%% file: main.tex
\documentclass[11pt]{article}

\usepackage{wermarxiv}

\author{
Gabriel Arpino\thanks{Department of Statistics, Harvard University, and Department of Statistics and Data Science, Yale University, \texttt{gabrielarpino@fas.harvard.edu}.} \and
Ramji Venkataramanan\thanks{Department of Engineering, University of Cambridge, \texttt{rv285@cam.ac.uk}.}}

\begin{document}

\title{Inferring Change Points in Regression via \\ Sample Weighting}

\date{\today}
\maketitle
\begin{abstract}
We study the problem of  identifying change points in high-dimensional  generalized linear models, and  propose an approach based on sample-weighted empirical risk minimization. Our method,  \WeightedERM{},  encodes priors on the change
points via sample weights, to obtain weighted versions of standard estimators  such as M-estimators and  maximum-likelihood estimators.  Under standard assumptions on the data, we obtain a precise asymptotic characterization of the performance of our method for general Gaussian designs, in the high-dimensional limit where the number of samples and covariate dimension grow proportionally. We show how this characterization can be used to efficiently construct a posterior distribution over change points. Numerical experiments on both simulated and real data illustrate the efficacy of \WeightedERM{} compared to existing approaches, demonstrating that sample weights constructed  with weakly informative priors can yield accurate change point estimators. Our method is implemented as an open-source package, \texttt{weightederm}, available in Python and R.
\end{abstract}
\thispagestyle{empty}

{\small
\setlength{\parskip}{0pt}
\tableofcontents
}
\addtocontents{toc}{\protect\thispagestyle{empty}}
\thispagestyle{empty}
\clearpage
\setcounter{page}{1}

\section{Introduction} \label{sec:introduction}

Heterogeneity is a common characteristic of large, high-dimensional datasets. A simple form of heterogeneity
in data ordered by time is a change in the data generating mechanism at certain unknown time instants.
For example, a model predicting a country’s economic output from monthly survey data may be affected
by small changes in government policy or in import/export regulations of other countries. A key question is: how
can we fit accurate statistical models for high-dimensional data in the presence of such \emph{change points}?
Change points are the unknown time instants where there is a change in the underlying generative mechanism.

Models with change points have been studied in several statistical contexts, such as the detection of changes in: signal means \citep{wang_univariate_2020,liu2021minimax}; covariance structures \citep{cho2015multiple,wang_optimal_2021b}; graphs \citep{bhattacharjee_change_2020,fan2018approximate}; dynamic networks \citep{wang_optimal_2021a}; functionals \citep{madrid_padilla_optimal_2022}; distributions \citep{londschien_random_2023}; and data streams of various formats in the online setting \citep{Romano23online,moen_general_2025}. Change point models have found application in a range of areas including genomics \citep{braun2000multiple}, neuroscience \citep{AstonKirch12}, finance \citep{andreou2002detecting} and economics  \citep{he_leveraging_2022}.

We consider (offline) change point estimation in the setting of high-dimensional generalized linear models (GLMs). In this setting, we are given a sequence of data
$(y_i, \x_i) \in \reals \times \reals^p$ from the model
\begin{align}
    y_i = q(\x^\top_i \bbeta^{(i)}, \varepsilon_i), \; \text{for } i \in [n]. \label{eq:model}
\end{align}
Here, $\bbeta^{(i)} \in \reals^p$ is the unknown regression vector (or signal) for the $i$th sample,  $\x_i \in \reals^p$ is the covariate vector, $\varepsilon_i$ is additive noise, and $q: \reals \times \reals \to \reals$ is a known function. We denote the unknown change points, i.e., the sample indices where the regression vector changes,  by $\eta_1, \ldots, \eta_{L^*-1}$. Specifically, we have
\begin{align}
   1 = \eta_0 < \eta_1 < \dots < \eta_{L^*} = n+1, \label{eq:cp_constraint}
\end{align}
with $\bbeta^{(i)} \neq \bbeta^{(i -1)}$ if and only if $i \in \{ \eta_{\ell}\}_{\ell = 1}^{L^*-1}$ $(i=2,\ldots,n)$.  We note that $L^*$ is the number of distinct signals in the sequence $\{\bbeta^{(i)}\}_{i=1}^n$, and $(L^*-1)$ is the number of change points. The number of change points is not known, but we assume that an upper bound $L$ on the value of $L^*$ is available.  The goal is to estimate the change point locations as well as the $L^*$ signals, and to quantify our uncertainty around these estimates.

The model \eqref{eq:model} covers many widely studied regression models including linear, logistic, quantile, and Poisson regression. In recent years, there has been considerable work on detecting change points in high-dimensional linear models, which we now review, along with related work for logistic regression and other GLMs.

\paragraph{Linear regression with change points}

In this model, the data $(y_i, \x_i) \in \reals \times \reals^p$ are generated as:
\begin{align}
y_i = \x_i^\top \bbeta^{(\Psi_i)} + \varepsilon_i, \quad i = 1, \dots, n. \label{eq:linear-chgpt-model}
\end{align}
This corresponds to model \eqref{eq:model} with $q(z, v) := z + v$, and when $L^* = 1$, it reduces to standard linear regression.

Linear regression with change points in the high-dimensional regime, where the dimension $p \gtrsim n$, has been studied in a number of papers, e.g. \citep{lee2016lasso, leonardi2016computationally,kaul2019efficient,rinaldo_localizing_2021,xu_change-point_2024, li_divide_2023, bai2023unified, liu2024change}. Most of these works consider sparse signals where the number of non-zero components  of  $\bbeta^{(\Psi_i)} \in \reals^p$ is $o(p)$, and analyze procedures which combine the LASSO estimator (or a variant) with a partitioning technique, e.g., dynamic programming. \citet{gao_sparse_2022} assume sparsity on the difference between signals across a change point, and \citet{cho_detection_2025} consider general non-sparse signals. The recent works of \citet{liu2024change,cho2026} study change point detection in high-dimensional linear models with heavy-tailed errors and outliers.

\paragraph{Logistic regression with change points}
For the logistic model, defining $\zeta(z) := \log{(1 + e^z)}$,  the data $(y_i, \x_i) \in \{0, 1\} \times \reals^p$ are generated according to:
\begin{align}
    \P\left[y_i = 1 \middle| \x_i^\top \bbeta^{(\Psi_i)} \right] = \frac{e^{\x_i^\top \bbeta^{(\Psi_i)}}}{1 + e^{\x_i^\top \bbeta^{(\Psi_i)}}} = \zeta'(\x_i^\top \bbeta^{(\Psi_i)}), \quad i = 1, \dots, n. \label{eq:logistic-chgpt-model_AMP}
\end{align}
We may view this as an instance of \eqref{eq:model} with $\varepsilon_1, \dots, \varepsilon_n \distas{i.i.d} U[0, 1]$ and $q(z, v) = \ind\{v \leq \zeta'(z) \}$, so that $y_i = q(\x_i^\top \bbeta^{(\Psi_i)}, \varepsilon_i) := \ind\{\varepsilon_i \leq \zeta'(\x_i^\top \bbeta^{(\Psi_i)})\}$ for $i \in \{1,\dots,n\}$. When $L^* = 1$, model \eqref{eq:logistic-chgpt-model_AMP} reduces to the standard logistic model.

Logistic regression with change points has been used in epidemiology to model the relationship between the continuous exposure variable and disease risk \citep{pastorbarriuso_transition_2003}. It has also  been used in medicine to identify relevant immune response biomarkers in patients with potentially infectious diseases \citep{fong_change_2015}.

\citet{hofrichter} studied  change point detection in generic low-dimensional GLMs, and  \citet{wang_efficient_2023} recently proposed a method for detecting change points in high-dimensional GLMs with sparse regression vectors.
 Their estimator, which combines an $\ell_1$-penalized estimator with a partitioning technique, is shown to be consistent assuming $s$-sparse signals with $s = o(\sqrt{n} / \log{p})$.

\paragraph{Challenges in the high-dimensional setting} Existing procedures such as \citep{li_divide_2023, gao_sparse_2022, cho_detection_2025, wang_efficient_2023} for detecting change points in high-dimensional linear and generalized linear models can incorporate structural assumptions on the regression vectors such as sparsity or sparse differences, but they are not equipped to exploit prior information on the change point locations. For example, we may wish to use prior information such as: ``a change point is more likely to occur between samples $\frac{n}{3}$ and $\frac{2n}{3}$ than outside this range".

The recent work \citep{arpino_2025_jmlr} proposed an Approximate Message Passing (AMP) based estimator for detecting change points in high-dimensional GLMs, which can be tailored to take advantage of priors on both signals and change points. However, the AMP estimator can be difficult to tailor when the signal prior isn't   known, and its asymptotic performance guarantees are based on i.i.d. Gaussian design assumptions. Bayesian approaches to change point detection have also been studied in  works such as \citep{fearnhead2006exact, lungu2022changepoint}, however they  mainly focus  on low-dimensional time-series.

\paragraph{Main contributions}
\begin{enumerate}
    \item We propose \WeightedERM{}, a sample-weighting procedure for estimating the number and locations of multiple change points in high-dimensional generalized linear models. Our method, described in Section \ref{sec:methodology}, applies careful sample weighting to standard estimators based on empirical risk minimization (ERM), such as least squares or other maximum-likelihood estimators.      We show how to construct such sample weight patterns given prior information on change point locations, and demonstrate that even seemingly uninformative priors yield accurate estimators.
	\item We rigorously characterize the performance of \WeightedERM{} for general Gaussian designs, in the asymptotic regime where the sample size grows proportionally with the covariate dimension. This characterization (Theorem \ref{thm:ERM_general_characterization_pres} and Proposition \ref{prop:ERM_hausdorff_characterization_pres}     in Section \ref{sec:asymptotic-characterization}) is in terms of the solution of a set of nonlinear equations, which can be numerically solved under mild assumptions.  To highlight the key ideas, we first present the result for the linear model with least squares loss, and use it to compute the asymptotic error of the \WeightedERM{} changepoint estimates.

	\item We apply this asymptotic characterization to construct a posterior distribution over change point locations in Section \ref{sec:uncertainty-quantification}. We provide a theoretical convergence guarantee for  this posterior distribution (Proposition \ref{prop:pointwise_posterior_pres}) and experimental evidence that  it can be computed accurately and efficiently in finite dimension using  empirically estimated quantities.

\item     We validate the method's  performance  via the theory and numerical experiments in three settings, including M-estimation in the linear model and logistic regression (Section \ref{sec:synthetic-experiments}). In Section \ref{sec:real-data-experiments}, we demonstrate the   performance of our estimator on macroeconomic data  \citep{McCracken01102016} and of our posterior inference method on myocardial infarction data \citep{misc_myocardial_infarction_complications_579} .
\end{enumerate}

\begin{figure}[!t]
\begin{subfigure}[b]{.32\textwidth}
    \centering
        \raisebox{-1.65mm}{\includegraphics[width=\textwidth]{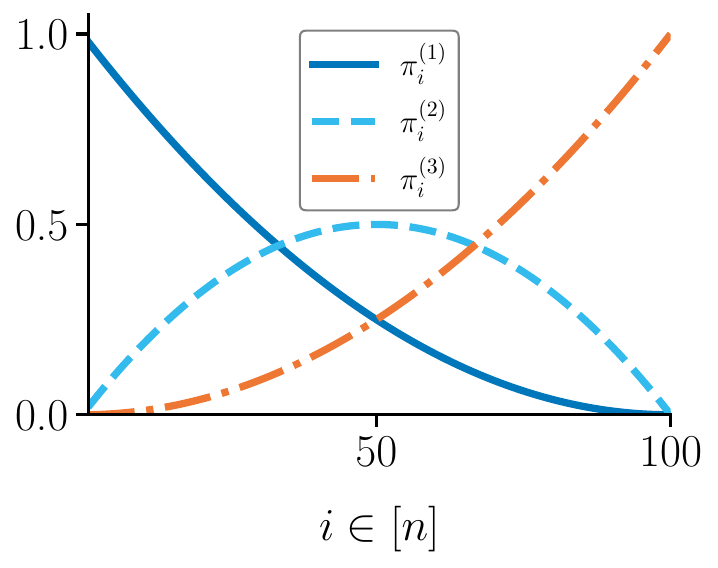}}
        \label{fig:marginal_L3}
\end{subfigure}
\begin{subfigure}[b]{.32\textwidth}
        \centering
        \resizebox{\textwidth}{!}{\input{figures/synthetic-final/M1a/location_asymmetric_errors.pgf}}
        \label{fig:M1a-location}
\end{subfigure}
\hfill
\begin{subfigure}[b]{.32\textwidth}
        \centering
        \resizebox{\textwidth}{!}{\input{figures/synthetic-final/M1a/runtime_asymmetric_errors.pgf}}
        \label{fig:M1a-runtime}
\end{subfigure}
\caption{\small Left: Sample weights $\pi_i^{(\ell)} = { {{i - 1} \choose {\ell - 1}} {{n - i} \choose {L - \ell}}}/{{{n - 1} \choose {L - 1}}}$ produced from a uniform location prior on two change points, with $L=3$, $n = 100$. Middle and right: \WeightedERM{} (\WERM{}) vs. other methods estimating exactly two change points in a sparse linear model. Dimension $p = 200$, noise $\varepsilon_i \distas{\text{i.i.d.}} \N(0, 0.6)$, covariates $\x_i \distas{\text{i.i.d.}} \N(0, \bSigma/n)$, $\Sigma_{i, j} = 0.2^{|i-j|}$, regression vector entries sampled independently according to $0.5 \N(0, \frac{n}{p}) + 0.5 \delta_0$. Error measured via the Hausdorff distance, see \eqref{eq:hausdorff-dist}. Error bars indicate  $25$th to $75$th percentile across $100$ trials. }
\label{fig:M1a}
\end{figure}

\paragraph{Example: Linear model with change points} \label{example:linear_model}
To explain the key ideas behind our method, let us first consider the linear model with change points described in \eqref{eq:linear-chgpt-model}. Assuming a maximum of $L-1$ change points $(\eta_1, \dots, \eta_{L-1})$,  \WeightedERM{}  begins by assigning \textit{weights} $\begin{bmatrix}
    \pi_i^{(1)}, \dots, \pi_i^{(L)}
\end{bmatrix}$ to each sample $i \in [n]$, which encode our priors on the change points. For example, suppose there are exactly two change points ($L = L^* = 3$), and we assume that all change point configurations compatible with these constraints are equally likely\label{page:lin-model-with-chgpts-ex}.
This induces a prior distribution over change point configurations. We can then compute the weights $\begin{bmatrix}
    \pi_i^{(1)}, \pi_i^{(2)}, \pi_i^{(3)}
\end{bmatrix}$ as follows for $i \in [n]$. For $\ell \in \{ 1, 2, 3\}$, let $\pi_i^{(\ell)}$ be the marginal prior probability of observing data generated from regression vector $\ell$ at index $i$. These weights are shown in Figure \ref{fig:M1a} (left panel). Having assigned weights to the samples, we estimate the $L=3$ regression vectors by solving the following weighted least squares problems:
\begin{align} \label{eq:data-fitting-step}
    \hat{\bbeta}^{(1)} = \argmin_{\tilde{\bbeta} \in \reals^p} \sum_{i = 1}^n \pi_i^{(1)} \left(y_i - \x_i^\top \tilde{\bbeta} \right)^2, \;\; \dots, \;\; \hat{\bbeta}^{(3)} = \argmin_{\tilde{\bbeta} \in \reals^p} \sum_{i = 1}^n \pi_i^{(L)} \left(y_i - \x_i^\top \tilde{\bbeta} \right)^2.
\end{align}
Our change point estimate $\hat{\eeta} = \begin{bmatrix}
    \hat{\eta}_1, \hat{\eta}_{2}
\end{bmatrix}$ is then constructed by finding the split in the data that minimizes the sum of squared residuals from the  least squares fits:
\begin{align} \label{eq:data-split}
\sum_{i = 1}^{\hat{\eta}_1-1} \left(y_i - \x_i^\top \hat{\bbeta}^{(1)} \right)^2 +
\sum_{i = \hat{\eta}_1}^{\hat{\eta}_2-1} \left(y_i - \x_i^\top \hat{\bbeta}^{(2)} \right)^2
+ \sum_{i = \hat{\eta}_{2}}^n \left(y_i - \x_i^\top \hat{\bbeta}^{(3)} \right)^2,
\end{align}
where the minimization is over all change point configurations that satisfy the constraints. Figure \ref{fig:M1a} shows the performance of this estimator on a sparse linear model with two change points at $2n/5$ and $7n/10$. We observe that \WeightedERM{} has smaller localisation error (Hausdorff distance) compared to state of the art methods for detecting change points in linear models (\MOSEG{} \citep{cho_high-dimensional_2024}, \McScan{} \citep{cho_detection_2025}), and competitive runtime.

We expand on this example in the next section,
and define \WeightedERM{} via a generic loss function for a generalized linear model.

\section{Methodology} \label{sec:methodology}
We begin with some notation, and then describe the \WeightedERM{} procedure for estimating the set of change points $\{\eta_{\ell}\}_{\ell \in [L^*-1]}$ in the GLM \eqref{eq:model}.
The covariate vectors are stacked to form the design matrix $\X :=
    \big[ \x_1^\top, \dots, \x_n^\top
\big]^\top \in \reals^{n \times p}$. Similarly let $\y := \big[
    y_1, \dots, y_n
\big]^\top \in \reals^n$, and $\bvarepsilon := \big[
    \varepsilon_1, \dots, \varepsilon_n
\big]^\top \in \reals^n$. We define the signal matrix $\B := \Big[
    \bbeta^{(1)}, \dots, \bbeta^{(L)}
\Big] \in \reals^{p \times L}$, with the last $L - L^*$ columns set to arbitrary vectors in $\reals^p$. We can then write the response $\y$ from model \eqref{eq:model} as:
\begin{align}
    \y = q(\X \B, \bPsi, \bvarepsilon) \in \reals^n, \label{eq:model_psi_AMP}
\end{align}
where $q$ is the known function in model \eqref{eq:model}, expanded to incorporate the \emph{signal configuration} vector $\bPsi$ and act row-wise on matrix inputs, with $(q(\X \B, \bPsi, \bvarepsilon))_i = q(\X \bbeta^{(\Psi_i)}, \varepsilon_i)$ for $i \in [n]$. The signal configuration vector is defined component-wise, with $\Psi_i$ being the index of the regression vector (signal)  corresponding to sample $i$, i.e. $\Psi_i := \sum_{\ell = 1}^{L^*} \ell \cdot \ind\{ \eta_{\ell-1} \leq i < \eta_{\ell} \}$ for $i \in [n]$.  We note the one-to-one mapping between valid change point configurations $\{\eta_1, \dots, \eta_{L^*-1}\}$ and signal configuration vectors $\bPsi$.  The set of valid signal configuration vectors is denoted by $\mathcal{X}$.

\paragraph{\WeightedERM{}}

Given observations $(\X, \y)$ and a base loss function $M: \reals \times \reals \to \reals$ tailored to the specifics of the model \eqref{eq:model}, our method solves $L$  weighted empirical risk minimization problems on the same set of observables $(\X, \y)$. Each of these minimizations, indexed by $\ell \in [L]$, is assigned a different set of sample weights $\big\{ \pi_i^{(\ell)} \big\}_{i \in [n]}$, and the corresponding optimization problem is:
\begin{align}
\begin{split}\label{eq:ERM_est}
    &\hat{\bbeta}^{(\ell)} \in \argmin_{{\bbeta}  \in \reals^p} \, \sum_{i = 1}^n  \pi_i^{(\ell)} M(\x_i^\top {\bbeta}, y_i) =: \argmin_{{\bbeta}  \in \reals^p} \; \cC^{(\ell)}({\bbeta}), \\
    &\hat{\ttheta}^{(\ell)} := \X \hat{\bbeta}^{(\ell)}.
\end{split}
\end{align}
This corresponds to the data fitting step \eqref{eq:data-fitting-step} for the special case of linear regression with change points.
Postulating a prior $\pi_{\bar{\bPsi}}$ for the signal configuration vector $\bPsi$,  we set $\pi_i^{(\ell)}$ to be the marginal prior probability of observing data generated from regression vector $\ell$ at time index $i$, for $\ell \in [L]$ and $i \in [n]$. In other words, for $\ell \in [L], i \in [n]$, we set:
\begin{align}
	\pi_i^{(\ell)} = \sum_{\bpsi : \psi_i = \ell} \pi_{\bar{\bPsi}}(\bpsi). \label{eq:pi_i_defn}
\end{align}
Through a range of examples, we  will demonstrate how even weakly informative priors $\pi_{\bar{\bPsi}}$ yield non-trivial sets of sample-weights $\{\pi_i^{(\ell)}\}_{i \in [n]}$ that favourably bias the empirical risk minimization problems \eqref{eq:ERM_est}. We emphasize that our method and theory do not require that the true $\bPsi$ be drawn according to $\pi_{\bar{\bPsi}}$. Rather,  the prior $\pi_{\bar{\bPsi}}$ allows us to encode any knowledge we may have about the change point locations, and use it to define the weights.

A natural choice for the loss function $M$ is the negative log-likelihood function of the model, or a convex relaxation. For example, for logistic regression we use $M(a,b) = \log{(1 + e^a)} - ba$. In Appendix \ref{sec:motivation-via-jensen}, we show how the sample-weighted estimator
in \eqref{eq:ERM_est} can be viewed as maximizing a relaxed version of the likelihood  of $\y$ given $(\X, \B)$,  assuming a prior $\pi_{\bar{\bPsi}}$ over change points.

We estimate the set of change points as follows, using the solutions of the  $L$ optimization problems  in \eqref{eq:ERM_est}:
\begin{align}
\begin{split}\label{eq:general_changepoint_estimator}
&\hat{\bPsi}(\hat{\ttheta}^{(1)}, \dots, \hat{\ttheta}^{(L)}; \y) \in \argmin_{\bpsi \in \mathcal{X}} \,  \sum_{i = 1}^n M(\hat{\theta}^{(\psi_i)}_i, y_i) + P(\bpsi) \, , \\
&\hat{\eeta}(\hat{\ttheta}^{(1)}, \dots, \hat{\ttheta}^{(L)}; \y) := U^{-1}\left( \hat{\bPsi}(\hat{\ttheta}^{(1)}, \dots, \hat{\ttheta}^{(L)}; \y)\right).
\end{split}
\end{align}
where $U$
denotes the one-to-one mapping between change point vectors $\eeta$ and signal configuration vectors $\bPsi$, given by $U(\eeta) = (U_i(\eeta))_{i \in [n]}, \, U_i(\eeta) := 1 + \sum_{\ell = 1}^{\|\eeta\|_0} \ind\{ \eta_\ell \leq i \}$. The function $P: [L]^n \to \reals$ is a chosen penalty on $\bpsi$,  e.g. one that penalizes the number of change points. For penalties of the form $P(\bpsi) = P(\max_i \psi_i)$ which depend only on the assumed number of change points, the optimization in \eqref{eq:general_changepoint_estimator} can be solved efficiently using dynamic programming. This then yields a total runtime for $\WeightedERM{}$ of $O(L n^2 + L \cdot \textsf{ERM}(n, p))$, where $\textsf{ERM}(n, p)$ is the runtime of the optimization problem \eqref{eq:ERM_est}. A key difference  from many existing methods for change point regression, such as \citep{li_divide_2023, cho_high-dimensional_2024, gao_sparse_2022, wang_efficient_2023}, is that \eqref{eq:general_changepoint_estimator}  scores candidate partitions using fixed fitted coefficients rather than refitting on every candidate segment. As a result, \WeightedERM{} has significantly lower computational cost, as illustrated in the numerical experiments in Section \ref{sec:synthetic-experiments}.  \label{page:runtime-statement}

The penalty function $P$ can be tuned to give an approximate threshold for any required significance level. Alternatively,  the number of change points can be estimated using data-driven methods such as cross-validation \citep{li_divide_2023, pein_cross-validation_2025}; we use this method  in our numerical experiments in Section \ref{sec:synthetic-experiments}.

The accuracy of a change point estimator is evaluated using the Hausdorff distance, a commonly used metric in this setting \citep{wang_high_2018, xu_change-point_2024, li_divide_2023}. The Hausdorff distance between two non-empty subsets $X,Y$  of $\reals$ is:
\begin{align}
	d_{H}(X, Y) = \max\Big\{ \sup_{x \in X} d(x, Y) , \,  \sup_{y \in Y} d(X, y) \Big\},  \label{eq:hausdorff-dist}
\end{align}
where $d(x, Y) := \min_{y \in Y} \|x - y\|_2$. The Hausdorff distance is a metric, and can be viewed as the largest of all distances from a point in $X$ to its closest point in $Y$ and vice versa. We interpret the Hausdorff distance between $\eeta$ and an estimate $\hat{\eeta}$ as the Hausdorff distance between the sets formed by their elements.

\begin{figure}[!t]
\begin{subfigure}[b]{.32\textwidth}
    \centering
        \raisebox{-1.65mm}{\includegraphics[width=\textwidth]{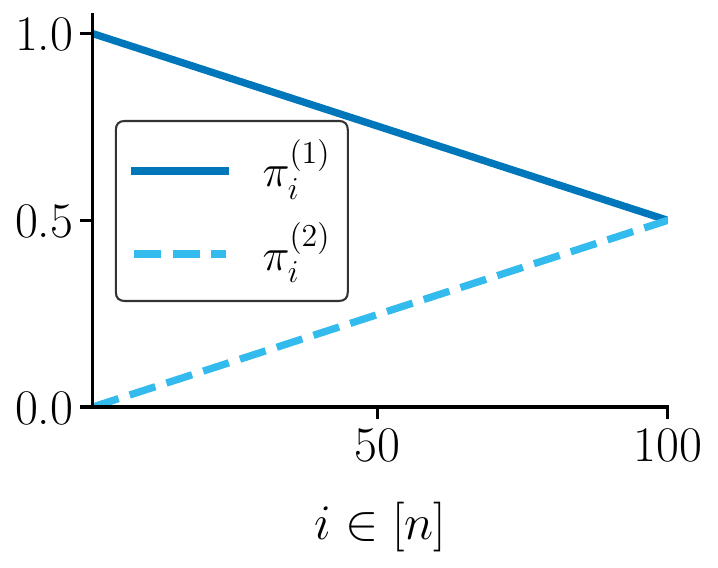}}
        \label{fig:logistic-unif-Lb-SE-match-est-2}
\end{subfigure}
\begin{subfigure}[b]{.32\textwidth}
    \centering
        \resizebox{\textwidth}{!}{\input{figures/SE_est_match/WERM_SE_match_est_logistic_Lub_L2_new5_loc.pgf}}
        \label{fig:logistic-unif-Lb-SE-match-est}
\end{subfigure}
\hfill
\begin{subfigure}[b]{.32\textwidth}
        \centering
        \resizebox{\textwidth}{!}{\input{figures/SE_est_match/WERM_SE_match_est_logistic_Lub_L2_new5_size.pgf}}
        \label{fig:logistic-unif-Lb-SE-match-size}
\end{subfigure}
\caption{\small Left: Sample weights used in the logistic model example on p.\pageref{page:logistic}. Middle and right: Theory (RHS of \eqref{eq:ERM_hausdorff_match_eqn}, \eqref{eq:ERM_size_match_eqn}) and method match in the setting of the logistic model example on p.\pageref{page:logistic}.
Error bars indicate the $25$-th to $75$-th percentiles across $15$ trials.}
\label{fig:logistic-example-match}
\end{figure}

We now provide an example to illustrate our methodology in the setting of logistic regression with change points.
\paragraph{Example: Logistic model with change points} \label{page:logistic}
Recall the definition of the logistic model  in \eqref{eq:logistic-chgpt-model_AMP} and
consider the problem of estimating the presence and location of a change point, when we know there to be at most one change point, i.e. $L=2$.  We postulate equal probability of there being one or no change points, and conditional on the presence of a change point, we assume that all change point locations are equally likely. An application of elementary set partition counting arguments to the formula in \eqref{eq:pi_i_defn} yields the  weights $\pi_i^{(\ell)} = \frac{1}{L} \sum_{k=\ell}^L { {{i - 1} \choose {\ell - 1}} {{n - i} \choose {k - \ell}}}/{{{n - 1} \choose {k - 1}}}$ for $\ell \in [L], i \in [n]$, displayed in Figure \ref{fig:logistic-example-match}.
We set $M(a, b)  = \log{(1 + e^a)} - ba$, the negative log-likelihood of the  logistic model, and  the penalty $P(\hat{\bpsi}) = P(\hat{\eeta})$ to $\log {{n-1} \choose {\|\hat{\eeta}\|_0}}$.  Figure \ref{fig:logistic-example-match} shows that the method correctly detects the presence of a change point (at $0.4n$)  for  $\frac{n}{p} \ge 5$, with growing accuracy as $n$ increases. Additional experiment details can be found in Appendix \ref{sec:logistic-example-additional-details}.

We provide further numerical results with synthetic data in Section \ref{sec:synthetic-experiments}, and evaluate our method on two real datasets in Section \ref{sec:real-data-experiments}, demonstrating that it can detect change points in macroeconomic and medical data that are consistent with findings in the literature.

\section{Asymptotic characterization} \label{sec:asymptotic-characterization}
In this section, we give a tight asymptotic characterization of the performance of \WeightedERM{} for general Gaussian designs.  We focus on the \emph{proportional} asymptotics regime where the number of samples $n$ and features $p$ grows proportionally. In this regime, the asymptotic statistical risk of convex penalized estimators has been  characterized for a number of linear and generalized linear models, e.g. \citep{Don09,donoho_high_2016, thrampoulidis2018precise,sur_modern_2019,mei_generalization_2022, liang_precise_2022,celentano2022fundamental,bu2023characterizing}. The asymptotic estimation error of general first-order iterative algorithms has also been precisely characterized in this regime \citep{celentano2020estimation, chandrasekher_sharp_2023}.
Although many of these works use i.i.d. Gaussian assumptions on the design matrix, the results can often be extended to Gaussian designs with general covariance, the setting we consider here \citep{zhao_asymptotic_2022, huang2022lasso}.
The asymptotics of estimators  in the proportional regime have also been studied under a rotational-invariance assumption on the design matrix \citep{li2023random, li2023spectrum, zhang2026orthogonallyinvariant}.

Although consistent estimation is generally not possible in the proportional asymptotics regime,
the properties of the random design can be used to obtain a sharp characterization of the distribution of the errors \citep{montanari2018mean}. This can then be used, for example, to construct accurate confidence intervals \citep{zhao_asymptotic_2022, celentano2023lasso, bellec2025observable}. The asymptotic results in this regime have been observed to closely match empirical performance even at moderate sample sizes, see e.g.  \citep{sur_modern_2019, liang_precise_2022}.

Existing results on high-dimensional change point regression usually assume signals that are $s$-sparse with $s$ sufficiently small, and propose change point estimators that are consistent when the separation between change points is at least of order $s \log p/\kappa^2$, where $\kappa$ is a constant determined by the separation between the signals \citep{wang_2021_statistically, li_divide_2023, wang_efficient_2023}.
In contrast, we do not assume signal sparsity that is sublinear in $n$, so the change point estimation error will not converge to zero unless $n/p \to \infty$. We therefore quantify our method's performance via precise asymptotics for the estimation error and the limiting posterior distribution.

\subsection{Preliminaries and notation} \label{sec:prelim_notation}
All vectors (even rows of matrices) are treated as column vectors unless otherwise stated.  We use $\| \boldsymbol{v} \|$ to denote the Euclidean norm of a vector $\boldsymbol{v}$. When referring to probability densities, we include probability mass functions, with integrals interpreted as sums when the distribution is discrete. For a real-valued multivariate function $f$,  $f'$ denotes the derivative with respect to the first argument, $\partial_2 f$  the derivative with respect to its second argument, and $\partial_{12} f$ the second derivative with respect to its first argument followed by the second. For two sequences (in $n$) of random variables $X_n, Y_n$, we write $X_n \overset{\P}{\simeq} Y_n$ when their difference converges in probability to $0$, that is, $\lim_{n \to \infty} \P(|X_n - Y_n| > \epsilon) = 0$ for any $\epsilon > 0$.

\subsection{Setting} \label{sec:setting}
\paragraph{Model assumptions} In the model \eqref{eq:model}, we consider
$n$ independent observations $(y_i, \x_i)_{i = 1}^n$, with the covariates $\x_i \in \reals^p$ following a multivariate normal distribution $\x_i \distas{} \N(\0, \bSigma / n)$.  We assume that the covariance $\bSigma \in \reals^{p \times p}$ is positive definite so that the model is identifiable, and let $\L \in \reals^{p \times p}$ be the lower triangular matrix from the unique Cholesky decomposition $\bSigma = \L \L^\top$.  We consider the high-dimensional regime where $n$ and $p$ both go to infinity such that $n / p \to \delta > 1$, with $L$ and $L^*$ fixed. As in \cite{zhao_asymptotic_2022}, we consider a scaling of the regression coefficient matrix obeying:
\begin{align} \label{eq:Gamma_as_limit}
	\cov(\x_i^\top \B) = \frac{1}{n} \B^\top \bSigma \B \to \bGamma \succ \0 \in \reals^{L \times L} \; \text{ as $n \to \infty$}.
\end{align}
This scaling keeps the ``signal-to-noise ratio'' in our problem fixed. The larger the diagonal values of $\bGamma$ are relative to positive off-diagonal terms, the easier it is to identify the presence of different signals in our dataset.

We assume that the noise vector $\bvarepsilon = (\varepsilon_i)_{i \in [n]}$, viewed as the first $n$ entries of the same infinite sequence $(\varepsilon_i)_{i \geq 1}$, is independent of the covariates $\{\x_i\}_{i \in [n]}$, and that the empirical distribution of the entries of $\bvarepsilon$ converges in Wasserstein-$2$ distance to a distribution $\P_{\bar{\varepsilon}}$ with finite second moment; see Appendix \ref{sec:preliminaries-for-proofs} for a precise definition.   This is a common assumption in the proportional asymptotics regime \citep{bayati_montanari_2012, javanmard_state-evolution_2013}, and includes as a special case $\bvarepsilon$ whose entries are i.i.d. samples from $\P_{\bar{\varepsilon}}$.

We assume that the link function $q$ in \eqref{eq:model} is Lipschitz.\footnote{The link function for logistic regression is not Lipschitz. In Appendix \ref{sec:logistic-guarantees}, we show that the theoretical guarantees still hold for the logistic model on the parameter region where the estimator exists, via a smoothing and stability argument.}
Lastly, as $n \to \infty$, we assume that the entries of the normalized change point vector $\eeta / n$ converge to constants $\alpha_1, \dots, \alpha_{L^*-1}$ such that $0 = \alpha_0 < \alpha_1 < \dots < \alpha_{L^*-1} < \alpha_{L^*} = 1$. This assumption is natural when the sample size $n$ and dimension $p$ grow proportionally, and the number of degrees of freedom in the signals also grows linearly in $p$.

The following mild assumption on the weight patterns $\{ \pi_i^{(\ell)} \}_{i=1}^n$ allows for a rigorous analysis of \eqref{eq:ERM_est} as $n, p \to \infty$.
\begin{assumption}[Regularity of weights]\label{asmpt:reg-weights}
For $\ell \in [L]$, the sequence $\{ \pi^{(\ell)}_i \}_{i \in [n]}$ is non-negative with $\pi^{(\ell)}_{\floor{nt} + 1} \xrightarrow[]{n \to \infty} \Phi^{(\ell)}(t)$ for $t \in [0, 1)$, for some bounded function $\Phi^{(\ell)}: [0, 1] \to \reals_{\geq 0}$ that is strictly positive for a.e. $t \in [0,1]$.
\end{assumption}
This assumption yields a natural interpretation of the weight pattern as the discretization of a bounded function on the unit interval. Importantly, this requirement is satisfied for uniform change point priors such as those used in the examples on p.\pageref{example:linear_model} (with $\Phi^{(\ell)}(t) = {{L-1} \choose {\ell - 1}} t^{\ell - 1} (1 - t)^{L - \ell}$) and p.\pageref{page:logistic} (with $\Phi^{(\ell)}(t) = \frac{1}{L} \sum_{k = \ell}^L {{k-1} \choose {\ell - 1}} t^{\ell - 1} (1 - t)^{k - \ell}$).

\subsection{Theory for the linear model with squared error loss}
\label{sec:linear-model-theory}
Before stating our results for the general case, we present the  \WeightedERM{}  theory for the simple setting of the linear model \eqref{eq:linear-chgpt-model} with isotropic design ($\bSigma = \I$) and squared loss $M(u, v) = (u - v)^2$.
The optimisation problem in \eqref{eq:ERM_est} is now a weighted least-squares problem that can be explicitly solved, for $\ell \in [L]$:
\begin{align}
    \label{eq:ls-wls}
    \hat{\bbeta}^{(\ell)} & = \argmin_{{\bbeta}  \in \reals^p} \, \sum_{i = 1}^n  \pi_i^{(\ell)} ( y_i - \x_i^\top {\bbeta})^2 =
     \big(\X^\top \boldsymbol{\Pi}^{(\ell)} \X\big)^{-1} \X^\top \boldsymbol{\Pi}^{(\ell)} \y, \\
     & \text{ where } \boldsymbol{\Pi}^{(\ell)} := \mathrm{diag}\big(\pi_1^{(\ell)}, \dots, \pi_n^{(\ell)}\big). \nonumber
\end{align}

We consider the model assumptions  in Section \ref{sec:setting}, with $\bGamma = \lim_{n \to \infty} \frac{1}{n} \B^{\top} \B$ invertible, and further assume that for $i \in [n]$, $\varepsilon_i \distas{i.i.d} \P_{\bar{\varepsilon}}$, with $\E \bar{\varepsilon} = 0, \E \bar{\varepsilon}^2 = \sigma^2$. Recalling the fractional change point limits $0= \alpha_0 < \alpha_1 < \ldots, < \alpha_{L^*-1} < \alpha_{L^*} =1$, we define the limiting configuration function $\bar\psi(t): [0,1] \to [L^*]$ as $\bar\psi(t) = r$, for $t \in [\alpha_{r-1}, \alpha_r)$; $\bar\psi(1)=L^*$.

The solutions $\{(\hat{\bbeta}^{(\ell)}, \,  \hat{\ttheta}^{(\ell)}= \X\hat{\bbeta}^{(\ell)}) \}_{\ell \in [L]}$  are collected to define the following matrices of the regression vector estimates and the fits:
\begin{align}
	&\hat{\B} := \begin{bmatrix}
		\hat{\bbeta}^{(1)} & \hat{\bbeta}^{(2)} & \dots & \hat{\bbeta}^{(L)}
	\end{bmatrix} \in \reals^{p \times L}, \label{eq:B-estimate-matrix} \\
    	&\hat{\TTheta} := \begin{bmatrix}
		\hat{\ttheta}^{(1)} & \hat{\ttheta}^{(2)} & \dots & \hat{\ttheta}^{(L)}
	\end{bmatrix} \in \reals^{n \times L}, \label{eq:Theta-estimate-matrix}
\end{align}

 The asymptotic behaviour of $(\hat{\B}, \hat{\TTheta})$ is described by scalars $b^{(\ell)} > 0$, vectors $\blambda^{(\ell)} \in \reals^L$ for $\ell \in [L]$, and a covariance matrix $\bK = (\kappa_{\ell, \ell'})_{\ell, \ell' \in [L]}$. Writing $\bLambda := [\blambda^{(1)}\ \cdots\ \blambda^{(L)}] \in \reals^{L \times L}$,   Proposition \ref{prop:ls-B-Theta} below shows that $[\B, \hat{\B}] \in \reals^{p \times 2L}$ has the asymptotic distributional representation $[\B, \B \bLambda + \sqrt{\delta} \G_{\hat{\B}}]$, where $\G_{\hat{\B}} \in \reals^{p \times L}$ has i.i.d. rows with $(\G_{\hat{\B}})_j \distas{\text{i.i.d.}} \N(\0_L, \bK)$ for $j \in [p]$. Thus $\bLambda$ captures the bias in the estimate $\hat{\B}$ and $\delta\bK$ is its covariance.

 The parameters $(b^{(\ell)})_{\ell \in [L]}, \bLambda, \bK  $ are sequentially computed  as follows. For $\ell \in [L]$, let $b^{(\ell)} >0$ be the solution of
\begin{align}
        1-\frac{1}{\delta}
    &= \lim_{n\to\infty}\, \frac{1}{n}\sum_{i=1}^n
       \frac{1}{1+2b^{(\ell)}\pi_i^{(\ell)}} \, ,
       \label{eq:ls-b-linear-model}
\end{align}
and write
\begin{align}
\omega^{(\ell)}_i := \frac{2b^{(\ell)} \pi_i^{(\ell)}}{1+2b^{(\ell)}\pi_i^{(\ell)}} \, \in [0, 1), \qquad i \in [n].
\label{eq:omega_ell_i}
\end{align}
Let $\e_{r}$ denote the $r$th standard basis vector of $\reals^L$. Then, for $\ell,\ell'\in [L]$,
\begin{align}
    \blambda^{(\ell)}
    &= \lim_{n\to\infty}\frac{\delta }{n}\sum_{i=1}^n
       \omega^{(\ell)}_i \,  \e_{\Psi_i},
       \label{eq:ls-lambda-linear-model}\\
    \kappa_{\ell,\ell'}
    &= \frac{1}{1 - s_{\ell,\ell'}} \left[ \lim_{n\to\infty}\frac{\delta}{n}\sum_{i=1}^n
  \omega^{(\ell)}_i  \omega^{(\ell')}_i \,
       (\blambda^{(\ell)}-\e_{\Psi_i})^\top\bGamma
       (\blambda^{(\ell')}-\e_{\Psi_i}) \; + \; s_{\ell, \ell'}\, \sigma^2 \right],
       \label{eq:ls-kappa-linear-model} \\
       & \text{ where } \  s_{\ell,\ell'} := \lim_{n\to\infty}\frac{\delta}{n}\sum_{i=1}^n
  \omega^{(\ell)}_i  \omega^{(\ell')}_i. \label{eq:ls_sll}
\end{align}
The limits in \eqref{eq:ls-b-linear-model}-\eqref{eq:ls_sll} exist and can be succinctly represented in terms of scalar integrals using Assumption \ref{asmpt:reg-weights} and the limiting configuration function $\bar \psi(t)$. Writing $\omega^{(\ell)}(t) := 2 b^{(\ell)} \Phi^{(\ell)}(t)/(1 + 2 b^{(\ell)} \Phi^{(\ell)}(t))$, we have
\begin{equation}
    \begin{split}
        \int_0^1 \frac{\de t}{1 + 2 b^{(\ell)} \Phi^{(\ell)}(t)} = 1 - \frac{1}{\delta}, \qquad
s_{\ell, \ell'} = \delta \int_0^1 \omega^{(\ell)}(t)\, \omega^{(\ell')}(t)\, \de t, \\
\lambda^{(\ell)}_r = \delta \int_{\alpha_{r-1}}^{\alpha_r} \omega^{(\ell)}(t)\, \de t, \ r \in [L^*], \quad \text{  and } \quad \lambda^{(\ell)}_r = 0, \ r > L^*.
    \end{split}
    \label{eq:ls-integrals}
\end{equation}
Similarly, the limit in \eqref{eq:ls-kappa-linear-model} is
\[
\delta \int_0^1 \omega^{(\ell)}(t)\, \omega^{(\ell')}(t)\, \big(\blambda^{(\ell)}-\e_{\bar\psi(t)}\big)^\top\bGamma\big(\blambda^{(\ell')}-\e_{\bar\psi(t)}\big)\, \de t.
\] All the limiting parameters therefore depend on the model only through $(\delta, \bGamma, \sigma^2, \alpha_1, \dots, \alpha_{L^*-1})$, and on the method only through the weight profiles $\Phi^{(\ell)}$.

Our asymptotic characterizations of the matrix estimates $\hat{\B}$ and $\hat{\TTheta}$ are described in terms of pseudo-Lipschitz functions of order $r$ for some finite $r > 1$.  Essentially, these are continuous functions which do not grow faster than $\|x\|^r$; see Appendix \ref{sec:preliminaries-for-proofs} for the precise definition.

\begin{proposition} \label{prop:ls-B-Theta}
Consider the linear model setting above, and let $(b^{(\ell)}, \blambda^{(\ell)}, \kappa_{\ell,\ell'})_{\ell,\ell' \in [L]}$ be given by \eqref{eq:ls-b-linear-model}--\eqref{eq:ls_sll}. Let $\Z, \G_{\TTheta} \in \reals^{n \times L}$ and $\G_{\hat{\B}} \in \reals^{p \times L}$ be mutually independent, with i.i.d.\ rows $\Z_i \sim \N(\0_L, \bGamma)$, $(\G_{\TTheta})_i \sim \N(\0_L, \bK)$ for $i \in [n]$, and $(\G_{\hat{\B}})_j \sim \N(\0_L, \bK)$ for $j \in [p]$. Define $\tilde{\y} \in \reals^n$ and $\tilde{\TTheta} \in \reals^{n \times L}$ by
\begin{align}
    \label{eq:ls-Theta-tilde}
    \tilde{y}_i := Z_{i, \Psi_i} + \varepsilon_i, \qquad
    \tilde{\Theta}_{i, \ell} := \big(1 - \omega_i^{(\ell)}\big)\big(\Z_i^\top \blambda^{(\ell)} + (\G_{\TTheta})_{i, \ell}\big) + \omega_i^{(\ell)}\, \tilde{y}_i, \\
    \hspace{2in} i \in [n], \ \ell \in [L]. \nonumber
\end{align}
Then for any sequences of uniformly pseudo-Lipschitz functions $\varphi_{p}(\cdot \,; \B) : \reals^{p \times L} \to \reals$ and $\varphi_{n}(\cdot \,; \bPsi, \bvarepsilon) : \reals^{n \times 2L} \to \reals$,
\begin{align}
    \varphi_p(\hat{\B} ; \B) &\stackrel{\P}{\simeq} \E \, \varphi_p\big(\B \bLambda + \sqrt{\delta}\, \G_{\hat{\B}} ; \B\big),
    \label{eq:ls-B-rep} \\
    \varphi_n(\hat{\TTheta}, \X \B ; \bPsi, \bvarepsilon) &\stackrel{\P}{\simeq} \E\, \varphi_n\big(\tilde{\TTheta}, \Z ; \bPsi, \bvarepsilon\big),
    \label{eq:ls-Theta-rep}
\end{align}
where the expectations are over $(\Z, \G_{\TTheta}, \G_{\hat{\B}})$. In particular, for $\ell, r \in [L]$,
\begin{align}
    \frac1p\|\hat{\bbeta}^{(\ell)}-\bbeta^{(r)}\|^2
    \xrightarrow{\P}
    \delta\left[(\blambda^{(\ell)}-\e_r)^\top\bGamma
       (\blambda^{(\ell)}-\e_r)+\kappa_{\ell,\ell}\right].
       \label{eq:ls-mse}
\end{align}
\end{proposition}
The proposition is a special case of Theorem \ref{thm:ERM_general_characterization_pres} in Section \ref{sec:characterization-nonlinear-eqns}, and a proof is  given in Appendix \ref{subsec:ls-B-Theta_proof}.

The quantities $\blambda^{(\ell)}, \kappa_{\ell, \ell'}$ and $(\omega^{(\ell)}_i)_{i \in [n]}$ all have clear statistical interpretations. By \eqref{eq:ls-B-rep}, $\hat{\bbeta}^{(\ell)}$ behaves like $\sum_{r} \lambda_r^{(\ell)} \bbeta^{(r)}$ plus an independent   $\N(0, \kappa_{\ell, \ell} \I)$ noise vector. The $L$ regression estimates are correlated since they are all computed using the data $(\X, \y)$, and the covariance between $\hat{\bbeta}^{(\ell)}$ and $\hat{\bbeta}^{(\ell')}$ is $\delta \kappa_{\ell, \ell'}$. The quantity $(\omega^{(\ell)}_i)_{i \in [n]}$ in \eqref{eq:omega_ell_i} is the asymptotic \emph{leverage} of the observation $y_i$ in the $\ell$th fit. The leverage $h_i^{(\ell)} := \partial \hat{\theta}_i^{(\ell)}/\partial y_i$ measures the sensitivity of the $i$th component of the fit to $y_i$ \citep{hoaglin1978hat}. From \eqref{eq:ls-wls}, it follows that $ h_i^{(\ell)} =  \pi_i \x_i^{\top} \big(\X^\top \boldsymbol{\Pi}^{(\ell)} \X\big)^{-1}  \x_i$. Denoting the Hessian of the objective function in \eqref{eq:ls-wls} by $\Q$, we have
\[
\Q =  2\pi_i^{(\ell)}\x_i \x_i^{\top}   +  \Q_{\backslash i} \, , \quad  \text{ where } \ \Q_{\backslash i} := 2\sum_{j \ne i} \pi_j^{(\ell)}\x_j \x_j^{\top},
\]
 and using the Sherman-Morrison identity for $\Q^{-1}$, we obtain
 \begin{equation}
     h_i^{(\ell)} = \frac{2\pi_i^{(\ell)} a_i} {1 + 2\pi_i^{(\ell)} a_i}, \quad \text{ where } \ a_i:= \x_i^\top \Q_{\backslash i}^{-1} \x_i
 \end{equation}
 Since $\x_i \sim \N(\0, \I/n)$ is independent of $\Q_{\backslash i}$, the quadratic form $a_i$ concentrates around $\frac{1}{n}\mathrm{tr}({\Q_{\backslash i}}^{-1}) \approx \frac{1}{n}\mathrm{tr}(\Q^{-1})$, which converges to a constant, which can be identified as $b^{(\ell)}$  \citep{elkaroui2013robust, elkaroui2018impact}. Indeed, leverages always sum to the number of parameters \citep{hoaglin1978hat}, so we must have $\sum_i h_i^{(\ell)} = \mathrm{tr}\big(\X(\X^\top \boldsymbol{\Pi}^{(\ell)}\X)^{-1}\X^\top \boldsymbol{\Pi}^{(\ell)}\big) = p$. Then, letting  $b^{(\ell)} := \lim_{n \to \infty} \frac{1}{n}\mathrm{tr}(\Q^{-1})$ gives $h_i^{(\ell)} \approx \omega_i^{(\ell)}$, and the leverage-sum condition is $\frac{1}{n}\sum_i \omega_i^{(\ell)} \to \frac{1}{\delta}$, which is exactly \eqref{eq:ls-b-linear-model}. Thus, samples with large weight have leverage close to one and are fitted closely, while samples with weight near zero have little influence on the fit.

 By \eqref{eq:ls-Theta-tilde}, the fitted value $\hat{\theta}_i^{(\ell)}$ behaves like a convex combination of the observation $y_i$ and the prediction $\Z_i^\top\blambda^{(\ell)} + (\G_{\TTheta})_{i,\ell}$, with weight equal to the leverage. The random variable $\Z_i^\top\blambda^{(\ell)} + (\G_{\TTheta})_{i,\ell}$ can be viewed as the distributional limit of the \emph{leave-one-out} prediction $\x_i^\top \hat{\bbeta}^{(\ell)}_{-i}$, where $\hat{\bbeta}^{(\ell)}_{-i}$ is the fit with sample $i$ removed. Indeed,  since $\hat{\bbeta}^{(\ell)}_{-i}$ is independent of $\x_i$ and behaves like $\B\blambda^{(\ell)} + \sqrt{\delta}\, (\G_{\hat{\B}})_{:,\ell}$, we have $\x_i^\top \hat{\bbeta}^{(\ell)}_{-i} \approx \Z_i^\top \blambda^{(\ell)} + (\G_{\TTheta})_{i,\ell}$. Consequently, the residuals behave like
    \begin{align}
    \tilde{y}_i - \tilde{\Theta}_{i,\ell} = \big(1 - \omega_i^{(\ell)}\big) \Big[ \Z_i^\top \big(\e_{\Psi_i} - \blambda^{(\ell)}\big) + \varepsilon_i - (\G_{\TTheta})_{i,\ell} \Big],
    \label{eq:ls-residuals}
    \end{align}
    which is the asymptotic form of the classical identity $y_i - \x_i^\top \hat{\bbeta}^{(\ell)} = (1 - h_i^{(\ell)})(y_i - \x_i^\top \hat{\bbeta}^{(\ell)}_{-i})$ \citep{allen1974relationship}. A similar identity is used in \cite{rad2020scalable} to analyze the out-of-sample prediction error in high dimensions. Within the bracket in \eqref{eq:ls-residuals}, the first term captures the mismatch between the signal that generated sample $i$ and the linear combination of signals targeted by the $\ell$th fit, while the second and third terms correspond to the noise and the estimation error, respectively.

For general design covariance $\bSigma = \L \L^{\top}$, the limiting parameters depend on $\bSigma$ only through $\bGamma = \lim_n \B^\top \bSigma \B / n$, so the formulas in \eqref{eq:ls-b-linear-model}-\eqref{eq:ls-integrals} are unchanged. In this setting, Proposition \ref{prop:ls-B-Theta} still holds, with the only change being $\G_{\hat{\B}}$ in \eqref{eq:ls-B-rep} is replaced by $\L^{-\top}\G_{\hat{\B}}$ (Theorem \ref{thm:ERM_general_characterization_pres}). The representation \eqref{eq:ls-Theta-rep} of the fits is unchanged, and  the change point results below also hold unchanged.

\paragraph{Characterizing the change point error}
We now consider estimating the change points when their number $L^* - 1 = L - 1$ is known. This is \eqref{eq:general_changepoint_estimator} with $P \equiv 0$ and $\mathcal{X}$ restricted to configurations with exactly $L - 1$ change points. As in \eqref{eq:data-split}, the estimator minimizes the total squared residual when each segment is fitted by its own weighted least squares estimate:
\begin{align}
    \hat{\eeta} \in \argmin_{1 = k_0 < k_1 < \cdots < k_{L-1} < k_L = n+1} \; \sum_{\ell = 1}^{L} \; \sum_{i = k_{\ell - 1}}^{k_\ell - 1} \big(y_i - \hat{\theta}_i^{(\ell)}\big)^2,
    \label{eq:ls-split}
\end{align}
where the estimated signal configuration $\hat\Psi_i = \ell$ for $\hat\eta_{\ell - 1} \le i < \hat \eta_\ell$. To analyse \eqref{eq:ls-split}, we evaluate the total squared residual for an arbitrary candidate set of \emph{fractional} change points
\[
\tau \in \mathcal{T}_L := \big\{ (\tau_1, \dots, \tau_{L-1}) : 0 = \tau_0 < \tau_1 < \cdots < \tau_{L-1} < \tau_L = 1 \big\},
\]
which assigns sample $i$ to segment $\ell$ when $\tau_{\ell-1} < i/n \le \tau_\ell$. The corresponding residual is
\begin{align}
    R_n(\tau) := \frac{1}{n} \sum_{\ell=1}^{L} \sum_{i=1}^n \ind\Big\{ \tau_{\ell-1} < \tfrac{i}{n} \le \tau_\ell \Big\} \big(y_i - \hat{\theta}_i^{(\ell)}\big)^2.
    \label{eq:ls-Rn}
\end{align}
The estimator \eqref{eq:ls-split} minimizes $R_n$ over the grid $\tau_\ell = (k_\ell - 1)/n$, so that $\hat{\eeta}/n = \hat{\tau} + 1/n$ for a minimizer $\hat{\tau}$ of $R_n$ over this grid.

Proposition \ref{prop:ls-changepoints} below shows that for any candidate set $\tau$ of fractional change points, the residual $R_n$ converges to $\sum_{\ell=1}^L  \int_{\tau_{\ell - 1}}^{\tau_\ell} v_\ell(t)\, \de t$, where  $v_\ell(t)$ are the \emph{expected squared residual profiles} defined as
\begin{align}
    v_\ell(t) := \frac{(\e_{\bar{\psi}(t)} - \blambda^{(\ell)})^\top \bGamma\, (\e_{\bar{\psi}(t)} - \blambda^{(\ell)}) + \kappa_{\ell, \ell} + \sigma^2}{\big(1 + 2 b^{(\ell)} \Phi^{(\ell)}(t)\big)^2}, \qquad t \in [0, 1], \; \ell \in [L].
    \label{eq:ls-v}
\end{align}
By \eqref{eq:ls-residuals}, $v_\ell(i/n)$ is the limiting mean squared residual of the $\ell$th fit at sample $i$. We note that each $v_\ell(t)$ is piecewise continuous, with jumps at the change points, where $\bar{\psi}(t)$ changes.  In between the change points $v_\ell(t)$ varies smoothly through the weight profile $\Phi^{(\ell)}(t)$.
\begin{proposition} \label{prop:ls-changepoints}
Consider the setting of Proposition \ref{prop:ls-B-Theta}, and define
\begin{align}
    F(\tau) := \sum_{\ell = 1}^{L} \int_{\tau_{\ell - 1}}^{\tau_\ell} v_\ell(t)\, \de t, \qquad \tau \in \mathcal{T}_L.
    \label{eq:ls-F}
\end{align}
\begin{enumerate}
    \item[(a)]
    $\sup_{\tau \in \mathcal{T}_L} |R_n(\tau) - F(\tau)| \xrightarrow{\P} 0$. That is, the total squared residual \eqref{eq:ls-Rn} converges uniformly over candidate change points.
    \item[(b)] Suppose $L = L^*$ and that $F$ has a unique minimizer $\tau^* = (\tau^*_1, \dots, \tau^*_{L-1})$ over $\mathcal{T}_L$. Then the estimator \eqref{eq:ls-split} satisfies
    \begin{align}
    \frac{\hat{\eta}_j}{n} \xrightarrow{\P} \tau_j^*, \quad j \in [L-1], \quad \text{and} \quad
    \frac{1}{n} d_H(\hat{\eeta}, \eeta) \xrightarrow{\P} d_H\big(\{\tau_j^*\}_{j \in [L-1]}, \{\alpha_j\}_{j \in [L-1]}\big),
    \label{eq:ls-cp-limit}
    \end{align}
    where $0< \alpha_1 < \dots < \alpha_{L^*-1} < 1$ are the true fractional change points.
\end{enumerate}
\end{proposition}
The proof is given in Appendix \ref{subsec:ls-changepoints}.
Proposition \ref{prop:ls-changepoints} says that asymptotically, the estimator divides the interval $[0, 1]$ among the $L$ fits (in order) so as to minimize the total expected squared residual $F$. This gives a simple condition for the normalized change point error to vanish. For every $\tau$, we have $F(\tau) \ge \int_0^1 \min_\ell v_\ell(t)\, \de t$. If the correct fit has the smallest expected squared residual throughout its own segment, that is, $v_{\bar{\psi}(t)}(t) < v_\ell(t)$ for all $\ell \neq \bar{\psi}(t)$ and almost every $t$, then equality holds only at $\tau = \alpha$, so $\tau^* = \alpha$ and $d_H(\hat{\eeta}, \eeta) = o_{\P}(n)$. When the condition fails, $\tau^*$ can differ from $\alpha$, and the normalized error converges to the non-zero constant $d_H(\tau^*, \alpha)$. In this case the estimator is asymptotically biased, by an amount determined by $(\delta, \bGamma, \sigma^2, \alpha)$ and the weight profiles alone.

\begin{figure}[!t]
\centering
\includegraphics[width=\textwidth]{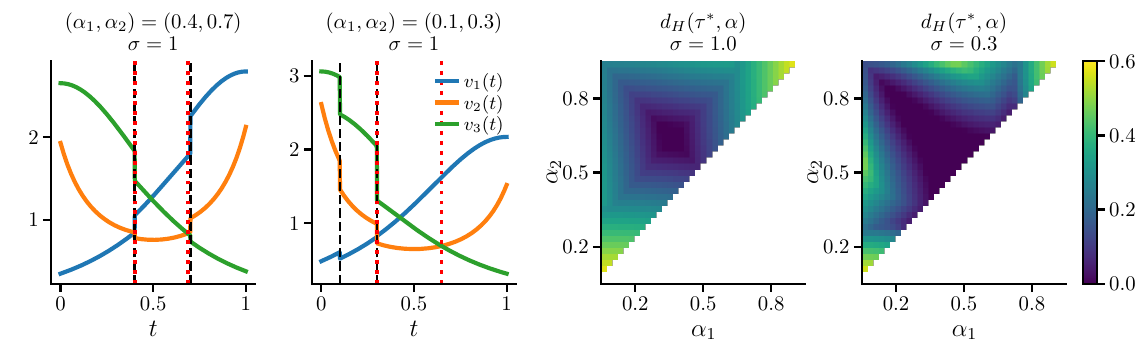}
\caption{\small {Linear model with $L = L^* = 3$, uniform-prior weights, $\delta = 4$ and $\Gamma_{r,s} = 0.5^{|r-s|}$. Two left panels ($\sigma = 1$): expected squared residual profiles $v_\ell(t)$ from \eqref{eq:ls-v}, true change points $\alpha$ (black dashed) and the limits $\tau^*$ of the estimated change points from Proposition \ref{prop:ls-changepoints} (red dotted). Two right panels: limiting normalized Hausdorff error $d_H(\tau^*, \alpha)$ as a function of the true change points $(\alpha_1, \alpha_2)$, for $\sigma = 1$ and $\sigma = 0.3$.}}
\label{fig:ls-limit-1x4}
\end{figure}

\paragraph{Example: asymptotic bias with a uniform prior} \label{sec:asymptotic-bias-example}
Figure \ref{fig:ls-limit-1x4} illustrates Proposition \ref{prop:ls-changepoints} for $L = L^* = 3$ with the uniform-prior weights of Figure \ref{fig:M1a}, whose profiles are $\Phi^{(1)}(t) = (1-t)^2$, $\Phi^{(2)}(t) = 2t(1-t)$ and $\Phi^{(3)}(t) = t^2$. We take $\delta = 4$ and $\Gamma_{r, s} = 0.5^{|r - s|}$. For $(\alpha_1, \alpha_2) = (0.4, 0.7)$ and $\sigma = 1$ (first panel), the correct fit has the smallest profile in each segment except just before $\alpha_2$, where $v_3$ falls below $v_2$. Hence $\tau^* = (0.400, 0.686)$: the first change point is located without asymptotic bias and the second with a small bias. For $(\alpha_1, \alpha_2) = (0.1, 0.3)$ (second panel), the first fit has the smallest profile on the whole of the first two segments, because the uniform prior gives early samples large  weights for the first fit. The estimator therefore merges the first two segments and places a spurious change point near $0.65$: $\tau^* = (0.300, 0.649)$. In simulations with $n = 4000$ and $p = 1000$, the estimated change points averaged over $10$ trials were $(0.395, 0.676)$ and $(0.308, 0.649)$ for the two configurations, in close agreement with $\tau^*$. The heat maps show $d_H(\tau^*, \alpha)$ over all $(\alpha_1, \alpha_2)$ for $\sigma = 1$ and $\sigma = 0.3$. With the uniform prior, the change points are located without asymptotic bias  in a region around $(1/3, 2/3)$, where the weight profiles cross, and this region grows as the noise level falls. A different prior changes the weight profiles $\Phi^{(\ell)}$, and hence this region.

\subsection{Theory for the general case} \label{sec:characterization-nonlinear-eqns}

In this section, we present theory for \WeightedERM{} in the general setting of GLMs with change points.
We make the following assumptions on the loss $M$ used in our estimator \eqref{eq:ERM_est}:
\begin{assumption}[Regularity of the loss]\label{asmpt:reg-loss}
\leavevmode\par
\begin{enumerate}
\item $M$ is \textit{strictly convex} and \textit{smooth}: it is continuously differentiable with absolutely continuous derivative $M'$ having an a.e. continuous second derivative $M'' > 0$ that is bounded above by a constant. \label{cond:strict-cvx-smooth}
\item For $\ell \in [L]$, we have $\lim_{p \to \infty} \P\left[ \nabla^2 \cC^{(\ell)}(\bbeta) \succeq \gamma(\|\bbeta\| /
   \sqrt{p}) \cdot \I, \;\;\; \forall \bbeta \in \reals^p \right] = 1$ for some non-increasing continuous function $0 < \gamma(\cdot) < 1$ independent of $n$, where $\cC^{(\ell)}$ is defined in \eqref{eq:ERM_est}.\label{cond:likelihood_curvature}
\item The following higher order derivatives exist and are bounded above and below by a constant: $M''', \partial_{12} M, \partial_2 M''$. Moreover, we have that $(M'')^2 - M' \cdot M''' > 0$. \label{eq:loss-asmpt-log-conc}
\item $b M''(\prox_{b M(\cdot, v)}(u), v) \to \infty \text{ as } b \to \infty$. \label{eq:loss-asmpt-b-M-to-inf}
\item
The solution to \eqref{eq:ERM_est} exists for $\ell \in [L]$ with probability tending to $1$ as $n \to \infty$  (over the randomness of the design matrix $\X$). \label{eq:existence-of-estimator}
\end{enumerate}
\end{assumption}
Assumption~\ref{asmpt:reg-loss}(\ref{cond:strict-cvx-smooth}) is standard in the literature on characterising risk in the regime where $n$ grows proportionally with $p$ \citep{donoho_high_2016}. The ``likelihood curvature'' Assumption~\ref{asmpt:reg-loss}(\ref{cond:likelihood_curvature}) \citep{sur_modern_2019} controls the decay of the Hessian of $\cC^{(\ell)}$ as the magnitude of the input increases, and is akin to a local strong convexity condition. It is satisfied for many widely used loss functions including the squared, logistic, and Poisson losses, and is satisfied for the Huber loss in part of the domain. Assumption~\ref{asmpt:reg-loss}(\ref{eq:loss-asmpt-log-conc}) imposes certain regularity conditions on the higher order derivatives of $M$ which are satisfied by common loss functions such as the squared loss and the logistic loss and, under additional technical steps, can be relaxed to admit the non-differentiable Huber loss and the Poisson loss.
Assumption~\ref{asmpt:reg-loss}(\ref{eq:loss-asmpt-b-M-to-inf}), moreover, prevents the decay of $M''$ to zero as $b$ grows.

Assumptions \ref{asmpt:reg-loss}(\ref{cond:likelihood_curvature}), \ref{asmpt:reg-loss}(\ref{eq:loss-asmpt-log-conc}), and \ref{asmpt:reg-loss}(\ref{eq:loss-asmpt-b-M-to-inf}) are standard in the study of logistic regression in the proportional asymptotic regime \citep{sur_modern_2019, sur_likelihood_2019}, and are satisfied by other important loss functions such as squared loss and the Huber loss. While Assumption~\ref{asmpt:reg-loss}(\ref{eq:existence-of-estimator}) is always satisfied for the squared and Huber losses, it is not trivial for important losses such as the logistic loss. \cite{sur_modern_2019}  precisely quantify the $(\delta, \bGamma)$ parameter regime in which the maximum likelihood estimator for the (homogeneous) logistic model does not exist. For the logistic model with change points, we identify a sufficient $(\delta,\bGamma)$ parameter region for existence of the corresponding \WeightedERM{} estimator (Proposition \ref{prop:logistic_mle_existence} in Appendix \ref{sec:logistic}).

\paragraph{Characterization via nonlinear equations}

We recall the  signal estimate matrix $\hat{\B} \in \reals^{p \times L}$ and the fit matrix $\hat{\TTheta} \in \reals^{n \times L}$ from \eqref{eq:B-estimate-matrix}-\eqref{eq:Theta-estimate-matrix}. As in the case of the linear model, the asymptotic behaviour of $(\hat{\B}, \hat{\TTheta})$ is described by  positive scalars $(b^{(\ell)})_{\ell \in [L]}$,  a bias matrix $\bLambda = [\blambda^{(1)}, \ldots, \blambda^{(L)} ] \in \reals^{L \times L}$, and a covariance matrix $\bK = (\kappa_{\ell, \ell'})_{\ell, \ell' \in [L]}$. These parameters, which depend on the GLM and the loss function $M$, are given by the solution of a set of nonlinear equations, given in \eqref{eq:b-fp-implicit}--\eqref{eq:cross-cov-fp-implicit} (Appendix \ref{sec:nonlinear_equations}).

Our main theoretical result, Theorem \ref{thm:ERM_general_characterization_pres} below,
shows that $[\B, \hat{\B}] \in \reals^{p \times 2L}$ has the asymptotic distributional representation $[\B, \B \bLambda + \sqrt{\delta} \L^{-\top} \G_{\hat{\B}}]$, where $\L$ is the lower Cholesky factor of $\bSigma$. The matrix $\G_{\hat{\B}} \in \reals^{p \times L}$ has i.i.d. rows with $(\G_{\hat{\B}})_j \distas{\text{i.i.d.}} \N(\0_L, \bK)$ for $j \in [p]$.
Similarly,  $[\X \B, \hat{\TTheta}]$ has the asymptotic distributional representation $[\Z, f(\Z \bLambda + \G_{\TTheta}; \bPsi, \bvarepsilon) ]$, for a certain nonlinear function $f$ (see \eqref{eq:SE_ERM_Theta}). Here, $\Z, \G_{\TTheta} \in \reals^{n \times L}$ are independent matrices with rows  $\Z_i \distas{\text{i.i.d.}} \N(\0_L, \bGamma)$  and $(\G_{\TTheta})_i \distas{\text{i.i.d.}} \N(\0_L, \bK)$, for $i \in [n]$.
Thus, $\bLambda$ encodes the bias of estimators in \eqref{eq:ERM_est} and $\bK$ their covariance structure.

The \textit{proximal operator} is central to our analysis. Given a convex, lower semi-continuous function $f: \reals \to \reals$ and  $\varrho >0$, the proximal operator is defined as:
\begin{align}
\prox_{\varrho f}(z) := \argmin_{t \in \reals} \left\{ \varrho f(t) + (t - z)^2 / 2\right\}, \qquad z \in \reals. \label{eq:prox_op_first_defn}
\end{align}
Intuitively, the mapping $z \mapsto \prox_{\varrho f}(z)$ produces a minimizer of $f$ close to $z$ with the proximity of the minimizer to $z$ controlled by the parameter $\varrho$.
We need the following matrix variant of the proximal operator in \eqref{eq:prox_op_first_defn}, denoted by $\bprox_{ \b \bpi M(\cdot, \tilde{\y})} : \reals^{n \times L} \to \reals^{n \times L}$. For $\tilde{\y} \in \reals^n, \tilde{\TTheta} \in \reals^{n \times L}, \b \in \reals^L$,  and $i \in [n], \ell \in [L]$, define:
\begin{align} \label{eq:bold-prox-def}
	&\left(\bprox_{\b \bpi M(\cdot, \tilde{\y})}(\tilde{\TTheta})\right)_{i, \ell} :=
		\prox_{ b_{\ell} \pi^{(\ell)}_i M(\cdot, \tilde{y}_i)}(\tilde{\TTheta}_{i, \ell}).
\end{align}
For example, for the squared loss $M(u,v) = (u -v)^2$,  $$\prox_{b_{\ell} \pi^{(\ell)}_i M(\cdot, \tilde{y}_i)}(\tilde{\TTheta}_{i, \ell}) = \frac{(\tilde{\TTheta}_{i, \ell} + 2 b_{\ell} \pi^{(\ell)}_i \tilde{y}_i)}{1 + 2 b_{\ell} \pi^{(\ell)}_i}.$$
\begin{theorem} \label{thm:ERM_general_characterization_pres}
Consider the setting described in Section \ref{sec:setting} and recall the matrices $\bGamma, \bLambda, \bK$ defined via \eqref{eq:Gamma_as_limit} and \eqref{eq:b-fp-implicit}--\eqref{eq:cross-cov-fp-implicit}, and the  lower-triangular Cholesky factor $\L$ for the covariance $\bSigma$. Further assume that a solution to equations \eqref{eq:b-fp-implicit}--\eqref{eq:cross-cov-fp-implicit} exists. Then, {$\b$, $\bLambda$ and $\bK$ are unique}, and for any sequence of uniformly pseudo-Lipschitz functions $\varphi_{n}(\cdot \;; \bPsi, \bvarepsilon) : \reals^{n \times 2L} \to \reals$, $\varphi_{p}(\cdot \; ; \B) : \reals^{p \times L} \to \reals$:
   \begin{align}
    &\varphi_n(\hat{\TTheta}, \X \B ; \bPsi, \bvarepsilon) \stackrel{\P}{\simeq} \E \left\{ \varphi_n \left( \bprox_{\b \bpi M(\cdot, q(\Z, \bPsi, \bvarepsilon))}\left(\Z \bLambda + \G_{\TTheta}\right), \Z ; \bPsi, \bvarepsilon \right) \right\} \label{eq:SE_ERM_Theta}, \\
    &\varphi_p(\hat{\B} ; \B) \stackrel{\P}{\simeq} \E \{ \varphi_p(\B \bLambda + \sqrt{\delta} \L^{-\top} \G_{\hat{\B}} ; \B)\} \label{eq:SE_ERM_B},
    \end{align}
where, independently over $i \in [n]$, we have $\Z_i \distas{} \N(\0_L, \bGamma)$, $(\G_{\TTheta})_i \distas{} \N(\0_L, \bK)$, and independently over $j \in [p]$, we have $(\G_{\hat{\B}})_j \distas{} \N(\0_L, \bK)$.

\end{theorem}
The proof is given in Appendix \ref{sec:proof-of-main-results}, and we give a brief outline at the end of this section. Theorem \ref{thm:ERM_general_characterization_pres} states that the behaviour of any pseudo-Lipschitz function of the matrix estimates $\hat{\TTheta}, \hat{\B}$ converges to a deterministic quantity, namely an expectation over the random matrices $\Z, \G_{\TTheta}, \G_{\hat{\B}}$ involving the solution of the coupled system of equations \eqref{eq:b-fp-implicit}--\eqref{eq:cross-cov-fp-implicit}. We prove the existence of a solution to \eqref{eq:b-fp-implicit}--\eqref{eq:cross-cov-fp-implicit} under mild additional assumptions in Proposition \ref{prop:existence-of-fixed-point-solution} (Appendix \ref{sec:proof-of-main-results}).  In practice, the existence of a solution  can be checked numerically.

Theorem \ref{thm:ERM_general_characterization_pres} allows us to evaluate performance metrics such as the mean squared error between the signal matrix $\B$ and the estimate $\hat{\B}$. Taking $\varphi_p(\hat{\B}, \B) = \|\hat{\B} - \B\|_F^2 / p$ leads to:
\[\|\hat{\B} - \B\|_F^2 / p \stackrel{\P}{\simeq} \E[\|\B \bLambda + \sqrt{\delta} \L^{-\top} \G_{\hat{\B}} - \B\|_F^2] / p,\]
where the limiting value of the RHS (as $p \to \infty$) can be precisely computed under suitable assumptions discussed near the end of this section. In the homogeneous case of $L=1$, the result  recovers known results for high-dimensional asymptotics of M-estimation \citep{donoho_high_2016} and logistic regression \citep{sur_modern_2019}

We recall the one-to-one mapping $U$ between a set of change points $\eeta$ and signal configuration vectors $\bPsi$ defined via \eqref{eq:general_changepoint_estimator}, and highlight that the change point estimator $\hat{\bPsi}(\hat{\TTheta}; \y)$ from \eqref{eq:general_changepoint_estimator} takes on a similar form to $\varphi_n$ in Theorem \ref{thm:ERM_general_characterization_pres}. This allows us to precisely evaluate performance metrics of interest for well-behaved estimators of this form, such as the scaled Hausdorff distance between the estimated and true change points, and the number of detected change points.
\begin{proposition} \label{prop:ERM_hausdorff_characterization_pres}
    Consider the setting of Theorem \ref{thm:ERM_general_characterization_pres}, and recall the   Hausdorff distance $d_H$ in \eqref{eq:hausdorff-dist}. Let $\hat{\eeta}(\hat{\TTheta}, \y)$ be an estimator for the set of change points  such that $(\V, \tilde{\y}) \mapsto U(\hat{\eeta}(\V, \tilde{\y}))$ is uniformly pseudo-Lipschitz. Then,
    \begin{align} \label{eq:ERM_hausdorff_match_eqn}
    & \frac{1}{n} d_H\left(\hat{\eeta} \big(\hat{\TTheta}; \y \big), \eeta \right)
    \stackrel{\P}{\simeq} \frac{1}{n} \E \; d_H\left(\hat{\eeta}\left( \bprox_{\b \bpi M(\cdot, q(\Z, \bPsi, \bvarepsilon))}\big( \Z \bLambda + \G_{\TTheta}\big) ; q(\Z, \bPsi, \bvarepsilon) \right), \eeta \right).
    \end{align}
    Moreover, if $(\V, \tilde{\y}) \mapsto \|\hat{\eeta}(\V, \tilde{\y})\|_0$ is uniformly pseudo-Lipschitz, then the number of estimated change points, denoted by $\|\hat{\eeta}(\hat{\TTheta}, \y) \|_0$, satisfies:
    \begin{align} \label{eq:ERM_size_match_eqn}
        \left\| \hat{\eeta}\big(\hat{\TTheta}, \y\big) \right\|_0 \stackrel{\P}{\simeq} \E \left\| \hat{\eeta}\left( \bprox_{\b \bpi M(\cdot, q(\Z, \bPsi, \bvarepsilon))} \big( \Z \bLambda + \G_{\TTheta} \big), q(\Z, \bPsi, \bvarepsilon) \right) \right\|_0,
    \end{align}
where, independently over $i \in [n]$, we have $\Z_i \distas{} \N(\0_L, \bGamma)$, $(\G_{\TTheta})_i \distas{} \N(\0_L, \bK)$.
\end{proposition}
The proof, presented in Appendix \ref{sec:proof-of-main-results}, involves showing that $\frac{1}{n}d_H\Big(\hat{\eeta} \big(\hat{\TTheta}; \y \big), \eeta \Big)$ is uniformly pseudo-Lipschitz and then applying Theorem \ref{thm:ERM_general_characterization_pres}. For any ground truth change point set $\eeta$, Proposition \ref{prop:ERM_hausdorff_characterization_pres} precisely characterizes the asymptotic Hausdorff distance and size errors for a large class of estimators of the form \eqref{eq:general_changepoint_estimator}. Figure \ref{fig:logistic-example-match} shows the normalized Hausdorff error and the number of change points for logistic regression with a change point. We observe a close match between the empirical performance and the theory (left and right sides of \eqref{eq:ERM_hausdorff_match_eqn} and \eqref{eq:ERM_size_match_eqn}, respectively). We emphasize, and evidence empirically in Appendix \ref{sec:E-L-upper-bound}, that the theoretical results of Theorem \ref{thm:ERM_general_characterization_pres} and Proposition \ref{prop:ERM_hausdorff_characterization_pres} hold for any $L \geq 1$, regardless of the actual number of change points $(L^* - 1)$. \label{page:post-theory-remark}

\paragraph{Computing theoretical quantities and predictions}
The fixed point equations \eqref{eq:b-fp-implicit}--\eqref{eq:cross-cov-fp-implicit} and the theoretical predictions on the RHS of \eqref{eq:SE_ERM_Theta}, \eqref{eq:ERM_hausdorff_match_eqn}, \eqref{eq:ERM_size_match_eqn} can be computed with minimal problem-specific knowledge  under reasonable assumptions. The explicit dependence of these quantities  on the noise vector $\bvarepsilon$ can be replaced by expectations over a scalar random variable $\bar{\varepsilon}$
under the assumptions in Section \ref{sec:setting}.
This can be proved using an  argument similar to that in the proof of Proposition \ref{prop:conv-nonlin-eqns} in Appendix \ref{sec:variants-nonlin-eqns}, and experimental evidence  is provided  via Figures
\ref{fig:logistic-example-match}, \ref{fig:posterior-sq-loss}, and \ref{fig:alt-signal-est-match-app} (in Appendix \ref{sec:lin-model-alternating-signals-example}), where the empirical results and the theoretical predictions  were produced using independent copies of $\bvarepsilon$ with the same limiting distribution. Finally, the signal strength matrix $\bGamma$ can be estimated using a procedure described at the end of Section \ref{sec:uncertainty-quantification}.

The explicit dependence of the theoretical quantities on the signal configuration vector $\bPsi$ is fundamental. Since the entries of $\bPsi$ change value only at a finite number of change  points, the quantities in question will depend on the limiting fractional values of these change points. This is consistent with recent change point regression literature, where the limiting distribution of the change point estimators in \cite{xu_change-point_2024} is shown to be a function of the data generating mechanism.

\paragraph{Proof outline for Theorem \ref{thm:ERM_general_characterization_pres}} We first prove that equations \eqref{eq:b-fp-implicit}--\eqref{eq:cross-cov-fp-implicit} are well-defined. The proof of Theorem \ref{thm:ERM_general_characterization_pres} then leverages the equivalence between estimator \eqref{eq:ERM_est} and a certain non-separable Approximate Message Passing (AMP) iteration  in the limit where $n \to \infty$ followed by $t \to \infty$, where $t$ is the iteration number (Lemma \ref{lemma:ERM_frobenius_characterization}). This equivalence allows us to translate the asymptotic characterization of AMP in \citep{arpino_2025_jmlr} to an asymptotic characterization of the Weighted ERM estimator \eqref{eq:ERM_est} in terms of quantities $(\bLambda, \bK, \b)$ defined via \eqref{eq:b-fp-implicit}--\eqref{eq:cross-cov-fp-implicit}.
\section{Uncertainty quantification} \label{sec:uncertainty-quantification}

Given a set of estimators $\{(\hat{\bbeta}^{(\ell)}, \hat{\ttheta}^{(\ell)} )\}_{\ell \in [L]}$ from \eqref{eq:ERM_est}, we apply the asymptotic characterization of Theorem \ref{thm:ERM_general_characterization_pres} to construct a posterior distribution over signal configuration vectors $\bPsi$, or equivalently, over change point vectors $\eeta$.

First, let $\ell \in [L]$, and let $\m^{\prime}\left( \hat{\ttheta}^{(\ell)}, \y \right) \in \reals^n$ denote a vector with components $\left(\m^{\prime}\left( \hat{\ttheta}^{(\ell)}, \y  \right) \right)_i = \pi^{(\ell)}_i M'(\hat{\theta}_i^{(\ell)}, y_i)$ for $i \in [n]$.
We modify the estimate $\hat{\ttheta}^{(\ell)}$ from \eqref{eq:ERM_est} as follows, using a rescaled gradient ascent step:
\begin{align}
    \ttheta^{\adj, (\ell)} := \hat{\ttheta}^{(\ell)} + \hat{b}^{(\ell)} \m'(\hat{\ttheta}^{(\ell)}, \y), \label{eq:theta-deb-defn}
\end{align}
where $\hat{b}^{(\ell)} > 0$ is the solution to the empirical version of fixed point equation \eqref{eq:b-fp-implicit}:
\begin{align} \label{eq:b-hat-eqn}
   1 - \frac{1}{\delta} = \frac{1}{n} \sum_{i=1}^n  \left(1 + \hat{b}^{(\ell)} \pi^{(\ell)}_i M''(\hat{\theta}^{(\ell)}_i, y_i)\right)^{-1}.
\end{align}
We call $\TTheta^{\adj} := \big[
    \ttheta^{\adj, (1)} \dots \ttheta^{\adj, (L)}
\big]$ the adjusted estimate of $\TTheta := \big[
   \ttheta^{(1)}  \dots  \ttheta^{(L)} \big]$. The following theorem shows that $\TTheta^{\adj}$ is asymptotically characterized by a random variable $\Z \bLambda + \G_{\TTheta}$, in contrast with $\hat{\TTheta}$ whose asymptotic behaviour is characterized by $\bprox_{\b \bpi M(\cdot, q(\Z, \bPsi, \bvarepsilon))}\left(\Z \bLambda + \G_{\TTheta}\right)$ in Theorem \ref{thm:ERM_general_characterization_pres}. The simpler characterization of $\TTheta^{\adj}$ allows us to construct a computable likelihood function using $(\TTheta^{\adj}, \y)$, enabling valid inference over
   $\bPsi$.
\begin{theorem} \label{thm:characterization_of_debiased_est_pres}
Consider the setting of Theorem \ref{thm:ERM_general_characterization_pres}. Then, for any sequence of uniformly pseudo-Lipschitz functions $\varphi_{n}(\cdot \;; \bPsi, \bvarepsilon) : \reals^{n \times 2L} \to \reals$:
    \begin{align}
    &\varphi_n(\TTheta^{\adj}, \X \B ; \bPsi, \bvarepsilon) \stackrel{\P}{\simeq} \E \left\{ \varphi_n\left(\Z \bLambda + \G_{\TTheta}, \Z ; \bPsi, \bvarepsilon \right) \right\} \label{eq:SE_ERM_Theta_deb},
    \end{align}
where independently over $i \in [n]$, we have $\Z_i \distas{} \N(\0_L, \bGamma)$ and $(\G_{\TTheta})_i \distas{} \N(\0_L, \bK)$.
\end{theorem}
The proof of Theorem \ref{thm:characterization_of_debiased_est_pres} is given in Appendix \ref{subsec:debiased_est_posterior_proofs}. The random variable $\Z \bLambda + \G_{\TTheta}$, combined with an observation of the form $q(\Z, \bPsi, \bar{\bvarepsilon})$, gives a recipe for constructing a posterior distribution over $\bPsi$. Recalling that $\bPsi$ is the unknown ground-truth signal  configuration vector (deterministic), we can consider a probability mass function $\pi_{\bar{\bPsi}}$ over a postulated random variable $\bar{\bPsi}$, and view $\bpsi$ as a realization of $\bar{\bPsi}$. Using the prior
$\pi_{\bar{\bPsi}}$, the posterior is:
\begin{align}
p(\bpsi | \V, \u) := p_{\bar{\bPsi} | \Z \bLambda + \G_{\TTheta}, \; q(\Z, \bar{\bPsi}, \bar{\bvarepsilon})}(\bpsi | \V, \u) &= \frac{  \pi_{\bar{\bPsi}}(\bpsi) \cL(\V, \u | \bpsi)}{\sum_{\tilde{\bpsi}} \pi_{\bar{\bPsi}}(\tilde{\bpsi}) \cL(\V, \u | \tilde{\bpsi})}, \label{eq:posterior_density_2_pres}
\end{align}
where $\V \in \reals^{n \times L}$, $\u \in \reals^n$. Here, $\cL(\cdot, \cdot | \bpsi)$ is the likelihood of $(\Z \bLambda + \G_{\TTheta}, q(\Z, \bar{\bPsi}, \bar{\bvarepsilon}))$ given $\bar{\bPsi} = \bpsi \in \mathcal{X}$, where the matrices $\bLambda, \bK$ associated to $\Z \bLambda + \G_{\TTheta}$ are computed as in \eqref{eq:b-fp-implicit}--\eqref{eq:cross-cov-fp-implicit} with $\bPsi$ replaced by $\bpsi$. As described in \cite[Appendix C.$3$]{arpino_2025_jmlr}, the likelihood $\cL$ can be computed in closed form for the linear model under the assumption of additive and independent Gaussian noise, while for the logistic model, the likelihood can be well approximated.
Since Theorem \ref{thm:characterization_of_debiased_est_pres} states that $(\TTheta^{\adj}, \y)$ converges in a specific sense to $(\Z \bLambda + \G_{\TTheta}, q(\Z, \bPsi, \bvarepsilon))$, we can obtain an uncertainty estimate over $\bPsi$ by plugging in $(\TTheta^{\adj}, \y)$ for $(\V, \u)$ in \eqref{eq:posterior_density_2_pres}. It then follows from our theory that this uncertainty estimate converges point-wise in probability to a faithful posterior distribution over the change points in the high-dimensional limit.
\begin{proposition}\label{prop:pointwise_posterior_pres}
Under the setting described in Section \ref{sec:setting}, assume that $(\V, \u) \mapsto p(\cdot | \V, \u)$ is uniformly pseudo-Lipschitz.  Then, for $\bpsi \in \mathcal{X}$:
\begin{align}
   p(\bpsi | \TTheta^{\adj}, \y) \stackrel{\P}{\simeq}  p(\bpsi | \Z \bLambda + \G_{\TTheta}, q(\Z, \bPsi, {\bvarepsilon})),\label{eq:posterior_convergence_pres}
\end{align}
where independently over $i \in [n]$, we have $\Z_i \distas{} \N(\0_L, \bGamma)$ and $(\G_{\TTheta})_i \distas{} \N(\0_L, \bK)$.
\end{proposition}
The proof  is given in Appendix \ref{subsec:debiased_est_posterior_proofs}. Given $\bpsi$ and ground-truth variables $\bPsi, \bvarepsilon$, the RHS of \eqref{eq:posterior_convergence_pres} can be computed by sampling $(\Z \bLambda + \G_{\TTheta}, \Z)$, where the matrices $\bLambda, \bK$ associated to $\Z \bLambda + \G_{\TTheta}$ are computed exactly as in \eqref{eq:b-fp-implicit}--\eqref{eq:cross-cov-fp-implicit} (using $\bPsi$, not $\bpsi$), and then evaluating $p(\bpsi | \V, \u)$ with $(\V, \u)= (\Z \bLambda + \G_{\TTheta}, \, q( \Z , \bPsi, \bvarepsilon)$.

\begin{figure}[!tp]
    \centering
    \begin{subfigure}[b]{0.99\textwidth}
        \centering
        \includegraphics[width=0.99\textwidth]{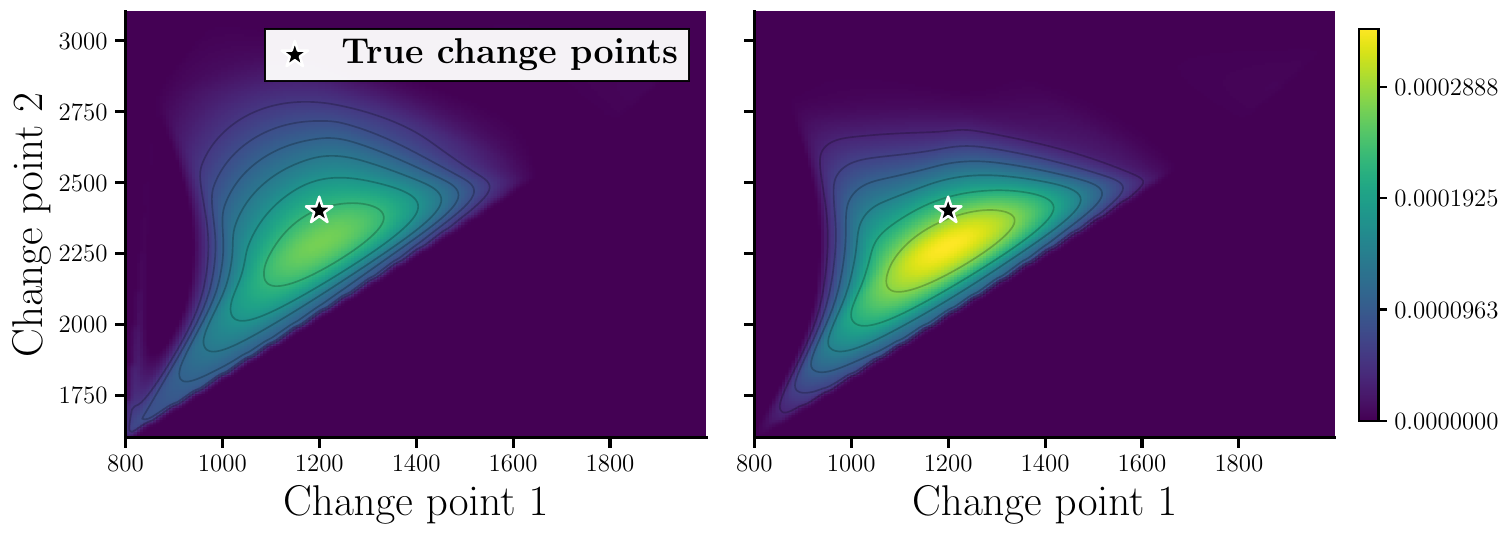}
    \end{subfigure}
    \caption{\small Posterior distributions produced from \WeightedERM{} (left) and theory in \eqref{eq:posterior_convergence_pres} (right),     downsampled to a grid of $13$ points, averaged over $40$ trials, and smoothed using a unit Gaussian kernel. Linear model with two change points at $0.3n, 0.6n$, with $n=4000$, $p=1000$.
  }
    \label{fig:posterior-sq-loss}
\end{figure}

In Figure \ref{fig:posterior-sq-loss},
we compare the posterior distribution produced via our method and that dictated by theory, for a linear model with exactly two change points at $0.3n$, $0.6n$, with other experiment details given in Appendix \ref{sec:posterior-example-additional-details}. The plots on the left and right correspond to left and right sides of \eqref{eq:posterior_convergence_pres}, respectively.
We observe a close match between the posterior produced by our method and the theoretical version in Proposition \ref{prop:pointwise_posterior_pres}.
The explicit parametric dependence on $\bvarepsilon$ on the right sides of \eqref{eq:SE_ERM_Theta_deb} and \eqref{eq:posterior_convergence_pres} can be removed via the arguments presented at the end of Section \ref{sec:asymptotic-characterization}.

\paragraph{Estimating the signal strength matrix} \label{para:estimating-signal-strength-matrix}
The likelihood function $\cL$ depends on the unknown signal strength matrix $\bGamma$ from \eqref{eq:Gamma_as_limit}. We now describe a simple procedure for estimating $\bGamma$ from data, after which the likelihood function $\cL$ can be computed using only empirical quantities $(\hat{\TTheta}, \y)$ and the noise distribution $\P_{\bar{\varepsilon}}$. First, suppose the true signal configuration vector $\bPsi$ were known. Applying Theorem \ref{thm:characterization_of_debiased_est_pres} with $\varphi_n(\TTheta^{\adj}, \X\B; \bPsi, \bvarepsilon) := \left( \frac{1}{n} (\TTheta^{\adj})^\top (\TTheta^{\adj}) \right)_{[\ell, \ell']}$ for $\ell, \ell' \in [L]$, we obtain:
\begin{align}
    \frac{1}{n} (\TTheta^{\adj})^\top (\TTheta^{\adj}) \stackrel{\P}{\simeq} \bLambda^\top \bGamma \bLambda + \bK. \label{eq:theta-deb-fp-eqn}
\end{align}
Hence \eqref{eq:theta-deb-fp-eqn} can be incorporated into equations \eqref{eq:b-fp-implicit}--\eqref{eq:cross-cov-fp-implicit} (with sample weights set to \eqref{eq:pi_i_defn}) to form a set of $\frac{3}{2} (L + L^2) + L(L+1)$ equations in the same number of unknowns, allowing
one to solve for $(\bGamma, \bLambda, \bK, \b)$. Our procedure consists of replacing $\bPsi$ in  \eqref{eq:b-fp-implicit}--\eqref{eq:cross-cov-fp-implicit} with the empirical estimate $\hat{\bPsi}(\hat{\TTheta}, \y)$ from \eqref{eq:general_changepoint_estimator},  and using the recipe above to produce an empirical estimate $\hat{\bGamma}$ of $\bGamma$.  In Section \ref{sec:myocardial} we use this data-driven method described to estimate the signal strength matrix $\bGamma$, and consequently construct a posterior distribution over change points in myocardial infarction data \citep{misc_myocardial_infarction_complications_579}.
\section{Numerical experiments on synthetic data} \label{sec:synthetic-experiments}
We evaluate the empirical performance of \WeightedERM{} alongside current state of the art  change point regression methods in three different settings presented below.
Code for reproducing the experiments is available in the \texttt{WeightedERM-reference} package \citep{arpino_2026_weightederm_reference}. A more robust and flexible implementation intended for practical use is available via the \texttt{weightederm} package in Python \citep{arpino_2026_weightederm_python} and R \citep{arpino_2026_weightederm_R}. In Appendix \ref{sec:additional-experiments-and-examples}, we provide additional experiments investigating the sensitivity of \WeightedERM{} to the misspecification of $L$ (Appendix \ref{sec:E-L-upper-bound}), to the change point prior (Appendix \ref{sec:sensitivity-change-point-prior}), to the approach for selecting number of change points (Appendix \ref{sec:investigating-count-selection}), as well as a numerical comparison to the AMP estimator for isotropic covariates \citep{arpino_2025_jmlr} (Appendix \ref{sec:AMP-comparison}).

For each experiment in this section, we fix $p = 200$ and run $100$ trials, in each trial independently generating the regression coefficient matrix $\B$, the covariates $\x_i$, and the noise $\varepsilon_i$ for $i \in [n]$. We compare existing methods with the unpenalized \WeightedERM{} estimator \eqref{eq:ERM_est}, denoted \WERM{} in the plots, and its $\ell_1$-penalized variant, \WeightedERML{}, denoted \WERML{}. The latter adds the penalty $\lambda \sqrt{2(\log p)/n} \| \bbeta \|_1$ to \eqref{eq:ERM_est} to promote sparsity and enable a fairer comparison with competing methods, most of which assume sparse signals. Setting $\lambda=0$ recovers \WeightedERM{}. Further implementation details can be found in Appendix \ref{sec:numerical-implementation-details}.

\label{par:CV}
Rather than tuning the penalty $P$ in \eqref{eq:general_changepoint_estimator} to control the number of change points, we jointly select $(\hat{L},\hat{\lambda})$ by cross-validation. Given an upper bound $L$ on the number of signals, for each candidate $\hat{L}\in \{ 1,\dots,L \}$ and $\hat{\lambda}\geq0$, we apply \eqref{eq:ERM_est}--\eqref{eq:general_changepoint_estimator} with $P=0$ and restrict the weights to configurations with exactly $\hat{L}-1$ change points. These weights are obtained as marginals of $\pi_{\bar{\bPsi}}$ in \eqref{eq:pi_i_defn}, conditional on $\hat{L}-1$ change points. Via the dynamic programming method for solving \eqref{eq:general_changepoint_estimator}, this yields a candidate change point vector $\hat{\eeta}^{(\hat{L})}\in[n]^{\hat{L}-1}$ for each $\hat{L}$.
We then select the optimal $\hat{\lambda}$ (when applicable) and the optimal number of change points $\hat{L}-1$, together with the corresponding vector $\hat{\eeta}^{(\hat{L})}$, using the absolute-error cross-validation procedure of \citep[Appendix A]{pein_cross-validation_2025}. By default, we use $2$ folds for \WeightedERML{} and $10$ folds for \WeightedERM{}. We show in Appendices \ref{sec:investigating-count-selection}, \ref{sec:E-L-upper-bound}, \ref{sec:AMP-comparison} that the empirical performance of cross-validation is comparable to using \eqref{eq:general_changepoint_estimator} with an appropriate penalty $P(\bpsi)$, but cross-validation is a  robust practical alternative that is more easily tunable via the interpretable `number of folds' parameter \citep{pein_cross-validation_2025}.

\subsection{Linear model with sparse signals and heavy-tailed noise} \label{sec:lin-model-sparse-signals}
We consider  robust estimation with Huber loss in the linear model \eqref{eq:linear-chgpt-model} with  heavy-tailed noise. We set the number of change points to be two, and fix their locations at $0.35n, 0.65n$. For $i \in [n]$, we sample $\varepsilon_i \distas{\text{i.i.d.}} \sqrt{0.1}\,T(4)$,  a scaled Student's t-distribution with four degrees of freedom, and use $\x_i \distas{\text{i.i.d.}} \N(\0_p, \bSigma/n)$ where $\Sigma_{k, j} = 0.4^{|k-j|}$ for $k, j \in [p]$. The regression vectors are sparse, with entries independently sampled according to $0.3 \N(0, \delta) + 0.7 \delta_0$. We evaluate \WeightedERML{} with $L = 4$, i.e. the method considers $0$ to $3$ change points, and take $M$ to be the Huber loss:
\begin{align}\label{eq:Huber-loss}
M^{\text{Huber}}_{\tau}(u, v) =
\begin{cases}
\displaystyle \frac{1}{2} (u-v)^2, & \text{if } |u-v| \le \tau, \\[8pt]
\displaystyle \tau \bigl(|u-v| - \tfrac{1}{2}\,\tau\bigr), & \text{if } |u-v| > \tau,
\end{cases}
\end{align}
with $\tau$ set to the standard value of $1.35$. We select the sample weights according to a uniform prior over change point locations where change points are at least $n/20$ away from each other and from end points.
We run \McScan{}, \DPDU{} \citep{xu_change-point_2024}, and \MOSEG{} exactly as described in the multiple change point scenarios  in \cite[Section 4.2.1]{cho_detection_2025}, and \DCDP{} using the cross-validation implementation provided in \cite{li_divide_2023}, with default parameters.

Figure \ref{fig:M1b} displays the location error, the predicted number of change points, and the runtime of each method, where the estimation error of a size zero change point estimate is ignored.  \WeightedERML{}  performs favourably compared to competing methods, particularly as the sample proportion $\delta = n/p$ grows, achieving near-perfect estimation for $\delta \geq 5$. Moreover, the runtime is comparable to those of \DCDP{} and \MOSEG{}.  Figure \ref{fig:M1b-hist} shows a histogram of predicted change points across $100$ trials for each method. \WeightedERML{} has a stronger concentration of estimates around the true change point locations compared to other methods.
\begin{figure}[!tp]
    \centering
    \begin{subfigure}[b]{0.32\textwidth}
        \centering
        \resizebox{\textwidth}{!}{\input{figures/synthetic-final/M1/location.pgf}}
        \label{fig:M1b-location}
    \end{subfigure}
    \hfill
    \begin{subfigure}[b]{0.32\textwidth}
        \centering
        \resizebox{\textwidth}{!}{\input{figures/synthetic-final/M1/size.pgf}}
        \label{fig:M1b-size}
    \end{subfigure}
    \hfill
    \begin{subfigure}[b]{0.32\textwidth}
        \centering
        \resizebox{\textwidth}{!}{\input{figures/synthetic-final/M1/runtime.pgf}}
        \label{fig:M1b-runtime}
    \end{subfigure}
    \caption{\small Comparison against \McScan{}, \DCDP{}, \DPDU{}, \MOSEG{} in the setting of two change points in the linear model with heavy-tailed noise, where regression vector entries are sampled independently from $0.3 \N(0, \delta) + 0.7 \delta_0$ and the change point prior assumes these are at least $n/20$ apart but otherwise uniformly distributed. Error bars indicate the $25$-th to $75$-th percentiles across $100$ trials.}
    \label{fig:M1b}
\end{figure}
\begin{figure}[!tp]
    \centering
    \begin{subfigure}[b]{0.9\textwidth}
        \centering
        \resizebox{\textwidth}{!}{\input{figures/synthetic-final/M1/chgpt_histograms.pgf}}
    \end{subfigure}
    \caption{\small Estimated change point locations (across 100 trials) for sparse linear model with heavy-tailed noise. Histograms show the distribution of predictions across methods for varying sampling ratios $\delta = n/p$ ($p=200$ fixed). Grey regions indicate the Gaussian kernel density estimate with bandwidth selected via $5$-fold cross-validation. True change points are shown as black dashed lines.}
    \label{fig:M1b-hist}
\end{figure}

\subsection{Linear model with sparse signal differences} \label{sec:lin-model-sparse-signal-diffs}
Next, we consider the linear model \eqref{eq:linear-chgpt-model} where the change in the signals across a change point (i.e. $\bbeta^{(\eta_\ell-1)} - \bbeta^{(\eta_\ell)}$) is  sparse. The \charcoal{} method  \cite{gao_sparse_2022} is designed for this setting, and \McScan{} \citep{cho_detection_2025} offers asymptotic performance guarantees under such an assumption. We fix two ground truth change points at $0.2n$ and $0.5n$ and for $i \in [n]$ sample $\varepsilon_i \distas{\text{i.i.d.}} \N(0, 0.1)$, $\x_i \distas{\text{i.i.d.}} \N(0, \I/n)$. We sample differentially-sparse regression vectors as follows:
\begin{align}
&\beta^{(1)}_j \distas{\text{i.i.d.}} \N(0, 8) \nonumber\\
&\beta^{(\ell)}_j =
\left\{
    \begin{array}{lr}
        \beta^{(\ell - 1)}_j, & \text{with probability } 1 - p_s\\
        \nu(\beta^{(\ell-1)}_j + w_\ell), & \text{with probability } p_s
    \end{array}
\right. , \quad \ell \in \{2, \dots, L\},
\label{eq:sparse_diff_prior_WERM}
\end{align}
where $p_s = 0.3$ and $w_\ell \distas{\text{i.i.d.}} \N(0, 400 \delta)$ create the sparse change between adjacent signals, and $\nu := \sqrt{\frac{8}{8 + 400 \delta}}$ is a rescaling factor that ensures uniform signal magnitude: $\E[(\beta_j^{(1)})^2]=\dots =\E[(\beta_j^{(L)})^2] =8$. In other words, two adjacent regression vectors differ in approximately a third of their coefficients by a Gaussian perturbation.
We evaluate \WeightedERM{} with $L = 4$, i.e. the method considers at most three change points, and take $M$ to be the squared error loss. Given a number of change points, \WeightedERM{} assumes a uniform prior distribution over change point locations such that they are at least $n/40$ apart and away from end points. We run \charcoal{} as described in the multiple change point scenario in \cite[Section $4.2.1$]{cho_detection_2025} for $n/p \leq 8.5$ due to prohibitive runtime, and run \McScan{} in \texttt{auto} mode. Figure \ref{fig:M2} shows the location error, number of change point estimates, and runtimes of all  the methods. The location error of \WeightedERM{} is consistently smaller than that of \charcoal{}, and smaller than that of \McScan{} for $n/p \geq 7.5$. In terms of the number of change points, \WeightedERM{} improves on \McScan{} and \charcoal{}. We note that \McScan{} and \WeightedERM{} have much smaller runtime than \charcoal{}. The histograms of change point estimates in Figure \ref{fig:M2-hist} indicate stronger concentration of the \WeightedERM{} estimates around the truth, compared to \charcoal{} and \McScan{}. We emphasize that \WeightedERM{} does not assume or exploit the sparse difference signal structure in the regression vectors. An additional comparison of \WeightedERM{} and \McScan{} for dense independent Gaussian signals is provided in Appendix \ref{sec:dense-gaussian-comparison}.
\begin{figure}[!tp]
    \centering
    \begin{subfigure}[b]{0.32\textwidth}
        \centering
        \resizebox{\textwidth}{!}{\input{figures/synthetic-final/M2/location.pgf}}
        \label{fig:M2-location}
    \end{subfigure}
    \hfill
    \begin{subfigure}[b]{0.32\textwidth}
        \centering
        \resizebox{\textwidth}{!}{\input{figures/synthetic-final/M2/size.pgf}}
        \label{fig:M2-size}
    \end{subfigure}
    \hfill
    \begin{subfigure}[b]{0.32\textwidth}
        \centering
        \resizebox{\textwidth}{!}{\input{figures/synthetic-final/M2/runtime.pgf}}
        \label{fig:M2-runtime}
    \end{subfigure}
    \caption{\small Comparison between \WeightedERM{} (\WERM{}) and other methods for a  linear model with sparse signal differences and two change points. Error bars indicate the $25$-th to $75$-th percentiles across $100$ trials.}
    \label{fig:M2}
\end{figure}
\begin{figure}[!tp]
    \centering
    \begin{subfigure}[b]{0.6\textwidth}
        \centering
        \resizebox{\textwidth}{!}{\input{figures/synthetic-final/M2/chgpt_histograms.pgf}}
    \end{subfigure}
    \caption{\small Estimated change point locations across 100 trials for the linear model with sparse signal differences. Histograms show the distribution of predictions across methods for varying sampling ratios $\delta = n/p$ ($p=200$ fixed). Grey regions indicate the Gaussian kernel density estimate with bandwidth selected via $5$-fold cross-validation. True change points are shown as black dashed lines. }
    \label{fig:M2-hist}
\end{figure}
\subsection{Logistic model with sparse signals} \label{sec:logistic-model-sparse-signals-exp}
Finally, we consider change point estimation in the logistic model \eqref{eq:logistic-chgpt-model_AMP} when the regression vectors are sparse. We fix two change points at $n/3$, $8n/15$, and sample $\x_i \distas{\text{i.i.d.}}  \N(0, \bSigma / n)$ for $i \in [n]$, with $\bSigma$ generated as follows. We pick an orthogonal matrix $\U$ uniformly at random, and eigenvalues $d_1, \dots, d_p$ i.i.d. from a chi-squared distribution with $10$ degrees of freedom. We then define $\boldsymbol{S} = \U^\top \D \U$, where $\D = \text{diag}(d_1, \dots, d_p)$. The covariance matrix $\bSigma$ is then obtained by dividing each entry $S_{kj}$ of $\boldsymbol{S}$ by $\sqrt{S_{kk} S_{jj}}$, for $k, j \in [p]$. We fix $n/p = \delta = 6.0$ and sample regression coefficients independently according to $0.1 \N(0, \delta \kappa^2) + 0.9 \delta_0$ while varying the signal strength parameter $\kappa^2$ from $10$ to $2000$. We evaluate \WeightedERML{} with $L = 4$ (at most three change points), and take $M$ to be the logistic loss $(a, b) \mapsto \log(1 + e^a) - ba$.
We recall that \WeightedERML{} incorporates a penalty term $\lambda \sqrt{2 (\log{p}) / n} \|\bbeta\|_1$ in \eqref{eq:ERM_est} in order to promote sparsity, comparable to the $\ell_1$ penalty term used in the \GLMBSA{} implementation \citep{wang_efficient_2023}. Moreover, we use cross-validation with the logistic loss as outlined at the start of Section \ref{sec:synthetic-experiments} to select the number of change points and the $\ell_1$ penalty constant in \WeightedERML{}, and to select the $\ell_1$ penalty constant used in \GLMBSA{} (initialized with the default parameter).
Given a number of change points, \WeightedERM{} assumes a uniform prior  over change point locations such that they are at least $n/10$ apart and away from end points. We run the \texttt{bsa\_chgpt} function from the \GLMBSA{} implementation (obtained from the authors) which allows us to pass in the maximum number of change points $L-1 = 3$ while using default parameters. In Figure \ref{fig:M3b}, we plot the location estimation error, the predicted number of change points, and the runtime for both methods. \WeightedERML{} produces favourable estimates to those of \GLMBSA{}, particularly in terms of location error and runtime.
\begin{figure}[!tp]
    \centering
    \begin{subfigure}[b]{0.32\textwidth}
        \centering
        \resizebox{\textwidth}{!}{\input{figures/synthetic-final/M3/delta6/location_boxplots.pgf}}
        \label{fig:M3b-location}
    \end{subfigure}
    \hfill
    \begin{subfigure}[b]{0.32\textwidth}
        \centering
        \resizebox{\textwidth}{!}{\input{figures/synthetic-final/M3/delta6/size_asymmetric_errors.pgf}}
        \label{fig:M3b-size}
    \end{subfigure}
    \hfill
    \begin{subfigure}[b]{0.32\textwidth}
        \centering
        \resizebox{\textwidth}{!}{\input{figures/synthetic-final/M3/delta6/runtime_boxplots.pgf}}
        \label{fig:M3b-runtime}
    \end{subfigure}
    \caption{\small Comparison between \WeightedERM{} (\WERM{}) and other methods for a logistic model with sparse signals and two change points. Error bars in the middle plot indicate the $25$-th to $75$-th percentiles across $100$ trials. Metrics on the left and right plots are jittered in dots with a low intensity, and the overall performance is summarized as a boxplot. }
    \label{fig:M3b}
\end{figure}
\section{Real data applications} \label{sec:real-data-experiments}
\subsection{Myocardial infarction data} \label{sec:myocardial}
We consider the myocardial infarction (MI) complications dataset studied in \citep{arpino_2025_jmlr}, originally obtained from \citep{misc_myocardial_infarction_complications_579}, containing the medical information of $n = 1700$  patients (samples) aged $26$-$92$ with MI complications.
Each sample has $p = 111$ medical features such as age, sex, heredity, and the presence of diabetes. The dataset also contains $12$ binary response variables for each patient relating to the state of the patient's overall heart health, indicating the presence of complications such as `Atrial Fibrillation' and `Chronic Heart Failure' (CHF). We investigate the relation between the binary CHF response  variable and the features of each patient, using the logistic model \eqref{eq:logistic-chgpt-model_AMP}.
We consider the presence of at most one change point ($L = 2$), and perform preprocessing identical  to that in  \cite[Section 4.3]{arpino_2025_jmlr}.
We use the data-driven approach outlined at the end of Section \ref{sec:uncertainty-quantification} for estimating the signal strength matrix $\bGamma$, with $M$ set to the squared error loss for numerical stability (see Appendix \ref{sec:numerical-challenges-in-the-logistic-model} for a discussion on numerical challenges in the logistic model, and Appendix \ref{sec:numerical-implementation-details} for additional numerical implementation details).
We obtain the rounded estimate $\hat{\bGamma} = \operatorname{diag}(0.0043,0.0040)$. Solving the fixed point equations \eqref{eq:b-fp-implicit}--\eqref{eq:cross-cov-fp-implicit} at this estimate gives an error with Frobenius norm $0.013$ for \eqref{eq:theta-deb-fp-eqn}.

\begin{figure}[!tp]
    \centering
    \begin{subfigure}[b]{0.8\textwidth}
        \centering
        \includegraphics[width=\dimexpr\textwidth-2\fboxrule\relax]{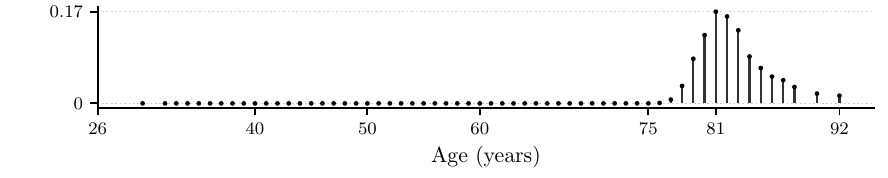}
    \end{subfigure}
    \caption{\small Posterior distribution over a single change point in myocardial infarction data, referenced in Section \ref{sec:myocardial}.}
    \label{fig:myocardial_chgpts}
\end{figure}

We set $M$ to the logistic loss $(a, b) \mapsto \log(1 + e^a) - ba$, and place a uniform prior over the number of change points ($0$ or $1$, each with probability $\frac{1}{2}$), and a uniform distribution on the location conditional on a change point being present.
This yields sample weights identical to those in Figure \ref{fig:logistic-example-match}.
Using \eqref{eq:posterior_convergence_pres}, we compute the estimated posterior $p(\bpsi| \TTheta^{\adj}, \y)$, which assigns probability $0.607$ to the presence of a change point. Conditioning on a single change point, Figure \ref{fig:myocardial_chgpts} shows the posterior over its location. The mass is concentrated between ages $75$ and $92$, peaking at $81$, locating a possible change in how CHF relates to features such as heredity, sex, and diabetes within this range. This aligns with prior medical findings: patients with CHF aged over $75$ are more often female and exhibit fewer cardiovascular morbidities and risk factors than those aged $55$ or younger \citep{Azad2014-ky, tromp_age_2021}. The posterior in Figure \ref{fig:myocardial_chgpts} resembles a shifted version of that in \citep{arpino_2025_jmlr}, which was derived using an Approximate Message Passing estimator with a Gaussian prior on regression coefficients. Conditional on a change point, both results place it later in life, consistent with the aforementioned medical literature.
In Appendix \ref{sec:myocardial-sensitivity} we investigate the sensitivity of the posterior probability of a change point to perturbations of $\hat{\bGamma}$.
\subsection{Macroeconomic data} \label{sec:FRED}
The FRED-MD database is a large collection of monthly macroeconomic data maintained by the Federal Reserve Bank of St. Louis \citep{McCracken01102016}. We consider the task (investigated in \citep{cho_detection_2025}) of detecting change points in FRED-MD data within the period from January $1960$ to May $2024$, for a factor-augmented forecasting model for the growth rate of the industrial production total index (INDPRO). We perform the same preprocessing steps as  \cite{cho_detection_2025}, yielding a dataset with $p = 114$ features, $6$ of which are explicitly constructed factors, and $n = 773$ monthly samples.

We evaluate both the \WeightedERM{} and \WeightedERML{} methods with base loss $M$ set to the Huber loss \eqref{eq:Huber-loss}, with maximum number of change points $L-1$ set to $4$, and with a uniform prior over change point locations given a number of change points.
\WeightedERM{} yields no change points, while \WeightedERML{} yields two change points in
September $1980$ and in June $2009$.
With possible estimation bias, the first change point can be seen to follow the impactful `Volcker shock' of $1979$ \citep{GOODFRIEND2005981}, and the second change point follows the
$2008$ global financial crisis. These are comparable to the February $1974$ and May $2010$ change points detected by \McScan{} in \citep{cho_detection_2025}. \cite{cho_detection_2025} identify an additional change point at May $2019$ where two adjacent $114$-dimensional regression vectors reportedly differ by only two entries. We highlight this as an interesting finding and consider the adaptation of our method to highly sparse signal differences as a direction for future work.
\section{Discussion} \label{sec:discussion}

In this work, we proposed  \WeightedERM{}, which modifies standard estimators via sample weighting to identify change points in high-dimensional GLMs. Its strong performance against state of the art methods, even with weakly informative priors on the change points, was demonstrated via  empirical studies on synthetic and real data. We established precise asymptotic performance guarantees for general Gaussian designs, and used the theory to construct an efficiently computable posterior distribution over change point locations.

Our theory has a few limitations in that it does not fully cover the range of scenarios in which $\WeightedERM{}$ is  effective. It is of interest to extend the theoretical guarantees to $\WeightedERM{}$ with regression penalties such as $\ell_1$-regularization to promote sparsity, and with cross-validation as presented in Section \ref{sec:synthetic-experiments}. Other generalizations, which require  further development of the approximate message passing proof technique, include  allowing for the number of change points to grow with $n$, and different scaling regimes of $n$ vs. $p$.

Moreover, the theoretical framework may be extended to broader classes of random covariates, including rotationally invariant designs, which can be analysed via variants of approximate message passing (AMP) and their connections to convex optimization \citep{ li2023random, li2023spectrum, zhang2026orthogonallyinvariant}. Another important direction is to handle temporally dependent data, where the rows of $\X$ form a high-dimensional time series. Related settings have recently been studied in high-dimensional change point regression \citep{xu_change-point_2024} and in AMP-based analyses of linear regression \citep{tieplova_information_2026}.
\label{page:theory-growing-L}

\section{Acknowledgements}
Part of this work was done while G. Arpino was a doctoral student at the University of Cambridge, where he was supported in part by a Cambridge Trust Scholarship and by a  Cambridge Philosophical Society Research Studentship.
R. Venkataramanan was supported in part by an EPSRC Mathematical Sciences Small Grant.

\section*{Statement on AI use}
In the revised version of this paper, we used Claude Opus 5.5  and ChatGPT Astra 6 to formulate and prove Proposition \ref{prop:ls-changepoints} and the associated example illustrating the asymptotic  bias; to prove the validity of the theoretical guarantees for the logistic model presented in Appendix \ref{sec:logistic-guarantees}; to develop code for figures in Appendix \ref{sec:additional-experiments-and-examples}; and for  proofreading. The first version of this paper used earlier versions of ChatGPT and Claude to convert the reference code to Python and R packages.

{\small{
\bibliographystyle{plainnat}
\bibliography{reference}
}}

\newpage

\appendix

\section{\WeightedERM{} via likelihood relaxation} \label{sec:motivation-via-jensen}
We derive the \WeightedERM{} estimator described in Section \ref{sec:methodology} via a relaxation of the likelihood function using Jensen's inequality. Given data $(\X, \y)$ and a prior over signal configuration vectors $\pi_{\bar{\bPsi}}$, the likelihood   $\sp(\y | \X, \B) = \sum_{\bpsi} \pi_{\bar{\bPsi}}(\bpsi) \sp(\y | \X, \B, \bpsi)$. It is natural to consider maximum-likelihood estimators for $\B$ taking on the form:
\begin{align}
    \hat{\B} = \argmin_{\tilde{\B} \in \reals^{p \times L}} -\log \sp(\y | \X, \tilde{\B}). \label{eq:max-likelihood-motivation}
\end{align}
Assuming separability of the likelihood across columns of $\B$ and rows of $(\X, \y)$ conditional on $\bpsi$, i.e. $\sp(\y | \X, \tilde{\B}, \bpsi) = \Pi_{i = 1}^n \sp(y_i | \x_i, \tilde{\B}_{[:, \psi_i]})$, we apply Jensen's inequality to obtain:
\begin{align*}
    -\log \sp(\y | \X, \tilde{\B})
    &= -\log \sum_{\bpsi} \pi_{\bar{\bPsi}}(\bpsi) \Pi_{i=1}^n \sp(y_i | \x_i, \tilde{\B}_{[:, \psi_i]}) \\
    & \leq \sum_{\bpsi} \pi_{\bar{\bPsi}}(\bpsi) \sum_{i=1}^n \left\{ - \log \sp(y_i | \x_i, \tilde{\B}_{[:, \psi_i]}) \right\} \\
    & = \sum_{\bpsi} \pi_{\bar{\bPsi}}(\bpsi) \sum_{\ell = 1}^L \sum_{i=1}^n \left\{ - \log \sp(y_i | \x_i, \tilde{\B}_{[:, \psi_i]}) \ind_{\ell = \psi_i} \right\} \\
    &= \sum_{\ell = 1}^L \sum_{i = 1}^n \left( \sum_{\bpsi} \pi_{\bar{\bPsi}}(\bpsi) \left\{ - \log \sp(y_i | \x_i, \tilde{\B}_{[:, \psi_i]}) \ind_{\ell = \psi_i}\right\} \right) \\
    &= \sum_{\ell = 1}^L \sum_{i = 1}^n \left( \sum_{\bpsi: \psi_i = \ell} \pi_{\bar{\bPsi}}(\bpsi) \right) \left\{ - \log \sp(y_i | \x_i, \tilde{\B}_{[:, \ell]}) \right\}.
\end{align*}
We then associate $\left( \sum_{\bpsi: \psi_i = \ell} \pi_{\bar{\bPsi}}(\bpsi) \right)$ with $\pi_i^{(\ell)}$ as per \eqref{eq:pi_i_defn} and associate $M(\x_i^\top \bbeta, y_i)$ with $- \log \sp(y_i | \x_i, \tilde{\B}_{[:, \ell]}, \ell)$ to obtain our proposed set of estimators \eqref{eq:ERM_est}. This argument frames step \eqref{eq:ERM_est} as a relaxation of the maximum likelihood estimator for $\B$ in the presence of a prior distribution over change points, and resembles existing mean field and variational methods for inference in the presence of complex dependence structure \citep{wainwright_graphical_2008}.
\section{Preliminaries for proofs} \label{sec:preliminaries-for-proofs}

\paragraph{Convergence of empirical distributions}
Given a matrix $\A \in \reals^{p \times L}$, the empirical distribution of the rows of $\A$ is the measure $\mu_p(A) := \frac{1}{p}|\{ \A_j \in A: j \in [p]\}|$ for any measurable set $A \subseteq \reals^L$. We say that the empirical distribution of the entries of $\A$ converges weakly to distribution $\P_{\bar{\A}}$ if $ \frac{1}{p}\sum_{j=1}^p f(\A_j) \to \E_{\P_{\bar{\A}}}[f(\bar{\A})]$ as $p \to \infty$ for all bounded continuous functions $f: \reals^L \to \reals$. We say that the empirical distribution of the rows of $\A$ converges in Wasserstein-$2$ distance to distribution $\P_{\bar{\A}}$, or $d_2(\mu_p, \P_{\bar{\A}}) \to 0$, if there exists a sequence of couplings $\Pi_p$ with marginals $\mu_p$ and $\P_{\bar{\A}}$ respectively, so that if $(W_p, W) \distas{} \Pi_p$, then $\E[(W_p - W)^2] \to 0$ as $p \to \infty$.

\paragraph{\textbf{Pseudo-Lipschitz Functions}}
Our results are stated in terms of uniformly \textit{pseudo-Lipschitz} functions \citep{berthier_state-evolution_2019}. For $C > 0$ and $r \in [1, \infty)$, let $\PL_{n, m, q}(r, C)$ be the set of functions $\phi: \reals^{n \times q} \to \reals^{m \times q}$ such that for all $\x, \tilde{\x} \in \reals^{n\times q}$, we have
\begin{align}
\frac{\|\phi(\x) - \phi(\tilde{\x})\|_F}{\sqrt{m}} \leq C\left(1 + \left(\frac{\|\x\|_F}{\sqrt{n}}\right)^{r - 1} + \left(\frac{\|\tilde{\x}\|_F}{\sqrt{n}}\right)^{r - 1}\right)\frac{\|\x - \tilde{\x} \|_F}{\sqrt{n}}.
\label{eq:PL_def}
\end{align}
Note that $\cup_{C > 0} \PL_{n, m, q}(r_1, C) \subseteq \cup_{C > 0} \PL_{n, m, q}(r_2, C)$ for any $1 \leq r_1 \leq r_2$. A function $\phi \in \PL_{n, m, q}(r, C)$ is called pseudo-Lipschitz of order $r$. A family of pseudo-Lipschitz functions is said to be \textit{uniformly} pseudo-Lipschitz if all functions of the family are pseudo-Lipschitz with the same order $r$ and the same constant $C$.  For $\x, \y \in \reals^n$, the mean squared error $\phi(\x,\y)= (\x - \y)^\top (\x - \y) /n$ and the normalized  squared correlation $\phi(\x, \y)=|\x^\top \y |/n$ are examples of uniformly pseudo-Lipschitz functions of order $2$.

\subsection{Properties of proximal operators} \label{sec:prop-prox-op}
We collect some mathematical identities for proximal operators involving $M$  used in our proofs, which  hold under Assumption~\ref{asmpt:reg-loss}(\ref{cond:strict-cvx-smooth}). Recalling the definition of the proximal operator in \eqref{eq:prox_op_first_defn}, we will use the associated proximal operator for $M$:
\begin{align}
\label{eq:proxM}
   \prox_{b M(\cdot, v)}(u) := \argmin_{t \in \reals} \{ b M(t, v) + (t - u)^2/2\},
\end{align}
for $u, v \in \reals$. For $b = 0$, we define $\prox_{b M(\cdot, v)}(u) := u$.
\begin{remark} \label{rmk:prox-implicit-function-theorem-and-b}
By the Implicit Function Theorem \cite[Theorem 9.28]{rudin1976principles}, the proximal operator in \eqref{eq:proxM} is a differentiable function of $u,b$ whenever $1 + b M''(\prox_{b M(\cdot, v)}(u), v) \neq 0$, which is satisfied by convexity of $M$. Together with Proposition \ref{prop:prox-b-to-0} below, this implies that $\prox_{b M(\cdot, v)}(u)$ is continuous with respect to $b \in [0, \infty)$ whenever $M$ is convex.
\end{remark}

By differentiating the proximal objective, we notice that the proximal operator satisfies:
\begin{align}
    b M'(\prox_{b M(\cdot, v)}(u), v) + \prox_{b M(\cdot, v)}(u) - u = 0, \label{eq:prox_id_1}
\end{align}
which yields our first identity:
\begin{align}
   \prox_{b M(\cdot, v)}(u) - u = - b M'(\prox_{b M(\cdot, v)}(u), v). \label{eq:prox_id_2}
\end{align}
Differentiating \eqref{eq:prox_id_1}, we obtain:
\begin{align*}
   b M''(\prox_{b M(\cdot, v)}(u), v) \cdot \prox'_{bM(\cdot, v)}(u) + \prox'_{bM(\cdot, v)}(u) - 1 = 0,
\end{align*}
which yields the identity:
\begin{align}
    \prox'_{bM(\cdot, v)}(u) = \left(1 + b M''(\prox_{bM(\cdot, v)}(u), v) \right)^{-1}. \label{eq:prox_id_3}
\end{align}
Applying a derivative $\partial_b$ with respect to $b$ and the chain rule to \eqref{eq:prox_id_2}, we obtain the identity:
\begin{align}
    \partial_b \prox_{b M(\cdot, v)}(u) = -\frac{M'(\prox_{bM(\cdot, v)}(u), v)}{1 + b M''(\prox_{bM(\cdot, v)}(u), v)}. \label{eq:prox_id_4}
\end{align}

Letting $v := s(u)$ in \eqref{eq:prox_id_1} for some function $s: \reals \to \reals$, letting $\partial_2$ denote the partial derivative with respect to the second argument, and taking the derivative $\partial_u$ with respect to $u \in \reals$ we obtain:
\begin{align*}
   &b M''(\prox_{bM(\cdot, s(u))}(u), s(u)) \partial_u\prox_{b M(\cdot, s(u))}(u) \\
   &\quad + b \partial_2 M' (\prox_{b M(\cdot, s(u))}(u), s(u)) s'(u) + \partial_u \prox_{b M(\cdot, s(u))}(u) - 1 = 0,
\end{align*}
yielding the identity:
\begin{align}
    \partial_u \prox_{b M(\cdot, s(u))}(u) = \frac{1 - b \partial_2 M'(\prox_{bM(\cdot, s(u))}(u), s(u)) s'(u)}{1 + b M''(\prox_{b M(\cdot, s(u))}(u), s(u))}. \label{eq:prox_id_5}
\end{align}
\begin{proposition} \label{prop:prox-b-to-0}
Let $u, v \in \reals$ and assume $M(\cdot, v) : \reals \to \reals$ is convex and differentiable.
Then,
\begin{align*}
    \lim_{b \to 0^+} \prox_{b M(\cdot, v)}(u) = u.
\end{align*}
\end{proposition}
\begin{proof}
Let $x_b := \prox_{b M(\cdot, v)}(u)$, and notice that by optimality with respect to the proximal objective, the following inequality holds:
\begin{align*}
   b M(x_b, v) + (x_b - u)^2 / 2 \leq b M(u, v).
\end{align*}
By convexity of $M(\cdot, v)$, we have that:
\begin{align*}
    M(u, v) - M(x_b, v) \leq M'(u, v) (u - x_b).
\end{align*}
Combining the above inequalities, we obtain:
\begin{align*}
    (x_b - u)^2 / 2 \leq b M'(u, v) (u - x_b),
\end{align*}
which gives
   $ |x_b - u| \leq \sqrt{2} b |M'(u, v)|$.
Letting $b \to 0^+$, we see that  $x_b = \prox_{b M(\cdot, v)}(u) \to u$.
\end{proof}
\begin{proposition} \label{prop:prox-deriv-wrt-y}
Let $u, v \in \reals, b \in \reals_{>0}$ and assume $M: \reals \times \reals \to \reals$ is twice differentiable with continuous second derivative $M'' \geq 0$ and continuous cross derivative $\partial_{12} M$. We have that:
\begin{align*}
    \partial_v \prox_{b M(\cdot, v)}(u) = -\frac{b \partial_{12} M(\prox_{b M(\cdot, v)}(u), v)}{b M''(\prox_{b M(\cdot, v)}(u), v) + 1},
\end{align*}
and consequently, $v \mapsto \prox_{b M(\cdot, v)}(u)$ is Lipschitz continuous with Lipschitz constant
\[\sup_{v \in \reals}  \, \left|\frac{b \partial_{12} M(\prox_{b M(\cdot, v)}(u), v)}{b M''(\prox_{b M(\cdot, v)}(u), v) + 1} \right|.\]
\end{proposition}
\begin{proof}
Having fixed $u \in \reals, b \in \reals_{> 0}$, we abbreviate $\prox_{b M(\cdot, v)}(u)$ as $p(v)$. First, for $v_1, v_2 \in \reals$, notice that by \eqref{eq:prox_id_2} we have:
\begin{align*}
    |M'(p(v_1), v_1) - M'(p(v_2), v_2)| = \frac{1}{b}|p(v_1) - p(v_2)|.
\end{align*}
Define the objective function $J(x, v) = bM(x, v) + \frac{1}{2} (x - u)^2$ associated with the proximal operator, with first, second, and cross derivatives given by:
\begin{align*}
    &J'(x, v) = b M'(x, v) + (x - u), \\
    &J''(x, v) = b M''(x, v) + 1, \\
    &\partial_{12} J(x, v) = b \partial_{12} M(x, v).
\end{align*}
By the implicit function theorem \cite[Theorem 9.28]{rudin1976principles}, we have:
\begin{align*}
    p'(v) = -(J''(p(v), v))^{-1} \partial_{12} J(p(v), v) = -\frac{b \partial_{12} M(p(v), v)}{b M''(p(v), v) + 1}.
\end{align*}
The proposition statement follows by taking the absolute value on both sides and by applying the mean value theorem.
\end{proof}
\begin{proposition} \label{prop:M_prime_ub}
Let $u, v \in \reals, b \in \reals_{>0}$ and assume $M: \reals \times \reals \to \reals$ is twice differentiable with continuous second derivative $M'' \geq 0$ such that $M'' \leq B_{11}$ and continuous cross derivative $\partial_{12} M$ such that $|\partial_{12} M| \leq B_{12}$, for some non-negative constants $B_{11}, B_{12}$. We have that:
\begin{align}
    \left| M'\left( \prox_{b M(\cdot, v)}(u), v \right)\right| \leq |M'(0, 0)| + B_{11}|u| + B_{12}|v|.
\end{align}
\end{proposition}
\begin{proof}
By the fundamental theorem of calculus, we have that:
\begin{align*}
    &M'\left( \prox_{b M(\cdot, v)}(u), v \right) \\
    &= M'(0, 0) \\
    &\qquad + \int_{0}^1 \left( M''(t \prox_{b M(\cdot, v)}(u), tv) \prox_{b M(\cdot, v)}(u) + \partial_{12} M(t \prox_{b M(\cdot, v)}(u), tv) v \right) dt.
\end{align*}
Further applying \eqref{eq:prox_id_2} we obtain:
\begin{align*}
    &M'\left( \prox_{b M(\cdot, v)}(u), v \right)
    = M'(0, 0) + \int_{0}^1 \left( M''(t \prox_{b M(\cdot, v)}(u), tv) (u
 \right. \\
  &\hspace{2.5cm} \left. - \,  b M'(\prox_{b M(\cdot, v)}(u), v)) + \partial_{12} M(t \prox_{b M(\cdot, v)}(u), tv) v \right) dt.
\end{align*}
We rearrange to further obtain the equation:
\begin{align*}
    &M'\left( \prox_{b M(\cdot, v)}(u), v \right)
    = \left(1 + b \int_{0}^1 M''(t \prox_{b M(\cdot, v)}(u), tv) dt \right)^{-1} \cdot \Big( M'(0, 0)
    \\
    &\hspace{2cm}  \left. +  \int_{0}^1 \left( M''(t \prox_{b M(\cdot, v)}(u), tv) u + \partial_{12} M(t \prox_{b M(\cdot, v)}(u), tv) v \right) dt \right),
\end{align*}
which, after recalling that $b \geq 0, M'' \geq 0$ and after applying the triangle inequality, leads to the desired upper bound.
\end{proof}
{
\section{Proofs for linear model with squared loss} \label{sec:proof-linear-model}}

Throughout, $M(u, v) = (u - v)^2$, $q(z, v) = z + v$ and $\bSigma = \I$. For a sequence $(f_i)_{i \ge 1}$ we repeatedly use the identity
\begin{align}
    \frac{1}{n}\sum_{i=1}^n f_i = \int_0^1 f_{\floor{nt}+1}\, \de t,
    \label{eq:ls-riemann}
\end{align}
so that, by dominated convergence, $\frac{1}{n}\sum_{i} g(\pi_i^{(\ell)}, \Psi_i, i/n) \to \int_0^1 g(\Phi^{(\ell)}(t), \bar\psi(t), t)\, \de t$ for any bounded $g$ that is continuous in its first argument, using Assumption \ref{asmpt:reg-weights} and $\Psi_{\floor{nt}+1} = \bar\psi(t)$ for all $t \in [0,1)\setminus\{\alpha_1, \dots, \alpha_{L^*-1}\}$ and $n$ large enough. We also use the weak law of large numbers for weighted sums: if $c_{n,i}$ are deterministic with $|c_{n,i}| \le C$, then $\frac{1}{n}\sum_i c_{n,i}(\varepsilon_i^2 - \sigma^2) \xrightarrow{\P} 0$, which follows from the weak law of large numbers for triangular arrays \citep[Section~2.2]{durrett2019probability}, since  $\E\bar\varepsilon^2 < \infty$.

\subsection{Proof of Proposition \ref{prop:ls-B-Theta}} \label{subsec:ls-B-Theta_proof}
We first check that the assumptions of Theorem \ref{thm:ERM_general_characterization_pres} hold. The link $q(z, v) = z + v$ is Lipschitz. For Assumption \ref{asmpt:reg-loss}: $M'' = 2$, $M''' = \partial_2 M'' = 0$ and $\partial_{12} M = -2$ are bounded, $(M'')^2 - M' M''' = 4 > 0$, and $b M'' = 2b \to \infty$, giving parts (\ref{cond:strict-cvx-smooth}), (\ref{eq:loss-asmpt-log-conc}) and (\ref{eq:loss-asmpt-b-M-to-inf}). For part (\ref{cond:likelihood_curvature}), we  note that $\nabla^2 \cC^{(\ell)} = 2\X^\top \boldsymbol{\Pi}^{(\ell)}\X$ does not depend on $\bbeta$. Choose $c > 0$ such that the set $S_c := \{t : \Phi^{(\ell)}(t) \geq 2c\}$ has length greater than $1/\delta$, which is possible since $\Phi^{(\ell)} > 0$ almost everywhere. Then $\nabla^2 \cC^{(\ell)} \succeq 2c \sum_{i : \pi_i^{(\ell)} \geq c} \x_i \x_i^\top$, and the number of such $i$ exceeds $(1 + \epsilon) p$ for some $\epsilon > 0$ and all large $n$. Hence, the smallest eigenvalue is bounded away from zero with probability tending to one \citep{bai1993limit}. This also gives part (\ref{eq:existence-of-estimator}).

Next, we show that \eqref{eq:ls-b-linear-model}--\eqref{eq:ls_sll} solve the general equations \eqref{eq:b-fp-implicit}--\eqref{eq:cross-cov-fp-implicit}. For $c > 0$,
\begin{align}
    \prox_{c M(\cdot, y)}(z) = \frac{z + 2 c y}{1 + 2 c},
    \qquad
    M'\big(\prox_{c M(\cdot, y)}(z), y\big) = \frac{2 (z - y)}{1 + 2c}.
    \label{eq:ls-prox}
\end{align}
In \eqref{eq:sM-def} we have $c = b^{(\ell)} \pi_i^{(\ell)}$, $z = \Z_i^\top \blambda^{(\ell)} + w_i^{(\ell)}$ and $y = q(\Z_i, \Psi_i, \varepsilon_i) = \Z_i^\top \e_{\Psi_i} + \varepsilon_i$. Hence, with $\omega_i^{(\ell)}$ as in \eqref{eq:omega_ell_i},
\begin{align*}
    (\sM_i^{(\ell)})'' = 2 \pi_i^{(\ell)}, \qquad
    (\sM_i^{(\ell)})' = \frac{\omega_i^{(\ell)}}{b^{(\ell)}} \Big( \Z_i^\top (\blambda^{(\ell)} - \e_{\Psi_i}) + w_i^{(\ell)} - \varepsilon_i \Big).
\end{align*}
Since $(\sM_i^{(\ell)})''$ is deterministic, \eqref{eq:b-fp-implicit} is \eqref{eq:ls-b-linear-model}. Since $\Z_i$ has mean zero and is independent of $w_i^{(\ell)}$, $\E[\Z_i (\sM_i^{(\ell)})'] = \frac{\omega_i^{(\ell)}}{b^{(\ell)}} \bGamma (\blambda^{(\ell)} - \e_{\Psi_i})$. As $\bGamma$ is invertible, \eqref{eq:mu-fp-implicit} is equivalent to $\blambda^{(\ell)} \lim_n \frac{1}{n}\sum_i \omega_i^{(\ell)} = \lim_n \frac{1}{n}\sum_i \omega_i^{(\ell)} \e_{\Psi_i}$. By \eqref{eq:ls-b-linear-model}, $\frac{1}{n}\sum_i \omega_i^{(\ell)} = 1 - \frac{1}{n}\sum_i (1 + 2b^{(\ell)}\pi_i^{(\ell)})^{-1} \to 1/\delta$, which gives \eqref{eq:ls-lambda-linear-model}. Finally, $\Z_i$ and $(w_i^{(\ell)}, w_i^{(\ell')})$ are independent and centred, with $\E[w_i^{(\ell)} w_i^{(\ell')}] = \kappa_{\ell,\ell'}$, so
\begin{align*}
    b^{(\ell)} b^{(\ell')}\, \E\big[(\sM_i^{(\ell)})' (\sM_i^{(\ell')})'\big] = \omega_i^{(\ell)}\omega_i^{(\ell')} \Big[ (\blambda^{(\ell)}-\e_{\Psi_i})^\top\bGamma(\blambda^{(\ell')}-\e_{\Psi_i}) + \kappa_{\ell,\ell'} + \varepsilon_i^2 \Big].
\end{align*}
Averaging over $i$ and using the weak law of large numbers for weighted sums to replace $\varepsilon_i^2$ by $\sigma^2$, \eqref{eq:sigma-fp-implicit} and \eqref{eq:cross-cov-fp-implicit} become the linear equation $\kappa_{\ell,\ell'} = A_{\ell,\ell'} + s_{\ell,\ell'}(\kappa_{\ell,\ell'} + \sigma^2)$, where $A_{\ell,\ell'}$ is the limit in \eqref{eq:ls-kappa-linear-model} and $s_{\ell,\ell'}$ is given by \eqref{eq:ls_sll}. Since $s_{\ell,\ell'} < 1$, its unique solution is \eqref{eq:ls-kappa-linear-model}. The limits exist by \eqref{eq:ls-riemann}, and by Proposition \ref{prop:uniqueness_of_fixed_points} this is the unique solution of the general equations.

Theorem \ref{thm:ERM_general_characterization_pres} with $\L = \I$ now gives \eqref{eq:ls-B-rep}. For \eqref{eq:ls-Theta-rep}, by \eqref{eq:bold-prox-def} and \eqref{eq:ls-prox}, the $(i,\ell)$ entry of $\bprox_{\b \bpi M(\cdot, q(\Z, \bPsi, \bvarepsilon))}(\Z\bLambda + \G_{\TTheta})$ is
\[
\frac{\Z_i^\top\blambda^{(\ell)} + (\G_{\TTheta})_{i,\ell} + 2b^{(\ell)}\pi_i^{(\ell)}\tilde{y}_i}{1 + 2b^{(\ell)}\pi_i^{(\ell)}}
= \big(1-\omega_i^{(\ell)}\big)\big(\Z_i^\top\blambda^{(\ell)} + (\G_{\TTheta})_{i,\ell}\big) + \omega_i^{(\ell)}\tilde{y}_i = \tilde{\Theta}_{i,\ell}.
\]
Therefore \eqref{eq:SE_ERM_Theta} reduces to \eqref{eq:ls-Theta-rep} in the linear model setting. For \eqref{eq:ls-mse}, apply \eqref{eq:ls-B-rep} with the order-2 pseudo-Lipschitz function $\varphi_p(\hat{\B}; \B) = \frac{1}{p}\|\hat{\bbeta}^{(\ell)} - \bbeta^{(r)}\|^2$. Since $\G_{\hat{\B}}$ is centred and independent of $\B$,
\[
\E\, \frac{1}{p}\big\|\B(\blambda^{(\ell)} - \e_r) + \sqrt{\delta}\, (\G_{\hat{\B}})_{:,\ell}\big\|^2 = \frac{n}{p}\, (\blambda^{(\ell)} - \e_r)^\top \frac{\B^\top \B}{n} (\blambda^{(\ell)} - \e_r) + \delta\, \kappa_{\ell,\ell} \, ,
\]
which converges to the right side of \eqref{eq:ls-mse} since $n/p \to \delta$ and $\B^\top\B/n \to \bGamma$. \hfill $\square$

\subsection{Proof of Proposition \ref{prop:ls-changepoints}}
\label{subsec:ls-changepoints}
\emph{Part (a)}: We first show pointwise convergence, i.e. $R_n(\tau) \xrightarrow{\P} F(\tau)$ for each $\tau \in \mathcal{T}_L$, and then prove the uniform convergence claim of part (a). Fix $\tau \in \mathcal{T}_L$. Since $y_i = (\X\B)_{i,\Psi_i} + \varepsilon_i$, we have $R_n(\tau) = \varphi_n(\hat{\TTheta}, \X\B; \bPsi, \bvarepsilon)$ with
\[
\varphi_n(\TTheta, \Z; \bPsi, \bvarepsilon) := \frac{1}{n}\sum_{i=1}^n \sum_{\ell=1}^L \ind\{\tau_{\ell-1} < i/n \le \tau_\ell\}\, u_{i,\ell}^2, \qquad u_{i,\ell} := Z_{i,\Psi_i} + \varepsilon_i - \Theta_{i,\ell}.
\]
We check that $\varphi_n$ is uniformly pseudo-Lipschitz of order $2$ in the sense of \eqref{eq:PL_def}. Let $\boldsymbol{a} = (\TTheta, \Z)$ and $\boldsymbol{a}' = (\TTheta', \Z')$ be two points of $\reals^{n \times 2L}$, with residuals $u_{i,\ell}$ and $u'_{i,\ell}$. The Cauchy--Schwarz inequality then gives
\begin{equation}
|\varphi_n(\boldsymbol{a}) - \varphi_n(\boldsymbol{a}')| \le \frac{1}{n} \sum_{i, \ell} |u_{i,\ell} - u'_{i,\ell}|\, |u_{i,\ell} + u'_{i,\ell}| \le \frac{1}{n}\, \|\u - \u'\|_F \big( \|\u\|_F + \|\u'\|_F \big).
\label{eq:ls-pl-step}
\end{equation}
Since $u_{i,\ell} - u'_{i,\ell} = (Z - Z')_{i, \Psi_i} - (\Theta - \Theta')_{i,\ell}$ we have
\[
\|\u - \u'\|_F^2 \le 2L\, \|\Z - \Z'\|_F^2 + 2\, \|\TTheta - \TTheta'\|_F^2 \le 2L\, \|\boldsymbol{a} - \boldsymbol{a}'\|_F^2 .
\]
Similarly, writing $u_{i,\ell} = (Z_{i,\Psi_i} - \Theta_{i,\ell}) + \varepsilon_i$, we have $\|\u\|_F \le \sqrt{2L}\, \|\boldsymbol{a}\|_F + \sqrt{L}\, \|\bvarepsilon\|$, and likewise for $\u'$. Substituting into \eqref{eq:ls-pl-step},
\[
|\varphi_n(\boldsymbol{a}) - \varphi_n(\boldsymbol{a}')| \le C \left( 1 + \frac{\|\boldsymbol{a}\|_F}{\sqrt{n}} + \frac{\|\boldsymbol{a}'\|_F}{\sqrt{n}} \right) \frac{\|\boldsymbol{a} - \boldsymbol{a}'\|_F}{\sqrt{n}},
\quad C := 2\sqrt{2}\, L \max\Big\{ 1, \sup_n \frac{\|\bvarepsilon\|}{\sqrt{n}} \Big\}.
\]
The constant $C$ does not depend on either $n$ or  $\tau$ (the weights $c_{i,\ell}$ enter only through $|c_{i,\ell}| \le 1$), and it is finite because the Wasserstein-$2$ convergence of the empirical distribution of $\bvarepsilon$ (Section \ref{sec:setting}) implies $\|\bvarepsilon\|^2/n \to \sigma^2$. Hence $\{\varphi_n\}$ is uniformly pseudo-Lipschitz of order $2$.

By \eqref{eq:ls-Theta-rep}, $R_n(\tau) \stackrel{\P}{\simeq} \E\, \varphi_n(\tilde{\TTheta}, \Z; \bPsi, \bvarepsilon)$. Here $\tilde{y}_i - \tilde{\Theta}_{i,\ell}$ is given by \eqref{eq:ls-residuals}; since $\Z_i$ and $(\G_{\TTheta})_i$ are centred and independent,
\[
\E\, \varphi_n(\tilde{\TTheta}, \Z; \bPsi, \bvarepsilon) = \frac{1}{n}\sum_{i=1}^n\sum_{\ell=1}^L c_{i,\ell}\, \frac{(\e_{\Psi_i}-\blambda^{(\ell)})^\top\bGamma(\e_{\Psi_i}-\blambda^{(\ell)}) + \kappa_{\ell,\ell} + \varepsilon_i^2}{\big(1 + 2b^{(\ell)}\pi_i^{(\ell)}\big)^2}.
\]
We use the weak law of large numbers for weighted sums to replace the $\varepsilon_i^2$ by $\sigma^2$ (at a cost of $o_{\P}(1)$). Then \eqref{eq:ls-riemann} shows that the right-hand side converges to $F(\tau)$. Hence $R_n(\tau) \xrightarrow{\P} F(\tau)$ for each $\tau \in \mathcal{T}_L$.

We now show that this convergence is uniform over $\tau \in    \mathcal{T}_L$. For $\ell \in [L]$ and $t \in [0, 1]$, let $P_\ell(t) := \frac{1}{n}\sum_{i \le nt} (y_i - \hat{\theta}_i^{(\ell)})^2$ and $\bar{P}_\ell(t) := \int_0^t v_\ell(s)\, \de s$. Then
\[
R_n(\tau) = \sum_{\ell=1}^L \big[P_\ell(\tau_\ell) - P_\ell(\tau_{\ell-1})\big], \qquad F(\tau) = \sum_{\ell=1}^L \big[\bar{P}_\ell(\tau_\ell) - \bar{P}_\ell(\tau_{\ell-1})\big],
\]
so $\sup_{\tau}|R_n(\tau) - F(\tau)| \le 2\sum_\ell \sup_t |P_\ell(t) - \bar{P}_\ell(t)|$. Applying the pointwise convergence result with a single segment $(0, t]$ shows that $P_\ell(t) \xrightarrow{\P} \bar{P}_\ell(t)$ for each $t$. Both $P_\ell$ and $\bar{P}_\ell$ are non-decreasing, and $\bar{P}_\ell$ is continuous because $v_\ell$ is bounded. Given $\gamma > 0$, choose $0 = t_0 < t_1 < \dots < t_m = 1$ with $\bar{P}_\ell(t_{k+1}) - \bar{P}_\ell(t_k) \le \gamma$ for all $k$. For $t \in [t_k, t_{k+1}]$, monotonicity gives
\[
P_\ell(t_k) - \bar{P}_\ell(t_k) - \gamma \;\le\; P_\ell(t) - \bar{P}_\ell(t) \;\le\; P_\ell(t_{k+1}) - \bar{P}_\ell(t_{k+1}) + \gamma,
\]
so $\sup_t |P_\ell(t) - \bar{P}_\ell(t)| \le \max_{k} |P_\ell(t_k) - \bar{P}_\ell(t_k)| + \gamma$. The maximum is over finitely many terms and each one converges to zero in probability. Since $\gamma$ is arbitrary, this proves part (a).

\emph{Part (b)}: Let $\mathcal{G}_n \subset \mathcal{T}_L$ be the set of $\tau$ with $\tau_\ell = (k_\ell - 1)/n$ for integers $1 = k_0 < k_1 < \dots < k_{L-1} < k_L = n+1$. By \eqref{eq:ls-split}, $\hat{\tau} := \hat{\eeta}/n - 1/n$ minimizes $R_n$ over $\mathcal{G}_n$. Let $\mathcal{O}$ be any open neighbourhood of $\tau^*$. Since $F$ is continuous on the compact set $\mathcal{T}_L$ with unique minimizer $\tau^*$, we have $\gamma := \inf_{\tau \in \mathcal{T}_L \setminus \mathcal{O}} F(\tau) - F(\tau^*) > 0$. Moreover, $F$ is Lipschitz, since $v_\ell$ is bounded, and $\mathcal{G}_n$ contains a point $\tau^{(n)}$ with $\|\tau^{(n)} - \tau^*\|_\infty \le L/n$, so $F(\tau^{(n)}) < F(\tau^*) + \gamma/3$ for $n$ large. On the event $\{\sup_\tau |R_n - F| < \gamma/3\}$, whose probability tends to one by part (a), we then have $R_n(\hat{\tau}) \le R_n(\tau^{(n)}) < F(\tau^*) + 2\gamma/3$, while $R_n(\tau) > F(\tau) - \gamma/3 \ge F(\tau^*) + 2\gamma/3$ for every $\tau \notin \mathcal{O}$. Hence $\hat{\tau} \in \mathcal{O}$ with probability tending to one, that is, $\hat{\tau} \xrightarrow{\P} \tau^*$, and so $\hat{\eeta}/n \xrightarrow{\P} \tau^*$. Finally, $\frac{1}{n}d_H(\hat{\eeta}, \eeta) = d_H(\hat{\eeta}/n, \eeta/n)$, the Hausdorff distance is $1$-Lipschitz in each argument with respect to the maximum norm, and $\eeta/n \to \alpha$, which gives \eqref{eq:ls-cp-limit}. \hfill $\square$

\section{Nonlinear equations defining asymptotic \WeightedERM{} parameters} \label{sec:nonlinear_equations}
In this section, we specify the equations whose solution yields the scalars $( b^{(\ell)})_{\ell \in [L]}$ and the two $L \times L$ matrices $\bLambda = \begin{bmatrix}
    	\blambda^{(1)} & \blambda^{(2)} & \dots & \blambda^{(L)}
    \end{bmatrix}$ and $\K = (\kappa_{\ell, \ell'})_{\ell, \ell' \in [L]}$.
For $\ell \in [L]$, the $\ell$-th diagonal entry of $\bK$, $\kappa_{\ell, \ell} \in \reals$, and the $\ell$-th column of $\bLambda$, $\blambda^{(\ell)} \in \reals^L$, are defined as components of the unique triple $(\blambda^{(\ell)}, b^{(\ell)}, \kappa_{\ell, \ell})$ that solves the following system of nonlinear equations, for $b^{(\ell)} > 0$:
\begin{align}
    1 - \frac{1}{\delta} &= \lim_{n \to \infty} \frac{1}{n} \sum_{i=1}^n \E\left[ \left(1 + b^{(\ell)} ({\sM}_i^{(\ell)})'' \right)^{-1} \right], \label{eq:b-fp-implicit} \\
    \0_L &= \lim_{n \to \infty} \frac{1}{n} \sum_{i = 1}^n \E\left[\Z_i ({\sM}_i^{(\ell)})' \right], \label{eq:mu-fp-implicit} \\
    \kappa_{\ell, \ell} &= \lim_{n \to \infty} \frac{\delta \left(b^{(\ell)}\right)^2}{n} \sum_{i = 1}^n \E \left[\left( ({\sM}_i^{(\ell)})' \right)^2 \right], \label{eq:sigma-fp-implicit}
\end{align}
where we have used the shorthand:
\begin{align}
\begin{split} \label{eq:sM-def}
    &(\sM_i^{(\ell)})' := \pi_i^{(\ell)} M' \left(\prox_{b^{(\ell)} \pi_i^{(\ell)} M(\cdot, q(\Z_i, \Psi_i, \varepsilon_i))}(\Z_i^{\top} \blambda^{(\ell)} + w^{(\ell)}_i), q(\Z_i, \Psi_i, \varepsilon_i) \right), \\
    &(\sM_i^{(\ell)})'' := \pi_i^{(\ell)} M''\left(\prox_{b^{(\ell)} \pi_i^{(\ell)} M(\cdot, q(\Z_i, \Psi_i, \varepsilon_i))}(\Z_i^{\top} \blambda^{(\ell)} + w^{(\ell)}_i), q(\Z_i, \Psi_i, \varepsilon_i) \right).
\end{split}
\end{align}

The expectations in \eqref{eq:b-fp-implicit}--\eqref{eq:sigma-fp-implicit} are taken over the mutually independent random variables $\Z_i \distas{\text{i.i.d.}} \N(\0_L, \bGamma)$ and $w_i^{(\ell)} \distas{\text{i.i.d.}} \N(0, \kappa_{\ell, \ell})$ for $\ell \in [L], i \in [n]$. For $\ell \neq \ell' \in [L]$, the cross-covariance term $\kappa_{\ell, \ell'}$ is defined as the solution to the following equation:
\begin{align}
    \kappa_{\ell, \ell'} &= \lim_{n \to \infty} \frac{\delta b^{(\ell)} b^{(\ell')}}{n} \sum_{i = 1}^n \E \left[\left( \sM^{(\ell)}_i\right)' \left( \sM^{(\ell')}_i \right)'\right], \label{eq:cross-cov-fp-implicit}
\end{align}
where the expectation in \eqref{eq:cross-cov-fp-implicit} is taken over $\Z_i \distas{\text{i.i.d.}} \N(\0_L, \bGamma)$ and
$(w_i^{(\ell)}, w_i^{(\ell')}) \distas{\text{i.i.d.}} \N\left(\0_2, \begin{bmatrix}
    \kappa_{\ell, \ell} & \kappa_{\ell, \ell'} \\
    \kappa_{\ell, \ell'} & \kappa_{\ell', \ell'}
\end{bmatrix} \right)$, for $i \in [n]$.
 We observe that, once $(\blambda^{(l)}, b^{(l)}, \kappa_{l, l})$ are fixed for $l \in \{\ell, \ell'\}$, \eqref{eq:cross-cov-fp-implicit} is a self-contained expression in the variable $\kappa_{\ell, \ell'}$.

In summary, \eqref{eq:b-fp-implicit}--\eqref{eq:sigma-fp-implicit} form a set of self-contained fixed point equations in the variables ($\blambda^{(\ell)}, \kappa_{\ell, \ell}, b^{(\ell)}$), parametrized by $(\delta, \bGamma)$, which serve to characterize any single estimator $(\hat{\ttheta}^{(\ell)}, \hat{\bbeta}^{(\ell)})$, for $\ell \in [L]$.  The full set of equations \eqref{eq:b-fp-implicit}--\eqref{eq:cross-cov-fp-implicit} in the variables $\left(\bLambda, \bK, \b \right)$ where $\b := \big[
   b^{(1)} \dots b^{(L)} \big]$, in turn, forms a set of $\frac{3}{2} (L + L^2)$ fixed point equations that characterize the full joint asymptotic distribution of the $L$ estimators in $\hat{\B}$ and $\hat{\TTheta}$.

At first glance, it is not evident whether the limits in \eqref{eq:b-fp-implicit}--\eqref{eq:cross-cov-fp-implicit} exist, and if so, whether a solution to the system of equations exists and is unique. We prove in Appendix \ref{sec:proof-existence-of-limits-in-fixed-point-equations} that the limits in \eqref{eq:b-fp-implicit}--\eqref{eq:cross-cov-fp-implicit} exist under the aforementioned assumptions, and moreover, that any solution to \eqref{eq:b-fp-implicit}--\eqref{eq:cross-cov-fp-implicit} is unique. We prove the existence of solutions to \eqref{eq:b-fp-implicit}--\eqref{eq:cross-cov-fp-implicit} under mild additional assumptions in Proposition \ref{prop:existence-of-fixed-point-solution} (Appendix \ref{sec:proof-of-main-results}).

\subsection{Alternative formulation of nonlinear equations}\label{sec:variants-nonlin-eqns}

In this section, we present an equivalent formulation of equations \eqref{eq:b-fp-implicit}--\eqref{eq:cross-cov-fp-implicit} that will be used in our proofs. This formulation replaces each of the noise variables $(\varepsilon_i)_{i \in [n]}$ with a random variable $\bar{\varepsilon} \distas{} \P_{\bar{\varepsilon}}$, where $\P_{\bar{\varepsilon}}$ is  the limiting distribution of the entries of $\bvarepsilon$, as specified in Section \ref{sec:setting}. Let $\bar{y}_i := q(\Z_i, \Psi_i, \bar{\varepsilon})$ for $i \in [n]$.  We recall and extend the notation in \eqref{eq:sM-def}, and define for $\ell \in [L]$:
\begin{align}
\begin{split}\label{eq:sM-def-proof}
&(\sM_i^{(\ell)})' := \pi_i^{(\ell)} M' \left(\prox_{b^{(\ell)} \pi^{(\ell)}_i M(\cdot, q(\Z_i, \Psi_i, \varepsilon_i))}(\Z_i^{\top} \blambda^{(\ell)} + w^{(\ell)}_i), q(\Z_i, \Psi_i, \varepsilon_i) \right), \\
&(\sM_i^{(\ell)})'' := \pi^{(\ell)}_i M'' \left(\prox_{b^{(\ell)} \pi^{(\ell)}_i M(\cdot, q(\Z_i, \Psi_i, \varepsilon_i))}(\Z_i^{\top} \blambda^{(\ell)} + w^{(\ell)}_i), q(\Z_i, \Psi_i, \varepsilon_i) \right), \\
&(\sbM_i^{(\ell)})' := \pi^{(\ell)}_i M' \left(\prox_{b^{(\ell)} \pi^{(\ell)}_i M(\cdot, \bar{y}_i)}(\Z_i^{\top} \blambda^{(\ell)} + w^{(\ell)}_i), \bar{y}_i \right), \\
&(\sbM_i^{(\ell)})'' := \pi^{(\ell)}_i M''\left(\prox_{b^{(\ell)} \pi^{(\ell)}_i M(\cdot, \bar{y}_i)}(\Z_i^{\top} \blambda^{(\ell)} + w^{(\ell)}_i), \bar{y}_i \right), \\
&(\sbM_i^{(\ell)})''' := \pi^{(\ell)}_i M'''\left(\prox_{b^{(\ell)} \pi^{(\ell)}_i M(\cdot, \bar{y}_i)}(\Z_i^{\top} \blambda^{(\ell)} + w^{(\ell)}_i), \bar{y}_i \right), \\
&\partial_{12} \sbM_i^{(\ell)} := \pi^{(\ell)}_i \partial_{12} M\left(\prox_{b^{(\ell)} \pi^{(\ell)}_i M(\cdot, \bar{y}_i)}(\Z_i^{\top} \blambda^{(\ell)} + w^{(\ell)}_i), \bar{y}_i \right),
\end{split}
\end{align}
where, independently over $i \in [n]$, we have $\Z_i \distas{} \N(\0_L, \bGamma), w_i^{(\ell)} \distas{} \N(0, \kappa_{\ell, \ell})$ as per the definition in Section \ref{sec:characterization-nonlinear-eqns}.

We state a couple of  technical lemmas used to derive the alternative formulation in \eqref{eq:b-fp-implicit-bar}-\eqref{eq:cross-cov-fp-implicit-bar}.
\begin{lemma} \label{lemma:h-conv-to-Eh-gen}
Consider the model assumptions in Section~\ref{sec:setting} and suppose that Assumption~\ref{asmpt:reg-weights} holds.  Let
$h : \mathbb{R} \times [\underline{\pi}, \bar{\pi}] \times \mathcal{S} \to \mathbb{R}$,
where $\mathcal{S}$ is a finite set, be a function satisfying the following for some constant $C > 0$:
\begin{enumerate}[(a)]
  \item[(a)]  \begin{align} \label{eq:h-conv-to-Eh-cond-1}
      \sup_{\pi \in [\underline{\pi},\bar{\pi}], \psi \in \mathcal{S}} |h(0, \pi, \psi)| \leq C.
  \end{align}
  \item[(b)] For all $\varepsilon^{(1)},\varepsilon^{(2)} \in \mathbb{R}$,
        $\pi \in [\underline{\pi},\bar{\pi}]$, $\psi \in \mathcal{S}$,
        \begin{align} \label{eq:h-conv-to-Eh-cond-2}
          \left|h(\varepsilon^{(1)},\pi,\psi)-h(\varepsilon^{(2)},\pi,\psi) \right|
          \leq C \left(1+|\varepsilon^{(1)}|+|\varepsilon^{(2)}| \right)|\varepsilon^{(1)}-\varepsilon^{(2)}|.
        \end{align}
  \item[(c)] For all
        $\pi^{(1)},\pi^{(2)} \in [\underline{\pi},\bar{\pi}]$, $\varepsilon \in \mathbb{R}$, $\psi \in \mathcal{S}$,
        \begin{align} \label{eq:h-conv-to-Eh-cond-3}
          \left|h(\varepsilon,\pi^{(1)},\psi)-h(\varepsilon,\pi^{(2)},\psi) \right|
          \leq C(1+|\varepsilon|^2) |\pi^{(1)}-\pi^{(2)}|.
        \end{align}
\end{enumerate}
Then, we have that as $n\to \infty$,
\begin{align}\label{eq:gen-noise-replacement}
  \left|\frac{1}{n}\sum_{i=1}^n h(\varepsilon_i,\pi_i,\Psi_i)
        -\frac{1}{n}\sum_{i=1}^n \mathbb{E}[h(\bar{\varepsilon},\pi_i,\Psi_i)]\right| \to 0.
\end{align}
\end{lemma}
\begin{proof}
Let $A_m(f):=m^{-1}\sum_{i=1}^m f(\varepsilon_i)$ for $m\geq1$, and set $A_0(f):=0$. Since the noise vector $\bvarepsilon$ is a prefix of the same sequence $(\varepsilon_i)_{i \geq 1}$, Wasserstein-$2$ convergence implies, for every bounded continuous $f$ and $0\leq a<b\leq1$,
\begin{align*}
\frac1n\sum_{i=\floor{an}+1}^{\floor{bn}}f(\varepsilon_i)
&=\frac{\floor{bn}}n A_{\floor{bn}}(f)-\frac{\floor{an}}n A_{\floor{an}}(f)\\
&\longrightarrow (b-a)\E[f(\bar\varepsilon)].
\end{align*}
When $\floor{an}=0$, the second term is zero by the definition of $A_0$. The same holds for $f(\varepsilon)=\varepsilon^2$, so the noise empirical distribution on every interval of asymptotically positive length converges in Wasserstein-$2$ to $\P_{\bar\varepsilon}$.

First truncate the noise using $T_R(e):=\max\{-R,\min\{e,R\}\}$, and write $h_R(e,\pi,\psi):=h(T_R(e),\pi,\psi)$. By \eqref{eq:h-conv-to-Eh-cond-1} and \eqref{eq:h-conv-to-Eh-cond-2}, $|h_R|\leq C_R$ for a constant $C_R$ independent of $n,i$. Define $\widetilde\Psi_i:=\Psi_{\eta_{r-1}}$ when $\floor{n\alpha_{r-1}}<i\leq\floor{n\alpha_r}$, for $r\in[L^*]$. Since $\eta_r/n\to\alpha_r$,
\[
\frac{\#\{i:\Psi_i\neq\widetilde\Psi_i\}}{n}
\leq\sum_{r=1}^{L^*-1}\frac{|\eta_r-1-\floor{n\alpha_r}|}{n}\longrightarrow0.
\]
Replacing $\Psi_i$ by $\widetilde\Psi_i$ therefore changes each of the two averages in \eqref{eq:gen-noise-replacement}, with $h$ replaced by $h_R$, by at most $2C_R\#\{i:\Psi_i\neq\widetilde\Psi_i\}/n$, which converges to zero.

By \eqref{eq:pi_i_defn}, $0\leq\pi_i\leq1$. Define $P_n(t):=\pi_{\floor{nt}+1}$ for $t\in[0,1)$. Assumption~\ref{asmpt:reg-weights} and the dominated convergence theorem give
\[
\int_0^1|P_n(t)-\Phi(t)|\,dt\longrightarrow0.
\]
For $m\geq1$, choose $\Phi_m$ with values in $[\underline\pi,\bar\pi]$, constant on finitely many intervals, such that $\int_0^1|\Phi_m(t)-\Phi(t)|\,dt\leq1/m$, and set
\[
\widetilde\pi_{i,m}:=n\int_{(i-1)/n}^{i/n}\Phi_m(t)\,dt.
\]
Then
\[
\frac1n\sum_{i=1}^n|\pi_i-\widetilde\pi_{i,m}|
\leq\int_0^1|P_n(t)-\Phi_m(t)|\,dt
\leq\int_0^1|P_n(t)-\Phi(t)|\,dt+\frac1m.
\]
By \eqref{eq:h-conv-to-Eh-cond-3}, replacing $\pi_i$ by $\widetilde\pi_{i,m}$ changes each of the two averages in \eqref{eq:gen-noise-replacement}, with $h$ replaced by $h_R$, by at most $C(1+R^2)n^{-1}\sum_i|\pi_i-\widetilde\pi_{i,m}|$. For fixed $m$, these weights and $\widetilde\Psi_i$ are constant on finitely many blocks, apart from finitely many boundary cells. Applying the preceding interval convergence to $f(e)=h_R(e,\pi,\psi)$ on each block proves the claim for these weights. Taking $n\to\infty$ and then $m\to\infty$ proves it for the original weights and truncated noise.

Finally, \eqref{eq:h-conv-to-Eh-cond-1} and \eqref{eq:h-conv-to-Eh-cond-2} bound the truncation error by a constant times
\[
\frac1n\sum_{i=1}^n(1+\varepsilon_i^2)\ind\{|\varepsilon_i|>R\}
+\E[(1+\bar\varepsilon^2)\ind\{|\bar\varepsilon|>R\}].
\]
Its limit superior tends to zero as $R\to\infty$ by Wasserstein-$2$ convergence. This proves the result.
\end{proof}
\begin{lemma}\label{lemma:bounded-deriv-expect}
Consider the model assumptions in Section \ref{sec:setting}, and further suppose Assumptions \ref{asmpt:reg-loss}(\ref{cond:strict-cvx-smooth}), \ref{asmpt:reg-loss}(\ref{eq:loss-asmpt-log-conc}) hold. Then, for $\ell \in [L], i\in[n], \varepsilon \in \reals, \vartheta \in (0, \infty)$, we have that
\begin{align*}
 \E\left[ M'\left( \prox_{b \vartheta M(\cdot, q(\Z_1, \Psi_i, \varepsilon))}({\Z_1^\top} \blambda^{(\ell)} + w^{(\ell)}_1), q(\Z_1, \Psi_i, \varepsilon) \right)^2 \right] \leq C_{h01} + C_{h02} \varepsilon^2,
\end{align*}
for some constants $C_{h01}, C_{h02} > 0$.
\end{lemma}
\begin{proof}
\begin{align}
   &{\E\left[ M'\left( \prox_{b \vartheta M(\cdot, q(\Z_1, \Psi_i, \varepsilon))}({\Z_1^\top} \blambda^{(\ell)} + w^{(\ell)}_1), q(\Z_1, \Psi_i, \varepsilon) \right)^2 \right]} \notag \\
    &\leq {\E\left[ 3|M'(0, 0)|^2 + 3 B_{11}^2 |{\Z_1^\top} \blambda^{(\ell)} + w_1^{(\ell)}|^2 + 3 B_{12}^2 |q(\Z_1, \Psi_i, \varepsilon)|^2 \right]} \label{eq:h-0-1}\\
    &\leq {\E\left[ 3|M'(0, 0)|^2 + 3 B_{11}^2 |{\Z_1^\top} \blambda^{(\ell)} + w_1^{(\ell)}|^2 + 3 B_{12}^2 C_{\text{qLip}} (3\|\Z_1\|^2_2 + 3L^2 + 3|\varepsilon|^2) \right]} \label{eq:h-0-2} \\
    &\leq C_{h01} + C_{h02} \varepsilon^2, \label{eq:h-0-3}
\end{align}
where \eqref{eq:h-0-1} follows from Proposition \ref{prop:M_prime_ub} with $B_{11} := \sup_{a_1, a_2} M''(a_1, a_2)$ and $B_{12} := \sup_{a_1, a_2} \partial_{12} M(a_1, a_2)$ together with the Cauchy-Schwarz inequality, \eqref{eq:h-0-2} follows from the Lipschitz assumption on $q$ with Lipschitz constant $C_{\text{qLip}} > 0$, \eqref{eq:h-0-3} follows for large enough constants $C_{h01}, C_{h02} > 0$.
\end{proof}
\begin{proposition} \label{prop:conv-nonlin-eqns}
For  $\ell \in [L], i \in [n]$, recall $(\sM_i^{(\ell)})', (\sbM_i^{(\ell)})', (\sM_i^{(\ell)})'', (\sbM_i^{(\ell)})''$  defined in \eqref{eq:sM-def-proof}.
Consider the model assumptions in Section \ref{sec:setting}, and further suppose Assumptions \ref{asmpt:reg-loss}(\ref{cond:strict-cvx-smooth}), \ref{asmpt:reg-loss}(\ref{eq:loss-asmpt-log-conc}), \ref{asmpt:reg-weights} hold.
We then have that, as $n \to \infty$ and for $\ell, \ell' \in [L]$:
\begin{align}
    &\left| \frac{1}{n} \sum_{i = 1}^n \E\left[ (1 + b^{(\ell)} (\sM^{(\ell)}_i)'')^{-1} \right] - \frac{1}{n} \sum_{i = 1}^{n} \E\left[ (1 + b^{(\ell)} (\sbM^{(\ell)}_i)'')^{-1} \right] \right| \to 0, \label{eq:fp-asymp-b} \\
    &\left\| \frac{1}{n} \sum_{i = 1}^n \E\left[\Z_i (\sM_i^{(\ell)})' \right] - \frac{1}{n} \sum_{i = 1}^n \E\left[\Z_i (\sbM_i^{(\ell)})' \right] \right\|_2 \to 0, \label{eq:fp-asymp-mu} \\
    & \left| \frac{1}{n} \sum_{i=1}^n \E \left[(\sM_i^{(\ell)})' (\sM_i^{(\ell')})'\right] - \frac{1}{n} \sum_{i=1}^n \E \left[(\sbM_i^{(\ell)})' (\sbM_i^{(\ell')})'\right] \right| \to 0, \label{eq:fp-asymp-sigma}
\end{align}
where the expectation is taken over $\bar{\varepsilon}$, $\Z_i$, $w_i^{(\ell)}$, $w_i^{(\ell')}$, where we recall that for $i \in [n], \ell \in [L]$, we have $\Z_i \distas{i.i.d} \N(0, \bGamma)$ and $w^{(\ell)}_i \distas{i.i.d} \N(0, \kappa_{\ell, \ell})$ independent of $\bar{\varepsilon} \distas{} \P_{\bar{\varepsilon}}$ as defined in Section \ref{sec:characterization-nonlinear-eqns}.
\end{proposition}
The proof is given below. In light of Proposition \ref{prop:conv-nonlin-eqns}, the original equations \eqref{eq:b-fp-implicit}--\eqref{eq:cross-cov-fp-implicit} given by
\begin{align*}
    1 - \frac{1}{\delta} &= \lim_{n \to \infty} \frac{1}{n} \sum_{i=1}^n \E\left[ \left(1 + b^{(\ell)} ({\sM}_i^{(\ell)})'' \right)^{-1} \right], \\
    \0_L &= \lim_{n \to \infty} \frac{1}{n} \sum_{i = 1}^n \E\left[\Z_i ({\sM}_i^{(\ell)})' \right],\\
    \kappa_{\ell, \ell'} &= \lim_{n \to \infty} \frac{\delta b^{(\ell)} b^{(\ell')}}{n} \sum_{i = 1}^n \E \left[\left( \sM^{(\ell)}_i\right)' \left( \sM^{(\ell')}_i \right)'\right],
\end{align*}
can be rewritten with the parametric dependence on $\bvarepsilon$ replaced by an expectation over $\bar{\varepsilon}$:
\begin{align}
    1 - \frac{1}{\delta} &= \lim_{n \to \infty} \frac{1}{n} \sum_{i=1}^n \E\left[ \left(1 + b^{(\ell)} (\sbM_i^{(\ell)})'' \right)^{-1} \right], \label{eq:b-fp-implicit-bar} \\
    \0_L &= \lim_{n \to \infty} \frac{1}{n} \sum_{i = 1}^n \E\left[\Z_i (\sbM_i^{(\ell)})' \right] \label{eq:mu-fp-implicit-bar}, \\
    \kappa_{\ell, \ell'} &= \lim_{n \to \infty} \frac{\delta b^{(\ell)} b^{(\ell')}}{n} \sum_{i = 1}^n \E \left[(\sbM_i^{(\ell)})' (\sbM_i^{(\ell')})'\right] \label{eq:cross-cov-fp-implicit-bar}.
\end{align}
In Appendix \ref{sec:proof-existence-of-limits-in-fixed-point-equations}, we prove that the limits in equations \eqref{eq:b-fp-implicit-bar}--\eqref{eq:cross-cov-fp-implicit-bar} are well defined. Propositions \ref{prop:conv-nonlin-eqns}--\ref{prop:steins-lemma-fp} then yield the existence of the limits in all three variants of equations presented in this section. In Appendix \ref{sec:uniqueness_of_solns}, we prove that any solution to equations \eqref{eq:b-fp-implicit-bar}--\eqref{eq:cross-cov-fp-implicit-bar} is unique, and in Appendix \ref{sec:proof-ERM_frobenius_characterization} we use equations \eqref{eq:b-fp-implicit-bar}--\eqref{eq:cross-cov-fp-implicit-bar} to prove a Cauchy property of non-separable Approximate Message Passing iterates that is essential for proving Theorems \ref{thm:ERM_general_characterization_pres}--\ref{thm:characterization_of_debiased_est_pres}. \label{eq:existence-of-limits-argument}
\begin{proof}[Proof of Proposition \ref{prop:conv-nonlin-eqns}.]
Fix $\ell, \ell' \in [L]$. Define the quantities:
\begin{align*}
    B_{11} := \sup_{u, v} M''(u, v), \;\; B_{12} := {\sup_{u,v}|\partial_{12}M(u,v)|}, \;\; \bar{\pi} := \sup_i \pi_i, \;\; \underline{\pi} := \inf_i \pi_i.
\end{align*}
In what follows, we drop the superscript $^{(\ell)}$ when it is clear from context, especially
for \eqref{eq:fp-asymp-b}, \eqref{eq:fp-asymp-mu}. We will only prove \eqref{eq:fp-asymp-sigma} for $\ell = \ell'$, the case $\ell \neq \ell'$ can be proven similarly.
It follows from Lemma \ref{lemma:h-conv-to-Eh-gen} that if $f: \reals \times \reals \times [L^*]$ satisfies \eqref{eq:h-conv-to-Eh-cond-1}--\eqref{eq:h-conv-to-Eh-cond-3}, then, as $n \to \infty$,
\begin{align} \label{eq:h-form}
    \left|\frac{1}{n} \sum_{i=1}^n f(\varepsilon_i, \pi_i, \Psi_i) - \frac{1}{n} \sum_{i=1}^n \E[f(\bar{\varepsilon}, \pi_i, \Psi_i)] \right| \to 0.
\end{align}
Hence, our proof will consist of showing that \eqref{eq:fp-asymp-b}--\eqref{eq:fp-asymp-sigma} take the form \eqref{eq:h-form} with $f: \reals \times \reals \times [L^*]$ satisfying \eqref{eq:h-conv-to-Eh-cond-1}--\eqref{eq:h-conv-to-Eh-cond-3}, after which the result follows from Lemma \ref{lemma:h-conv-to-Eh-gen}.
Define:
\begin{align*}
    &\sa = \sa(\varepsilon_i, \pi_i, \Psi_i) := \prox_{b \pi_i M(\cdot, q(\Z_1, \Psi_{i}, \varepsilon_i))}({\Z_1^\top} \blambda^{(\ell)} + w^{(\ell)}_1), \\
    &\sy = \sy(\varepsilon_i, \pi_i, \Psi_i) := q(\Z_1, \Psi_{i}, \varepsilon_i).
\end{align*}
We note that the left-hand side quantity in \eqref{eq:fp-asymp-b} can be written as $\big|\frac{1}{n} \sum_{i=1}^n f(\varepsilon_i, \pi_i, \Psi_i) - \frac{1}{n} \sum_{i=1}^n \E[f(\bar{\varepsilon}, \pi_i, \Psi_i)] \big|$ with
\begin{align}
    f(\varepsilon_i, \pi_i, \Psi_i) = \E[(1 + b \pi_i M''(\sa, \sy))^{-1}], \label{eq:f-choice-1}
\end{align}
where we recall that, for $i \in [n]$ and $\tilde{\ell} \in [L]$, $(\Z_i, w_i^{(\tilde{\ell})})$ have the same distribution as $(\Z_1, w_1^{(\tilde{\ell})})$. Note that $f$ in \eqref{eq:f-choice-1} is bounded above by $1$, and hence by the dominated convergence theorem we have that
\begin{align}
    &|\partial_{\varepsilon_i} f(\varepsilon_i, \pi_i, \Psi_i)| \notag \\
    &= \left|\E\left[ \left(\frac{b^2 \pi_i M'''(\sa, \sy) \cdot \pi_i \partial_{12} M(\sa, \sy)}{(b \pi_i M''(\sa, \sy) + 1)^3} - \frac{b \pi_i \partial_2  M''(\sa, \sy)}{(b \pi_i M''(\sa, \sy) + 1)^2} \right) \cdot \partial_{\varepsilon_i} q(\Z_i, \Psi_i, \varepsilon_i) \right]\right| \notag \\
    &\leq \E\left[ \left(\left| \frac{b^2 \pi_i M'''(\sa, \sy) \cdot \pi_i \partial_{12} M(\sa, \sy)}{(b \pi_i M''(\sa, \sy) + 1)^3}\right| + \left|\frac{b \pi_i \partial_2  M''(\sa, \sy)}{(b \pi_i M''(\sa, \sy) + 1)^2}\right| \right) \cdot \left|\partial_{\varepsilon_i} q(\Z_i, \Psi_i, \varepsilon_i)\right| \right] \notag \\
    &\leq C_{\text{qLip}} \, \E\left[ b^2 \pi_i |M'''(\sa, \sy)| \cdot |\partial_{12} M(\sa, \sy)| + b \pi_i |\partial_2 M''(\sa, \sy)| \right] \notag\\
    &\leq C_{h11} \bar{\pi} \max\{b^2, b\}, \label{eq:f-partial-varepsilon-1}
\end{align}
for some constant $C_{h11} > 0$, where the second inequality holds because $q$ is Lipschitz. {
Similarly, by \eqref{eq:prox_id_4},
\begin{align}
\partial_{\pi_i}f
&=\E\left[-\frac{bM''(\sa,\sy)}{(1+b\pi_iM''(\sa,\sy))^2}
+\frac{b^2\pi_iM'''(\sa,\sy)M'(\sa,\sy)}{(1+b\pi_iM''(\sa,\sy))^3}\right],\label{eq:f-partial-pi-1-1}\\
|\partial_{\pi_i}f|&\leq C(1+|\varepsilon_i|),\label{eq:f-partial-pi-1-4}
\end{align}
for a constant $C$ independent of $n,i$, by Lemma \ref{lemma:bounded-deriv-expect}. Together with $|f|\leq1$ and \eqref{eq:f-partial-varepsilon-1}, absolute continuity verifies \eqref{eq:h-conv-to-Eh-cond-1} to \eqref{eq:h-conv-to-Eh-cond-3}. Since $q$ in \eqref{eq:model} is Lipschitz by Section \ref{sec:setting}, its derivative with respect to the noise argument in \eqref{eq:f-partial-varepsilon-1} and the calculations below is understood for Lebesgue-almost every $\varepsilon_i$.
}

Moving onto equation \eqref{eq:fp-asymp-mu}, notice that it suffices to prove that
\begin{align}
    & \left|\frac{1}{n} \sum_{i = 1}^n \E\left[Z_{i, \ell'} (\sM_i^{(\ell)})' \right] - \frac{1}{n} \sum_{i = 1}^n \E\left[Z_{i, \ell'} (\sbM_i^{(\ell)})' \right]\right| \to 0, \label{eq:fp-asymp-mu-sufficient-cond}
\end{align}
for $\ell, \ell' \in [L]$. Let $i \in [n]$ and define $f^{(2)}$ as follows:
\begin{align}
   f^{(2)}(\varepsilon_i, \pi_i, \Psi_i) &:= \E\left[ Z_{1, \ell'} \pi_i M'\left( \prox_{b \pi_i M(\cdot, q(\Z_1, \Psi_i, \varepsilon_i))}({\Z_1^\top} \blambda^{(\ell)} + w^{(\ell)}_1), q(\Z_1, \Psi_i, \varepsilon_i) \right)\right] \notag \\
   &= \E\left[ Z_{1, \ell'} \pi_i M'\left( \sa, \sy \right)\right], \label{eq:h-choice-2}
\end{align}
and notice that the left-hand side of \eqref{eq:fp-asymp-mu-sufficient-cond} can be written as:
\begin{align}
    & \left|\frac{1}{n} \sum_{i = 1}^n f^{(2)}(\varepsilon_i, \pi_i, \Psi_i) - \frac{1}{n} \sum_{i = 1}^n \E\left[ f^{(2)}(\bar{\varepsilon}, \pi_i, \Psi_i) \right] \right| \to 0, \label{eq:fp-asymp-mu-sufficient-cond-with-f}
\end{align}
since for $i \in [n], \tilde{\ell} \in [L]$, $(\Z_i, w_i^{(\tilde{\ell})})$ have the same distribution as $(\Z_1, w_1^{(\tilde{\ell})})$. One can {bound $f^{(2)}$} via the following steps:
\begin{align}
   |f^{(2)}(\varepsilon_i, \pi_i, \Psi_i)| & = {\left|\E[Z_{1,\ell'}\pi_i M'(\sa,\sy)]\right|} \notag \\
   &\leq \pi_i \sqrt{\E\left[ Z_{1, \ell'}^2 \right]} \sqrt{\E\left[ M'( \sa, \sy)^2 \right]} \label{eq:f-2-1}\\
    &\leq \bar{\pi} (C_{h21} + C_{h22} |\varepsilon_i|) \label{eq:f-2-4}
\end{align}
where \eqref{eq:f-2-1} follows from the Cauchy-Schwarz inequality,
\eqref{eq:f-2-4} follows from Lemma \ref{lemma:bounded-deriv-expect} for some positive constants $C_{h21}, C_{h22}$. {
We then have, by the chain rule and \eqref{eq:prox_id_4},
\begin{align}
\partial_{\varepsilon_i} f^{(2)}
&=\E\left[Z_{1,\ell'}\frac{\pi_i\partial_{12}M(\sa,\sy)}{1+b\pi_iM''(\sa,\sy)}\,
\partial_{\varepsilon_i}q(\Z_1,\Psi_i,\varepsilon_i)\right],\label{eq:f-2-2-1}\\
\partial_{\pi_i} f^{(2)}
&=\E\left[\frac{Z_{1,\ell'}M'(\sa,\sy)}{1+b\pi_iM''(\sa,\sy)}\right].\label{eq:h-2-vartheta-2}
\end{align}
Thus $|\partial_{\varepsilon_i}f^{(2)}|\leq C$ and $|\partial_{\pi_i}f^{(2)}|\leq C(1+|\varepsilon_i|)$ by Cauchy-Schwarz and Lemma \ref{lemma:bounded-deriv-expect}. Together with \eqref{eq:f-2-4}, these verify \eqref{eq:h-conv-to-Eh-cond-1} to \eqref{eq:h-conv-to-Eh-cond-3}.
}

Moving onto \eqref{eq:fp-asymp-sigma} for $\ell = \ell'$, notice that it can be written as
\begin{align}
    & \left| \frac{1}{n} \sum_{i=1}^n f^{(3)}(\varepsilon_i, \pi_i, \Psi_i) - \frac{1}{n} \sum_{i=1}^n \E [ f^{(3)}(\bar{\varepsilon}, \pi_i, \Psi_i) ] \right| \to 0,
\end{align}
where
\begin{align}
   {f^{(3)}(\varepsilon_i,\pi_i,\Psi_i)} := (\pi^{(\ell)}_i)^2 \E \left[ M'(\sa, \sy)^2 \right].
\end{align}
We note $f^{(3)}$ satisfies the following upper bound:
\begin{align}
    &|f^{(3)}(\varepsilon_i, \pi_i, \Psi_i)|
 \leq \bar{\pi}^2 \E\left[ \left| M'(\sa, \sy)^2 \right| \right]
    \leq C_{h3} \bar{\pi}^2 (C_{h21} + C_{h22} |\varepsilon_i|)^2,
    \label{eq:f-3-0-1}
\end{align}
where \eqref{eq:f-3-0-1} follows from similar computations to \eqref{eq:f-2-1}--\eqref{eq:f-2-4} for some positive constant $C_{h3}$. {
Moreover, the chain rule gives
\begin{align}
\partial_{\pi_i}f^{(3)}
&=2\pi_i\E\left[\frac{M'(\sa,\sy)^2}{1+b\pi_iM''(\sa,\sy)}\right],\label{eq:f-3-1-1}\\
\partial_{\varepsilon_i}f^{(3)}
&=2\pi_i^2\E\left[\frac{M'(\sa,\sy)\partial_{12}M(\sa,\sy)}{1+b\pi_iM''(\sa,\sy)}\,
\partial_{\varepsilon_i}q(\Z_1,\Psi_i,\varepsilon_i)\right].\label{eq:f-3-2-1}
\end{align}
Consequently, $|\partial_{\pi_i}f^{(3)}|\leq C(1+\varepsilon_i^2)$ and $|\partial_{\varepsilon_i}f^{(3)}|\leq C(1+|\varepsilon_i|)$. These bounds and \eqref{eq:f-3-0-1} verify \eqref{eq:h-conv-to-Eh-cond-1} to \eqref{eq:h-conv-to-Eh-cond-3}. Lemma \ref{lemma:h-conv-to-Eh-gen} therefore proves \eqref{eq:fp-asymp-sigma} for $\ell=\ell'$.

For $\ell\neq\ell'$, the product rule and Cauchy-Schwarz verify \eqref{eq:h-conv-to-Eh-cond-1} and \eqref{eq:h-conv-to-Eh-cond-2} for $\E[(\sM_i^{(\ell)})'(\sM_i^{(\ell')})']$ in \eqref{eq:fp-asymp-sigma}, uniformly in its two weights. Replacing its two weights by $u,v\in[0,1]$ changes this expectation by at most
\[
C(1+\varepsilon_i^2)\left(|\pi_i^{(\ell)}-u|+|\pi_i^{(\ell')}-v|\right).
\]
Apply the proof of Lemma \ref{lemma:h-conv-to-Eh-gen}, approximating each weight sequence separately and using the combined interval endpoints. For each average in \eqref{eq:fp-asymp-sigma}, after truncating the noise to $[-R,R]$, the combined approximation error has limit superior at most $2C(1+R^2)/m$ as $n\to\infty$. Taking $m\to\infty$ and then $R\to\infty$ proves \eqref{eq:fp-asymp-sigma} for $\ell\neq\ell'$.
}
\end{proof}
\section{Proofs of main results}\label{sec:proof-of-main-results}
In this section, we give the proofs of Theorem \ref{thm:ERM_general_characterization_pres}, Proposition \ref{prop:ERM_hausdorff_characterization_pres}, Theorem \ref{thm:characterization_of_debiased_est_pres}, and Proposition \ref{prop:pointwise_posterior_pres}. These results are proved using  the theory of non-separable Approximate Message Passing (AMP) algorithms \citep{berthier_state-evolution_2019, gerbelot_graph-based_2023}, and generalize the techniques used to prove similar asymptotic characterizations for M-estimation and logistic regression in homogeneous regression models without change points \citep{donoho_high_2016, sur_modern_2019}. At a high level, our proofs make use of the non-separable AMP results from \citep{arpino_2025_jmlr} stating that a certain iterative algorithm operating on $(\X, \y)$ can be succinctly characterized via random variables whose parameters are given by a low-dimensional deterministic recursion, called \textit{state evolution}, as $n, p \to \infty$. We then construct a specific variant of this iterative algorithm that converges to the minimizer of \eqref{eq:ERM_est} as the number of iterations grows, allowing us to translate the asymptotic characterization of the iterative algorithm into an asymptotic characterization of \eqref{eq:ERM_est}. Importantly, our results rely on equations \eqref{eq:b-fp-implicit}--\eqref{eq:cross-cov-fp-implicit} being well-defined, which is guaranteed by the following lemma.
\begin{lemma}\label{lemma:general-pointwise-conv}
Consider the model assumptions in Section \ref{sec:setting} and further suppose Assumptions \ref{asmpt:reg-loss}(\ref{cond:strict-cvx-smooth}), \ref{asmpt:reg-loss}(\ref{eq:loss-asmpt-log-conc}), \ref{asmpt:reg-weights} (described in the same section) hold. Then, the limits in \eqref{eq:b-fp-implicit}--\eqref{eq:cross-cov-fp-implicit} exist.
\end{lemma}
Lemma \ref{lemma:general-pointwise-conv} is proved in Appendix \ref{sec:proof-existence-of-limits-in-fixed-point-equations}, and the existence of the limits in \eqref{eq:b-fp-implicit}--\eqref{eq:cross-cov-fp-implicit} is assumed going forward. Our main results will be in terms of solutions to equations \eqref{eq:b-fp-implicit}--\eqref{eq:cross-cov-fp-implicit}. The existence of solutions to \eqref{eq:b-fp-implicit}--\eqref{eq:cross-cov-fp-implicit} is guaranteed under additional technical assumptions, as stated in the following proposition.
\begin{proposition} \label{prop:existence-of-fixed-point-solution}
Assume the setting outlined in Section \ref{sec:setting}. Further assume that $q$ is differentiable with respect to its first argument, and that $M$ is strongly convex.
Then, we have that equations \eqref{eq:b-fp-implicit}--\eqref{eq:cross-cov-fp-implicit} admit a solution within a compact subset of $\reals^{L + 3}$.
\end{proposition}
We note that the existence of solutions to \eqref{eq:b-fp-implicit}--\eqref{eq:cross-cov-fp-implicit} was numerically confirmed for the experiments in Sections \ref{sec:synthetic-experiments} and \ref{sec:real-data-experiments}. The proof of Proposition \ref{prop:existence-of-fixed-point-solution} is provided in Appendix \ref{sec:existence-of-solns}.

Moreover, any solution to equations \eqref{eq:b-fp-implicit}--\eqref{eq:cross-cov-fp-implicit} is unique by the following proposition, which we prove in Appendix \ref{sec:uniqueness_of_solns}:
\begin{proposition} \label{prop:uniqueness_of_fixed_points}
Consider the setting described in Section \ref{sec:setting}. Then {any solution $(\b,\bLambda,\bK)$ to equations \eqref{eq:b-fp-implicit}--\eqref{eq:cross-cov-fp-implicit}, with $\b>0$ and $\bK=\bK^\top\succeq0$, is unique}.
\end{proposition}

Our asymptotic characterization allows for dependence among covariates, i.e., $\bSigma \neq \I$.
The following lemma allows us to reduce the analysis for general covariance  to the isotropic setting via a simple change of variables argument \citep{zhao_asymptotic_2022}.

\begin{lemma} \label{lem:covariance-reduction_pres}
Let $\L \in \reals^{p \times p}$ be the lower triangular matrix obtained from the unique Cholesky decomposition $\bSigma = \L \L^\top$.  For $\bbeta \in \reals^p$, consider the empirical risk function ${\cC}^{(\ell)}(\bbeta; \X, \y) = \cC^{(\ell)}(\bbeta)$ for $\ell \in [L]$, where $\cC^{(\ell)}$ is defined in \eqref{eq:ERM_est} and $\y = q(\X \B, \bPsi, \bvarepsilon)$ follows the model definition in \eqref{eq:model}. If $\hat{\bbeta}$ is a minimizer of ${\cC}^{(\ell)}(\cdot \;; \X, \y)$, then $\L^\top \hat{\bbeta}$ is a minimizer of ${\cC}^{(\ell)}(\cdot \; ; \tilde{\X}, \y)$, where $\tilde{\X} := \X \L^{-\top}$ is an isotropic design matrix.
\end{lemma}
\begin{proof}
Let $\ell \in [L]$. From \eqref{eq:ERM_est}, note that ${\cC^{(\ell)}}(\bbeta; \X, \y)$ depends on the rows of $\X$, $\x_i$ for $i \in [n]$, and $\bbeta$ only through their inner product. Therefore,
\begin{align}
    {\cC}^{(\ell)}(\bbeta; \X, \y) = {\cC}^{(\ell)}(\L^\top \bbeta; \X \L^{-\top}, \y), \label{eq:tilde_C}
\end{align}
for $\bbeta \in \reals^p$. Hence, if $\hat{\bbeta}$ is the minimizer of the original empirical loss function on the LHS of \eqref{eq:tilde_C}, then $\L^\top \bbeta$ is the minimizer of the same empirical loss function defined using the isotropic design matrix $\tilde{\X}$ and with true regression coefficient matrix given by $\L^\top \B$.
\end{proof}
The reparametrization  in Lemma \ref{lem:covariance-reduction_pres} allows us to work with isotropic covariates $\tilde{\X} := \X \L^{-\top}$ and ground truth signal $\L^\top \B$ when constructing our AMP iteration, and hence to obtain asymptotic guarantees via the non-separable AMP theory in \citep{arpino_2025_jmlr}. Moreover, this reparametrization does not alter either the linearly transformed signal matrix $\X \B$ or $\y$, nor the estimate matrix $\hat{\TTheta}$ in \eqref{eq:Theta-estimate-matrix}.

\subsection{Proof of Theorem \ref{thm:ERM_general_characterization_pres} and Proposition \ref{prop:ERM_hausdorff_characterization_pres}}
The proof of Theorem \ref{thm:ERM_general_characterization_pres} relies on two main results.  The first result, Lemma \ref{lemma:SE-matrix-amp}, states that a certain matrix-valued AMP iteration admits an asymptotic characterization parametrized by the solution to a deterministic, $L$ dimensional recursion. The proof of this result is a reduction to the AMP result in \citep{arpino_2025_jmlr}, with a specific choice of ``denoiser'' function. The second result, Lemma \ref{lemma:ERM_frobenius_characterization}, shows that with this choice of denoiser, the AMP iteration converges to the solution of the optimization problem in \eqref{eq:ERM_est}

\paragraph{AMP iteration and its asymptotic characterization}
We  begin with some notation. For a function $G^t: \R^{n\times L}\times \R^{n}\to \R^{n\times L}$, let $\partial_i{G^t_i}$ denote the $L\times L$ Jacobian of $G^t_i$ w.r.t. the $i$th row of its first input.
Let $\odot$ denote element-wise multiplication, and let superscript $^\odot$ denote element-wise power.
We recall the notation $\bprox_{\tilde{\b} \bpi M(\cdot, \tilde{\y})}(\tilde{\TTheta})$ defined in \eqref{eq:bold-prox-def}, and the
random variables $\Z_i, \G_{\TTheta}$ and the deterministic quantities $(\bLambda, \bK, \b)$ defined in Section \ref{sec:characterization-nonlinear-eqns}. We let $\1_L$ denote the $L$-dimensional all-ones vector, and for a matrix $\A \in \reals^{L \times L}$, we let $\diag(\A) \in \reals^L$ denote the $L$-dimensional vector such that $(\diag(\A))_i = A_{i, i}$ for $i \in [L]$.

We now define the matrix-valued AMP iteration in question. We start with {initialization $\B^0\in\reals^{p\times L}$ independent of $\tilde\X$}, and define $F^0: \B^0 \mapsto \B^0$, {$\bar G^0(\TTheta^0,\y):=\bprox_{\b\bpi M(\cdot,\y)}(\TTheta^0)$, $G_i^0:=(\bar G_i^0-\TTheta_i^0)\odot\b^{\odot(-1)}$, $\A^0:=\operatorname{diag}(\b)$}, $\hat{\B}^0 := F^0(\B^0)$, $\hat{\bR}^{-1}:=\0_{n\times L}$.
For $t \geq 0$, the AMP algorithm then computes:
\begin{align}
\begin{split}
\label{eq:matrix-valued-amp-1}
    &\TTheta^{t} = \tilde{\X} \hat{\B}^t -  \hat{\bR}^{t-1} (\A^t)^\top\,, \; \; \; \hat{\bR}^t = G^t\left(\TTheta^t, \y\right) \,, \;\;\; \hat{\TTheta}^t = \bar{G}^t(\TTheta^t, \y)  \\
    &\B^{t+1} = \tilde{\X}^\top \hat{\bR}^t -  \hat{\B}^{t} (\C^t)^\top\,, \; \; \; \hat{\B}^{t+1} = F^{t+1}\left(\B^{t+1}\right), \, \\
\end{split}
\end{align}
where $G^t: \R^{n\times L}\times \R^{n}\to \R^{n\times L}$ and $F^{t+1}: \R^{p\times L}\to \R^{p\times L}$ are
functions specified via the following recursive procedure.
Given $G^{t}$, we define the matrix $\C^{t}$ as follows:
\begin{align} \label{eq:matrix-amp-C-def}
    \C^{t} &= \lim_{n \to \infty} \frac{1}{n} \sum_{i=1}^n \E \, \partial_i{G^{t}_i}\left({\bLambda^\top\Z_i} + (\G_{\TTheta})_i, q(\Z_i, \Psi_i, \varepsilon_i)\right) \in \reals^{L \times L},
\end{align}
and define the successive $F^{t+1}$ function as follows:
\begin{align} \label{eq:matrix-amp-F-def}
    &F_j^{t+1}(\B_j^{t+1})
    = - \B_j^{t+1} \odot \diag(\C^t)^{\odot (-1)} \in \reals^L
\end{align}
for $j \in [p]$. We then define $\A^{t+1}$ as follows:
\begin{align} \label{eq:matrix-amp-A-t+1-diag}
    \A^{t+1} = - \frac{1}{\delta} \begin{bmatrix}
        1/C^t_{1, 1} & \cdots & 0 \\
        0 & \ddots & 0 \\
        0 & \cdots & 1/C^t_{L, L}
    \end{bmatrix} \in \reals^{L \times L},
\end{align}
for which $G^{t+1}$ is defined as:
\begin{align} \label{eq:matrix-amp-G-def}
    &\bar{G}^{t+1}(\TTheta^{t+1}, \y) = \bprox_{\diag(\A^{t+1}) \bpi M(\cdot, \y)}(\TTheta^{t+1}) \in \reals^{n \times L} \\
    &G_i^{t+1}(\TTheta^{t+1}_i, y_i) := \left(\bar{G}_i^{t+1}(\TTheta^{t+1}_i, y_i) - \TTheta^{t+1}_i\right) \odot \diag(\A^{t+1})^{\odot (-1)} \in \reals^{L},
\end{align}
for $i \in [n]$. We note that this yields a diagonal representation for $\C^{t+1}$ following formula \eqref{eq:matrix-amp-C-def}, where for $k, l \in [L]$ {with $k\neq l$} we have that $C^{t+1}_{k,l} = 0$ and
\begin{align}
   {C^{t+1}_{l,l}=\lim_{n\to\infty}\frac1n\sum_{i=1}^n\E\left[\left(\prox'_{A^{t+1}_{l,l}\pi_i^{(l)}M(\cdot,q(\Z_i,\Psi_i,\varepsilon_i))}(\Z_i^\top\blambda^{(l)}+w_i^{(l)})-1\right)/A^{t+1}_{l,l}\right].} \label{eq:matrix-valued-amp-C-diagonal-representation-L}
\end{align}
A similar diagonal representation holds for $\C^0$ by plugging $G^0$ into \eqref{eq:matrix-amp-C-def}.

We now state the convergence result. For $s, r \geq 0$, we define the random matrix $\G_{\TTheta}^s \in \reals^{n \times L}$ where, for $i \in [n]$, we let $(\G_{\TTheta}^s)_i \distas{\text{i.i.d.}} \N(\0_L, \bK)$ with $\cov((\G_{\TTheta}^r)_i, (\G_{\TTheta}^s)_i) = \bK^{(r, s)}$. For $s \geq 0$, we let $\G^s_{\hat{\B}} := (\G_{\TTheta}^s)_{[:p, :]}$ be a truncation of $\G_{\TTheta}^s$, where $\bK$ is defined in Appendix \ref{sec:nonlinear_equations}. The interaction matrices $\bK^{(r, s)}$ are defined recursively as follows:
\begin{align}
    &(\bK^{r + 1, s+1})_{[\ell, \ell']} \notag \\
    &= \lim_{n \to \infty} \frac{\delta b^{(\ell)} b^{(\ell')}}{n} \sum_{i = 1}^n \notag \\
    & \hspace{0.9cm} \E\left[\pi_i^{(\ell)} M' \left( \prox_{b^{(\ell)} \pi_i^{(\ell)} M(\cdot, q(\Z_i, \Psi_i, \varepsilon_i))}(\Z_i^{\top} \blambda^{(\ell)} + {w_i^{(\ell), r}}), q(\Z_i, \Psi_i, \varepsilon_i)) \right) \right. \notag \\
    &\left. \hspace{0.9cm} \cdot \, \pi_i^{(\ell')} M' \left( \prox_{b^{(\ell')} \pi_i^{(\ell')} M(\cdot, q(\Z_i, \Psi_i, \varepsilon_i))}(\Z^\top_i \blambda^{(\ell')} + {w_i^{(\ell'), s}}), q(\Z_i, \Psi_i, \varepsilon_i)) \right) \right], \label{eq:interaction-matrices}
\end{align}
where independently across $i \in [n]$ and for $\ell, \ell' \in [L]$, $s, r \geq 0$, we have ${w_i^{(\ell), r}} \distas{} \N(0, \kappa_{\ell, \ell})$, ${w_i^{(\ell'), s}} \distas{} \N(0, {\kappa_{\ell', \ell'}})$ and $\cov({w_i^{(\ell), r}}, {w_i^{(\ell'), s}}) = {(\bK^{r, s})}_{[\ell, \ell']}$. {The existence of the limit in \eqref{eq:interaction-matrices} follows by the averaging argument in the proof of Lemma \ref{lemma:general-pointwise-conv}, with Proposition \ref{prop:conv-nonlin-eqns} replacing $\varepsilon_i$ by $\bar{\varepsilon}$. Lemma \ref{lemma:prop-of-cross-cov-simple} gives continuity in the Gaussian covariance. For the initialization in Lemma \ref{lemma:SE-matrix-amp}, $\bK^{0,0}=\bK${ and $\bK^{0,s}=\bK^{s,0}=\0$ for $s\geq1$}.}
\begin{lemma} \label{lemma:SE-matrix-amp}
    Consider the setting in Section \ref{sec:setting} and the matrix-valued AMP algorithm in \eqref{eq:matrix-valued-amp-1}, with initialization $\hat{\B}^0 := \L^\top \B \bLambda + \sqrt{\delta} \G^{det}$, where $(\G^{det})_j$ are independent samples from a $\N(0, \bK)$ density, for $j \in [p]$, {independent of $(\X,\bvarepsilon)$}. For $t \geq 0$, and for any sequence of uniformly pseudo-Lipschitz functions $\varphi_{n}(\cdot \;; \bPsi, \bvarepsilon) : \reals^{n \times (L(t+2))} \to \reals$, $\varphi_{p}(\cdot \; ; \L^\top \B) : \reals^{p \times (L(t+1))} \to \reals$:
    \begin{align}
    &\varphi_n(\TTheta^0, \dots, \TTheta^t, \X \B ; \bPsi, \bvarepsilon) \stackrel{\P}{\simeq} \E \{ \varphi_n(\Z \bLambda + \G_{\TTheta}^0, \dots, \Z \bLambda + \G_{\TTheta}^t, \Z; \bPsi, \bvarepsilon)\} \label{eq:SE_Theta}, \\
    &\varphi_p(\hat{\B}^1, \dots, \hat{\B}^{t+1} ; \L^\top \B) \stackrel{\P}{\simeq} \E\{ \varphi_p(\L^\top \B \bLambda + \sqrt{\delta} \G_{\B}^1, \dots, \L^\top \B \bLambda + \sqrt{\delta} \G_{\B}^{t+1} ; \L^\top \B)\} \label{eq:SE_B},
    \end{align}
    as $n, p \to \infty$ with $n/p \to \delta$, where $\bLambda, \bK$ are solutions to \eqref{eq:b-fp-implicit}--\eqref{eq:cross-cov-fp-implicit}.
\end{lemma}
The proof, presented in Appendix \ref{sec:proof-SE-matrix-amp-red}, is an application of \cite[Theorem 1]{arpino_2025_jmlr} together with the chosen initialization, which guarantees that the asymptotic behaviour of the AMP iteration stays the same for all $t \geq 0$.
\begin{remark}
The expression in \eqref{eq:SE_B} differs from that in \cite[Theorem 1]{arpino_2025_jmlr} in that it is a convergence result for the iterates $\hat{\B}^t$ rather than $\B^t$, as is usual in the AMP literature \citep{gerbelot_graph-based_2023, berthier_state-evolution_2019, feng_unifying_2022}. This is because $F^{t+1}$ as defined in \eqref{eq:matrix-amp-F-def} produces a matrix whose $l$-th column is a scalar multiple of the $l$-th input column, and hence $\hat{\B}^t = {F^{t}(\B^t)}$ can be computed in closed form.
Moreover, Lemma \ref{lemma:ERM_frobenius_characterization} below shows that $\hat{\B}^t$  tracks the asymptotic behaviour of $\L^\top \hat{\B}$, where $\hat{\B} =[\hat{\bbeta}^{(1)}, \ldots, \hat{\bbeta}^{(L)}]$ is the set of signal estimates produced by \WeightedERM{} in \eqref{eq:ERM_est}.
\end{remark}

\paragraph{AMP tracks \WeightedERM{}} The second key step in the proof of Theorem \ref{thm:ERM_general_characterization_pres} is the following convergence result relating AMP and the \WeightedERM{} estimator \eqref{eq:ERM_est}.
\begin{lemma} \label{lemma:ERM_frobenius_characterization}
    Consider the setting in Section \ref{sec:setting}. Then, for any $\epsilon > 0$, we have:
    \begin{align}
        &\lim_{t \to \infty} \lim_{p \to \infty} \P\left[\left| \frac{1}{p} \| \L^\top \hat{\B} -  \hat{\B}^t \|^2_F \right| > \epsilon \right] = \lim_{t \to \infty} \lim_{n \to \infty} \P\left[\left| \frac{1}{n} \| \hat{\TTheta} -  \hat{\TTheta}^t \|^2_F \right| > \epsilon \right] = 0,
    \end{align}
    where $\hat{\TTheta}$, $\hat{\B}$ are defined in \eqref{eq:Theta-estimate-matrix}-\eqref{eq:B-estimate-matrix} and
    $\hat{\TTheta}^t,\hat{\B}^t$ are the AMP iterates defined in  \eqref{eq:matrix-valued-amp-1}.
\end{lemma}
The proof of Lemma \ref{lemma:ERM_frobenius_characterization}, given in Appendix \ref{sec:proof-ERM_frobenius_characterization}, invokes Taylor's theorem and the curvature assumption on the objective, Assumption~\ref{asmpt:reg-loss}(\ref{cond:likelihood_curvature}), to prove that the gradient of $\cC^{(\ell)}$ in \eqref{eq:ERM_est} is bounded above by the difference of successive AMP iterates. The result is then obtained by applying a Cauchy convergence property of the iterates of the non-separable AMP in \eqref{eq:matrix-valued-amp-1}
(Lemma \ref{lemma:cauchy_t_plus_h}), which may be of independent interest.

We are now ready to prove Theorem \ref{thm:ERM_general_characterization_pres} and Proposition \ref{prop:ERM_hausdorff_characterization_pres}.

\begin{proof}[Proof of Theorem \ref{thm:ERM_general_characterization_pres}.]
The uniqueness of the solution to \eqref{eq:b-fp-implicit}--\eqref{eq:cross-cov-fp-implicit} follows from Proposition \ref{prop:uniqueness_of_fixed_points}.
The proof of the convergence results \eqref{eq:SE_ERM_B}-\eqref{eq:SE_ERM_Theta} relies on the pseudo-Lipschitz property of the proximal operator, which follows from  \eqref{eq:prox_id_3} and Assumption \ref{asmpt:reg-loss}(\ref{cond:strict-cvx-smooth}).
We have that, for $\epsilon > 0$:
\begin{align*}
&\limsup_{n \to \infty} \P\left[ \left| \varphi_n(\hat{\TTheta}, \X \B ; \bPsi, \bvarepsilon) - \right. \right. \\
&\hspace{2cm} \left. \left.  \E \left\{ \varphi_n\left(\bprox_{\b \bpi M(\cdot, q(\Z, \bPsi, \bvarepsilon))}\left(\Z \bLambda + \G_{\TTheta}\right), \Z ; \bPsi, \bvarepsilon \right) \right\} \right| > \epsilon\right]  \\
&  \leq \limsup_{t \to \infty} \; \limsup_{n \to \infty} \; (K_{t, n} + J_{t, n}),
\end{align*}
where,
\begin{align}
&K_{t, n} := \P\left[ \left| \varphi_n\left(\hat{\TTheta}, \X\B ; \bPsi, \bvarepsilon \right) - \varphi_n\left(\hat{\TTheta}^t, \X \B ; \bPsi, \bvarepsilon\right) \right| > \epsilon/2 \right], \nonumber \\
&J_{t, n} := \P\left[ \left|\varphi_n\left(\hat{\TTheta}^t, \X \B ; \bPsi, \bvarepsilon\right) \right. \right. \nonumber \\
&\hspace{2cm}\left. \left. - \E \left\{ \varphi_n\left(\bprox_{\b \bpi M(\cdot, q(\Z, \bPsi, \bvarepsilon))}\left(\Z \bLambda + \G_{\TTheta}\right), \Z ; \bPsi, \bvarepsilon \right) \right\}\right| > \epsilon/2 \right]. \nonumber
\end{align}
From state evolution of the stationary matrix-valued AMP (Lemma \ref{lemma:SE-matrix-amp}), and from the pseudo-Lipschitz hypothesis on $\varphi_n$ (noting that $(\tilde{\TTheta}, \tilde{\y}) \mapsto \bprox_{\b \bpi M(\cdot, \tilde{\y})}(\tilde{\TTheta})$ is pseudo-Lipschitz as a consequence of \eqref{eq:prox_id_3}, Assumption \ref{asmpt:reg-loss}(\ref{cond:strict-cvx-smooth}), Assumption \ref{asmpt:reg-weights}), we have that $\lim_{n \to \infty} J_{t, n} = 0$ for any $t \geq 0$. We now  prove that $\lim_{t \to \infty} \lim_{n \to \infty} K_{t, n} = 0$. We note that by pseudo-Lipschitzness of $\varphi_{n}(\cdot \;; \bPsi, \bvarepsilon)$ we have that:
\begin{align}
   &\left| \varphi_n\left(\hat{\TTheta}, \X \B ; \bPsi, \bvarepsilon \right) - \varphi_n\left(\hat{\TTheta}^t, \X \B ; \bPsi, \bvarepsilon\right) \right| \notag \\
   &\leq L \left(1 + \left(\frac{\|\hat{\TTheta}\|_F}{\sqrt{n}} \right)^{r-1} + \left( \frac{\|\hat{\TTheta}^t\|_2}{\sqrt{n}} \right)^{r - 1} \right) \frac{\|\hat{\TTheta} - \hat{\TTheta}^t \|_F}{\sqrt{n}}. \label{eq:PL_phi_n}
\end{align}
By existence of the estimator in question, and by the fact that $\lim_{n \to \infty} \|\X \L^{-\top}\|_2 < \infty$ almost surely \citep{Anderson_Guionnet_Zeitouni_2009}, we have that $\|\hat{\TTheta}\|_F / \sqrt{n} < \infty$ with high probability. Moreover, from the definition of the AMP iterates in \eqref{eq:matrix-valued-amp-1}--\eqref{eq:matrix-amp-F-def}, we have that:
\begin{align} \label{eq:TTheta_hat_bound}
    \|\hat{\TTheta}^{t+1}\|_F / \sqrt{n} &= \|\X \L^{-\top} \hat{\B}^{t + 1} + \TTheta^t - \TTheta^{t + 1} \|_F / \sqrt{n} \nonumber \\
    &\leq \left( \| \X \L^{-\top} \|_2 \|\hat{\B}^{t + 1} \|_F + \|\TTheta^t\|_F + \|\TTheta^{t + 1} \|_F \right) / \sqrt{n}.
\end{align}
It follows from Lemma \ref{lemma:SE-matrix-amp} that $\|\hat{\B}^t\|_F^2 / p \stackrel{\P}{\simeq} \delta \|\TTheta^t\|_F^2 / n \stackrel{\P}{\simeq} \delta (\bLambda)^\top \bGamma \bLambda + \delta \bK$, and hence the RHS of \eqref{eq:TTheta_hat_bound} is bounded with high probability by observing that $\linebreak[4] \lim_{n \to \infty} \|\X \L^{-\top}\|_2 < \infty$ almost surely \citep{Anderson_Guionnet_Zeitouni_2009}. These observations when applied to \eqref{eq:PL_phi_n}, together with Lemma \ref{lemma:ERM_frobenius_characterization}, yield:
\begin{align*}
&\lim_{t \to \infty} \lim_{n \to \infty} K_{t, n} \\
&= \lim_{t \to \infty} \lim_{n \to \infty} \P\left[ \left| \varphi_n\left(\hat{\TTheta}, \X\B ; \bPsi, \bvarepsilon \right) - \varphi_n\left(\hat{\TTheta}^t, \X \B ; \bPsi, \bvarepsilon\right) \right| > \epsilon/2 \right] \\
&= 0.
\end{align*}
This proves \eqref{eq:SE_ERM_Theta}.
Similarly, \eqref{eq:SE_ERM_B} follows from the pseudo-Lipschitz assumption on $\varphi_p$, Lemma \ref{lemma:ERM_frobenius_characterization}, boundedness of $\|\hat{\B}\|^2_F / n$ via the existence of the estimator in question, boundedness of $\|\hat{\B}^t\|_F^2 / n$ via Lemma \ref{lemma:SE-matrix-amp} and, finally, Lemma \ref{lemma:SE-matrix-amp} applied to the pseudo-Lipschitz function $(\A^{(1)}, \A^{(2)}) \mapsto \varphi_p( \L^{-\top} \A^{(1)}; \L^{-\top} \A^{(2)})$.
\end{proof}
\begin{proof}[Proof of Proposition \ref{prop:ERM_hausdorff_characterization_pres}.]
     We use the following lemma, which states that the Hausdorff distance defined in \eqref{eq:hausdorff-dist} preserves the pseudo-Lipschitz property:
\begin{lemma} \label{lemma:hausdorff_PL}
    Let $\hat{\eeta}(\hat{\TTheta}^t, \y)$ be an estimator such that \[(\V, \tilde{\y}) \mapsto U(\hat{\eeta}(\V, \tilde{\y})) \] is uniformly pseudo-Lipschitz, where we recall that  $U$
is the one-to-one mapping between change point vectors $\hat{\eeta}$ and signal configuration vectors $\hat{\bPsi}$. Then $(\V, \tilde{\y}) \mapsto d_H(\eeta, \hat{\eeta}(\V, \tilde{\y})) / n$ is uniformly pseudo-Lipschitz.
\end{lemma}
The proof of Lemma \ref{lemma:hausdorff_PL} is identical to the first part of the proof of  \cite[Proposition $3$]{arpino_2025_jmlr}.  The next lemma shows that under a Lipschitz link $q$, if $(\V, \tilde{\y}) \mapsto U(\hat{\eeta}(\V, \tilde{\y}))$ is uniformly pseudo-Lipschitz so is the map $(\V, \z) \mapsto U(\hat{\eeta}(\V, q(\z, \bPsi, \bvarepsilon)))$.

\begin{lemma}\label{lemma:unc-implies-comp}
Suppose the link $q$ is $L_q$-Lipschitz in its first argument, uniformly in
$\varepsilon$, and that $c_0 := \sup_n \|q(\0, \bPsi, \bvarepsilon)\| / \sqrt{n} < \infty$.
Let $f_n(\cdot\,; \bPsi, \bvarepsilon) : \reals^{n \times (L+1)} \to \reals^{{m}}$
be uniformly pseudo-Lipschitz of order $r$ with constant $C$. Then
\[
  (\V, \z) \;\longmapsto\; f_n\big(\V,\, q(\z, \bPsi, \bvarepsilon);\, \bPsi, \bvarepsilon\big)
\]
is uniformly pseudo-Lipschitz of the same order $r$, with a constant depending
only on $C$, {$r$,} $L_q$ and $c_0$.
\end{lemma}
\begin{proof}
Write $g(\z) := q(\z, \bPsi, \bvarepsilon) \in \reals^n$, so that
$g(\z)_i = q(z_{i, \Psi_i}, \varepsilon_i)$ for $i \in [n]$. Since each row of
$\z$ contributes exactly one entry and $q(\cdot, \varepsilon)$ is
$L_q$-Lipschitz,
\begin{equation} \label{eq:g-lip}
  \|g(\z) - g(\tilde{\z})\|^2
  = \sum_{i=1}^n \big| q(z_{i, \Psi_i}, \varepsilon_i) - q(\tilde{z}_{i, \Psi_i}, \varepsilon_i) \big|^2
  \le L_q^2 \sum_{i=1}^n \big| z_{i, \Psi_i} - \tilde{z}_{i, \Psi_i} \big|^2
  \le L_q^2 \, \|\z - \tilde{\z}\|_F^2 ,
\end{equation}
so $g$ is $L_q$-Lipschitz
to
$(\reals^n, \|\cdot\|)$, and $\|g(\z)\| \le \|g(\0)\| + L_q \|\z\|_F$.

Stacking the arguments as $[\V, g(\z)] \in \reals^{n \times (L+1)}$, by
\eqref{eq:g-lip}, we have
\begin{equation} \label{eq:stack-lip}
  \big\| [\V, g(\z)] - [\tilde{\V}, g(\tilde{\z})] \big\|_F^2
  = \|\V - \tilde{\V}\|_F^2 + \|g(\z) - g(\tilde{\z})\|^2
  \le (1 + L_q^2) \, \big\| [\V, \z] - [\tilde{\V}, \tilde{\z}] \big\|_F^2 .
\end{equation}
Likewise, using $\|g(\z)\| \le \|g(\0)\| + L_q\|\z\|_F$ and
$\|g(\0)\|/\sqrt{n} \le c_0$ \, ,
\begin{equation} \label{eq:stack-norm}
  \frac{\big\| [\V, g(\z)] \big\|_F}{\sqrt{n}}
  \;\le\; c_0 + (1 + L_q)\, \frac{\big\| [\V, \z] \big\|_F}{\sqrt{n}} .
\end{equation}
Applying the pseudo-Lipschitz bound \eqref{eq:PL_def} for $f_n$ at the points
$[\V, g(\z)]$ and $[\tilde{\V}, g(\tilde{\z})]$, and then substituting
\eqref{eq:stack-lip} and \eqref{eq:stack-norm}, gives
\begin{align*}
 &  \frac{\big\| f_n(\V, g(\z)) - f_n(\tilde{\V}, g(\tilde{\z})) \big\|}{\sqrt{{m}}} \\
&   \;\le\; C' \left( 1 + \Big( \tfrac{\|[\V, \z]\|_F}{\sqrt{n}} \Big)^{r-1}
                      + \Big( \tfrac{\|[\tilde{\V}, \tilde{\z}]\|_F}{\sqrt{n}} \Big)^{r-1} \right)
            \frac{\big\| [\V, \z] - [\tilde{\V}, \tilde{\z}] \big\|_F}{\sqrt{n}} \, ,
\end{align*}
for a constant $C'$ depending only on $C$, $r$, $L_q$ and $c_0$, where we used
$(a + b)^{r-1} \le 2^{r-1}(a^{r-1} + b^{r-1})$ to expand \eqref{eq:stack-norm}.
Since $L_q$ and $c_0$ do not depend on $n$, the family is uniformly
pseudo-Lipschitz. Taking $f_n = U \circ \hat{\eeta}$ gives the claim.
\end{proof}
The proof of Proposition \ref{prop:ERM_hausdorff_characterization_pres} is then a direct application of Theorem \ref{thm:ERM_general_characterization_pres} and {Lemmas \ref{lemma:hausdorff_PL}--\ref{lemma:unc-implies-comp}}. {The bound on $c_0$ follows from the Lipschitz assumption on $q$ and the bounded empirical second moment of $\bvarepsilon$ in Section \ref{sec:setting}.}
\end{proof}
\subsection{Proof of Theorem \ref{thm:characterization_of_debiased_est_pres} and Proposition \ref{prop:pointwise_posterior_pres}}
\label{subsec:debiased_est_posterior_proofs}
The proof of Theorem \ref{thm:characterization_of_debiased_est_pres} applies the same high-level techniques used to prove Theorem \ref{thm:ERM_general_characterization_pres}, with added details relating to the adjusted estimate $\TTheta^{\adj}$. We first have the following lemma, which shows that $\hat{b}^{(\ell)}$ in \eqref{eq:b-hat-eqn} and $b^{(\ell)}$ in \eqref{eq:b-fp-implicit} are asymptotically equivalent, for $\ell \in [L]$.
\begin{lemma} \label{lemma:b-hat-conv}
Let $\tilde{b} \in \reals_{\geq 0}$ and let $\ell \in [L]$ with superscripts omitted. Then, we have that:
\begin{align*}
    \frac{1}{n} \sum_{i=1}^n  \left(1 + \tilde{b} \pi_i M''(\hat{\theta}_i, y_i)\right)^{-1} \stackrel{\P}{\simeq} \frac{1}{n} \sum_{i=1}^n \E\left[ \left(1 + \tilde{b} \sM_i'' \right)^{-1} \right],
\end{align*}
where the expectation is taken over $\Z_i \distas{} \N(\0_L, \bGamma)$ and $w_i \distas{} \N(0, \kappa_{\ell, \ell})$ independently over $i \in [n]$ as defined in \eqref{eq:sM-def}.
\end{lemma}
\begin{proof}
For $\u, \v \in \reals^n$, the family of functions \[\tilde{\varphi}_n(\u, \v) := \frac{1}{n} \sum_{i=1}^n  \left(1 + \tilde{b} \pi_i M''(a_i, d_i)\right)^{-1}\] indexed by $n \in \naturals$ is uniformly Lipschitz by the mean value theorem argument used in Lemma \ref{lemma:equicontinuous_h_i}. For $\ell \in [L]$, the result then follows by applying Theorem \ref{thm:ERM_general_characterization_pres} to $\varphi_n(\hat{\TTheta}, \X \B; \bPsi, \bvarepsilon) := \tilde{\varphi}_n\left( \hat{\TTheta}_{[:, \ell]}, q(\X\B, \bPsi, \bvarepsilon) \right) = \tilde{\varphi}_n\left( \hat{\ttheta}, \y\right)$.
\end{proof}
\begin{proof}[Proof of Theorem \ref{thm:characterization_of_debiased_est_pres}.]
For $\ell \in [L]$, we omit the superscripts when these can be inferred from context. In light of Lemma \ref{lemma:b-hat-conv}, it suffices to prove the theorem statement for $\ttheta^{\adj}$ in \eqref{eq:theta-deb-defn} with $\hat{b}$ replaced by $b$. We have that, for $\epsilon > 0$:
\begin{align*}
&\limsup_{n \to \infty} \P\left[ \left| \varphi_n(\TTheta^{\adj}, \X \B ; \bPsi, \bvarepsilon) - \E \left\{ \varphi_n\left(\Z \bLambda + \G_{\TTheta}, \Z ; \bPsi, \bvarepsilon \right) \right\} \right| > \epsilon\right]  \\
& \hspace{4cm} \leq \limsup_{t \to \infty} \; \limsup_{n \to \infty} \; (K_{t, n} + J_{t, n}),
\end{align*}
where,
\begin{align}
&K_{t, n} := \P\left[ \left| \varphi_n\left({\TTheta}^{\adj}, \X\B ; \bPsi, \bvarepsilon \right) - \varphi_n\left({\TTheta}^t, \X \B ; \bPsi, \bvarepsilon\right) \right| > \epsilon/2 \right], \\
&J_{t, n} := \P\left[ \left|\varphi_n\left({\TTheta}^t, \X \B ; \bPsi, \bvarepsilon\right) - \E \left\{ \varphi_n\left(\Z \bLambda + \G_{\TTheta}, \Z ; \bPsi, \bvarepsilon \right) \right\}\right| > \epsilon/2 \right].
\end{align}
From state evolution of the stationary matrix-valued AMP (Lemma \ref{lemma:SE-matrix-amp}), and from the pseudo-Lipschitz hypothesis on $\varphi_n$, we have that $\lim_{n \to \infty} J_{t, n} = 0$ for any $t \geq 0$. We now  prove that $\lim_{t \to \infty} \lim_{n \to \infty} K_{t, n} = 0$. We note that by pseudo-Lipschitzness of $\varphi_{n}(\cdot \;; \bPsi, \bvarepsilon)$ we have that:
\begin{align}
   &\left| \varphi_n\left({\TTheta}^{\adj}, \X \B ; \bPsi, \bvarepsilon \right) - \varphi_n\left({\TTheta}^t, \X \B ; \bPsi, \bvarepsilon\right) \right| \notag \\
   &\leq L \left(1 + \left(\frac{\|\TTheta^{\adj}\|_F}{\sqrt{n}} \right)^{r-1} + \left( \frac{\|{\TTheta}^t\|_2}{\sqrt{n}} \right)^{r - 1} \right) \frac{\|{\TTheta}^{\adj} - {\TTheta}^t \|_F}{\sqrt{n}}.
\end{align}
By Lemma \ref{lemma:SE-matrix-amp}, we have that $\|\TTheta^t\|^2_F / n \stackrel{\P}{\simeq} (\bLambda)^\top \bGamma \bLambda + \bK$. We proceed to bound $\frac{\|\TTheta^{\adj}\|_F}{\sqrt{n}}$ and $\frac{\|{\TTheta}^{\adj} - {\TTheta}^t \|_F}{\sqrt{n}}$. Noting that
  $  \| \TTheta^{\adj} - \TTheta^t \|^2_F = \sum_{\ell = 1}^L \|\ttheta^{(\ell), \adj} - \ttheta^{(\ell), t} \|_2^2$,
 it suffices to bound $\frac{\|\ttheta^{\adj}\|_2}{\sqrt{n}}$ and $\frac{\|{\ttheta}^{\adj} - {\ttheta}^t \|_F}{\sqrt{n}}$ (dropping the superscript $\ell$ for brevity). Applying the triangle inequality and the assumption that $\sup_{u, v} M''(u, v) < \infty$, we obtain:
\begin{align*}
    \|\ttheta^{\adj} \|_2 & \leq \| \hat{\ttheta} \|_2 + b \|\m'(\hat{\ttheta}, \y)\|_2 \\
    &\leq \| \hat{\ttheta} \|_2 + b \tilde{C} \|\hat{\ttheta}\|_2 + b \|\m'(\0_n, \y) \|_2,
\end{align*}
for some large enough constant $\tilde{C} < \infty$. Boundedness of $\| \hat{\ttheta} \|^2_2 / n$ then follows directly from Theorem \ref{thm:ERM_general_characterization_pres}, and hence we have proven that $\frac{\|\TTheta^{\adj}\|_F}{\sqrt{n}}$ is bounded. Next, we let $\prox_{b \m(\cdot, \y)}(\ttheta^{t-1})$ denote a vector such that $(\prox_{b \m(\cdot, \y)}(\ttheta^{t-1}))_i = \prox_{b \pi_i M(\cdot, y_i)}(\theta^{t-1}_i)$ for $i \in [n]$. We expand $\ttheta^{\adj}, \ttheta^t$ and apply proximal identity \eqref{eq:prox_id_2} together with the definition of $g$ in \eqref{eq:den_choice} to obtain:
\begin{align*}
   \| \ttheta^{\adj} - \ttheta^t \|_2
   &\leq \left\| \hat{\ttheta} + b \m'(\hat{\ttheta}, \y) - \tilde{\X} \hat{\bzeta}^t + b g^{t-1}(\ttheta^{t-1}, \y) \right\|_2 \\
   &\leq \left\| \hat{\ttheta} - \tilde{\X} \hat{\bzeta}^t \right\|_2 + b \left\| \m'(\hat{\ttheta}, y) - \m'\left(\prox_{b \m(\cdot, \y)}(\ttheta^{t-1}), \y \right) \right\|_2,
\end{align*}
where we recall $\tilde{\X} := \X \L^{-\top}$ as per Lemma \ref{lem:covariance-reduction_pres}.
We then apply the definition $\hat{\ttheta} = \X \hat{\bbeta}$ and the assumption that $\sup_{u, v} \sup_{i \in [n]} \pi_i M''(u, v) < \infty$ to obtain the upper bound:
\begin{align*}
   \| \ttheta^{\adj} - \ttheta^t \|_2
   &\leq \|\tilde{\X} \|_{op} \|\L^\top \hat{\bbeta} - \hat{\bzeta}^t\|_2 + b \tilde{C} \left\| \hat{\ttheta} - \prox_{b \m(\cdot, \y)}(\ttheta^{t-1})\right\|_2, \\
   &= \|\tilde{\X} \|_{op} \|\L^\top \hat{\bbeta} - \hat{\bzeta}^t\|_2 + b \tilde{C} \left\| \hat{\ttheta} - \hat{\bvartheta}^t\right\|_2,
\end{align*}
where $\tilde{C} < \infty$ is a large enough constant. By Lemma \ref{lemma:ERM_frobenius_characterization} and by the fact that $\lim_{n \to \infty} \|\tilde{\X}\|_2 < \infty$ almost surely \citep{Anderson_Guionnet_Zeitouni_2009}, we then have that $\| \TTheta^{\adj} - \TTheta^t \|^2_F / n \to 0$ in probability, and the result is proved.
\end{proof}
\begin{proof}[Proof of Proposition \ref{prop:pointwise_posterior_pres}.]
Let $\varphi_n(\V, \z ; \bPsi, \bvarepsilon) := p(\bpsi | \V, q(\z, \bPsi, \bvarepsilon))$. By assumption, $\varphi_n$ is uniformly pseudo-Lipschitz with respect to $(\V, \z)$. Applying Theorem \ref{thm:characterization_of_debiased_est_pres} to $\varphi_n$, we obtain:
\begin{align*}
    p(\bpsi | \TTheta^{\adj}, \y) = p(\bpsi | \TTheta^{\adj}, q(\X \B, \bPsi, \bvarepsilon)) \stackrel{\P}{\simeq}  \E \, [p(\bpsi | \Z \bLambda + \G_{\TTheta}, q(\Z, \bPsi, \bvarepsilon))] =: \mu_p(\bpsi).
\end{align*}
Similarly, we apply \citep[Lemma 19]{gerbelot_graph-based_2023}, a concentration result regarding pseudo-Lipschitz functions of Gaussian random variables, to $\varphi_n(\Z \bLambda + \G_{\TTheta}, \Z; \bPsi, \bvarepsilon)$, to obtain $p(\bpsi | \Z \bLambda + \G_{\TTheta}, q(\Z, \bPsi, {\bvarepsilon})) \stackrel{\P}{\simeq} \mu_p(\bpsi)$.
Combining the aforementioned results, we obtain that for any $\epsilon >0$:
\begin{align*}
    &\P[ | p(\bpsi | \TTheta^{\adj}, \y) - p(\bpsi | \Z \bLambda + \G_{\TTheta}, q(\Z, \bPsi, {\bvarepsilon}))| > \epsilon] \\
    &\leq \P[|p(\bpsi | \TTheta^{\adj}, \y) -  \mu_p(\bpsi)| > \epsilon/2] \\
    &\hspace{4cm} + \P[|p(\bpsi | \Z \bLambda + \G_{\TTheta}, q(\Z, \bPsi, {\bvarepsilon})) - \mu_p(\bpsi)| > \epsilon/2]
    \\
    & \stackrel{n \to \infty}{\longrightarrow} 0.
\end{align*}
\end{proof}
\subsection{Proof of Lemma \ref{lemma:SE-matrix-amp}}\label{sec:proof-SE-matrix-amp-red}
We first define a vector-valued AMP iteration that will facilitate the proof of Lemma \ref{lemma:SE-matrix-amp}.
\paragraph{Vector-valued AMP iteration} Let $\ell \in [L]$. Recall the isotropic design matrix $\tilde{\X}$ defined in Lemma \ref{lem:covariance-reduction_pres}. Starting with an initializer $\hat{\bzeta}^0\in\reals^{p}$ and defining {$b_0:=b^{(\ell)}$, $\hat\s^{-1}:=\0_n$, $f^0(w):=w$,}  for $t \geq 0$ the iteration reads:
\begin{align}
\begin{split}
\label{eq:vector-valued-amp}
    &\ttheta^{t} = \tilde{\X} \hat{\bzeta}^t - {b_t} \hat{\s}^{t-1}\,, \; \; \; \hat{\s}^t = g^{(\ell), t}\left(\ttheta^t, \y\right) \, , \; \; \; \hat{\bvartheta}^t := \bar{g}^{(\ell), t}(\ttheta^t, \y), \,\\
    &\bbeta^{t+1} = \tilde{\X}^\top \hat{\s}^t -  c_t \hat{\bzeta}^t\,, \; \; \; \hat{\bzeta}^{t+1} = f^{t+1}\left(\bbeta^{t+1}\right), \, \\
\end{split}
\end{align}
where the denoiser functions $g^{(\ell), t}: \R^{n}\times \R^{n}\to \R^{n}$ and $f^t: \R \to \R$ are used to define the quantities ${b}_t, {c}_t$ as follows:
\begin{align}
\begin{split}\label{eq:c_b}
    &c_t := \lim_{n \to \infty} n^{-1} \sum_{i = 1}^n \E (g_i^{(\ell), t})'(\Z_i^{\top} \blambda^{(\ell)} + w_i^{(\ell)}, q(\Z_i, \Psi_i, \varepsilon_i){)}, \\
    &b_{t + 1} := \lim_{n \to \infty} n^{-1} \sum_{j = 1}^p (f^{t+1})'(\beta_j^{t + 1}),
\end{split}
\end{align}
where primes $'$ denote derivatives with respect to their first argument. The denoisers are defined as:
\begin{align}
    \bar{g}_i^{(\ell), t}(u, v) := \prox_{b_t \pi_i M(\cdot, v)}(u)\, , g_i^{(\ell), t}(u, v) := (\bar{g}_i^{(\ell), t}(u, v) - u) / b_t \, , f^{t+1}(w) := - w/c_t \label{eq:den_choice}
\end{align}
for $i \in [n]$, where $\prox_{\eta M}(z) := \argmin_{t \in \reals} \left\{ \eta M(t) + (t - z)^2 / 2\right\}$. We note that the definition of $f^{t+1}(w)$ and $b_{t+1}$ above implies $f^{t+1}(w) = \delta b_{t+1} w$.

The above vector-valued AMP iteration has the important property that it tracks the $\ell$-th column of the iterates in the matrix-valued AMP iteration previously defined in \eqref{eq:matrix-valued-amp-1}. Indeed, the diagonal representation of $\A^{t+1}$ in \eqref{eq:matrix-amp-A-t+1-diag} and of $\C^{t+1}$ in \eqref{eq:matrix-valued-amp-C-diagonal-representation-L} when combined with \eqref{eq:den_choice} yield:
\begin{align} \label{eq:matrix-amp-F-G-other-forms}
\begin{split}
    &F_j^{t+1}(\B_j^{t+1}) = \begin{bmatrix}
        {f^{t+1}((B_j^{t+1})_1)} & \dots & {f^{t+1}((B_j^{t+1})_L)}
    \end{bmatrix}, \\
    &\bar{G}_i^t(\TTheta^t_i, y_i) = \begin{bmatrix}
        \bar{g}_i^{(1), t}((\Theta^t_i)_1, y_i) & \dots & \bar{g}_i^{(L), t}((\Theta^t_i)_L, y_i)
    \end{bmatrix}, \\
    &G_i^t(\TTheta^t_i, y_i) = \begin{bmatrix}
        g_i^{(1), t}((\Theta^t_i)_1, y_i) & \dots & g_i^{(L), t}((\Theta^t_i)_L, y_i)
    \end{bmatrix}, \\
\end{split}
\end{align}
where $f^t, g^{(\ell), t}, \bar{g}^{(\ell), t}$ are defined in \eqref{eq:den_choice}, {with $f^t$ chosen for the corresponding column}. The following result verifies that the columns of the iterates of the matrix-valued AMP in \eqref{eq:matrix-valued-amp-1} are tracked by the vector-valued AMP iterates in \eqref{eq:vector-valued-amp}.
\begin{proposition} \label{prop:matrix-vector-AMP-red}
    Let $t \geq 0, \tilde{\ell} \in [L]$ and $(\bbeta^{t+1}, \ttheta^{t}, \hat{\bzeta}^{t+1}, \hat{\bvartheta}^{t}, \hat{\s}^t)$ be the iterates in \eqref{eq:vector-valued-amp} with $\ell = \tilde{\ell}$ {and $\hat\bzeta^0=\hat\B^0_{[:,\tilde\ell]}$}. Consider the matrix-valued AMP iterates in \eqref{eq:matrix-valued-amp-1}. We have that:
    \begin{align*}
        \B^{t+1}_{[:, \tilde{\ell}]} = \bbeta^{t + 1}, \;\;\; \TTheta^t_{[:, \tilde{\ell}]} = \ttheta^t, \;\;\; \hat{\B}^{t + 1}_{[:, \tilde{\ell}]} = \hat{\bzeta}^{t + 1}, \;\;\; \hat{\TTheta}^t_{[:, \tilde{\ell}]} = \hat{\bvartheta}^t, \;\;\; \hat{\bR}^t_{[:, \tilde{\ell}]} = {\hat{\s}^t}.
    \end{align*}
    In other words, the $\tilde{\ell}$-th column of the iterates \eqref{eq:matrix-valued-amp-1} corresponds to $(\bbeta^{t+1}, \ttheta^{t}, \allowbreak \hat{\bzeta}^{t+1}, \allowbreak \hat{\bvartheta}^{t}, \allowbreak \hat{\s}^t)$.
\end{proposition}
\begin{proof}
Plugging \eqref{eq:matrix-amp-F-def} and \eqref{eq:matrix-amp-G-def} into \eqref{eq:matrix-valued-amp-1} and noticing that $\A^t, \C^t$ are diagonal matrices, we obtain the following recursion:
\begin{align*}
\begin{split}
    &\TTheta^{t} = \begin{bmatrix} \tilde{\X} f^t(\B^t_{[:, 1]}) & \dots & \tilde{\X} f^t(\B^t_{[:, L]}) \end{bmatrix} \\
    &\hspace{2cm}-  \begin{bmatrix}
    \hat{\bR}_{[:, 1]}^{t-1} \cdot {b_t^{(1)}} & \dots & \hat{\bR}_{[:, L]}^{t-1} \cdot {b_t^{(L)}}
    \end{bmatrix}, \\
    &\hat{\bR}^t = G^t\left(\TTheta^t, \y\right), \\
    &\B^{t+1} = \begin{bmatrix}
        \tilde{\X}^\top g^{(1), t}(\TTheta^t_{[:, 1]}, \y) & \dots & \tilde{\X}^\top g^{(L), t}(\TTheta^t_{[:, L]}, \y)
    \end{bmatrix} \\
    &\hspace{2cm} - \left[
        \hat{\B}_{[:, 1]}^t \cdot {c_t^{(1)}} \; \dots \right. \\
    &\hspace{3cm} \left. \hat{\B}_{[:, L]}^t \cdot {c_t^{(L)}}
    \right], \\
    &\hat{\B}^{t+1} = F^{t+1}\left(\B^{t+1}\right),
\end{split}
\end{align*}
from which the result follows.
\end{proof}

We also show that it is possible to isolate the variables $\blambda^{(\ell)}$ for $\ell \in [L]$ on the LHS of equations \eqref{eq:b-fp-implicit}--\eqref{eq:cross-cov-fp-implicit}. This is necessary for the proof of Lemma \ref{lemma:SE-matrix-amp} below, and follows from a Stein's Lemma argument.
\begin{proposition}\label{prop:steins-lemma-fp}
Recall the notation in \eqref{eq:sM-def-proof}. Consider the model assumptions outlined in Section \ref{sec:setting} and suppose Assumption \ref{asmpt:reg-loss}(\ref{eq:loss-asmpt-log-conc}) holds. Moreover, for $i \in [n]$, assume $q$ is differentiable with respect to its first argument. For $\ell \in [L]$, equations \eqref{eq:b-fp-implicit}--\eqref{eq:cross-cov-fp-implicit} can then be rearranged to yield the {equivalent representation}:
\begin{align}
    1 - \frac{1}{\delta} &= \lim_{n \to \infty} \frac{1}{n} \sum_{i=1}^n \E\left[ \left(1 + b^{(\ell)} ({\sM}_i^{(\ell)})'' \right)^{-1} \right],
    \label{eq:b-fp-form}
    \\
    \blambda^{(\ell)} &= \lim_{n \to \infty} \frac{\delta b^{(\ell)}}{n} \sum_{i = 1}^n \E\left[ - \left( \frac{{\partial_{12}} {\sM}^{(\ell)}_i}{b^{(\ell)} ({\sM}^{(\ell)}_i)'' + 1} \right) \nabla_{\Z_i} q(\Z_i, \Psi_i, \varepsilon_i) \right],
    \label{eq:mu-fp-form}
    \\
    \kappa_{\ell, \ell'} &= \lim_{n \to \infty} \frac{\delta b^{(\ell)} b^{(\ell')}}{n} \sum_{i = 1}^n \E \left[ ({\sM}_i^{(\ell)})' ({\sM}_i^{(\ell')})' \right]
    \label{eq:cross-cov-fp-form}
    .
\end{align}
Similarly, equations \eqref{eq:b-fp-implicit-bar}--\eqref{eq:cross-cov-fp-implicit-bar} yield the equivalent representation:
\begin{align}
    1 - \frac{1}{\delta} &= \lim_{n \to \infty} \frac{1}{n} \sum_{i=1}^n \E\left[ \left(1 + b^{(\ell)} ({\sbM}_i^{(\ell)})'' \right)^{-1} \right],
    \label{eq:b-fp-form-bar}
    \\
    \blambda^{(\ell)} &= \lim_{n \to \infty} \frac{\delta b^{(\ell)}}{n} \sum_{i = 1}^n \E\left[ - \left( \frac{{\partial_{12}} {\sbM}^{(\ell)}_i}{b^{(\ell)} ({\sbM}^{(\ell)}_i)'' + 1} \right) \nabla_{\Z_i} q(\Z_i, \Psi_i, \bar{\varepsilon}) \right],
    \label{eq:mu-fp-form-bar}
    \\
    \kappa_{\ell, \ell'} &= \lim_{n \to \infty} \frac{\delta b^{(\ell)} b^{(\ell')}}{n} \sum_{i = 1}^n \E \left[ ({\sbM}_i^{(\ell)})' ({\sbM}_i^{(\ell')})' \right]
    \label{eq:cross-cov-fp-form-bar}
    .
\end{align}

\end{proposition}
\begin{proof}
We first derive equations \eqref{eq:b-fp-form}--\eqref{eq:cross-cov-fp-form}, and note that the only differing equation between \eqref{eq:b-fp-implicit}--\eqref{eq:cross-cov-fp-implicit} and \eqref{eq:b-fp-form}--\eqref{eq:cross-cov-fp-form} is the middle equation \eqref{eq:mu-fp-form}. Let $\ell \in [L]$ with superscripts omitted for notational convenience and, for $i \in [n]$, let $v_{\TTheta, i} := {\Z_i^\top} \blambda + w_i$. For $i \in [n], b > 0, \z \in \reals^{L}, u \in \reals, \Psi_i \in [L], \varepsilon_i \in \reals$, let $\tilde{g}_i(\z, u, \Psi_i, \varepsilon_i) := (\prox_{b \pi_i M(\cdot, q(\z, \Psi_i, \varepsilon_i))}(u) - u) / b$. Note that, by \eqref{eq:prox_id_2}, we have that $\tilde{g}_i(\Z_i, v_{\TTheta, i}, \Psi_i, \varepsilon_i) = -\sM'_i$. Also note that, by \eqref{eq:prox_id_3}, we have:
\begin{align}
    \frac{\delta b}{n} \sum_{i = 1}^n \blambda \E\left[ \nabla_{v_{\TTheta, i}} \tilde{g}_i(\Z_i, v_{\TTheta, i}, \Psi_i, \varepsilon_i) \right] = \blambda \left(\frac{\delta}{n} \sum_{i = 1}^n  \E\left[ (1 + b \sM_i'')^{-1} \right] - \delta\right). \label{id:steins-arg-1}
\end{align}
Moreover, by Proposition \ref{prop:prox-deriv-wrt-y} and by the chain rule, we have that:
\begin{align}
    \frac{\delta b}{n} \sum_{i = 1}^n \E[\nabla_{\Z_i} \tilde{g}_i(\Z_i, v_{\TTheta, i}, \Psi_i, \varepsilon_i)] = \frac{\delta b}{n} \sum_{i = 1}^n \E\left[ - \left( \frac{{\partial_{12}} {\sM}_i}{b ({\sM}_i)'' + 1} \right) \nabla_{\Z_i} q(\Z_i, \Psi_i, \varepsilon_i) \right]. \label{id:steins-arg-2}
\end{align}
We apply a generalized version of Stein's Lemma   \citep[Lemma 15]{berthier_state-evolution_2019}, where for $\x \distas{} \N(\0_d, \tilde{\bSigma})$ and $h: \reals^d \to \reals$, we have that $\E[\x h(\x)] = \tilde{\bSigma} \E[\nabla_{\x} h(\x)]$. Applying \eqref{eq:prox_id_2} we obtain:
\begin{align*}
&\E\left[ \Z_i (-\sM'_i) \right] \\
&= \left(\E\left[ \begin{bmatrix}
    \Z_i \\ v_{\TTheta, i}
\end{bmatrix} \tilde{g}_i(\Z_i, \v_{\TTheta, i}, \Psi_i, \varepsilon_i)\right]\right)_{[1:L]} \\
&= \left( \cov(\Z_i, v_{\TTheta, i}) \E\left[\nabla_{(\Z_i, v_{\TTheta, i})} \tilde{g}_i(\Z_i, \v_{\TTheta, i}, \Psi_i, \varepsilon_i)\right]\right)_{[1:L]} \\
&= \cov(\Z_i) \E[\nabla_{\Z_i} \tilde{g}_i(\Z_i, v_{\TTheta, i}, \Psi_i, \varepsilon_i)] + \E[v_{\TTheta, i} \Z_i] \E\left[ \nabla_{v_{\TTheta, i}} \tilde{g}_i(\Z_i, v_{\TTheta, i}, \Psi_i, \varepsilon_i) \right] \\
&= \bGamma \E[\nabla_{\Z_i} \tilde{g}_i(\Z_i, v_{\TTheta, i}, \Psi_i, \varepsilon_i)] + \bGamma \blambda \E\left[ \nabla_{v_{\TTheta, i}} \tilde{g}_i(\Z_i, v_{\TTheta, i}, \Psi_i, \varepsilon_i) \right].
\end{align*}
{Since $\bGamma$ is positive definite by \eqref{eq:Gamma_as_limit}, it is invertible, and rearranging gives:}
\begin{align*}
   &\frac{\delta b}{n} \sum_{i = 1}^n \E[\nabla_{\Z_i} \tilde{g}_i(\Z_i, v_{\TTheta, i}, \Psi_i, \varepsilon_i)] \\
   &= \delta b \bGamma^{-1} \frac{1}{n} \sum_{i = 1}^n \E[\Z_i (-\sM_i')] - \frac{\delta b}{n} \sum_{i = 1}^n \blambda \E\left[ \nabla_{v_{\TTheta, i}} \tilde{g}_i(\Z_i, v_{\TTheta, i}, \Psi_i, \varepsilon_i) \right].
\end{align*}
Plugging in identities \eqref{id:steins-arg-1}--\eqref{id:steins-arg-2} into the above equality, we obtain:
\begin{align*}
   &\frac{\delta b}{n} \sum_{i = 1}^n \E\left[ - \left( \frac{{\partial_{12}} {\sM}_i}{b ({\sM}_i)'' + 1} \right) \nabla_{\Z_i} q(\Z_i, \Psi_i, \varepsilon_i) \right] \\
   &= \delta b \bGamma^{-1} \frac{1}{n} \sum_{i = 1}^n \E[\Z_i (-\sM_i')] - \blambda \left(\frac{\delta}{n} \sum_{i = 1}^n  \E\left[ (1 + b \sM_i'')^{-1} \right] - \delta\right).
\end{align*}
{Taking limits and plugging in equations \eqref{eq:b-fp-implicit}--\eqref{eq:mu-fp-implicit}, we obtain}
\begin{align*}
   &{\lim_{n\to\infty}}\frac{\delta b}{n} \sum_{i = 1}^n \E\left[ - \left( \frac{{\partial_{12}} {\sM}_i}{b ({\sM}_i)'' + 1} \right) \nabla_{\Z_i} q(\Z_i, \Psi_i, \varepsilon_i) \right] = 0 - \blambda \left( -1 \right).
\end{align*}
Equations \eqref{eq:b-fp-form-bar}--\eqref{eq:cross-cov-fp-form-bar} can be derived similarly.
\end{proof}
We are now in a position to prove Lemma \ref{lemma:SE-matrix-amp}. The proof references properties of the proximal operator outlined in Appendix \ref{sec:prop-prox-op}.
\begin{proof}[Proof of Lemma \ref{lemma:SE-matrix-amp}.]

The proof uses a reduction to the setting of \citep[Theorem 1]{arpino_2025_jmlr} which provides an asymptotic characterization of the matrix-valued AMP \eqref{eq:matrix-valued-amp-1}, with $\C^t$ in \eqref{eq:matrix-amp-C-def} substituted for an empirical variant. The empirical versions of $\C^t, \A^t$ used in \citep[Theorem 1]{arpino_2025_jmlr} can be replaced with the deterministic versions in \eqref{eq:matrix-amp-C-def} and \eqref{eq:matrix-amp-A-t+1-diag} via an argument similar to \cite[Corollary 2]{berthier_state-evolution_2019}. We first restate their result for general denoiser functions $G^t: \reals^{n \times L} \times \reals^n \to \reals^{n \times L}$ and $F^t: \reals^{p \times L} \to \reals^{p \times L}$, noting that the matrix $\tilde{\X} := \X \L^{-\top}$ in our manuscript corresponds to the isotropic covariate matrix $\X$ used in \citep{arpino_2025_jmlr}.

In the high-dimensional limit as $n,p\to\infty$ (with $n/p \to \delta$),  the empirical distributions of $\TTheta^t$ and $\B^{t+1}$ are quantified through the random variables $\V_{\TTheta}^t$ and $\V_{\B}^{t+1}$ respectively, where
\begin{align}
&\V_{\TTheta}^t := \Z \bGamma^{-1} \bnu^t_{\TTheta} + \G^t_{\TTheta} \,  \in \reals^{n \times L}, \label{eq:V_TTheta}\\
&\V_{\B}^{t+1} := \L^\top \B \bnu^{t+1}_{\B} + \G^{t+1}_{\B} \, \in \reals^{p \times L}. \label{eq:V_B}
\end{align}
The random matrices $\Z$, $\bG_{\TTheta}^t$, and $\bG_{\B}^{t+1} $ are independent of $\X$, and have i.i.d. rows following a Gaussian distribution. Namely, for $i \in [n]$ we have $\Z_i \distas{\text{i.i.d.}} \N(\0_L, \bGamma)$. For $i \in [n]$ and $s, r \geq 0$, $(\G_{\TTheta}^t)_i \distas{\text{i.i.d.}} \N(\0_L, \bkappa_{\TTheta}^{t, t})$ with $\cov((\G_{\TTheta}^r)_i, (\G_{\TTheta}^s)_i) = \bkappa_{\TTheta}^{r, s}$. Similarly, for $j \in [p]$, $(\G_{\B}^t)_j \distas{\text{i.i.d.}} \N(\0_L, \bkappa_{\B}^{t, t})$ with $\cov((\G_{\B}^r)_j, (\G_{\B}^s)_j) = \bkappa_{\B}^{r, s}$. The $L\times L$ deterministic matrices $\bnu_{\TTheta}^t, \bkappa_{\TTheta}^{r,s}, \bnu_{\B}^{t},$ and $\bkappa_{\B}^{r,s}$ are defined below via the \emph{state evolution} recursion.

Given an initializer $\hat{\B}^0$ for the iteration \eqref{eq:matrix-valued-amp-1}, {the state evolution is initialized by}
{
\begin{align}
\bGamma&:=\lim_{p\to\infty}\frac{(\L^\top\B)^\top\L^\top\B}{\delta p},\qquad
\bnu_{\TTheta}^0:=\lim_{p\to\infty}\frac{(\L^\top\B)^\top\hat\B^0}{\delta p},\notag\\
\bkappa_{\TTheta}^{0,0}&:=\lim_{p\to\infty}\frac{(\mathbf E^0)^\top\mathbf E^0}{\delta p},\qquad
\mathbf E^0:=\hat\B^0-\L^\top\B\bGamma^{-1}\bnu_{\TTheta}^0.\label{eq:kappa_theta_0}
\end{align}
}
Note that the definition of $\bGamma$ above corresponds to that in \eqref{eq:Gamma_as_limit}.
Let $\tilde{G}_i^t(\Z, \V_{\TTheta}^t, \bPsi, \bvarepsilon) := G_i^t(\V_{\TTheta}^t, q(\Z, \bPsi, \bvarepsilon))$ and let $\partial_{1i} \tilde{G}_i^t$ be the partial derivative (Jacobian) w.r.t. the $i$th row of the first argument.  Then, the state evolution matrices are defined recursively as follows:
\begin{align}
&{\bnu}_{\B}^{t+1} := \lim_{n \to \infty} \frac{1}{n} \E\left[\sum_{i=1}^n \partial_{1i} \tilde{G}_i^t(\Z, \V_{\TTheta}^t, \bPsi, \bvarepsilon) \right], \label{eq:nu_B_SE}\\
&\bkappa_{\B}^{s+1, t+1} := \lim_{n \to \infty} \frac{1}{n} \E\left[G^s\left( \V_{\TTheta}^s, q(\Z, \bPsi, \bvarepsilon)\right)^\top G^t\left(\V_{\TTheta}^t, q(\Z, \bPsi, \bvarepsilon)\right)  \right], \label{eq:kappa_B_SE} \\
&\bnu^{t+1}_{\TTheta} := \frac{1}{\delta} \lim_{p \to \infty} \frac{1}{p} \E\left[ \left( \L^\top \B \right)^\top F^{t+1}(\V_{\B}^{t+1}) \right], \label{eq:nu_theta_SE}\\
&\bkappa_{\TTheta}^{s+1, t+1} \nonumber \\
&:= \frac{1}{\delta} \lim_{p \to \infty} \frac{1}{p} \E\left[\left(F^{s+1}(\V_{\B}^{s+1}) - \L^\top \B \bGamma^{-1} \bnu^{s+1}_{\TTheta}\right)^\top \left(F^{t+1}(\V_{\B}^{t+1}) - \L^\top \B \bGamma^{-1} \bnu^{t+1}_{\TTheta} \right) \right]. \label{eq:kappa_theta_SE}
\end{align}
The expectations above are taken with respect to $ \Z, \V_{\TTheta}^t, \V_{\TTheta}^s, \V_{\B}^{t+1}$ and $\V_{\B}^{s+1}$, and depend on  $g^t$, $f^t$, $\B$, $\bvarepsilon$, and $\bPsi$. Under assumptions (A1)--(A3) outlined in \cite{arpino_2025_jmlr}, \cite[Theorem 1]{arpino_2025_jmlr} states that for $t \geq 0$ and any sequence of uniformly pseudo-Lipschitz functions $\varphi_{p}(\cdot \; ; \L^\top \B) : \reals^{p \times (L(t+1))} \to \reals$ and $\varphi_{n}(\cdot \;; \bPsi, \bvarepsilon) : \reals^{n \times (L(t+2))} \to \reals$,
\begin{align}
&\varphi_n(\TTheta^0, \dots, \TTheta^t, \X \B ; \bPsi, \bvarepsilon) \stackrel{\P}{\simeq} \E_{\V_{\TTheta}^0, \dots, \V_{\TTheta}^t, \Z} \{ \varphi_n(\V_{\TTheta}^0, \dots, \V_{\TTheta}^t, \Z ; \bPsi, \bvarepsilon)\} \label{eq:SE_Theta_thm1-proof}, \\
&\varphi_p(\B^1, \dots, \B^{t+1} ; \L^\top \B) \stackrel{\P}{\simeq} \E_{\V_{\B}^1, \dots, \V_{\B}^{t+1}} \{ \varphi_p(\V_{\B}^1, \dots, \V_{\B}^{t+1} ; \L^\top \B) \} \label{eq:SE_B_thm1-proof},
\end{align}
as $n, p \to \infty$ with $n/p \to \delta$, where the random variables $\Z, \V_{\TTheta}^t$ and $\V_{\B}^{t+1}$ are defined in \eqref{eq:V_TTheta}, \eqref{eq:V_B}.

Specifying the denoisers $F^t, G^t$ as in \eqref{eq:matrix-amp-F-def} and \eqref{eq:matrix-amp-G-def}, and noting the equivalent representation using $f^{t}, g^{(\ell), t}, \tilde{g}^{(\ell), t}$ in \eqref{eq:matrix-amp-F-G-other-forms}, the columns of the state evolution matrices \eqref{eq:nu_B_SE}, \eqref{eq:nu_theta_SE} take on the form:
\begin{align}
    &(\bnu_{\B}^{t + 1})_{[:, \ell]} = \lim_{n \to \infty} \frac{1}{n} \sum_{i = 1}^n \E \nabla_{\Z_i} \tilde{g}_i^{(\ell), t}(\Z_i, (\V_{\TTheta}^t)_{[i, \ell]}, \Psi_i, \varepsilon_i), \label{eq:mu-expansion}\\
    &(\bnu_{\TTheta}^{t + 1})_{[:, \ell]} = \lim_{p \to \infty} \frac{1}{\delta p} \delta b_{t+1}^{(\ell)} \E[\left( \L^\top \B\right)^\top (\V_{\B}^{t+1})_{[:, \ell]}] =  \bGamma \cdot \delta b_{t+1}^{(\ell)} (\bnu_{\B}^{t + 1})_{[:, \ell]}, \label{eq:alpha-expansion}
\end{align}
for $\ell \in [L]$, where $\tilde{g}^{(\ell), t}$ is defined in \eqref{eq:den_choice}.
Moreover, specifying the denoisers $F^t, G^t$ as in \eqref{eq:matrix-amp-F-def} and \eqref{eq:matrix-amp-G-def}, the entries of the state evolution matrix \eqref{eq:kappa_B_SE} take on the form:
\begin{align}
    &(\bkappa_{\B}^{t + 1, s+1})_{[\ell, \ell']} \notag \\
    &\quad = \lim_{n \to \infty} \frac{1}{n} \sum_{i = 1}^n \E\left[g_i^{(\ell), t}((\V_{\TTheta}^t)_{[i, \ell]}, q(\Z_i, \Psi_i, \varepsilon_i)) g_i^{(\ell'), s}((\V_{\TTheta}^s)_{[i, \ell']}, q(\Z_i, \Psi_i, \varepsilon_i)) \right], \label{eq:sigma_ell_ellprime_def}
\end{align}
for $\ell, \ell' \in [L]$. Similarly, specifying the denoisers $F^t, G^t$ as in \eqref{eq:matrix-amp-F-def} and \eqref{eq:matrix-amp-G-def}, the entries of the state evolution matrix \eqref{eq:kappa_theta_SE} take on the form:
\begin{align}
\begin{split} \label{eq:omega-expansion}
    \kappa_{\ell, \ell'}^{t+1, s+1}
    &:= (\bkappa_{\TTheta}^{t + 1, s+1})_{[\ell, \ell']} \\
    &= \lim_{p \to \infty} \frac{1}{\delta p} \E\left[ \left( \delta b^{(\ell)}_{t+1} (\V_{\B}^{t+1})_{[:, \ell]} - \L^\top \B \bGamma^{-1} (\bnu_{\TTheta}^{t+1})_{[:, \ell]} \right)^\top \right. \\
    & \hspace{5em} \left. \left( \delta b^{(\ell')}_{s+1} (\V_{\B}^{s+1})_{[:, \ell']} - \L^\top \B \bGamma^{-1} (\bnu_{\TTheta}^{s+1})_{[:, \ell']} \right)\right] \\
    &= \lim_{p \to \infty} \frac{1}{p} \delta b_{t+1}^{(\ell)} b_{s+1}^{(\ell')} \E\left[(\V_{\B}^{t+1})_{[:, \ell]}^\top (\V_{\B}^{s+1})_{[:, \ell']} \right] - {((\bnu_{\TTheta}^{t+1})_{[:, \ell]})^\top} \bGamma^{-1} (\bnu_{\TTheta}^{s+1})_{[:, \ell']} \\
    &= \delta b_{t+1}^{(\ell)} b_{s+1}^{(\ell')} (\bkappa_{\B}^{t + 1, s+1})_{[\ell, \ell']},
\end{split}
\end{align}
for $\ell, \ell' \in [L]$, where the fourth equality follows from the definition of $\V_{\B}^{t+1}$ in \eqref{eq:V_B}, the relation between $\bnu_{\B}^{t+1}$ and $\bnu_{\TTheta}^{t+1}$ established in \eqref{eq:alpha-expansion}, and the definition of $\sigma_{(\ell, \ell'), t+1}$ in \eqref{eq:sigma_ell_ellprime_def}. We note that, applying proximal identity \eqref{eq:prox_id_2} to \eqref{eq:omega-expansion} yields expression \eqref{eq:interaction-matrices} relevant to the theorem statement.
Next, note that by the definition of $\bnu_{\TTheta}^{t+1}$, and the expansion \eqref{eq:alpha-expansion}, the $\ell$-th column $\blambda^{(\ell), t+1}$ of $\bGamma^{-1} {\bnu_{\TTheta}}^{t+1}$ satisfies the recursive relation:

\begin{align}
   \blambda^{(\ell), t+1} &:= (\bGamma^{-1} {\bnu_{\TTheta}}^{t+1})_{[:, \ell]} \notag\\
    &= \lim_{n \to \infty} \frac{{\delta b_{t+1}^{(\ell)}}}{n} \sum_{i = 1}^n \E \nabla_{\Z_i} \tilde{g}_i^{(\ell), t}(\Z_i, \Z_i^{\top} \blambda^{(\ell), t} + (\G_{\TTheta}^{t})_{[i, \ell]}, \Psi_i, \varepsilon_i). \label{eq:bmu-recursion}
\end{align}
Similarly, the entries of ${\bkappa}_{\TTheta}^{t+1}$ satisfy the recursive relation:
\begin{align}
    \kappa^{t+1}_{\ell, \ell'} = \lim_{n \to \infty} \frac{\delta b_{t+1}^{(\ell)} b_{t+1}^{(\ell')}}{n} &
     \sum_{i = 1}^n \E \left[g_i^{(\ell), t}\left(\Z_i^{\top} \blambda^{(\ell), t} + (\G^t_{\TTheta})_{[i, \ell]}, q(\Z_i, \Psi_i, \varepsilon_i) \right) \right. \nonumber \\
    &\hspace{2cm} \left. g_i^{(\ell'), t}\left({\Z_i^\top \blambda^{(\ell'), t} + (\G^t_{\TTheta})_{[i, \ell']}}, q(\Z_i, \Psi_i, \varepsilon_i{)} \right)\right]. \label{eq:sigma-recursion}
\end{align}
Moreover, by the diagonal representation of $\C^t, \A^t$ in \eqref{eq:matrix-valued-amp-C-diagonal-representation-L} and \eqref{eq:matrix-amp-A-t+1-diag}, we have that:
\begin{align*}
    &C^t_{[\ell, \ell]} = {\lim_{n\to\infty}}\frac{1}{n} \sum_{i=1}^n \E\left[ (g_i^{(\ell), t})' ((\V_{\TTheta}^t)_{[i, \ell]}, q(\Z_i, \Psi_i, \varepsilon_i)) \right], \\
    &b^{(\ell)}_{t+1} = A^{t+1}_{[\ell, \ell]} = -\frac{1}{\delta C^t_{[\ell, \ell]}},
\end{align*}
which yield the following expression for $b^{(\ell)}_{t+1}$:
\begin{align}
    &b^{(\ell)}_{t+1} = -\frac{1}{\delta} \left({\lim_{n\to\infty}}\frac{1}{n} \sum_{i=1}^n \E\left[ (g_i^{(\ell), t})' ((\V_{\TTheta}^t)_{[i, \ell]}, q(\Z_i, \Psi_i, \varepsilon_i)) \right] \right)^{-1}. \label{eq:b-recursion}
\end{align}
It then follows from the proximal identity \eqref{eq:prox_id_2} that solutions to the nonlinear equations \eqref{eq:b-fp-form}--\eqref{eq:cross-cov-fp-form} are fixed points of the recursions \eqref{eq:bmu-recursion}--\eqref{eq:b-recursion}.

We now argue that, with denoisers $F^t, G^t$ specified as in \eqref{eq:matrix-amp-F-def} and \eqref{eq:matrix-amp-G-def}, assumptions (A1)--(A3) in \citep{arpino_2025_jmlr} are satisfied. Indeed, due to the choice of initialization $\B^0$ and by assumption \eqref{eq:Gamma_as_limit}, assumption (A1) is satisfied. Moreover, assumption (A3) follows from Lemma \ref{lemma:general-pointwise-conv}. We note that, by using the deterministic definition of $\C^t, \A^t$ in \eqref{eq:matrix-valued-amp-C-diagonal-representation-L} and \eqref{eq:matrix-amp-A-t+1-diag} rather than the empirical version used in \citep[Theorem 1]{arpino_2025_jmlr}, the uniform pseudo-Lipschitz assumption on the Jacobians of $f_j^t, \tilde{g}_i^t$ in assumption (A2) is not required.
The remainder of assumption (A2) in \citep{arpino_2025_jmlr} then follows from the fact that for $\ell, \psi \in [L], \varepsilon \in \reals$ with superscripts omitted, the proximal operator $(u, z) \mapsto \prox_{b \pi_i M(\cdot, q(z, \psi, \varepsilon))}(u)$ is pseudo-Lipschitz whenever $M: \reals \times \reals \to \reals$ is lower semicontinuous and convex and $q$ is pseudo-Lipschitz. This can be derived using the mean value theorem, Proposition \ref{prop:prox-deriv-wrt-y}, and Assumption \ref{asmpt:reg-loss}(\ref{eq:loss-asmpt-log-conc}).

Initializing as in the theorem statement, it follows from \citep[Lemma 19]{gerbelot_graph-based_2023} that $\TTheta^0 = \X \L^{-\top} \hat{\B}^0$ admits the following asymptotic characterization for any sequence of uniformly pseudo-Lipschitz functions $\varphi_n(\cdot; \bPsi, \bvarepsilon): \reals^{n \times L} \to \reals$:
\begin{align*}
    \varphi_n(\TTheta^0, \X\B; \bPsi, \bvarepsilon) \stackrel{\P}{\simeq} \E\varphi_n(\Z {\bGamma^{-1}}\bnu_{\TTheta}^0 + \G_{\TTheta}^0, \Z; \bPsi, \bvarepsilon),
\end{align*}
where $\bnu_{\TTheta}^0 = \bGamma \bLambda, \bkappa_{\TTheta}^{0, 0} = \bK$. Propagating this through the state evolution equations \eqref{eq:nu_B_SE}--\eqref{eq:kappa_theta_SE} and consequently through the recursions \eqref{eq:bmu-recursion}--\eqref{eq:sigma-recursion}, we observe that the recursions are stationary. That is, we initialize \eqref{eq:bmu-recursion}--\eqref{eq:b-recursion} at their fixed points, and we can write $\bGamma^{-1} \bnu_{\TTheta}^t = \bLambda$ and $\bkappa_{\TTheta}^{t, t} = \bK$ for all $t \geq 0$. Moreover, by linearity of the denoiser $F := F^t$ in \eqref{eq:matrix-amp-F-def} and by the relations \eqref{eq:alpha-expansion}, \eqref{eq:sigma_ell_ellprime_def}, it follows that:
\begin{align*}
   \left(F(\V_{\B}^{t+1})\right)_j \distas{} \N({(\L^\top \B \bGamma^{-1} \bnu_{\TTheta}^{t+1})_j}, \delta \bkappa_{\TTheta}^{t+1, t+1})
\end{align*}
independently over $j \in [p]$, with $\cov\left(\left(F(\V_{\B}^{s})\right)_j, \left(F(\V_{\B}^{r})\right)_j \right) = \delta \bkappa_{\TTheta}^{s, r}$ for $s, r \geq 0$. It follows that for any sequence of uniformly pseudo-Lipschitz functions $\varphi_p(\cdot; \L^\top \B): \reals^{p \to L} \to \reals$ and $t \geq 1$:
\begin{align*}
    \varphi_p(\hat{\B}^t; \L^\top \B)
    &= \varphi_p(F(\B^t); \L^\top \B) \stackrel{\P}{\simeq} \E \varphi_p(F(\L^\top \B \bnu_{\B}^t + \G_{\B}^t); \L^\top \B) \\
    &= \E \varphi_p(\L^\top \B \bLambda + \sqrt{\delta} \G_{\B, \TTheta}^t; \L^\top \B),
\end{align*}
where $\G_{\B, \TTheta}^t := (\G_{\TTheta}^t)_{[:p, :]}$ is a truncation of $\G_{\TTheta}^t$. Applying  \citep[Theorem $1$]{arpino_2025_jmlr}, the result follows.
\end{proof}
\subsection{Proof of Lemma \ref{lemma:ERM_frobenius_characterization}}\label{sec:proof-ERM_frobenius_characterization}
We begin by establishing an asymptotic Cauchy  convergence result for the AMP iterates \eqref{eq:matrix-valued-amp-1}.
\begin{lemma}[Asymptotic Cauchy convergence of AMP iterates.] \label{lemma:cauchy_t_plus_h}
Let $\ell \in [L]$ and omit the superscript $^{(\ell)}$. Consider the setting in Section \ref{sec:setting}, {with the fixed-point initialization of Lemma \ref{lemma:SE-matrix-amp}}, and assume $b > 0$.
Let $(\bbeta^{t+1}, \ttheta^{t}, \hat{\bzeta}^{t+1}, \hat{\bvartheta}^{t}, \hat{\s}^t)$ be the iterates in \eqref{eq:vector-valued-amp}. Then for any fixed $h \in \naturals$:
     \begin{align*}
        &\lim_{t \to\infty} \lim_{n \to \infty} \frac{1}{p} \| \hat{\bzeta}^{t + h} - \hat{\bzeta}^t \|_2^2 \stackrel{\P}{\simeq} 0, \qquad \lim_{t \to\infty} \lim_{n \to \infty} \frac{1}{n} \| \ttheta^{t + h} - \ttheta^{t} \|_2^2 \stackrel{\P}{\simeq} 0.
     \end{align*}
\end{lemma}
The proof of Lemma \ref{lemma:cauchy_t_plus_h} relies on the following technical lemma.
\begin{lemma} \label{lemma:prop-of-cross-cov-simple}
Let $\ell, \ell' \in [L], i \in [n]$. {Fix $b^{(\ell)},b^{(\ell')}>0$ and $\blambda^{(\ell)},\blambda^{(\ell')}\in\reals^L$.} Consider the function:
\begin{align}
    & \sh_i  = \sh_i(\bSigma_{\xi}) = \sh_i(\tau_1, \tau_{1, 2}, \tau_2) = \E\left[ \pi_i^{(\ell)} M'(\prox_{{b^{(\ell)}} \pi_i^{(\ell)} M(\cdot, \bar{y}_i)}(\Z_i^{\top} \blambda^{(\ell)} + \xi_1), \bar{y}_i) \right. \notag \\
    &\hspace{4.5cm} \left. \cdot  {\pi_i^{(\ell')}} M'(\prox_{{b^{(\ell')}} \pi_i^{(\ell')} M(\cdot, \bar{y}_i)}({\Z_i^\top} \blambda^{(\ell')} + \xi_2), \bar{y}_i)\right], \label{eq:cross-cov-eqn}
\end{align}
where, for $i \in [n]$, $\bar{y}_i = q(\Z_i, \Psi_i, \bar{\varepsilon})$, and the expectation is taken with respect to the following set of mutually independent random variables:
\begin{itemize}
    \item A Gaussian vector $(\xi_1, \xi_2) \distas{} \N\left(\0_2, \bSigma_{\xi}\right)$, where $\bSigma_{\xi} \succeq 0$ with $(\Sigma_{\xi})_{[1, 1]} = \tau_1, (\Sigma_{\xi})_{[2, 2]} = \tau_2, (\Sigma_{\xi})_{[1, 2]} = (\Sigma_{\xi})_{[2, 1]} = \tau_{1, 2}$,
    \item $\Z_i \distas{\text{i.i.d.}} \N(\0_L, \bGamma)$,
    \item $\bar{\varepsilon} \distas{} \P_{\bar{\varepsilon}}$.
\end{itemize}
{For $R>0$, define the compact covariance set}
{
\[
\mathcal{K}_R := \left\{
\begin{pmatrix}
\tau_1 & \tau_{1,2} \\
\tau_{1,2} & \tau_2
\end{pmatrix}
: 0\leq\tau_1,\tau_2\leq R,\quad \tau_{1,2}^2\leq\tau_1\tau_2
\right\}.
\]
}
{Thus, $\mathcal{K}_R$ contains all positive-semidefinite covariance matrices above with variances at most $R$, including singular matrices.}
Then, under the model assumptions outlined in Section \ref{sec:setting} and under Assumption \ref{asmpt:reg-loss}(\ref{cond:strict-cvx-smooth}){, (\ref{eq:loss-asmpt-log-conc})}, the following holds:
\begin{enumerate}
    \item $\sh_i$ is {uniformly bounded over $n,i$ and $\bSigma_{\xi}\in\mathcal{K}_R$, for every $R>0$} \label{eq:sh1}
    \item $\sh_i$ is continuous \label{eq:sh2}
    \item The family $\{\sh_i\}_{i = 1}^\infty$ is equicontinuous with respect to $\tau_1, \tau_{1, 2}, \tau_2$ {on $\mathcal{K}_R$, for every $R>0$}. \label{eq:sh3}
\end{enumerate}
\end{lemma}
\begin{proof}
Let $i \in [n]$, and let
\begin{align*}
    &\cM' := \pi_i^{(\ell)} M'\left(\prox_{{b^{(\ell)}} \pi_i^{(\ell)} M(\cdot, \bar{y}_i)}(\Z_i^{\top} \blambda^{(\ell)} + \xi_1), \bar{y}_i\right), \\
    &\tilde{\cM}' := \pi_i^{(\ell')} M'\left(\prox_{{b^{(\ell')}} \pi_i^{(\ell')} M(\cdot, \bar{y}_i)}({\Z_i^\top} \blambda^{(\ell')} + \xi_2), \bar{y}_i\right).
\end{align*}
Result \eqref{eq:sh1} can be shown using the Cauchy-Schwarz inequality:
\begin{align*}
    {|\sh_i| = |\E[\cM' \tilde{\cM}']|} \leq \sqrt{\E[(\cM')^2] \E[(\tilde{\cM}')^2]},
\end{align*}
together with Proposition \ref{prop:M_prime_ub} and the Lipschitz assumption on $q$ and the bounded second moment assumption on $\bar{\varepsilon}$. {For \eqref{eq:sh2} and \eqref{eq:sh3}, couple the Gaussian inputs as $\bSigma_\xi^{1/2}U$ and $\widetilde{\bSigma}_\xi^{1/2}U$, using the same $U\sim\N(\0_2,\I_2)$, $\Z_i$ and $\bar\varepsilon$. The quantities $\cM'$ and $\tilde{\cM}'$ are uniformly Lipschitz in $\xi_1$ and $\xi_2$, respectively, by \eqref{eq:prox_id_3}. Their uniformly bounded second moments on $\mathcal{K}_R$ and Cauchy-Schwarz give}
{
\[
 |\sh_i(\bSigma_\xi)-\sh_i(\widetilde{\bSigma}_\xi)|
 \le C\|\bSigma_\xi^{1/2}-\widetilde{\bSigma}_\xi^{1/2}\|_F,
\]
}
{where $C=C_R$ is independent of $n,i$. Continuity of the positive-semidefinite square root proves both claims, including at singular covariances. For use in Lemma \ref{lemma:cauchy_t_plus_h}, we also record a derivative bound in the positive-definite interior.} We fix $\tau_1, \tau_2 \in (0, \infty)$ and let $\tau_a, \tau_b \in {(-\sqrt{\tau_1 \tau_2}, \sqrt{\tau_1 \tau_2})}$, noting that $\bSigma_{\xi} \succeq 0$ implies $\tau_1 \tau_2 \geq \tau_{1, 2}^2$. Applying the mean value theorem, we obtain:
\begin{align} \label{eq:h-mvt-step}
    &|\sh_i(\tau_1, \tau_a, \tau_2) - \sh_i(\tau_1, \tau_b, \tau_2) | = \left| \partial_{2} \sh_i(\tau_1, \tilde{\tau}_{1, 2}, \tau_2) \right| \cdot |\tau_a - \tau_b|,
\end{align}
for some $\tilde{\tau}_{1, 2} \in (-\sqrt{\tau_1 \tau_2}, \sqrt{\tau_1 \tau_2})$. We now proceed to bound the derivative term uniformly over $i \in \naturals$. {Define
\begin{align*}
\cM''&:=\pi_i^{(\ell)}M''\left(\prox_{b^{(\ell)}\pi_i^{(\ell)}M(\cdot,\bar y_i)}(\Z_i^\top\blambda^{(\ell)}+\xi_1),\bar y_i\right),\\
\tilde{\cM}''&:=\pi_i^{(\ell')}M''\left(\prox_{b^{(\ell')}\pi_i^{(\ell')}M(\cdot,\bar y_i)}(\Z_i^\top\blambda^{(\ell')}+\xi_2),\bar y_i\right).
\end{align*}
} By Price's Theorem \citep{prices_theorem} and by the proximal identity \eqref{eq:prox_id_3}, we have:
\begin{align}
\begin{split} \label{eq:mvt-bound-deriv-step}
    |\partial_{\tau_{1, 2}} \E[\cM' \tilde{\cM}']|
    &= \left|\E\left[ \partial_{\xi_1} \cM' \cdot \partial_{\xi_2} \tilde{\cM}' \right] \right| \\
    &= \frac{1}{b^{(\ell)} b^{(\ell')}} \left|\E\left[ \left(\frac{b^{(\ell)} \cM''}{1 + b^{(\ell)} \cM''} \right) \left( \frac{b^{(\ell')}\tilde{\cM}''}{1 + b^{(\ell')} \tilde{\cM}''} \right) \right] \right| \\
    &\leq \frac{1}{b^{(\ell)} b^{(\ell')}},
\end{split}
\end{align}
{The resulting Lipschitz bound in $\tau_{1,2}$ extends to the endpoints by \eqref{eq:sh2}.}
\end{proof}
\begin{proof}[Proof of Lemma \ref{lemma:cauchy_t_plus_h}.]
By Lemma \ref{lemma:SE-matrix-amp}, we have that:
\begin{align}
\begin{split} \label{eq:theta_cauchy_SE}
&\frac{1}{n} \| \ttheta^{t + h} - \ttheta^{t} \|_2^2 \stackrel{\P}{\simeq} \frac{1}{n} \E \| (\G_{\TTheta}^{t+h})_{[:, \ell]} - (\G_{\TTheta}^{t})_{[:, \ell]}\|^2_2 = 2 (\kappa - \kappa^{(t+h, t)}), \\
& \frac{1}{p} \| \hat{\bzeta}^{t + h} - \hat{\bzeta}^t \|_2^2 \stackrel{\P}{\simeq} \frac{\delta}{p} \E \| (\G_{\B}^{t+h})_{[:, \ell]} - (\G_{\B}^t)_{[:, \ell]} \|_2^2 = 2 \delta (\kappa - \kappa^{(t+h, t)}),
\end{split}
\end{align}
where we abbreviate $\kappa := \bK^{(t, t)}_{[\ell, \ell]} = \bK_{[\ell, \ell]}$, $ \kappa^{(t+h, t)} := \bK^{(t+h, t)}_{[\ell, \ell]}$, with $\bK^{(r+1, s+1)}$ defined in \eqref{eq:interaction-matrices} for $s, r \geq 0$. Note that $g_i(u, v)$ takes the form
\begin{align*}
    g_i(u, v) := - \pi_i M'(\prox_{b \pi_i M(\cdot, v)}(u), v),
\end{align*}
as per \eqref{eq:den_choice} and  \eqref{eq:prox_id_2}. {If $\kappa=0$, \eqref{eq:theta_cauchy_SE} proves the claim. Otherwise, $\kappa>0$.} Letting $q_t := \kappa^{t+h, t} / \kappa$  and recalling \eqref{eq:interaction-matrices}, we obtain the recursion:
\begin{align*}
    &{q_{t+1} = \sH(q_t),\quad \sH:=\lim_{n\to\infty}\sH_n}, \\
    &\sH_n(q) = \frac{\delta b^2}{\kappa n} \sum_{i = 1}^n \E\left[ g_i({\Z_i^\top} \blambda + \sqrt{\kappa} \xi_1, \bar{y}_i) g_i({\Z_i^\top} \blambda + \sqrt{\kappa} \xi_2, {\bar{y}_i})\right],
\end{align*}
where the expectation is taken with respect to the centered Gaussian vector $(\xi_1, \xi_2)$ with $\E[\xi_1^2] = \E[\xi_2^2] = 1$ and $\E[\xi_1 \xi_2] = q$, independent of $\Z_i \distas{\text{i.i.d.}} \N(0, \bGamma)$ {and $\bar\varepsilon$} for all $i \in [n]$. {We will prove that $q_t\to1$. For each $q\in[-1,1]$, define
\[
\sh_q(\pi,r):=\E\left[g_\pi(\Z^\top\blambda+\sqrt\kappa\xi_1,\bar y^{(r)})
g_\pi(\Z^\top\blambda+\sqrt\kappa\xi_2,\bar y^{(r)})\right],
\]
where $g_\pi$ is $g_i$ above with $\pi_i=\pi$, $\bar y^{(r)}:=q(\Z,r,\bar\varepsilon)$, and $(\Z,\bar\varepsilon,\xi_1,\xi_2)$ has the distribution just specified. By Proposition \ref{prop:M_prime_ub}, Remark \ref{rmk:prox-implicit-function-theorem-and-b}, and the dominated convergence theorem, $\sh_q(\pi,r)$ is bounded and continuous in $\pi\in[0,1]$ for each $r\in[L^*]$. Assumption \ref{asmpt:reg-weights} gives $\pi_{\floor{nt}+1}\to\Phi(t)$. Applying the integral calculation in the proof of Lemma \ref{lemma:general-pointwise-conv} therefore yields
\[
\lim_{n\to\infty}\sH_n(q)
=\frac{\delta b^2}{\kappa}\sum_{r=1}^{L^*}\int_{\alpha_{r-1}}^{\alpha_r}\sh_q(\Phi(t),r)\,dt.
\]
Proposition \ref{prop:conv-nonlin-eqns} justifies the use of $\bar{\varepsilon}$ in this recursion.}

We prove that $\lim_n \sH_n(1) = 1$ and that $\lim_n \sH_n$ is a contraction, after which the result follows by Banach's contraction mapping principle \cite[Theorem $9.23$]{rudin1976principles}.  Note that, for $q = 1$, we have $\xi_1 = \xi_2 =: \xi \distas{} \N(0, 1)$, and hence:
\begin{align*}
    \lim_n \sH_n(1) = \lim_n \frac{\delta b^2}{\kappa n} \sum_{i = 1}^n \E\left[ g_i({\Z_i^\top} \blambda + \sqrt{\kappa} \xi, \bar{y}_i)^2\right],
\end{align*}
which is equal to $1$ since $b, \kappa, \blambda$ satisfy the set of nonlinear equations \eqref{eq:b-fp-implicit-bar}--\eqref{eq:cross-cov-fp-implicit-bar}.

{We next prove that $\sH$ is a contraction on $[-1,1]$.} First, define the quantity
\begin{align*}
\sr(i, \tilde{\xi}) := \frac{b \pi_i M''(\prox_{b \pi_i M(\cdot, \bar{y}_i)}({\Z_i^\top} \blambda + \sqrt{\kappa} \tilde{\xi}), \bar{y}_i)}{1 + b \pi_i M''(\prox_{b \pi_i M(\cdot, \bar{y}_i)}({\Z_i^\top} \blambda + \sqrt{\kappa} \tilde{\xi}), \bar{y}_i)},
\end{align*}
{and note that $0\le\sr(i,\xi)\le\rho:=bB_{11}/(1+bB_{11})<1$, where $B_{11}:=\sup_{u,v}M''(u,v)$ and $\pi_i\le1$ by \eqref{eq:pi_i_defn}. Thus}
{
\begin{align}
 \frac{\delta}{n}\sum_{i=1}^n\E[\sr(i,\xi_1)^2]
 \le \rho\,\frac{\delta}{n}\sum_{i=1}^n\E[\sr(i,\xi_1)]
 =:\rho D_n. \label{eq:r_lb_3}
\end{align}
}
{For $-1<q<1$, we then have}
\begin{align}
   {\sH_n'(q)} &= \frac{\delta b^2}{\kappa n} {\sum_{i=1}^n} \E\left[ \partial_{\xi_1} g_i({\Z_i^\top} \blambda + \sqrt{\kappa} \xi_1, \bar{y}_i) \partial_{\xi_2} g_i({\Z_i^\top} \blambda + \sqrt{\kappa} \xi_2, \bar{y}_i) \right] \label{eq:price-deriv-0} \\
   &= \frac{\delta}{n} \sum_{i=1}^n \E\left[ \sr(i, \xi_1) \sr(i, \xi_2) \right] \label{eq:price-deriv-1}\\
   &\leq \sqrt{\frac{\delta}{n} \sum_{i=1}^n \E[\sr(i, \xi_1)^2]} \sqrt{\frac{\delta}{n} \sum_{i=1}^n \E[\sr(i, \xi_2)^2]} \label{eq:price-deriv-2}\\
   &{\leq \rho D_n,} \label{eq:price-deriv-3}
\end{align}
where \eqref{eq:price-deriv-0} follows from {Lemma \ref{lemma:prop-of-cross-cov-simple}, specifically \eqref{eq:mvt-bound-deriv-step}}, \eqref{eq:price-deriv-1} follows from \eqref{eq:prox_id_3} and the chain rule, \eqref{eq:price-deriv-2} follows from the Cauchy-Schwarz inequality, \eqref{eq:price-deriv-3} follows from \eqref{eq:r_lb_3}. Moreover, $\lim_n D_n = 1$ due to $b, \kappa, \blambda$ satisfying \eqref{eq:b-fp-implicit-bar} (can be verified from \eqref{eq:b-fp-implicit-bar} by noting that $1 -  \frac{b \sbM_i''}{1 + b \sbM_i''} = (1 + b \sbM_i'')^{-1} $).
{The derivative is nonnegative, and continuity from Lemma \ref{lemma:prop-of-cross-cov-simple} extends its Lipschitz bound to the endpoints. For $q_1,q_2\in[-1,1]$, we therefore have}
{
\begin{align}
 |\sH(q_1)-\sH(q_2)|
 &=\lim_n|\sH_n(q_1)-\sH_n(q_2)| \notag\\
 &\leq\lim_n\rho D_n|q_1-q_2| \label{eq:contract-1}\\
 &=\rho|q_1-q_2|. \label{eq:contract-2}
\end{align}
}
{Finally, Cauchy-Schwarz gives $|\sH_n(q)|\le\sH_n(1)$, so $\sH$ maps $[-1,1]$ into itself. Since $\sH(1)=1$, Banach's contraction mapping principle gives $q_t\to1$, and \eqref{eq:theta_cauchy_SE} proves the claim.}
\end{proof}
We are now in a position to prove Lemma \ref{lemma:ERM_frobenius_characterization}.
\begin{proof}[Proof of Lemma \ref{lemma:ERM_frobenius_characterization}.]
We follow the proof technique of \cite[Theorem 4.1]{donoho_high_2016}, and omit the superscript $\ell \in [L]$ for brevity. Throughout this proof, we say an event occurs `with high probability' if it occurs with probability tending to $1$ as $p \to \infty$. Notice that, by construction,
\begin{align*}
    \| \L^\top \hat{\B} - \hat{\B}^t \|^2_F = \sum_{\ell = 1}^L \|\L^\top \hat{\bbeta}^{(\ell)} - \hat{\bbeta}^{(\ell), t} \|^2_2,
\end{align*}
and hence for the first part of the result it suffices to show that $\frac{1}{p} \|\L^\top \hat{\bbeta}^{(\ell)} - \hat{\bbeta}^{(\ell), t} \|^2_2 \to 0$ for each $\ell \in [L]$. We recall the empirical risk function $\cC$ defined in \eqref{eq:ERM_est} and the iterates $\hat{\bzeta}^t, \hat{\bvartheta}^t$ from the vector-valued AMP iteration \eqref{eq:vector-valued-amp}. We note that, according to Proposition \ref{prop:matrix-vector-AMP-red}, the iterates $\hat{\bzeta}^t, \hat{\bvartheta}^t$ track the columns of the matrix-valued AMP iterates $\hat{\B}^t, \hat{\TTheta}^t$ in \eqref{eq:matrix-valued-amp-1}.

Recall the definition of $\cC$ from \eqref{eq:ERM_est}, and note that $\cC(\hat{\bbeta})$ with original covariates $\bx_i$ equals $\cC(\L^\top \hat{\bbeta})$ with isotropic covariates $\L^{-T} \bx_i$. From Taylor's theorem we have that for $t \geq 0$ and for some $\lambda \in (0, 1)$,
\begin{align*}
   \cC(\L^\top \hat{\bbeta})
   &= \cC(\hat{\bzeta}^t) + \left( \nabla\cC(\hat{\bzeta}^t) \right)^\top \left(\L^\top \hat{\bbeta} - \hat{\bzeta}^t \right) \\
   &\hspace{2cm} + \frac{1}{2} \left( \L^\top \hat{\bbeta} - \hat{\bzeta}^t \right)^\top \nabla^2 \cC\left( \hat{\bzeta}^t + \lambda (\L^\top \hat{\bbeta} - \hat{\bzeta}^t)\right) \left( \L^\top \hat{\bbeta} - \hat{\bzeta}^t \right).
\end{align*}
We bound the Hessian of the empirical risk via Assumption \ref{asmpt:reg-loss}(\ref{cond:likelihood_curvature}) to obtain:
\begin{align*}
   & \cC(\hat{\bzeta}^t)  \geq \cC(\L^\top \hat{\bbeta}) \\
   &\geq \cC(\hat{\bzeta}^t) + \left( \nabla\cC(\hat{\bzeta}^t) \right)^\top \left(\L^\top \hat{\bbeta} - \hat{\bzeta}^t \right) + \frac{\gamma\left(\max\left\{ \|\L^\top \hat{\bbeta}\|, \|\hat{\bzeta}^t\| \right\} / \sqrt{p} \right)}{2} \|\L^\top \hat{\bbeta} - \hat{\bzeta}^t \|_2^2
\end{align*}
with high probability, where $0 < \gamma(\cdot) < 1$ is a non-increasing continuous function independent of $n$. It follows from Lemma \ref{lemma:SE-matrix-amp} that $\|\hat{\B}^t\|_F^2 / p \stackrel{\P}{\simeq} \delta (\blambda)^\top \bGamma \blambda + \bK$, and hence that $\|\hat{\bzeta}^t\|_2^2 / p < \infty$ with high probability for $\ell \in [L]$. Moreover, by existence of the estimator in question, we have that $\|\L^\top \hat{\bbeta}\|_2^2 / p < \infty$ with high probability. Applying Cauchy-Schwarz, we then obtain:
\begin{align}
    \|\L^\top \hat{\bbeta} - \hat{\bzeta}^t \|_2 \leq \frac{2}{c_0} \|\nabla  \cC(\hat{\bzeta}^t) \|_2.
\end{align}
for some constant $c_0 > 0$.

We will now prove that, for any $\epsilon > 0$, $\lim_{t \to \infty} \lim_{n \to \infty} \P( \frac{1}{p} \|\nabla  \cC(\hat{\bzeta}^t) \|_2^2 > \epsilon) = 0$. Recall that the columns of the matrix-valued AMP \eqref{eq:matrix-valued-amp-1} track the vector-valued AMP iterates \eqref{eq:vector-valued-amp}, with the denoiser $g^t$ defined in \eqref{eq:den_choice}. With the chosen initialization in Lemma \ref{lemma:SE-matrix-amp}, the scalar terms $c_t, b_{t+1}$ defined in \eqref{eq:c_b} are initialized at their fixed points, meaning $b_{t+1} = b$ where $b$ is a solution to \eqref{eq:b-fp-implicit} for $t \geq 0$. The functions $g^t, \bar{g}^t$ defined in \eqref{eq:den_choice} are therefore also stationary, and we denote $g^t$ as $g$ and $\bar{g}^t$ as $\bar{g}$ for $t \geq 0$.
Recalling the definition of $g$, we have $g(u, v) = \frac{1}{b} \left(\bar{g}(u, v) - u\right)$. Plugging this definition of $g$ into \eqref{eq:vector-valued-amp} and rearranging, we obtain:
\begin{align*}
    \hat{\bvartheta}^t = \bar{g}(\ttheta^t, \y) = \tilde{\X}  \hat{\bzeta}^{t + 1} + \ttheta^t - \ttheta^{t + 1}.
\end{align*}
For $\a, \b \in \reals^n, \ell \in [L]$, let $\m^{\prime}\left( \a, \b \right) \in \reals^n$ denote the vector such that $\left(\m^{\prime}\left( \a, \b \right) \right)_i = \pi^{(\ell)}_i M'(a_i, b_i)$ for $i \in [n]$.
Applying proximal identity \eqref{eq:prox_id_2}, a standard property of the proximal operator yielding $g_i(u, v) = - \pi_i M^{\prime}\left( \prox_{b \pi_i M(\cdot, v)}(u), v \right)$ for $i \in [n]$, together with the AMP equations \eqref{eq:vector-valued-amp}, we obtain:
\begin{align*}
    \frac{1}{\delta b} \left(\hat{\bzeta}^{t + 1} - \hat{\bzeta}^t \right) &= - \tilde{\X}^\top \m^{\prime}\left(\hat{\bvartheta}^t, \y \right) \\
    &= - \tilde{\X}^\top \m^{\prime}\left(\tilde{\X} \hat{\bzeta}^{t + 1} + \ttheta^t - \ttheta^{t + 1}, \y \right).
\end{align*}
Using the above, we write
\begin{align*}
\|\nabla  \cC(\hat{\bzeta}^t) \|_2
&= \left\| - \tilde{\X}^\top \m^{\prime}\left(\tilde{\X} \hat{\bzeta}^{t + 1}, \y \right) \right\|_2 \\
&= \left\|- \tilde{\X}^\top \m^{\prime}\left(\tilde{\X} \hat{\bzeta}^{t + 1}, \y \right) \right. \\
&\hspace{2cm} \left. + \tilde{\X}^{\top} \m^{\prime}\left(\tilde{\X} \hat{\bzeta}^{t + 1} + \ttheta^t - \ttheta^{t + 1}, \y \right) + \frac{1}{\delta b} (\hat{\bzeta}^{t + 1} - \hat{\bzeta}^t)  \right\|_2\\
&\leq \frac{1}{\delta b} \| \hat{\bzeta}^{t + 1} - \hat{\bzeta}^t \|_2 + \tilde{C} \| \tilde{\X} \|_2 \| \ttheta^t - \ttheta^{t + 1} \|_2.
\end{align*}
where $\tilde{C} := \sup_{v, u \in \reals, i \in [n]} \pi_i M^{\prime \prime} (u, v) < \infty$ by Assumptions \ref{asmpt:reg-loss}(\ref{cond:strict-cvx-smooth}), and \ref{asmpt:reg-weights}.
Applying Lemma \ref{lemma:cauchy_t_plus_h} and that $\lim_{n \to \infty} \| \tilde{\X} \|_2 <\infty$ almost surely \citep{Anderson_Guionnet_Zeitouni_2009}, we have that for any $\epsilon > 0$,
\begin{align*}
    \lim_{t \to \infty} \lim_{n \to \infty} \P\left( \frac{1}{p} \|\nabla  \cC(\hat{\bzeta}^t) \|_2^2 > \epsilon \right) = 0.
\end{align*}
The remaining result concerning $\hat{\TTheta}$ and $\hat{\TTheta}^t$ follows similarly by noticing that,
\begin{align*}
    \|\hat{\ttheta}^{(\ell)} - \hat{\ttheta}^{(\ell), t} \|_2 &= \| \tilde{\X} \hat{\bbeta}^{(\ell)} - \left( \tilde{\X} \hat{\bbeta}^{(\ell), t + 1} + \ttheta^{(\ell), t} - \ttheta^{(\ell), t + 1}\right) \|_2 \\
    &\leq \| \tilde{\X} \hat{\bbeta}^{(\ell)} - \tilde{\X} \hat{\bbeta}^{(\ell), t + 1} \|_2 + \| \ttheta^{(\ell), t} - \ttheta^{(\ell), t + 1} \|_2 \\
    &\leq \|\tilde{\X}\|_2 \| \hat{\bbeta}^{(\ell)} - \hat{\bbeta}^{(\ell), t + 1} \|_2 + \| \ttheta^{(\ell), t} - \ttheta^{(\ell), t + 1} \|_2.
\end{align*}
Applying the arguments above together with Lemma \ref{lemma:cauchy_t_plus_h}, and noting that $\allowbreak \lim_{n \to \infty} \allowbreak \|\tilde{\X}\|_2 \allowbreak < \infty$ almost surely, we obtain the result.
\end{proof}
\section{Proof of existence of limits in fixed point equations} \label{sec:proof-existence-of-limits-in-fixed-point-equations}
We will use the following lemma to prove the existence of limits in  \eqref{eq:b-fp-implicit}--\eqref{eq:cross-cov-fp-implicit}.
\begin{lemma} \label{lemma:conv-of-chgpts}
Consider the model assumptions in Section \ref{sec:setting}, and for $i \in [n]$ let $f_i: \reals^c \times \reals \to \reals^c$ be bounded where $c \in \naturals$. Then for any $\A \in \reals^{n \times L}$ with rows denoted $\a_i$ for $i \in [n]$, we have
\begin{align*}
 &    \lim_{n \to \infty} \left\| \frac{1}{n} \sum_{i=1}^n f_i(\a_i, \Psi_i) - \sum_{\ell = 1}^{L^*} \frac{1}{n} \sum_{i = \floor{\alpha_{\ell-1} n} + 1}^{\floor{\alpha_{\ell} n}} f_i(\a_i, \Psi_{\eta_{\ell-1}}) \right\|  \\
  & =   \lim_{n \to \infty} \left\| \frac{1}{n} \sum_{i=1}^n f_i(\a_i, \Psi_i) - \sum_{\ell = 1}^{L^*} \frac{1}{n} \sum_{i = \floor{\alpha_{\ell-1} n} + 1}^{\floor{\alpha_{\ell} n}} f_i(\a_i, \Psi_{i}) \right\| = 0.
\end{align*}
\end{lemma}
\begin{proof}
Let $\ell \in [L^*]$. Due to the convergence assumption on $\eeta/n$, we have that $\eta_{\ell-1} = \floor{\alpha_{\ell-1} n} + 1 + r_n$ and $\eta_{\ell} = \floor{\alpha_{\ell} n} + 1 + s_n$, where $r_n, s_n = o(n)$. Moreover, due to the structure of $\bPsi$, we have that $\Psi_i = \Psi_{\eta_{\ell-1}}$ for $i \in [\eta_{\ell-1}, \eta_{\ell})$, and hence that $\frac{1}{n} \sum_{i=\eta_{\ell - 1}}^{\eta_{\ell} - 1} f_i(\a_i, \Psi_i) = \frac{1}{n} \sum_{i=\eta_{\ell - 1}}^{\eta_{\ell} -1} f_i(\a_i, \Psi_{\eta_{\ell-1}})$. This yields:
\begin{align}
    \frac{1}{n} \sum_{i=1}^n f_i(\a_i, \Psi_i) = \sum_{\ell = 1}^{L^*} \frac{1}{n} \sum_{i=\eta_{\ell-1}}^{\eta_{\ell} - 1} f_i(\a_i, \Psi_i) = \sum_{\ell = 1}^{L^*} \frac{1}{n} \sum_{i=\eta_{\ell-1}}^{\eta_{\ell} - 1} f_i(\a_i, \Psi_{\eta_{\ell-1}}). \label{eq:relabeling-eq}
\end{align}
We then have that:
\begin{align*}
    &\left\|\frac{1}{n} \sum_{i=\eta_{\ell - 1}}^{\eta_{\ell}-1} f_i(\a_i, \Psi_{\eta_{\ell-1}}) - \frac{1}{n} \sum_{i = \floor{\alpha_{\ell-1} n} + 1}^{\floor{\alpha_{\ell} n}} f_i(\a_i, \Psi_{\eta_{\ell-1}}) \right\| \\
    &\leq M |s_n| / n + M |r_n| / n \to 0,
\end{align*}
where $M \in \reals_{\geq 0}$ is such that $\|f_i\| \leq M$, for $i \in [n]$. The result then follows from \eqref{eq:relabeling-eq}.
\end{proof}
The existence of limits in \eqref{eq:b-fp-implicit}--\eqref{eq:cross-cov-fp-implicit} then follows from a simple argument involving the dominated convergence theorem.
\begin{proof}[Proof of Lemma \ref{lemma:general-pointwise-conv}.]
We first prove that the limits in \eqref{eq:b-fp-implicit-bar}--\eqref{eq:cross-cov-fp-implicit-bar} exist, after which the result follows by Proposition \ref{prop:conv-nonlin-eqns}. We argue that the following terms are bounded:
\begin{enumerate}
    \item $\frac{1}{n} \sum_{i=1}^n \E[(1 + b^{(\ell)} \sbM_i'')^{-1}]$, \label{eq:bound-qty-1}
    \item $\frac{1}{n} \sum_{i = 1}^n \E[\Z_i \sbM_i']$, \label{eq:bound-qty-2}
    \item $\frac{1}{n} \sum_{i = 1}^n \E \left[ (\sbM_i^{(\ell)})' (\sbM_i^{(\ell')})' \right]$ \label{eq:bound-qty-3},
\end{enumerate}
for $\ell, \ell' \in [L]$. The first is bounded above by $1$ due to $b^{(\ell)} > 0$ and Assumption \ref{asmpt:reg-loss}(\ref{cond:strict-cvx-smooth}). Quantities (\ref{eq:bound-qty-2})--(\ref{eq:bound-qty-3}) can be bounded by applying the Cauchy-Schwarz inequality together with Proposition \ref{prop:M_prime_ub}.
The smoothness assumption on $M$ together with Remark \ref{rmk:prox-implicit-function-theorem-and-b} on the continuity of the proximal operator implies that each term involved in the sums (\ref{eq:bound-qty-1})--(\ref{eq:bound-qty-3}) is continuous with respect to $b^{(\ell)} \in \reals_{\geq 0}$ and $\pi_i \in \reals_{\geq 0}$.

It then suffices to prove that
\begin{equation*}
        \lim_{n \to \infty} \frac{1}{n} \sum_{i=1}^n \sh(\a, \pi_{i}, \Psi_i)
\end{equation*}
exists for some $c \in \naturals$, where $\sh: \reals^c \times \reals \times \reals \to \reals^c$ is a bounded and continuous function with respect to its second argument, and for all $\a \in \reals^c$. From Lemma \ref{lemma:conv-of-chgpts}, we have that
\begin{align*}
    \frac{1}{n} \sum_{i=1}^n \sh(\a, \pi_{i}, \Psi_i) \to \sum_{\ell = 1}^{L^*} \frac{1}{n} \sum_{i = \floor{\alpha_{\ell-1} n} + 1}^{\floor{\alpha_{\ell} n}} \sh(\a, \pi_i, \Psi_{\eta_{\ell-1}}),
\end{align*}
and hence it suffices to prove that
\begin{align*}
    D_{n, \ell}(\a) := \lim_{n \to \infty} \frac{1}{n} \sum_{i = \floor{\alpha_{\ell-1} n} + 1}^{\floor{\alpha_{\ell} n}} \sh(\a, \pi_i, \Psi_{\eta_{\ell-1}})
\end{align*}
exists, for $\ell \in [L^*]$. We express the above sum as follows:
\begin{align*}
    D_{n, \ell}(\a)
    &= \sum_{j = \floor{\alpha_{\ell - 1}n}}^{\floor{\alpha_{\ell} n} - 1} \int_{\frac{j}{n}}^{\frac{j+1}{n}} \sh(\a, \pi_{j + 1}, \Psi_{\eta_{\ell-1}}) dt
    = \int_{\frac{\floor{\alpha_{\ell-1}n}}{n}}^{\frac{\floor{\alpha_{\ell}n}}{n}} \sh(\a, \pi_{\floor{nt} + 1}, \Psi_{\eta_{\ell-1}}) dt.
\end{align*}
Recalling $\Phi(t)$ from Assumption~\ref{asmpt:reg-weights}, by the bounded property of $\sh$ we obtain
\begin{align*}
    &\left|D_{n, \ell}(\a) - \int_{\alpha_{\ell - 1}}^{\alpha_{\ell}} \sh(\a, \Phi(t), \Psi_{\eta_{\ell-1}}) \right| \\
    &\leq \left|D_{n, \ell}(\a) - \int_{\alpha_{\ell - 1}}^{\alpha_{\ell}} \sh(\a, \pi_{\floor{nt} + 1}, \Psi_{\eta_{\ell-1}}) \right| \\
    &\hspace{2cm} + \left|\int_{\alpha_{\ell - 1}}^{\alpha_{\ell}} \sh(\a, \pi_{\floor{nt} + 1}, \Psi_{\eta_{\ell-1}}) - \int_{\alpha_{\ell - 1}}^{\alpha_{\ell}} \sh(\a, \Phi(t), \Psi_{\eta_{\ell-1}}) \right| \\
    &\leq \left|\int_{\left[\frac{\floor{\alpha_{\ell-1}n}}{n}, \alpha_{\ell-1}\right) \cup \left[\frac{\floor{\alpha_{\ell}n}}{n}, \alpha_{\ell}\right)} |\sh(\a, \pi_{\floor{nt} + 1}, \Psi_{\eta_{\ell-1}})| dt \right| + r_n \\
    &\leq M \left( \left|\frac{\floor{\alpha_{\ell-1}n}}{n} - \alpha_{\ell-1}\right| + \left|\frac{\floor{\alpha_{\ell}n}}{n} - \alpha_{\ell} \right|\right) + r_n \\
    &\leq \frac{2M}{n} + r_n,
\end{align*}
where $M>0$ is a constant such that $|\sh| \leq M$, and {$r_n\to0$ as $n\to\infty$} due to the dominated convergence theorem. Hence, $\lim_{n \to \infty} D_{n, \ell}(\a) = \int_{\alpha_{\ell - 1}}^{\alpha_\ell} h(\a, \Phi(t), \Psi_{\eta_{\ell - 1}})$ and the result follows.
\end{proof}
\section{Proof of existence and uniqueness of solutions to  \eqref{eq:b-fp-implicit}--\eqref{eq:cross-cov-fp-implicit}}
\subsection{Proof of Proposition \ref{prop:existence-of-fixed-point-solution} (existence of solution) } \label{sec:existence-of-solns}
By Proposition \ref{prop:conv-nonlin-eqns}, it suffices instead to prove the existence of solutions to equations \eqref{eq:b-fp-implicit-bar}--\eqref{eq:cross-cov-fp-implicit-bar} which we recall do not carry an explicit dependence on $\bvarepsilon$. Applying Proposition \ref{prop:steins-lemma-fp}, this is equivalent to proving the existence of solutions to the following set of equations, for $\ell, \ell' \in [L]$:
\begin{align}
    1 - \frac{1}{\delta} &= \lim_{n \to \infty} \frac{1}{n} \sum_{i=1}^n \E\left[ \left(1 + b^{(\ell)} (\sbM_i^{(\ell)})'' \right)^{-1} \right], \label{eq:b-fp-form-bar-2} \\
    \blambda^{(\ell)} &= \lim_{n \to \infty} \frac{\delta b^{(\ell)}}{n} \sum_{i = 1}^n \E\left[ - \left( \frac{{\partial_{12}} \sbM^{(\ell)}_i}{b^{(\ell)} (\sbM^{(\ell)}_i)'' + 1} \right) \nabla_{\Z_i} q(\Z_i, \Psi_i, \bar{\varepsilon}) \right], \label{eq:mu-fp-form-bar-2} \\
    \kappa_{\ell, \ell'} &= \lim_{n \to \infty} \frac{\delta b^{(\ell)} b^{(\ell')}}{n} \sum_{i = 1}^n \E \left[ (\sbM_i^{(\ell)})' (\sbM_i^{(\ell')})' \right] \label{eq:cross-cov-fp-form-bar-2}.
\end{align}
We show in Lemma \ref{lemma:continuity-of-fp-eqns} that the quantities on the right in  \eqref{eq:b-fp-form-bar-2}--\eqref{eq:cross-cov-fp-form-bar-2} are continuous with respect to the  parameters $(\blambda^{(\ell)}, b^{(\ell)}, \kappa_{\ell, \ell}, \kappa_{\ell, \ell'})$, and then use Schauder's Fixed Point Theorem  \citep[Theorem 1]{Franklin1980} to deduce the existence of solutions to equations \eqref{eq:b-fp-form-bar-2}--\eqref{eq:cross-cov-fp-form-bar-2}.  We begin with the following technical lemma.
\begin{lemma} \label{lemma:equicontinuous_h_i}
Let $\ell \in [L], i \in [n], \bar{b} \in \reals_{>0}$. Consider the model assumptions  in Section \ref{sec:setting} and further suppose Assumptions \ref{asmpt:reg-loss}(\ref{cond:strict-cvx-smooth}), \ref{asmpt:reg-loss}(\ref{eq:loss-asmpt-log-conc}), and \ref{asmpt:reg-weights} hold. Then, recalling that $\bar{y}_i := q(\Z_i, \Psi_i, \bar{\varepsilon})$, the family of functions
\begin{align*}
       h_i(b) := \E\left[ \left(1 + b \pi^{(\ell)}_i M''\left(\prox_{b \pi^{(\ell)}_i M(\cdot, \bar{y}_i)}(\Z_i^{\top} \blambda^{(\ell)} + w^{(\ell)}_i), \bar{y}_i \right) \right)^{-1} \right],
\end{align*}
for $i \in [n]$, is equicontinuous with respect to $b \in [0, \bar{b}]$, as per  \cite[Definition $7.22$]{rudin1976principles}.
\end{lemma}
\begin{proof}
Let $i \in [n]$ and omit the superscript $^{(\ell)}$ for brevity. We first note that by proximal identity \eqref{eq:prox_id_3}, $h_i(b)$ is the expectation of the derivative of a proximal operator. By Remark \ref{rmk:prox-implicit-function-theorem-and-b} and by continuity of $M''$, we have that the following function is continuous with respect to $b \in [0, \bar{b}]$:
\begin{align*}
    \tilde{h}_i(b; z, \bar{y}) := (1 + b \pi_i M''(\prox_{b \pi_i M(\cdot, \bar{y})}(z), \bar{y}))^{-1},
\end{align*}
where $z, \bar{y} \in \reals$. Moreover, by Remark \ref{rmk:prox-implicit-function-theorem-and-b} and differentiability of $x \mapsto (1+x)$ and $\pi_i M''$, $\tilde{h}_i$ is differentiable with respect to $b \in (0, \bar{b})$.
From the mean-value theorem, we then have the following, for $b_1, b_2 \in [0, \bar{b}]$:
\begin{align*}
    &\left|\E[\tilde{h}_i(b_1; {\Z_i^\top} \blambda + w_i, \bar{y}_i)] - \E[\tilde{h}_i(b_2; {\Z_i^\top} \blambda + w_i, \bar{y}_i)] \right| \\
    &= |\E \partial_b \tilde{h}_i(\tilde{b}; {\Z_i^\top} \blambda + w_i, \bar{y}_i)| \cdot |b_1 - b_2|,
\end{align*}
for some $\tilde{b} \in (0, \bar{b})$. We now seek to bound the derivative term uniformly over $i \in [n]$ and $b \in [0, \bar{b}]$. Using \eqref{eq:prox_id_4}, we expand the derivative term as
\begin{align*}
   \partial_b (1 + b \sbM_i'')^{-1} &= - (1 + b \sbM_i'')^{-2} \left(\sbM_i'' + b \sbM_i''' \cdot \partial_b \prox_{b \pi_i M(\cdot, \bar{y}_i)}({\Z_i^\top} \blambda + w_i) \right)\\
   &= - \frac{\sbM_i''}{(1 + b \sbM_i'')^2} + \frac{b \sbM_i'''}{(1 + b \sbM_i'')^2} \cdot \frac{\sbM_i'}{(1 + b \sbM_i'')} \\
   &= \frac{b\left( \sbM_i'''\sbM_i' - (\sbM_i'')^2 \right) - \sbM_i''}{(1 + b \sbM_i'')^3}.
\end{align*}
Note that, by the Lipschitz property of $q$, we have that:
\begin{align}
    \left| q(\Z_i, \Psi_i, \bar{\varepsilon}) \right| \leq |q(0, 0, 0)| + C_q \frac{\sqrt{\|\Z_i\|^2_2 + \Psi^2_i + |\bar{\varepsilon}|^2}}{\sqrt{L + 2}}, \label{eq:q_PL_bound}
\end{align}
for some constant $C_q > 0$. We then obtain the bound:
\begin{align}
    &|\E \partial_b \tilde{h}_i(\tilde{b}; {\Z^\top} \blambda + w, \bar{y})| \\
    &\leq \E |\partial_b \tilde{h}_i(\tilde{b}; {\Z^\top} \blambda + w, \bar{y}_i)| \notag \\
    &= \E \left| \partial_b (1 + b \sbM_i'')^{-1} \right| \notag \\
    &\leq \E \left| \frac{b\left( \sbM_i'''\sbM_i' - (\sbM_i'')^2 \right)}{(1 + b \sbM_i'')^3} \right| + \E \left| \frac{\sbM_i''}{(1 + b \sbM_i'')^3} \right|\notag \\
    &\leq b \E [|\sbM_i'''| |\sbM_i'|]  + b \E[ (\sbM_i'')^2]+ \E \left| \sbM_i'' \right|\notag \\
    &\leq b B_{111} \E \big[|M'(0, 0, 0)| + B_{11} |{\Z_i^\top} \blambda + w_i| + B_{12}|q(\Z_i, \Psi_i, \bar{\varepsilon})| \big] + b B_{11}^2 + B_{11} \label{eq:E_partial_bound_1} \\
    &\leq \bar{b} \tilde{c} + B_{11} \label{eq:E_partial_bound_2}
\end{align}
for some positive constants $B_{111}, B_{11}, B_{12}, \tilde{c} > 0$, where \eqref{eq:E_partial_bound_1} follows from Proposition \ref{prop:M_prime_ub}, and \eqref{eq:E_partial_bound_2} follows from \eqref{eq:q_PL_bound} and from the fact that $\Z_i, \bar{\varepsilon}$ are random variables with bounded second moment as per the model assumptions in Section \ref{sec:setting}. Since the above argument holds for any fixed $\bar{b} > 0$, the result follows.
\end{proof}
We are now in a position to prove that the RHS of equations \eqref{eq:b-fp-form-bar-2}--\eqref{eq:cross-cov-fp-form-bar-2} are continuous with respect to the parameters of interest.
\begin{lemma}\label{lemma:continuity-of-fp-eqns}
Consider the model assumptions  in Section \ref{sec:setting} and further suppose Assumptions \ref{asmpt:reg-loss}(\ref{cond:strict-cvx-smooth}), \ref{asmpt:reg-loss}(\ref{eq:loss-asmpt-log-conc}),  \ref{asmpt:reg-weights} hold.
{Then, for $\ell,\ell'\in[L]$, the RHS of equations \eqref{eq:b-fp-form-bar}--\eqref{eq:cross-cov-fp-form-bar} are continuous in $(\b,\bLambda,\bK)$ on every compact set with $\b\geq0$ and $\bK=\bK^\top\succeq0$. In particular, $\kappa_{\ell,\ell'}$ may be negative, but must satisfy $\kappa_{\ell,\ell'}^2\leq\kappa_{\ell,\ell}\kappa_{\ell',\ell'}$.}
\end{lemma}
\begin{proof}
We let $\ell, \ell' \in [L]$ and drop the superscript $^{(\ell)} \in [L]$ for notational convenience when it can be inferred from context. Since the averaging operation preserves equicontinuity, it suffices to prove equicontinuity of the following equations for $i \in [n]$:
\begin{enumerate}
    \item $\sh_{1i}(b, \kappa^{1/2}_{(\ell, \ell)}, \blambda) := \E\left[ \left(1 + b (\sbM_i)'' \right)^{-1} \right]$, \label{eq:cont-eq1}
    \item $\sh_{2i}(b, \kappa^{1/2}_{(\ell, \ell)}, \blambda) := \E\left[ \Z_i \sbM_i' \right]$, \label{eq:cont-eq2}
    \item $\sh_{3i}\left(b, \kappa^{1/2}_{(\ell, \ell)}, {\kappa_{\ell,\ell'}}, \blambda^{(\ell)}\right) := \E \left[ (\sbM_i^{(\ell)})' (\sbM_i^{(\ell')})' \right]$. \label{eq:cont-eq3}
\end{enumerate}
The RHS of (\ref{eq:cont-eq1}) is bounded above by $1$ due to $b^{(\ell)} > 0$ and Assumption \ref{asmpt:reg-loss}(\ref{cond:strict-cvx-smooth}). The RHS of equations (\ref{eq:cont-eq2})--(\ref{eq:cont-eq3}) can be bounded by applying the Cauchy-Schwarz inequality together with Proposition \ref{prop:M_prime_ub}.
By the dominated convergence theorem, smoothness and {convexity} of $M$, and Remark \ref{rmk:prox-implicit-function-theorem-and-b}, (\ref{eq:cont-eq1})--(\ref{eq:cont-eq3}) are continuous with respect to the parameters in question. We proceed by proving equicontinuity of each equation with respect to each parameter via the mean value theorem, and then applying the Arzel\`a-Ascoli theorem to obtain the result. {Continuity in the covariance matrix is obtained by the Gaussian coupling in Lemma \ref{lemma:prop-of-cross-cov-simple}, including at singular matrices.}

We first let $\tilde{\ell} \in [L]$ and list the following partial derivatives, obtained via identities \eqref{eq:prox_id_3}, \eqref{eq:prox_id_4}:
\begin{align}
\begin{aligned} \label{eq:M'-M''-derivs}
    &\partial_b \sbM_i' = - \frac{\sbM_i'}{1 + b \sbM_i''} \cdot \sbM_i'', \ \quad
    \partial_{\lambda_{\tilde{\ell}}} \sbM_i' = \frac{Z_{i, \tilde{\ell}}}{1 + b \sbM_i''} \cdot \sbM_i'', \  \quad
    \partial_{\lambda_{\tilde{\ell}}} \sbM_i'' = \frac{Z_{i, \tilde{\ell}}}{1 + b \sbM_i''} \cdot \sbM_i''', \\
    &\partial_{{\kappa^{1/2}_{(\ell, \ell)}}} \sbM_i' = \frac{\tilde{w}_i}{1 + b \sbM_i''} \cdot \sbM_i'', \qquad
    \partial_{{\kappa_{(\ell, \ell)}}^{1/2}} \sbM_i'' = \frac{\tilde{w}_i}{1 + b \sbM_i''} \cdot \sbM_i''',
\end{aligned}
\end{align}
where $\tilde{w}_i \distas{\text{i.i.d.}} \N(0, 1)$ for $i \in [n]$.

Equicontinuity of the family $\{\sh_{1i}\}_{i \in [n]}$ on the domain $[0, \bar{b}]$ follows from Lemma \ref{lemma:equicontinuous_h_i}. Further, we apply the mean value theorem and the dominated convergence theorem to obtain, for $\blambda^{(a)}, \blambda^{(b)} \in \reals^L$ and $i \in [n]$:
\begin{align*}
    &\left|\sh_{1i}(\cdot, \cdot, \blambda^{(a)}) - \sh_{1i}(\cdot, \cdot, \blambda^{(b)}) \right| \\
    &\leq \sum_{\tilde{\ell} = 1}^L \E\left| \partial_{\lambda_{\tilde{\ell}}} (1 + b (\sbM_i)'')^{-1} \right| \left| \lambda_{\tilde{\ell}}^{(a)} - \lambda_{\tilde{\ell}}^{(b)} \right| \\
    & \le \sum_{\tilde{\ell} = 1}^L \left( \E  \max_{\lambda_{\tilde{\ell}} \in (\lambda_{\tilde{\ell}}^{(a)},\lambda_{\tilde{\ell}}^{(b)})} \left| \frac{b \sbM_i'''}{(1 + b \sbM_i'')^3} Z_{i, \tilde{\ell}} \right| \right) \cdot \left| \lambda_{\tilde{\ell}}^{(a)} - \lambda_{\tilde{\ell}}^{(b)} \right|, \\
    &\leq \sum_{\tilde{\ell} = 1}^L \left( \E\left[ \left( \max_{\lambda_{\tilde{\ell}} \in (\lambda_{\tilde{\ell}}^{(a)},\lambda_{\tilde{\ell}}^{(b)})}  \left| \frac{b \sbM_i'''}{(1 + b \sbM_i'')^3} \right|\right)^2 \right]^{1/2} \cdot \E[Z^2_{i, \tilde{\ell}}]^{1/2} \right) \cdot \left| \lambda_{\tilde{\ell}}^{(a)} - \lambda_{\tilde{\ell}}^{(b)} \right|,
\end{align*}
where the first term is uniformly bounded over $i \in [n]$ by assumption. Next, we observe that, for $\kappa_{(a)}^{1/2}, \kappa_{(b)}^{1/2}$ and $i \in [n]$:
\begin{align*}
     &\left|\sh_{1i}(\cdot, \kappa^{1/2}_{(a)}, \cdot) - \sh_{1i}(\cdot, \kappa^{1/2}_{(b)}, \cdot) \right| \\
     &\leq \left( \E \max_{\kappa_{\ell, \ell}^{1/2} \in (\kappa^{1/2}_{(a)}, \kappa^{1/2}_{(b)})} \left| \frac{b \sbM_i'''}{(1 + b \sbM_i'')^3} \tilde{w}_i \right| \right) \left|\kappa^{1/2}_{(a)} - \kappa^{1/2}_{(b)} \right| \\
     &\leq  \E\left[  \left( \max_{\kappa_{\ell, \ell}^{1/2} \in (\kappa^{1/2}_{(a)}, \kappa^{1/2}_{(b)})} \left| \frac{b \sbM_i'''}{(1 + b \sbM_i'')^3} \right| \right)^2 \right]^{1/2} \E\left[\tilde{w}^2_i\right]^{1/2}  \left|\kappa^{1/2}_{(a)} - \kappa^{1/2}_{(b)} \right|,
\end{align*}
where the first term is uniformly bounded over $i \in [n]$ by assumption.

{For $\sh_{3i}$, apply the Gaussian coupling in Lemma \ref{lemma:prop-of-cross-cov-simple} to the covariance matrix of $(w_i^{(\ell)},w_i^{(\ell')})$. The quantities in the product defining $\sh_{3i}$ are the weighted derivatives $(\sbM_i^{(\ell)})'$ and $(\sbM_i^{(\ell')})'$ from \eqref{eq:sM-def-proof}. On any compact set $\mathcal D$ in the statement, Proposition \ref{prop:M_prime_ub} gives
\[
\sup_{(\b,\bLambda,\bK)\in\mathcal D}\ \sup_{n,\,i\in[n],\,\ell\in[L]}
\E\left[((\sbM_i^{(\ell)})')^2\right]<\infty.
\]
Together with \eqref{eq:M'-M''-derivs} and Cauchy-Schwarz, this proves joint equicontinuity of (\ref{eq:cont-eq1}) to (\ref{eq:cont-eq3}) on $\mathcal D$. The bounds also hold at $b=0$, since $\pi_iM''/(1+b\pi_iM'')\leq B_{11}$, where $B_{11}:=\sup_{u,v}M''(u,v)<\infty$.

The right hand side of \eqref{eq:mu-fp-form-bar}, which defines $\blambda^{(\ell)}$, is treated similarly: the numerator involves $\partial_{12}M$, whose derivative in its first argument is $\partial_2M''$. Both are bounded by Assumption \ref{asmpt:reg-loss}(\ref{eq:loss-asmpt-log-conc}), and $\|\nabla_{\Z_i}q\|_2\leq C_{\text{qLip}}$ almost everywhere. Thus its summands are also equicontinuous on $\mathcal D$. Averaging preserves these bounds, and the pointwise limits exist by the averaging argument in Lemma \ref{lemma:general-pointwise-conv}. The Arzel\`a-Ascoli theorem \citep[Theorem 7.25]{rudin1976principles} yields a uniformly convergent subsequence on $\mathcal D$. Its limit equals the pointwise limit and is continuous, proving the claim.
}
\end{proof}
We now apply the above continuity result together with Schauder's Fixed Point Theorem  \citep[Theorem 1]{Franklin1980} to deduce the existence of solutions to equations \eqref{eq:b-fp-form-bar-2}--\eqref{eq:cross-cov-fp-form-bar-2} within a compact set.
\begin{proof}[Proof of Proposition \ref{prop:existence-of-fixed-point-solution}.]
We will prove the existence of a solution to \eqref{eq:b-fp-form-bar-2}--\eqref{eq:cross-cov-fp-form-bar-2} for $\ell=\ell' \in [L]$, and will omit the superscript $^{(\ell)}$. {We treat the off-diagonal covariances at the end of the proof.}

Define the quantities:
\begin{align*}
    B_{11} := \sup_{u, v} M''(u, v), \;\; B_{12} := {\sup_{u, v} |\partial_{12} M(u, v)|}, \;\; {\bar{\pi} := \sup_{n,\,i\in[n]} \pi_i \leq 1.}
\end{align*}
\begin{align*}
    \varpi_n^{(2)}(b, \blambda, \kappa) &:= \frac{\delta b}{n} \sum_{i = 1}^n \E\left[ - \left( \frac{{\partial_{12}} \sbM_i}{b (\sbM_i)'' + 1} \right) \nabla_{\Z_i} q(\Z_i, \Psi_i, \bar{\varepsilon}) \right], \\
    \varpi^{(2)}(b, \blambda, \kappa) &:= \lim_{n \to \infty} \varpi^{(2)}_n(b, \blambda, \kappa), \\
    \varpi_n^{(3)}(b, \blambda, \kappa) &:= \frac{\delta b^2}{n} \sum_{i = 1}^n \E \left[ (\sbM'_i)^2 \right], \\
    \varpi^{(3)}(b, \blambda, \kappa) &:= \lim_{n \to \infty} \varpi^{(3)}_n(b, \blambda, \kappa),
\end{align*}
and let $b = \varpi_1^{-1}(\blambda, \kappa)$ denote the unique solution to \eqref{eq:b-fp-form-bar-2}; {the integral representation \eqref{eq:A-limit-integral}, the strict decrease in \eqref{eq:deriv-wrt-b-simple}, and the bounds below guarantee its existence and uniqueness. Its continuity follows from Lemma \ref{lemma:continuity-of-fp-eqns}, uniqueness, and the uniform bounds $0<\underline b\leq b\leq\bar b<\infty$ below.} We recall that the quantities $\varpi^{(2)}(b, \blambda, \kappa)$, $\varpi^{(3)}(b, \blambda, \kappa)$ are well-defined (the corresponding limits exist) by the arguments in Appendix \ref{sec:variants-nonlin-eqns}, p.\pageref{eq:existence-of-limits-argument}. We note that continuity of $\lim_n \varpi_n^{(3)}$ follows from Lemma \ref{lemma:continuity-of-fp-eqns}, and a similar argument to that in Lemma \ref{lemma:continuity-of-fp-eqns} together with Proposition \ref{prop:steins-lemma-fp} and the assumptions in the proposition statement can be used to show that $\lim_n \varpi_n^{(3)}$ is continuous.

The proof of existence involves showing that, for $(b, \blambda, \kappa)$ in a compact set $B$ of the form
\begin{align}
    B := [\underline{b}, \bar{b}] \times [\underline{\blambda}, \bar{\blambda}] \times [0, \bar{\kappa}], \label{eq:B-set}
\end{align}
{we have that $(b_*, \varpi^{(2)}(b_*, \blambda, \kappa), \varpi^{(3)}(b_*, \blambda, \kappa)) \in B$, where $b_*:=\varpi_1^{-1}(\blambda,\kappa)$.}
The existence of a fixed point then follows by the Schauder fixed-point theorem \citep[Theorem $1$]{Franklin1980}.

We first construct the set $[\underline{\blambda}, \bar{\blambda}]$, and restate equation \eqref{eq:mu-fp-form-bar-2} as $\blambda = \lim_n \varpi_n^{(2)}(b, \blambda, \kappa)$.
We have that
\begin{align}
    &\left\|\frac{\delta b}{n} \sum_{i = 1}^n \E\left[ - \left( \frac{{\partial_{12}} \sbM_i}{b \sbM_i'' + 1}\right) \nabla_{\Z_i} q(\Z_i, \Psi_i, \bar{\varepsilon})\right] \right\|_2 \notag \\
    &\leq \frac{\delta b}{n} \sum_{i = 1}^n \E\left[ \left|\left( \frac{{\partial_{12}} \sbM_i}{b \sbM_i'' + 1}\right) \right| \left\|\nabla_{\Z_i} q(\Z_i, \Psi_i, \bar{\varepsilon}) \right\|_2\right]  \notag \\
    &\leq \delta b \cdot B_{12} \cdot \sqrt{C_{PL2}} \label{eq:ex-step-b}
\end{align}
where \eqref{eq:ex-step-b} follows for some constant ${C}_{\text{PL2}} > 0$ from Assumptions \ref{asmpt:reg-loss}(\ref{cond:strict-cvx-smooth}), \ref{asmpt:reg-loss}(\ref{eq:loss-asmpt-log-conc}), and the Lipschitz assumption on $q$. Therefore, setting
\begin{align}
    \bar{\blambda} := \delta \bar{b} B_{12} \sqrt{C_{PL2}} \1_L \;\; \text{and} \;\; \underline{\blambda} := - \delta \bar{b} B_{12} \sqrt{C_{PL2}} \1_L, \label{eq:set-lambdas}
\end{align}
we have that $\lim_n \varpi_n^{(2)}(\blambda) \in [\underline{\blambda}, \bar{\blambda}]$.

{
Turning to equation \eqref{eq:cross-cov-fp-form-bar-2}, let $b=b_*$ and define
\begin{align*}
    r_i(v) &:= v-\prox_{b\pi_iM(\cdot,\bar y_i)}(v), \qquad
    m_i := \argmin_u M(u,\bar y_i), \\
    \rho &:= \frac{\bar b\bar\pi B_{11}}{1+\bar b\bar\pi B_{11}}<1.
\end{align*}
By \eqref{eq:prox_id_3}, $0\leq r_i'(v)\leq\rho$, and $r_i(m_i)=0$. Hence $r_i(v)^2\leq\rho r_i(v)(v-m_i)$. Strong convexity gives $|m_i|\leq (|M'(0,0)|+B_{12}|\bar y_i|)/C_{\text{cvx}}$, where $M''\geq C_{\text{cvx}}>0$. The Lipschitz assumption on $q$ and the bounded second moment of $\bar\varepsilon$ therefore give a constant $C>0$ such that
\begin{align*}
    \sup_{n,\,i\in[n],\,\blambda\in[\underline{\blambda},\bar{\blambda}]}
    \E[(\Z_i^\top\blambda-m_i)^2]\leq C.
\end{align*}
Writing $v_i=\Z_i^\top\blambda+\sqrt\kappa\tilde w_i$, where $\tilde w_i\sim\N(0,1)$ is independent of $(\Z_i,\bar\varepsilon)$, we have $r_i(v_i)=b\sbM_i'$. Cauchy--Schwarz and Stein's lemma, conditional on $(\Z_i,\bar\varepsilon)$, yield
\begin{align*}
    \varpi_n^{(3)}(b,\blambda,\kappa)
    &\leq \frac{\delta\rho}{n}\sum_{i=1}^n\E[r_i(v_i)(\Z_i^\top\blambda-m_i)]
       +\frac{\delta\rho\kappa}{n}\sum_{i=1}^n\E[r_i'(v_i)] \\
    &\leq \rho\sqrt{\delta C\varpi_n^{(3)}(b,\blambda,\kappa)}
       +\frac{\delta\rho\kappa}{n}\sum_{i=1}^n\E[r_i'(v_i)].
\end{align*}
Since $b=b_*$ solves \eqref{eq:b-fp-form-bar-2}, $\lim_n n^{-1}\sum_i\E[r_i'(v_i)]=1/\delta$. Taking limits, we obtain
\begin{align*}
    \varpi^{(3)}(b,\blambda,\kappa)
    \leq \rho\sqrt{\delta C\varpi^{(3)}(b,\blambda,\kappa)}+\rho\kappa.
\end{align*}
Solving this quadratic inequality in $\sqrt{\varpi^{(3)}(b,\blambda,\kappa)}$ shows that $\varpi^{(3)}(b,\blambda,\kappa)\in[0,\bar\kappa]$ for $\kappa\in[0,\bar\kappa]$, upon setting
\begin{align}
    \bar\kappa := \frac{\delta C\rho^2}{(1-\rho)^2}. \label{eq:set-kappa-bar}
\end{align}
}
Moving on to constructing $[\underline{b}, \bar{b}]$, notice that if $b < \frac{1}{(\delta - 1) B_{11} \bar{\pi}}$, we have that
\begin{align*}
    \lim_{n \to \infty} \frac{1}{n} \sum_{i=1}^n \E\left[ (1 + b \sbM_i'')^{-1} \right] \geq (1 + b \bar{\pi} B_{11})^{-1} > 1 - 1/\delta.
\end{align*}
Hence,
\begin{align}
b \geq \frac{1}{(\delta - 1) B_{11} \bar{\pi}} =: \underline{b} \label{eq:set-underline-b}
\end{align}
is necessary for a solution to exist in \eqref{eq:b-fp-form-bar-2}. Moreover, since $M$ is strongly convex by assumption, there exists a positive constant $C_{\text{cvx}}$ such that $M'' \geq C_{\text{cvx}} > 0$. {By Assumption \ref{asmpt:reg-weights} and dominated convergence,
\begin{align*}
    \lim_{n\to\infty}\frac1n\sum_{i=1}^n\E[(1+b\sbM_i'')^{-1}]
    \leq \int_0^1\frac{dt}{1+bC_{\text{cvx}}\Phi(t)}\longrightarrow0
    \qquad (b\to\infty),
\end{align*}
since $\Phi>0$ almost everywhere. Choose $\bar b\geq\underline b$ large enough that
\begin{align}
    \int_0^1\frac{dt}{1+\bar b C_{\text{cvx}}\Phi(t)}<1-1/\delta. \label{eq:set-bar-b}
\end{align}
Then $b\leq\bar b$ is necessary for a solution to \eqref{eq:b-fp-form-bar-2}. Together with continuity, strict decrease in $b$, and the value $1$ at $b=0$, these bounds also establish the existence and uniqueness of $b_*$ used above.}

Finally, prescribing parameters \eqref{eq:set-bar-b}, \eqref{eq:set-underline-b}, \eqref{eq:set-kappa-bar}, \eqref{eq:set-lambdas} for the set $B$ in \eqref{eq:B-set} and applying Schauder's Fixed Point Theorem yields {a solution to the diagonal equations. For the full system, choose the bounds uniformly over the finitely many $\ell\in[L]$, and replace $[0,\bar\kappa]$ by the compact convex set $\{\bK=\bK^\top\succeq0:\kappa_{\ell,\ell}\leq\bar\kappa,\ \ell\in[L]\}$. The right hand side of \eqref{eq:cross-cov-fp-form-bar-2} is a limit of Gram matrices and hence is positive semidefinite, while the same argument bounds each diagonal entry by $\bar\kappa$. Thus the joint map also preserves this compact convex set, and Schauder's Fixed Point Theorem yields the result.}
\end{proof}
\subsection{Proof of Proposition \ref{prop:uniqueness_of_fixed_points} (uniqueness of solution)} \label{sec:uniqueness_of_solns}
We now prove that, under the setting described in Section \ref{sec:setting}, if a solution to equations \eqref{eq:b-fp-implicit}--\eqref{eq:cross-cov-fp-implicit} exists, then it is unique. By the equivalence established in Appendix \ref{sec:variants-nonlin-eqns}, we work with equations \eqref{eq:b-fp-implicit-bar}--\eqref{eq:cross-cov-fp-implicit-bar}. First we prove that, given $(\bLambda, \bK)$, $\b$ is unique, after which the result follows by a contradiction argument applying the AMP theory in Lemma \ref{lemma:SE-matrix-amp}.

The proof technique for this first step relies on the limits in \eqref{eq:b-fp-implicit-bar}--\eqref{eq:cross-cov-fp-implicit-bar} being well-defined, after which we apply equicontinuity with respect to $\b$ together with the Arzelà-Ascoli theorem to conclude that the RHS of equations \eqref{eq:b-fp-implicit-bar}--\eqref{eq:cross-cov-fp-implicit-bar} converges uniformly in $n$ and is hence continuous with respect to $\b$. Uniqueness then follows by a mean value theorem argument similar to those used to prove similar results for M-estimation and logistic regression in \cite{donoho_high_2016, sur_modern_2019}. We first prove uniform convergence.
\begin{lemma} \label{lemma:unif-conv-onsager-eq-cpt}
Consider the model assumptions  in Section \ref{sec:setting} and further suppose Assumptions \ref{asmpt:reg-loss}(\ref{cond:strict-cvx-smooth}), \ref{asmpt:reg-loss}(\ref{eq:loss-asmpt-log-conc}), \ref{asmpt:reg-weights} hold. For $\ell \in [L]$, the function:
\begin{align}
    A^{(\ell)}_n(b) = \frac{1}{n} \sum_{i=1}^n \E_{\Z_i, w^{(\ell)}_i, \bar{\varepsilon}} \left[ \left(1 + b (\sbM^{(\ell)}_i)''(b) \right)^{-1} \right] \label{eq:A_n_def}
\end{align}
converges uniformly as $n \to \infty$ for $b \in [0, \infty)$, where $(\sbM^{(\ell)}_i)'' = (\sbM^{(\ell)}_i)''(b)$ is defined in \eqref{eq:sM-def-proof}.
\end{lemma}
\begin{proof}
Let $\ell \in [L]$ and omit the superscripts $^{(\ell)}$ for notational convenience. We first prove that the result holds for $b$ in a compact domain $[0, \bar{b}]$, for $\bar{b} \in \reals_{> 0}$, and finally conclude by applying a truncation argument. By Lemma \ref{lemma:equicontinuous_h_i}, the family of functions $\{h(b; \pi_i)\}_{i \in \naturals}$  is equicontinuous on the domain $[0, \bar{b}]$, where:
\begin{align*}
       h(b; \pi_i) := \E\left[ (1 + b \pi_i M''(\prox_{b \pi_i M(\cdot, \bar{y}_i)}({\Z_i^\top} \blambda + w_i), \bar{y}_i))^{-1} \right].
\end{align*}
The sequence of averages $\{A_n\}_{n = 1}^\infty$ is also equicontinuous. Further noting that $|h(b; \pi_i)| \leq 1$, yields $|A_n| \leq 1$. Applying the Arzel\`a-Ascoli theorem \citep[Theorem 7.25]{rudin1976principles}, we obtain that $\{A_n\}_{n = 1}^\infty$ contains a uniformly convergent subsequence for $b \in [0, \bar{b}]$. Further, by Lemma \ref{lemma:general-pointwise-conv} we have that the sequence $\{A_n\}_{n = 1}^\infty$ converges pointwise to a limit.  Therefore, by a standard argument \cite[Exercise $7.16$]{rudin1976principles} the entire sequence converges uniformly for $b \in [0, \bar{b}]$, yielding:
\begin{align*}
    \lim_{n \to \infty} \sup_{b \in [0, \bar{b}]} |A_n(b) - \lim_{n \to \infty} A_n(b)| = 0.
\end{align*}
We now extend this result to prove that $\{A_n\}_{n = 1}^\infty$ converges uniformly within the unbounded domain $[0, \infty)$ via a truncation argument. By the assumption that, for all $t \in [0, 1]$, $\pi_{\lfloor n t \rfloor + 1} \to \Phi(t) \in \reals_{>0}$ as $n \to \infty$, we can choose $N$ large enough such that $\pi_i \geq \underline{\pi}$ and $\pi_i \leq \bar{\pi}$ for all $i \geq N$ and for some $\bar{\pi} \geq \underline{\pi} > 0$. We can then bound $h$ for $i \geq N$ as follows:
\begin{align*}
    &h(b; \pi_i) \\
    &\leq \E\left[\Big(1 + b \underline{\pi} \min_{\tilde{\pi} \in [\underline{\pi}, \bar{\pi}]} \min_{\tilde{\Psi} \in [L]} M''(\prox_{b \tilde{\pi} M(\cdot, q(\Z_i, \tilde{\Psi}, \bar{\varepsilon}))}({\Z_i^\top} \blambda + w_i), q(\Z_i, \tilde{\Psi}, \bar{\varepsilon})) \Big)^{-1} \right] \\
    &=: \phi(b) \\
    &\leq 1,
\end{align*}
where we note $\phi$ does not depend on $i, n \in \naturals$. We then have that, for $b > \bar{b}$ and for $n > N$:
\begin{align*}
    &\left|A_n(b) - \lim_{m \to \infty} A_m(b)\right| \\
    &= \left|\frac{1}{n} \sum_{i = 1}^n h(b; \pi_i) - \lim_{m \to \infty} \frac{1}{m} \sum_{j = 1}^m h(b; \pi_j) \right| \\
    &\leq \frac{1}{n} \sum_{i = 1}^N |h(b; \pi_i)| + \frac{1}{n} \sum_{i = N+1}^n |h(b; \pi_i)| + \limsup_{m \to \infty} \frac{1}{m} \sum_{j = 1}^m |h(b; \pi_j)| \\
    &\leq \frac{N}{n} \phi(b) + \phi(b) + \phi(b) \leq 3\phi(b),
\end{align*}
where $\phi(b) \to 0$ as $b \to \infty$ by the dominated convergence theorem and by the lower bound assumption on $M''$. For all $b \in [0, \infty)$, we have:
\begin{align*}
    &\left|A_n(b) - \lim_{m \to \infty} A_m(b)\right| \\
    &\leq \sup_{\tilde{b} \in [0, \bar{b}]} \left|A_n(\tilde{b}) - \lim_{m \to \infty} A_m(\tilde{b}) \right| + \sup_{\tilde{b} \in [\bar{b}, \infty)} \left|A_n(\tilde{b}) - \lim_{m \to \infty} A_m(\tilde{b}) \right|.
\end{align*}
Given $\epsilon > 0$, we choose $\bar{b}$ and $\tilde{N} > N$ large enough such that $\sup_{\tilde{b} \in [0, \bar{b}]} |A_n(\tilde{b}) - \lim_{m \to \infty} A_m(\tilde{b})| < \varepsilon/2$ and $\phi(b) < \varepsilon/6$ for $n > \tilde{N}$ and $b > \bar{b}$ to obtain the result.
\end{proof}
The following lemma proves the existence and uniqueness of $\b$, which will be used directly in our proof of Proposition \ref{prop:uniqueness_of_fixed_points} below.
\begin{lemma} \label{lemma:existence-uniqueness-onsager-solution}
Let $\ell \in [L]$ and consider the setting described in Section \ref{sec:setting}. Given $\blambda \in \reals^L, \kappa_{\ell, \ell} \in \reals_{\geq 0}$, a solution $b^{(\ell)}$ to equation \eqref{eq:b-fp-implicit-bar} exists and is unique in the domain $(0, \infty)$.
\end{lemma}
\begin{proof}
We prove the existence and uniqueness of $b^{(\ell)}$ for each $\ell \in [L]$, and omit the superscripts $^{(\ell)}$ for notational convenience. Recall the quantities defined in \eqref{eq:sM-def-proof} and \eqref{eq:A_n_def}:
\begin{align*}
   &\sbM''_i = \sbM''_i(b) := \pi_i M''(\prox_{b \pi_i M(\cdot, \bar{y}_i)}({\Z_i^\top} \blambda + w_i), \bar{y}_i), \\
    &A_n(b) := \frac{1}{n} \sum_{i = 1}^n \E_{\Z_i, w_i, \bar{\varepsilon}} \left[ (1 + b \sbM''_i(b))^{-1} \right],
\end{align*}
and let
\begin{align*}
   &\sbM'''_i = \sbM'''_i(b) := \pi_i M'''(\prox_{b \pi_i M(\cdot, \bar{y}_i)}({\Z_i^\top} \blambda + w_i), \bar{y}_i).
\end{align*}
We then restate \eqref{eq:b-fp-implicit-bar} as follows:
\begin{align}
    1 - \frac{1}{\delta} = \lim_{n \to \infty} A_n(b). \label{eq:b-fp-alternate-form}
\end{align}
Notice that $A_n$ is continuous with $|A_n| \leq 1$. Moreover, $A_n$ converges uniformly in $b \in [0, \infty)$ by Lemma \ref{lemma:unif-conv-onsager-eq-cpt}, yielding the existence of $\lim_n A_n =: A$ where $A: [0, \infty) \to \reals$ is continuous in $b$ by standard arguments \cite[Theorem 7.12]{rudin1976principles}.
Existence then follows by proving:
\begin{align}
    &\lim_{b \to 0} A(b) = 1 \label{eq:b-0-lim}\\
    &\lim_{b \to \infty} A(b) = 0. \label{eq:b-inf-lim}
\end{align}
By the dominated convergence theorem and by Assumption~\ref{asmpt:reg-loss}(\ref{eq:loss-asmpt-b-M-to-inf}), we have that $\lim_n \lim_{b \to \infty} A_n(b) = 0$. Moreover, due to uniform convergence, we have that $\lim_{b \to \bar{b}} \lim_n A_n(b) = \lim_n \lim_{b \to \bar{b}} A_n(b)$ for any limit point $\bar{b}$ of $(0, \infty)$  \cite[Theorem 7.11]{rudin1976principles}. Hence, we have that $\lim_{b \to \infty} A(b) = 0$. For the $b \to 0$ limit point, notice that
\begin{align*}
    A_n(b) = 1 - \frac{1}{n} \sum_{i = 1}^n \E\left[\frac{b \sbM''_i(b)}{1 + b \sbM''_i(b)} \right] =: 1 - B_n(b),
\end{align*}
and hence it suffices to show that $\lim_{b \to 0} B_n = 0$. Note that, for $b > 0$, $x \mapsto bx / (1 + bx)$ is strictly increasing in $x$ and concave for $bx > -1$. We have by Jensen's inequality that
\begin{align*}
    0 \leq B_n(b) \leq \frac{ b \sup_{\bar{b} > 0, i \in [n]} \E \sbM''_i(\bar{b})}{1 + b \sup_{\bar{b} > 0, i \in [n]} \E \sbM''_i(\bar{b})} \leq 1,
 \end{align*}
which implies that $B_n(b) \to 0$ as $b \to 0$ due to Assumption \ref{asmpt:reg-loss}(\ref{cond:strict-cvx-smooth}). By the preceding arguments and due to uniform convergence, we have proved \eqref{eq:b-0-lim} and hence the existence of a solution to \eqref{eq:b-fp-alternate-form}.

To show uniqueness, it suffices to prove that $A = \lim_n A_n$ is a strictly decreasing function of $b$. We first apply the chain rule and \eqref{eq:prox_id_4} to compute the following derivative:
\begin{align}
    \partial_b (1 + b \sbM''_i)^{-1} &= - \frac{1}{(1 + b \sbM_i'')^2} \left(\sbM_i'' + b \sbM_i''' \left(- \frac{\sbM_i'}{(1 + b \sbM_i'')} \right) \right) \notag\\
    &= \frac{b\left(\sbM_i''' \sbM_i' - (\sbM_i'')^2 \right) - \sbM_i''}{(1 + b \sbM_i'')^3} \label{eq:deriv-wrt-b-simple}\\
    &{\leq 0} \label{eq:deriv-wrt-b-neg}
\end{align}
where \eqref{eq:deriv-wrt-b-neg} follows from Assumptions \ref{asmpt:reg-loss}(\ref{cond:strict-cvx-smooth}) and \ref{asmpt:reg-loss}(\ref{eq:loss-asmpt-log-conc}). {For every positive weight, \eqref{eq:deriv-wrt-b-simple} is strictly negative. More explicitly, the integral calculation in the proof of Lemma \ref{lemma:general-pointwise-conv} gives
\begin{align}\label{eq:A-limit-integral}
A(b)&=\sum_{r=1}^{L^*}\int_{\alpha_{r-1}}^{\alpha_r}H(b,\Phi(t),r)\,dt,\notag\\
H(b,\pi,r)&:=\E\left[\left(1+b\pi M''\left(\prox_{b\pi M(\cdot,\bar y^{(r)})}(\Z^\top\blambda+w),\bar y^{(r)}\right)\right)^{-1}\right],
\end{align}
where $\bar y^{(r)}:=q(\Z,r,\bar\varepsilon)$ and $\Z\sim\N(0,\bGamma)$, $w\sim\N(0,\kappa_{\ell,\ell})$, $\bar\varepsilon\sim\P_{\bar\varepsilon}$ are independent. By \eqref{eq:deriv-wrt-b-simple}, $H(b_1,\pi,r)>H(b_2,\pi,r)$ for $0\leq b_1<b_2$ and $\pi>0$. Since $\Phi(t)>0$ almost everywhere by Assumption \ref{asmpt:reg-weights}, integrating in \eqref{eq:A-limit-integral} gives $A(b_1)>A(b_2)$. Thus $A$ is strictly decreasing, which proves uniqueness.
}
\end{proof}
\begin{proof}[Proof of Proposition \ref{prop:uniqueness_of_fixed_points}.]
The proof of uniqueness is analogous to \cite[Corollary 2]{sur_likelihood_2019}, and to \cite[Corollary 4.3]{donoho_high_2016}.

By Lemma \ref{lemma:SE-matrix-amp} combined with Lemma \ref{lemma:ERM_frobenius_characterization}, we have that
\begin{align*}
   \frac{1}{\delta p} \B^{\top} \bSigma \hat{\B} = \frac{1}{\delta p} \B^{\top} \L \L^\top \hat{\B} \stackrel{\P}{\simeq} \E\left[ \frac{1}{\delta p} \B^\top \bSigma \B \bLambda + \frac{\sqrt{\delta}}{\delta p} \B^\top \L \G \right] = \bGamma \bLambda,
\end{align*}
and hence we have that
\begin{align}
     {\bGamma^{-1}\frac{1}{\delta p}\B^\top\bSigma\hat\B\stackrel{\P}{\simeq}\bLambda}. \label{eq:uniqueness-of-nu}
\end{align}
Now, recall that $\hat{\B}$ satisfies $\|\hat{\B}\|_F < \infty$ with probability $\to 1$ as $p \to \infty$ by Assumption \ref{asmpt:reg-loss}(\ref{eq:existence-of-estimator}), and is hence unique by the curvature Assumption \ref{asmpt:reg-loss}(\ref{cond:likelihood_curvature}). We therefore have that the right hand side in \eqref{eq:uniqueness-of-nu} above must be unique. Similarly, we have that
\begin{align*}
   {\frac{1}{\delta p}(\L^\top\hat\B-\L^\top\B\bLambda)^\top(\L^\top\hat\B-\L^\top\B\bLambda)\stackrel{\P}{\simeq}\bK,}
\end{align*}
{Since $\hat\B$ and $\bLambda$ are unique, the left hand side has a unique limit, and hence $\bK$ is unique.} Then, Lemma \ref{lemma:existence-uniqueness-onsager-solution} establishes that $\b$ must also be unique.
\end{proof}
\section{Maximum likelihood estimation in the logistic model} \label{sec:logistic}
Let $\ell \in [L]$ and omit the superscripts. The logistic loss is defined as $M(u, v) = \zeta(u) - vu$, where $\zeta(z) := \log{(1 + e^z)}$. Plugging this into \eqref{eq:ERM_est}, we obtain:
\begin{align} \label{eq:logistic_mle}
    \hat{\bbeta} \in \argmin_{\tilde{\bbeta} \in \reals^p} \sum_{i = 1}^n \pi_i \left\{ \zeta(\x_i^\top \tilde{\bbeta}) - y_i \x_i^\top \tilde{\bbeta} \right\},
\end{align}
where $y_i = \ind\{ \varepsilon_i \leq \zeta'(\x_i^\top \bbeta)\}$ for $\varepsilon_1, \dots, \varepsilon_n \distas{\text{i.i.d.}} U[0, 1]$.
Plugging in $\pi_i M = \pi_i \left\{ \zeta(\x_i^\top \tilde{\bbeta}) - y_i \x_i^\top \tilde{\bbeta} \right\}$ into \eqref{eq:b-fp-implicit}--\eqref{eq:cross-cov-fp-implicit}, we obtain:
\begin{align}
    \begin{split}
        1 - \frac{1}{\delta} &= \lim_{n \to \infty} \frac{1}{n} \sum_{i = 1}^n \\
        &\qquad \E\left[\left({1 + b \pi_i \zeta''(\prox_{b \pi_i \zeta}({\Z_i^\top} \blambda + w_i + b \pi_i q(\Z_i, \Psi_i, \varepsilon_i))}\right)^{-1} \right] \\
        \0_L &= \lim_{n \to \infty} \frac{1}{n} \sum_{i = 1}^n \\
        &\qquad \E\left[ \Z_i \left(\pi_i q(\Z_i, \Psi_i, \varepsilon_i) - \pi_i \zeta'(\prox_{b \pi_i \zeta}({\Z_i^\top} \blambda + w_i + b \pi_i q(\Z_i, \Psi_i, \varepsilon_i))) \right) \right], \\
        \kappa_{\ell, \ell} &= \lim_{n \to \infty} \frac{{\delta b^2}}{n} \sum_{i = 1}^n \\
        &\qquad \E\left[ \left(\pi_i q(\Z_i, \Psi_i, \varepsilon_i) - \pi_i \zeta'(\prox_{b \pi_i \zeta}({\Z_i^\top} \blambda + w_i + b \pi_i q(\Z_i, \Psi_i, \varepsilon_i))) \right)^2\right].
    \end{split}
\end{align}
Letting $\pi_i = 1$ for $i \in [n]$, and $L = 1$, we recover the nonlinear equations (5) that characterize the logistic maximum likelihood estimator (MLE) in the homogeneous regression setting of \cite{sur_modern_2019}.
\subsection{Existence of MLE in the logistic model}

The works \citep{sur_modern_2019} and \citep{zhao_asymptotic_2022} considered the homogeneous ($L = 1, \pi_i = 1$) logistic observation model $y_i = \ind\{\varepsilon_i \leq \zeta'(\x_i^\top \bbeta)\}$ and proved that the logistic MLE \eqref{eq:logistic_mle} in this scenario exists with high probability if and only if $\gamma < g_{\text{MLE}}(\kappa)$, where $\gamma^2 := \lim_{p \to \infty} \Var{(\x_i^\top \bbeta)}$,
and $g_{\text{MLE}}$ is defined in \cite[Theorem 1]{sur_modern_2019}. In what follows, we prove that the logistic change point observation model $y_i = \ind\{\varepsilon_i \leq \zeta'(\x_i^\top \bbeta^{(\Psi_i)})\}$ {admits the following sufficient condition for existence.}
\begin{proposition}
 \label{prop:logistic_mle_existence}
   {Under Assumption~\ref{asmpt:reg-weights}, the logistic MLE} \eqref{eq:logistic_mle} operating for the model \eqref{eq:model} with $q: (a, b) \mapsto \ind\left\{ b \leq \zeta'(a) \right\}$ {exists with high probability for $\x_i\distas{\text{i.i.d.}}\N(\0_p,\I/n)$ if} $\lim_{p \to \infty} \Var(\x_i^\top \bbeta^{(\eta_{\ell -1})}) < g_{\text{MLE}}\left({\lim_{n\to\infty}\frac{p}{\eta_{\ell}-\eta_{\ell-1}}}\right)^2$ for all $\ell \in [L^*]$.
\end{proposition}
\begin{proof}
{For $\ell\in[L^*]$, suppose the inequality in the proposition holds and split the observations $(\y,\X)$ into homogeneous sections:
\[
I_\ell:=\{i\in[n]:\eta_{\ell-1}\leq i<\eta_\ell,\ \pi_i>0\},\qquad
\y^{(\ell)}:=\y_{I_\ell},\quad \X^{(\ell)}:=\X_{I_\ell,:}.
\]
{By Assumption~\ref{asmpt:reg-weights}, $(\eta_\ell-\eta_{\ell-1}-|I_\ell|)/n\to0$ as $n\to\infty$. Hence
\[
\frac{p}{|I_\ell|}\longrightarrow\frac{1}{\delta(\alpha_\ell-\alpha_{\ell-1})}
=\lim_{n\to\infty}\frac{p}{\eta_\ell-\eta_{\ell-1}},
\]
so the strict inequality in the proposition still holds after deleting zero-weight observations. Define}
\begin{align}\label{eq:logistic_obj}
\cL(\tilde{\bbeta};\y^{(\ell)},\X^{(\ell)})
&:=\sum_{i\in I_\ell}\left\{\zeta(\x_i^\top\tilde{\bbeta})-y_i\x_i^\top\tilde{\bbeta}\right\}\notag\\
&=\sum_{i\in I_\ell}\log\left(1+e^{-(2y_i-1)\x_i^\top\tilde{\bbeta}}\right).
\end{align}
By \cite[Theorem 1]{sur_modern_2019}, with high probability each retained section has a finite MLE and full column rank. On this event, for every unit vector $u$, some $i\in I_\ell$ satisfies $(2y_i-1)\x_i^\top u<0$: otherwise moving in direction $u$ would strictly decrease the loss, contradicting the existence of a minimizer. Continuity and compactness of the unit sphere therefore give
\[
c_{\ell,n}:=\min_{\|u\|_2=1}\max_{i\in I_\ell}\{-(2y_i-1)\x_i^\top u\}>0.
\]
Since $\log(1+e^a)\geq\max\{a,0\}$, we have $\cL(\tilde{\bbeta};\y^{(\ell)},\X^{(\ell)})\geq c_{\ell,n}\|\tilde{\bbeta}\|_2$. Letting $m_{\ell,n}:=\min_{i\in I_\ell}\pi_i>0$, we then obtain
\begin{align}\label{eq:logistic_obj_lb}
\sum_{i=1}^n\pi_i\left\{\zeta(\x_i^\top\tilde{\bbeta})-y_i\x_i^\top\tilde{\bbeta}\right\}
&\geq\sum_{\ell=1}^{L^*}m_{\ell,n}\cL(\tilde{\bbeta};\y^{(\ell)},\X^{(\ell)})\notag\\
&\geq\|\tilde{\bbeta}\|_2\sum_{\ell=1}^{L^*}m_{\ell,n}c_{\ell,n}\to\infty
\end{align}
as $\|\tilde{\bbeta}\|_2\to\infty$, for each fixed $n$. The weighted loss is continuous, so it attains its minimum on a sufficiently large closed ball. This proves existence of a finite weighted MLE with high probability.
}
\end{proof}
\subsection{Numerical challenges in the logistic model} \label{sec:numerical-challenges-in-the-logistic-model}
Proposition \ref{prop:logistic_mle_existence} specifies {a sufficient region where the logistic MLE \eqref{eq:logistic_mle} exists.
Outside this region, existence of the MLE is not guaranteed by the proposition. When no finite minimizer exists, the nonnegative loss can approach its infimum along a sequence of candidate optimizers with $\|\tilde{\bbeta}\|_2\to\infty$, which can make numerical optimization unstable.
This poses a challenge to estimating the signal strength matrix $\bGamma$ under the logistic model using the procedure in Section \ref{sec:uncertainty-quantification}, p.\pageref{para:estimating-signal-strength-matrix}. For each candidate $(\tilde\delta,\tilde{\bGamma})$, we check numerically whether equations \eqref{eq:b-fp-implicit} to \eqref{eq:cross-cov-fp-implicit} admit a solution. Any solution is unique by Proposition \ref{prop:uniqueness_of_fixed_points} when the corresponding logistic MLE \eqref{eq:logistic_mle} exists, but numerical solvers may fail to find a solution for some candidates. For the squared loss $a,b\mapsto(a-b)^2$, a least-squares minimizer always exists and the fixed point equations have explicit solutions (Section \ref{sec:linear-model-theory}), avoiding this numerical difficulty.}

\section{Additional experiments and details} \label{sec:additional-experiments-and-examples}
\subsection{{Numerical implementation details}} \label{sec:numerical-implementation-details}
For \WeightedERM{} and \WeightedERML{}, coefficient estimation used a direct weighted least-squares solver in the unpenalized least-squares setting, and an accelerated proximal
gradient solver (monotone FISTA with backtracking) otherwise. The iterative solver used a maximum of $50000$ iterations with convergence tolerance $10^{-5}$ for the Huber
loss experiment in Section \ref{sec:lin-model-sparse-signals}, $20000$ iterations with tolerance $10^{-6}$ for the sparse logistic experiment in Section \ref{sec:logistic-model-sparse-signals-exp}, and $10000$ iterations with tolerance $10^{-7}$ for the FRED-MD experiment in Section \ref{sec:FRED}. Convergence was checked using the relative
change in the coefficients and the residual in the first-order optimality conditions. The unpenalized least-squares fits used to produce Figure \ref{fig:M1a}, in the dense regression experiment in Appendix \ref{sec:dense-gaussian-comparison}, and in the differentially-sparse experiment in Section \ref{sec:lin-model-sparse-signal-diffs} did not require
iterative optimization. Cross-validation over the $\ell_1$ penalty used the three-point grid $\{\lambda^{\mathrm{def}}/10,\lambda^{\mathrm{def}},10\lambda^{\mathrm{def}}\}$, with
method-specific defaults $\lambda^{\mathrm{def}}$. For \WeightedERML{}, this corresponded to $\lambda=C\sqrt{2\log(p)/n_{\mathrm{train}}}$ with $C\in\{0.03,0.3,3\}$, where $n_{\mathrm{train}}$ is the
training sample size. The selected constant was retained when refitting on the full sample. For \GLMBSA{} in Section~\ref{sec:logistic-model-sparse-signals-exp}, the corresponding
penalty multiplier was selected from $\{0.018,0.18,1.8\}$, retaining the method's own penalty scaling.

For the Myocardial infarction data experiment in Section \ref{sec:myocardial}, the auxiliary least-squares fits and posterior logistic fits are unpenalized and have no intercept.  After removing observations with missing age and excluding age from the covariates, the analysis uses $n=1692$ and $p=110$.
For estimating $\hat{\bGamma}$, we initialize with $\hat{\bGamma}^{(0)}=\operatorname{diag}(G_{11},G_{22})$, where $\G= \hat{\B}^{\top}\hat{\B}/p \in \reals^{2 \times 2}$ is the outer product of the signal estimate matrix $\hat{\B}$ obtained from \eqref{eq:ERM_est} using unpenalized least-squares, the uniform change point prior and sample weights described in Section \ref{sec:myocardial}, and $L = 2$. We then execute $16$ iterations of equations \eqref{eq:b-fp-implicit}--\eqref{eq:cross-cov-fp-implicit} and \eqref{eq:theta-deb-fp-eqn} (as described on page \pageref{para:estimating-signal-strength-matrix}) to obtain $\hat{\bGamma}$.
The regression fits used to obtain \eqref{eq:theta-deb-defn} then use $16$ zero-start FISTA iterations.

Unless otherwise stated, trials with empty change point estimates are omitted from Hausdorff error plots as the Hausdorff error of an empty estimate is undefined. A Python implementation of the experiments in this manuscript is available in the WeightedERM-reference package
\citep{arpino_2026_weightederm_reference}.
\subsection{Additional details for logistic model example on p.\pageref{page:logistic}}\label{sec:logistic-example-additional-details}
We set dimension $p = 800$, true change point at $0.4n$. For $i \in [n]$,  $\varepsilon_i \distas{\text{i.i.d.}} U[0, 1]$, and  $\x_i \distas{\text{i.i.d.}} \N(0, \bSigma/n)$ where $\Sigma_{k, j} := 0.2^{|k-j|}$ for $k, j \in [p]$. Signals $(\beta_j^{(1)}, \beta_j^{(2)}) \distas{\text{i.i.d.}} \N\left(0, \delta \begin{bmatrix}
    1 & -0.3 \\
    -0.3 & 1
\end{bmatrix}\right)$, for $j \in [p]$.
\subsection{Additional details for posterior distribution example in Figure \ref{fig:posterior-sq-loss}} \label{sec:posterior-example-additional-details}
The prior $\pi_{\bar{\bPsi}}$ was set to be uniform over all two change point locations that are at least $n/5$ from each other and from the endpoints. The base loss $M$ was set to the squared error loss, and the weights were computed via \eqref{eq:pi_i_defn}.
For $j \in [p]$, we sampled $(\beta_j^{(1)}, \beta_j^{(2)}, \beta_j^{(3)}) \distas{\text{i.i.d.}} \N\left(\0_3, \delta \begin{bmatrix}
1 & 0.1 & 0 \\
0.1 & 1 & 0 \\
0 & 0 & 1
\end{bmatrix} \right)$. The noise  $\bvarepsilon$ was chosen to be standard Gaussian, and covariate vectors sampled as $\x_i \distas{\text{i.i.d.}} \N(\0_p, \bSigma/n)$, for $i \in [n]$, with $\Sigma_{k,j} = 0.1^{|k-j|}$.
\subsection{{Comparison with \McScan{} for dense Gaussian signals}} \label{sec:dense-gaussian-comparison}
We complement the sparse-signal-difference experiment in Section \ref{sec:lin-model-sparse-signal-diffs} with a comparison in which the regression signals are dense and independent. We consider the linear model \eqref{eq:linear-chgpt-model} with two change points at $0.35n$ and $0.65n$. We fix $p=200$ and vary the sampling ratio $\delta=n/p$ over $\{2,\ldots,9\}$. For $i\in[n]$, we sample $\x_i\distas{\text{i.i.d.}}\N(\0_p,\bSigma/n)$, where $\Sigma_{k,j}=0.4^{|k-j|}$, and independently sample the entries of each regression vector as $\beta_j^{(\ell)}\distas{\text{i.i.d.}}\N(0,\delta)$. The noise variables are independent $\N(0,0.1)$ random variables. Results are based on $100$ paired trials for every value of $\delta$, so both methods receive the same simulated data in each trial.

We run \McScan{} in \texttt{auto} mode. We run \WeightedERM{} using the $10$-fold absolute-error cross-validation procedure outlined on p.\pageref{par:CV} of Section \ref{sec:synthetic-experiments}, with $L = 4$, squared error loss $M(a, b) = (a-b)^2$, and no coefficient penalty. The covariate matrix is multiplied by $\sqrt{n}$ and the response is standardized before fitting both methods.

\begin{figure}[!tp]
    \centering
    \begin{subfigure}[b]{0.49\textwidth}
        \centering
        \includegraphics[width=\textwidth]{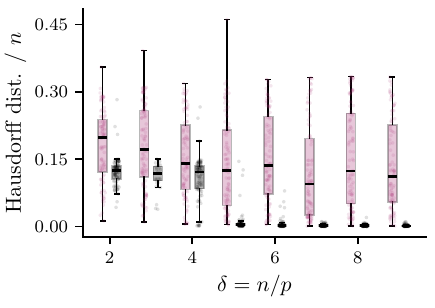}
        \label{fig:dense-gaussian-location}
    \end{subfigure}
    \hfill
    \begin{subfigure}[b]{0.49\textwidth}
        \centering
        \includegraphics[width=\textwidth]{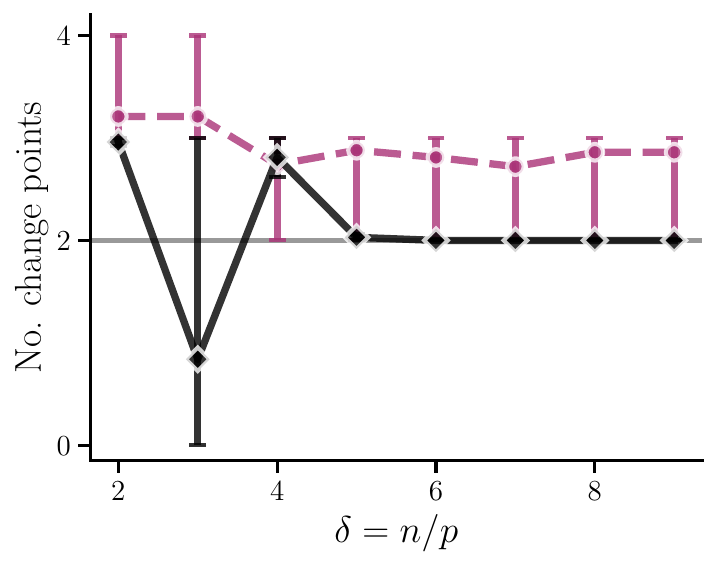}
        \label{fig:dense-gaussian-size}
    \end{subfigure}
    \caption{\small Comparison of \WeightedERM{} (\WERM{}, black) and \McScan{} (purple) with dense Gaussian signals and Gaussian noise. Left: normalized Hausdorff error, summarized by boxplots with jittered trials. Right: mean estimated number of change points over $100$ paired trials, with absolute mean-to-quartile error bars; the grey horizontal line marks the true value of two.}
    \label{fig:dense-gaussian-comparison}
\end{figure}

Figure \ref{fig:dense-gaussian-comparison} displays normalized Hausdorff location error and the estimated number of change points for both methods. For $\delta\geq5$, \WeightedERM{} has median normalized Hausdorff error below $0.005$, whereas the median error of \McScan{} remains above $0.09$. Over the same range, the median number of change points selected by \WeightedERM{} equals the true value of two, while \McScan{} has a median of three estimated change points for every displayed value of $\delta$.
\subsection{Example: Linear model with two change points and alternating signals}
\label{sec:lin-model-alternating-signals-example}

\begin{figure}[!tp]
\raisebox{2.65pt}{
\begin{subfigure}[c]{.49\textwidth}
    \centering
        \includegraphics[width=\textwidth]{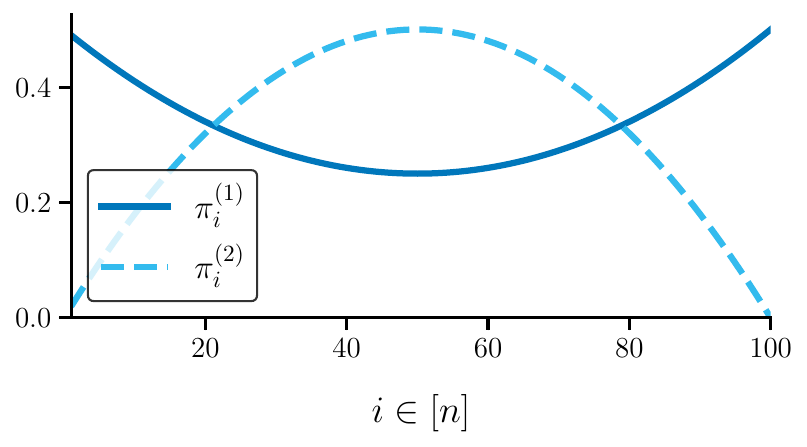}
        \caption{$\{\pi_i^{(\ell)}\}_{i \in [n], \ell \in [2]}$, $n = 100$.}
        \label{fig:marginal-two-state}
\end{subfigure}}
\hfill
\begin{subfigure}[c]{.49\textwidth}
        \centering
        \resizebox{\textwidth}{!}{\input{figures/TwoStatevsUnif/h_laplace_loc.pgf}}
        \caption{Uniform vs. alternating signal prior. Error bars indicate the $25$-th to $75$-th percentiles across $10$ trials.}
        \label{fig:unif-vs-two-state-estimation}
\end{subfigure}
\caption{Sample weights (left) and estimation performance (right) of \WeightedERM{} (\WERM{}) for linear model with two change points at $0.3n$ and $0.8n$ and alternating signals. }
\end{figure}

In this section, we consider a linear model with two change points at $0.3n$ and $0.8n$, with alternating signals. That is, the regression vector changes from $\bbeta^{(1)}$ to $\bbeta^{(2)}$ at $0.3n$, and it changes back to $\bbeta^{(1)}$  at $0.8n$. We compare the performance of \WeightedERM{} with two sets of weights.  The first  set of weights is the one used in the example on p.\pageref{page:lin-model-with-chgpts-ex} (shown in Figure \ref{fig:M1a}, left), obtained from the uniform prior that assigns equal  probability to all configurations with exactly two change points  ($L = L^* = 3$). The second set of weights is obtained from  an `alternating signal' prior, where we assume that exactly two change points are generated by two alternating regression vectors. In other words, the prior encodes the assumption that the first and third segments of the data correspond to the same signal $\bbeta^{(1)}$. This prior yields a uniform signal configuration prior $\pi_{\bar{\Psi}}$ over the set of all binary sequences of length $n$ with exactly one positive jump followed by one negative jump. The resulting sets of weights $\pi_i^{(1)} = 0.5 { {{i - 1} \choose {1 - 1}} {{n - i} \choose {L - 1}}}/{{{n - 1} \choose {L - 1}}} + 0.5 { {{i - 1} \choose {3 - 1}} {{n - i} \choose {L - 3}}}/{{{n- 1} \choose {L - 1}}}$ and $\pi_i^{(2)} = { {{i - 1} \choose {2 - 1}} {{n - i} \choose {L - 2}}}/{{{n - 1} \choose {L - 1}}}$, derived from \eqref{eq:pi_i_defn}, are displayed in Figure \ref{fig:marginal-two-state}.

\begin{figure}[!t]
    \centering
    \begin{subfigure}[b]{0.49\textwidth}
        \centering
        \resizebox{\textwidth}{!}{\input{figures/SE_est_match/h_laplace_SEmatch.pgf}}
        \label{fig:alt-signal-est-match-location-app}
    \end{subfigure}
    \caption{Theory and method match for the alternating signals example in Appendix \ref{sec:lin-model-alternating-signals-example}. Error bars indicate the $25$-th to $75$-th percentiles across $10$ trials.}
    \label{fig:alt-signal-est-match-app}
\end{figure}

We fix $p = 800$, and for $j \in [p]$, sample $(\beta_i^{(1)}, \beta_i^{(2)}) \distas{\text{i.i.d.}} \N\left(0, 0.3 {\delta} \begin{bmatrix}
   1 & 0.1\\
   0.1 & 1
\end{bmatrix}\right)$, $\x_i \distas{\text{i.i.d.}} \N(0, \bSigma/n)$ where $\Sigma_{k, j} = 0.1^{|k-j|}$ for $k, j \in [p]$, and $\varepsilon_i$ independently from a centred Laplace distribution with scale $\sqrt{0.1}$. We set the base loss $M$ in \eqref{eq:ERM_est} to be the Huber loss \eqref{eq:Huber-loss}. Omitting the penalty function $P$, we observe the estimation performance in Figure \ref{fig:unif-vs-two-state-estimation}. The figure shows the superior performance of the alternating signal prior over the uniform prior
as we vary $\delta = n/p$, and highlights the impact of the prior on estimation performance.

\subsection{{Investigating procedures for selecting number of change points}} \label{sec:investigating-count-selection}
\paragraph{Linear model with sparse signal differences.}
We compare cross-validation with penalized partition search for selecting the number of change points in the linear model with sparse signal differences described in Section \ref{sec:lin-model-sparse-signal-diffs}. We use $p=200$, $100$ paired trials, change points at $0.20n$ and $0.50n$, covariates $\N(0,\I_p/n)$, noise $\N(0,0.1)$, and the coefficient distribution in \eqref{eq:sparse_diff_prior_WERM} with $p_s=0.3$. The methods use the same simulated data within each trial. Both WERM procedures use squared error loss, no coefficient penalty and at most $L=4$ segments. Their fixed-count fits share weights obtained from a uniform location prior conditional on the count. The minimum separation between successive change points is $\Delta=\max\{\lceil n/50\rceil,\lfloor n/40\rfloor,1\}$, applied also to training-fold sample sizes.

\textbf{WERM+CV} selects the number of segments by exhaustive $10$-fold cross-validation using absolute held-out error, then refits on all observations. \textbf{WERM+P} compares the full-data fixed-count fits using the summed squared-error objective plus
\begin{align}
    P(\bpsi)=\gamma\log\binom{n-1}{K -1},\qquad K:= \max_i \psi_i \in\{1,\ldots,L\},\label{eq:search-penalty-experiment}
\end{align}
where $K := \max_i \psi_i$ is the number of segments in $\bpsi$ and $\gamma=6$ is fixed. We include \McScan{} in automatic count-selection mode and \charcoal{} with threshold $10$ and interval seed $42$. Due to its runtime, \charcoal{} is evaluated only for $\delta\leq9$, with $100$ additional trials at $\delta=2.5$ to retain its original plotting grid. Four R failures, two each at $\delta=5.5$ and $8.5$, are excluded from charcoal summaries; valid empty estimates remain in count summaries but have undefined localization error.

\begin{figure}[!tp]
    \centering
    \begin{subfigure}[t]{0.32\textwidth}
        \centering
        \includegraphics[width=\textwidth]{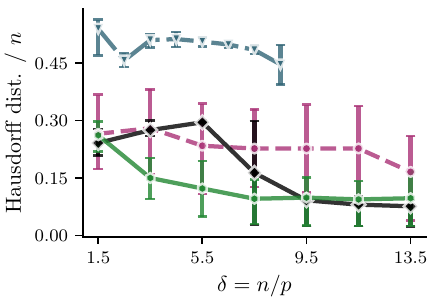}
    \end{subfigure}
    \begin{subfigure}[t]{0.32\textwidth}
        \centering
        \includegraphics[width=\textwidth]{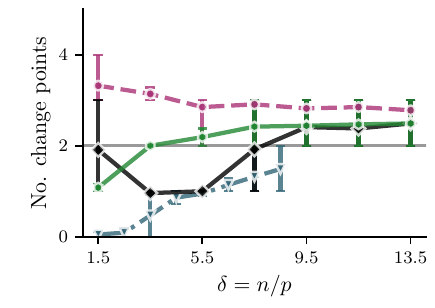}
    \end{subfigure}
    \begin{subfigure}[t]{0.32\textwidth}
        \centering
        \includegraphics[width=\textwidth]{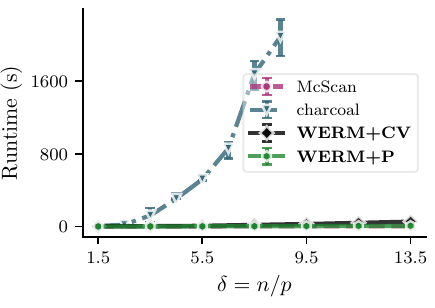}
    \end{subfigure}
    \caption{Count-selection comparison from Appendix \ref{sec:investigating-count-selection}: \textbf{WERM+CV}, \textbf{WERM+P}, \McScan{} and \charcoal{}, over $100$ paired trials. Curves show means; bars show absolute mean-to-quartile deviations.}
    \label{fig:M2-count-selection-comparison}
\end{figure}

Figure \ref{fig:M2-count-selection-comparison} shows that the penalty improves mean conditional localization relative to cross-validation at $\delta=3.5,5.5,7.5$, while cross-validation has the smaller mean error at $\delta=9.5,11.5,13.5$. The penalized search is faster in these comparisons. Both WERM timings exclude their shared preparation of full-data and fold-specific marginal weights, performed once per trial. Among the $1000$ paired WERM comparisons, the procedures select the same count in $322$ trials and return identical locations in every such trial.

\subsection{{Sensitivity to the upper bound $L$}} \label{sec:sensitivity-number-of-change-points}\label{sec:E-L-upper-bound}
We study the sensitivity of \textbf{WERM+CV} and \textbf{WERM+P} presented in Appendix \ref{sec:investigating-count-selection} to the upper bound $L$ on the number of signals or segments in the linear model \eqref{eq:linear-chgpt-model}, in the setting of three true change points ($L^*=4$). We fix $p=100$, $n=1000$, and independently sample $\x_i\sim\N(0,\I_p/n)$ and $\varepsilon_i\sim\N(0,0.5^2)$ for $i\in[n]$. We draw the initial coefficient vector $\bbeta^{(1)}$ from $\N(0,(n/p)\I_p)$ and successive independent jumps $\bbeta^{(\ell)}-\bbeta^{(\ell-1)}$ from $\N(0,1.5^2(n/p)\I_p)$ for $\ell\in\{2,3,4\}$. We run $100$ trials, using the same generated data for both methods and each upper bound $L\in\{1,\ldots,6\}$ within a trial.

Both methods fit \WeightedERM{} with base loss $M(a,b)=\tfrac12(a-b)^2$ and no regression coefficient penalty. For each upper bound, \textbf{WERM+CV} selects the number of segments from $\{1,\ldots,L\}$ using $10$-fold cross-validation as described in Section \ref{sec:synthetic-experiments} (using $P=0$) with a uniform location prior conditional on the number of segments. \textbf{WERM+P} instead uses a uniform prior on the number of segments in $\{1,\ldots,L\}$, conditionally uniform locations, and the penalty \eqref{eq:search-penalty-experiment} with $\gamma=6$ fixed. We impose a minimum separation $\Delta=n/10$ (top row) or $\Delta=n/20$ (bottom row) between successive change points, scaled to the training sample size within each fold.

\begin{figure}[!tp]
    \centering
    \begin{subfigure}[t]{0.325\textwidth}
        \centering
        \includegraphics[width=\textwidth]{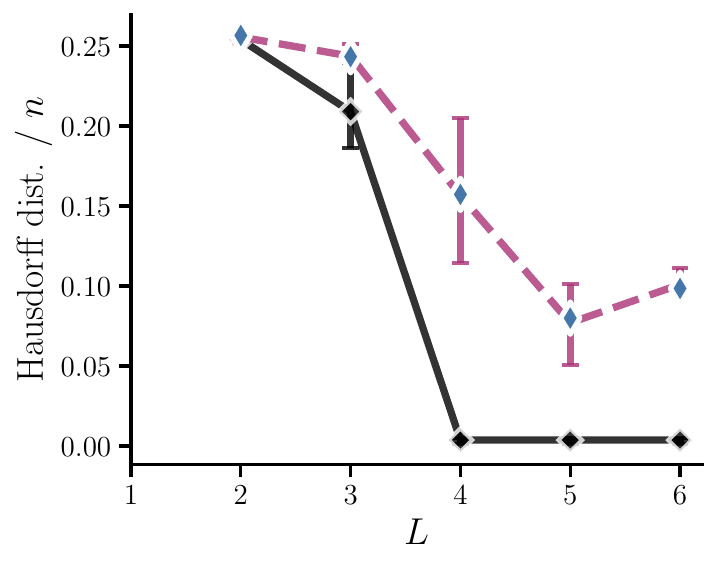}
    \end{subfigure}\hfill
    \begin{subfigure}[t]{0.325\textwidth}
        \centering
        \includegraphics[width=\textwidth]{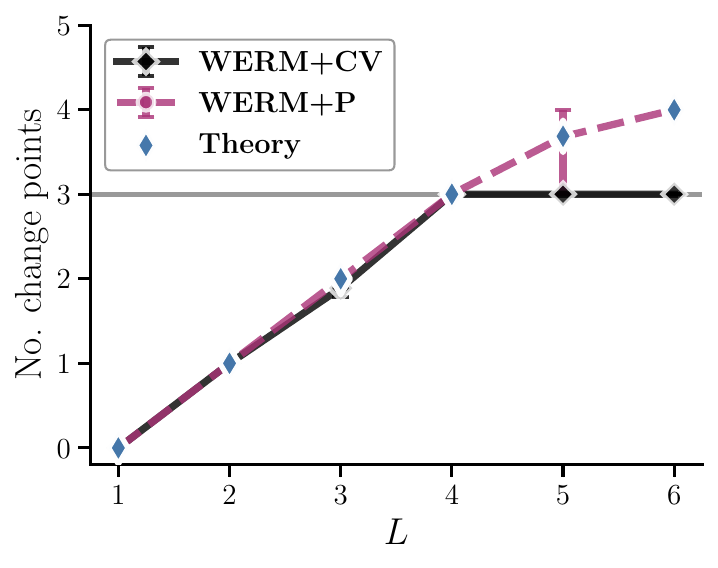}
    \end{subfigure}\hfill
    \begin{subfigure}[t]{0.325\textwidth}
        \centering
        \includegraphics[width=\textwidth]{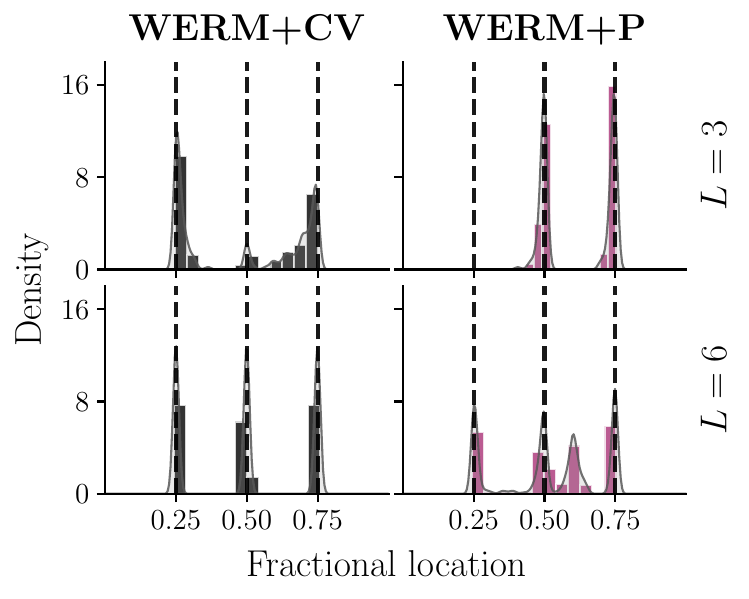}
    \end{subfigure}
    \par\medskip
    \begin{subfigure}[t]{0.325\textwidth}
        \centering
        \includegraphics[width=\textwidth]{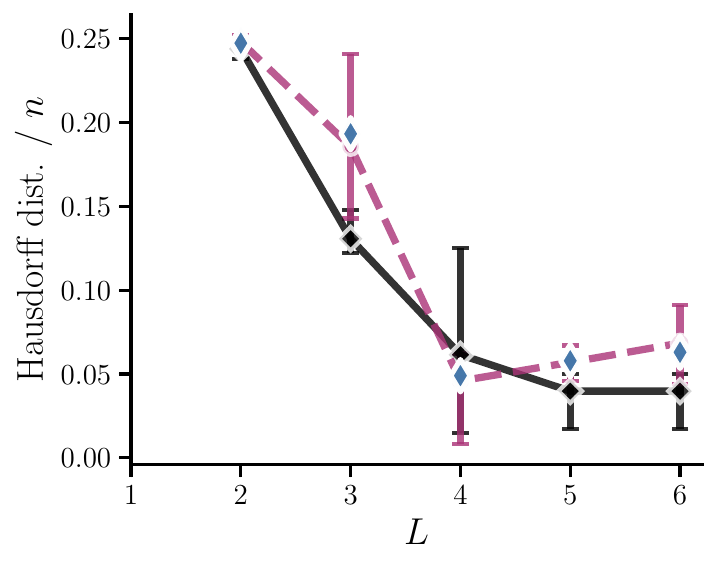}
    \end{subfigure}\hfill
    \begin{subfigure}[t]{0.325\textwidth}
        \centering
        \includegraphics[width=\textwidth]{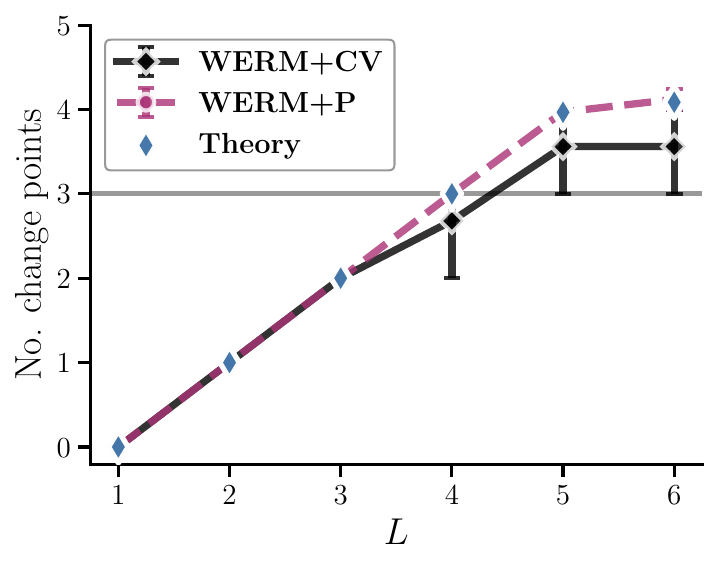}
    \end{subfigure}\hfill
    \begin{subfigure}[t]{0.325\textwidth}
        \centering
        \includegraphics[width=\textwidth]{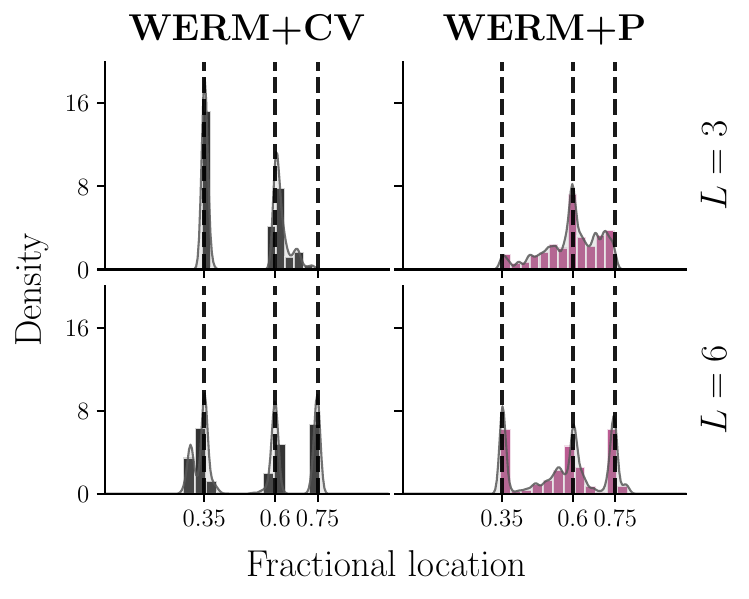}
    \end{subfigure}
    \caption{Sensitivity to the upper bound $L$ with $L^*=4$, $n=1000$ and $p=100$, as detailed in Appendix \ref{sec:sensitivity-number-of-change-points}. Top row: true change points $(0.25n,0.50n,0.75n)$ and minimum separation $\Delta_n=n/10$. Bottom row: true change points $(0.35n,0.60n,0.75n)$ and $\Delta_n=n/20$. Left: normalized Hausdorff error. Middle: selected number of change points. Curves show means over $100$ paired trials; error bars show first and third quartiles. Blue diamonds show Theory for \textbf{WERM+P}, obtained from \eqref{eq:b-fp-implicit}--\eqref{eq:cross-cov-fp-implicit} and Proposition \ref{prop:ERM_hausdorff_characterization_pres}. Right: histograms of change point estimates, with Gaussian kernel density estimates in gray using five-fold bandwidth selection. Dashed lines mark the true locations.}
    \label{fig:E-L-location}
\end{figure}

Figure \ref{fig:E-L-location} reports the normalized Hausdorff error, selected number of change points, and histograms of change point estimates, for two different ground truth change point locations of $(0.25n, 0.50n, 0.75n)$ (top) and $(0.35n, 0.60n, 0.75n)$ (bottom). In both cases and for both methods, we observe that $L > L^*$ does not yield reduced change point estimation accuracy. When $L < L^*$, the estimates are accurate with respect to a subset of true change points.
The theoretical predictions for \textbf{WERM+P}, obtained from \eqref{eq:b-fp-implicit}--\eqref{eq:cross-cov-fp-implicit} and Proposition \ref{prop:ERM_hausdorff_characterization_pres}, closely track the performance measures.
\subsection{{Sensitivity to change point prior}}\label{sec:sensitivity-change-point-prior}
\begin{figure}[!tp]
    \centering
    \includegraphics[width=0.90\textwidth]{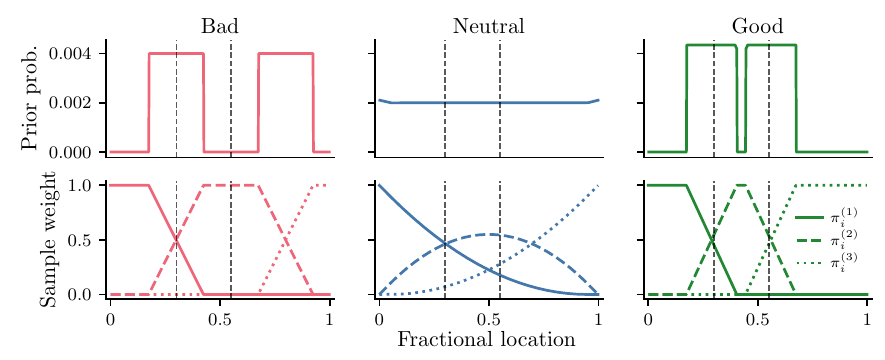}
    \par\medskip
    \begin{subfigure}[c]{0.30\textwidth}
        \centering
        \includegraphics[width=\textwidth]{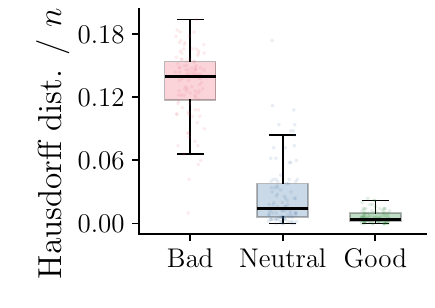}
    \end{subfigure}\hfill
    \begin{subfigure}[c]{0.66\textwidth}
        \centering
        \includegraphics[width=\textwidth]{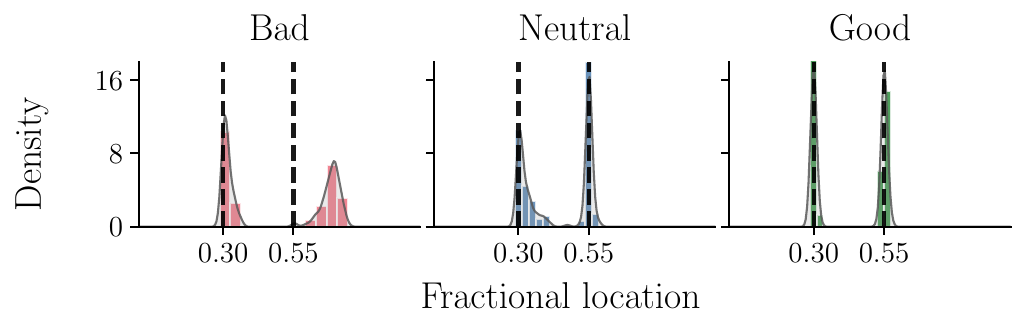}
    \end{subfigure}
    \caption{Sensitivity to the change point prior in Appendix \ref{sec:sensitivity-change-point-prior}. Top: prior probabilities. Middle: sample weights $\pi_i^{(\ell)}$. Bottom: normalized Hausdorff error (left) and estimated fractional locations (right), over $100$ paired trials. Dashed lines mark the true locations $0.30$ and $0.55$; gray curves show Gaussian kernel density estimates, with densities clipped at $18$.}
    \label{fig:E-PRIOR-priors}\label{fig:E-PRIOR-location}
\end{figure}

We investigate the effect of change point prior misspecification on estimation performance in the setting of a linear model \eqref{eq:linear-chgpt-model} with two true change points at $(0.30n,0.55n)$. We use $100$ paired trials with $n=500$, $p=100$, independent covariates $\x_i\sim\N(0,\I_p/n)$ and noise $\varepsilon_i\sim\N(0,0.5^2)$. The initial coefficient vector is drawn from $\N(0,(n/p)\I_p)$, and successive independent jumps from $\N(0,1.5^2(n/p)\I_p)$. We run unpenalized \WeightedERM{} with $M(a,b)=\tfrac12(a-b)^2$, without cross-validation, while keeping the working number of segments $\hat{L} = 3$ fixed, and set a minimum change point separation of $\Delta = 25$. With the number of segments $\hat{L} = 3$ fixed, the penalty $P$ in \eqref{eq:general_changepoint_estimator} does not play a role. Each change point location prior is uniform over a set of admissible ordered change point pairs $(c_1,c_2)$ with $1\leq c_1<c_2\leq n-1$ and $c_2-c_1\geq25$. The neutral prior includes every admissible pair. The bad prior restricts $(c_1/n,c_2/n)$ to $[0.175,0.425)\times[0.675,0.925)$, placing its second-change support away from the truth. The good prior restricts it to $[0.175,0.405)\times[0.445,0.675)$, covering both true locations while leaving a $0.04$ gap between its two support windows. The resulting prior probability of observing a change point at a given fractional location is displayed at the top of Figure \ref{fig:E-PRIOR-priors}. The joint configuration priors are then marginalized as per step \eqref{eq:pi_i_defn} of the \WeightedERM{} methodology to obtain sample weights $\{ \pi_i^{(\ell)} \}_{i \in [n], \ell \in [\hat{L}]}$ used in the weighted least squares fits, these sample weight patterns are displayed in the {middle row} of Figure \ref{fig:E-PRIOR-priors}.
The estimation error is presented in the {bottom-left panel} of Figure \ref{fig:E-PRIOR-location}, where the median normalized Hausdorff error reads as $0.14$ under the bad prior, $0.014$ under the neutral prior and $0.004$ under the good prior. The {bottom-right panel} displays histograms of estimated locations, where the bad prior's estimates need not remain inside its support windows because the change point search procedure as described in \eqref{eq:general_changepoint_estimator} does not restrict the final change point search to locations with non-zero prior mass. We note this as a design choice to mitigate prior misspecification.
\subsection{{Myocardial infarction data: sensitivity to signal strength estimate}}
\label{sec:myocardial-sensitivity}
We investigate the sensitivity of the posterior probability of one change point to the estimated signal strength matrix $\hat{\bGamma} \in \reals^{2 \times 2}$ in the Myocardial infarction real data application outlined in Section \ref{sec:myocardial}. {We replace $\hat{\bGamma}$ by $(1+t)\hat{\bGamma}$ for $t\in\{-0.2,-0.1,0,0.1,0.2\}$, holding the data and adjusted fitted values fixed while recomputing the fixed point quantities and normalized posterior. Table \ref{tab:myocardial-rho-sensitivity} reports probabilities from $0.497$ to $0.713$, compared with $0.607$ at the unperturbed estimate.}
\begin{table}[!tp]
    \centering
    \caption{\small {Posterior probability of one change point in the myocardial infarction sensitivity experiment outlined in Appendix \ref{sec:myocardial-sensitivity}, under the scalar perturbation $(1+t)\hat{\bGamma}$.}}
    \label{tab:myocardial-rho-sensitivity}
    \begin{tabular}{crrrrr}
        \hline
        $t$ & $-20\%$ & $-10\%$ & $0\%$ & $+10\%$ & $+20\%$ \\
        \hline
        Probability of one change point & 0.497 & 0.552 & 0.607 & 0.661 & 0.713 \\
        \hline
    \end{tabular}
\end{table}

\subsection{{Comparison with AMP}}\label{sec:AMP-comparison}
We compare \WeightedERM{} with the AMP estimator of \citet{arpino_2025_jmlr} in the linear model \eqref{eq:linear-chgpt-model}, and highlight that the AMP estimator is designed for the setting of identity (or known) covariance. We use the AMP designed for $\bSigma = \I_p$ in what follows. We fix $p=200$ and run $100$ paired trials at each $\delta=n/p\in\{1.5,3.5,7.5,11.5\}$, independently sampling $\beta_j^{(\ell)}\sim\N(0,\delta)$, $\varepsilon_i\sim\N(0,0.1)$ and $\x_i\sim\N(\0_p,\bSigma/n)$ for $i \in [n], j \in [p]$. We consider three covariances: \emph{isotropic}, $\bSigma=\I_p$; \emph{AR(1)}, $\Sigma_{kj}=0.4^{|k-j|}$ for $k, j \in [p]$; and the \emph{random orthogonal} covariance used in Section \ref{sec:logistic-model-sparse-signals-exp}, fixed across trials. The two true change points are $(0.25n,0.50n)$ or $(0.45n,0.75n)$.

\begin{figure}[!tp]
    \centering
    \includegraphics[width=0.325\textwidth]{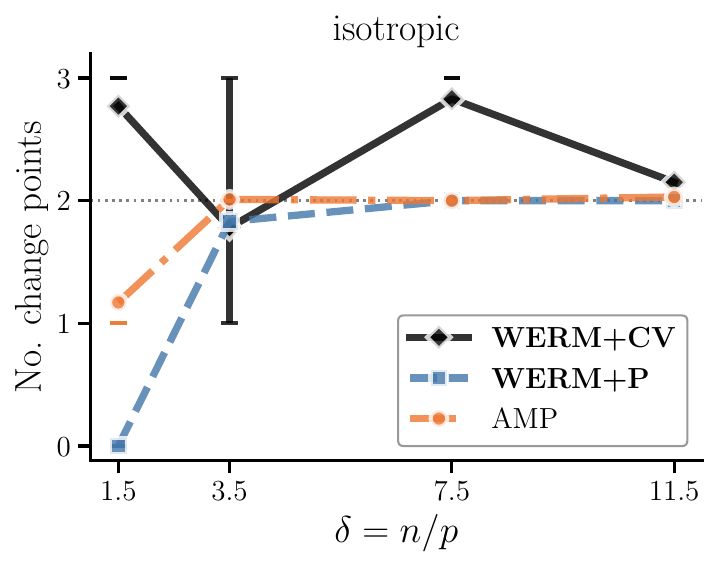}\hfill
    \includegraphics[width=0.325\textwidth]{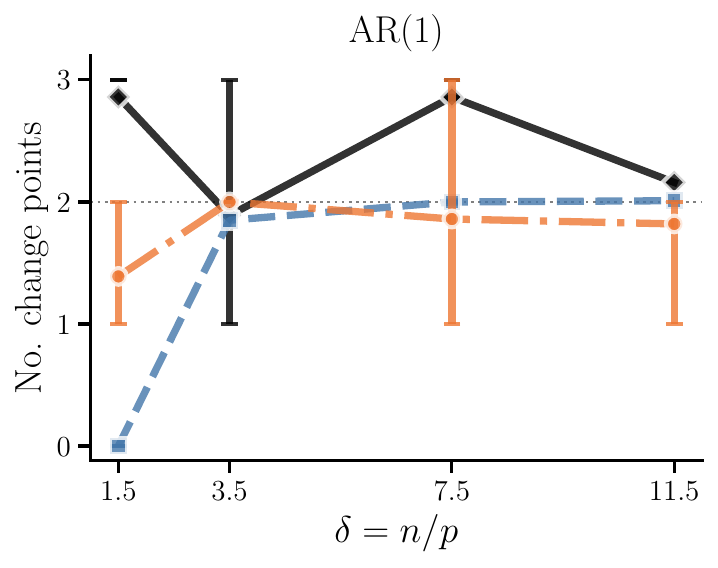}\hfill
    \includegraphics[width=0.325\textwidth]{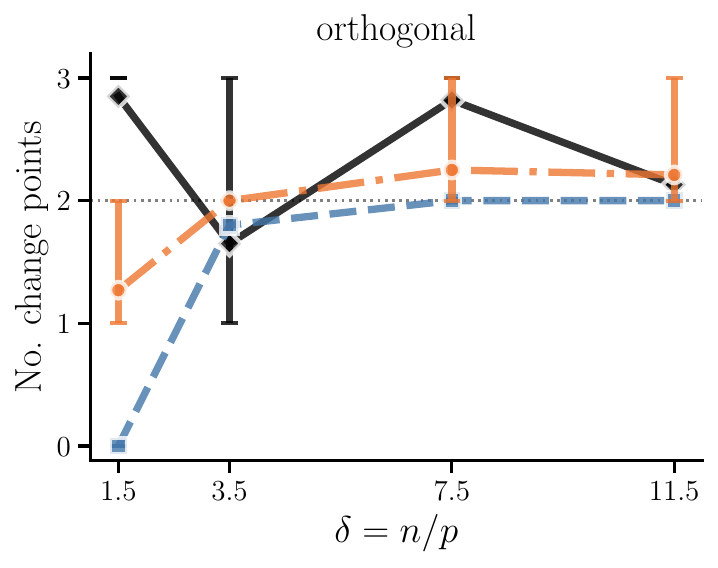}
    \par\smallskip
    \includegraphics[width=\textwidth]{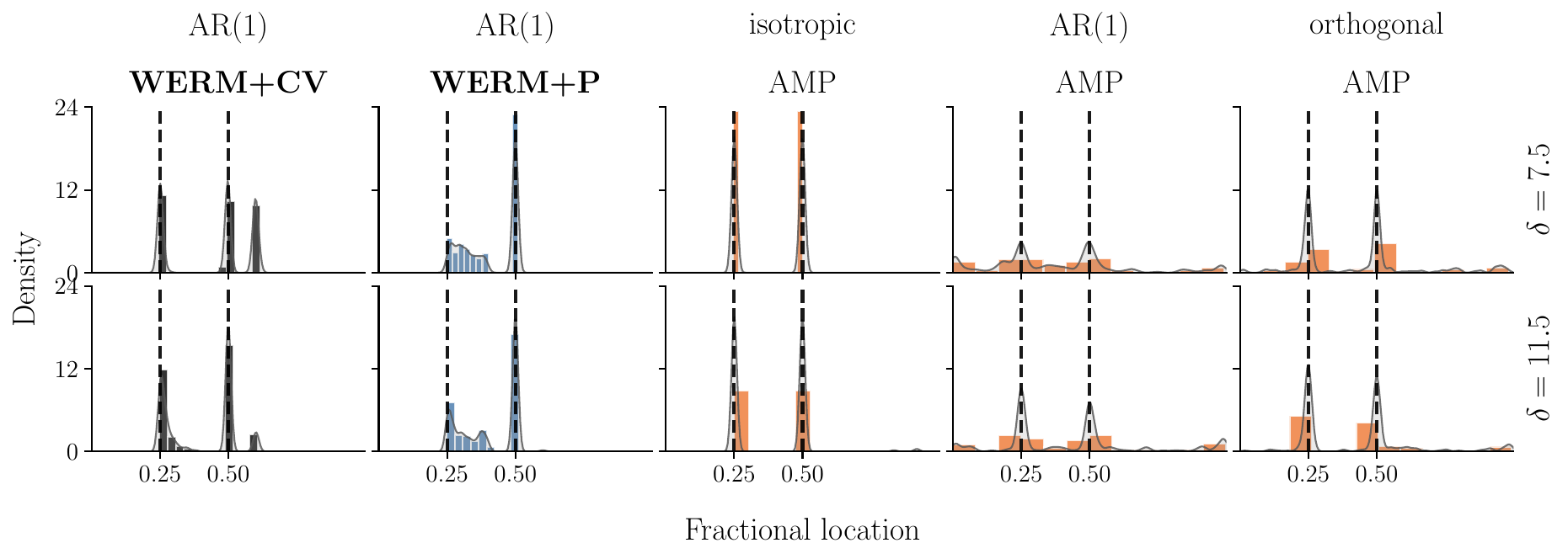}
    \caption{\small Comparison with AMP from Appendix \ref{sec:AMP-comparison} for true change points $(0.25n,0.50n)$. Top: mean estimated counts and first/third quartiles over $100$ trials, for isotropic, AR(1) and random orthogonal designs. Bottom: histogram of change point estimates for the different methods under different designs; \textbf{WERM+CV} and \textbf{WERM+P} shown only under AR(1) as other designs yielded nearly identical histograms. Gray curves are Gaussian kernel density estimates; dashed lines mark the true locations.}
    \label{fig:AMP-comparison-original}
\end{figure}

\begin{figure}[!tp]
    \centering
    \includegraphics[width=0.325\textwidth]{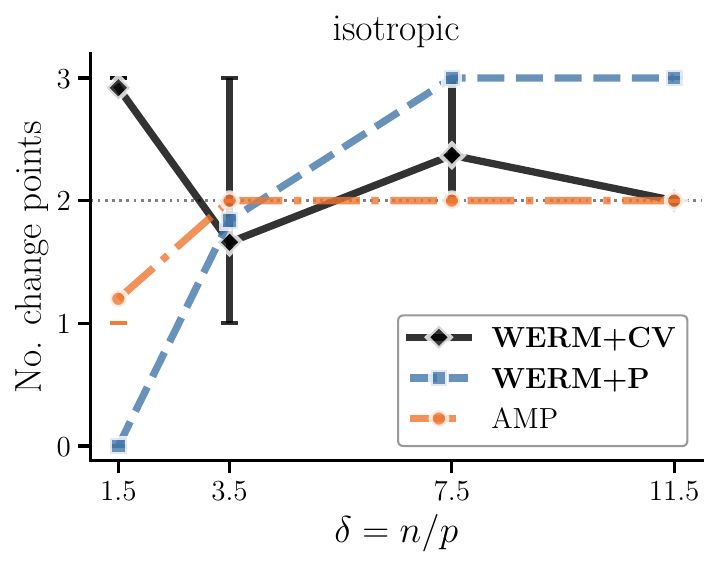}\hfill
    \includegraphics[width=0.325\textwidth]{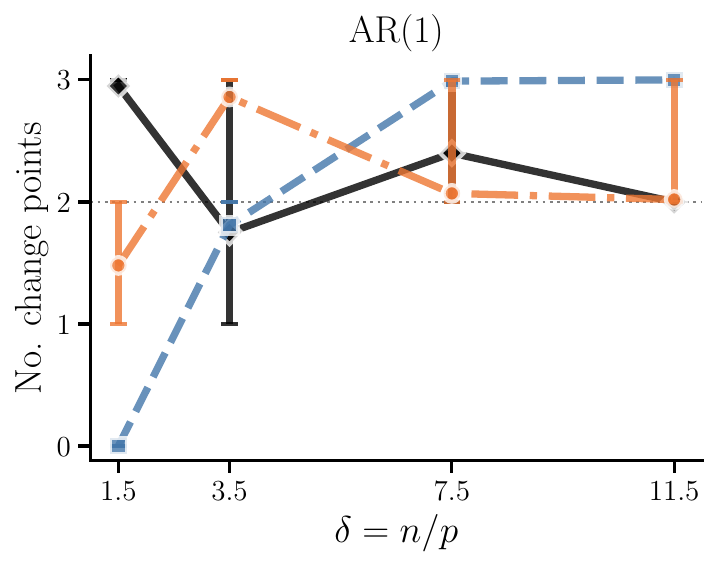}\hfill
    \includegraphics[width=0.325\textwidth]{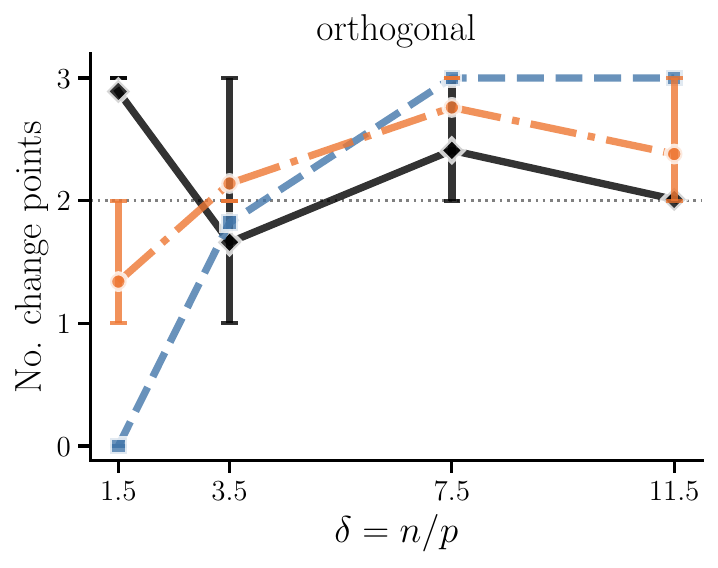}
    \par\smallskip
    \includegraphics[width=\textwidth]{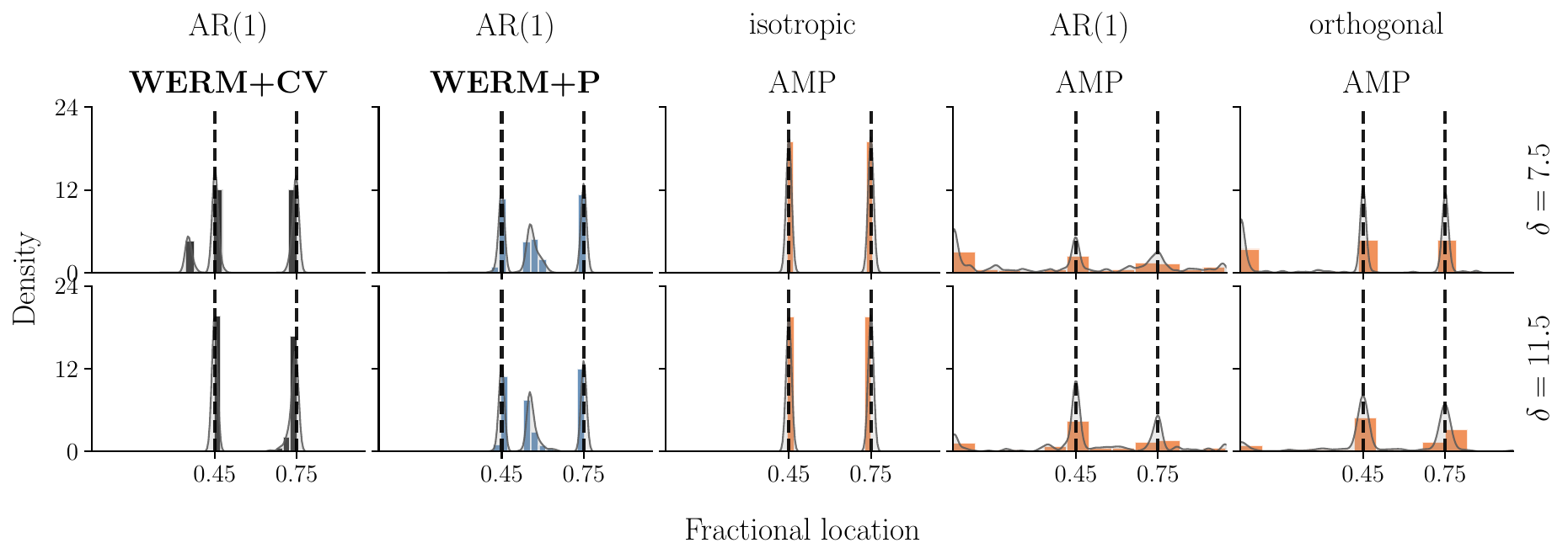}
    \caption{\small Comparison with AMP from Appendix \ref{sec:AMP-comparison} for true change points $(0.45n,0.75n)$. Top: mean estimated counts and first/third quartiles over $100$ trials, for isotropic, AR(1) and orthogonal designs. Bottom: histogram of change point estimates for the different methods under different designs; \textbf{WERM+CV} and \textbf{WERM+P} shown only under AR(1) as other designs yielded nearly identical histograms. Gray curves are Gaussian kernel density estimates; dashed lines mark the true locations.}
    \label{fig:AMP-comparison-relocated}
\end{figure}

All methods allow zero to three change points ($L=4$), assuming a uniform prior on change point locations with minimum separation $\Delta = n/10$ between successive change points and end points.
\textbf{WERM+CV} uses the least squares loss and $10$-fold absolute-error cross-validation as introduced in Section \ref{sec:synthetic-experiments}. \textbf{WERM+P} uses least squares loss and selects change points based on \eqref{eq:general_changepoint_estimator} with the penalty used  in Appendix \ref{sec:investigating-count-selection}: $P(\bpsi) = 6 \log\binom{n-1}{K-1},\; K=\max_i \psi_i$.
We run $10$ iterations of the AMP algorithm, and highlight that AMP uses a Gaussian denoiser which incorporates exact knowledge of the generating signal prior and noise variance.  Despite this advantage, Figures \ref{fig:AMP-comparison-original}--\ref{fig:AMP-comparison-relocated} show that the location estimates of AMP are significantly more dispersed under the AR(1) and the random orthogonal designs than those of \textbf{WERM+CV} and \textbf{WERM+P}, while yielding comparable estimates on the number of change points. This is consistent with existing evidence that Approximate Message Passing methods can fail to converge on zero-mean Gaussian designs with correlated columns (see Fig. 1(c) in \cite{VilaAdaptiveDamping2015} and Theorem 2 in \cite{RanganOnTheConv2019}).

\clearpage
{

\section{Theoretical guarantees for logistic regression with change points}
\label{sec:logistic-guarantees}}

Recall from Section \ref{sec:introduction} that the logistic model is an instance of \eqref{eq:model} with $\varepsilon_1, \dots, \varepsilon_n \distas{\text{i.i.d.}} U[0,1]$ and
\begin{equation} \label{eq:logistic-link}
    q(z, \varepsilon) = \ind\{\varepsilon \le \zeta'(z)\},
    \qquad \text{so that} \qquad
    y_i = \ind\{\varepsilon_i \le \zeta'(\x_i^\top \bbeta^{(\Psi_i)})\}, \quad i \in [n].
\end{equation}
The link \eqref{eq:logistic-link} is not Lipschitz, and so does not satisfy the model assumptions of Section \ref{sec:setting}. In this appendix we show that the conclusions of Theorems \ref{thm:ERM_general_characterization_pres} and \ref{thm:characterization_of_debiased_est_pres} and of Proposition \ref{prop:ERM_hausdorff_characterization_pres} still hold for the logistic model, on the region of $(\delta, \bGamma)$ for which the weighted estimator \eqref{eq:logistic_mle} exists (Proposition \ref{prop:logistic_mle_existence}).

The assumption that $q$ is Lipschitz is used in the proofs of Section \ref{sec:asymptotic-characterization} in four places:
\begin{enumerate}[(a)]
    \item \label{item:lip-equicont} {the moment bounds and noise replacement argument in Lemmas \ref{lemma:bounded-deriv-expect} and \ref{lemma:h-conv-to-Eh-gen}, and the equicontinuity arguments in Lemmas \ref{lemma:prop-of-cross-cov-simple}, \ref{lemma:equicontinuous_h_i}, and \ref{lemma:continuity-of-fp-eqns}.}
    \item \label{item:lip-existence} the proof that a solution to \eqref{eq:b-fp-implicit}--\eqref{eq:cross-cov-fp-implicit} exists (Appendix \ref{sec:existence-of-solns});
    \item \label{item:lip-conv} the convergence of the finite-$n$ nonlinear equations to their limits (Appendix \ref{sec:proof-existence-of-limits-in-fixed-point-equations});
    \item \label{item:lip-amp} the verification of assumption (A2) of \citet{arpino_2025_jmlr}, namely that the denoiser $(u, z) \mapsto \prox_{b \pi_i M(\cdot,\, q(z, \psi, \varepsilon))}(u)$ is pseudo-Lipschitz (Appendix \ref{sec:proof-ERM_frobenius_characterization}).
\end{enumerate}
In \eqref{item:lip-equicont}--\eqref{item:lip-conv}, $q$ appears only inside an expectation over $(\Z, \bvarepsilon)$, and the Lipschitz property is used only to bound quantities of the form $\E[ (q(z_1, \psi, \bar\varepsilon) - q(z_2, \psi, \bar\varepsilon))^2 ]$. These three uses are therefore covered by the weaker Assumption \ref{asmpt:avg-lipschitz} below, which the logistic link satisfies. Only \eqref{item:lip-amp} requires a bound that holds for each realization of $\varepsilon$, which we circumvent  using a smoothing argument in Appendix \ref{sec:logistic-smoothing}.

\subsection{An averaged Lipschitz condition satisfied by the logistic link}
\label{sec:logistic-avg-lip}

\begin{assumption}[Mean-square continuity of the link] \label{asmpt:avg-lipschitz}
There exist constants $C_q > 0$ and $\rho \in (0, 1]$ such that, for all $z_1, z_2 \in \reals$ and $\psi \in [L]$,
\begin{equation} \label{eq:avg-lipschitz}
    \E_{\bar{\varepsilon}} \left[ \big( q(z_1, \psi, \bar{\varepsilon}) - q(z_2, \psi, \bar{\varepsilon}) \big)^2 \right]
    \; \le \; C_q \, |z_1 - z_2|^{2\rho},
\end{equation}
and $\sup_{z, \psi} \E_{\bar{\varepsilon}}[ q(z, \psi, \bar{\varepsilon})^2 ] < \infty$.
\end{assumption}

Any Lipschitz link satisfies \eqref{eq:avg-lipschitz} with $\rho = 1$. The logistic link satisfies it with $\rho = 1/2$: since $q \in \{0,1\}$ and $\bar\varepsilon \distas{} U[0,1]$,
\begin{equation} \label{eq:logistic-avg-lip-computation}
\begin{split}
    \E_{\bar{\varepsilon}}\left[ \big( q(z_1, \bar{\varepsilon}) - q(z_2, \bar{\varepsilon}) \big)^2 \right]
    &= \E_{\bar{\varepsilon}}\left| \ind\{\bar\varepsilon \le \zeta'(z_1)\} - \ind\{\bar\varepsilon \le \zeta'(z_2)\} \right| \\
    & = \left| \zeta'(z_1) - \zeta'(z_2) \right|
    \le \tfrac{1}{4} |z_1 - z_2|,
    \end{split}
\end{equation}
using $\sup_z \zeta''(z) = 1/4$.
Replacing the Lipschitz hypothesis on $q$ in Section \ref{sec:setting} by Assumption \ref{asmpt:avg-lipschitz} leaves the arguments \eqref{item:lip-equicont}--\eqref{item:lip-conv} unchanged up to constants, with each occurrence of $C_{\text{qLip}} |z_1 - z_2|$ replaced by $\sqrt{C_q} |z_1 - z_2|^{\rho}$ and each subsequent application of the Cauchy--Schwarz inequality carried out in $L^2(\P_{\bar\varepsilon})$.

To summarize, the link $q$ enters \eqref{eq:b-fp-implicit}--\eqref{eq:cross-cov-fp-implicit} only through expectations over $(\Z_i, \varepsilon_i)$. Since the logistic $q$ is bounded and its discontinuity set $\{(z, \varepsilon) : \varepsilon = \zeta'(z)\}$ has zero measure under the joint law of $(\Z_i, \varepsilon_i)$, the system \eqref{eq:b-fp-implicit}--\eqref{eq:cross-cov-fp-implicit} is well defined for the logistic link, and Propositions \ref{prop:existence-of-fixed-point-solution} and \ref{prop:uniqueness_of_fixed_points} apply under Assumption \ref{asmpt:avg-lipschitz}.  The Lipschitz continuity of $q$ is required in the method of proof of Theorem \ref{thm:ERM_general_characterization_pres}, via \eqref{item:lip-amp}, which we now address, by showing that the solutions of \WeightedERM{} and the parameters characterizing its asymptotic performance can be approximated to arbitrary accuracy by a model which uses a smoothed version of $q$.

\subsection{Smoothed link and a coupling}
\label{sec:logistic-smoothing}

To handle \eqref{item:lip-amp} we introduce a one-parameter family of Lipschitz links that agree with $q$ outside a vanishing set. Let $f: \reals \to \reals$ be the smooth function
\begin{equation*}
    f(x) = \begin{cases} e^{-1/x}, & x > 0, \\ 0, & x \le 0, \end{cases}
\end{equation*}
and, for $\xi \in (0, 1)$, define
\begin{equation} \label{eq:smooth-ramp}
    g_{\xi}(x) = \frac{f(x/\xi)}{f(x/\xi) + f(1 - x/\xi)}, \qquad x \in \reals.
\end{equation}
The function $g_\xi$ is smooth, non-decreasing, takes values in $[0,1]$, satisfies $g_\xi(x) = 0$ for $x \le 0$ and $g_\xi(x) = 1$ for $x \ge \xi$, and is Lipschitz with constant $C_\xi = \Theta(1/\xi)$. Consequently
\begin{equation} \label{eq:smoothed-link}
    q_{\xi}(z, \varepsilon) := g_{\xi}\big( \zeta'(z) - \varepsilon \big)
\end{equation}
is Lipschitz jointly in $(z, \varepsilon)$ with constant $C_\xi \max\{1, \sup_z \zeta''(z)\} = \Theta(1/\xi)$. Hence, for each fixed $\xi > 0$, the model $y_i^{\xi} = q_\xi(\x_i^\top \bbeta^{(\Psi_i)}, \varepsilon_i)$ satisfies all the model assumptions of Section \ref{sec:setting}, and Theorems \ref{thm:ERM_general_characterization_pres}, \ref{thm:characterization_of_debiased_est_pres} and Proposition \ref{prop:ERM_hausdorff_characterization_pres} apply to it {provided their remaining hypotheses, including fixed-point existence, hold}, with limiting parameters $(\bLambda_\xi, \bK_\xi)$ given by the unique solution of \eqref{eq:b-fp-implicit}--\eqref{eq:cross-cov-fp-implicit} with $q_\xi$ in place of $q$.

Throughout we couple the two models on a single probability space, where they have the same covariates $\{\x_i\}$, the noise $\{\varepsilon_i\}$, the configuration $\bPsi$ and the signals $\B$, and only the responses differ. Write $\hat{\B}, \hat{\TTheta}$ for the solutions of \eqref{eq:logistic_mle} computed from $\y$, and $\hat{\B}_\xi, \hat{\TTheta}_\xi$ for those computed from $\y^\xi$.

\begin{lemma}\label{lemma:logistic-coupling}
Let $S_\xi := \{ i \in [n] : y_i^{\xi} \neq y_i \}$. Then $|y_i^\xi - y_i| \le 1$ for all $i$, and
\begin{equation} \label{eq:coupling-fraction}
    \lim_{n \to \infty} \frac{|S_\xi|}{n} \;\le\; \xi \quad \text{almost surely,}
    \quad\text{hence}\quad
    \limsup_{n \to \infty} \frac{1}{n} \big\| \y^{\xi} - \y \big\|^2 \;\le\; \xi \quad \text{almost surely.}
\end{equation}
\end{lemma}

\begin{proof}
By construction $q_\xi(z, \varepsilon) = q(z, \varepsilon)$ whenever $\zeta'(z) - \varepsilon \notin (0, \xi)$, so $i \in S_\xi$ implies $\varepsilon_i \in (\zeta'(z_i) - \xi, \zeta'(z_i))$ where $z_i = \x_i^\top \bbeta^{(\Psi_i)}$. Since $\varepsilon_i \distas{} U[0,1]$ is independent of $\x_i$, we have $\P(i \in S_\xi \mid \x_i) \le \xi$ for every $i$, and the events $\{i \in S_\xi\}$ are independent across $i$. The first claim follows from the strong law of large numbers, and the second follows since $|y_i^\xi - y_i| \le \ind\{i \in S_\xi\}$.
\end{proof}

\subsection{Stability of the weighted logistic estimator}
\label{sec:logistic-stability}

The transfer from $q_\xi$ to $q$ builds on the observation that the objective function in \eqref{eq:logistic_mle} is affine in the response. Writing $\cC^{(\ell)}$ and $\cC^{(\ell)}_\xi$ for the objectives in \eqref{eq:logistic_mle} formed from $\y$ and $\y^\xi$ respectively, we have
\begin{equation} \label{eq:objective-linear-perturbation}
    \cC^{(\ell)}_{\xi}(\bbeta) = \cC^{(\ell)}(\bbeta) - \big\langle \v^{(\ell)}, \bbeta \big\rangle,
    \qquad
    \v^{(\ell)} := \sum_{i=1}^n \pi_i^{(\ell)} \big( y_i^{\xi} - y_i \big) \x_i \; \in \reals^p ,
\end{equation}
so that the perturbation is linear in $\bbeta$ and, in particular, $\nabla^2 \cC^{(\ell)}_\xi \equiv \nabla^2 \cC^{(\ell)}$. The two objectives therefore share the curvature bound of Assumption \ref{asmpt:reg-loss}(\ref{cond:likelihood_curvature}).

\begin{lemma} \label{lemma:logistic-stability}
Suppose Assumptions \ref{asmpt:reg-weights} and \ref{asmpt:reg-loss} hold, that $\|\bSigma\|_{\mathrm{op}}$ is bounded uniformly in $p$, and that $(\delta, \bGamma)$ lies strictly inside the existence region of Proposition \ref{prop:logistic_mle_existence}. Let $R < \infty$ be such that $\max_{\ell \in [L]} \|\hat{\bbeta}^{(\ell)}\| \vee \|\hat{\bbeta}^{(\ell)}_\xi\| \le R \sqrt{p}$ with probability tending to one.
Then there is a constant $C = C(R, \delta, C_{\X})$, independent of
$\xi$ and $n$, where $C_{\X}$ is the high-probability bound on
$\|\X\|_{\mathrm{op}}$ obtained in \eqref{eq:chevet-bound}, such that with probability tending to one
\begin{equation} \label{eq:stability-bound}
    \frac{1}{\sqrt{p}} \big\| \hat{\B}_{\xi} - \hat{\B} \big\|_F \;\le\; C \sqrt{\xi},
    \qquad
    \frac{1}{\sqrt{n}} \big\| \hat{\TTheta}_{\xi} - \hat{\TTheta} \big\|_F \;\le\; C \sqrt{\xi} .
\end{equation}
\end{lemma}

\begin{proof}
Fix $\ell \in [L]$ and drop the superscript. On the event $\{\|\hat{\bbeta}\| \vee \|\hat{\bbeta}_\xi\| \le R\sqrt{p}\}$, Assumption \ref{asmpt:reg-loss}(\ref{cond:likelihood_curvature}) gives that $\cC$ is $\gamma(R)$-strongly convex on the segment joining $\hat{\bbeta}$ and $\hat{\bbeta}_\xi$, with $\gamma(R) > 0$. Since $\nabla \cC(\hat{\bbeta}) = \0$ and, by \eqref{eq:objective-linear-perturbation}, $\nabla \cC(\hat{\bbeta}_\xi) = \nabla \cC_\xi(\hat{\bbeta}_\xi) + \v = \v$,
\begin{equation*}
    \gamma(R) \big\| \hat{\bbeta}_\xi - \hat{\bbeta} \big\|^2
    \;\le\; \big\langle \nabla \cC(\hat{\bbeta}_\xi) - \nabla \cC(\hat{\bbeta}), \, \hat{\bbeta}_\xi - \hat{\bbeta} \big\rangle
    \;=\; \big\langle \v, \, \hat{\bbeta}_\xi - \hat{\bbeta} \big\rangle
    \;\le\; \|\v\| \, \big\| \hat{\bbeta}_\xi - \hat{\bbeta} \big\| ,
\end{equation*}
whence $\|\hat{\bbeta}_\xi - \hat{\bbeta}\| \le \|\v\| / \gamma(R)$. Writing $\v = \X^\top \u$ with $u_i = \pi_i (y_i^\xi - y_i)$ supported on $S_\xi$, we have $|u_i| \le 1$, since by \eqref{eq:pi_i_defn} each $\pi^{(\ell)}_i$ is a marginal prior probability and hence lies in $[0,1]$, and $|y_i^\xi - y_i| \le 1$. Lemma \ref{lemma:logistic-coupling} then gives $\|\u\| \le \sqrt{|S_\xi|} \le \sqrt{2 \xi n}$ with probability tending to one. Under the scaling $\x_i \distas{} \N(\0, \bSigma/n)$ we may write $\X = n^{-1/2} \G \bSigma^{1/2}$ with $\G \in \reals^{n \times p}$ having i.i.d.\ $\N(0,1)$ entries. Chevet's inequality \citep[Thm.~8.7.1]{Vershynin_2018} then gives
\begin{equation} \label{eq:chevet-bound}
    \E \big\| \X \big\|_{\mathrm{op}}
    \;\le\; \sqrt{\frac{\tr(\bSigma)}{n}} \; + \; \|\bSigma\|_{\mathrm{op}}^{1/2}.
\end{equation}
Since $\G \mapsto \|n^{-1/2}\G\bSigma^{1/2}\|_{\mathrm{op}}$ is Lipschitz with respect to the Frobenius norm with constant $\|\bSigma\|_{\mathrm{op}}^{1/2}/\sqrt{n}$, the concentration inequality for Lipschitz functions of standard Gaussians \citep[Thm.~5.2.2]{Vershynin_2018} gives $\P\big(\|\X\|_{\mathrm{op}} \ge \E\|\X\|_{\mathrm{op}} + s\big) \le \exp\big(-n s^2 / (2\|\bSigma\|_{\mathrm{op}})\big)$. Hence $\|\X\|_{\mathrm{op}} \le C_{\X} < \infty$ with probability tending to one, and
so $\|\v\| \le \|\X\|_{\mathrm{op}} \|\u\| \le C' \sqrt{\xi n}$. Dividing by $\sqrt{p}$ and using $n/p \to \delta$ gives the first bound in \eqref{eq:stability-bound} after taking a maximum over the $L$ (fixed) values of $\ell$. The second bound in \eqref{eq:stability-bound} follows from $\hat{\ttheta}_\xi - \hat{\ttheta} = \X(\hat{\bbeta}_\xi - \hat{\bbeta})$ and $\|\X\|_{\mathrm{op}} = O(1)$.
\end{proof}

\subsection{Transfer of the asymptotic characterization}
\label{sec:logistic-transfer}

\begin{lemma}[Continuity of the limiting parameters] \label{lemma:logistic-fp-continuity}
Under the hypotheses of Lemma \ref{lemma:logistic-stability}, {suppose the fixed points for the logistic link and for all sufficiently small $\xi>0$ exist. Then} $(\bLambda_\xi, \bK_\xi) \to (\bLambda, \bK)$ as $\xi \to 0$, where $(\bLambda, \bK)$ is the unique solution of \eqref{eq:b-fp-implicit}--\eqref{eq:cross-cov-fp-implicit} for the logistic link \eqref{eq:logistic-link}.
\end{lemma}

\begin{proof}
By the arguments in Appendix \ref{sec:logistic-avg-lip}, the system is well defined at $\xi = 0$. By \eqref{eq:coupling-fraction}, $\E[(q_\xi(\bar Z, \bar\varepsilon) - q(\bar Z, \bar\varepsilon))^2] \le \xi$, so $q_\xi \to q$ in $L^2(\P_{\bar Z, \bar\varepsilon})$. The maps defining the right-hand sides of \eqref{eq:b-fp-implicit}--\eqref{eq:cross-cov-fp-implicit} depend on the link only through such $L^2$ quantities, and are Lipschitz in this dependence uniformly over compact subsets of the $(\blambda, b, \kappa)$ domain, by the $1$-Lipschitz property of $\prox_{b\pi\zeta}$ together with Assumptions \ref{asmpt:reg-loss}(\ref{cond:strict-cvx-smooth}) and \ref{asmpt:reg-loss}(\ref{eq:loss-asmpt-log-conc}). Hence the solution maps converge uniformly on compact subsets. Combined with uniqueness of the limiting solution and the bounds of Appendix \ref{sec:existence-of-solns}, every subsequential limit of $(\bLambda_\xi, \bK_\xi)$ solves the $\xi = 0$ system, giving the claim.
\end{proof}

\begin{theorem}[Asymptotic characterization for the logistic model] \label{thm:logistic-characterization}
Consider the logistic change point model \eqref{eq:logistic-link} with the weighted logistic estimator \eqref{eq:logistic_mle}. Suppose Assumptions \ref{asmpt:reg-loss} and \ref{asmpt:reg-weights} hold, that $\|\bSigma\|_{\mathrm{op}}$ is bounded uniformly in $p$, and that $(\delta, \bGamma)$ lies strictly inside the {sufficient existence region} of Proposition \ref{prop:logistic_mle_existence}. {Suppose also that the logistic fixed point and the smoothed fixed points for all sufficiently small $\xi>0$ exist.} Then the conclusions \eqref{eq:SE_ERM_Theta} and \eqref{eq:SE_ERM_B} of Theorem \ref{thm:ERM_general_characterization_pres} hold, with $(\bLambda, \bK)$ the unique solution of \eqref{eq:b-fp-implicit}--\eqref{eq:cross-cov-fp-implicit} for the logistic link. The same holds for Theorem \ref{thm:characterization_of_debiased_est_pres} and, provided the map $(\V, \tilde{\y}) \mapsto U(\hat{\eeta}(\V, \tilde{\y}))$ is uniformly pseudo-Lipschitz, for Proposition \ref{prop:ERM_hausdorff_characterization_pres}. {The count conclusion \eqref{eq:ERM_size_match_eqn} additionally requires its corresponding assumption on the count map.}
\end{theorem}

\begin{proof}
Fix a sequence of uniformly pseudo-Lipschitz functions $\varphi_n(\cdot\,; \bPsi, \bvarepsilon)$ of order $r = 2$ and constant $C_\varphi$, and abbreviate $A_n := \varphi_n(\hat{\TTheta}, \X\B; \bPsi, \bvarepsilon)$, $A_n^\xi := \varphi_n(\hat{\TTheta}_\xi, \X\B; \bPsi, \bvarepsilon)$, and let $\mathcal{E}_\xi$ denote the right-hand side of \eqref{eq:SE_ERM_Theta} computed with $(q_\xi, \bLambda_\xi, \bK_\xi)$, and $\mathcal{E}_0$ the same expression computed with $(q, \bLambda, \bK)$. Then
\begin{equation} \label{eq:three-epsilon}
    \big| A_n - \mathcal{E}_0 \big|
    \;\le\; \underbrace{\big| A_n - A_n^{\xi} \big|}_{\mathrm{(I)}}
    \;+\; \underbrace{\big| A_n^{\xi} - \mathcal{E}_{\xi} \big|}_{\mathrm{(II)}}
    \;+\; \underbrace{\big| \mathcal{E}_{\xi} - \mathcal{E}_0 \big|}_{\mathrm{(III)}} .
\end{equation}
For (I), the pseudo-Lipschitz property of order $2$ gives
$|A_n - A_n^\xi| \le C_\varphi \big(1 + \tfrac{1}{\sqrt n}\|\hat{\TTheta}\|_F + \tfrac{1}{\sqrt n}\|\hat{\TTheta}_\xi\|_F + \tfrac{1}{\sqrt n}\|\X\B\|_F\big) \cdot \tfrac{1}{\sqrt n}\|\hat{\TTheta}_\xi - \hat{\TTheta}\|_F$;
the term within brackets is $O_\P(1)$ by \eqref{eq:Gamma_as_limit} and the a priori bound of Lemma \ref{lemma:logistic-stability}, and $
\tfrac{1}{\sqrt n}\|\hat{\TTheta}_\xi - \hat{\TTheta}\|_F$ is $O_\P(\sqrt\xi)$ by \eqref{eq:stability-bound}. Hence $\limsup_n \mathrm{(I)} \le C \sqrt{\xi}$ with probability tending to one. Term (II) converges to zero in probability for each fixed $\xi > 0$, by Theorem \ref{thm:ERM_general_characterization_pres} applied to the smoothed model, which satisfies all assumptions of Section \ref{sec:setting}. Term (III) is deterministic and converges to zero as $\xi \to 0$ by Lemma \ref{lemma:logistic-fp-continuity}, the $L^2$ convergence $q_\xi \to q$, and the pseudo-Lipschitz property of $\varphi_n$ and of the proximal operator. Taking $n \to \infty$ and then $\xi \to 0$ in \eqref{eq:three-epsilon} gives \eqref{eq:SE_ERM_Theta}; the argument for \eqref{eq:SE_ERM_B} is identical, using the first bound in \eqref{eq:stability-bound} and $\varphi_p$ in place of $\varphi_n$. The statement for Theorem \ref{thm:characterization_of_debiased_est_pres} follows in the same way. For Proposition \ref{prop:ERM_hausdorff_characterization_pres}, by {Lemma \ref{lemma:hausdorff_PL}} $(\V, \tilde{\y}) \mapsto d_H(\eeta, \hat{\eeta}(\V, \tilde{\y}))/n$ is uniformly pseudo-Lipschitz in both arguments; applying \eqref{eq:three-epsilon} with $\varphi_n(\V, \tilde{\y}) = d_H(\eeta, \hat{\eeta}(\V, \tilde{\y}))/n$ and controlling the response argument by $\tfrac{1}{\sqrt n}\|\y^\xi - \y\| = O_\P(\sqrt\xi)$ from Lemma \ref{lemma:logistic-coupling} gives \eqref{eq:ERM_hausdorff_match_eqn}, and likewise \eqref{eq:ERM_size_match_eqn} {under its assumption on the count map}.
\end{proof}

\end{document}

%% file: figures/synthetic-final/M1a/location_asymmetric_errors.pgf
\begingroup%
\makeatletter%
\begin{pgfpicture}%
\pgfpathrectangle{\pgfpointorigin}{\pgfqpoint{4.743111in}{3.749100in}}%
\pgfusepath{use as bounding box, clip}%
\begin{pgfscope}%
\pgfsetbuttcap%
\pgfsetmiterjoin%
\definecolor{currentfill}{rgb}{1.000000,1.000000,1.000000}%
\pgfsetfillcolor{currentfill}%
\pgfsetlinewidth{0.000000pt}%
\definecolor{currentstroke}{rgb}{1.000000,1.000000,1.000000}%
\pgfsetstrokecolor{currentstroke}%
\pgfsetdash{}{0pt}%
\pgfpathmoveto{\pgfqpoint{0.000000in}{0.000000in}}%
\pgfpathlineto{\pgfqpoint{4.743111in}{0.000000in}}%
\pgfpathlineto{\pgfqpoint{4.743111in}{3.749100in}}%
\pgfpathlineto{\pgfqpoint{0.000000in}{3.749100in}}%
\pgfpathlineto{\pgfqpoint{0.000000in}{0.000000in}}%
\pgfpathclose%
\pgfusepath{fill}%
\end{pgfscope}%
\begin{pgfscope}%
\pgfsetbuttcap%
\pgfsetmiterjoin%
\definecolor{currentfill}{rgb}{1.000000,1.000000,1.000000}%
\pgfsetfillcolor{currentfill}%
\pgfsetlinewidth{0.000000pt}%
\definecolor{currentstroke}{rgb}{0.000000,0.000000,0.000000}%
\pgfsetstrokecolor{currentstroke}%
\pgfsetstrokeopacity{0.000000}%
\pgfsetdash{}{0pt}%
\pgfpathmoveto{\pgfqpoint{0.911181in}{0.831623in}}%
\pgfpathlineto{\pgfqpoint{4.643111in}{0.831623in}}%
\pgfpathlineto{\pgfqpoint{4.643111in}{3.649100in}}%
\pgfpathlineto{\pgfqpoint{0.911181in}{3.649100in}}%
\pgfpathlineto{\pgfqpoint{0.911181in}{0.831623in}}%
\pgfpathclose%
\pgfusepath{fill}%
\end{pgfscope}%
\begin{pgfscope}%
\pgfsetbuttcap%
\pgfsetroundjoin%
\definecolor{currentfill}{rgb}{0.000000,0.000000,0.000000}%
\pgfsetfillcolor{currentfill}%
\pgfsetlinewidth{1.505625pt}%
\definecolor{currentstroke}{rgb}{0.000000,0.000000,0.000000}%
\pgfsetstrokecolor{currentstroke}%
\pgfsetdash{}{0pt}%
\pgfsys@defobject{currentmarker}{\pgfqpoint{0.000000in}{-0.083333in}}{\pgfqpoint{0.000000in}{0.000000in}}{%
\pgfpathmoveto{\pgfqpoint{0.000000in}{0.000000in}}%
\pgfpathlineto{\pgfqpoint{0.000000in}{-0.083333in}}%
\pgfusepath{stroke,fill}%
}%
\begin{pgfscope}%
\pgfsys@transformshift{1.080814in}{0.831623in}%
\pgfsys@useobject{currentmarker}{}%
\end{pgfscope}%
\end{pgfscope}%
\begin{pgfscope}%
\definecolor{textcolor}{rgb}{0.000000,0.000000,0.000000}%
\pgfsetstrokecolor{textcolor}%
\pgfsetfillcolor{textcolor}%
\pgftext[x=1.080814in,y=0.699679in,,top]{\color{textcolor}{\rmfamily\fontsize{22.000000}{26.400000}\selectfont\catcode`\^=\active\def^{\ifmmode\sp\else\^{}\fi}\catcode`\%=\active\def
\end{pgfscope}%
\begin{pgfscope}%
\pgfsetbuttcap%
\pgfsetroundjoin%
\definecolor{currentfill}{rgb}{0.000000,0.000000,0.000000}%
\pgfsetfillcolor{currentfill}%
\pgfsetlinewidth{1.505625pt}%
\definecolor{currentstroke}{rgb}{0.000000,0.000000,0.000000}%
\pgfsetstrokecolor{currentstroke}%
\pgfsetdash{}{0pt}%
\pgfsys@defobject{currentmarker}{\pgfqpoint{0.000000in}{-0.083333in}}{\pgfqpoint{0.000000in}{0.000000in}}{%
\pgfpathmoveto{\pgfqpoint{0.000000in}{0.000000in}}%
\pgfpathlineto{\pgfqpoint{0.000000in}{-0.083333in}}%
\pgfusepath{stroke,fill}%
}%
\begin{pgfscope}%
\pgfsys@transformshift{2.777146in}{0.831623in}%
\pgfsys@useobject{currentmarker}{}%
\end{pgfscope}%
\end{pgfscope}%
\begin{pgfscope}%
\definecolor{textcolor}{rgb}{0.000000,0.000000,0.000000}%
\pgfsetstrokecolor{textcolor}%
\pgfsetfillcolor{textcolor}%
\pgftext[x=2.777146in,y=0.699679in,,top]{\color{textcolor}{\rmfamily\fontsize{22.000000}{26.400000}\selectfont\catcode`\^=\active\def^{\ifmmode\sp\else\^{}\fi}\catcode`\%=\active\def
\end{pgfscope}%
\begin{pgfscope}%
\pgfsetbuttcap%
\pgfsetroundjoin%
\definecolor{currentfill}{rgb}{0.000000,0.000000,0.000000}%
\pgfsetfillcolor{currentfill}%
\pgfsetlinewidth{1.505625pt}%
\definecolor{currentstroke}{rgb}{0.000000,0.000000,0.000000}%
\pgfsetstrokecolor{currentstroke}%
\pgfsetdash{}{0pt}%
\pgfsys@defobject{currentmarker}{\pgfqpoint{0.000000in}{-0.083333in}}{\pgfqpoint{0.000000in}{0.000000in}}{%
\pgfpathmoveto{\pgfqpoint{0.000000in}{0.000000in}}%
\pgfpathlineto{\pgfqpoint{0.000000in}{-0.083333in}}%
\pgfusepath{stroke,fill}%
}%
\begin{pgfscope}%
\pgfsys@transformshift{4.473478in}{0.831623in}%
\pgfsys@useobject{currentmarker}{}%
\end{pgfscope}%
\end{pgfscope}%
\begin{pgfscope}%
\definecolor{textcolor}{rgb}{0.000000,0.000000,0.000000}%
\pgfsetstrokecolor{textcolor}%
\pgfsetfillcolor{textcolor}%
\pgftext[x=4.473478in,y=0.699679in,,top]{\color{textcolor}{\rmfamily\fontsize{22.000000}{26.400000}\selectfont\catcode`\^=\active\def^{\ifmmode\sp\else\^{}\fi}\catcode`\%=\active\def
\end{pgfscope}%
\begin{pgfscope}%
\definecolor{textcolor}{rgb}{0.000000,0.000000,0.000000}%
\pgfsetstrokecolor{textcolor}%
\pgfsetfillcolor{textcolor}%
\pgftext[x=2.777146in,y=0.388056in,,top]{\color{textcolor}{\rmfamily\fontsize{22.000000}{26.400000}\selectfont\catcode`\^=\active\def^{\ifmmode\sp\else\^{}\fi}\catcode`\%=\active\def
\end{pgfscope}%
\begin{pgfscope}%
\pgfsetbuttcap%
\pgfsetroundjoin%
\definecolor{currentfill}{rgb}{0.000000,0.000000,0.000000}%
\pgfsetfillcolor{currentfill}%
\pgfsetlinewidth{1.505625pt}%
\definecolor{currentstroke}{rgb}{0.000000,0.000000,0.000000}%
\pgfsetstrokecolor{currentstroke}%
\pgfsetdash{}{0pt}%
\pgfsys@defobject{currentmarker}{\pgfqpoint{-0.083333in}{0.000000in}}{\pgfqpoint{-0.000000in}{0.000000in}}{%
\pgfpathmoveto{\pgfqpoint{-0.000000in}{0.000000in}}%
\pgfpathlineto{\pgfqpoint{-0.083333in}{0.000000in}}%
\pgfusepath{stroke,fill}%
}%
\begin{pgfscope}%
\pgfsys@transformshift{0.911181in}{0.831623in}%
\pgfsys@useobject{currentmarker}{}%
\end{pgfscope}%
\end{pgfscope}%
\begin{pgfscope}%
\definecolor{textcolor}{rgb}{0.000000,0.000000,0.000000}%
\pgfsetstrokecolor{textcolor}%
\pgfsetfillcolor{textcolor}%
\pgftext[x=0.443111in, y=0.731604in, left, base]{\color{textcolor}{\rmfamily\fontsize{22.000000}{26.400000}\selectfont\catcode`\^=\active\def^{\ifmmode\sp\else\^{}\fi}\catcode`\%=\active\def
\end{pgfscope}%
\begin{pgfscope}%
\pgfsetbuttcap%
\pgfsetroundjoin%
\definecolor{currentfill}{rgb}{0.000000,0.000000,0.000000}%
\pgfsetfillcolor{currentfill}%
\pgfsetlinewidth{1.505625pt}%
\definecolor{currentstroke}{rgb}{0.000000,0.000000,0.000000}%
\pgfsetstrokecolor{currentstroke}%
\pgfsetdash{}{0pt}%
\pgfsys@defobject{currentmarker}{\pgfqpoint{-0.083333in}{0.000000in}}{\pgfqpoint{-0.000000in}{0.000000in}}{%
\pgfpathmoveto{\pgfqpoint{-0.000000in}{0.000000in}}%
\pgfpathlineto{\pgfqpoint{-0.083333in}{0.000000in}}%
\pgfusepath{stroke,fill}%
}%
\begin{pgfscope}%
\pgfsys@transformshift{0.911181in}{1.712085in}%
\pgfsys@useobject{currentmarker}{}%
\end{pgfscope}%
\end{pgfscope}%
\begin{pgfscope}%
\definecolor{textcolor}{rgb}{0.000000,0.000000,0.000000}%
\pgfsetstrokecolor{textcolor}%
\pgfsetfillcolor{textcolor}%
\pgftext[x=0.443111in, y=1.612065in, left, base]{\color{textcolor}{\rmfamily\fontsize{22.000000}{26.400000}\selectfont\catcode`\^=\active\def^{\ifmmode\sp\else\^{}\fi}\catcode`\%=\active\def
\end{pgfscope}%
\begin{pgfscope}%
\pgfsetbuttcap%
\pgfsetroundjoin%
\definecolor{currentfill}{rgb}{0.000000,0.000000,0.000000}%
\pgfsetfillcolor{currentfill}%
\pgfsetlinewidth{1.505625pt}%
\definecolor{currentstroke}{rgb}{0.000000,0.000000,0.000000}%
\pgfsetstrokecolor{currentstroke}%
\pgfsetdash{}{0pt}%
\pgfsys@defobject{currentmarker}{\pgfqpoint{-0.083333in}{0.000000in}}{\pgfqpoint{-0.000000in}{0.000000in}}{%
\pgfpathmoveto{\pgfqpoint{-0.000000in}{0.000000in}}%
\pgfpathlineto{\pgfqpoint{-0.083333in}{0.000000in}}%
\pgfusepath{stroke,fill}%
}%
\begin{pgfscope}%
\pgfsys@transformshift{0.911181in}{2.592546in}%
\pgfsys@useobject{currentmarker}{}%
\end{pgfscope}%
\end{pgfscope}%
\begin{pgfscope}%
\definecolor{textcolor}{rgb}{0.000000,0.000000,0.000000}%
\pgfsetstrokecolor{textcolor}%
\pgfsetfillcolor{textcolor}%
\pgftext[x=0.443111in, y=2.492527in, left, base]{\color{textcolor}{\rmfamily\fontsize{22.000000}{26.400000}\selectfont\catcode`\^=\active\def^{\ifmmode\sp\else\^{}\fi}\catcode`\%=\active\def
\end{pgfscope}%
\begin{pgfscope}%
\definecolor{textcolor}{rgb}{0.000000,0.000000,0.000000}%
\pgfsetstrokecolor{textcolor}%
\pgfsetfillcolor{textcolor}%
\pgftext[x=0.387555in,y=2.240361in,,bottom,rotate=90.000000]{\color{textcolor}{\rmfamily\fontsize{22.000000}{26.400000}\selectfont\catcode`\^=\active\def^{\ifmmode\sp\else\^{}\fi}\catcode`\%=\active\def
\end{pgfscope}%
\begin{pgfscope}%
\pgfpathrectangle{\pgfqpoint{0.911181in}{0.831623in}}{\pgfqpoint{3.731930in}{2.817477in}}%
\pgfusepath{clip}%
\pgfsetbuttcap%
\pgfsetroundjoin%
\pgfsetlinewidth{3.513125pt}%
\definecolor{currentstroke}{rgb}{0.666667,0.200000,0.466667}%
\pgfsetstrokecolor{currentstroke}%
\pgfsetstrokeopacity{0.800000}%
\pgfsetdash{}{0pt}%
\pgfpathmoveto{\pgfqpoint{1.080814in}{1.904686in}}%
\pgfpathlineto{\pgfqpoint{1.080814in}{3.428984in}}%
\pgfusepath{stroke}%
\end{pgfscope}%
\begin{pgfscope}%
\pgfpathrectangle{\pgfqpoint{0.911181in}{0.831623in}}{\pgfqpoint{3.731930in}{2.817477in}}%
\pgfusepath{clip}%
\pgfsetbuttcap%
\pgfsetroundjoin%
\pgfsetlinewidth{3.513125pt}%
\definecolor{currentstroke}{rgb}{0.666667,0.200000,0.466667}%
\pgfsetstrokecolor{currentstroke}%
\pgfsetstrokeopacity{0.800000}%
\pgfsetdash{}{0pt}%
\pgfpathmoveto{\pgfqpoint{1.928980in}{1.671730in}}%
\pgfpathlineto{\pgfqpoint{1.928980in}{3.285909in}}%
\pgfusepath{stroke}%
\end{pgfscope}%
\begin{pgfscope}%
\pgfpathrectangle{\pgfqpoint{0.911181in}{0.831623in}}{\pgfqpoint{3.731930in}{2.817477in}}%
\pgfusepath{clip}%
\pgfsetbuttcap%
\pgfsetroundjoin%
\pgfsetlinewidth{3.513125pt}%
\definecolor{currentstroke}{rgb}{0.666667,0.200000,0.466667}%
\pgfsetstrokecolor{currentstroke}%
\pgfsetstrokeopacity{0.800000}%
\pgfsetdash{}{0pt}%
\pgfpathmoveto{\pgfqpoint{2.777146in}{1.513981in}}%
\pgfpathlineto{\pgfqpoint{2.777146in}{3.156592in}}%
\pgfusepath{stroke}%
\end{pgfscope}%
\begin{pgfscope}%
\pgfpathrectangle{\pgfqpoint{0.911181in}{0.831623in}}{\pgfqpoint{3.731930in}{2.817477in}}%
\pgfusepath{clip}%
\pgfsetbuttcap%
\pgfsetroundjoin%
\pgfsetlinewidth{3.513125pt}%
\definecolor{currentstroke}{rgb}{0.666667,0.200000,0.466667}%
\pgfsetstrokecolor{currentstroke}%
\pgfsetstrokeopacity{0.800000}%
\pgfsetdash{}{0pt}%
\pgfpathmoveto{\pgfqpoint{3.625312in}{1.366504in}}%
\pgfpathlineto{\pgfqpoint{3.625312in}{3.292513in}}%
\pgfusepath{stroke}%
\end{pgfscope}%
\begin{pgfscope}%
\pgfpathrectangle{\pgfqpoint{0.911181in}{0.831623in}}{\pgfqpoint{3.731930in}{2.817477in}}%
\pgfusepath{clip}%
\pgfsetbuttcap%
\pgfsetroundjoin%
\pgfsetlinewidth{3.513125pt}%
\definecolor{currentstroke}{rgb}{0.666667,0.200000,0.466667}%
\pgfsetstrokecolor{currentstroke}%
\pgfsetstrokeopacity{0.800000}%
\pgfsetdash{}{0pt}%
\pgfpathmoveto{\pgfqpoint{4.473478in}{1.281025in}}%
\pgfpathlineto{\pgfqpoint{4.473478in}{3.434487in}}%
\pgfusepath{stroke}%
\end{pgfscope}%
\begin{pgfscope}%
\pgfpathrectangle{\pgfqpoint{0.911181in}{0.831623in}}{\pgfqpoint{3.731930in}{2.817477in}}%
\pgfusepath{clip}%
\pgfsetbuttcap%
\pgfsetroundjoin%
\definecolor{currentfill}{rgb}{0.666667,0.200000,0.466667}%
\pgfsetfillcolor{currentfill}%
\pgfsetfillopacity{0.800000}%
\pgfsetlinewidth{2.007500pt}%
\definecolor{currentstroke}{rgb}{0.666667,0.200000,0.466667}%
\pgfsetstrokecolor{currentstroke}%
\pgfsetstrokeopacity{0.800000}%
\pgfsetdash{}{0pt}%
\pgfsys@defobject{currentmarker}{\pgfqpoint{-0.055556in}{-0.000000in}}{\pgfqpoint{0.055556in}{0.000000in}}{%
\pgfpathmoveto{\pgfqpoint{0.055556in}{-0.000000in}}%
\pgfpathlineto{\pgfqpoint{-0.055556in}{0.000000in}}%
\pgfusepath{stroke,fill}%
}%
\begin{pgfscope}%
\pgfsys@transformshift{1.080814in}{1.904686in}%
\pgfsys@useobject{currentmarker}{}%
\end{pgfscope}%
\begin{pgfscope}%
\pgfsys@transformshift{1.928980in}{1.671730in}%
\pgfsys@useobject{currentmarker}{}%
\end{pgfscope}%
\begin{pgfscope}%
\pgfsys@transformshift{2.777146in}{1.513981in}%
\pgfsys@useobject{currentmarker}{}%
\end{pgfscope}%
\begin{pgfscope}%
\pgfsys@transformshift{3.625312in}{1.366504in}%
\pgfsys@useobject{currentmarker}{}%
\end{pgfscope}%
\begin{pgfscope}%
\pgfsys@transformshift{4.473478in}{1.281025in}%
\pgfsys@useobject{currentmarker}{}%
\end{pgfscope}%
\end{pgfscope}%
\begin{pgfscope}%
\pgfpathrectangle{\pgfqpoint{0.911181in}{0.831623in}}{\pgfqpoint{3.731930in}{2.817477in}}%
\pgfusepath{clip}%
\pgfsetbuttcap%
\pgfsetroundjoin%
\definecolor{currentfill}{rgb}{0.666667,0.200000,0.466667}%
\pgfsetfillcolor{currentfill}%
\pgfsetfillopacity{0.800000}%
\pgfsetlinewidth{2.007500pt}%
\definecolor{currentstroke}{rgb}{0.666667,0.200000,0.466667}%
\pgfsetstrokecolor{currentstroke}%
\pgfsetstrokeopacity{0.800000}%
\pgfsetdash{}{0pt}%
\pgfsys@defobject{currentmarker}{\pgfqpoint{-0.055556in}{-0.000000in}}{\pgfqpoint{0.055556in}{0.000000in}}{%
\pgfpathmoveto{\pgfqpoint{0.055556in}{-0.000000in}}%
\pgfpathlineto{\pgfqpoint{-0.055556in}{0.000000in}}%
\pgfusepath{stroke,fill}%
}%
\begin{pgfscope}%
\pgfsys@transformshift{1.080814in}{3.428984in}%
\pgfsys@useobject{currentmarker}{}%
\end{pgfscope}%
\begin{pgfscope}%
\pgfsys@transformshift{1.928980in}{3.285909in}%
\pgfsys@useobject{currentmarker}{}%
\end{pgfscope}%
\begin{pgfscope}%
\pgfsys@transformshift{2.777146in}{3.156592in}%
\pgfsys@useobject{currentmarker}{}%
\end{pgfscope}%
\begin{pgfscope}%
\pgfsys@transformshift{3.625312in}{3.292513in}%
\pgfsys@useobject{currentmarker}{}%
\end{pgfscope}%
\begin{pgfscope}%
\pgfsys@transformshift{4.473478in}{3.434487in}%
\pgfsys@useobject{currentmarker}{}%
\end{pgfscope}%
\end{pgfscope}%
\begin{pgfscope}%
\pgfpathrectangle{\pgfqpoint{0.911181in}{0.831623in}}{\pgfqpoint{3.731930in}{2.817477in}}%
\pgfusepath{clip}%
\pgfsetbuttcap%
\pgfsetroundjoin%
\pgfsetlinewidth{3.513125pt}%
\definecolor{currentstroke}{rgb}{0.133333,0.533333,0.200000}%
\pgfsetstrokecolor{currentstroke}%
\pgfsetstrokeopacity{0.800000}%
\pgfsetdash{}{0pt}%
\pgfpathmoveto{\pgfqpoint{1.080814in}{1.376409in}}%
\pgfpathlineto{\pgfqpoint{1.080814in}{2.372431in}}%
\pgfusepath{stroke}%
\end{pgfscope}%
\begin{pgfscope}%
\pgfpathrectangle{\pgfqpoint{0.911181in}{0.831623in}}{\pgfqpoint{3.731930in}{2.817477in}}%
\pgfusepath{clip}%
\pgfsetbuttcap%
\pgfsetroundjoin%
\pgfsetlinewidth{3.513125pt}%
\definecolor{currentstroke}{rgb}{0.133333,0.533333,0.200000}%
\pgfsetstrokecolor{currentstroke}%
\pgfsetstrokeopacity{0.800000}%
\pgfsetdash{}{0pt}%
\pgfpathmoveto{\pgfqpoint{1.928980in}{1.528655in}}%
\pgfpathlineto{\pgfqpoint{1.928980in}{3.175852in}}%
\pgfusepath{stroke}%
\end{pgfscope}%
\begin{pgfscope}%
\pgfpathrectangle{\pgfqpoint{0.911181in}{0.831623in}}{\pgfqpoint{3.731930in}{2.817477in}}%
\pgfusepath{clip}%
\pgfsetbuttcap%
\pgfsetroundjoin%
\pgfsetlinewidth{3.513125pt}%
\definecolor{currentstroke}{rgb}{0.133333,0.533333,0.200000}%
\pgfsetstrokecolor{currentstroke}%
\pgfsetstrokeopacity{0.800000}%
\pgfsetdash{}{0pt}%
\pgfpathmoveto{\pgfqpoint{2.777146in}{1.929449in}}%
\pgfpathlineto{\pgfqpoint{2.777146in}{3.464753in}}%
\pgfusepath{stroke}%
\end{pgfscope}%
\begin{pgfscope}%
\pgfpathrectangle{\pgfqpoint{0.911181in}{0.831623in}}{\pgfqpoint{3.731930in}{2.817477in}}%
\pgfusepath{clip}%
\pgfsetbuttcap%
\pgfsetroundjoin%
\pgfsetlinewidth{3.513125pt}%
\definecolor{currentstroke}{rgb}{0.133333,0.533333,0.200000}%
\pgfsetstrokecolor{currentstroke}%
\pgfsetstrokeopacity{0.800000}%
\pgfsetdash{}{0pt}%
\pgfpathmoveto{\pgfqpoint{3.625312in}{1.789125in}}%
\pgfpathlineto{\pgfqpoint{3.625312in}{3.457599in}}%
\pgfusepath{stroke}%
\end{pgfscope}%
\begin{pgfscope}%
\pgfpathrectangle{\pgfqpoint{0.911181in}{0.831623in}}{\pgfqpoint{3.731930in}{2.817477in}}%
\pgfusepath{clip}%
\pgfsetbuttcap%
\pgfsetroundjoin%
\pgfsetlinewidth{3.513125pt}%
\definecolor{currentstroke}{rgb}{0.133333,0.533333,0.200000}%
\pgfsetstrokecolor{currentstroke}%
\pgfsetstrokeopacity{0.800000}%
\pgfsetdash{}{0pt}%
\pgfpathmoveto{\pgfqpoint{4.473478in}{2.300893in}}%
\pgfpathlineto{\pgfqpoint{4.473478in}{3.652433in}}%
\pgfusepath{stroke}%
\end{pgfscope}%
\begin{pgfscope}%
\pgfpathrectangle{\pgfqpoint{0.911181in}{0.831623in}}{\pgfqpoint{3.731930in}{2.817477in}}%
\pgfusepath{clip}%
\pgfsetbuttcap%
\pgfsetroundjoin%
\definecolor{currentfill}{rgb}{0.133333,0.533333,0.200000}%
\pgfsetfillcolor{currentfill}%
\pgfsetfillopacity{0.800000}%
\pgfsetlinewidth{2.007500pt}%
\definecolor{currentstroke}{rgb}{0.133333,0.533333,0.200000}%
\pgfsetstrokecolor{currentstroke}%
\pgfsetstrokeopacity{0.800000}%
\pgfsetdash{}{0pt}%
\pgfsys@defobject{currentmarker}{\pgfqpoint{-0.055556in}{-0.000000in}}{\pgfqpoint{0.055556in}{0.000000in}}{%
\pgfpathmoveto{\pgfqpoint{0.055556in}{-0.000000in}}%
\pgfpathlineto{\pgfqpoint{-0.055556in}{0.000000in}}%
\pgfusepath{stroke,fill}%
}%
\begin{pgfscope}%
\pgfsys@transformshift{1.080814in}{1.376409in}%
\pgfsys@useobject{currentmarker}{}%
\end{pgfscope}%
\begin{pgfscope}%
\pgfsys@transformshift{1.928980in}{1.528655in}%
\pgfsys@useobject{currentmarker}{}%
\end{pgfscope}%
\begin{pgfscope}%
\pgfsys@transformshift{2.777146in}{1.929449in}%
\pgfsys@useobject{currentmarker}{}%
\end{pgfscope}%
\begin{pgfscope}%
\pgfsys@transformshift{3.625312in}{1.789125in}%
\pgfsys@useobject{currentmarker}{}%
\end{pgfscope}%
\begin{pgfscope}%
\pgfsys@transformshift{4.473478in}{2.300893in}%
\pgfsys@useobject{currentmarker}{}%
\end{pgfscope}%
\end{pgfscope}%
\begin{pgfscope}%
\pgfpathrectangle{\pgfqpoint{0.911181in}{0.831623in}}{\pgfqpoint{3.731930in}{2.817477in}}%
\pgfusepath{clip}%
\pgfsetbuttcap%
\pgfsetroundjoin%
\definecolor{currentfill}{rgb}{0.133333,0.533333,0.200000}%
\pgfsetfillcolor{currentfill}%
\pgfsetfillopacity{0.800000}%
\pgfsetlinewidth{2.007500pt}%
\definecolor{currentstroke}{rgb}{0.133333,0.533333,0.200000}%
\pgfsetstrokecolor{currentstroke}%
\pgfsetstrokeopacity{0.800000}%
\pgfsetdash{}{0pt}%
\pgfsys@defobject{currentmarker}{\pgfqpoint{-0.055556in}{-0.000000in}}{\pgfqpoint{0.055556in}{0.000000in}}{%
\pgfpathmoveto{\pgfqpoint{0.055556in}{-0.000000in}}%
\pgfpathlineto{\pgfqpoint{-0.055556in}{0.000000in}}%
\pgfusepath{stroke,fill}%
}%
\begin{pgfscope}%
\pgfsys@transformshift{1.080814in}{2.372431in}%
\pgfsys@useobject{currentmarker}{}%
\end{pgfscope}%
\begin{pgfscope}%
\pgfsys@transformshift{1.928980in}{3.175852in}%
\pgfsys@useobject{currentmarker}{}%
\end{pgfscope}%
\begin{pgfscope}%
\pgfsys@transformshift{2.777146in}{3.464753in}%
\pgfsys@useobject{currentmarker}{}%
\end{pgfscope}%
\begin{pgfscope}%
\pgfsys@transformshift{3.625312in}{3.457599in}%
\pgfsys@useobject{currentmarker}{}%
\end{pgfscope}%
\begin{pgfscope}%
\pgfsys@transformshift{4.473478in}{3.667443in}%
\pgfsys@useobject{currentmarker}{}%
\end{pgfscope}%
\end{pgfscope}%
\begin{pgfscope}%
\pgfpathrectangle{\pgfqpoint{0.911181in}{0.831623in}}{\pgfqpoint{3.731930in}{2.817477in}}%
\pgfusepath{clip}%
\pgfsetbuttcap%
\pgfsetroundjoin%
\pgfsetlinewidth{3.513125pt}%
\definecolor{currentstroke}{rgb}{0.000000,0.000000,0.000000}%
\pgfsetstrokecolor{currentstroke}%
\pgfsetstrokeopacity{0.800000}%
\pgfsetdash{}{0pt}%
\pgfpathmoveto{\pgfqpoint{1.080814in}{1.403923in}}%
\pgfpathlineto{\pgfqpoint{1.080814in}{1.948709in}}%
\pgfusepath{stroke}%
\end{pgfscope}%
\begin{pgfscope}%
\pgfpathrectangle{\pgfqpoint{0.911181in}{0.831623in}}{\pgfqpoint{3.731930in}{2.817477in}}%
\pgfusepath{clip}%
\pgfsetbuttcap%
\pgfsetroundjoin%
\pgfsetlinewidth{3.513125pt}%
\definecolor{currentstroke}{rgb}{0.000000,0.000000,0.000000}%
\pgfsetstrokecolor{currentstroke}%
\pgfsetstrokeopacity{0.800000}%
\pgfsetdash{}{0pt}%
\pgfpathmoveto{\pgfqpoint{1.928980in}{1.088424in}}%
\pgfpathlineto{\pgfqpoint{1.928980in}{1.466289in}}%
\pgfusepath{stroke}%
\end{pgfscope}%
\begin{pgfscope}%
\pgfpathrectangle{\pgfqpoint{0.911181in}{0.831623in}}{\pgfqpoint{3.731930in}{2.817477in}}%
\pgfusepath{clip}%
\pgfsetbuttcap%
\pgfsetroundjoin%
\pgfsetlinewidth{3.513125pt}%
\definecolor{currentstroke}{rgb}{0.000000,0.000000,0.000000}%
\pgfsetstrokecolor{currentstroke}%
\pgfsetstrokeopacity{0.800000}%
\pgfsetdash{}{0pt}%
\pgfpathmoveto{\pgfqpoint{2.777146in}{1.004964in}}%
\pgfpathlineto{\pgfqpoint{2.777146in}{1.326883in}}%
\pgfusepath{stroke}%
\end{pgfscope}%
\begin{pgfscope}%
\pgfpathrectangle{\pgfqpoint{0.911181in}{0.831623in}}{\pgfqpoint{3.731930in}{2.817477in}}%
\pgfusepath{clip}%
\pgfsetbuttcap%
\pgfsetroundjoin%
\pgfsetlinewidth{3.513125pt}%
\definecolor{currentstroke}{rgb}{0.000000,0.000000,0.000000}%
\pgfsetstrokecolor{currentstroke}%
\pgfsetstrokeopacity{0.800000}%
\pgfsetdash{}{0pt}%
\pgfpathmoveto{\pgfqpoint{3.625312in}{0.919669in}}%
\pgfpathlineto{\pgfqpoint{3.625312in}{1.210222in}}%
\pgfusepath{stroke}%
\end{pgfscope}%
\begin{pgfscope}%
\pgfpathrectangle{\pgfqpoint{0.911181in}{0.831623in}}{\pgfqpoint{3.731930in}{2.817477in}}%
\pgfusepath{clip}%
\pgfsetbuttcap%
\pgfsetroundjoin%
\pgfsetlinewidth{3.513125pt}%
\definecolor{currentstroke}{rgb}{0.000000,0.000000,0.000000}%
\pgfsetstrokecolor{currentstroke}%
\pgfsetstrokeopacity{0.800000}%
\pgfsetdash{}{0pt}%
\pgfpathmoveto{\pgfqpoint{4.473478in}{0.895824in}}%
\pgfpathlineto{\pgfqpoint{4.473478in}{1.084756in}}%
\pgfusepath{stroke}%
\end{pgfscope}%
\begin{pgfscope}%
\pgfpathrectangle{\pgfqpoint{0.911181in}{0.831623in}}{\pgfqpoint{3.731930in}{2.817477in}}%
\pgfusepath{clip}%
\pgfsetbuttcap%
\pgfsetroundjoin%
\definecolor{currentfill}{rgb}{0.000000,0.000000,0.000000}%
\pgfsetfillcolor{currentfill}%
\pgfsetfillopacity{0.800000}%
\pgfsetlinewidth{2.007500pt}%
\definecolor{currentstroke}{rgb}{0.000000,0.000000,0.000000}%
\pgfsetstrokecolor{currentstroke}%
\pgfsetstrokeopacity{0.800000}%
\pgfsetdash{}{0pt}%
\pgfsys@defobject{currentmarker}{\pgfqpoint{-0.055556in}{-0.000000in}}{\pgfqpoint{0.055556in}{0.000000in}}{%
\pgfpathmoveto{\pgfqpoint{0.055556in}{-0.000000in}}%
\pgfpathlineto{\pgfqpoint{-0.055556in}{0.000000in}}%
\pgfusepath{stroke,fill}%
}%
\begin{pgfscope}%
\pgfsys@transformshift{1.080814in}{1.403923in}%
\pgfsys@useobject{currentmarker}{}%
\end{pgfscope}%
\begin{pgfscope}%
\pgfsys@transformshift{1.928980in}{1.088424in}%
\pgfsys@useobject{currentmarker}{}%
\end{pgfscope}%
\begin{pgfscope}%
\pgfsys@transformshift{2.777146in}{1.004964in}%
\pgfsys@useobject{currentmarker}{}%
\end{pgfscope}%
\begin{pgfscope}%
\pgfsys@transformshift{3.625312in}{0.919669in}%
\pgfsys@useobject{currentmarker}{}%
\end{pgfscope}%
\begin{pgfscope}%
\pgfsys@transformshift{4.473478in}{0.895824in}%
\pgfsys@useobject{currentmarker}{}%
\end{pgfscope}%
\end{pgfscope}%
\begin{pgfscope}%
\pgfpathrectangle{\pgfqpoint{0.911181in}{0.831623in}}{\pgfqpoint{3.731930in}{2.817477in}}%
\pgfusepath{clip}%
\pgfsetbuttcap%
\pgfsetroundjoin%
\definecolor{currentfill}{rgb}{0.000000,0.000000,0.000000}%
\pgfsetfillcolor{currentfill}%
\pgfsetfillopacity{0.800000}%
\pgfsetlinewidth{2.007500pt}%
\definecolor{currentstroke}{rgb}{0.000000,0.000000,0.000000}%
\pgfsetstrokecolor{currentstroke}%
\pgfsetstrokeopacity{0.800000}%
\pgfsetdash{}{0pt}%
\pgfsys@defobject{currentmarker}{\pgfqpoint{-0.055556in}{-0.000000in}}{\pgfqpoint{0.055556in}{0.000000in}}{%
\pgfpathmoveto{\pgfqpoint{0.055556in}{-0.000000in}}%
\pgfpathlineto{\pgfqpoint{-0.055556in}{0.000000in}}%
\pgfusepath{stroke,fill}%
}%
\begin{pgfscope}%
\pgfsys@transformshift{1.080814in}{1.948709in}%
\pgfsys@useobject{currentmarker}{}%
\end{pgfscope}%
\begin{pgfscope}%
\pgfsys@transformshift{1.928980in}{1.466289in}%
\pgfsys@useobject{currentmarker}{}%
\end{pgfscope}%
\begin{pgfscope}%
\pgfsys@transformshift{2.777146in}{1.326883in}%
\pgfsys@useobject{currentmarker}{}%
\end{pgfscope}%
\begin{pgfscope}%
\pgfsys@transformshift{3.625312in}{1.210222in}%
\pgfsys@useobject{currentmarker}{}%
\end{pgfscope}%
\begin{pgfscope}%
\pgfsys@transformshift{4.473478in}{1.084756in}%
\pgfsys@useobject{currentmarker}{}%
\end{pgfscope}%
\end{pgfscope}%
\begin{pgfscope}%
\pgfpathrectangle{\pgfqpoint{0.911181in}{0.831623in}}{\pgfqpoint{3.731930in}{2.817477in}}%
\pgfusepath{clip}%
\pgfsetrectcap%
\pgfsetroundjoin%
\pgfsetlinewidth{3.513125pt}%
\definecolor{currentstroke}{rgb}{0.666667,0.200000,0.466667}%
\pgfsetstrokecolor{currentstroke}%
\pgfsetstrokeopacity{0.800000}%
\pgfsetdash{}{0pt}%
\pgfpathmoveto{\pgfqpoint{1.080814in}{2.723735in}}%
\pgfpathlineto{\pgfqpoint{1.928980in}{2.453580in}}%
\pgfpathlineto{\pgfqpoint{2.777146in}{2.429771in}}%
\pgfpathlineto{\pgfqpoint{3.625312in}{2.274964in}}%
\pgfpathlineto{\pgfqpoint{4.473478in}{2.305002in}}%
\pgfusepath{stroke}%
\end{pgfscope}%
\begin{pgfscope}%
\pgfpathrectangle{\pgfqpoint{0.911181in}{0.831623in}}{\pgfqpoint{3.731930in}{2.817477in}}%
\pgfusepath{clip}%
\pgfsetbuttcap%
\pgfsetroundjoin%
\definecolor{currentfill}{rgb}{0.666667,0.200000,0.466667}%
\pgfsetfillcolor{currentfill}%
\pgfsetfillopacity{0.800000}%
\pgfsetlinewidth{2.007500pt}%
\definecolor{currentstroke}{rgb}{1.000000,1.000000,1.000000}%
\pgfsetstrokecolor{currentstroke}%
\pgfsetstrokeopacity{0.800000}%
\pgfsetdash{}{0pt}%
\pgfsys@defobject{currentmarker}{\pgfqpoint{-0.055556in}{-0.055556in}}{\pgfqpoint{0.055556in}{0.055556in}}{%
\pgfpathmoveto{\pgfqpoint{0.000000in}{-0.055556in}}%
\pgfpathcurveto{\pgfqpoint{0.014734in}{-0.055556in}}{\pgfqpoint{0.028866in}{-0.049702in}}{\pgfqpoint{0.039284in}{-0.039284in}}%
\pgfpathcurveto{\pgfqpoint{0.049702in}{-0.028866in}}{\pgfqpoint{0.055556in}{-0.014734in}}{\pgfqpoint{0.055556in}{0.000000in}}%
\pgfpathcurveto{\pgfqpoint{0.055556in}{0.014734in}}{\pgfqpoint{0.049702in}{0.028866in}}{\pgfqpoint{0.039284in}{0.039284in}}%
\pgfpathcurveto{\pgfqpoint{0.028866in}{0.049702in}}{\pgfqpoint{0.014734in}{0.055556in}}{\pgfqpoint{0.000000in}{0.055556in}}%
\pgfpathcurveto{\pgfqpoint{-0.014734in}{0.055556in}}{\pgfqpoint{-0.028866in}{0.049702in}}{\pgfqpoint{-0.039284in}{0.039284in}}%
\pgfpathcurveto{\pgfqpoint{-0.049702in}{0.028866in}}{\pgfqpoint{-0.055556in}{0.014734in}}{\pgfqpoint{-0.055556in}{0.000000in}}%
\pgfpathcurveto{\pgfqpoint{-0.055556in}{-0.014734in}}{\pgfqpoint{-0.049702in}{-0.028866in}}{\pgfqpoint{-0.039284in}{-0.039284in}}%
\pgfpathcurveto{\pgfqpoint{-0.028866in}{-0.049702in}}{\pgfqpoint{-0.014734in}{-0.055556in}}{\pgfqpoint{0.000000in}{-0.055556in}}%
\pgfpathlineto{\pgfqpoint{0.000000in}{-0.055556in}}%
\pgfpathclose%
\pgfusepath{stroke,fill}%
}%
\begin{pgfscope}%
\pgfsys@transformshift{1.080814in}{2.723735in}%
\pgfsys@useobject{currentmarker}{}%
\end{pgfscope}%
\begin{pgfscope}%
\pgfsys@transformshift{1.928980in}{2.453580in}%
\pgfsys@useobject{currentmarker}{}%
\end{pgfscope}%
\begin{pgfscope}%
\pgfsys@transformshift{2.777146in}{2.429771in}%
\pgfsys@useobject{currentmarker}{}%
\end{pgfscope}%
\begin{pgfscope}%
\pgfsys@transformshift{3.625312in}{2.274964in}%
\pgfsys@useobject{currentmarker}{}%
\end{pgfscope}%
\begin{pgfscope}%
\pgfsys@transformshift{4.473478in}{2.305002in}%
\pgfsys@useobject{currentmarker}{}%
\end{pgfscope}%
\end{pgfscope}%
\begin{pgfscope}%
\pgfpathrectangle{\pgfqpoint{0.911181in}{0.831623in}}{\pgfqpoint{3.731930in}{2.817477in}}%
\pgfusepath{clip}%
\pgfsetbuttcap%
\pgfsetroundjoin%
\pgfsetlinewidth{3.513125pt}%
\definecolor{currentstroke}{rgb}{0.133333,0.533333,0.200000}%
\pgfsetstrokecolor{currentstroke}%
\pgfsetstrokeopacity{0.800000}%
\pgfsetdash{{12.950000pt}{5.600000pt}}{0.000000pt}%
\pgfpathmoveto{\pgfqpoint{1.080814in}{1.913710in}}%
\pgfpathlineto{\pgfqpoint{1.928980in}{2.332810in}}%
\pgfpathlineto{\pgfqpoint{2.777146in}{2.637229in}}%
\pgfpathlineto{\pgfqpoint{3.625312in}{2.673020in}}%
\pgfpathlineto{\pgfqpoint{4.473478in}{3.006510in}}%
\pgfusepath{stroke}%
\end{pgfscope}%
\begin{pgfscope}%
\pgfpathrectangle{\pgfqpoint{0.911181in}{0.831623in}}{\pgfqpoint{3.731930in}{2.817477in}}%
\pgfusepath{clip}%
\pgfsetbuttcap%
\pgfsetmiterjoin%
\definecolor{currentfill}{rgb}{0.133333,0.533333,0.200000}%
\pgfsetfillcolor{currentfill}%
\pgfsetfillopacity{0.800000}%
\pgfsetlinewidth{2.007500pt}%
\definecolor{currentstroke}{rgb}{1.000000,1.000000,1.000000}%
\pgfsetstrokecolor{currentstroke}%
\pgfsetstrokeopacity{0.800000}%
\pgfsetdash{}{0pt}%
\pgfsys@defobject{currentmarker}{\pgfqpoint{-0.055556in}{-0.055556in}}{\pgfqpoint{0.055556in}{0.055556in}}{%
\pgfpathmoveto{\pgfqpoint{-0.055556in}{-0.055556in}}%
\pgfpathlineto{\pgfqpoint{0.055556in}{-0.055556in}}%
\pgfpathlineto{\pgfqpoint{0.055556in}{0.055556in}}%
\pgfpathlineto{\pgfqpoint{-0.055556in}{0.055556in}}%
\pgfpathlineto{\pgfqpoint{-0.055556in}{-0.055556in}}%
\pgfpathclose%
\pgfusepath{stroke,fill}%
}%
\begin{pgfscope}%
\pgfsys@transformshift{1.080814in}{1.913710in}%
\pgfsys@useobject{currentmarker}{}%
\end{pgfscope}%
\begin{pgfscope}%
\pgfsys@transformshift{1.928980in}{2.332810in}%
\pgfsys@useobject{currentmarker}{}%
\end{pgfscope}%
\begin{pgfscope}%
\pgfsys@transformshift{2.777146in}{2.637229in}%
\pgfsys@useobject{currentmarker}{}%
\end{pgfscope}%
\begin{pgfscope}%
\pgfsys@transformshift{3.625312in}{2.673020in}%
\pgfsys@useobject{currentmarker}{}%
\end{pgfscope}%
\begin{pgfscope}%
\pgfsys@transformshift{4.473478in}{3.006510in}%
\pgfsys@useobject{currentmarker}{}%
\end{pgfscope}%
\end{pgfscope}%
\begin{pgfscope}%
\pgfpathrectangle{\pgfqpoint{0.911181in}{0.831623in}}{\pgfqpoint{3.731930in}{2.817477in}}%
\pgfusepath{clip}%
\pgfsetbuttcap%
\pgfsetroundjoin%
\pgfsetlinewidth{3.513125pt}%
\definecolor{currentstroke}{rgb}{0.000000,0.000000,0.000000}%
\pgfsetstrokecolor{currentstroke}%
\pgfsetstrokeopacity{0.800000}%
\pgfsetdash{{22.400000pt}{5.600000pt}{3.500000pt}{5.600000pt}}{0.000000pt}%
\pgfpathmoveto{\pgfqpoint{1.080814in}{1.675766in}}%
\pgfpathlineto{\pgfqpoint{1.928980in}{1.304431in}}%
\pgfpathlineto{\pgfqpoint{2.777146in}{1.184248in}}%
\pgfpathlineto{\pgfqpoint{3.625312in}{1.099372in}}%
\pgfpathlineto{\pgfqpoint{4.473478in}{1.020922in}}%
\pgfusepath{stroke}%
\end{pgfscope}%
\begin{pgfscope}%
\pgfpathrectangle{\pgfqpoint{0.911181in}{0.831623in}}{\pgfqpoint{3.731930in}{2.817477in}}%
\pgfusepath{clip}%
\pgfsetbuttcap%
\pgfsetmiterjoin%
\definecolor{currentfill}{rgb}{0.000000,0.000000,0.000000}%
\pgfsetfillcolor{currentfill}%
\pgfsetfillopacity{0.800000}%
\pgfsetlinewidth{2.007500pt}%
\definecolor{currentstroke}{rgb}{1.000000,1.000000,1.000000}%
\pgfsetstrokecolor{currentstroke}%
\pgfsetstrokeopacity{0.800000}%
\pgfsetdash{}{0pt}%
\pgfsys@defobject{currentmarker}{\pgfqpoint{-0.078567in}{-0.078567in}}{\pgfqpoint{0.078567in}{0.078567in}}{%
\pgfpathmoveto{\pgfqpoint{-0.000000in}{-0.078567in}}%
\pgfpathlineto{\pgfqpoint{0.078567in}{0.000000in}}%
\pgfpathlineto{\pgfqpoint{0.000000in}{0.078567in}}%
\pgfpathlineto{\pgfqpoint{-0.078567in}{0.000000in}}%
\pgfpathlineto{\pgfqpoint{-0.000000in}{-0.078567in}}%
\pgfpathclose%
\pgfusepath{stroke,fill}%
}%
\begin{pgfscope}%
\pgfsys@transformshift{1.080814in}{1.675766in}%
\pgfsys@useobject{currentmarker}{}%
\end{pgfscope}%
\begin{pgfscope}%
\pgfsys@transformshift{1.928980in}{1.304431in}%
\pgfsys@useobject{currentmarker}{}%
\end{pgfscope}%
\begin{pgfscope}%
\pgfsys@transformshift{2.777146in}{1.184248in}%
\pgfsys@useobject{currentmarker}{}%
\end{pgfscope}%
\begin{pgfscope}%
\pgfsys@transformshift{3.625312in}{1.099372in}%
\pgfsys@useobject{currentmarker}{}%
\end{pgfscope}%
\begin{pgfscope}%
\pgfsys@transformshift{4.473478in}{1.020922in}%
\pgfsys@useobject{currentmarker}{}%
\end{pgfscope}%
\end{pgfscope}%
\begin{pgfscope}%
\pgfsetrectcap%
\pgfsetmiterjoin%
\pgfsetlinewidth{1.505625pt}%
\definecolor{currentstroke}{rgb}{0.000000,0.000000,0.000000}%
\pgfsetstrokecolor{currentstroke}%
\pgfsetdash{}{0pt}%
\pgfpathmoveto{\pgfqpoint{0.911181in}{0.831623in}}%
\pgfpathlineto{\pgfqpoint{0.911181in}{3.649100in}}%
\pgfusepath{stroke}%
\end{pgfscope}%
\begin{pgfscope}%
\pgfsetrectcap%
\pgfsetmiterjoin%
\pgfsetlinewidth{1.505625pt}%
\definecolor{currentstroke}{rgb}{0.000000,0.000000,0.000000}%
\pgfsetstrokecolor{currentstroke}%
\pgfsetdash{}{0pt}%
\pgfpathmoveto{\pgfqpoint{0.911181in}{0.831623in}}%
\pgfpathlineto{\pgfqpoint{4.643111in}{0.831623in}}%
\pgfusepath{stroke}%
\end{pgfscope}%
\end{pgfpicture}%
\makeatother%
\endgroup%

%% file: figures/synthetic-final/M1a/runtime_asymmetric_errors.pgf
\begingroup%
\makeatletter%
\begin{pgfpicture}%
\pgfpathrectangle{\pgfpointorigin}{\pgfqpoint{4.747194in}{3.743420in}}%
\pgfusepath{use as bounding box, clip}%
\begin{pgfscope}%
\pgfsetbuttcap%
\pgfsetmiterjoin%
\definecolor{currentfill}{rgb}{1.000000,1.000000,1.000000}%
\pgfsetfillcolor{currentfill}%
\pgfsetlinewidth{0.000000pt}%
\definecolor{currentstroke}{rgb}{1.000000,1.000000,1.000000}%
\pgfsetstrokecolor{currentstroke}%
\pgfsetdash{}{0pt}%
\pgfpathmoveto{\pgfqpoint{0.000000in}{0.000000in}}%
\pgfpathlineto{\pgfqpoint{4.747194in}{0.000000in}}%
\pgfpathlineto{\pgfqpoint{4.747194in}{3.743420in}}%
\pgfpathlineto{\pgfqpoint{0.000000in}{3.743420in}}%
\pgfpathlineto{\pgfqpoint{0.000000in}{0.000000in}}%
\pgfpathclose%
\pgfusepath{fill}%
\end{pgfscope}%
\begin{pgfscope}%
\pgfsetbuttcap%
\pgfsetmiterjoin%
\definecolor{currentfill}{rgb}{1.000000,1.000000,1.000000}%
\pgfsetfillcolor{currentfill}%
\pgfsetlinewidth{0.000000pt}%
\definecolor{currentstroke}{rgb}{0.000000,0.000000,0.000000}%
\pgfsetstrokecolor{currentstroke}%
\pgfsetstrokeopacity{0.000000}%
\pgfsetdash{}{0pt}%
\pgfpathmoveto{\pgfqpoint{0.839270in}{0.831623in}}%
\pgfpathlineto{\pgfqpoint{4.647194in}{0.831623in}}%
\pgfpathlineto{\pgfqpoint{4.647194in}{3.543400in}}%
\pgfpathlineto{\pgfqpoint{0.839270in}{3.543400in}}%
\pgfpathlineto{\pgfqpoint{0.839270in}{0.831623in}}%
\pgfpathclose%
\pgfusepath{fill}%
\end{pgfscope}%
\begin{pgfscope}%
\pgfsetbuttcap%
\pgfsetroundjoin%
\definecolor{currentfill}{rgb}{0.000000,0.000000,0.000000}%
\pgfsetfillcolor{currentfill}%
\pgfsetlinewidth{1.505625pt}%
\definecolor{currentstroke}{rgb}{0.000000,0.000000,0.000000}%
\pgfsetstrokecolor{currentstroke}%
\pgfsetdash{}{0pt}%
\pgfsys@defobject{currentmarker}{\pgfqpoint{0.000000in}{-0.083333in}}{\pgfqpoint{0.000000in}{0.000000in}}{%
\pgfpathmoveto{\pgfqpoint{0.000000in}{0.000000in}}%
\pgfpathlineto{\pgfqpoint{0.000000in}{-0.083333in}}%
\pgfusepath{stroke,fill}%
}%
\begin{pgfscope}%
\pgfsys@transformshift{1.012358in}{0.831623in}%
\pgfsys@useobject{currentmarker}{}%
\end{pgfscope}%
\end{pgfscope}%
\begin{pgfscope}%
\definecolor{textcolor}{rgb}{0.000000,0.000000,0.000000}%
\pgfsetstrokecolor{textcolor}%
\pgfsetfillcolor{textcolor}%
\pgftext[x=1.012358in,y=0.699679in,,top]{\color{textcolor}{\rmfamily\fontsize{22.000000}{26.400000}\selectfont\catcode`\^=\active\def^{\ifmmode\sp\else\^{}\fi}\catcode`\%=\active\def
\end{pgfscope}%
\begin{pgfscope}%
\pgfsetbuttcap%
\pgfsetroundjoin%
\definecolor{currentfill}{rgb}{0.000000,0.000000,0.000000}%
\pgfsetfillcolor{currentfill}%
\pgfsetlinewidth{1.505625pt}%
\definecolor{currentstroke}{rgb}{0.000000,0.000000,0.000000}%
\pgfsetstrokecolor{currentstroke}%
\pgfsetdash{}{0pt}%
\pgfsys@defobject{currentmarker}{\pgfqpoint{0.000000in}{-0.083333in}}{\pgfqpoint{0.000000in}{0.000000in}}{%
\pgfpathmoveto{\pgfqpoint{0.000000in}{0.000000in}}%
\pgfpathlineto{\pgfqpoint{0.000000in}{-0.083333in}}%
\pgfusepath{stroke,fill}%
}%
\begin{pgfscope}%
\pgfsys@transformshift{2.743232in}{0.831623in}%
\pgfsys@useobject{currentmarker}{}%
\end{pgfscope}%
\end{pgfscope}%
\begin{pgfscope}%
\definecolor{textcolor}{rgb}{0.000000,0.000000,0.000000}%
\pgfsetstrokecolor{textcolor}%
\pgfsetfillcolor{textcolor}%
\pgftext[x=2.743232in,y=0.699679in,,top]{\color{textcolor}{\rmfamily\fontsize{22.000000}{26.400000}\selectfont\catcode`\^=\active\def^{\ifmmode\sp\else\^{}\fi}\catcode`\%=\active\def
\end{pgfscope}%
\begin{pgfscope}%
\pgfsetbuttcap%
\pgfsetroundjoin%
\definecolor{currentfill}{rgb}{0.000000,0.000000,0.000000}%
\pgfsetfillcolor{currentfill}%
\pgfsetlinewidth{1.505625pt}%
\definecolor{currentstroke}{rgb}{0.000000,0.000000,0.000000}%
\pgfsetstrokecolor{currentstroke}%
\pgfsetdash{}{0pt}%
\pgfsys@defobject{currentmarker}{\pgfqpoint{0.000000in}{-0.083333in}}{\pgfqpoint{0.000000in}{0.000000in}}{%
\pgfpathmoveto{\pgfqpoint{0.000000in}{0.000000in}}%
\pgfpathlineto{\pgfqpoint{0.000000in}{-0.083333in}}%
\pgfusepath{stroke,fill}%
}%
\begin{pgfscope}%
\pgfsys@transformshift{4.474107in}{0.831623in}%
\pgfsys@useobject{currentmarker}{}%
\end{pgfscope}%
\end{pgfscope}%
\begin{pgfscope}%
\definecolor{textcolor}{rgb}{0.000000,0.000000,0.000000}%
\pgfsetstrokecolor{textcolor}%
\pgfsetfillcolor{textcolor}%
\pgftext[x=4.474107in,y=0.699679in,,top]{\color{textcolor}{\rmfamily\fontsize{22.000000}{26.400000}\selectfont\catcode`\^=\active\def^{\ifmmode\sp\else\^{}\fi}\catcode`\%=\active\def
\end{pgfscope}%
\begin{pgfscope}%
\definecolor{textcolor}{rgb}{0.000000,0.000000,0.000000}%
\pgfsetstrokecolor{textcolor}%
\pgfsetfillcolor{textcolor}%
\pgftext[x=2.743232in,y=0.388056in,,top]{\color{textcolor}{\rmfamily\fontsize{22.000000}{26.400000}\selectfont\catcode`\^=\active\def^{\ifmmode\sp\else\^{}\fi}\catcode`\%=\active\def
\end{pgfscope}%
\begin{pgfscope}%
\pgfsetbuttcap%
\pgfsetroundjoin%
\definecolor{currentfill}{rgb}{0.000000,0.000000,0.000000}%
\pgfsetfillcolor{currentfill}%
\pgfsetlinewidth{1.505625pt}%
\definecolor{currentstroke}{rgb}{0.000000,0.000000,0.000000}%
\pgfsetstrokecolor{currentstroke}%
\pgfsetdash{}{0pt}%
\pgfsys@defobject{currentmarker}{\pgfqpoint{-0.083333in}{0.000000in}}{\pgfqpoint{-0.000000in}{0.000000in}}{%
\pgfpathmoveto{\pgfqpoint{-0.000000in}{0.000000in}}%
\pgfpathlineto{\pgfqpoint{-0.083333in}{0.000000in}}%
\pgfusepath{stroke,fill}%
}%
\begin{pgfscope}%
\pgfsys@transformshift{0.839270in}{0.831623in}%
\pgfsys@useobject{currentmarker}{}%
\end{pgfscope}%
\end{pgfscope}%
\begin{pgfscope}%
\definecolor{textcolor}{rgb}{0.000000,0.000000,0.000000}%
\pgfsetstrokecolor{textcolor}%
\pgfsetfillcolor{textcolor}%
\pgftext[x=0.575218in, y=0.731604in, left, base]{\color{textcolor}{\rmfamily\fontsize{22.000000}{26.400000}\selectfont\catcode`\^=\active\def^{\ifmmode\sp\else\^{}\fi}\catcode`\%=\active\def
\end{pgfscope}%
\begin{pgfscope}%
\pgfsetbuttcap%
\pgfsetroundjoin%
\definecolor{currentfill}{rgb}{0.000000,0.000000,0.000000}%
\pgfsetfillcolor{currentfill}%
\pgfsetlinewidth{1.505625pt}%
\definecolor{currentstroke}{rgb}{0.000000,0.000000,0.000000}%
\pgfsetstrokecolor{currentstroke}%
\pgfsetdash{}{0pt}%
\pgfsys@defobject{currentmarker}{\pgfqpoint{-0.083333in}{0.000000in}}{\pgfqpoint{-0.000000in}{0.000000in}}{%
\pgfpathmoveto{\pgfqpoint{-0.000000in}{0.000000in}}%
\pgfpathlineto{\pgfqpoint{-0.083333in}{0.000000in}}%
\pgfusepath{stroke,fill}%
}%
\begin{pgfscope}%
\pgfsys@transformshift{0.839270in}{2.187512in}%
\pgfsys@useobject{currentmarker}{}%
\end{pgfscope}%
\end{pgfscope}%
\begin{pgfscope}%
\definecolor{textcolor}{rgb}{0.000000,0.000000,0.000000}%
\pgfsetstrokecolor{textcolor}%
\pgfsetfillcolor{textcolor}%
\pgftext[x=0.443111in, y=2.087493in, left, base]{\color{textcolor}{\rmfamily\fontsize{22.000000}{26.400000}\selectfont\catcode`\^=\active\def^{\ifmmode\sp\else\^{}\fi}\catcode`\%=\active\def
\end{pgfscope}%
\begin{pgfscope}%
\pgfsetbuttcap%
\pgfsetroundjoin%
\definecolor{currentfill}{rgb}{0.000000,0.000000,0.000000}%
\pgfsetfillcolor{currentfill}%
\pgfsetlinewidth{1.505625pt}%
\definecolor{currentstroke}{rgb}{0.000000,0.000000,0.000000}%
\pgfsetstrokecolor{currentstroke}%
\pgfsetdash{}{0pt}%
\pgfsys@defobject{currentmarker}{\pgfqpoint{-0.083333in}{0.000000in}}{\pgfqpoint{-0.000000in}{0.000000in}}{%
\pgfpathmoveto{\pgfqpoint{-0.000000in}{0.000000in}}%
\pgfpathlineto{\pgfqpoint{-0.083333in}{0.000000in}}%
\pgfusepath{stroke,fill}%
}%
\begin{pgfscope}%
\pgfsys@transformshift{0.839270in}{3.543400in}%
\pgfsys@useobject{currentmarker}{}%
\end{pgfscope}%
\end{pgfscope}%
\begin{pgfscope}%
\definecolor{textcolor}{rgb}{0.000000,0.000000,0.000000}%
\pgfsetstrokecolor{textcolor}%
\pgfsetfillcolor{textcolor}%
\pgftext[x=0.443111in, y=3.443381in, left, base]{\color{textcolor}{\rmfamily\fontsize{22.000000}{26.400000}\selectfont\catcode`\^=\active\def^{\ifmmode\sp\else\^{}\fi}\catcode`\%=\active\def
\end{pgfscope}%
\begin{pgfscope}%
\definecolor{textcolor}{rgb}{0.000000,0.000000,0.000000}%
\pgfsetstrokecolor{textcolor}%
\pgfsetfillcolor{textcolor}%
\pgftext[x=0.387555in,y=2.187512in,,bottom,rotate=90.000000]{\color{textcolor}{\rmfamily\fontsize{22.000000}{26.400000}\selectfont\catcode`\^=\active\def^{\ifmmode\sp\else\^{}\fi}\catcode`\%=\active\def
\end{pgfscope}%
\begin{pgfscope}%
\pgfpathrectangle{\pgfqpoint{0.839270in}{0.831623in}}{\pgfqpoint{3.807924in}{2.711777in}}%
\pgfusepath{clip}%
\pgfsetbuttcap%
\pgfsetroundjoin%
\pgfsetlinewidth{3.513125pt}%
\definecolor{currentstroke}{rgb}{0.666667,0.200000,0.466667}%
\pgfsetstrokecolor{currentstroke}%
\pgfsetdash{}{0pt}%
\pgfpathmoveto{\pgfqpoint{1.012358in}{0.835171in}}%
\pgfpathlineto{\pgfqpoint{1.012358in}{0.837058in}}%
\pgfusepath{stroke}%
\end{pgfscope}%
\begin{pgfscope}%
\pgfpathrectangle{\pgfqpoint{0.839270in}{0.831623in}}{\pgfqpoint{3.807924in}{2.711777in}}%
\pgfusepath{clip}%
\pgfsetbuttcap%
\pgfsetroundjoin%
\pgfsetlinewidth{3.513125pt}%
\definecolor{currentstroke}{rgb}{0.666667,0.200000,0.466667}%
\pgfsetstrokecolor{currentstroke}%
\pgfsetdash{}{0pt}%
\pgfpathmoveto{\pgfqpoint{1.877795in}{0.836977in}}%
\pgfpathlineto{\pgfqpoint{1.877795in}{0.838981in}}%
\pgfusepath{stroke}%
\end{pgfscope}%
\begin{pgfscope}%
\pgfpathrectangle{\pgfqpoint{0.839270in}{0.831623in}}{\pgfqpoint{3.807924in}{2.711777in}}%
\pgfusepath{clip}%
\pgfsetbuttcap%
\pgfsetroundjoin%
\pgfsetlinewidth{3.513125pt}%
\definecolor{currentstroke}{rgb}{0.666667,0.200000,0.466667}%
\pgfsetstrokecolor{currentstroke}%
\pgfsetdash{}{0pt}%
\pgfpathmoveto{\pgfqpoint{2.743232in}{0.838664in}}%
\pgfpathlineto{\pgfqpoint{2.743232in}{0.841942in}}%
\pgfusepath{stroke}%
\end{pgfscope}%
\begin{pgfscope}%
\pgfpathrectangle{\pgfqpoint{0.839270in}{0.831623in}}{\pgfqpoint{3.807924in}{2.711777in}}%
\pgfusepath{clip}%
\pgfsetbuttcap%
\pgfsetroundjoin%
\pgfsetlinewidth{3.513125pt}%
\definecolor{currentstroke}{rgb}{0.666667,0.200000,0.466667}%
\pgfsetstrokecolor{currentstroke}%
\pgfsetdash{}{0pt}%
\pgfpathmoveto{\pgfqpoint{3.608670in}{0.842926in}}%
\pgfpathlineto{\pgfqpoint{3.608670in}{0.846899in}}%
\pgfusepath{stroke}%
\end{pgfscope}%
\begin{pgfscope}%
\pgfpathrectangle{\pgfqpoint{0.839270in}{0.831623in}}{\pgfqpoint{3.807924in}{2.711777in}}%
\pgfusepath{clip}%
\pgfsetbuttcap%
\pgfsetroundjoin%
\pgfsetlinewidth{3.513125pt}%
\definecolor{currentstroke}{rgb}{0.666667,0.200000,0.466667}%
\pgfsetstrokecolor{currentstroke}%
\pgfsetdash{}{0pt}%
\pgfpathmoveto{\pgfqpoint{4.474107in}{0.846149in}}%
\pgfpathlineto{\pgfqpoint{4.474107in}{0.854164in}}%
\pgfusepath{stroke}%
\end{pgfscope}%
\begin{pgfscope}%
\pgfpathrectangle{\pgfqpoint{0.839270in}{0.831623in}}{\pgfqpoint{3.807924in}{2.711777in}}%
\pgfusepath{clip}%
\pgfsetbuttcap%
\pgfsetroundjoin%
\definecolor{currentfill}{rgb}{0.666667,0.200000,0.466667}%
\pgfsetfillcolor{currentfill}%
\pgfsetlinewidth{2.007500pt}%
\definecolor{currentstroke}{rgb}{0.666667,0.200000,0.466667}%
\pgfsetstrokecolor{currentstroke}%
\pgfsetdash{}{0pt}%
\pgfsys@defobject{currentmarker}{\pgfqpoint{-0.055556in}{-0.000000in}}{\pgfqpoint{0.055556in}{0.000000in}}{%
\pgfpathmoveto{\pgfqpoint{0.055556in}{-0.000000in}}%
\pgfpathlineto{\pgfqpoint{-0.055556in}{0.000000in}}%
\pgfusepath{stroke,fill}%
}%
\begin{pgfscope}%
\pgfsys@transformshift{1.012358in}{0.835171in}%
\pgfsys@useobject{currentmarker}{}%
\end{pgfscope}%
\begin{pgfscope}%
\pgfsys@transformshift{1.877795in}{0.836977in}%
\pgfsys@useobject{currentmarker}{}%
\end{pgfscope}%
\begin{pgfscope}%
\pgfsys@transformshift{2.743232in}{0.838664in}%
\pgfsys@useobject{currentmarker}{}%
\end{pgfscope}%
\begin{pgfscope}%
\pgfsys@transformshift{3.608670in}{0.842926in}%
\pgfsys@useobject{currentmarker}{}%
\end{pgfscope}%
\begin{pgfscope}%
\pgfsys@transformshift{4.474107in}{0.846149in}%
\pgfsys@useobject{currentmarker}{}%
\end{pgfscope}%
\end{pgfscope}%
\begin{pgfscope}%
\pgfpathrectangle{\pgfqpoint{0.839270in}{0.831623in}}{\pgfqpoint{3.807924in}{2.711777in}}%
\pgfusepath{clip}%
\pgfsetbuttcap%
\pgfsetroundjoin%
\definecolor{currentfill}{rgb}{0.666667,0.200000,0.466667}%
\pgfsetfillcolor{currentfill}%
\pgfsetlinewidth{2.007500pt}%
\definecolor{currentstroke}{rgb}{0.666667,0.200000,0.466667}%
\pgfsetstrokecolor{currentstroke}%
\pgfsetdash{}{0pt}%
\pgfsys@defobject{currentmarker}{\pgfqpoint{-0.055556in}{-0.000000in}}{\pgfqpoint{0.055556in}{0.000000in}}{%
\pgfpathmoveto{\pgfqpoint{0.055556in}{-0.000000in}}%
\pgfpathlineto{\pgfqpoint{-0.055556in}{0.000000in}}%
\pgfusepath{stroke,fill}%
}%
\begin{pgfscope}%
\pgfsys@transformshift{1.012358in}{0.837058in}%
\pgfsys@useobject{currentmarker}{}%
\end{pgfscope}%
\begin{pgfscope}%
\pgfsys@transformshift{1.877795in}{0.838981in}%
\pgfsys@useobject{currentmarker}{}%
\end{pgfscope}%
\begin{pgfscope}%
\pgfsys@transformshift{2.743232in}{0.841942in}%
\pgfsys@useobject{currentmarker}{}%
\end{pgfscope}%
\begin{pgfscope}%
\pgfsys@transformshift{3.608670in}{0.846899in}%
\pgfsys@useobject{currentmarker}{}%
\end{pgfscope}%
\begin{pgfscope}%
\pgfsys@transformshift{4.474107in}{0.854164in}%
\pgfsys@useobject{currentmarker}{}%
\end{pgfscope}%
\end{pgfscope}%
\begin{pgfscope}%
\pgfpathrectangle{\pgfqpoint{0.839270in}{0.831623in}}{\pgfqpoint{3.807924in}{2.711777in}}%
\pgfusepath{clip}%
\pgfsetbuttcap%
\pgfsetroundjoin%
\pgfsetlinewidth{3.513125pt}%
\definecolor{currentstroke}{rgb}{0.133333,0.533333,0.200000}%
\pgfsetstrokecolor{currentstroke}%
\pgfsetdash{}{0pt}%
\pgfpathmoveto{\pgfqpoint{1.012358in}{1.102973in}}%
\pgfpathlineto{\pgfqpoint{1.012358in}{1.189291in}}%
\pgfusepath{stroke}%
\end{pgfscope}%
\begin{pgfscope}%
\pgfpathrectangle{\pgfqpoint{0.839270in}{0.831623in}}{\pgfqpoint{3.807924in}{2.711777in}}%
\pgfusepath{clip}%
\pgfsetbuttcap%
\pgfsetroundjoin%
\pgfsetlinewidth{3.513125pt}%
\definecolor{currentstroke}{rgb}{0.133333,0.533333,0.200000}%
\pgfsetstrokecolor{currentstroke}%
\pgfsetdash{}{0pt}%
\pgfpathmoveto{\pgfqpoint{1.877795in}{1.282290in}}%
\pgfpathlineto{\pgfqpoint{1.877795in}{1.447106in}}%
\pgfusepath{stroke}%
\end{pgfscope}%
\begin{pgfscope}%
\pgfpathrectangle{\pgfqpoint{0.839270in}{0.831623in}}{\pgfqpoint{3.807924in}{2.711777in}}%
\pgfusepath{clip}%
\pgfsetbuttcap%
\pgfsetroundjoin%
\pgfsetlinewidth{3.513125pt}%
\definecolor{currentstroke}{rgb}{0.133333,0.533333,0.200000}%
\pgfsetstrokecolor{currentstroke}%
\pgfsetdash{}{0pt}%
\pgfpathmoveto{\pgfqpoint{2.743232in}{1.422936in}}%
\pgfpathlineto{\pgfqpoint{2.743232in}{1.543820in}}%
\pgfusepath{stroke}%
\end{pgfscope}%
\begin{pgfscope}%
\pgfpathrectangle{\pgfqpoint{0.839270in}{0.831623in}}{\pgfqpoint{3.807924in}{2.711777in}}%
\pgfusepath{clip}%
\pgfsetbuttcap%
\pgfsetroundjoin%
\pgfsetlinewidth{3.513125pt}%
\definecolor{currentstroke}{rgb}{0.133333,0.533333,0.200000}%
\pgfsetstrokecolor{currentstroke}%
\pgfsetdash{}{0pt}%
\pgfpathmoveto{\pgfqpoint{3.608670in}{1.873454in}}%
\pgfpathlineto{\pgfqpoint{3.608670in}{2.065841in}}%
\pgfusepath{stroke}%
\end{pgfscope}%
\begin{pgfscope}%
\pgfpathrectangle{\pgfqpoint{0.839270in}{0.831623in}}{\pgfqpoint{3.807924in}{2.711777in}}%
\pgfusepath{clip}%
\pgfsetbuttcap%
\pgfsetroundjoin%
\pgfsetlinewidth{3.513125pt}%
\definecolor{currentstroke}{rgb}{0.133333,0.533333,0.200000}%
\pgfsetstrokecolor{currentstroke}%
\pgfsetdash{}{0pt}%
\pgfpathmoveto{\pgfqpoint{4.474107in}{2.035513in}}%
\pgfpathlineto{\pgfqpoint{4.474107in}{2.295317in}}%
\pgfusepath{stroke}%
\end{pgfscope}%
\begin{pgfscope}%
\pgfpathrectangle{\pgfqpoint{0.839270in}{0.831623in}}{\pgfqpoint{3.807924in}{2.711777in}}%
\pgfusepath{clip}%
\pgfsetbuttcap%
\pgfsetroundjoin%
\definecolor{currentfill}{rgb}{0.133333,0.533333,0.200000}%
\pgfsetfillcolor{currentfill}%
\pgfsetlinewidth{2.007500pt}%
\definecolor{currentstroke}{rgb}{0.133333,0.533333,0.200000}%
\pgfsetstrokecolor{currentstroke}%
\pgfsetdash{}{0pt}%
\pgfsys@defobject{currentmarker}{\pgfqpoint{-0.055556in}{-0.000000in}}{\pgfqpoint{0.055556in}{0.000000in}}{%
\pgfpathmoveto{\pgfqpoint{0.055556in}{-0.000000in}}%
\pgfpathlineto{\pgfqpoint{-0.055556in}{0.000000in}}%
\pgfusepath{stroke,fill}%
}%
\begin{pgfscope}%
\pgfsys@transformshift{1.012358in}{1.102973in}%
\pgfsys@useobject{currentmarker}{}%
\end{pgfscope}%
\begin{pgfscope}%
\pgfsys@transformshift{1.877795in}{1.282290in}%
\pgfsys@useobject{currentmarker}{}%
\end{pgfscope}%
\begin{pgfscope}%
\pgfsys@transformshift{2.743232in}{1.422936in}%
\pgfsys@useobject{currentmarker}{}%
\end{pgfscope}%
\begin{pgfscope}%
\pgfsys@transformshift{3.608670in}{1.873454in}%
\pgfsys@useobject{currentmarker}{}%
\end{pgfscope}%
\begin{pgfscope}%
\pgfsys@transformshift{4.474107in}{2.035513in}%
\pgfsys@useobject{currentmarker}{}%
\end{pgfscope}%
\end{pgfscope}%
\begin{pgfscope}%
\pgfpathrectangle{\pgfqpoint{0.839270in}{0.831623in}}{\pgfqpoint{3.807924in}{2.711777in}}%
\pgfusepath{clip}%
\pgfsetbuttcap%
\pgfsetroundjoin%
\definecolor{currentfill}{rgb}{0.133333,0.533333,0.200000}%
\pgfsetfillcolor{currentfill}%
\pgfsetlinewidth{2.007500pt}%
\definecolor{currentstroke}{rgb}{0.133333,0.533333,0.200000}%
\pgfsetstrokecolor{currentstroke}%
\pgfsetdash{}{0pt}%
\pgfsys@defobject{currentmarker}{\pgfqpoint{-0.055556in}{-0.000000in}}{\pgfqpoint{0.055556in}{0.000000in}}{%
\pgfpathmoveto{\pgfqpoint{0.055556in}{-0.000000in}}%
\pgfpathlineto{\pgfqpoint{-0.055556in}{0.000000in}}%
\pgfusepath{stroke,fill}%
}%
\begin{pgfscope}%
\pgfsys@transformshift{1.012358in}{1.189291in}%
\pgfsys@useobject{currentmarker}{}%
\end{pgfscope}%
\begin{pgfscope}%
\pgfsys@transformshift{1.877795in}{1.447106in}%
\pgfsys@useobject{currentmarker}{}%
\end{pgfscope}%
\begin{pgfscope}%
\pgfsys@transformshift{2.743232in}{1.543820in}%
\pgfsys@useobject{currentmarker}{}%
\end{pgfscope}%
\begin{pgfscope}%
\pgfsys@transformshift{3.608670in}{2.065841in}%
\pgfsys@useobject{currentmarker}{}%
\end{pgfscope}%
\begin{pgfscope}%
\pgfsys@transformshift{4.474107in}{2.295317in}%
\pgfsys@useobject{currentmarker}{}%
\end{pgfscope}%
\end{pgfscope}%
\begin{pgfscope}%
\pgfpathrectangle{\pgfqpoint{0.839270in}{0.831623in}}{\pgfqpoint{3.807924in}{2.711777in}}%
\pgfusepath{clip}%
\pgfsetbuttcap%
\pgfsetroundjoin%
\pgfsetlinewidth{3.513125pt}%
\definecolor{currentstroke}{rgb}{0.000000,0.000000,0.000000}%
\pgfsetstrokecolor{currentstroke}%
\pgfsetdash{}{0pt}%
\pgfpathmoveto{\pgfqpoint{1.012358in}{0.834485in}}%
\pgfpathlineto{\pgfqpoint{1.012358in}{0.837026in}}%
\pgfusepath{stroke}%
\end{pgfscope}%
\begin{pgfscope}%
\pgfpathrectangle{\pgfqpoint{0.839270in}{0.831623in}}{\pgfqpoint{3.807924in}{2.711777in}}%
\pgfusepath{clip}%
\pgfsetbuttcap%
\pgfsetroundjoin%
\pgfsetlinewidth{3.513125pt}%
\definecolor{currentstroke}{rgb}{0.000000,0.000000,0.000000}%
\pgfsetstrokecolor{currentstroke}%
\pgfsetdash{}{0pt}%
\pgfpathmoveto{\pgfqpoint{1.877795in}{0.837380in}}%
\pgfpathlineto{\pgfqpoint{1.877795in}{0.841983in}}%
\pgfusepath{stroke}%
\end{pgfscope}%
\begin{pgfscope}%
\pgfpathrectangle{\pgfqpoint{0.839270in}{0.831623in}}{\pgfqpoint{3.807924in}{2.711777in}}%
\pgfusepath{clip}%
\pgfsetbuttcap%
\pgfsetroundjoin%
\pgfsetlinewidth{3.513125pt}%
\definecolor{currentstroke}{rgb}{0.000000,0.000000,0.000000}%
\pgfsetstrokecolor{currentstroke}%
\pgfsetdash{}{0pt}%
\pgfpathmoveto{\pgfqpoint{2.743232in}{0.841298in}}%
\pgfpathlineto{\pgfqpoint{2.743232in}{0.847052in}}%
\pgfusepath{stroke}%
\end{pgfscope}%
\begin{pgfscope}%
\pgfpathrectangle{\pgfqpoint{0.839270in}{0.831623in}}{\pgfqpoint{3.807924in}{2.711777in}}%
\pgfusepath{clip}%
\pgfsetbuttcap%
\pgfsetroundjoin%
\pgfsetlinewidth{3.513125pt}%
\definecolor{currentstroke}{rgb}{0.000000,0.000000,0.000000}%
\pgfsetstrokecolor{currentstroke}%
\pgfsetdash{}{0pt}%
\pgfpathmoveto{\pgfqpoint{3.608670in}{0.849253in}}%
\pgfpathlineto{\pgfqpoint{3.608670in}{0.857702in}}%
\pgfusepath{stroke}%
\end{pgfscope}%
\begin{pgfscope}%
\pgfpathrectangle{\pgfqpoint{0.839270in}{0.831623in}}{\pgfqpoint{3.807924in}{2.711777in}}%
\pgfusepath{clip}%
\pgfsetbuttcap%
\pgfsetroundjoin%
\pgfsetlinewidth{3.513125pt}%
\definecolor{currentstroke}{rgb}{0.000000,0.000000,0.000000}%
\pgfsetstrokecolor{currentstroke}%
\pgfsetdash{}{0pt}%
\pgfpathmoveto{\pgfqpoint{4.474107in}{0.852834in}}%
\pgfpathlineto{\pgfqpoint{4.474107in}{0.867224in}}%
\pgfusepath{stroke}%
\end{pgfscope}%
\begin{pgfscope}%
\pgfpathrectangle{\pgfqpoint{0.839270in}{0.831623in}}{\pgfqpoint{3.807924in}{2.711777in}}%
\pgfusepath{clip}%
\pgfsetbuttcap%
\pgfsetroundjoin%
\definecolor{currentfill}{rgb}{0.000000,0.000000,0.000000}%
\pgfsetfillcolor{currentfill}%
\pgfsetlinewidth{2.007500pt}%
\definecolor{currentstroke}{rgb}{0.000000,0.000000,0.000000}%
\pgfsetstrokecolor{currentstroke}%
\pgfsetdash{}{0pt}%
\pgfsys@defobject{currentmarker}{\pgfqpoint{-0.055556in}{-0.000000in}}{\pgfqpoint{0.055556in}{0.000000in}}{%
\pgfpathmoveto{\pgfqpoint{0.055556in}{-0.000000in}}%
\pgfpathlineto{\pgfqpoint{-0.055556in}{0.000000in}}%
\pgfusepath{stroke,fill}%
}%
\begin{pgfscope}%
\pgfsys@transformshift{1.012358in}{0.834485in}%
\pgfsys@useobject{currentmarker}{}%
\end{pgfscope}%
\begin{pgfscope}%
\pgfsys@transformshift{1.877795in}{0.837380in}%
\pgfsys@useobject{currentmarker}{}%
\end{pgfscope}%
\begin{pgfscope}%
\pgfsys@transformshift{2.743232in}{0.841298in}%
\pgfsys@useobject{currentmarker}{}%
\end{pgfscope}%
\begin{pgfscope}%
\pgfsys@transformshift{3.608670in}{0.849253in}%
\pgfsys@useobject{currentmarker}{}%
\end{pgfscope}%
\begin{pgfscope}%
\pgfsys@transformshift{4.474107in}{0.852834in}%
\pgfsys@useobject{currentmarker}{}%
\end{pgfscope}%
\end{pgfscope}%
\begin{pgfscope}%
\pgfpathrectangle{\pgfqpoint{0.839270in}{0.831623in}}{\pgfqpoint{3.807924in}{2.711777in}}%
\pgfusepath{clip}%
\pgfsetbuttcap%
\pgfsetroundjoin%
\definecolor{currentfill}{rgb}{0.000000,0.000000,0.000000}%
\pgfsetfillcolor{currentfill}%
\pgfsetlinewidth{2.007500pt}%
\definecolor{currentstroke}{rgb}{0.000000,0.000000,0.000000}%
\pgfsetstrokecolor{currentstroke}%
\pgfsetdash{}{0pt}%
\pgfsys@defobject{currentmarker}{\pgfqpoint{-0.055556in}{-0.000000in}}{\pgfqpoint{0.055556in}{0.000000in}}{%
\pgfpathmoveto{\pgfqpoint{0.055556in}{-0.000000in}}%
\pgfpathlineto{\pgfqpoint{-0.055556in}{0.000000in}}%
\pgfusepath{stroke,fill}%
}%
\begin{pgfscope}%
\pgfsys@transformshift{1.012358in}{0.837026in}%
\pgfsys@useobject{currentmarker}{}%
\end{pgfscope}%
\begin{pgfscope}%
\pgfsys@transformshift{1.877795in}{0.841983in}%
\pgfsys@useobject{currentmarker}{}%
\end{pgfscope}%
\begin{pgfscope}%
\pgfsys@transformshift{2.743232in}{0.847052in}%
\pgfsys@useobject{currentmarker}{}%
\end{pgfscope}%
\begin{pgfscope}%
\pgfsys@transformshift{3.608670in}{0.857702in}%
\pgfsys@useobject{currentmarker}{}%
\end{pgfscope}%
\begin{pgfscope}%
\pgfsys@transformshift{4.474107in}{0.867224in}%
\pgfsys@useobject{currentmarker}{}%
\end{pgfscope}%
\end{pgfscope}%
\begin{pgfscope}%
\pgfpathrectangle{\pgfqpoint{0.839270in}{0.831623in}}{\pgfqpoint{3.807924in}{2.711777in}}%
\pgfusepath{clip}%
\pgfsetrectcap%
\pgfsetroundjoin%
\pgfsetlinewidth{3.513125pt}%
\definecolor{currentstroke}{rgb}{0.666667,0.200000,0.466667}%
\pgfsetstrokecolor{currentstroke}%
\pgfsetdash{}{0pt}%
\pgfpathmoveto{\pgfqpoint{1.012358in}{0.836645in}}%
\pgfpathlineto{\pgfqpoint{1.877795in}{0.838914in}}%
\pgfpathlineto{\pgfqpoint{2.743232in}{0.840943in}}%
\pgfpathlineto{\pgfqpoint{3.608670in}{0.845901in}}%
\pgfpathlineto{\pgfqpoint{4.474107in}{0.849704in}}%
\pgfusepath{stroke}%
\end{pgfscope}%
\begin{pgfscope}%
\pgfpathrectangle{\pgfqpoint{0.839270in}{0.831623in}}{\pgfqpoint{3.807924in}{2.711777in}}%
\pgfusepath{clip}%
\pgfsetbuttcap%
\pgfsetroundjoin%
\definecolor{currentfill}{rgb}{0.666667,0.200000,0.466667}%
\pgfsetfillcolor{currentfill}%
\pgfsetlinewidth{2.007500pt}%
\definecolor{currentstroke}{rgb}{1.000000,1.000000,1.000000}%
\pgfsetstrokecolor{currentstroke}%
\pgfsetdash{}{0pt}%
\pgfsys@defobject{currentmarker}{\pgfqpoint{-0.055556in}{-0.055556in}}{\pgfqpoint{0.055556in}{0.055556in}}{%
\pgfpathmoveto{\pgfqpoint{0.000000in}{-0.055556in}}%
\pgfpathcurveto{\pgfqpoint{0.014734in}{-0.055556in}}{\pgfqpoint{0.028866in}{-0.049702in}}{\pgfqpoint{0.039284in}{-0.039284in}}%
\pgfpathcurveto{\pgfqpoint{0.049702in}{-0.028866in}}{\pgfqpoint{0.055556in}{-0.014734in}}{\pgfqpoint{0.055556in}{0.000000in}}%
\pgfpathcurveto{\pgfqpoint{0.055556in}{0.014734in}}{\pgfqpoint{0.049702in}{0.028866in}}{\pgfqpoint{0.039284in}{0.039284in}}%
\pgfpathcurveto{\pgfqpoint{0.028866in}{0.049702in}}{\pgfqpoint{0.014734in}{0.055556in}}{\pgfqpoint{0.000000in}{0.055556in}}%
\pgfpathcurveto{\pgfqpoint{-0.014734in}{0.055556in}}{\pgfqpoint{-0.028866in}{0.049702in}}{\pgfqpoint{-0.039284in}{0.039284in}}%
\pgfpathcurveto{\pgfqpoint{-0.049702in}{0.028866in}}{\pgfqpoint{-0.055556in}{0.014734in}}{\pgfqpoint{-0.055556in}{0.000000in}}%
\pgfpathcurveto{\pgfqpoint{-0.055556in}{-0.014734in}}{\pgfqpoint{-0.049702in}{-0.028866in}}{\pgfqpoint{-0.039284in}{-0.039284in}}%
\pgfpathcurveto{\pgfqpoint{-0.028866in}{-0.049702in}}{\pgfqpoint{-0.014734in}{-0.055556in}}{\pgfqpoint{0.000000in}{-0.055556in}}%
\pgfpathlineto{\pgfqpoint{0.000000in}{-0.055556in}}%
\pgfpathclose%
\pgfusepath{stroke,fill}%
}%
\begin{pgfscope}%
\pgfsys@transformshift{1.012358in}{0.836645in}%
\pgfsys@useobject{currentmarker}{}%
\end{pgfscope}%
\begin{pgfscope}%
\pgfsys@transformshift{1.877795in}{0.838914in}%
\pgfsys@useobject{currentmarker}{}%
\end{pgfscope}%
\begin{pgfscope}%
\pgfsys@transformshift{2.743232in}{0.840943in}%
\pgfsys@useobject{currentmarker}{}%
\end{pgfscope}%
\begin{pgfscope}%
\pgfsys@transformshift{3.608670in}{0.845901in}%
\pgfsys@useobject{currentmarker}{}%
\end{pgfscope}%
\begin{pgfscope}%
\pgfsys@transformshift{4.474107in}{0.849704in}%
\pgfsys@useobject{currentmarker}{}%
\end{pgfscope}%
\end{pgfscope}%
\begin{pgfscope}%
\pgfpathrectangle{\pgfqpoint{0.839270in}{0.831623in}}{\pgfqpoint{3.807924in}{2.711777in}}%
\pgfusepath{clip}%
\pgfsetbuttcap%
\pgfsetroundjoin%
\pgfsetlinewidth{3.513125pt}%
\definecolor{currentstroke}{rgb}{0.133333,0.533333,0.200000}%
\pgfsetstrokecolor{currentstroke}%
\pgfsetdash{{12.950000pt}{5.600000pt}}{0.000000pt}%
\pgfpathmoveto{\pgfqpoint{1.012358in}{1.145601in}}%
\pgfpathlineto{\pgfqpoint{1.877795in}{1.359634in}}%
\pgfpathlineto{\pgfqpoint{2.743232in}{1.499520in}}%
\pgfpathlineto{\pgfqpoint{3.608670in}{1.957909in}}%
\pgfpathlineto{\pgfqpoint{4.474107in}{2.184674in}}%
\pgfusepath{stroke}%
\end{pgfscope}%
\begin{pgfscope}%
\pgfpathrectangle{\pgfqpoint{0.839270in}{0.831623in}}{\pgfqpoint{3.807924in}{2.711777in}}%
\pgfusepath{clip}%
\pgfsetbuttcap%
\pgfsetmiterjoin%
\definecolor{currentfill}{rgb}{0.133333,0.533333,0.200000}%
\pgfsetfillcolor{currentfill}%
\pgfsetlinewidth{2.007500pt}%
\definecolor{currentstroke}{rgb}{1.000000,1.000000,1.000000}%
\pgfsetstrokecolor{currentstroke}%
\pgfsetdash{}{0pt}%
\pgfsys@defobject{currentmarker}{\pgfqpoint{-0.055556in}{-0.055556in}}{\pgfqpoint{0.055556in}{0.055556in}}{%
\pgfpathmoveto{\pgfqpoint{-0.055556in}{-0.055556in}}%
\pgfpathlineto{\pgfqpoint{0.055556in}{-0.055556in}}%
\pgfpathlineto{\pgfqpoint{0.055556in}{0.055556in}}%
\pgfpathlineto{\pgfqpoint{-0.055556in}{0.055556in}}%
\pgfpathlineto{\pgfqpoint{-0.055556in}{-0.055556in}}%
\pgfpathclose%
\pgfusepath{stroke,fill}%
}%
\begin{pgfscope}%
\pgfsys@transformshift{1.012358in}{1.145601in}%
\pgfsys@useobject{currentmarker}{}%
\end{pgfscope}%
\begin{pgfscope}%
\pgfsys@transformshift{1.877795in}{1.359634in}%
\pgfsys@useobject{currentmarker}{}%
\end{pgfscope}%
\begin{pgfscope}%
\pgfsys@transformshift{2.743232in}{1.499520in}%
\pgfsys@useobject{currentmarker}{}%
\end{pgfscope}%
\begin{pgfscope}%
\pgfsys@transformshift{3.608670in}{1.957909in}%
\pgfsys@useobject{currentmarker}{}%
\end{pgfscope}%
\begin{pgfscope}%
\pgfsys@transformshift{4.474107in}{2.184674in}%
\pgfsys@useobject{currentmarker}{}%
\end{pgfscope}%
\end{pgfscope}%
\begin{pgfscope}%
\pgfpathrectangle{\pgfqpoint{0.839270in}{0.831623in}}{\pgfqpoint{3.807924in}{2.711777in}}%
\pgfusepath{clip}%
\pgfsetbuttcap%
\pgfsetroundjoin%
\pgfsetlinewidth{3.513125pt}%
\definecolor{currentstroke}{rgb}{0.000000,0.000000,0.000000}%
\pgfsetstrokecolor{currentstroke}%
\pgfsetdash{{22.400000pt}{5.600000pt}{3.500000pt}{5.600000pt}}{0.000000pt}%
\pgfpathmoveto{\pgfqpoint{1.012358in}{0.836021in}}%
\pgfpathlineto{\pgfqpoint{1.877795in}{0.839867in}}%
\pgfpathlineto{\pgfqpoint{2.743232in}{0.845138in}}%
\pgfpathlineto{\pgfqpoint{3.608670in}{0.854180in}}%
\pgfpathlineto{\pgfqpoint{4.474107in}{0.860893in}}%
\pgfusepath{stroke}%
\end{pgfscope}%
\begin{pgfscope}%
\pgfpathrectangle{\pgfqpoint{0.839270in}{0.831623in}}{\pgfqpoint{3.807924in}{2.711777in}}%
\pgfusepath{clip}%
\pgfsetbuttcap%
\pgfsetmiterjoin%
\definecolor{currentfill}{rgb}{0.000000,0.000000,0.000000}%
\pgfsetfillcolor{currentfill}%
\pgfsetlinewidth{2.007500pt}%
\definecolor{currentstroke}{rgb}{1.000000,1.000000,1.000000}%
\pgfsetstrokecolor{currentstroke}%
\pgfsetdash{}{0pt}%
\pgfsys@defobject{currentmarker}{\pgfqpoint{-0.078567in}{-0.078567in}}{\pgfqpoint{0.078567in}{0.078567in}}{%
\pgfpathmoveto{\pgfqpoint{-0.000000in}{-0.078567in}}%
\pgfpathlineto{\pgfqpoint{0.078567in}{0.000000in}}%
\pgfpathlineto{\pgfqpoint{0.000000in}{0.078567in}}%
\pgfpathlineto{\pgfqpoint{-0.078567in}{0.000000in}}%
\pgfpathlineto{\pgfqpoint{-0.000000in}{-0.078567in}}%
\pgfpathclose%
\pgfusepath{stroke,fill}%
}%
\begin{pgfscope}%
\pgfsys@transformshift{1.012358in}{0.836021in}%
\pgfsys@useobject{currentmarker}{}%
\end{pgfscope}%
\begin{pgfscope}%
\pgfsys@transformshift{1.877795in}{0.839867in}%
\pgfsys@useobject{currentmarker}{}%
\end{pgfscope}%
\begin{pgfscope}%
\pgfsys@transformshift{2.743232in}{0.845138in}%
\pgfsys@useobject{currentmarker}{}%
\end{pgfscope}%
\begin{pgfscope}%
\pgfsys@transformshift{3.608670in}{0.854180in}%
\pgfsys@useobject{currentmarker}{}%
\end{pgfscope}%
\begin{pgfscope}%
\pgfsys@transformshift{4.474107in}{0.860893in}%
\pgfsys@useobject{currentmarker}{}%
\end{pgfscope}%
\end{pgfscope}%
\begin{pgfscope}%
\pgfsetrectcap%
\pgfsetmiterjoin%
\pgfsetlinewidth{1.505625pt}%
\definecolor{currentstroke}{rgb}{0.000000,0.000000,0.000000}%
\pgfsetstrokecolor{currentstroke}%
\pgfsetdash{}{0pt}%
\pgfpathmoveto{\pgfqpoint{0.839270in}{0.831623in}}%
\pgfpathlineto{\pgfqpoint{0.839270in}{3.543400in}}%
\pgfusepath{stroke}%
\end{pgfscope}%
\begin{pgfscope}%
\pgfsetrectcap%
\pgfsetmiterjoin%
\pgfsetlinewidth{1.505625pt}%
\definecolor{currentstroke}{rgb}{0.000000,0.000000,0.000000}%
\pgfsetstrokecolor{currentstroke}%
\pgfsetdash{}{0pt}%
\pgfpathmoveto{\pgfqpoint{0.839270in}{0.831623in}}%
\pgfpathlineto{\pgfqpoint{4.647194in}{0.831623in}}%
\pgfusepath{stroke}%
\end{pgfscope}%
\begin{pgfscope}%
\pgfsetbuttcap%
\pgfsetmiterjoin%
\definecolor{currentfill}{rgb}{1.000000,1.000000,1.000000}%
\pgfsetfillcolor{currentfill}%
\pgfsetfillopacity{0.400000}%
\pgfsetlinewidth{1.003750pt}%
\definecolor{currentstroke}{rgb}{0.000000,0.000000,0.000000}%
\pgfsetstrokecolor{currentstroke}%
\pgfsetstrokeopacity{0.400000}%
\pgfsetdash{}{0pt}%
\pgfpathmoveto{\pgfqpoint{1.033715in}{2.136309in}}%
\pgfpathlineto{\pgfqpoint{3.019149in}{2.136309in}}%
\pgfpathquadraticcurveto{\pgfqpoint{3.074705in}{2.136309in}}{\pgfqpoint{3.074705in}{2.191864in}}%
\pgfpathlineto{\pgfqpoint{3.074705in}{3.348956in}}%
\pgfpathquadraticcurveto{\pgfqpoint{3.074705in}{3.404512in}}{\pgfqpoint{3.019149in}{3.404512in}}%
\pgfpathlineto{\pgfqpoint{1.033715in}{3.404512in}}%
\pgfpathquadraticcurveto{\pgfqpoint{0.978159in}{3.404512in}}{\pgfqpoint{0.978159in}{3.348956in}}%
\pgfpathlineto{\pgfqpoint{0.978159in}{2.191864in}}%
\pgfpathquadraticcurveto{\pgfqpoint{0.978159in}{2.136309in}}{\pgfqpoint{1.033715in}{2.136309in}}%
\pgfpathlineto{\pgfqpoint{1.033715in}{2.136309in}}%
\pgfpathclose%
\pgfusepath{stroke,fill}%
\end{pgfscope}%
\begin{pgfscope}%
\pgfsetbuttcap%
\pgfsetroundjoin%
\pgfsetlinewidth{3.513125pt}%
\definecolor{currentstroke}{rgb}{0.666667,0.200000,0.466667}%
\pgfsetstrokecolor{currentstroke}%
\pgfsetdash{}{0pt}%
\pgfpathmoveto{\pgfqpoint{1.367048in}{3.051695in}}%
\pgfpathlineto{\pgfqpoint{1.367048in}{3.329473in}}%
\pgfusepath{stroke}%
\end{pgfscope}%
\begin{pgfscope}%
\pgfsetbuttcap%
\pgfsetroundjoin%
\definecolor{currentfill}{rgb}{0.666667,0.200000,0.466667}%
\pgfsetfillcolor{currentfill}%
\pgfsetlinewidth{2.007500pt}%
\definecolor{currentstroke}{rgb}{0.666667,0.200000,0.466667}%
\pgfsetstrokecolor{currentstroke}%
\pgfsetdash{}{0pt}%
\pgfsys@defobject{currentmarker}{\pgfqpoint{-0.055556in}{-0.000000in}}{\pgfqpoint{0.055556in}{0.000000in}}{%
\pgfpathmoveto{\pgfqpoint{0.055556in}{-0.000000in}}%
\pgfpathlineto{\pgfqpoint{-0.055556in}{0.000000in}}%
\pgfusepath{stroke,fill}%
}%
\begin{pgfscope}%
\pgfsys@transformshift{1.367048in}{3.051695in}%
\pgfsys@useobject{currentmarker}{}%
\end{pgfscope}%
\end{pgfscope}%
\begin{pgfscope}%
\pgfsetbuttcap%
\pgfsetroundjoin%
\definecolor{currentfill}{rgb}{0.666667,0.200000,0.466667}%
\pgfsetfillcolor{currentfill}%
\pgfsetlinewidth{2.007500pt}%
\definecolor{currentstroke}{rgb}{0.666667,0.200000,0.466667}%
\pgfsetstrokecolor{currentstroke}%
\pgfsetdash{}{0pt}%
\pgfsys@defobject{currentmarker}{\pgfqpoint{-0.055556in}{-0.000000in}}{\pgfqpoint{0.055556in}{0.000000in}}{%
\pgfpathmoveto{\pgfqpoint{0.055556in}{-0.000000in}}%
\pgfpathlineto{\pgfqpoint{-0.055556in}{0.000000in}}%
\pgfusepath{stroke,fill}%
}%
\begin{pgfscope}%
\pgfsys@transformshift{1.367048in}{3.329473in}%
\pgfsys@useobject{currentmarker}{}%
\end{pgfscope}%
\end{pgfscope}%
\begin{pgfscope}%
\pgfsetrectcap%
\pgfsetroundjoin%
\pgfsetlinewidth{3.513125pt}%
\definecolor{currentstroke}{rgb}{0.666667,0.200000,0.466667}%
\pgfsetstrokecolor{currentstroke}%
\pgfsetdash{}{0pt}%
\pgfpathmoveto{\pgfqpoint{1.089270in}{3.190584in}}%
\pgfpathlineto{\pgfqpoint{1.644826in}{3.190584in}}%
\pgfusepath{stroke}%
\end{pgfscope}%
\begin{pgfscope}%
\pgfsetbuttcap%
\pgfsetroundjoin%
\definecolor{currentfill}{rgb}{0.666667,0.200000,0.466667}%
\pgfsetfillcolor{currentfill}%
\pgfsetlinewidth{2.007500pt}%
\definecolor{currentstroke}{rgb}{1.000000,1.000000,1.000000}%
\pgfsetstrokecolor{currentstroke}%
\pgfsetdash{}{0pt}%
\pgfsys@defobject{currentmarker}{\pgfqpoint{-0.055556in}{-0.055556in}}{\pgfqpoint{0.055556in}{0.055556in}}{%
\pgfpathmoveto{\pgfqpoint{0.000000in}{-0.055556in}}%
\pgfpathcurveto{\pgfqpoint{0.014734in}{-0.055556in}}{\pgfqpoint{0.028866in}{-0.049702in}}{\pgfqpoint{0.039284in}{-0.039284in}}%
\pgfpathcurveto{\pgfqpoint{0.049702in}{-0.028866in}}{\pgfqpoint{0.055556in}{-0.014734in}}{\pgfqpoint{0.055556in}{0.000000in}}%
\pgfpathcurveto{\pgfqpoint{0.055556in}{0.014734in}}{\pgfqpoint{0.049702in}{0.028866in}}{\pgfqpoint{0.039284in}{0.039284in}}%
\pgfpathcurveto{\pgfqpoint{0.028866in}{0.049702in}}{\pgfqpoint{0.014734in}{0.055556in}}{\pgfqpoint{0.000000in}{0.055556in}}%
\pgfpathcurveto{\pgfqpoint{-0.014734in}{0.055556in}}{\pgfqpoint{-0.028866in}{0.049702in}}{\pgfqpoint{-0.039284in}{0.039284in}}%
\pgfpathcurveto{\pgfqpoint{-0.049702in}{0.028866in}}{\pgfqpoint{-0.055556in}{0.014734in}}{\pgfqpoint{-0.055556in}{0.000000in}}%
\pgfpathcurveto{\pgfqpoint{-0.055556in}{-0.014734in}}{\pgfqpoint{-0.049702in}{-0.028866in}}{\pgfqpoint{-0.039284in}{-0.039284in}}%
\pgfpathcurveto{\pgfqpoint{-0.028866in}{-0.049702in}}{\pgfqpoint{-0.014734in}{-0.055556in}}{\pgfqpoint{0.000000in}{-0.055556in}}%
\pgfpathlineto{\pgfqpoint{0.000000in}{-0.055556in}}%
\pgfpathclose%
\pgfusepath{stroke,fill}%
}%
\begin{pgfscope}%
\pgfsys@transformshift{1.367048in}{3.190584in}%
\pgfsys@useobject{currentmarker}{}%
\end{pgfscope}%
\end{pgfscope}%
\begin{pgfscope}%
\definecolor{textcolor}{rgb}{0.000000,0.000000,0.000000}%
\pgfsetstrokecolor{textcolor}%
\pgfsetfillcolor{textcolor}%
\pgftext[x=1.867048in,y=3.093362in,left,base]{\color{textcolor}{\rmfamily\fontsize{20.000000}{24.000000}\selectfont\catcode`\^=\active\def^{\ifmmode\sp\else\^{}\fi}\catcode`\%=\active\def
\end{pgfscope}%
\begin{pgfscope}%
\pgfsetbuttcap%
\pgfsetroundjoin%
\pgfsetlinewidth{3.513125pt}%
\definecolor{currentstroke}{rgb}{0.133333,0.533333,0.200000}%
\pgfsetstrokecolor{currentstroke}%
\pgfsetdash{}{0pt}%
\pgfpathmoveto{\pgfqpoint{1.367048in}{2.656739in}}%
\pgfpathlineto{\pgfqpoint{1.367048in}{2.934517in}}%
\pgfusepath{stroke}%
\end{pgfscope}%
\begin{pgfscope}%
\pgfsetbuttcap%
\pgfsetroundjoin%
\definecolor{currentfill}{rgb}{0.133333,0.533333,0.200000}%
\pgfsetfillcolor{currentfill}%
\pgfsetlinewidth{2.007500pt}%
\definecolor{currentstroke}{rgb}{0.133333,0.533333,0.200000}%
\pgfsetstrokecolor{currentstroke}%
\pgfsetdash{}{0pt}%
\pgfsys@defobject{currentmarker}{\pgfqpoint{-0.055556in}{-0.000000in}}{\pgfqpoint{0.055556in}{0.000000in}}{%
\pgfpathmoveto{\pgfqpoint{0.055556in}{-0.000000in}}%
\pgfpathlineto{\pgfqpoint{-0.055556in}{0.000000in}}%
\pgfusepath{stroke,fill}%
}%
\begin{pgfscope}%
\pgfsys@transformshift{1.367048in}{2.656739in}%
\pgfsys@useobject{currentmarker}{}%
\end{pgfscope}%
\end{pgfscope}%
\begin{pgfscope}%
\pgfsetbuttcap%
\pgfsetroundjoin%
\definecolor{currentfill}{rgb}{0.133333,0.533333,0.200000}%
\pgfsetfillcolor{currentfill}%
\pgfsetlinewidth{2.007500pt}%
\definecolor{currentstroke}{rgb}{0.133333,0.533333,0.200000}%
\pgfsetstrokecolor{currentstroke}%
\pgfsetdash{}{0pt}%
\pgfsys@defobject{currentmarker}{\pgfqpoint{-0.055556in}{-0.000000in}}{\pgfqpoint{0.055556in}{0.000000in}}{%
\pgfpathmoveto{\pgfqpoint{0.055556in}{-0.000000in}}%
\pgfpathlineto{\pgfqpoint{-0.055556in}{0.000000in}}%
\pgfusepath{stroke,fill}%
}%
\begin{pgfscope}%
\pgfsys@transformshift{1.367048in}{2.934517in}%
\pgfsys@useobject{currentmarker}{}%
\end{pgfscope}%
\end{pgfscope}%
\begin{pgfscope}%
\pgfsetbuttcap%
\pgfsetroundjoin%
\pgfsetlinewidth{3.513125pt}%
\definecolor{currentstroke}{rgb}{0.133333,0.533333,0.200000}%
\pgfsetstrokecolor{currentstroke}%
\pgfsetdash{{12.950000pt}{5.600000pt}}{0.000000pt}%
\pgfpathmoveto{\pgfqpoint{1.089270in}{2.795628in}}%
\pgfpathlineto{\pgfqpoint{1.644826in}{2.795628in}}%
\pgfusepath{stroke}%
\end{pgfscope}%
\begin{pgfscope}%
\pgfsetbuttcap%
\pgfsetmiterjoin%
\definecolor{currentfill}{rgb}{0.133333,0.533333,0.200000}%
\pgfsetfillcolor{currentfill}%
\pgfsetlinewidth{2.007500pt}%
\definecolor{currentstroke}{rgb}{1.000000,1.000000,1.000000}%
\pgfsetstrokecolor{currentstroke}%
\pgfsetdash{}{0pt}%
\pgfsys@defobject{currentmarker}{\pgfqpoint{-0.055556in}{-0.055556in}}{\pgfqpoint{0.055556in}{0.055556in}}{%
\pgfpathmoveto{\pgfqpoint{-0.055556in}{-0.055556in}}%
\pgfpathlineto{\pgfqpoint{0.055556in}{-0.055556in}}%
\pgfpathlineto{\pgfqpoint{0.055556in}{0.055556in}}%
\pgfpathlineto{\pgfqpoint{-0.055556in}{0.055556in}}%
\pgfpathlineto{\pgfqpoint{-0.055556in}{-0.055556in}}%
\pgfpathclose%
\pgfusepath{stroke,fill}%
}%
\begin{pgfscope}%
\pgfsys@transformshift{1.367048in}{2.795628in}%
\pgfsys@useobject{currentmarker}{}%
\end{pgfscope}%
\end{pgfscope}%
\begin{pgfscope}%
\definecolor{textcolor}{rgb}{0.000000,0.000000,0.000000}%
\pgfsetstrokecolor{textcolor}%
\pgfsetfillcolor{textcolor}%
\pgftext[x=1.867048in,y=2.698405in,left,base]{\color{textcolor}{\rmfamily\fontsize{20.000000}{24.000000}\selectfont\catcode`\^=\active\def^{\ifmmode\sp\else\^{}\fi}\catcode`\%=\active\def
\end{pgfscope}%
\begin{pgfscope}%
\pgfsetbuttcap%
\pgfsetroundjoin%
\pgfsetlinewidth{3.513125pt}%
\definecolor{currentstroke}{rgb}{0.000000,0.000000,0.000000}%
\pgfsetstrokecolor{currentstroke}%
\pgfsetdash{}{0pt}%
\pgfpathmoveto{\pgfqpoint{1.367048in}{2.261782in}}%
\pgfpathlineto{\pgfqpoint{1.367048in}{2.539560in}}%
\pgfusepath{stroke}%
\end{pgfscope}%
\begin{pgfscope}%
\pgfsetbuttcap%
\pgfsetroundjoin%
\definecolor{currentfill}{rgb}{0.000000,0.000000,0.000000}%
\pgfsetfillcolor{currentfill}%
\pgfsetlinewidth{2.007500pt}%
\definecolor{currentstroke}{rgb}{0.000000,0.000000,0.000000}%
\pgfsetstrokecolor{currentstroke}%
\pgfsetdash{}{0pt}%
\pgfsys@defobject{currentmarker}{\pgfqpoint{-0.055556in}{-0.000000in}}{\pgfqpoint{0.055556in}{0.000000in}}{%
\pgfpathmoveto{\pgfqpoint{0.055556in}{-0.000000in}}%
\pgfpathlineto{\pgfqpoint{-0.055556in}{0.000000in}}%
\pgfusepath{stroke,fill}%
}%
\begin{pgfscope}%
\pgfsys@transformshift{1.367048in}{2.261782in}%
\pgfsys@useobject{currentmarker}{}%
\end{pgfscope}%
\end{pgfscope}%
\begin{pgfscope}%
\pgfsetbuttcap%
\pgfsetroundjoin%
\definecolor{currentfill}{rgb}{0.000000,0.000000,0.000000}%
\pgfsetfillcolor{currentfill}%
\pgfsetlinewidth{2.007500pt}%
\definecolor{currentstroke}{rgb}{0.000000,0.000000,0.000000}%
\pgfsetstrokecolor{currentstroke}%
\pgfsetdash{}{0pt}%
\pgfsys@defobject{currentmarker}{\pgfqpoint{-0.055556in}{-0.000000in}}{\pgfqpoint{0.055556in}{0.000000in}}{%
\pgfpathmoveto{\pgfqpoint{0.055556in}{-0.000000in}}%
\pgfpathlineto{\pgfqpoint{-0.055556in}{0.000000in}}%
\pgfusepath{stroke,fill}%
}%
\begin{pgfscope}%
\pgfsys@transformshift{1.367048in}{2.539560in}%
\pgfsys@useobject{currentmarker}{}%
\end{pgfscope}%
\end{pgfscope}%
\begin{pgfscope}%
\pgfsetbuttcap%
\pgfsetroundjoin%
\pgfsetlinewidth{3.513125pt}%
\definecolor{currentstroke}{rgb}{0.000000,0.000000,0.000000}%
\pgfsetstrokecolor{currentstroke}%
\pgfsetdash{{22.400000pt}{5.600000pt}{3.500000pt}{5.600000pt}}{0.000000pt}%
\pgfpathmoveto{\pgfqpoint{1.089270in}{2.400671in}}%
\pgfpathlineto{\pgfqpoint{1.644826in}{2.400671in}}%
\pgfusepath{stroke}%
\end{pgfscope}%
\begin{pgfscope}%
\pgfsetbuttcap%
\pgfsetmiterjoin%
\definecolor{currentfill}{rgb}{0.000000,0.000000,0.000000}%
\pgfsetfillcolor{currentfill}%
\pgfsetlinewidth{2.007500pt}%
\definecolor{currentstroke}{rgb}{1.000000,1.000000,1.000000}%
\pgfsetstrokecolor{currentstroke}%
\pgfsetdash{}{0pt}%
\pgfsys@defobject{currentmarker}{\pgfqpoint{-0.078567in}{-0.078567in}}{\pgfqpoint{0.078567in}{0.078567in}}{%
\pgfpathmoveto{\pgfqpoint{-0.000000in}{-0.078567in}}%
\pgfpathlineto{\pgfqpoint{0.078567in}{0.000000in}}%
\pgfpathlineto{\pgfqpoint{0.000000in}{0.078567in}}%
\pgfpathlineto{\pgfqpoint{-0.078567in}{0.000000in}}%
\pgfpathlineto{\pgfqpoint{-0.000000in}{-0.078567in}}%
\pgfpathclose%
\pgfusepath{stroke,fill}%
}%
\begin{pgfscope}%
\pgfsys@transformshift{1.367048in}{2.400671in}%
\pgfsys@useobject{currentmarker}{}%
\end{pgfscope}%
\end{pgfscope}%
\begin{pgfscope}%
\definecolor{textcolor}{rgb}{0.000000,0.000000,0.000000}%
\pgfsetstrokecolor{textcolor}%
\pgfsetfillcolor{textcolor}%
\pgftext[x=1.867048in,y=2.303449in,left,base]{\color{textcolor}{\rmfamily\fontsize{20.000000}{24.000000}\selectfont\catcode`\^=\active\def^{\ifmmode\sp\else\^{}\fi}\catcode`\%=\active\def
\end{pgfscope}%
\end{pgfpicture}%
\makeatother%
\endgroup%

%% file: figures/SE_est_match/WERM_SE_match_est_logistic_Lub_L2_new5_loc.pgf
\begingroup%
\makeatletter%
\begin{pgfpicture}%
\pgfpathrectangle{\pgfpointorigin}{\pgfqpoint{4.737079in}{3.749100in}}%
\pgfusepath{use as bounding box, clip}%
\begin{pgfscope}%
\pgfsetbuttcap%
\pgfsetmiterjoin%
\definecolor{currentfill}{rgb}{1.000000,1.000000,1.000000}%
\pgfsetfillcolor{currentfill}%
\pgfsetlinewidth{0.000000pt}%
\definecolor{currentstroke}{rgb}{1.000000,1.000000,1.000000}%
\pgfsetstrokecolor{currentstroke}%
\pgfsetdash{}{0pt}%
\pgfpathmoveto{\pgfqpoint{0.000000in}{0.000000in}}%
\pgfpathlineto{\pgfqpoint{4.737079in}{0.000000in}}%
\pgfpathlineto{\pgfqpoint{4.737079in}{3.749100in}}%
\pgfpathlineto{\pgfqpoint{0.000000in}{3.749100in}}%
\pgfpathlineto{\pgfqpoint{0.000000in}{0.000000in}}%
\pgfpathclose%
\pgfusepath{fill}%
\end{pgfscope}%
\begin{pgfscope}%
\pgfsetbuttcap%
\pgfsetmiterjoin%
\definecolor{currentfill}{rgb}{1.000000,1.000000,1.000000}%
\pgfsetfillcolor{currentfill}%
\pgfsetlinewidth{0.000000pt}%
\definecolor{currentstroke}{rgb}{0.000000,0.000000,0.000000}%
\pgfsetstrokecolor{currentstroke}%
\pgfsetstrokeopacity{0.000000}%
\pgfsetdash{}{0pt}%
\pgfpathmoveto{\pgfqpoint{1.043289in}{0.831623in}}%
\pgfpathlineto{\pgfqpoint{4.637079in}{0.831623in}}%
\pgfpathlineto{\pgfqpoint{4.637079in}{3.649100in}}%
\pgfpathlineto{\pgfqpoint{1.043289in}{3.649100in}}%
\pgfpathlineto{\pgfqpoint{1.043289in}{0.831623in}}%
\pgfpathclose%
\pgfusepath{fill}%
\end{pgfscope}%
\begin{pgfscope}%
\pgfpathrectangle{\pgfqpoint{1.043289in}{0.831623in}}{\pgfqpoint{3.593791in}{2.817477in}}%
\pgfusepath{clip}%
\pgfsetbuttcap%
\pgfsetroundjoin%
\pgfsetlinewidth{3.513125pt}%
\definecolor{currentstroke}{rgb}{0.000000,0.000000,0.000000}%
\pgfsetstrokecolor{currentstroke}%
\pgfsetstrokeopacity{0.800000}%
\pgfsetdash{}{0pt}%
\pgfpathmoveto{\pgfqpoint{1.206643in}{1.795880in}}%
\pgfpathlineto{\pgfqpoint{1.206643in}{2.417668in}}%
\pgfusepath{stroke}%
\end{pgfscope}%
\begin{pgfscope}%
\pgfpathrectangle{\pgfqpoint{1.043289in}{0.831623in}}{\pgfqpoint{3.593791in}{2.817477in}}%
\pgfusepath{clip}%
\pgfsetbuttcap%
\pgfsetroundjoin%
\pgfsetlinewidth{3.513125pt}%
\definecolor{currentstroke}{rgb}{0.000000,0.000000,0.000000}%
\pgfsetstrokecolor{currentstroke}%
\pgfsetstrokeopacity{0.800000}%
\pgfsetdash{}{0pt}%
\pgfpathmoveto{\pgfqpoint{2.023413in}{1.327515in}}%
\pgfpathlineto{\pgfqpoint{2.023413in}{2.023788in}}%
\pgfusepath{stroke}%
\end{pgfscope}%
\begin{pgfscope}%
\pgfpathrectangle{\pgfqpoint{1.043289in}{0.831623in}}{\pgfqpoint{3.593791in}{2.817477in}}%
\pgfusepath{clip}%
\pgfsetbuttcap%
\pgfsetroundjoin%
\pgfsetlinewidth{3.513125pt}%
\definecolor{currentstroke}{rgb}{0.000000,0.000000,0.000000}%
\pgfsetstrokecolor{currentstroke}%
\pgfsetstrokeopacity{0.800000}%
\pgfsetdash{}{0pt}%
\pgfpathmoveto{\pgfqpoint{2.840184in}{1.013788in}}%
\pgfpathlineto{\pgfqpoint{2.840184in}{2.030439in}}%
\pgfusepath{stroke}%
\end{pgfscope}%
\begin{pgfscope}%
\pgfpathrectangle{\pgfqpoint{1.043289in}{0.831623in}}{\pgfqpoint{3.593791in}{2.817477in}}%
\pgfusepath{clip}%
\pgfsetbuttcap%
\pgfsetroundjoin%
\pgfsetlinewidth{3.513125pt}%
\definecolor{currentstroke}{rgb}{0.000000,0.000000,0.000000}%
\pgfsetstrokecolor{currentstroke}%
\pgfsetstrokeopacity{0.800000}%
\pgfsetdash{}{0pt}%
\pgfpathmoveto{\pgfqpoint{3.656954in}{1.025932in}}%
\pgfpathlineto{\pgfqpoint{3.656954in}{1.675652in}}%
\pgfusepath{stroke}%
\end{pgfscope}%
\begin{pgfscope}%
\pgfpathrectangle{\pgfqpoint{1.043289in}{0.831623in}}{\pgfqpoint{3.593791in}{2.817477in}}%
\pgfusepath{clip}%
\pgfsetbuttcap%
\pgfsetroundjoin%
\pgfsetlinewidth{3.513125pt}%
\definecolor{currentstroke}{rgb}{0.000000,0.000000,0.000000}%
\pgfsetstrokecolor{currentstroke}%
\pgfsetstrokeopacity{0.800000}%
\pgfsetdash{}{0pt}%
\pgfpathmoveto{\pgfqpoint{4.473725in}{0.973307in}}%
\pgfpathlineto{\pgfqpoint{4.473725in}{1.658785in}}%
\pgfusepath{stroke}%
\end{pgfscope}%
\begin{pgfscope}%
\pgfpathrectangle{\pgfqpoint{1.043289in}{0.831623in}}{\pgfqpoint{3.593791in}{2.817477in}}%
\pgfusepath{clip}%
\pgfsetbuttcap%
\pgfsetroundjoin%
\definecolor{currentfill}{rgb}{0.000000,0.000000,0.000000}%
\pgfsetfillcolor{currentfill}%
\pgfsetfillopacity{0.800000}%
\pgfsetlinewidth{2.007500pt}%
\definecolor{currentstroke}{rgb}{0.000000,0.000000,0.000000}%
\pgfsetstrokecolor{currentstroke}%
\pgfsetstrokeopacity{0.800000}%
\pgfsetdash{}{0pt}%
\pgfsys@defobject{currentmarker}{\pgfqpoint{-0.055556in}{-0.000000in}}{\pgfqpoint{0.055556in}{0.000000in}}{%
\pgfpathmoveto{\pgfqpoint{0.055556in}{-0.000000in}}%
\pgfpathlineto{\pgfqpoint{-0.055556in}{0.000000in}}%
\pgfusepath{stroke,fill}%
}%
\begin{pgfscope}%
\pgfsys@transformshift{1.206643in}{1.795880in}%
\pgfsys@useobject{currentmarker}{}%
\end{pgfscope}%
\begin{pgfscope}%
\pgfsys@transformshift{2.023413in}{1.327515in}%
\pgfsys@useobject{currentmarker}{}%
\end{pgfscope}%
\begin{pgfscope}%
\pgfsys@transformshift{2.840184in}{1.013788in}%
\pgfsys@useobject{currentmarker}{}%
\end{pgfscope}%
\begin{pgfscope}%
\pgfsys@transformshift{3.656954in}{1.025932in}%
\pgfsys@useobject{currentmarker}{}%
\end{pgfscope}%
\begin{pgfscope}%
\pgfsys@transformshift{4.473725in}{0.973307in}%
\pgfsys@useobject{currentmarker}{}%
\end{pgfscope}%
\end{pgfscope}%
\begin{pgfscope}%
\pgfpathrectangle{\pgfqpoint{1.043289in}{0.831623in}}{\pgfqpoint{3.593791in}{2.817477in}}%
\pgfusepath{clip}%
\pgfsetbuttcap%
\pgfsetroundjoin%
\definecolor{currentfill}{rgb}{0.000000,0.000000,0.000000}%
\pgfsetfillcolor{currentfill}%
\pgfsetfillopacity{0.800000}%
\pgfsetlinewidth{2.007500pt}%
\definecolor{currentstroke}{rgb}{0.000000,0.000000,0.000000}%
\pgfsetstrokecolor{currentstroke}%
\pgfsetstrokeopacity{0.800000}%
\pgfsetdash{}{0pt}%
\pgfsys@defobject{currentmarker}{\pgfqpoint{-0.055556in}{-0.000000in}}{\pgfqpoint{0.055556in}{0.000000in}}{%
\pgfpathmoveto{\pgfqpoint{0.055556in}{-0.000000in}}%
\pgfpathlineto{\pgfqpoint{-0.055556in}{0.000000in}}%
\pgfusepath{stroke,fill}%
}%
\begin{pgfscope}%
\pgfsys@transformshift{1.206643in}{2.417668in}%
\pgfsys@useobject{currentmarker}{}%
\end{pgfscope}%
\begin{pgfscope}%
\pgfsys@transformshift{2.023413in}{2.023788in}%
\pgfsys@useobject{currentmarker}{}%
\end{pgfscope}%
\begin{pgfscope}%
\pgfsys@transformshift{2.840184in}{2.030439in}%
\pgfsys@useobject{currentmarker}{}%
\end{pgfscope}%
\begin{pgfscope}%
\pgfsys@transformshift{3.656954in}{1.675652in}%
\pgfsys@useobject{currentmarker}{}%
\end{pgfscope}%
\begin{pgfscope}%
\pgfsys@transformshift{4.473725in}{1.658785in}%
\pgfsys@useobject{currentmarker}{}%
\end{pgfscope}%
\end{pgfscope}%
\begin{pgfscope}%
\pgfpathrectangle{\pgfqpoint{1.043289in}{0.831623in}}{\pgfqpoint{3.593791in}{2.817477in}}%
\pgfusepath{clip}%
\pgfsetrectcap%
\pgfsetroundjoin%
\pgfsetlinewidth{3.513125pt}%
\definecolor{currentstroke}{rgb}{0.000000,0.000000,0.000000}%
\pgfsetstrokecolor{currentstroke}%
\pgfsetstrokeopacity{0.800000}%
\pgfsetdash{}{0pt}%
\pgfpathmoveto{\pgfqpoint{1.206643in}{2.022412in}}%
\pgfpathlineto{\pgfqpoint{2.023413in}{1.723284in}}%
\pgfpathlineto{\pgfqpoint{2.840184in}{1.610015in}}%
\pgfpathlineto{\pgfqpoint{3.656954in}{1.374676in}}%
\pgfpathlineto{\pgfqpoint{4.473725in}{1.351579in}}%
\pgfusepath{stroke}%
\end{pgfscope}%
\begin{pgfscope}%
\pgfpathrectangle{\pgfqpoint{1.043289in}{0.831623in}}{\pgfqpoint{3.593791in}{2.817477in}}%
\pgfusepath{clip}%
\pgfsetbuttcap%
\pgfsetroundjoin%
\definecolor{currentfill}{rgb}{0.000000,0.000000,0.000000}%
\pgfsetfillcolor{currentfill}%
\pgfsetfillopacity{0.800000}%
\pgfsetlinewidth{2.007500pt}%
\definecolor{currentstroke}{rgb}{1.000000,1.000000,1.000000}%
\pgfsetstrokecolor{currentstroke}%
\pgfsetstrokeopacity{0.800000}%
\pgfsetdash{}{0pt}%
\pgfsys@defobject{currentmarker}{\pgfqpoint{-0.055556in}{-0.055556in}}{\pgfqpoint{0.055556in}{0.055556in}}{%
\pgfpathmoveto{\pgfqpoint{0.000000in}{-0.055556in}}%
\pgfpathcurveto{\pgfqpoint{0.014734in}{-0.055556in}}{\pgfqpoint{0.028866in}{-0.049702in}}{\pgfqpoint{0.039284in}{-0.039284in}}%
\pgfpathcurveto{\pgfqpoint{0.049702in}{-0.028866in}}{\pgfqpoint{0.055556in}{-0.014734in}}{\pgfqpoint{0.055556in}{0.000000in}}%
\pgfpathcurveto{\pgfqpoint{0.055556in}{0.014734in}}{\pgfqpoint{0.049702in}{0.028866in}}{\pgfqpoint{0.039284in}{0.039284in}}%
\pgfpathcurveto{\pgfqpoint{0.028866in}{0.049702in}}{\pgfqpoint{0.014734in}{0.055556in}}{\pgfqpoint{0.000000in}{0.055556in}}%
\pgfpathcurveto{\pgfqpoint{-0.014734in}{0.055556in}}{\pgfqpoint{-0.028866in}{0.049702in}}{\pgfqpoint{-0.039284in}{0.039284in}}%
\pgfpathcurveto{\pgfqpoint{-0.049702in}{0.028866in}}{\pgfqpoint{-0.055556in}{0.014734in}}{\pgfqpoint{-0.055556in}{0.000000in}}%
\pgfpathcurveto{\pgfqpoint{-0.055556in}{-0.014734in}}{\pgfqpoint{-0.049702in}{-0.028866in}}{\pgfqpoint{-0.039284in}{-0.039284in}}%
\pgfpathcurveto{\pgfqpoint{-0.028866in}{-0.049702in}}{\pgfqpoint{-0.014734in}{-0.055556in}}{\pgfqpoint{0.000000in}{-0.055556in}}%
\pgfpathlineto{\pgfqpoint{0.000000in}{-0.055556in}}%
\pgfpathclose%
\pgfusepath{stroke,fill}%
}%
\begin{pgfscope}%
\pgfsys@transformshift{1.206643in}{2.022412in}%
\pgfsys@useobject{currentmarker}{}%
\end{pgfscope}%
\begin{pgfscope}%
\pgfsys@transformshift{2.023413in}{1.723284in}%
\pgfsys@useobject{currentmarker}{}%
\end{pgfscope}%
\begin{pgfscope}%
\pgfsys@transformshift{2.840184in}{1.610015in}%
\pgfsys@useobject{currentmarker}{}%
\end{pgfscope}%
\begin{pgfscope}%
\pgfsys@transformshift{3.656954in}{1.374676in}%
\pgfsys@useobject{currentmarker}{}%
\end{pgfscope}%
\begin{pgfscope}%
\pgfsys@transformshift{4.473725in}{1.351579in}%
\pgfsys@useobject{currentmarker}{}%
\end{pgfscope}%
\end{pgfscope}%
\begin{pgfscope}%
\pgfpathrectangle{\pgfqpoint{1.043289in}{0.831623in}}{\pgfqpoint{3.593791in}{2.817477in}}%
\pgfusepath{clip}%
\pgfsetbuttcap%
\pgfsetroundjoin%
\definecolor{currentfill}{rgb}{0.200000,0.733333,0.933333}%
\pgfsetfillcolor{currentfill}%
\pgfsetfillopacity{0.999000}%
\pgfsetlinewidth{2.509375pt}%
\definecolor{currentstroke}{rgb}{1.000000,1.000000,1.000000}%
\pgfsetstrokecolor{currentstroke}%
\pgfsetstrokeopacity{0.999000}%
\pgfsetdash{}{0pt}%
\pgfsys@defobject{currentmarker}{\pgfqpoint{-0.064550in}{-0.107583in}}{\pgfqpoint{0.064550in}{0.107583in}}{%
\pgfpathmoveto{\pgfqpoint{-0.000000in}{-0.107583in}}%
\pgfpathlineto{\pgfqpoint{0.064550in}{0.000000in}}%
\pgfpathlineto{\pgfqpoint{0.000000in}{0.107583in}}%
\pgfpathlineto{\pgfqpoint{-0.064550in}{0.000000in}}%
\pgfpathlineto{\pgfqpoint{-0.000000in}{-0.107583in}}%
\pgfpathclose%
\pgfusepath{stroke,fill}%
}%
\begin{pgfscope}%
\pgfsys@transformshift{1.206643in}{2.068722in}%
\pgfsys@useobject{currentmarker}{}%
\end{pgfscope}%
\begin{pgfscope}%
\pgfsys@transformshift{2.023413in}{1.835552in}%
\pgfsys@useobject{currentmarker}{}%
\end{pgfscope}%
\begin{pgfscope}%
\pgfsys@transformshift{2.840184in}{1.489960in}%
\pgfsys@useobject{currentmarker}{}%
\end{pgfscope}%
\begin{pgfscope}%
\pgfsys@transformshift{3.656954in}{1.250197in}%
\pgfsys@useobject{currentmarker}{}%
\end{pgfscope}%
\begin{pgfscope}%
\pgfsys@transformshift{4.473725in}{1.270257in}%
\pgfsys@useobject{currentmarker}{}%
\end{pgfscope}%
\end{pgfscope}%
\begin{pgfscope}%
\pgfsetbuttcap%
\pgfsetroundjoin%
\definecolor{currentfill}{rgb}{0.000000,0.000000,0.000000}%
\pgfsetfillcolor{currentfill}%
\pgfsetlinewidth{1.505625pt}%
\definecolor{currentstroke}{rgb}{0.000000,0.000000,0.000000}%
\pgfsetstrokecolor{currentstroke}%
\pgfsetdash{}{0pt}%
\pgfsys@defobject{currentmarker}{\pgfqpoint{0.000000in}{-0.083333in}}{\pgfqpoint{0.000000in}{0.000000in}}{%
\pgfpathmoveto{\pgfqpoint{0.000000in}{0.000000in}}%
\pgfpathlineto{\pgfqpoint{0.000000in}{-0.083333in}}%
\pgfusepath{stroke,fill}%
}%
\begin{pgfscope}%
\pgfsys@transformshift{1.206643in}{0.831623in}%
\pgfsys@useobject{currentmarker}{}%
\end{pgfscope}%
\end{pgfscope}%
\begin{pgfscope}%
\definecolor{textcolor}{rgb}{0.000000,0.000000,0.000000}%
\pgfsetstrokecolor{textcolor}%
\pgfsetfillcolor{textcolor}%
\pgftext[x=1.206643in,y=0.699679in,,top]{\color{textcolor}{\rmfamily\fontsize{22.000000}{26.400000}\selectfont\catcode`\^=\active\def^{\ifmmode\sp\else\^{}\fi}\catcode`\%=\active\def
\end{pgfscope}%
\begin{pgfscope}%
\pgfsetbuttcap%
\pgfsetroundjoin%
\definecolor{currentfill}{rgb}{0.000000,0.000000,0.000000}%
\pgfsetfillcolor{currentfill}%
\pgfsetlinewidth{1.505625pt}%
\definecolor{currentstroke}{rgb}{0.000000,0.000000,0.000000}%
\pgfsetstrokecolor{currentstroke}%
\pgfsetdash{}{0pt}%
\pgfsys@defobject{currentmarker}{\pgfqpoint{0.000000in}{-0.083333in}}{\pgfqpoint{0.000000in}{0.000000in}}{%
\pgfpathmoveto{\pgfqpoint{0.000000in}{0.000000in}}%
\pgfpathlineto{\pgfqpoint{0.000000in}{-0.083333in}}%
\pgfusepath{stroke,fill}%
}%
\begin{pgfscope}%
\pgfsys@transformshift{2.840184in}{0.831623in}%
\pgfsys@useobject{currentmarker}{}%
\end{pgfscope}%
\end{pgfscope}%
\begin{pgfscope}%
\definecolor{textcolor}{rgb}{0.000000,0.000000,0.000000}%
\pgfsetstrokecolor{textcolor}%
\pgfsetfillcolor{textcolor}%
\pgftext[x=2.840184in,y=0.699679in,,top]{\color{textcolor}{\rmfamily\fontsize{22.000000}{26.400000}\selectfont\catcode`\^=\active\def^{\ifmmode\sp\else\^{}\fi}\catcode`\%=\active\def
\end{pgfscope}%
\begin{pgfscope}%
\pgfsetbuttcap%
\pgfsetroundjoin%
\definecolor{currentfill}{rgb}{0.000000,0.000000,0.000000}%
\pgfsetfillcolor{currentfill}%
\pgfsetlinewidth{1.505625pt}%
\definecolor{currentstroke}{rgb}{0.000000,0.000000,0.000000}%
\pgfsetstrokecolor{currentstroke}%
\pgfsetdash{}{0pt}%
\pgfsys@defobject{currentmarker}{\pgfqpoint{0.000000in}{-0.083333in}}{\pgfqpoint{0.000000in}{0.000000in}}{%
\pgfpathmoveto{\pgfqpoint{0.000000in}{0.000000in}}%
\pgfpathlineto{\pgfqpoint{0.000000in}{-0.083333in}}%
\pgfusepath{stroke,fill}%
}%
\begin{pgfscope}%
\pgfsys@transformshift{4.473725in}{0.831623in}%
\pgfsys@useobject{currentmarker}{}%
\end{pgfscope}%
\end{pgfscope}%
\begin{pgfscope}%
\definecolor{textcolor}{rgb}{0.000000,0.000000,0.000000}%
\pgfsetstrokecolor{textcolor}%
\pgfsetfillcolor{textcolor}%
\pgftext[x=4.473725in,y=0.699679in,,top]{\color{textcolor}{\rmfamily\fontsize{22.000000}{26.400000}\selectfont\catcode`\^=\active\def^{\ifmmode\sp\else\^{}\fi}\catcode`\%=\active\def
\end{pgfscope}%
\begin{pgfscope}%
\definecolor{textcolor}{rgb}{0.000000,0.000000,0.000000}%
\pgfsetstrokecolor{textcolor}%
\pgfsetfillcolor{textcolor}%
\pgftext[x=2.840184in,y=0.388056in,,top]{\color{textcolor}{\rmfamily\fontsize{22.000000}{26.400000}\selectfont\catcode`\^=\active\def^{\ifmmode\sp\else\^{}\fi}\catcode`\%=\active\def
\end{pgfscope}%
\begin{pgfscope}%
\pgfsetbuttcap%
\pgfsetroundjoin%
\definecolor{currentfill}{rgb}{0.000000,0.000000,0.000000}%
\pgfsetfillcolor{currentfill}%
\pgfsetlinewidth{1.505625pt}%
\definecolor{currentstroke}{rgb}{0.000000,0.000000,0.000000}%
\pgfsetstrokecolor{currentstroke}%
\pgfsetdash{}{0pt}%
\pgfsys@defobject{currentmarker}{\pgfqpoint{-0.083333in}{0.000000in}}{\pgfqpoint{-0.000000in}{0.000000in}}{%
\pgfpathmoveto{\pgfqpoint{-0.000000in}{0.000000in}}%
\pgfpathlineto{\pgfqpoint{-0.083333in}{0.000000in}}%
\pgfusepath{stroke,fill}%
}%
\begin{pgfscope}%
\pgfsys@transformshift{1.043289in}{0.831623in}%
\pgfsys@useobject{currentmarker}{}%
\end{pgfscope}%
\end{pgfscope}%
\begin{pgfscope}%
\definecolor{textcolor}{rgb}{0.000000,0.000000,0.000000}%
\pgfsetstrokecolor{textcolor}%
\pgfsetfillcolor{textcolor}%
\pgftext[x=0.443111in, y=0.731604in, left, base]{\color{textcolor}{\rmfamily\fontsize{22.000000}{26.400000}\selectfont\catcode`\^=\active\def^{\ifmmode\sp\else\^{}\fi}\catcode`\%=\active\def
\end{pgfscope}%
\begin{pgfscope}%
\pgfsetbuttcap%
\pgfsetroundjoin%
\definecolor{currentfill}{rgb}{0.000000,0.000000,0.000000}%
\pgfsetfillcolor{currentfill}%
\pgfsetlinewidth{1.505625pt}%
\definecolor{currentstroke}{rgb}{0.000000,0.000000,0.000000}%
\pgfsetstrokecolor{currentstroke}%
\pgfsetdash{}{0pt}%
\pgfsys@defobject{currentmarker}{\pgfqpoint{-0.083333in}{0.000000in}}{\pgfqpoint{-0.000000in}{0.000000in}}{%
\pgfpathmoveto{\pgfqpoint{-0.000000in}{0.000000in}}%
\pgfpathlineto{\pgfqpoint{-0.083333in}{0.000000in}}%
\pgfusepath{stroke,fill}%
}%
\begin{pgfscope}%
\pgfsys@transformshift{1.043289in}{1.803167in}%
\pgfsys@useobject{currentmarker}{}%
\end{pgfscope}%
\end{pgfscope}%
\begin{pgfscope}%
\definecolor{textcolor}{rgb}{0.000000,0.000000,0.000000}%
\pgfsetstrokecolor{textcolor}%
\pgfsetfillcolor{textcolor}%
\pgftext[x=0.443111in, y=1.703148in, left, base]{\color{textcolor}{\rmfamily\fontsize{22.000000}{26.400000}\selectfont\catcode`\^=\active\def^{\ifmmode\sp\else\^{}\fi}\catcode`\%=\active\def
\end{pgfscope}%
\begin{pgfscope}%
\pgfsetbuttcap%
\pgfsetroundjoin%
\definecolor{currentfill}{rgb}{0.000000,0.000000,0.000000}%
\pgfsetfillcolor{currentfill}%
\pgfsetlinewidth{1.505625pt}%
\definecolor{currentstroke}{rgb}{0.000000,0.000000,0.000000}%
\pgfsetstrokecolor{currentstroke}%
\pgfsetdash{}{0pt}%
\pgfsys@defobject{currentmarker}{\pgfqpoint{-0.083333in}{0.000000in}}{\pgfqpoint{-0.000000in}{0.000000in}}{%
\pgfpathmoveto{\pgfqpoint{-0.000000in}{0.000000in}}%
\pgfpathlineto{\pgfqpoint{-0.083333in}{0.000000in}}%
\pgfusepath{stroke,fill}%
}%
\begin{pgfscope}%
\pgfsys@transformshift{1.043289in}{2.774710in}%
\pgfsys@useobject{currentmarker}{}%
\end{pgfscope}%
\end{pgfscope}%
\begin{pgfscope}%
\definecolor{textcolor}{rgb}{0.000000,0.000000,0.000000}%
\pgfsetstrokecolor{textcolor}%
\pgfsetfillcolor{textcolor}%
\pgftext[x=0.443111in, y=2.674691in, left, base]{\color{textcolor}{\rmfamily\fontsize{22.000000}{26.400000}\selectfont\catcode`\^=\active\def^{\ifmmode\sp\else\^{}\fi}\catcode`\%=\active\def
\end{pgfscope}%
\begin{pgfscope}%
\definecolor{textcolor}{rgb}{0.000000,0.000000,0.000000}%
\pgfsetstrokecolor{textcolor}%
\pgfsetfillcolor{textcolor}%
\pgftext[x=0.387555in,y=2.240361in,,bottom,rotate=90.000000]{\color{textcolor}{\rmfamily\fontsize{22.000000}{26.400000}\selectfont\catcode`\^=\active\def^{\ifmmode\sp\else\^{}\fi}\catcode`\%=\active\def
\end{pgfscope}%
\begin{pgfscope}%
\pgfsetrectcap%
\pgfsetmiterjoin%
\pgfsetlinewidth{1.505625pt}%
\definecolor{currentstroke}{rgb}{0.000000,0.000000,0.000000}%
\pgfsetstrokecolor{currentstroke}%
\pgfsetdash{}{0pt}%
\pgfpathmoveto{\pgfqpoint{1.043289in}{0.831623in}}%
\pgfpathlineto{\pgfqpoint{1.043289in}{3.649100in}}%
\pgfusepath{stroke}%
\end{pgfscope}%
\begin{pgfscope}%
\pgfsetrectcap%
\pgfsetmiterjoin%
\pgfsetlinewidth{1.505625pt}%
\definecolor{currentstroke}{rgb}{0.000000,0.000000,0.000000}%
\pgfsetstrokecolor{currentstroke}%
\pgfsetdash{}{0pt}%
\pgfpathmoveto{\pgfqpoint{1.043289in}{0.831623in}}%
\pgfpathlineto{\pgfqpoint{4.637079in}{0.831623in}}%
\pgfusepath{stroke}%
\end{pgfscope}%
\begin{pgfscope}%
\pgfsetbuttcap%
\pgfsetmiterjoin%
\definecolor{currentfill}{rgb}{1.000000,1.000000,1.000000}%
\pgfsetfillcolor{currentfill}%
\pgfsetfillopacity{0.400000}%
\pgfsetlinewidth{1.003750pt}%
\definecolor{currentstroke}{rgb}{0.000000,0.000000,0.000000}%
\pgfsetstrokecolor{currentstroke}%
\pgfsetstrokeopacity{0.400000}%
\pgfsetdash{}{0pt}%
\pgfpathmoveto{\pgfqpoint{2.021647in}{2.434734in}}%
\pgfpathlineto{\pgfqpoint{4.403746in}{2.434734in}}%
\pgfpathquadraticcurveto{\pgfqpoint{4.470412in}{2.434734in}}{\pgfqpoint{4.470412in}{2.501401in}}%
\pgfpathlineto{\pgfqpoint{4.470412in}{3.415766in}}%
\pgfpathquadraticcurveto{\pgfqpoint{4.470412in}{3.482433in}}{\pgfqpoint{4.403746in}{3.482433in}}%
\pgfpathlineto{\pgfqpoint{2.021647in}{3.482433in}}%
\pgfpathquadraticcurveto{\pgfqpoint{1.954981in}{3.482433in}}{\pgfqpoint{1.954981in}{3.415766in}}%
\pgfpathlineto{\pgfqpoint{1.954981in}{2.501401in}}%
\pgfpathquadraticcurveto{\pgfqpoint{1.954981in}{2.434734in}}{\pgfqpoint{2.021647in}{2.434734in}}%
\pgfpathlineto{\pgfqpoint{2.021647in}{2.434734in}}%
\pgfpathclose%
\pgfusepath{stroke,fill}%
\end{pgfscope}%
\begin{pgfscope}%
\pgfsetbuttcap%
\pgfsetroundjoin%
\definecolor{currentfill}{rgb}{0.200000,0.733333,0.933333}%
\pgfsetfillcolor{currentfill}%
\pgfsetfillopacity{0.999000}%
\pgfsetlinewidth{2.509375pt}%
\definecolor{currentstroke}{rgb}{1.000000,1.000000,1.000000}%
\pgfsetstrokecolor{currentstroke}%
\pgfsetstrokeopacity{0.999000}%
\pgfsetdash{}{0pt}%
\pgfsys@defobject{currentmarker}{\pgfqpoint{-0.064550in}{-0.107583in}}{\pgfqpoint{0.064550in}{0.107583in}}{%
\pgfpathmoveto{\pgfqpoint{-0.000000in}{-0.107583in}}%
\pgfpathlineto{\pgfqpoint{0.064550in}{0.000000in}}%
\pgfpathlineto{\pgfqpoint{0.000000in}{0.107583in}}%
\pgfpathlineto{\pgfqpoint{-0.064550in}{0.000000in}}%
\pgfpathlineto{\pgfqpoint{-0.000000in}{-0.107583in}}%
\pgfpathclose%
\pgfusepath{stroke,fill}%
}%
\begin{pgfscope}%
\pgfsys@transformshift{2.421647in}{3.196630in}%
\pgfsys@useobject{currentmarker}{}%
\end{pgfscope}%
\end{pgfscope}%
\begin{pgfscope}%
\definecolor{textcolor}{rgb}{0.000000,0.000000,0.000000}%
\pgfsetstrokecolor{textcolor}%
\pgfsetfillcolor{textcolor}%
\pgftext[x=3.021647in,y=3.109130in,left,base]{\color{textcolor}{\rmfamily\fontsize{24.000000}{28.800000}\selectfont\catcode`\^=\active\def^{\ifmmode\sp\else\^{}\fi}\catcode`\%=\active\def
\end{pgfscope}%
\begin{pgfscope}%
\pgfsetbuttcap%
\pgfsetroundjoin%
\pgfsetlinewidth{3.513125pt}%
\definecolor{currentstroke}{rgb}{0.000000,0.000000,0.000000}%
\pgfsetstrokecolor{currentstroke}%
\pgfsetstrokeopacity{0.800000}%
\pgfsetdash{}{0pt}%
\pgfpathmoveto{\pgfqpoint{2.421647in}{2.585281in}}%
\pgfpathlineto{\pgfqpoint{2.421647in}{2.918614in}}%
\pgfusepath{stroke}%
\end{pgfscope}%
\begin{pgfscope}%
\pgfsetbuttcap%
\pgfsetroundjoin%
\definecolor{currentfill}{rgb}{0.000000,0.000000,0.000000}%
\pgfsetfillcolor{currentfill}%
\pgfsetfillopacity{0.800000}%
\pgfsetlinewidth{2.007500pt}%
\definecolor{currentstroke}{rgb}{0.000000,0.000000,0.000000}%
\pgfsetstrokecolor{currentstroke}%
\pgfsetstrokeopacity{0.800000}%
\pgfsetdash{}{0pt}%
\pgfsys@defobject{currentmarker}{\pgfqpoint{-0.055556in}{-0.000000in}}{\pgfqpoint{0.055556in}{0.000000in}}{%
\pgfpathmoveto{\pgfqpoint{0.055556in}{-0.000000in}}%
\pgfpathlineto{\pgfqpoint{-0.055556in}{0.000000in}}%
\pgfusepath{stroke,fill}%
}%
\begin{pgfscope}%
\pgfsys@transformshift{2.421647in}{2.585281in}%
\pgfsys@useobject{currentmarker}{}%
\end{pgfscope}%
\end{pgfscope}%
\begin{pgfscope}%
\pgfsetbuttcap%
\pgfsetroundjoin%
\definecolor{currentfill}{rgb}{0.000000,0.000000,0.000000}%
\pgfsetfillcolor{currentfill}%
\pgfsetfillopacity{0.800000}%
\pgfsetlinewidth{2.007500pt}%
\definecolor{currentstroke}{rgb}{0.000000,0.000000,0.000000}%
\pgfsetstrokecolor{currentstroke}%
\pgfsetstrokeopacity{0.800000}%
\pgfsetdash{}{0pt}%
\pgfsys@defobject{currentmarker}{\pgfqpoint{-0.055556in}{-0.000000in}}{\pgfqpoint{0.055556in}{0.000000in}}{%
\pgfpathmoveto{\pgfqpoint{0.055556in}{-0.000000in}}%
\pgfpathlineto{\pgfqpoint{-0.055556in}{0.000000in}}%
\pgfusepath{stroke,fill}%
}%
\begin{pgfscope}%
\pgfsys@transformshift{2.421647in}{2.918614in}%
\pgfsys@useobject{currentmarker}{}%
\end{pgfscope}%
\end{pgfscope}%
\begin{pgfscope}%
\pgfsetrectcap%
\pgfsetroundjoin%
\pgfsetlinewidth{3.513125pt}%
\definecolor{currentstroke}{rgb}{0.000000,0.000000,0.000000}%
\pgfsetstrokecolor{currentstroke}%
\pgfsetstrokeopacity{0.800000}%
\pgfsetdash{}{0pt}%
\pgfpathmoveto{\pgfqpoint{2.088314in}{2.751948in}}%
\pgfpathlineto{\pgfqpoint{2.754981in}{2.751948in}}%
\pgfusepath{stroke}%
\end{pgfscope}%
\begin{pgfscope}%
\pgfsetbuttcap%
\pgfsetroundjoin%
\definecolor{currentfill}{rgb}{0.000000,0.000000,0.000000}%
\pgfsetfillcolor{currentfill}%
\pgfsetfillopacity{0.800000}%
\pgfsetlinewidth{2.007500pt}%
\definecolor{currentstroke}{rgb}{1.000000,1.000000,1.000000}%
\pgfsetstrokecolor{currentstroke}%
\pgfsetstrokeopacity{0.800000}%
\pgfsetdash{}{0pt}%
\pgfsys@defobject{currentmarker}{\pgfqpoint{-0.055556in}{-0.055556in}}{\pgfqpoint{0.055556in}{0.055556in}}{%
\pgfpathmoveto{\pgfqpoint{0.000000in}{-0.055556in}}%
\pgfpathcurveto{\pgfqpoint{0.014734in}{-0.055556in}}{\pgfqpoint{0.028866in}{-0.049702in}}{\pgfqpoint{0.039284in}{-0.039284in}}%
\pgfpathcurveto{\pgfqpoint{0.049702in}{-0.028866in}}{\pgfqpoint{0.055556in}{-0.014734in}}{\pgfqpoint{0.055556in}{0.000000in}}%
\pgfpathcurveto{\pgfqpoint{0.055556in}{0.014734in}}{\pgfqpoint{0.049702in}{0.028866in}}{\pgfqpoint{0.039284in}{0.039284in}}%
\pgfpathcurveto{\pgfqpoint{0.028866in}{0.049702in}}{\pgfqpoint{0.014734in}{0.055556in}}{\pgfqpoint{0.000000in}{0.055556in}}%
\pgfpathcurveto{\pgfqpoint{-0.014734in}{0.055556in}}{\pgfqpoint{-0.028866in}{0.049702in}}{\pgfqpoint{-0.039284in}{0.039284in}}%
\pgfpathcurveto{\pgfqpoint{-0.049702in}{0.028866in}}{\pgfqpoint{-0.055556in}{0.014734in}}{\pgfqpoint{-0.055556in}{0.000000in}}%
\pgfpathcurveto{\pgfqpoint{-0.055556in}{-0.014734in}}{\pgfqpoint{-0.049702in}{-0.028866in}}{\pgfqpoint{-0.039284in}{-0.039284in}}%
\pgfpathcurveto{\pgfqpoint{-0.028866in}{-0.049702in}}{\pgfqpoint{-0.014734in}{-0.055556in}}{\pgfqpoint{0.000000in}{-0.055556in}}%
\pgfpathlineto{\pgfqpoint{0.000000in}{-0.055556in}}%
\pgfpathclose%
\pgfusepath{stroke,fill}%
}%
\begin{pgfscope}%
\pgfsys@transformshift{2.421647in}{2.751948in}%
\pgfsys@useobject{currentmarker}{}%
\end{pgfscope}%
\end{pgfscope}%
\begin{pgfscope}%
\definecolor{textcolor}{rgb}{0.000000,0.000000,0.000000}%
\pgfsetstrokecolor{textcolor}%
\pgfsetfillcolor{textcolor}%
\pgftext[x=3.021647in,y=2.635281in,left,base]{\color{textcolor}{\rmfamily\fontsize{24.000000}{28.800000}\selectfont\catcode`\^=\active\def^{\ifmmode\sp\else\^{}\fi}\catcode`\%=\active\def
\end{pgfscope}%
\end{pgfpicture}%
\makeatother%
\endgroup%

%% file: figures/SE_est_match/WERM_SE_match_est_logistic_Lub_L2_new5_size.pgf
\begingroup%
\makeatletter%
\begin{pgfpicture}%
\pgfpathrectangle{\pgfpointorigin}{\pgfqpoint{4.757956in}{3.743420in}}%
\pgfusepath{use as bounding box, clip}%
\begin{pgfscope}%
\pgfsetbuttcap%
\pgfsetmiterjoin%
\definecolor{currentfill}{rgb}{1.000000,1.000000,1.000000}%
\pgfsetfillcolor{currentfill}%
\pgfsetlinewidth{0.000000pt}%
\definecolor{currentstroke}{rgb}{1.000000,1.000000,1.000000}%
\pgfsetstrokecolor{currentstroke}%
\pgfsetdash{}{0pt}%
\pgfpathmoveto{\pgfqpoint{0.000000in}{0.000000in}}%
\pgfpathlineto{\pgfqpoint{4.757956in}{0.000000in}}%
\pgfpathlineto{\pgfqpoint{4.757956in}{3.743420in}}%
\pgfpathlineto{\pgfqpoint{0.000000in}{3.743420in}}%
\pgfpathlineto{\pgfqpoint{0.000000in}{0.000000in}}%
\pgfpathclose%
\pgfusepath{fill}%
\end{pgfscope}%
\begin{pgfscope}%
\pgfsetbuttcap%
\pgfsetmiterjoin%
\definecolor{currentfill}{rgb}{1.000000,1.000000,1.000000}%
\pgfsetfillcolor{currentfill}%
\pgfsetlinewidth{0.000000pt}%
\definecolor{currentstroke}{rgb}{0.000000,0.000000,0.000000}%
\pgfsetstrokecolor{currentstroke}%
\pgfsetstrokeopacity{0.000000}%
\pgfsetdash{}{0pt}%
\pgfpathmoveto{\pgfqpoint{0.675675in}{0.831623in}}%
\pgfpathlineto{\pgfqpoint{4.657956in}{0.831623in}}%
\pgfpathlineto{\pgfqpoint{4.657956in}{3.543400in}}%
\pgfpathlineto{\pgfqpoint{0.675675in}{3.543400in}}%
\pgfpathlineto{\pgfqpoint{0.675675in}{0.831623in}}%
\pgfpathclose%
\pgfusepath{fill}%
\end{pgfscope}%
\begin{pgfscope}%
\pgfpathrectangle{\pgfqpoint{0.675675in}{0.831623in}}{\pgfqpoint{3.982281in}{2.711777in}}%
\pgfusepath{clip}%
\pgfsetbuttcap%
\pgfsetroundjoin%
\pgfsetlinewidth{3.513125pt}%
\definecolor{currentstroke}{rgb}{0.000000,0.000000,0.000000}%
\pgfsetstrokecolor{currentstroke}%
\pgfsetstrokeopacity{0.800000}%
\pgfsetdash{}{0pt}%
\pgfpathmoveto{\pgfqpoint{0.856688in}{2.187512in}}%
\pgfpathlineto{\pgfqpoint{0.856688in}{2.187512in}}%
\pgfusepath{stroke}%
\end{pgfscope}%
\begin{pgfscope}%
\pgfpathrectangle{\pgfqpoint{0.675675in}{0.831623in}}{\pgfqpoint{3.982281in}{2.711777in}}%
\pgfusepath{clip}%
\pgfsetbuttcap%
\pgfsetroundjoin%
\pgfsetlinewidth{3.513125pt}%
\definecolor{currentstroke}{rgb}{0.000000,0.000000,0.000000}%
\pgfsetstrokecolor{currentstroke}%
\pgfsetstrokeopacity{0.800000}%
\pgfsetdash{}{0pt}%
\pgfpathmoveto{\pgfqpoint{1.761752in}{2.187512in}}%
\pgfpathlineto{\pgfqpoint{1.761752in}{2.187512in}}%
\pgfusepath{stroke}%
\end{pgfscope}%
\begin{pgfscope}%
\pgfpathrectangle{\pgfqpoint{0.675675in}{0.831623in}}{\pgfqpoint{3.982281in}{2.711777in}}%
\pgfusepath{clip}%
\pgfsetbuttcap%
\pgfsetroundjoin%
\pgfsetlinewidth{3.513125pt}%
\definecolor{currentstroke}{rgb}{0.000000,0.000000,0.000000}%
\pgfsetstrokecolor{currentstroke}%
\pgfsetstrokeopacity{0.800000}%
\pgfsetdash{}{0pt}%
\pgfpathmoveto{\pgfqpoint{2.666815in}{2.187512in}}%
\pgfpathlineto{\pgfqpoint{2.666815in}{2.187512in}}%
\pgfusepath{stroke}%
\end{pgfscope}%
\begin{pgfscope}%
\pgfpathrectangle{\pgfqpoint{0.675675in}{0.831623in}}{\pgfqpoint{3.982281in}{2.711777in}}%
\pgfusepath{clip}%
\pgfsetbuttcap%
\pgfsetroundjoin%
\pgfsetlinewidth{3.513125pt}%
\definecolor{currentstroke}{rgb}{0.000000,0.000000,0.000000}%
\pgfsetstrokecolor{currentstroke}%
\pgfsetstrokeopacity{0.800000}%
\pgfsetdash{}{0pt}%
\pgfpathmoveto{\pgfqpoint{3.571879in}{2.187512in}}%
\pgfpathlineto{\pgfqpoint{3.571879in}{2.187512in}}%
\pgfusepath{stroke}%
\end{pgfscope}%
\begin{pgfscope}%
\pgfpathrectangle{\pgfqpoint{0.675675in}{0.831623in}}{\pgfqpoint{3.982281in}{2.711777in}}%
\pgfusepath{clip}%
\pgfsetbuttcap%
\pgfsetroundjoin%
\pgfsetlinewidth{3.513125pt}%
\definecolor{currentstroke}{rgb}{0.000000,0.000000,0.000000}%
\pgfsetstrokecolor{currentstroke}%
\pgfsetstrokeopacity{0.800000}%
\pgfsetdash{}{0pt}%
\pgfpathmoveto{\pgfqpoint{4.476943in}{2.187512in}}%
\pgfpathlineto{\pgfqpoint{4.476943in}{2.187512in}}%
\pgfusepath{stroke}%
\end{pgfscope}%
\begin{pgfscope}%
\pgfpathrectangle{\pgfqpoint{0.675675in}{0.831623in}}{\pgfqpoint{3.982281in}{2.711777in}}%
\pgfusepath{clip}%
\pgfsetbuttcap%
\pgfsetroundjoin%
\definecolor{currentfill}{rgb}{0.000000,0.000000,0.000000}%
\pgfsetfillcolor{currentfill}%
\pgfsetfillopacity{0.800000}%
\pgfsetlinewidth{2.007500pt}%
\definecolor{currentstroke}{rgb}{0.000000,0.000000,0.000000}%
\pgfsetstrokecolor{currentstroke}%
\pgfsetstrokeopacity{0.800000}%
\pgfsetdash{}{0pt}%
\pgfsys@defobject{currentmarker}{\pgfqpoint{-0.055556in}{-0.000000in}}{\pgfqpoint{0.055556in}{0.000000in}}{%
\pgfpathmoveto{\pgfqpoint{0.055556in}{-0.000000in}}%
\pgfpathlineto{\pgfqpoint{-0.055556in}{0.000000in}}%
\pgfusepath{stroke,fill}%
}%
\begin{pgfscope}%
\pgfsys@transformshift{0.856688in}{2.187512in}%
\pgfsys@useobject{currentmarker}{}%
\end{pgfscope}%
\begin{pgfscope}%
\pgfsys@transformshift{1.761752in}{2.187512in}%
\pgfsys@useobject{currentmarker}{}%
\end{pgfscope}%
\begin{pgfscope}%
\pgfsys@transformshift{2.666815in}{2.187512in}%
\pgfsys@useobject{currentmarker}{}%
\end{pgfscope}%
\begin{pgfscope}%
\pgfsys@transformshift{3.571879in}{2.187512in}%
\pgfsys@useobject{currentmarker}{}%
\end{pgfscope}%
\begin{pgfscope}%
\pgfsys@transformshift{4.476943in}{2.187512in}%
\pgfsys@useobject{currentmarker}{}%
\end{pgfscope}%
\end{pgfscope}%
\begin{pgfscope}%
\pgfpathrectangle{\pgfqpoint{0.675675in}{0.831623in}}{\pgfqpoint{3.982281in}{2.711777in}}%
\pgfusepath{clip}%
\pgfsetbuttcap%
\pgfsetroundjoin%
\definecolor{currentfill}{rgb}{0.000000,0.000000,0.000000}%
\pgfsetfillcolor{currentfill}%
\pgfsetfillopacity{0.800000}%
\pgfsetlinewidth{2.007500pt}%
\definecolor{currentstroke}{rgb}{0.000000,0.000000,0.000000}%
\pgfsetstrokecolor{currentstroke}%
\pgfsetstrokeopacity{0.800000}%
\pgfsetdash{}{0pt}%
\pgfsys@defobject{currentmarker}{\pgfqpoint{-0.055556in}{-0.000000in}}{\pgfqpoint{0.055556in}{0.000000in}}{%
\pgfpathmoveto{\pgfqpoint{0.055556in}{-0.000000in}}%
\pgfpathlineto{\pgfqpoint{-0.055556in}{0.000000in}}%
\pgfusepath{stroke,fill}%
}%
\begin{pgfscope}%
\pgfsys@transformshift{0.856688in}{2.187512in}%
\pgfsys@useobject{currentmarker}{}%
\end{pgfscope}%
\begin{pgfscope}%
\pgfsys@transformshift{1.761752in}{2.187512in}%
\pgfsys@useobject{currentmarker}{}%
\end{pgfscope}%
\begin{pgfscope}%
\pgfsys@transformshift{2.666815in}{2.187512in}%
\pgfsys@useobject{currentmarker}{}%
\end{pgfscope}%
\begin{pgfscope}%
\pgfsys@transformshift{3.571879in}{2.187512in}%
\pgfsys@useobject{currentmarker}{}%
\end{pgfscope}%
\begin{pgfscope}%
\pgfsys@transformshift{4.476943in}{2.187512in}%
\pgfsys@useobject{currentmarker}{}%
\end{pgfscope}%
\end{pgfscope}%
\begin{pgfscope}%
\pgfpathrectangle{\pgfqpoint{0.675675in}{0.831623in}}{\pgfqpoint{3.982281in}{2.711777in}}%
\pgfusepath{clip}%
\pgfsetrectcap%
\pgfsetroundjoin%
\pgfsetlinewidth{3.513125pt}%
\definecolor{currentstroke}{rgb}{0.000000,0.000000,0.000000}%
\pgfsetstrokecolor{currentstroke}%
\pgfsetstrokeopacity{0.800000}%
\pgfsetdash{}{0pt}%
\pgfpathmoveto{\pgfqpoint{0.856688in}{2.187512in}}%
\pgfpathlineto{\pgfqpoint{1.761752in}{2.187512in}}%
\pgfpathlineto{\pgfqpoint{2.666815in}{2.187512in}}%
\pgfpathlineto{\pgfqpoint{3.571879in}{2.187512in}}%
\pgfpathlineto{\pgfqpoint{4.476943in}{2.187512in}}%
\pgfusepath{stroke}%
\end{pgfscope}%
\begin{pgfscope}%
\pgfpathrectangle{\pgfqpoint{0.675675in}{0.831623in}}{\pgfqpoint{3.982281in}{2.711777in}}%
\pgfusepath{clip}%
\pgfsetbuttcap%
\pgfsetroundjoin%
\definecolor{currentfill}{rgb}{0.000000,0.000000,0.000000}%
\pgfsetfillcolor{currentfill}%
\pgfsetfillopacity{0.800000}%
\pgfsetlinewidth{2.007500pt}%
\definecolor{currentstroke}{rgb}{0.000000,0.000000,0.000000}%
\pgfsetstrokecolor{currentstroke}%
\pgfsetstrokeopacity{0.800000}%
\pgfsetdash{}{0pt}%
\pgfsys@defobject{currentmarker}{\pgfqpoint{-0.055556in}{-0.055556in}}{\pgfqpoint{0.055556in}{0.055556in}}{%
\pgfpathmoveto{\pgfqpoint{0.000000in}{-0.055556in}}%
\pgfpathcurveto{\pgfqpoint{0.014734in}{-0.055556in}}{\pgfqpoint{0.028866in}{-0.049702in}}{\pgfqpoint{0.039284in}{-0.039284in}}%
\pgfpathcurveto{\pgfqpoint{0.049702in}{-0.028866in}}{\pgfqpoint{0.055556in}{-0.014734in}}{\pgfqpoint{0.055556in}{0.000000in}}%
\pgfpathcurveto{\pgfqpoint{0.055556in}{0.014734in}}{\pgfqpoint{0.049702in}{0.028866in}}{\pgfqpoint{0.039284in}{0.039284in}}%
\pgfpathcurveto{\pgfqpoint{0.028866in}{0.049702in}}{\pgfqpoint{0.014734in}{0.055556in}}{\pgfqpoint{0.000000in}{0.055556in}}%
\pgfpathcurveto{\pgfqpoint{-0.014734in}{0.055556in}}{\pgfqpoint{-0.028866in}{0.049702in}}{\pgfqpoint{-0.039284in}{0.039284in}}%
\pgfpathcurveto{\pgfqpoint{-0.049702in}{0.028866in}}{\pgfqpoint{-0.055556in}{0.014734in}}{\pgfqpoint{-0.055556in}{0.000000in}}%
\pgfpathcurveto{\pgfqpoint{-0.055556in}{-0.014734in}}{\pgfqpoint{-0.049702in}{-0.028866in}}{\pgfqpoint{-0.039284in}{-0.039284in}}%
\pgfpathcurveto{\pgfqpoint{-0.028866in}{-0.049702in}}{\pgfqpoint{-0.014734in}{-0.055556in}}{\pgfqpoint{0.000000in}{-0.055556in}}%
\pgfpathlineto{\pgfqpoint{0.000000in}{-0.055556in}}%
\pgfpathclose%
\pgfusepath{stroke,fill}%
}%
\begin{pgfscope}%
\pgfsys@transformshift{0.856688in}{2.187512in}%
\pgfsys@useobject{currentmarker}{}%
\end{pgfscope}%
\begin{pgfscope}%
\pgfsys@transformshift{1.761752in}{2.187512in}%
\pgfsys@useobject{currentmarker}{}%
\end{pgfscope}%
\begin{pgfscope}%
\pgfsys@transformshift{2.666815in}{2.187512in}%
\pgfsys@useobject{currentmarker}{}%
\end{pgfscope}%
\begin{pgfscope}%
\pgfsys@transformshift{3.571879in}{2.187512in}%
\pgfsys@useobject{currentmarker}{}%
\end{pgfscope}%
\begin{pgfscope}%
\pgfsys@transformshift{4.476943in}{2.187512in}%
\pgfsys@useobject{currentmarker}{}%
\end{pgfscope}%
\end{pgfscope}%
\begin{pgfscope}%
\pgfpathrectangle{\pgfqpoint{0.675675in}{0.831623in}}{\pgfqpoint{3.982281in}{2.711777in}}%
\pgfusepath{clip}%
\pgfsetbuttcap%
\pgfsetroundjoin%
\definecolor{currentfill}{rgb}{0.200000,0.733333,0.933333}%
\pgfsetfillcolor{currentfill}%
\pgfsetfillopacity{0.990000}%
\pgfsetlinewidth{2.509375pt}%
\definecolor{currentstroke}{rgb}{1.000000,1.000000,1.000000}%
\pgfsetstrokecolor{currentstroke}%
\pgfsetstrokeopacity{0.990000}%
\pgfsetdash{}{0pt}%
\pgfsys@defobject{currentmarker}{\pgfqpoint{-0.064550in}{-0.107583in}}{\pgfqpoint{0.064550in}{0.107583in}}{%
\pgfpathmoveto{\pgfqpoint{-0.000000in}{-0.107583in}}%
\pgfpathlineto{\pgfqpoint{0.064550in}{0.000000in}}%
\pgfpathlineto{\pgfqpoint{0.000000in}{0.107583in}}%
\pgfpathlineto{\pgfqpoint{-0.064550in}{0.000000in}}%
\pgfpathlineto{\pgfqpoint{-0.000000in}{-0.107583in}}%
\pgfpathclose%
\pgfusepath{stroke,fill}%
}%
\begin{pgfscope}%
\pgfsys@transformshift{0.856688in}{2.187512in}%
\pgfsys@useobject{currentmarker}{}%
\end{pgfscope}%
\begin{pgfscope}%
\pgfsys@transformshift{1.761752in}{2.187512in}%
\pgfsys@useobject{currentmarker}{}%
\end{pgfscope}%
\begin{pgfscope}%
\pgfsys@transformshift{2.666815in}{2.187512in}%
\pgfsys@useobject{currentmarker}{}%
\end{pgfscope}%
\begin{pgfscope}%
\pgfsys@transformshift{3.571879in}{2.187512in}%
\pgfsys@useobject{currentmarker}{}%
\end{pgfscope}%
\begin{pgfscope}%
\pgfsys@transformshift{4.476943in}{2.187512in}%
\pgfsys@useobject{currentmarker}{}%
\end{pgfscope}%
\end{pgfscope}%
\begin{pgfscope}%
\pgfsetbuttcap%
\pgfsetroundjoin%
\definecolor{currentfill}{rgb}{0.000000,0.000000,0.000000}%
\pgfsetfillcolor{currentfill}%
\pgfsetlinewidth{1.505625pt}%
\definecolor{currentstroke}{rgb}{0.000000,0.000000,0.000000}%
\pgfsetstrokecolor{currentstroke}%
\pgfsetdash{}{0pt}%
\pgfsys@defobject{currentmarker}{\pgfqpoint{0.000000in}{-0.083333in}}{\pgfqpoint{0.000000in}{0.000000in}}{%
\pgfpathmoveto{\pgfqpoint{0.000000in}{0.000000in}}%
\pgfpathlineto{\pgfqpoint{0.000000in}{-0.083333in}}%
\pgfusepath{stroke,fill}%
}%
\begin{pgfscope}%
\pgfsys@transformshift{0.856688in}{0.831623in}%
\pgfsys@useobject{currentmarker}{}%
\end{pgfscope}%
\end{pgfscope}%
\begin{pgfscope}%
\definecolor{textcolor}{rgb}{0.000000,0.000000,0.000000}%
\pgfsetstrokecolor{textcolor}%
\pgfsetfillcolor{textcolor}%
\pgftext[x=0.856688in,y=0.699679in,,top]{\color{textcolor}{\rmfamily\fontsize{22.000000}{26.400000}\selectfont\catcode`\^=\active\def^{\ifmmode\sp\else\^{}\fi}\catcode`\%=\active\def
\end{pgfscope}%
\begin{pgfscope}%
\pgfsetbuttcap%
\pgfsetroundjoin%
\definecolor{currentfill}{rgb}{0.000000,0.000000,0.000000}%
\pgfsetfillcolor{currentfill}%
\pgfsetlinewidth{1.505625pt}%
\definecolor{currentstroke}{rgb}{0.000000,0.000000,0.000000}%
\pgfsetstrokecolor{currentstroke}%
\pgfsetdash{}{0pt}%
\pgfsys@defobject{currentmarker}{\pgfqpoint{0.000000in}{-0.083333in}}{\pgfqpoint{0.000000in}{0.000000in}}{%
\pgfpathmoveto{\pgfqpoint{0.000000in}{0.000000in}}%
\pgfpathlineto{\pgfqpoint{0.000000in}{-0.083333in}}%
\pgfusepath{stroke,fill}%
}%
\begin{pgfscope}%
\pgfsys@transformshift{2.666815in}{0.831623in}%
\pgfsys@useobject{currentmarker}{}%
\end{pgfscope}%
\end{pgfscope}%
\begin{pgfscope}%
\definecolor{textcolor}{rgb}{0.000000,0.000000,0.000000}%
\pgfsetstrokecolor{textcolor}%
\pgfsetfillcolor{textcolor}%
\pgftext[x=2.666815in,y=0.699679in,,top]{\color{textcolor}{\rmfamily\fontsize{22.000000}{26.400000}\selectfont\catcode`\^=\active\def^{\ifmmode\sp\else\^{}\fi}\catcode`\%=\active\def
\end{pgfscope}%
\begin{pgfscope}%
\pgfsetbuttcap%
\pgfsetroundjoin%
\definecolor{currentfill}{rgb}{0.000000,0.000000,0.000000}%
\pgfsetfillcolor{currentfill}%
\pgfsetlinewidth{1.505625pt}%
\definecolor{currentstroke}{rgb}{0.000000,0.000000,0.000000}%
\pgfsetstrokecolor{currentstroke}%
\pgfsetdash{}{0pt}%
\pgfsys@defobject{currentmarker}{\pgfqpoint{0.000000in}{-0.083333in}}{\pgfqpoint{0.000000in}{0.000000in}}{%
\pgfpathmoveto{\pgfqpoint{0.000000in}{0.000000in}}%
\pgfpathlineto{\pgfqpoint{0.000000in}{-0.083333in}}%
\pgfusepath{stroke,fill}%
}%
\begin{pgfscope}%
\pgfsys@transformshift{4.476943in}{0.831623in}%
\pgfsys@useobject{currentmarker}{}%
\end{pgfscope}%
\end{pgfscope}%
\begin{pgfscope}%
\definecolor{textcolor}{rgb}{0.000000,0.000000,0.000000}%
\pgfsetstrokecolor{textcolor}%
\pgfsetfillcolor{textcolor}%
\pgftext[x=4.476943in,y=0.699679in,,top]{\color{textcolor}{\rmfamily\fontsize{22.000000}{26.400000}\selectfont\catcode`\^=\active\def^{\ifmmode\sp\else\^{}\fi}\catcode`\%=\active\def
\end{pgfscope}%
\begin{pgfscope}%
\definecolor{textcolor}{rgb}{0.000000,0.000000,0.000000}%
\pgfsetstrokecolor{textcolor}%
\pgfsetfillcolor{textcolor}%
\pgftext[x=2.666815in,y=0.388056in,,top]{\color{textcolor}{\rmfamily\fontsize{22.000000}{26.400000}\selectfont\catcode`\^=\active\def^{\ifmmode\sp\else\^{}\fi}\catcode`\%=\active\def
\end{pgfscope}%
\begin{pgfscope}%
\pgfsetbuttcap%
\pgfsetroundjoin%
\definecolor{currentfill}{rgb}{0.000000,0.000000,0.000000}%
\pgfsetfillcolor{currentfill}%
\pgfsetlinewidth{1.505625pt}%
\definecolor{currentstroke}{rgb}{0.000000,0.000000,0.000000}%
\pgfsetstrokecolor{currentstroke}%
\pgfsetdash{}{0pt}%
\pgfsys@defobject{currentmarker}{\pgfqpoint{-0.083333in}{0.000000in}}{\pgfqpoint{-0.000000in}{0.000000in}}{%
\pgfpathmoveto{\pgfqpoint{-0.000000in}{0.000000in}}%
\pgfpathlineto{\pgfqpoint{-0.083333in}{0.000000in}}%
\pgfusepath{stroke,fill}%
}%
\begin{pgfscope}%
\pgfsys@transformshift{0.675675in}{0.831623in}%
\pgfsys@useobject{currentmarker}{}%
\end{pgfscope}%
\end{pgfscope}%
\begin{pgfscope}%
\definecolor{textcolor}{rgb}{0.000000,0.000000,0.000000}%
\pgfsetstrokecolor{textcolor}%
\pgfsetfillcolor{textcolor}%
\pgftext[x=0.411623in, y=0.731604in, left, base]{\color{textcolor}{\rmfamily\fontsize{22.000000}{26.400000}\selectfont\catcode`\^=\active\def^{\ifmmode\sp\else\^{}\fi}\catcode`\%=\active\def
\end{pgfscope}%
\begin{pgfscope}%
\pgfsetbuttcap%
\pgfsetroundjoin%
\definecolor{currentfill}{rgb}{0.000000,0.000000,0.000000}%
\pgfsetfillcolor{currentfill}%
\pgfsetlinewidth{1.505625pt}%
\definecolor{currentstroke}{rgb}{0.000000,0.000000,0.000000}%
\pgfsetstrokecolor{currentstroke}%
\pgfsetdash{}{0pt}%
\pgfsys@defobject{currentmarker}{\pgfqpoint{-0.083333in}{0.000000in}}{\pgfqpoint{-0.000000in}{0.000000in}}{%
\pgfpathmoveto{\pgfqpoint{-0.000000in}{0.000000in}}%
\pgfpathlineto{\pgfqpoint{-0.083333in}{0.000000in}}%
\pgfusepath{stroke,fill}%
}%
\begin{pgfscope}%
\pgfsys@transformshift{0.675675in}{2.187512in}%
\pgfsys@useobject{currentmarker}{}%
\end{pgfscope}%
\end{pgfscope}%
\begin{pgfscope}%
\definecolor{textcolor}{rgb}{0.000000,0.000000,0.000000}%
\pgfsetstrokecolor{textcolor}%
\pgfsetfillcolor{textcolor}%
\pgftext[x=0.411623in, y=2.087493in, left, base]{\color{textcolor}{\rmfamily\fontsize{22.000000}{26.400000}\selectfont\catcode`\^=\active\def^{\ifmmode\sp\else\^{}\fi}\catcode`\%=\active\def
\end{pgfscope}%
\begin{pgfscope}%
\pgfsetbuttcap%
\pgfsetroundjoin%
\definecolor{currentfill}{rgb}{0.000000,0.000000,0.000000}%
\pgfsetfillcolor{currentfill}%
\pgfsetlinewidth{1.505625pt}%
\definecolor{currentstroke}{rgb}{0.000000,0.000000,0.000000}%
\pgfsetstrokecolor{currentstroke}%
\pgfsetdash{}{0pt}%
\pgfsys@defobject{currentmarker}{\pgfqpoint{-0.083333in}{0.000000in}}{\pgfqpoint{-0.000000in}{0.000000in}}{%
\pgfpathmoveto{\pgfqpoint{-0.000000in}{0.000000in}}%
\pgfpathlineto{\pgfqpoint{-0.083333in}{0.000000in}}%
\pgfusepath{stroke,fill}%
}%
\begin{pgfscope}%
\pgfsys@transformshift{0.675675in}{3.543400in}%
\pgfsys@useobject{currentmarker}{}%
\end{pgfscope}%
\end{pgfscope}%
\begin{pgfscope}%
\definecolor{textcolor}{rgb}{0.000000,0.000000,0.000000}%
\pgfsetstrokecolor{textcolor}%
\pgfsetfillcolor{textcolor}%
\pgftext[x=0.411623in, y=3.443381in, left, base]{\color{textcolor}{\rmfamily\fontsize{22.000000}{26.400000}\selectfont\catcode`\^=\active\def^{\ifmmode\sp\else\^{}\fi}\catcode`\%=\active\def
\end{pgfscope}%
\begin{pgfscope}%
\definecolor{textcolor}{rgb}{0.000000,0.000000,0.000000}%
\pgfsetstrokecolor{textcolor}%
\pgfsetfillcolor{textcolor}%
\pgftext[x=0.356068in,y=2.187512in,,bottom,rotate=90.000000]{\color{textcolor}{\rmfamily\fontsize{22.000000}{26.400000}\selectfont\catcode`\^=\active\def^{\ifmmode\sp\else\^{}\fi}\catcode`\%=\active\def
\end{pgfscope}%
\begin{pgfscope}%
\pgfsetrectcap%
\pgfsetmiterjoin%
\pgfsetlinewidth{1.505625pt}%
\definecolor{currentstroke}{rgb}{0.000000,0.000000,0.000000}%
\pgfsetstrokecolor{currentstroke}%
\pgfsetdash{}{0pt}%
\pgfpathmoveto{\pgfqpoint{0.675675in}{0.831623in}}%
\pgfpathlineto{\pgfqpoint{0.675675in}{3.543400in}}%
\pgfusepath{stroke}%
\end{pgfscope}%
\begin{pgfscope}%
\pgfsetrectcap%
\pgfsetmiterjoin%
\pgfsetlinewidth{1.505625pt}%
\definecolor{currentstroke}{rgb}{0.000000,0.000000,0.000000}%
\pgfsetstrokecolor{currentstroke}%
\pgfsetdash{}{0pt}%
\pgfpathmoveto{\pgfqpoint{0.675675in}{0.831623in}}%
\pgfpathlineto{\pgfqpoint{4.657956in}{0.831623in}}%
\pgfusepath{stroke}%
\end{pgfscope}%
\end{pgfpicture}%
\makeatother%
\endgroup%

%% file: figures/synthetic-final/M2/location.pgf
\begingroup%
\makeatletter%
\begin{pgfpicture}%
\pgfpathrectangle{\pgfpointorigin}{\pgfqpoint{4.740877in}{3.749100in}}%
\pgfusepath{use as bounding box, clip}%
\begin{pgfscope}%
\pgfsetbuttcap%
\pgfsetmiterjoin%
\definecolor{currentfill}{rgb}{1.000000,1.000000,1.000000}%
\pgfsetfillcolor{currentfill}%
\pgfsetlinewidth{0.000000pt}%
\definecolor{currentstroke}{rgb}{1.000000,1.000000,1.000000}%
\pgfsetstrokecolor{currentstroke}%
\pgfsetdash{}{0pt}%
\pgfpathmoveto{\pgfqpoint{0.000000in}{0.000000in}}%
\pgfpathlineto{\pgfqpoint{4.740877in}{0.000000in}}%
\pgfpathlineto{\pgfqpoint{4.740877in}{3.749100in}}%
\pgfpathlineto{\pgfqpoint{0.000000in}{3.749100in}}%
\pgfpathlineto{\pgfqpoint{0.000000in}{0.000000in}}%
\pgfpathclose%
\pgfusepath{fill}%
\end{pgfscope}%
\begin{pgfscope}%
\pgfsetbuttcap%
\pgfsetmiterjoin%
\definecolor{currentfill}{rgb}{1.000000,1.000000,1.000000}%
\pgfsetfillcolor{currentfill}%
\pgfsetlinewidth{0.000000pt}%
\definecolor{currentstroke}{rgb}{0.000000,0.000000,0.000000}%
\pgfsetstrokecolor{currentstroke}%
\pgfsetstrokeopacity{0.000000}%
\pgfsetdash{}{0pt}%
\pgfpathmoveto{\pgfqpoint{0.917618in}{0.831623in}}%
\pgfpathlineto{\pgfqpoint{4.569539in}{0.831623in}}%
\pgfpathlineto{\pgfqpoint{4.569539in}{3.649100in}}%
\pgfpathlineto{\pgfqpoint{0.917618in}{3.649100in}}%
\pgfpathlineto{\pgfqpoint{0.917618in}{0.831623in}}%
\pgfpathclose%
\pgfusepath{fill}%
\end{pgfscope}%
\begin{pgfscope}%
\pgfsetbuttcap%
\pgfsetroundjoin%
\definecolor{currentfill}{rgb}{0.000000,0.000000,0.000000}%
\pgfsetfillcolor{currentfill}%
\pgfsetlinewidth{1.505625pt}%
\definecolor{currentstroke}{rgb}{0.000000,0.000000,0.000000}%
\pgfsetstrokecolor{currentstroke}%
\pgfsetdash{}{0pt}%
\pgfsys@defobject{currentmarker}{\pgfqpoint{0.000000in}{-0.083333in}}{\pgfqpoint{0.000000in}{0.000000in}}{%
\pgfpathmoveto{\pgfqpoint{0.000000in}{0.000000in}}%
\pgfpathlineto{\pgfqpoint{0.000000in}{-0.083333in}}%
\pgfusepath{stroke,fill}%
}%
\begin{pgfscope}%
\pgfsys@transformshift{1.083614in}{0.831623in}%
\pgfsys@useobject{currentmarker}{}%
\end{pgfscope}%
\end{pgfscope}%
\begin{pgfscope}%
\definecolor{textcolor}{rgb}{0.000000,0.000000,0.000000}%
\pgfsetstrokecolor{textcolor}%
\pgfsetfillcolor{textcolor}%
\pgftext[x=1.083614in,y=0.699679in,,top]{\color{textcolor}{\rmfamily\fontsize{22.000000}{26.400000}\selectfont\catcode`\^=\active\def^{\ifmmode\sp\else\^{}\fi}\catcode`\%=\active\def
\end{pgfscope}%
\begin{pgfscope}%
\pgfsetbuttcap%
\pgfsetroundjoin%
\definecolor{currentfill}{rgb}{0.000000,0.000000,0.000000}%
\pgfsetfillcolor{currentfill}%
\pgfsetlinewidth{1.505625pt}%
\definecolor{currentstroke}{rgb}{0.000000,0.000000,0.000000}%
\pgfsetstrokecolor{currentstroke}%
\pgfsetdash{}{0pt}%
\pgfsys@defobject{currentmarker}{\pgfqpoint{0.000000in}{-0.083333in}}{\pgfqpoint{0.000000in}{0.000000in}}{%
\pgfpathmoveto{\pgfqpoint{0.000000in}{0.000000in}}%
\pgfpathlineto{\pgfqpoint{0.000000in}{-0.083333in}}%
\pgfusepath{stroke,fill}%
}%
\begin{pgfscope}%
\pgfsys@transformshift{2.190257in}{0.831623in}%
\pgfsys@useobject{currentmarker}{}%
\end{pgfscope}%
\end{pgfscope}%
\begin{pgfscope}%
\definecolor{textcolor}{rgb}{0.000000,0.000000,0.000000}%
\pgfsetstrokecolor{textcolor}%
\pgfsetfillcolor{textcolor}%
\pgftext[x=2.190257in,y=0.699679in,,top]{\color{textcolor}{\rmfamily\fontsize{22.000000}{26.400000}\selectfont\catcode`\^=\active\def^{\ifmmode\sp\else\^{}\fi}\catcode`\%=\active\def
\end{pgfscope}%
\begin{pgfscope}%
\pgfsetbuttcap%
\pgfsetroundjoin%
\definecolor{currentfill}{rgb}{0.000000,0.000000,0.000000}%
\pgfsetfillcolor{currentfill}%
\pgfsetlinewidth{1.505625pt}%
\definecolor{currentstroke}{rgb}{0.000000,0.000000,0.000000}%
\pgfsetstrokecolor{currentstroke}%
\pgfsetdash{}{0pt}%
\pgfsys@defobject{currentmarker}{\pgfqpoint{0.000000in}{-0.083333in}}{\pgfqpoint{0.000000in}{0.000000in}}{%
\pgfpathmoveto{\pgfqpoint{0.000000in}{0.000000in}}%
\pgfpathlineto{\pgfqpoint{0.000000in}{-0.083333in}}%
\pgfusepath{stroke,fill}%
}%
\begin{pgfscope}%
\pgfsys@transformshift{3.296900in}{0.831623in}%
\pgfsys@useobject{currentmarker}{}%
\end{pgfscope}%
\end{pgfscope}%
\begin{pgfscope}%
\definecolor{textcolor}{rgb}{0.000000,0.000000,0.000000}%
\pgfsetstrokecolor{textcolor}%
\pgfsetfillcolor{textcolor}%
\pgftext[x=3.296900in,y=0.699679in,,top]{\color{textcolor}{\rmfamily\fontsize{22.000000}{26.400000}\selectfont\catcode`\^=\active\def^{\ifmmode\sp\else\^{}\fi}\catcode`\%=\active\def
\end{pgfscope}%
\begin{pgfscope}%
\pgfsetbuttcap%
\pgfsetroundjoin%
\definecolor{currentfill}{rgb}{0.000000,0.000000,0.000000}%
\pgfsetfillcolor{currentfill}%
\pgfsetlinewidth{1.505625pt}%
\definecolor{currentstroke}{rgb}{0.000000,0.000000,0.000000}%
\pgfsetstrokecolor{currentstroke}%
\pgfsetdash{}{0pt}%
\pgfsys@defobject{currentmarker}{\pgfqpoint{0.000000in}{-0.083333in}}{\pgfqpoint{0.000000in}{0.000000in}}{%
\pgfpathmoveto{\pgfqpoint{0.000000in}{0.000000in}}%
\pgfpathlineto{\pgfqpoint{0.000000in}{-0.083333in}}%
\pgfusepath{stroke,fill}%
}%
\begin{pgfscope}%
\pgfsys@transformshift{4.403543in}{0.831623in}%
\pgfsys@useobject{currentmarker}{}%
\end{pgfscope}%
\end{pgfscope}%
\begin{pgfscope}%
\definecolor{textcolor}{rgb}{0.000000,0.000000,0.000000}%
\pgfsetstrokecolor{textcolor}%
\pgfsetfillcolor{textcolor}%
\pgftext[x=4.403543in,y=0.699679in,,top]{\color{textcolor}{\rmfamily\fontsize{22.000000}{26.400000}\selectfont\catcode`\^=\active\def^{\ifmmode\sp\else\^{}\fi}\catcode`\%=\active\def
\end{pgfscope}%
\begin{pgfscope}%
\definecolor{textcolor}{rgb}{0.000000,0.000000,0.000000}%
\pgfsetstrokecolor{textcolor}%
\pgfsetfillcolor{textcolor}%
\pgftext[x=2.743578in,y=0.388056in,,top]{\color{textcolor}{\rmfamily\fontsize{22.000000}{26.400000}\selectfont\catcode`\^=\active\def^{\ifmmode\sp\else\^{}\fi}\catcode`\%=\active\def
\end{pgfscope}%
\begin{pgfscope}%
\pgfsetbuttcap%
\pgfsetroundjoin%
\definecolor{currentfill}{rgb}{0.000000,0.000000,0.000000}%
\pgfsetfillcolor{currentfill}%
\pgfsetlinewidth{1.505625pt}%
\definecolor{currentstroke}{rgb}{0.000000,0.000000,0.000000}%
\pgfsetstrokecolor{currentstroke}%
\pgfsetdash{}{0pt}%
\pgfsys@defobject{currentmarker}{\pgfqpoint{-0.083333in}{0.000000in}}{\pgfqpoint{-0.000000in}{0.000000in}}{%
\pgfpathmoveto{\pgfqpoint{-0.000000in}{0.000000in}}%
\pgfpathlineto{\pgfqpoint{-0.083333in}{0.000000in}}%
\pgfusepath{stroke,fill}%
}%
\begin{pgfscope}%
\pgfsys@transformshift{0.917618in}{0.831623in}%
\pgfsys@useobject{currentmarker}{}%
\end{pgfscope}%
\end{pgfscope}%
\begin{pgfscope}%
\definecolor{textcolor}{rgb}{0.000000,0.000000,0.000000}%
\pgfsetstrokecolor{textcolor}%
\pgfsetfillcolor{textcolor}%
\pgftext[x=0.443111in, y=0.731604in, left, base]{\color{textcolor}{\rmfamily\fontsize{22.000000}{26.400000}\selectfont\catcode`\^=\active\def^{\ifmmode\sp\else\^{}\fi}\catcode`\%=\active\def
\end{pgfscope}%
\begin{pgfscope}%
\pgfsetbuttcap%
\pgfsetroundjoin%
\definecolor{currentfill}{rgb}{0.000000,0.000000,0.000000}%
\pgfsetfillcolor{currentfill}%
\pgfsetlinewidth{1.505625pt}%
\definecolor{currentstroke}{rgb}{0.000000,0.000000,0.000000}%
\pgfsetstrokecolor{currentstroke}%
\pgfsetdash{}{0pt}%
\pgfsys@defobject{currentmarker}{\pgfqpoint{-0.083333in}{0.000000in}}{\pgfqpoint{-0.000000in}{0.000000in}}{%
\pgfpathmoveto{\pgfqpoint{-0.000000in}{0.000000in}}%
\pgfpathlineto{\pgfqpoint{-0.083333in}{0.000000in}}%
\pgfusepath{stroke,fill}%
}%
\begin{pgfscope}%
\pgfsys@transformshift{0.917618in}{1.856160in}%
\pgfsys@useobject{currentmarker}{}%
\end{pgfscope}%
\end{pgfscope}%
\begin{pgfscope}%
\definecolor{textcolor}{rgb}{0.000000,0.000000,0.000000}%
\pgfsetstrokecolor{textcolor}%
\pgfsetfillcolor{textcolor}%
\pgftext[x=0.443111in, y=1.756141in, left, base]{\color{textcolor}{\rmfamily\fontsize{22.000000}{26.400000}\selectfont\catcode`\^=\active\def^{\ifmmode\sp\else\^{}\fi}\catcode`\%=\active\def
\end{pgfscope}%
\begin{pgfscope}%
\pgfsetbuttcap%
\pgfsetroundjoin%
\definecolor{currentfill}{rgb}{0.000000,0.000000,0.000000}%
\pgfsetfillcolor{currentfill}%
\pgfsetlinewidth{1.505625pt}%
\definecolor{currentstroke}{rgb}{0.000000,0.000000,0.000000}%
\pgfsetstrokecolor{currentstroke}%
\pgfsetdash{}{0pt}%
\pgfsys@defobject{currentmarker}{\pgfqpoint{-0.083333in}{0.000000in}}{\pgfqpoint{-0.000000in}{0.000000in}}{%
\pgfpathmoveto{\pgfqpoint{-0.000000in}{0.000000in}}%
\pgfpathlineto{\pgfqpoint{-0.083333in}{0.000000in}}%
\pgfusepath{stroke,fill}%
}%
\begin{pgfscope}%
\pgfsys@transformshift{0.917618in}{2.880697in}%
\pgfsys@useobject{currentmarker}{}%
\end{pgfscope}%
\end{pgfscope}%
\begin{pgfscope}%
\definecolor{textcolor}{rgb}{0.000000,0.000000,0.000000}%
\pgfsetstrokecolor{textcolor}%
\pgfsetfillcolor{textcolor}%
\pgftext[x=0.443111in, y=2.780678in, left, base]{\color{textcolor}{\rmfamily\fontsize{22.000000}{26.400000}\selectfont\catcode`\^=\active\def^{\ifmmode\sp\else\^{}\fi}\catcode`\%=\active\def
\end{pgfscope}%
\begin{pgfscope}%
\definecolor{textcolor}{rgb}{0.000000,0.000000,0.000000}%
\pgfsetstrokecolor{textcolor}%
\pgfsetfillcolor{textcolor}%
\pgftext[x=0.387555in,y=2.240361in,,bottom,rotate=90.000000]{\color{textcolor}{\rmfamily\fontsize{22.000000}{26.400000}\selectfont\catcode`\^=\active\def^{\ifmmode\sp\else\^{}\fi}\catcode`\%=\active\def
\end{pgfscope}%
\begin{pgfscope}%
\pgfpathrectangle{\pgfqpoint{0.917618in}{0.831623in}}{\pgfqpoint{3.651921in}{2.817477in}}%
\pgfusepath{clip}%
\pgfsetbuttcap%
\pgfsetroundjoin%
\pgfsetlinewidth{3.513125pt}%
\definecolor{currentstroke}{rgb}{0.666667,0.200000,0.466667}%
\pgfsetstrokecolor{currentstroke}%
\pgfsetstrokeopacity{0.800000}%
\pgfsetdash{}{0pt}%
\pgfpathmoveto{\pgfqpoint{1.083614in}{1.719555in}}%
\pgfpathlineto{\pgfqpoint{1.083614in}{2.714210in}}%
\pgfusepath{stroke}%
\end{pgfscope}%
\begin{pgfscope}%
\pgfpathrectangle{\pgfqpoint{0.917618in}{0.831623in}}{\pgfqpoint{3.651921in}{2.817477in}}%
\pgfusepath{clip}%
\pgfsetbuttcap%
\pgfsetroundjoin%
\pgfsetlinewidth{3.513125pt}%
\definecolor{currentstroke}{rgb}{0.666667,0.200000,0.466667}%
\pgfsetstrokecolor{currentstroke}%
\pgfsetstrokeopacity{0.800000}%
\pgfsetdash{}{0pt}%
\pgfpathmoveto{\pgfqpoint{1.636935in}{1.651253in}}%
\pgfpathlineto{\pgfqpoint{1.636935in}{2.783732in}}%
\pgfusepath{stroke}%
\end{pgfscope}%
\begin{pgfscope}%
\pgfpathrectangle{\pgfqpoint{0.917618in}{0.831623in}}{\pgfqpoint{3.651921in}{2.817477in}}%
\pgfusepath{clip}%
\pgfsetbuttcap%
\pgfsetroundjoin%
\pgfsetlinewidth{3.513125pt}%
\definecolor{currentstroke}{rgb}{0.666667,0.200000,0.466667}%
\pgfsetstrokecolor{currentstroke}%
\pgfsetstrokeopacity{0.800000}%
\pgfsetdash{}{0pt}%
\pgfpathmoveto{\pgfqpoint{2.190257in}{1.385805in}}%
\pgfpathlineto{\pgfqpoint{2.190257in}{2.595457in}}%
\pgfusepath{stroke}%
\end{pgfscope}%
\begin{pgfscope}%
\pgfpathrectangle{\pgfqpoint{0.917618in}{0.831623in}}{\pgfqpoint{3.651921in}{2.817477in}}%
\pgfusepath{clip}%
\pgfsetbuttcap%
\pgfsetroundjoin%
\pgfsetlinewidth{3.513125pt}%
\definecolor{currentstroke}{rgb}{0.666667,0.200000,0.466667}%
\pgfsetstrokecolor{currentstroke}%
\pgfsetstrokeopacity{0.800000}%
\pgfsetdash{}{0pt}%
\pgfpathmoveto{\pgfqpoint{2.743578in}{1.476228in}}%
\pgfpathlineto{\pgfqpoint{2.743578in}{2.516133in}}%
\pgfusepath{stroke}%
\end{pgfscope}%
\begin{pgfscope}%
\pgfpathrectangle{\pgfqpoint{0.917618in}{0.831623in}}{\pgfqpoint{3.651921in}{2.817477in}}%
\pgfusepath{clip}%
\pgfsetbuttcap%
\pgfsetroundjoin%
\pgfsetlinewidth{3.513125pt}%
\definecolor{currentstroke}{rgb}{0.666667,0.200000,0.466667}%
\pgfsetstrokecolor{currentstroke}%
\pgfsetstrokeopacity{0.800000}%
\pgfsetdash{}{0pt}%
\pgfpathmoveto{\pgfqpoint{3.296900in}{1.409273in}}%
\pgfpathlineto{\pgfqpoint{3.296900in}{2.580076in}}%
\pgfusepath{stroke}%
\end{pgfscope}%
\begin{pgfscope}%
\pgfpathrectangle{\pgfqpoint{0.917618in}{0.831623in}}{\pgfqpoint{3.651921in}{2.817477in}}%
\pgfusepath{clip}%
\pgfsetbuttcap%
\pgfsetroundjoin%
\pgfsetlinewidth{3.513125pt}%
\definecolor{currentstroke}{rgb}{0.666667,0.200000,0.466667}%
\pgfsetstrokecolor{currentstroke}%
\pgfsetstrokeopacity{0.800000}%
\pgfsetdash{}{0pt}%
\pgfpathmoveto{\pgfqpoint{3.850221in}{1.296563in}}%
\pgfpathlineto{\pgfqpoint{3.850221in}{2.560529in}}%
\pgfusepath{stroke}%
\end{pgfscope}%
\begin{pgfscope}%
\pgfpathrectangle{\pgfqpoint{0.917618in}{0.831623in}}{\pgfqpoint{3.651921in}{2.817477in}}%
\pgfusepath{clip}%
\pgfsetbuttcap%
\pgfsetroundjoin%
\pgfsetlinewidth{3.513125pt}%
\definecolor{currentstroke}{rgb}{0.666667,0.200000,0.466667}%
\pgfsetstrokecolor{currentstroke}%
\pgfsetstrokeopacity{0.800000}%
\pgfsetdash{}{0pt}%
\pgfpathmoveto{\pgfqpoint{4.403543in}{1.032262in}}%
\pgfpathlineto{\pgfqpoint{4.403543in}{2.159252in}}%
\pgfusepath{stroke}%
\end{pgfscope}%
\begin{pgfscope}%
\pgfpathrectangle{\pgfqpoint{0.917618in}{0.831623in}}{\pgfqpoint{3.651921in}{2.817477in}}%
\pgfusepath{clip}%
\pgfsetbuttcap%
\pgfsetroundjoin%
\definecolor{currentfill}{rgb}{0.666667,0.200000,0.466667}%
\pgfsetfillcolor{currentfill}%
\pgfsetfillopacity{0.800000}%
\pgfsetlinewidth{2.007500pt}%
\definecolor{currentstroke}{rgb}{0.666667,0.200000,0.466667}%
\pgfsetstrokecolor{currentstroke}%
\pgfsetstrokeopacity{0.800000}%
\pgfsetdash{}{0pt}%
\pgfsys@defobject{currentmarker}{\pgfqpoint{-0.055556in}{-0.000000in}}{\pgfqpoint{0.055556in}{0.000000in}}{%
\pgfpathmoveto{\pgfqpoint{0.055556in}{-0.000000in}}%
\pgfpathlineto{\pgfqpoint{-0.055556in}{0.000000in}}%
\pgfusepath{stroke,fill}%
}%
\begin{pgfscope}%
\pgfsys@transformshift{1.083614in}{1.719555in}%
\pgfsys@useobject{currentmarker}{}%
\end{pgfscope}%
\begin{pgfscope}%
\pgfsys@transformshift{1.636935in}{1.651253in}%
\pgfsys@useobject{currentmarker}{}%
\end{pgfscope}%
\begin{pgfscope}%
\pgfsys@transformshift{2.190257in}{1.385805in}%
\pgfsys@useobject{currentmarker}{}%
\end{pgfscope}%
\begin{pgfscope}%
\pgfsys@transformshift{2.743578in}{1.476228in}%
\pgfsys@useobject{currentmarker}{}%
\end{pgfscope}%
\begin{pgfscope}%
\pgfsys@transformshift{3.296900in}{1.409273in}%
\pgfsys@useobject{currentmarker}{}%
\end{pgfscope}%
\begin{pgfscope}%
\pgfsys@transformshift{3.850221in}{1.296563in}%
\pgfsys@useobject{currentmarker}{}%
\end{pgfscope}%
\begin{pgfscope}%
\pgfsys@transformshift{4.403543in}{1.032262in}%
\pgfsys@useobject{currentmarker}{}%
\end{pgfscope}%
\end{pgfscope}%
\begin{pgfscope}%
\pgfpathrectangle{\pgfqpoint{0.917618in}{0.831623in}}{\pgfqpoint{3.651921in}{2.817477in}}%
\pgfusepath{clip}%
\pgfsetbuttcap%
\pgfsetroundjoin%
\definecolor{currentfill}{rgb}{0.666667,0.200000,0.466667}%
\pgfsetfillcolor{currentfill}%
\pgfsetfillopacity{0.800000}%
\pgfsetlinewidth{2.007500pt}%
\definecolor{currentstroke}{rgb}{0.666667,0.200000,0.466667}%
\pgfsetstrokecolor{currentstroke}%
\pgfsetstrokeopacity{0.800000}%
\pgfsetdash{}{0pt}%
\pgfsys@defobject{currentmarker}{\pgfqpoint{-0.055556in}{-0.000000in}}{\pgfqpoint{0.055556in}{0.000000in}}{%
\pgfpathmoveto{\pgfqpoint{0.055556in}{-0.000000in}}%
\pgfpathlineto{\pgfqpoint{-0.055556in}{0.000000in}}%
\pgfusepath{stroke,fill}%
}%
\begin{pgfscope}%
\pgfsys@transformshift{1.083614in}{2.714210in}%
\pgfsys@useobject{currentmarker}{}%
\end{pgfscope}%
\begin{pgfscope}%
\pgfsys@transformshift{1.636935in}{2.783732in}%
\pgfsys@useobject{currentmarker}{}%
\end{pgfscope}%
\begin{pgfscope}%
\pgfsys@transformshift{2.190257in}{2.595457in}%
\pgfsys@useobject{currentmarker}{}%
\end{pgfscope}%
\begin{pgfscope}%
\pgfsys@transformshift{2.743578in}{2.516133in}%
\pgfsys@useobject{currentmarker}{}%
\end{pgfscope}%
\begin{pgfscope}%
\pgfsys@transformshift{3.296900in}{2.580076in}%
\pgfsys@useobject{currentmarker}{}%
\end{pgfscope}%
\begin{pgfscope}%
\pgfsys@transformshift{3.850221in}{2.560529in}%
\pgfsys@useobject{currentmarker}{}%
\end{pgfscope}%
\begin{pgfscope}%
\pgfsys@transformshift{4.403543in}{2.159252in}%
\pgfsys@useobject{currentmarker}{}%
\end{pgfscope}%
\end{pgfscope}%
\begin{pgfscope}%
\pgfpathrectangle{\pgfqpoint{0.917618in}{0.831623in}}{\pgfqpoint{3.651921in}{2.817477in}}%
\pgfusepath{clip}%
\pgfsetbuttcap%
\pgfsetroundjoin%
\pgfsetlinewidth{3.513125pt}%
\definecolor{currentstroke}{rgb}{0.200000,0.400000,0.466667}%
\pgfsetstrokecolor{currentstroke}%
\pgfsetstrokeopacity{0.800000}%
\pgfsetdash{}{0pt}%
\pgfpathmoveto{\pgfqpoint{1.083614in}{3.239285in}}%
\pgfpathlineto{\pgfqpoint{1.083614in}{3.652433in}}%
\pgfusepath{stroke}%
\end{pgfscope}%
\begin{pgfscope}%
\pgfpathrectangle{\pgfqpoint{0.917618in}{0.831623in}}{\pgfqpoint{3.651921in}{2.817477in}}%
\pgfusepath{clip}%
\pgfsetbuttcap%
\pgfsetroundjoin%
\pgfsetlinewidth{3.513125pt}%
\definecolor{currentstroke}{rgb}{0.200000,0.400000,0.466667}%
\pgfsetstrokecolor{currentstroke}%
\pgfsetstrokeopacity{0.800000}%
\pgfsetdash{}{0pt}%
\pgfpathmoveto{\pgfqpoint{1.360275in}{3.083043in}}%
\pgfpathlineto{\pgfqpoint{1.360275in}{3.264898in}}%
\pgfusepath{stroke}%
\end{pgfscope}%
\begin{pgfscope}%
\pgfpathrectangle{\pgfqpoint{0.917618in}{0.831623in}}{\pgfqpoint{3.651921in}{2.817477in}}%
\pgfusepath{clip}%
\pgfsetbuttcap%
\pgfsetroundjoin%
\pgfsetlinewidth{3.513125pt}%
\definecolor{currentstroke}{rgb}{0.200000,0.400000,0.466667}%
\pgfsetstrokecolor{currentstroke}%
\pgfsetstrokeopacity{0.800000}%
\pgfsetdash{}{0pt}%
\pgfpathmoveto{\pgfqpoint{1.636935in}{3.341739in}}%
\pgfpathlineto{\pgfqpoint{1.636935in}{3.524294in}}%
\pgfusepath{stroke}%
\end{pgfscope}%
\begin{pgfscope}%
\pgfpathrectangle{\pgfqpoint{0.917618in}{0.831623in}}{\pgfqpoint{3.651921in}{2.817477in}}%
\pgfusepath{clip}%
\pgfsetbuttcap%
\pgfsetroundjoin%
\pgfsetlinewidth{3.513125pt}%
\definecolor{currentstroke}{rgb}{0.200000,0.400000,0.466667}%
\pgfsetstrokecolor{currentstroke}%
\pgfsetstrokeopacity{0.800000}%
\pgfsetdash{}{0pt}%
\pgfpathmoveto{\pgfqpoint{1.913596in}{3.347431in}}%
\pgfpathlineto{\pgfqpoint{1.913596in}{3.548764in}}%
\pgfusepath{stroke}%
\end{pgfscope}%
\begin{pgfscope}%
\pgfpathrectangle{\pgfqpoint{0.917618in}{0.831623in}}{\pgfqpoint{3.651921in}{2.817477in}}%
\pgfusepath{clip}%
\pgfsetbuttcap%
\pgfsetroundjoin%
\pgfsetlinewidth{3.513125pt}%
\definecolor{currentstroke}{rgb}{0.200000,0.400000,0.466667}%
\pgfsetstrokecolor{currentstroke}%
\pgfsetstrokeopacity{0.800000}%
\pgfsetdash{}{0pt}%
\pgfpathmoveto{\pgfqpoint{2.190257in}{3.355710in}}%
\pgfpathlineto{\pgfqpoint{2.190257in}{3.476842in}}%
\pgfusepath{stroke}%
\end{pgfscope}%
\begin{pgfscope}%
\pgfpathrectangle{\pgfqpoint{0.917618in}{0.831623in}}{\pgfqpoint{3.651921in}{2.817477in}}%
\pgfusepath{clip}%
\pgfsetbuttcap%
\pgfsetroundjoin%
\pgfsetlinewidth{3.513125pt}%
\definecolor{currentstroke}{rgb}{0.200000,0.400000,0.466667}%
\pgfsetstrokecolor{currentstroke}%
\pgfsetstrokeopacity{0.800000}%
\pgfsetdash{}{0pt}%
\pgfpathmoveto{\pgfqpoint{2.466918in}{3.337798in}}%
\pgfpathlineto{\pgfqpoint{2.466918in}{3.396118in}}%
\pgfusepath{stroke}%
\end{pgfscope}%
\begin{pgfscope}%
\pgfpathrectangle{\pgfqpoint{0.917618in}{0.831623in}}{\pgfqpoint{3.651921in}{2.817477in}}%
\pgfusepath{clip}%
\pgfsetbuttcap%
\pgfsetroundjoin%
\pgfsetlinewidth{3.513125pt}%
\definecolor{currentstroke}{rgb}{0.200000,0.400000,0.466667}%
\pgfsetstrokecolor{currentstroke}%
\pgfsetstrokeopacity{0.800000}%
\pgfsetdash{}{0pt}%
\pgfpathmoveto{\pgfqpoint{2.743578in}{3.260629in}}%
\pgfpathlineto{\pgfqpoint{2.743578in}{3.375890in}}%
\pgfusepath{stroke}%
\end{pgfscope}%
\begin{pgfscope}%
\pgfpathrectangle{\pgfqpoint{0.917618in}{0.831623in}}{\pgfqpoint{3.651921in}{2.817477in}}%
\pgfusepath{clip}%
\pgfsetbuttcap%
\pgfsetroundjoin%
\pgfsetlinewidth{3.513125pt}%
\definecolor{currentstroke}{rgb}{0.200000,0.400000,0.466667}%
\pgfsetstrokecolor{currentstroke}%
\pgfsetstrokeopacity{0.800000}%
\pgfsetdash{}{0pt}%
\pgfpathmoveto{\pgfqpoint{3.020239in}{2.847550in}}%
\pgfpathlineto{\pgfqpoint{3.020239in}{3.377899in}}%
\pgfusepath{stroke}%
\end{pgfscope}%
\begin{pgfscope}%
\pgfpathrectangle{\pgfqpoint{0.917618in}{0.831623in}}{\pgfqpoint{3.651921in}{2.817477in}}%
\pgfusepath{clip}%
\pgfsetbuttcap%
\pgfsetroundjoin%
\pgfsetlinewidth{3.513125pt}%
\definecolor{currentstroke}{rgb}{0.200000,0.400000,0.466667}%
\pgfsetstrokecolor{currentstroke}%
\pgfsetstrokeopacity{0.800000}%
\pgfsetdash{}{0pt}%
\pgfusepath{stroke}%
\end{pgfscope}%
\begin{pgfscope}%
\pgfpathrectangle{\pgfqpoint{0.917618in}{0.831623in}}{\pgfqpoint{3.651921in}{2.817477in}}%
\pgfusepath{clip}%
\pgfsetbuttcap%
\pgfsetroundjoin%
\pgfsetlinewidth{3.513125pt}%
\definecolor{currentstroke}{rgb}{0.200000,0.400000,0.466667}%
\pgfsetstrokecolor{currentstroke}%
\pgfsetstrokeopacity{0.800000}%
\pgfsetdash{}{0pt}%
\pgfusepath{stroke}%
\end{pgfscope}%
\begin{pgfscope}%
\pgfpathrectangle{\pgfqpoint{0.917618in}{0.831623in}}{\pgfqpoint{3.651921in}{2.817477in}}%
\pgfusepath{clip}%
\pgfsetbuttcap%
\pgfsetroundjoin%
\pgfsetlinewidth{3.513125pt}%
\definecolor{currentstroke}{rgb}{0.200000,0.400000,0.466667}%
\pgfsetstrokecolor{currentstroke}%
\pgfsetstrokeopacity{0.800000}%
\pgfsetdash{}{0pt}%
\pgfusepath{stroke}%
\end{pgfscope}%
\begin{pgfscope}%
\pgfpathrectangle{\pgfqpoint{0.917618in}{0.831623in}}{\pgfqpoint{3.651921in}{2.817477in}}%
\pgfusepath{clip}%
\pgfsetbuttcap%
\pgfsetroundjoin%
\pgfsetlinewidth{3.513125pt}%
\definecolor{currentstroke}{rgb}{0.200000,0.400000,0.466667}%
\pgfsetstrokecolor{currentstroke}%
\pgfsetstrokeopacity{0.800000}%
\pgfsetdash{}{0pt}%
\pgfusepath{stroke}%
\end{pgfscope}%
\begin{pgfscope}%
\pgfpathrectangle{\pgfqpoint{0.917618in}{0.831623in}}{\pgfqpoint{3.651921in}{2.817477in}}%
\pgfusepath{clip}%
\pgfsetbuttcap%
\pgfsetroundjoin%
\pgfsetlinewidth{3.513125pt}%
\definecolor{currentstroke}{rgb}{0.200000,0.400000,0.466667}%
\pgfsetstrokecolor{currentstroke}%
\pgfsetstrokeopacity{0.800000}%
\pgfsetdash{}{0pt}%
\pgfusepath{stroke}%
\end{pgfscope}%
\begin{pgfscope}%
\pgfpathrectangle{\pgfqpoint{0.917618in}{0.831623in}}{\pgfqpoint{3.651921in}{2.817477in}}%
\pgfusepath{clip}%
\pgfsetbuttcap%
\pgfsetroundjoin%
\definecolor{currentfill}{rgb}{0.200000,0.400000,0.466667}%
\pgfsetfillcolor{currentfill}%
\pgfsetfillopacity{0.800000}%
\pgfsetlinewidth{2.007500pt}%
\definecolor{currentstroke}{rgb}{0.200000,0.400000,0.466667}%
\pgfsetstrokecolor{currentstroke}%
\pgfsetstrokeopacity{0.800000}%
\pgfsetdash{}{0pt}%
\pgfsys@defobject{currentmarker}{\pgfqpoint{-0.055556in}{-0.000000in}}{\pgfqpoint{0.055556in}{0.000000in}}{%
\pgfpathmoveto{\pgfqpoint{0.055556in}{-0.000000in}}%
\pgfpathlineto{\pgfqpoint{-0.055556in}{0.000000in}}%
\pgfusepath{stroke,fill}%
}%
\begin{pgfscope}%
\pgfsys@transformshift{1.083614in}{3.239285in}%
\pgfsys@useobject{currentmarker}{}%
\end{pgfscope}%
\begin{pgfscope}%
\pgfsys@transformshift{1.360275in}{3.083043in}%
\pgfsys@useobject{currentmarker}{}%
\end{pgfscope}%
\begin{pgfscope}%
\pgfsys@transformshift{1.636935in}{3.341739in}%
\pgfsys@useobject{currentmarker}{}%
\end{pgfscope}%
\begin{pgfscope}%
\pgfsys@transformshift{1.913596in}{3.347431in}%
\pgfsys@useobject{currentmarker}{}%
\end{pgfscope}%
\begin{pgfscope}%
\pgfsys@transformshift{2.190257in}{3.355710in}%
\pgfsys@useobject{currentmarker}{}%
\end{pgfscope}%
\begin{pgfscope}%
\pgfsys@transformshift{2.466918in}{3.337798in}%
\pgfsys@useobject{currentmarker}{}%
\end{pgfscope}%
\begin{pgfscope}%
\pgfsys@transformshift{2.743578in}{3.260629in}%
\pgfsys@useobject{currentmarker}{}%
\end{pgfscope}%
\begin{pgfscope}%
\pgfsys@transformshift{3.020239in}{2.847550in}%
\pgfsys@useobject{currentmarker}{}%
\end{pgfscope}%
\end{pgfscope}%
\begin{pgfscope}%
\pgfpathrectangle{\pgfqpoint{0.917618in}{0.831623in}}{\pgfqpoint{3.651921in}{2.817477in}}%
\pgfusepath{clip}%
\pgfsetbuttcap%
\pgfsetroundjoin%
\definecolor{currentfill}{rgb}{0.200000,0.400000,0.466667}%
\pgfsetfillcolor{currentfill}%
\pgfsetfillopacity{0.800000}%
\pgfsetlinewidth{2.007500pt}%
\definecolor{currentstroke}{rgb}{0.200000,0.400000,0.466667}%
\pgfsetstrokecolor{currentstroke}%
\pgfsetstrokeopacity{0.800000}%
\pgfsetdash{}{0pt}%
\pgfsys@defobject{currentmarker}{\pgfqpoint{-0.055556in}{-0.000000in}}{\pgfqpoint{0.055556in}{0.000000in}}{%
\pgfpathmoveto{\pgfqpoint{0.055556in}{-0.000000in}}%
\pgfpathlineto{\pgfqpoint{-0.055556in}{0.000000in}}%
\pgfusepath{stroke,fill}%
}%
\begin{pgfscope}%
\pgfsys@transformshift{1.083614in}{3.717402in}%
\pgfsys@useobject{currentmarker}{}%
\end{pgfscope}%
\begin{pgfscope}%
\pgfsys@transformshift{1.360275in}{3.264898in}%
\pgfsys@useobject{currentmarker}{}%
\end{pgfscope}%
\begin{pgfscope}%
\pgfsys@transformshift{1.636935in}{3.524294in}%
\pgfsys@useobject{currentmarker}{}%
\end{pgfscope}%
\begin{pgfscope}%
\pgfsys@transformshift{1.913596in}{3.548764in}%
\pgfsys@useobject{currentmarker}{}%
\end{pgfscope}%
\begin{pgfscope}%
\pgfsys@transformshift{2.190257in}{3.476842in}%
\pgfsys@useobject{currentmarker}{}%
\end{pgfscope}%
\begin{pgfscope}%
\pgfsys@transformshift{2.466918in}{3.396118in}%
\pgfsys@useobject{currentmarker}{}%
\end{pgfscope}%
\begin{pgfscope}%
\pgfsys@transformshift{2.743578in}{3.375890in}%
\pgfsys@useobject{currentmarker}{}%
\end{pgfscope}%
\begin{pgfscope}%
\pgfsys@transformshift{3.020239in}{3.377899in}%
\pgfsys@useobject{currentmarker}{}%
\end{pgfscope}%
\end{pgfscope}%
\begin{pgfscope}%
\pgfpathrectangle{\pgfqpoint{0.917618in}{0.831623in}}{\pgfqpoint{3.651921in}{2.817477in}}%
\pgfusepath{clip}%
\pgfsetbuttcap%
\pgfsetroundjoin%
\pgfsetlinewidth{3.513125pt}%
\definecolor{currentstroke}{rgb}{0.000000,0.000000,0.000000}%
\pgfsetstrokecolor{currentstroke}%
\pgfsetstrokeopacity{0.800000}%
\pgfsetdash{}{0pt}%
\pgfpathmoveto{\pgfqpoint{1.083614in}{1.890311in}}%
\pgfpathlineto{\pgfqpoint{1.083614in}{2.257437in}}%
\pgfusepath{stroke}%
\end{pgfscope}%
\begin{pgfscope}%
\pgfpathrectangle{\pgfqpoint{0.917618in}{0.831623in}}{\pgfqpoint{3.651921in}{2.817477in}}%
\pgfusepath{clip}%
\pgfsetbuttcap%
\pgfsetroundjoin%
\pgfsetlinewidth{3.513125pt}%
\definecolor{currentstroke}{rgb}{0.000000,0.000000,0.000000}%
\pgfsetstrokecolor{currentstroke}%
\pgfsetstrokeopacity{0.800000}%
\pgfsetdash{}{0pt}%
\pgfpathmoveto{\pgfqpoint{1.636935in}{2.163952in}}%
\pgfpathlineto{\pgfqpoint{1.636935in}{2.368429in}}%
\pgfusepath{stroke}%
\end{pgfscope}%
\begin{pgfscope}%
\pgfpathrectangle{\pgfqpoint{0.917618in}{0.831623in}}{\pgfqpoint{3.651921in}{2.817477in}}%
\pgfusepath{clip}%
\pgfsetbuttcap%
\pgfsetroundjoin%
\pgfsetlinewidth{3.513125pt}%
\definecolor{currentstroke}{rgb}{0.000000,0.000000,0.000000}%
\pgfsetstrokecolor{currentstroke}%
\pgfsetstrokeopacity{0.800000}%
\pgfsetdash{}{0pt}%
\pgfpathmoveto{\pgfqpoint{2.190257in}{2.339322in}}%
\pgfpathlineto{\pgfqpoint{2.190257in}{2.368429in}}%
\pgfusepath{stroke}%
\end{pgfscope}%
\begin{pgfscope}%
\pgfpathrectangle{\pgfqpoint{0.917618in}{0.831623in}}{\pgfqpoint{3.651921in}{2.817477in}}%
\pgfusepath{clip}%
\pgfsetbuttcap%
\pgfsetroundjoin%
\pgfsetlinewidth{3.513125pt}%
\definecolor{currentstroke}{rgb}{0.000000,0.000000,0.000000}%
\pgfsetstrokecolor{currentstroke}%
\pgfsetstrokeopacity{0.800000}%
\pgfsetdash{}{0pt}%
\pgfpathmoveto{\pgfqpoint{2.743578in}{0.974205in}}%
\pgfpathlineto{\pgfqpoint{2.743578in}{2.365013in}}%
\pgfusepath{stroke}%
\end{pgfscope}%
\begin{pgfscope}%
\pgfpathrectangle{\pgfqpoint{0.917618in}{0.831623in}}{\pgfqpoint{3.651921in}{2.817477in}}%
\pgfusepath{clip}%
\pgfsetbuttcap%
\pgfsetroundjoin%
\pgfsetlinewidth{3.513125pt}%
\definecolor{currentstroke}{rgb}{0.000000,0.000000,0.000000}%
\pgfsetstrokecolor{currentstroke}%
\pgfsetstrokeopacity{0.800000}%
\pgfsetdash{}{0pt}%
\pgfpathmoveto{\pgfqpoint{3.296900in}{0.960364in}}%
\pgfpathlineto{\pgfqpoint{3.296900in}{1.545429in}}%
\pgfusepath{stroke}%
\end{pgfscope}%
\begin{pgfscope}%
\pgfpathrectangle{\pgfqpoint{0.917618in}{0.831623in}}{\pgfqpoint{3.651921in}{2.817477in}}%
\pgfusepath{clip}%
\pgfsetbuttcap%
\pgfsetroundjoin%
\pgfsetlinewidth{3.513125pt}%
\definecolor{currentstroke}{rgb}{0.000000,0.000000,0.000000}%
\pgfsetstrokecolor{currentstroke}%
\pgfsetstrokeopacity{0.800000}%
\pgfsetdash{}{0pt}%
\pgfpathmoveto{\pgfqpoint{3.850221in}{0.953565in}}%
\pgfpathlineto{\pgfqpoint{3.850221in}{1.373870in}}%
\pgfusepath{stroke}%
\end{pgfscope}%
\begin{pgfscope}%
\pgfpathrectangle{\pgfqpoint{0.917618in}{0.831623in}}{\pgfqpoint{3.651921in}{2.817477in}}%
\pgfusepath{clip}%
\pgfsetbuttcap%
\pgfsetroundjoin%
\pgfsetlinewidth{3.513125pt}%
\definecolor{currentstroke}{rgb}{0.000000,0.000000,0.000000}%
\pgfsetstrokecolor{currentstroke}%
\pgfsetstrokeopacity{0.800000}%
\pgfsetdash{}{0pt}%
\pgfpathmoveto{\pgfqpoint{4.403543in}{0.953050in}}%
\pgfpathlineto{\pgfqpoint{4.403543in}{1.342013in}}%
\pgfusepath{stroke}%
\end{pgfscope}%
\begin{pgfscope}%
\pgfpathrectangle{\pgfqpoint{0.917618in}{0.831623in}}{\pgfqpoint{3.651921in}{2.817477in}}%
\pgfusepath{clip}%
\pgfsetbuttcap%
\pgfsetroundjoin%
\definecolor{currentfill}{rgb}{0.000000,0.000000,0.000000}%
\pgfsetfillcolor{currentfill}%
\pgfsetfillopacity{0.800000}%
\pgfsetlinewidth{2.007500pt}%
\definecolor{currentstroke}{rgb}{0.000000,0.000000,0.000000}%
\pgfsetstrokecolor{currentstroke}%
\pgfsetstrokeopacity{0.800000}%
\pgfsetdash{}{0pt}%
\pgfsys@defobject{currentmarker}{\pgfqpoint{-0.055556in}{-0.000000in}}{\pgfqpoint{0.055556in}{0.000000in}}{%
\pgfpathmoveto{\pgfqpoint{0.055556in}{-0.000000in}}%
\pgfpathlineto{\pgfqpoint{-0.055556in}{0.000000in}}%
\pgfusepath{stroke,fill}%
}%
\begin{pgfscope}%
\pgfsys@transformshift{1.083614in}{1.890311in}%
\pgfsys@useobject{currentmarker}{}%
\end{pgfscope}%
\begin{pgfscope}%
\pgfsys@transformshift{1.636935in}{2.163952in}%
\pgfsys@useobject{currentmarker}{}%
\end{pgfscope}%
\begin{pgfscope}%
\pgfsys@transformshift{2.190257in}{2.339322in}%
\pgfsys@useobject{currentmarker}{}%
\end{pgfscope}%
\begin{pgfscope}%
\pgfsys@transformshift{2.743578in}{0.974205in}%
\pgfsys@useobject{currentmarker}{}%
\end{pgfscope}%
\begin{pgfscope}%
\pgfsys@transformshift{3.296900in}{0.960364in}%
\pgfsys@useobject{currentmarker}{}%
\end{pgfscope}%
\begin{pgfscope}%
\pgfsys@transformshift{3.850221in}{0.953565in}%
\pgfsys@useobject{currentmarker}{}%
\end{pgfscope}%
\begin{pgfscope}%
\pgfsys@transformshift{4.403543in}{0.953050in}%
\pgfsys@useobject{currentmarker}{}%
\end{pgfscope}%
\end{pgfscope}%
\begin{pgfscope}%
\pgfpathrectangle{\pgfqpoint{0.917618in}{0.831623in}}{\pgfqpoint{3.651921in}{2.817477in}}%
\pgfusepath{clip}%
\pgfsetbuttcap%
\pgfsetroundjoin%
\definecolor{currentfill}{rgb}{0.000000,0.000000,0.000000}%
\pgfsetfillcolor{currentfill}%
\pgfsetfillopacity{0.800000}%
\pgfsetlinewidth{2.007500pt}%
\definecolor{currentstroke}{rgb}{0.000000,0.000000,0.000000}%
\pgfsetstrokecolor{currentstroke}%
\pgfsetstrokeopacity{0.800000}%
\pgfsetdash{}{0pt}%
\pgfsys@defobject{currentmarker}{\pgfqpoint{-0.055556in}{-0.000000in}}{\pgfqpoint{0.055556in}{0.000000in}}{%
\pgfpathmoveto{\pgfqpoint{0.055556in}{-0.000000in}}%
\pgfpathlineto{\pgfqpoint{-0.055556in}{0.000000in}}%
\pgfusepath{stroke,fill}%
}%
\begin{pgfscope}%
\pgfsys@transformshift{1.083614in}{2.257437in}%
\pgfsys@useobject{currentmarker}{}%
\end{pgfscope}%
\begin{pgfscope}%
\pgfsys@transformshift{1.636935in}{2.368429in}%
\pgfsys@useobject{currentmarker}{}%
\end{pgfscope}%
\begin{pgfscope}%
\pgfsys@transformshift{2.190257in}{2.368429in}%
\pgfsys@useobject{currentmarker}{}%
\end{pgfscope}%
\begin{pgfscope}%
\pgfsys@transformshift{2.743578in}{2.365013in}%
\pgfsys@useobject{currentmarker}{}%
\end{pgfscope}%
\begin{pgfscope}%
\pgfsys@transformshift{3.296900in}{1.545429in}%
\pgfsys@useobject{currentmarker}{}%
\end{pgfscope}%
\begin{pgfscope}%
\pgfsys@transformshift{3.850221in}{1.373870in}%
\pgfsys@useobject{currentmarker}{}%
\end{pgfscope}%
\begin{pgfscope}%
\pgfsys@transformshift{4.403543in}{1.342013in}%
\pgfsys@useobject{currentmarker}{}%
\end{pgfscope}%
\end{pgfscope}%
\begin{pgfscope}%
\pgfpathrectangle{\pgfqpoint{0.917618in}{0.831623in}}{\pgfqpoint{3.651921in}{2.817477in}}%
\pgfusepath{clip}%
\pgfsetbuttcap%
\pgfsetroundjoin%
\pgfsetlinewidth{3.513125pt}%
\definecolor{currentstroke}{rgb}{0.666667,0.200000,0.466667}%
\pgfsetstrokecolor{currentstroke}%
\pgfsetstrokeopacity{0.800000}%
\pgfsetdash{{12.950000pt}{5.600000pt}}{0.000000pt}%
\pgfpathmoveto{\pgfqpoint{1.083614in}{2.191184in}}%
\pgfpathlineto{\pgfqpoint{1.636935in}{2.265024in}}%
\pgfpathlineto{\pgfqpoint{2.190257in}{2.031216in}}%
\pgfpathlineto{\pgfqpoint{2.743578in}{1.993653in}}%
\pgfpathlineto{\pgfqpoint{3.296900in}{1.993610in}}%
\pgfpathlineto{\pgfqpoint{3.850221in}{1.994718in}}%
\pgfpathlineto{\pgfqpoint{4.403543in}{1.681590in}}%
\pgfusepath{stroke}%
\end{pgfscope}%
\begin{pgfscope}%
\pgfpathrectangle{\pgfqpoint{0.917618in}{0.831623in}}{\pgfqpoint{3.651921in}{2.817477in}}%
\pgfusepath{clip}%
\pgfsetbuttcap%
\pgfsetroundjoin%
\definecolor{currentfill}{rgb}{0.666667,0.200000,0.466667}%
\pgfsetfillcolor{currentfill}%
\pgfsetfillopacity{0.800000}%
\pgfsetlinewidth{2.007500pt}%
\definecolor{currentstroke}{rgb}{1.000000,1.000000,1.000000}%
\pgfsetstrokecolor{currentstroke}%
\pgfsetstrokeopacity{0.800000}%
\pgfsetdash{}{0pt}%
\pgfsys@defobject{currentmarker}{\pgfqpoint{-0.055556in}{-0.055556in}}{\pgfqpoint{0.055556in}{0.055556in}}{%
\pgfpathmoveto{\pgfqpoint{0.000000in}{-0.055556in}}%
\pgfpathcurveto{\pgfqpoint{0.014734in}{-0.055556in}}{\pgfqpoint{0.028866in}{-0.049702in}}{\pgfqpoint{0.039284in}{-0.039284in}}%
\pgfpathcurveto{\pgfqpoint{0.049702in}{-0.028866in}}{\pgfqpoint{0.055556in}{-0.014734in}}{\pgfqpoint{0.055556in}{0.000000in}}%
\pgfpathcurveto{\pgfqpoint{0.055556in}{0.014734in}}{\pgfqpoint{0.049702in}{0.028866in}}{\pgfqpoint{0.039284in}{0.039284in}}%
\pgfpathcurveto{\pgfqpoint{0.028866in}{0.049702in}}{\pgfqpoint{0.014734in}{0.055556in}}{\pgfqpoint{0.000000in}{0.055556in}}%
\pgfpathcurveto{\pgfqpoint{-0.014734in}{0.055556in}}{\pgfqpoint{-0.028866in}{0.049702in}}{\pgfqpoint{-0.039284in}{0.039284in}}%
\pgfpathcurveto{\pgfqpoint{-0.049702in}{0.028866in}}{\pgfqpoint{-0.055556in}{0.014734in}}{\pgfqpoint{-0.055556in}{0.000000in}}%
\pgfpathcurveto{\pgfqpoint{-0.055556in}{-0.014734in}}{\pgfqpoint{-0.049702in}{-0.028866in}}{\pgfqpoint{-0.039284in}{-0.039284in}}%
\pgfpathcurveto{\pgfqpoint{-0.028866in}{-0.049702in}}{\pgfqpoint{-0.014734in}{-0.055556in}}{\pgfqpoint{0.000000in}{-0.055556in}}%
\pgfpathlineto{\pgfqpoint{0.000000in}{-0.055556in}}%
\pgfpathclose%
\pgfusepath{stroke,fill}%
}%
\begin{pgfscope}%
\pgfsys@transformshift{1.083614in}{2.191184in}%
\pgfsys@useobject{currentmarker}{}%
\end{pgfscope}%
\begin{pgfscope}%
\pgfsys@transformshift{1.636935in}{2.265024in}%
\pgfsys@useobject{currentmarker}{}%
\end{pgfscope}%
\begin{pgfscope}%
\pgfsys@transformshift{2.190257in}{2.031216in}%
\pgfsys@useobject{currentmarker}{}%
\end{pgfscope}%
\begin{pgfscope}%
\pgfsys@transformshift{2.743578in}{1.993653in}%
\pgfsys@useobject{currentmarker}{}%
\end{pgfscope}%
\begin{pgfscope}%
\pgfsys@transformshift{3.296900in}{1.993610in}%
\pgfsys@useobject{currentmarker}{}%
\end{pgfscope}%
\begin{pgfscope}%
\pgfsys@transformshift{3.850221in}{1.994718in}%
\pgfsys@useobject{currentmarker}{}%
\end{pgfscope}%
\begin{pgfscope}%
\pgfsys@transformshift{4.403543in}{1.681590in}%
\pgfsys@useobject{currentmarker}{}%
\end{pgfscope}%
\end{pgfscope}%
\begin{pgfscope}%
\pgfpathrectangle{\pgfqpoint{0.917618in}{0.831623in}}{\pgfqpoint{3.651921in}{2.817477in}}%
\pgfusepath{clip}%
\pgfsetbuttcap%
\pgfsetroundjoin%
\pgfsetlinewidth{3.513125pt}%
\definecolor{currentstroke}{rgb}{0.200000,0.400000,0.466667}%
\pgfsetstrokecolor{currentstroke}%
\pgfsetstrokeopacity{0.800000}%
\pgfsetdash{{22.400000pt}{5.600000pt}{3.500000pt}{5.600000pt}}{0.000000pt}%
\pgfpathmoveto{\pgfqpoint{1.083614in}{3.594458in}}%
\pgfpathlineto{\pgfqpoint{1.360275in}{3.171409in}}%
\pgfpathlineto{\pgfqpoint{1.636935in}{3.443079in}}%
\pgfpathlineto{\pgfqpoint{1.913596in}{3.456635in}}%
\pgfpathlineto{\pgfqpoint{2.190257in}{3.423261in}}%
\pgfpathlineto{\pgfqpoint{2.466918in}{3.384690in}}%
\pgfpathlineto{\pgfqpoint{2.743578in}{3.310968in}}%
\pgfpathlineto{\pgfqpoint{3.020239in}{3.118198in}}%
\pgfusepath{stroke}%
\end{pgfscope}%
\begin{pgfscope}%
\pgfpathrectangle{\pgfqpoint{0.917618in}{0.831623in}}{\pgfqpoint{3.651921in}{2.817477in}}%
\pgfusepath{clip}%
\pgfsetbuttcap%
\pgfsetmiterjoin%
\definecolor{currentfill}{rgb}{0.200000,0.400000,0.466667}%
\pgfsetfillcolor{currentfill}%
\pgfsetfillopacity{0.800000}%
\pgfsetlinewidth{2.007500pt}%
\definecolor{currentstroke}{rgb}{1.000000,1.000000,1.000000}%
\pgfsetstrokecolor{currentstroke}%
\pgfsetstrokeopacity{0.800000}%
\pgfsetdash{}{0pt}%
\pgfsys@defobject{currentmarker}{\pgfqpoint{-0.055556in}{-0.055556in}}{\pgfqpoint{0.055556in}{0.055556in}}{%
\pgfpathmoveto{\pgfqpoint{-0.000000in}{-0.055556in}}%
\pgfpathlineto{\pgfqpoint{0.055556in}{0.055556in}}%
\pgfpathlineto{\pgfqpoint{-0.055556in}{0.055556in}}%
\pgfpathlineto{\pgfqpoint{-0.000000in}{-0.055556in}}%
\pgfpathclose%
\pgfusepath{stroke,fill}%
}%
\begin{pgfscope}%
\pgfsys@transformshift{1.083614in}{3.594458in}%
\pgfsys@useobject{currentmarker}{}%
\end{pgfscope}%
\begin{pgfscope}%
\pgfsys@transformshift{1.360275in}{3.171409in}%
\pgfsys@useobject{currentmarker}{}%
\end{pgfscope}%
\begin{pgfscope}%
\pgfsys@transformshift{1.636935in}{3.443079in}%
\pgfsys@useobject{currentmarker}{}%
\end{pgfscope}%
\begin{pgfscope}%
\pgfsys@transformshift{1.913596in}{3.456635in}%
\pgfsys@useobject{currentmarker}{}%
\end{pgfscope}%
\begin{pgfscope}%
\pgfsys@transformshift{2.190257in}{3.423261in}%
\pgfsys@useobject{currentmarker}{}%
\end{pgfscope}%
\begin{pgfscope}%
\pgfsys@transformshift{2.466918in}{3.384690in}%
\pgfsys@useobject{currentmarker}{}%
\end{pgfscope}%
\begin{pgfscope}%
\pgfsys@transformshift{2.743578in}{3.310968in}%
\pgfsys@useobject{currentmarker}{}%
\end{pgfscope}%
\begin{pgfscope}%
\pgfsys@transformshift{3.020239in}{3.118198in}%
\pgfsys@useobject{currentmarker}{}%
\end{pgfscope}%
\end{pgfscope}%
\begin{pgfscope}%
\pgfpathrectangle{\pgfqpoint{0.917618in}{0.831623in}}{\pgfqpoint{3.651921in}{2.817477in}}%
\pgfusepath{clip}%
\pgfsetrectcap%
\pgfsetroundjoin%
\pgfsetlinewidth{3.513125pt}%
\definecolor{currentstroke}{rgb}{0.000000,0.000000,0.000000}%
\pgfsetstrokecolor{currentstroke}%
\pgfsetstrokeopacity{0.800000}%
\pgfsetdash{}{0pt}%
\pgfpathmoveto{\pgfqpoint{1.083614in}{2.064662in}}%
\pgfpathlineto{\pgfqpoint{1.636935in}{2.240577in}}%
\pgfpathlineto{\pgfqpoint{2.190257in}{2.344725in}}%
\pgfpathlineto{\pgfqpoint{2.743578in}{1.669899in}}%
\pgfpathlineto{\pgfqpoint{3.296900in}{1.340737in}}%
\pgfpathlineto{\pgfqpoint{3.850221in}{1.263109in}}%
\pgfpathlineto{\pgfqpoint{4.403543in}{1.239787in}}%
\pgfusepath{stroke}%
\end{pgfscope}%
\begin{pgfscope}%
\pgfpathrectangle{\pgfqpoint{0.917618in}{0.831623in}}{\pgfqpoint{3.651921in}{2.817477in}}%
\pgfusepath{clip}%
\pgfsetbuttcap%
\pgfsetmiterjoin%
\definecolor{currentfill}{rgb}{0.000000,0.000000,0.000000}%
\pgfsetfillcolor{currentfill}%
\pgfsetfillopacity{0.800000}%
\pgfsetlinewidth{2.007500pt}%
\definecolor{currentstroke}{rgb}{1.000000,1.000000,1.000000}%
\pgfsetstrokecolor{currentstroke}%
\pgfsetstrokeopacity{0.800000}%
\pgfsetdash{}{0pt}%
\pgfsys@defobject{currentmarker}{\pgfqpoint{-0.078567in}{-0.078567in}}{\pgfqpoint{0.078567in}{0.078567in}}{%
\pgfpathmoveto{\pgfqpoint{-0.000000in}{-0.078567in}}%
\pgfpathlineto{\pgfqpoint{0.078567in}{0.000000in}}%
\pgfpathlineto{\pgfqpoint{0.000000in}{0.078567in}}%
\pgfpathlineto{\pgfqpoint{-0.078567in}{0.000000in}}%
\pgfpathlineto{\pgfqpoint{-0.000000in}{-0.078567in}}%
\pgfpathclose%
\pgfusepath{stroke,fill}%
}%
\begin{pgfscope}%
\pgfsys@transformshift{1.083614in}{2.064662in}%
\pgfsys@useobject{currentmarker}{}%
\end{pgfscope}%
\begin{pgfscope}%
\pgfsys@transformshift{1.636935in}{2.240577in}%
\pgfsys@useobject{currentmarker}{}%
\end{pgfscope}%
\begin{pgfscope}%
\pgfsys@transformshift{2.190257in}{2.344725in}%
\pgfsys@useobject{currentmarker}{}%
\end{pgfscope}%
\begin{pgfscope}%
\pgfsys@transformshift{2.743578in}{1.669899in}%
\pgfsys@useobject{currentmarker}{}%
\end{pgfscope}%
\begin{pgfscope}%
\pgfsys@transformshift{3.296900in}{1.340737in}%
\pgfsys@useobject{currentmarker}{}%
\end{pgfscope}%
\begin{pgfscope}%
\pgfsys@transformshift{3.850221in}{1.263109in}%
\pgfsys@useobject{currentmarker}{}%
\end{pgfscope}%
\begin{pgfscope}%
\pgfsys@transformshift{4.403543in}{1.239787in}%
\pgfsys@useobject{currentmarker}{}%
\end{pgfscope}%
\end{pgfscope}%
\begin{pgfscope}%
\pgfsetrectcap%
\pgfsetmiterjoin%
\pgfsetlinewidth{1.505625pt}%
\definecolor{currentstroke}{rgb}{0.000000,0.000000,0.000000}%
\pgfsetstrokecolor{currentstroke}%
\pgfsetdash{}{0pt}%
\pgfpathmoveto{\pgfqpoint{0.917618in}{0.831623in}}%
\pgfpathlineto{\pgfqpoint{0.917618in}{3.649100in}}%
\pgfusepath{stroke}%
\end{pgfscope}%
\begin{pgfscope}%
\pgfsetrectcap%
\pgfsetmiterjoin%
\pgfsetlinewidth{1.505625pt}%
\definecolor{currentstroke}{rgb}{0.000000,0.000000,0.000000}%
\pgfsetstrokecolor{currentstroke}%
\pgfsetdash{}{0pt}%
\pgfpathmoveto{\pgfqpoint{0.917618in}{0.831623in}}%
\pgfpathlineto{\pgfqpoint{4.569539in}{0.831623in}}%
\pgfusepath{stroke}%
\end{pgfscope}%
\end{pgfpicture}%
\makeatother%
\endgroup%

%% file: figures/synthetic-final/M2/size.pgf
\begingroup%
\makeatletter%
\begin{pgfpicture}%
\pgfpathrectangle{\pgfpointorigin}{\pgfqpoint{4.742995in}{3.643400in}}%
\pgfusepath{use as bounding box, clip}%
\begin{pgfscope}%
\pgfsetbuttcap%
\pgfsetmiterjoin%
\definecolor{currentfill}{rgb}{1.000000,1.000000,1.000000}%
\pgfsetfillcolor{currentfill}%
\pgfsetlinewidth{0.000000pt}%
\definecolor{currentstroke}{rgb}{1.000000,1.000000,1.000000}%
\pgfsetstrokecolor{currentstroke}%
\pgfsetdash{}{0pt}%
\pgfpathmoveto{\pgfqpoint{0.000000in}{0.000000in}}%
\pgfpathlineto{\pgfqpoint{4.742995in}{0.000000in}}%
\pgfpathlineto{\pgfqpoint{4.742995in}{3.643400in}}%
\pgfpathlineto{\pgfqpoint{0.000000in}{3.643400in}}%
\pgfpathlineto{\pgfqpoint{0.000000in}{0.000000in}}%
\pgfpathclose%
\pgfusepath{fill}%
\end{pgfscope}%
\begin{pgfscope}%
\pgfsetbuttcap%
\pgfsetmiterjoin%
\definecolor{currentfill}{rgb}{1.000000,1.000000,1.000000}%
\pgfsetfillcolor{currentfill}%
\pgfsetlinewidth{0.000000pt}%
\definecolor{currentstroke}{rgb}{0.000000,0.000000,0.000000}%
\pgfsetstrokecolor{currentstroke}%
\pgfsetstrokeopacity{0.000000}%
\pgfsetdash{}{0pt}%
\pgfpathmoveto{\pgfqpoint{0.675675in}{0.831623in}}%
\pgfpathlineto{\pgfqpoint{4.583279in}{0.831623in}}%
\pgfpathlineto{\pgfqpoint{4.583279in}{3.543400in}}%
\pgfpathlineto{\pgfqpoint{0.675675in}{3.543400in}}%
\pgfpathlineto{\pgfqpoint{0.675675in}{0.831623in}}%
\pgfpathclose%
\pgfusepath{fill}%
\end{pgfscope}%
\begin{pgfscope}%
\pgfsetbuttcap%
\pgfsetroundjoin%
\definecolor{currentfill}{rgb}{0.000000,0.000000,0.000000}%
\pgfsetfillcolor{currentfill}%
\pgfsetlinewidth{1.505625pt}%
\definecolor{currentstroke}{rgb}{0.000000,0.000000,0.000000}%
\pgfsetstrokecolor{currentstroke}%
\pgfsetdash{}{0pt}%
\pgfsys@defobject{currentmarker}{\pgfqpoint{0.000000in}{-0.083333in}}{\pgfqpoint{0.000000in}{0.000000in}}{%
\pgfpathmoveto{\pgfqpoint{0.000000in}{0.000000in}}%
\pgfpathlineto{\pgfqpoint{0.000000in}{-0.083333in}}%
\pgfusepath{stroke,fill}%
}%
\begin{pgfscope}%
\pgfsys@transformshift{0.853293in}{0.831623in}%
\pgfsys@useobject{currentmarker}{}%
\end{pgfscope}%
\end{pgfscope}%
\begin{pgfscope}%
\definecolor{textcolor}{rgb}{0.000000,0.000000,0.000000}%
\pgfsetstrokecolor{textcolor}%
\pgfsetfillcolor{textcolor}%
\pgftext[x=0.853293in,y=0.699679in,,top]{\color{textcolor}{\rmfamily\fontsize{22.000000}{26.400000}\selectfont\catcode`\^=\active\def^{\ifmmode\sp\else\^{}\fi}\catcode`\%=\active\def
\end{pgfscope}%
\begin{pgfscope}%
\pgfsetbuttcap%
\pgfsetroundjoin%
\definecolor{currentfill}{rgb}{0.000000,0.000000,0.000000}%
\pgfsetfillcolor{currentfill}%
\pgfsetlinewidth{1.505625pt}%
\definecolor{currentstroke}{rgb}{0.000000,0.000000,0.000000}%
\pgfsetstrokecolor{currentstroke}%
\pgfsetdash{}{0pt}%
\pgfsys@defobject{currentmarker}{\pgfqpoint{0.000000in}{-0.083333in}}{\pgfqpoint{0.000000in}{0.000000in}}{%
\pgfpathmoveto{\pgfqpoint{0.000000in}{0.000000in}}%
\pgfpathlineto{\pgfqpoint{0.000000in}{-0.083333in}}%
\pgfusepath{stroke,fill}%
}%
\begin{pgfscope}%
\pgfsys@transformshift{2.037416in}{0.831623in}%
\pgfsys@useobject{currentmarker}{}%
\end{pgfscope}%
\end{pgfscope}%
\begin{pgfscope}%
\definecolor{textcolor}{rgb}{0.000000,0.000000,0.000000}%
\pgfsetstrokecolor{textcolor}%
\pgfsetfillcolor{textcolor}%
\pgftext[x=2.037416in,y=0.699679in,,top]{\color{textcolor}{\rmfamily\fontsize{22.000000}{26.400000}\selectfont\catcode`\^=\active\def^{\ifmmode\sp\else\^{}\fi}\catcode`\%=\active\def
\end{pgfscope}%
\begin{pgfscope}%
\pgfsetbuttcap%
\pgfsetroundjoin%
\definecolor{currentfill}{rgb}{0.000000,0.000000,0.000000}%
\pgfsetfillcolor{currentfill}%
\pgfsetlinewidth{1.505625pt}%
\definecolor{currentstroke}{rgb}{0.000000,0.000000,0.000000}%
\pgfsetstrokecolor{currentstroke}%
\pgfsetdash{}{0pt}%
\pgfsys@defobject{currentmarker}{\pgfqpoint{0.000000in}{-0.083333in}}{\pgfqpoint{0.000000in}{0.000000in}}{%
\pgfpathmoveto{\pgfqpoint{0.000000in}{0.000000in}}%
\pgfpathlineto{\pgfqpoint{0.000000in}{-0.083333in}}%
\pgfusepath{stroke,fill}%
}%
\begin{pgfscope}%
\pgfsys@transformshift{3.221538in}{0.831623in}%
\pgfsys@useobject{currentmarker}{}%
\end{pgfscope}%
\end{pgfscope}%
\begin{pgfscope}%
\definecolor{textcolor}{rgb}{0.000000,0.000000,0.000000}%
\pgfsetstrokecolor{textcolor}%
\pgfsetfillcolor{textcolor}%
\pgftext[x=3.221538in,y=0.699679in,,top]{\color{textcolor}{\rmfamily\fontsize{22.000000}{26.400000}\selectfont\catcode`\^=\active\def^{\ifmmode\sp\else\^{}\fi}\catcode`\%=\active\def
\end{pgfscope}%
\begin{pgfscope}%
\pgfsetbuttcap%
\pgfsetroundjoin%
\definecolor{currentfill}{rgb}{0.000000,0.000000,0.000000}%
\pgfsetfillcolor{currentfill}%
\pgfsetlinewidth{1.505625pt}%
\definecolor{currentstroke}{rgb}{0.000000,0.000000,0.000000}%
\pgfsetstrokecolor{currentstroke}%
\pgfsetdash{}{0pt}%
\pgfsys@defobject{currentmarker}{\pgfqpoint{0.000000in}{-0.083333in}}{\pgfqpoint{0.000000in}{0.000000in}}{%
\pgfpathmoveto{\pgfqpoint{0.000000in}{0.000000in}}%
\pgfpathlineto{\pgfqpoint{0.000000in}{-0.083333in}}%
\pgfusepath{stroke,fill}%
}%
\begin{pgfscope}%
\pgfsys@transformshift{4.405661in}{0.831623in}%
\pgfsys@useobject{currentmarker}{}%
\end{pgfscope}%
\end{pgfscope}%
\begin{pgfscope}%
\definecolor{textcolor}{rgb}{0.000000,0.000000,0.000000}%
\pgfsetstrokecolor{textcolor}%
\pgfsetfillcolor{textcolor}%
\pgftext[x=4.405661in,y=0.699679in,,top]{\color{textcolor}{\rmfamily\fontsize{22.000000}{26.400000}\selectfont\catcode`\^=\active\def^{\ifmmode\sp\else\^{}\fi}\catcode`\%=\active\def
\end{pgfscope}%
\begin{pgfscope}%
\definecolor{textcolor}{rgb}{0.000000,0.000000,0.000000}%
\pgfsetstrokecolor{textcolor}%
\pgfsetfillcolor{textcolor}%
\pgftext[x=2.629477in,y=0.388056in,,top]{\color{textcolor}{\rmfamily\fontsize{22.000000}{26.400000}\selectfont\catcode`\^=\active\def^{\ifmmode\sp\else\^{}\fi}\catcode`\%=\active\def
\end{pgfscope}%
\begin{pgfscope}%
\pgfsetbuttcap%
\pgfsetroundjoin%
\definecolor{currentfill}{rgb}{0.000000,0.000000,0.000000}%
\pgfsetfillcolor{currentfill}%
\pgfsetlinewidth{1.505625pt}%
\definecolor{currentstroke}{rgb}{0.000000,0.000000,0.000000}%
\pgfsetstrokecolor{currentstroke}%
\pgfsetdash{}{0pt}%
\pgfsys@defobject{currentmarker}{\pgfqpoint{-0.083333in}{0.000000in}}{\pgfqpoint{-0.000000in}{0.000000in}}{%
\pgfpathmoveto{\pgfqpoint{-0.000000in}{0.000000in}}%
\pgfpathlineto{\pgfqpoint{-0.083333in}{0.000000in}}%
\pgfusepath{stroke,fill}%
}%
\begin{pgfscope}%
\pgfsys@transformshift{0.675675in}{0.831623in}%
\pgfsys@useobject{currentmarker}{}%
\end{pgfscope}%
\end{pgfscope}%
\begin{pgfscope}%
\definecolor{textcolor}{rgb}{0.000000,0.000000,0.000000}%
\pgfsetstrokecolor{textcolor}%
\pgfsetfillcolor{textcolor}%
\pgftext[x=0.411623in, y=0.731604in, left, base]{\color{textcolor}{\rmfamily\fontsize{22.000000}{26.400000}\selectfont\catcode`\^=\active\def^{\ifmmode\sp\else\^{}\fi}\catcode`\%=\active\def
\end{pgfscope}%
\begin{pgfscope}%
\pgfsetbuttcap%
\pgfsetroundjoin%
\definecolor{currentfill}{rgb}{0.000000,0.000000,0.000000}%
\pgfsetfillcolor{currentfill}%
\pgfsetlinewidth{1.505625pt}%
\definecolor{currentstroke}{rgb}{0.000000,0.000000,0.000000}%
\pgfsetstrokecolor{currentstroke}%
\pgfsetdash{}{0pt}%
\pgfsys@defobject{currentmarker}{\pgfqpoint{-0.083333in}{0.000000in}}{\pgfqpoint{-0.000000in}{0.000000in}}{%
\pgfpathmoveto{\pgfqpoint{-0.000000in}{0.000000in}}%
\pgfpathlineto{\pgfqpoint{-0.083333in}{0.000000in}}%
\pgfusepath{stroke,fill}%
}%
\begin{pgfscope}%
\pgfsys@transformshift{0.675675in}{1.916334in}%
\pgfsys@useobject{currentmarker}{}%
\end{pgfscope}%
\end{pgfscope}%
\begin{pgfscope}%
\definecolor{textcolor}{rgb}{0.000000,0.000000,0.000000}%
\pgfsetstrokecolor{textcolor}%
\pgfsetfillcolor{textcolor}%
\pgftext[x=0.411623in, y=1.816315in, left, base]{\color{textcolor}{\rmfamily\fontsize{22.000000}{26.400000}\selectfont\catcode`\^=\active\def^{\ifmmode\sp\else\^{}\fi}\catcode`\%=\active\def
\end{pgfscope}%
\begin{pgfscope}%
\pgfsetbuttcap%
\pgfsetroundjoin%
\definecolor{currentfill}{rgb}{0.000000,0.000000,0.000000}%
\pgfsetfillcolor{currentfill}%
\pgfsetlinewidth{1.505625pt}%
\definecolor{currentstroke}{rgb}{0.000000,0.000000,0.000000}%
\pgfsetstrokecolor{currentstroke}%
\pgfsetdash{}{0pt}%
\pgfsys@defobject{currentmarker}{\pgfqpoint{-0.083333in}{0.000000in}}{\pgfqpoint{-0.000000in}{0.000000in}}{%
\pgfpathmoveto{\pgfqpoint{-0.000000in}{0.000000in}}%
\pgfpathlineto{\pgfqpoint{-0.083333in}{0.000000in}}%
\pgfusepath{stroke,fill}%
}%
\begin{pgfscope}%
\pgfsys@transformshift{0.675675in}{3.001045in}%
\pgfsys@useobject{currentmarker}{}%
\end{pgfscope}%
\end{pgfscope}%
\begin{pgfscope}%
\definecolor{textcolor}{rgb}{0.000000,0.000000,0.000000}%
\pgfsetstrokecolor{textcolor}%
\pgfsetfillcolor{textcolor}%
\pgftext[x=0.411623in, y=2.901026in, left, base]{\color{textcolor}{\rmfamily\fontsize{22.000000}{26.400000}\selectfont\catcode`\^=\active\def^{\ifmmode\sp\else\^{}\fi}\catcode`\%=\active\def
\end{pgfscope}%
\begin{pgfscope}%
\definecolor{textcolor}{rgb}{0.000000,0.000000,0.000000}%
\pgfsetstrokecolor{textcolor}%
\pgfsetfillcolor{textcolor}%
\pgftext[x=0.356068in,y=2.187512in,,bottom,rotate=90.000000]{\color{textcolor}{\rmfamily\fontsize{22.000000}{26.400000}\selectfont\catcode`\^=\active\def^{\ifmmode\sp\else\^{}\fi}\catcode`\%=\active\def
\end{pgfscope}%
\begin{pgfscope}%
\pgfpathrectangle{\pgfqpoint{0.675675in}{0.831623in}}{\pgfqpoint{3.907604in}{2.711777in}}%
\pgfusepath{clip}%
\pgfsetbuttcap%
\pgfsetroundjoin%
\pgfsetlinewidth{3.513125pt}%
\definecolor{currentstroke}{rgb}{0.666667,0.200000,0.466667}%
\pgfsetstrokecolor{currentstroke}%
\pgfsetstrokeopacity{0.800000}%
\pgfsetdash{}{0pt}%
\pgfpathmoveto{\pgfqpoint{0.853293in}{2.458690in}}%
\pgfpathlineto{\pgfqpoint{0.853293in}{3.001045in}}%
\pgfusepath{stroke}%
\end{pgfscope}%
\begin{pgfscope}%
\pgfpathrectangle{\pgfqpoint{0.675675in}{0.831623in}}{\pgfqpoint{3.907604in}{2.711777in}}%
\pgfusepath{clip}%
\pgfsetbuttcap%
\pgfsetroundjoin%
\pgfsetlinewidth{3.513125pt}%
\definecolor{currentstroke}{rgb}{0.666667,0.200000,0.466667}%
\pgfsetstrokecolor{currentstroke}%
\pgfsetstrokeopacity{0.800000}%
\pgfsetdash{}{0pt}%
\pgfpathmoveto{\pgfqpoint{1.445355in}{2.458690in}}%
\pgfpathlineto{\pgfqpoint{1.445355in}{2.610549in}}%
\pgfusepath{stroke}%
\end{pgfscope}%
\begin{pgfscope}%
\pgfpathrectangle{\pgfqpoint{0.675675in}{0.831623in}}{\pgfqpoint{3.907604in}{2.711777in}}%
\pgfusepath{clip}%
\pgfsetbuttcap%
\pgfsetroundjoin%
\pgfsetlinewidth{3.513125pt}%
\definecolor{currentstroke}{rgb}{0.666667,0.200000,0.466667}%
\pgfsetstrokecolor{currentstroke}%
\pgfsetstrokeopacity{0.800000}%
\pgfsetdash{}{0pt}%
\pgfpathmoveto{\pgfqpoint{2.037416in}{1.916334in}}%
\pgfpathlineto{\pgfqpoint{2.037416in}{2.458690in}}%
\pgfusepath{stroke}%
\end{pgfscope}%
\begin{pgfscope}%
\pgfpathrectangle{\pgfqpoint{0.675675in}{0.831623in}}{\pgfqpoint{3.907604in}{2.711777in}}%
\pgfusepath{clip}%
\pgfsetbuttcap%
\pgfsetroundjoin%
\pgfsetlinewidth{3.513125pt}%
\definecolor{currentstroke}{rgb}{0.666667,0.200000,0.466667}%
\pgfsetstrokecolor{currentstroke}%
\pgfsetstrokeopacity{0.800000}%
\pgfsetdash{}{0pt}%
\pgfpathmoveto{\pgfqpoint{2.629477in}{1.916334in}}%
\pgfpathlineto{\pgfqpoint{2.629477in}{2.458690in}}%
\pgfusepath{stroke}%
\end{pgfscope}%
\begin{pgfscope}%
\pgfpathrectangle{\pgfqpoint{0.675675in}{0.831623in}}{\pgfqpoint{3.907604in}{2.711777in}}%
\pgfusepath{clip}%
\pgfsetbuttcap%
\pgfsetroundjoin%
\pgfsetlinewidth{3.513125pt}%
\definecolor{currentstroke}{rgb}{0.666667,0.200000,0.466667}%
\pgfsetstrokecolor{currentstroke}%
\pgfsetstrokeopacity{0.800000}%
\pgfsetdash{}{0pt}%
\pgfpathmoveto{\pgfqpoint{3.221538in}{1.916334in}}%
\pgfpathlineto{\pgfqpoint{3.221538in}{2.458690in}}%
\pgfusepath{stroke}%
\end{pgfscope}%
\begin{pgfscope}%
\pgfpathrectangle{\pgfqpoint{0.675675in}{0.831623in}}{\pgfqpoint{3.907604in}{2.711777in}}%
\pgfusepath{clip}%
\pgfsetbuttcap%
\pgfsetroundjoin%
\pgfsetlinewidth{3.513125pt}%
\definecolor{currentstroke}{rgb}{0.666667,0.200000,0.466667}%
\pgfsetstrokecolor{currentstroke}%
\pgfsetstrokeopacity{0.800000}%
\pgfsetdash{}{0pt}%
\pgfpathmoveto{\pgfqpoint{3.813599in}{1.916334in}}%
\pgfpathlineto{\pgfqpoint{3.813599in}{2.458690in}}%
\pgfusepath{stroke}%
\end{pgfscope}%
\begin{pgfscope}%
\pgfpathrectangle{\pgfqpoint{0.675675in}{0.831623in}}{\pgfqpoint{3.907604in}{2.711777in}}%
\pgfusepath{clip}%
\pgfsetbuttcap%
\pgfsetroundjoin%
\pgfsetlinewidth{3.513125pt}%
\definecolor{currentstroke}{rgb}{0.666667,0.200000,0.466667}%
\pgfsetstrokecolor{currentstroke}%
\pgfsetstrokeopacity{0.800000}%
\pgfsetdash{}{0pt}%
\pgfpathmoveto{\pgfqpoint{4.405661in}{1.916334in}}%
\pgfpathlineto{\pgfqpoint{4.405661in}{2.458690in}}%
\pgfusepath{stroke}%
\end{pgfscope}%
\begin{pgfscope}%
\pgfpathrectangle{\pgfqpoint{0.675675in}{0.831623in}}{\pgfqpoint{3.907604in}{2.711777in}}%
\pgfusepath{clip}%
\pgfsetbuttcap%
\pgfsetroundjoin%
\definecolor{currentfill}{rgb}{0.666667,0.200000,0.466667}%
\pgfsetfillcolor{currentfill}%
\pgfsetfillopacity{0.800000}%
\pgfsetlinewidth{2.007500pt}%
\definecolor{currentstroke}{rgb}{0.666667,0.200000,0.466667}%
\pgfsetstrokecolor{currentstroke}%
\pgfsetstrokeopacity{0.800000}%
\pgfsetdash{}{0pt}%
\pgfsys@defobject{currentmarker}{\pgfqpoint{-0.055556in}{-0.000000in}}{\pgfqpoint{0.055556in}{0.000000in}}{%
\pgfpathmoveto{\pgfqpoint{0.055556in}{-0.000000in}}%
\pgfpathlineto{\pgfqpoint{-0.055556in}{0.000000in}}%
\pgfusepath{stroke,fill}%
}%
\begin{pgfscope}%
\pgfsys@transformshift{0.853293in}{2.458690in}%
\pgfsys@useobject{currentmarker}{}%
\end{pgfscope}%
\begin{pgfscope}%
\pgfsys@transformshift{1.445355in}{2.458690in}%
\pgfsys@useobject{currentmarker}{}%
\end{pgfscope}%
\begin{pgfscope}%
\pgfsys@transformshift{2.037416in}{1.916334in}%
\pgfsys@useobject{currentmarker}{}%
\end{pgfscope}%
\begin{pgfscope}%
\pgfsys@transformshift{2.629477in}{1.916334in}%
\pgfsys@useobject{currentmarker}{}%
\end{pgfscope}%
\begin{pgfscope}%
\pgfsys@transformshift{3.221538in}{1.916334in}%
\pgfsys@useobject{currentmarker}{}%
\end{pgfscope}%
\begin{pgfscope}%
\pgfsys@transformshift{3.813599in}{1.916334in}%
\pgfsys@useobject{currentmarker}{}%
\end{pgfscope}%
\begin{pgfscope}%
\pgfsys@transformshift{4.405661in}{1.916334in}%
\pgfsys@useobject{currentmarker}{}%
\end{pgfscope}%
\end{pgfscope}%
\begin{pgfscope}%
\pgfpathrectangle{\pgfqpoint{0.675675in}{0.831623in}}{\pgfqpoint{3.907604in}{2.711777in}}%
\pgfusepath{clip}%
\pgfsetbuttcap%
\pgfsetroundjoin%
\definecolor{currentfill}{rgb}{0.666667,0.200000,0.466667}%
\pgfsetfillcolor{currentfill}%
\pgfsetfillopacity{0.800000}%
\pgfsetlinewidth{2.007500pt}%
\definecolor{currentstroke}{rgb}{0.666667,0.200000,0.466667}%
\pgfsetstrokecolor{currentstroke}%
\pgfsetstrokeopacity{0.800000}%
\pgfsetdash{}{0pt}%
\pgfsys@defobject{currentmarker}{\pgfqpoint{-0.055556in}{-0.000000in}}{\pgfqpoint{0.055556in}{0.000000in}}{%
\pgfpathmoveto{\pgfqpoint{0.055556in}{-0.000000in}}%
\pgfpathlineto{\pgfqpoint{-0.055556in}{0.000000in}}%
\pgfusepath{stroke,fill}%
}%
\begin{pgfscope}%
\pgfsys@transformshift{0.853293in}{3.001045in}%
\pgfsys@useobject{currentmarker}{}%
\end{pgfscope}%
\begin{pgfscope}%
\pgfsys@transformshift{1.445355in}{2.610549in}%
\pgfsys@useobject{currentmarker}{}%
\end{pgfscope}%
\begin{pgfscope}%
\pgfsys@transformshift{2.037416in}{2.458690in}%
\pgfsys@useobject{currentmarker}{}%
\end{pgfscope}%
\begin{pgfscope}%
\pgfsys@transformshift{2.629477in}{2.458690in}%
\pgfsys@useobject{currentmarker}{}%
\end{pgfscope}%
\begin{pgfscope}%
\pgfsys@transformshift{3.221538in}{2.458690in}%
\pgfsys@useobject{currentmarker}{}%
\end{pgfscope}%
\begin{pgfscope}%
\pgfsys@transformshift{3.813599in}{2.458690in}%
\pgfsys@useobject{currentmarker}{}%
\end{pgfscope}%
\begin{pgfscope}%
\pgfsys@transformshift{4.405661in}{2.458690in}%
\pgfsys@useobject{currentmarker}{}%
\end{pgfscope}%
\end{pgfscope}%
\begin{pgfscope}%
\pgfpathrectangle{\pgfqpoint{0.675675in}{0.831623in}}{\pgfqpoint{3.907604in}{2.711777in}}%
\pgfusepath{clip}%
\pgfsetbuttcap%
\pgfsetroundjoin%
\pgfsetlinewidth{3.513125pt}%
\definecolor{currentstroke}{rgb}{0.200000,0.400000,0.466667}%
\pgfsetstrokecolor{currentstroke}%
\pgfsetstrokeopacity{0.800000}%
\pgfsetdash{}{0pt}%
\pgfpathmoveto{\pgfqpoint{0.853293in}{0.831623in}}%
\pgfpathlineto{\pgfqpoint{0.853293in}{0.885859in}}%
\pgfusepath{stroke}%
\end{pgfscope}%
\begin{pgfscope}%
\pgfpathrectangle{\pgfqpoint{0.675675in}{0.831623in}}{\pgfqpoint{3.907604in}{2.711777in}}%
\pgfusepath{clip}%
\pgfsetbuttcap%
\pgfsetroundjoin%
\pgfsetlinewidth{3.513125pt}%
\definecolor{currentstroke}{rgb}{0.200000,0.400000,0.466667}%
\pgfsetstrokecolor{currentstroke}%
\pgfsetstrokeopacity{0.800000}%
\pgfsetdash{}{0pt}%
\pgfpathmoveto{\pgfqpoint{1.149324in}{0.831623in}}%
\pgfpathlineto{\pgfqpoint{1.149324in}{0.929247in}}%
\pgfusepath{stroke}%
\end{pgfscope}%
\begin{pgfscope}%
\pgfpathrectangle{\pgfqpoint{0.675675in}{0.831623in}}{\pgfqpoint{3.907604in}{2.711777in}}%
\pgfusepath{clip}%
\pgfsetbuttcap%
\pgfsetroundjoin%
\pgfsetlinewidth{3.513125pt}%
\definecolor{currentstroke}{rgb}{0.200000,0.400000,0.466667}%
\pgfsetstrokecolor{currentstroke}%
\pgfsetstrokeopacity{0.800000}%
\pgfsetdash{}{0pt}%
\pgfpathmoveto{\pgfqpoint{1.445355in}{0.831623in}}%
\pgfpathlineto{\pgfqpoint{1.445355in}{1.373979in}}%
\pgfusepath{stroke}%
\end{pgfscope}%
\begin{pgfscope}%
\pgfpathrectangle{\pgfqpoint{0.675675in}{0.831623in}}{\pgfqpoint{3.907604in}{2.711777in}}%
\pgfusepath{clip}%
\pgfsetbuttcap%
\pgfsetroundjoin%
\pgfsetlinewidth{3.513125pt}%
\definecolor{currentstroke}{rgb}{0.200000,0.400000,0.466667}%
\pgfsetstrokecolor{currentstroke}%
\pgfsetstrokeopacity{0.800000}%
\pgfsetdash{}{0pt}%
\pgfpathmoveto{\pgfqpoint{1.741385in}{1.222119in}}%
\pgfpathlineto{\pgfqpoint{1.741385in}{1.373979in}}%
\pgfusepath{stroke}%
\end{pgfscope}%
\begin{pgfscope}%
\pgfpathrectangle{\pgfqpoint{0.675675in}{0.831623in}}{\pgfqpoint{3.907604in}{2.711777in}}%
\pgfusepath{clip}%
\pgfsetbuttcap%
\pgfsetroundjoin%
\pgfsetlinewidth{3.513125pt}%
\definecolor{currentstroke}{rgb}{0.200000,0.400000,0.466667}%
\pgfsetstrokecolor{currentstroke}%
\pgfsetstrokeopacity{0.800000}%
\pgfsetdash{}{0pt}%
\pgfpathmoveto{\pgfqpoint{2.037416in}{1.298049in}}%
\pgfpathlineto{\pgfqpoint{2.037416in}{1.373979in}}%
\pgfusepath{stroke}%
\end{pgfscope}%
\begin{pgfscope}%
\pgfpathrectangle{\pgfqpoint{0.675675in}{0.831623in}}{\pgfqpoint{3.907604in}{2.711777in}}%
\pgfusepath{clip}%
\pgfsetbuttcap%
\pgfsetroundjoin%
\pgfsetlinewidth{3.513125pt}%
\definecolor{currentstroke}{rgb}{0.200000,0.400000,0.466667}%
\pgfsetstrokecolor{currentstroke}%
\pgfsetstrokeopacity{0.800000}%
\pgfsetdash{}{0pt}%
\pgfpathmoveto{\pgfqpoint{2.333446in}{1.373979in}}%
\pgfpathlineto{\pgfqpoint{2.333446in}{1.525838in}}%
\pgfusepath{stroke}%
\end{pgfscope}%
\begin{pgfscope}%
\pgfpathrectangle{\pgfqpoint{0.675675in}{0.831623in}}{\pgfqpoint{3.907604in}{2.711777in}}%
\pgfusepath{clip}%
\pgfsetbuttcap%
\pgfsetroundjoin%
\pgfsetlinewidth{3.513125pt}%
\definecolor{currentstroke}{rgb}{0.200000,0.400000,0.466667}%
\pgfsetstrokecolor{currentstroke}%
\pgfsetstrokeopacity{0.800000}%
\pgfsetdash{}{0pt}%
\pgfpathmoveto{\pgfqpoint{2.629477in}{1.373979in}}%
\pgfpathlineto{\pgfqpoint{2.629477in}{1.916334in}}%
\pgfusepath{stroke}%
\end{pgfscope}%
\begin{pgfscope}%
\pgfpathrectangle{\pgfqpoint{0.675675in}{0.831623in}}{\pgfqpoint{3.907604in}{2.711777in}}%
\pgfusepath{clip}%
\pgfsetbuttcap%
\pgfsetroundjoin%
\pgfsetlinewidth{3.513125pt}%
\definecolor{currentstroke}{rgb}{0.200000,0.400000,0.466667}%
\pgfsetstrokecolor{currentstroke}%
\pgfsetstrokeopacity{0.800000}%
\pgfsetdash{}{0pt}%
\pgfpathmoveto{\pgfqpoint{2.925508in}{1.373979in}}%
\pgfpathlineto{\pgfqpoint{2.925508in}{1.916334in}}%
\pgfusepath{stroke}%
\end{pgfscope}%
\begin{pgfscope}%
\pgfpathrectangle{\pgfqpoint{0.675675in}{0.831623in}}{\pgfqpoint{3.907604in}{2.711777in}}%
\pgfusepath{clip}%
\pgfsetbuttcap%
\pgfsetroundjoin%
\pgfsetlinewidth{3.513125pt}%
\definecolor{currentstroke}{rgb}{0.200000,0.400000,0.466667}%
\pgfsetstrokecolor{currentstroke}%
\pgfsetstrokeopacity{0.800000}%
\pgfsetdash{}{0pt}%
\pgfusepath{stroke}%
\end{pgfscope}%
\begin{pgfscope}%
\pgfpathrectangle{\pgfqpoint{0.675675in}{0.831623in}}{\pgfqpoint{3.907604in}{2.711777in}}%
\pgfusepath{clip}%
\pgfsetbuttcap%
\pgfsetroundjoin%
\pgfsetlinewidth{3.513125pt}%
\definecolor{currentstroke}{rgb}{0.200000,0.400000,0.466667}%
\pgfsetstrokecolor{currentstroke}%
\pgfsetstrokeopacity{0.800000}%
\pgfsetdash{}{0pt}%
\pgfusepath{stroke}%
\end{pgfscope}%
\begin{pgfscope}%
\pgfpathrectangle{\pgfqpoint{0.675675in}{0.831623in}}{\pgfqpoint{3.907604in}{2.711777in}}%
\pgfusepath{clip}%
\pgfsetbuttcap%
\pgfsetroundjoin%
\pgfsetlinewidth{3.513125pt}%
\definecolor{currentstroke}{rgb}{0.200000,0.400000,0.466667}%
\pgfsetstrokecolor{currentstroke}%
\pgfsetstrokeopacity{0.800000}%
\pgfsetdash{}{0pt}%
\pgfusepath{stroke}%
\end{pgfscope}%
\begin{pgfscope}%
\pgfpathrectangle{\pgfqpoint{0.675675in}{0.831623in}}{\pgfqpoint{3.907604in}{2.711777in}}%
\pgfusepath{clip}%
\pgfsetbuttcap%
\pgfsetroundjoin%
\pgfsetlinewidth{3.513125pt}%
\definecolor{currentstroke}{rgb}{0.200000,0.400000,0.466667}%
\pgfsetstrokecolor{currentstroke}%
\pgfsetstrokeopacity{0.800000}%
\pgfsetdash{}{0pt}%
\pgfusepath{stroke}%
\end{pgfscope}%
\begin{pgfscope}%
\pgfpathrectangle{\pgfqpoint{0.675675in}{0.831623in}}{\pgfqpoint{3.907604in}{2.711777in}}%
\pgfusepath{clip}%
\pgfsetbuttcap%
\pgfsetroundjoin%
\pgfsetlinewidth{3.513125pt}%
\definecolor{currentstroke}{rgb}{0.200000,0.400000,0.466667}%
\pgfsetstrokecolor{currentstroke}%
\pgfsetstrokeopacity{0.800000}%
\pgfsetdash{}{0pt}%
\pgfusepath{stroke}%
\end{pgfscope}%
\begin{pgfscope}%
\pgfpathrectangle{\pgfqpoint{0.675675in}{0.831623in}}{\pgfqpoint{3.907604in}{2.711777in}}%
\pgfusepath{clip}%
\pgfsetbuttcap%
\pgfsetroundjoin%
\definecolor{currentfill}{rgb}{0.200000,0.400000,0.466667}%
\pgfsetfillcolor{currentfill}%
\pgfsetfillopacity{0.800000}%
\pgfsetlinewidth{2.007500pt}%
\definecolor{currentstroke}{rgb}{0.200000,0.400000,0.466667}%
\pgfsetstrokecolor{currentstroke}%
\pgfsetstrokeopacity{0.800000}%
\pgfsetdash{}{0pt}%
\pgfsys@defobject{currentmarker}{\pgfqpoint{-0.055556in}{-0.000000in}}{\pgfqpoint{0.055556in}{0.000000in}}{%
\pgfpathmoveto{\pgfqpoint{0.055556in}{-0.000000in}}%
\pgfpathlineto{\pgfqpoint{-0.055556in}{0.000000in}}%
\pgfusepath{stroke,fill}%
}%
\begin{pgfscope}%
\pgfsys@transformshift{0.853293in}{0.831623in}%
\pgfsys@useobject{currentmarker}{}%
\end{pgfscope}%
\begin{pgfscope}%
\pgfsys@transformshift{1.149324in}{0.831623in}%
\pgfsys@useobject{currentmarker}{}%
\end{pgfscope}%
\begin{pgfscope}%
\pgfsys@transformshift{1.445355in}{0.831623in}%
\pgfsys@useobject{currentmarker}{}%
\end{pgfscope}%
\begin{pgfscope}%
\pgfsys@transformshift{1.741385in}{1.222119in}%
\pgfsys@useobject{currentmarker}{}%
\end{pgfscope}%
\begin{pgfscope}%
\pgfsys@transformshift{2.037416in}{1.298049in}%
\pgfsys@useobject{currentmarker}{}%
\end{pgfscope}%
\begin{pgfscope}%
\pgfsys@transformshift{2.333446in}{1.373979in}%
\pgfsys@useobject{currentmarker}{}%
\end{pgfscope}%
\begin{pgfscope}%
\pgfsys@transformshift{2.629477in}{1.373979in}%
\pgfsys@useobject{currentmarker}{}%
\end{pgfscope}%
\begin{pgfscope}%
\pgfsys@transformshift{2.925508in}{1.373979in}%
\pgfsys@useobject{currentmarker}{}%
\end{pgfscope}%
\end{pgfscope}%
\begin{pgfscope}%
\pgfpathrectangle{\pgfqpoint{0.675675in}{0.831623in}}{\pgfqpoint{3.907604in}{2.711777in}}%
\pgfusepath{clip}%
\pgfsetbuttcap%
\pgfsetroundjoin%
\definecolor{currentfill}{rgb}{0.200000,0.400000,0.466667}%
\pgfsetfillcolor{currentfill}%
\pgfsetfillopacity{0.800000}%
\pgfsetlinewidth{2.007500pt}%
\definecolor{currentstroke}{rgb}{0.200000,0.400000,0.466667}%
\pgfsetstrokecolor{currentstroke}%
\pgfsetstrokeopacity{0.800000}%
\pgfsetdash{}{0pt}%
\pgfsys@defobject{currentmarker}{\pgfqpoint{-0.055556in}{-0.000000in}}{\pgfqpoint{0.055556in}{0.000000in}}{%
\pgfpathmoveto{\pgfqpoint{0.055556in}{-0.000000in}}%
\pgfpathlineto{\pgfqpoint{-0.055556in}{0.000000in}}%
\pgfusepath{stroke,fill}%
}%
\begin{pgfscope}%
\pgfsys@transformshift{0.853293in}{0.885859in}%
\pgfsys@useobject{currentmarker}{}%
\end{pgfscope}%
\begin{pgfscope}%
\pgfsys@transformshift{1.149324in}{0.929247in}%
\pgfsys@useobject{currentmarker}{}%
\end{pgfscope}%
\begin{pgfscope}%
\pgfsys@transformshift{1.445355in}{1.373979in}%
\pgfsys@useobject{currentmarker}{}%
\end{pgfscope}%
\begin{pgfscope}%
\pgfsys@transformshift{1.741385in}{1.373979in}%
\pgfsys@useobject{currentmarker}{}%
\end{pgfscope}%
\begin{pgfscope}%
\pgfsys@transformshift{2.037416in}{1.373979in}%
\pgfsys@useobject{currentmarker}{}%
\end{pgfscope}%
\begin{pgfscope}%
\pgfsys@transformshift{2.333446in}{1.525838in}%
\pgfsys@useobject{currentmarker}{}%
\end{pgfscope}%
\begin{pgfscope}%
\pgfsys@transformshift{2.629477in}{1.916334in}%
\pgfsys@useobject{currentmarker}{}%
\end{pgfscope}%
\begin{pgfscope}%
\pgfsys@transformshift{2.925508in}{1.916334in}%
\pgfsys@useobject{currentmarker}{}%
\end{pgfscope}%
\end{pgfscope}%
\begin{pgfscope}%
\pgfpathrectangle{\pgfqpoint{0.675675in}{0.831623in}}{\pgfqpoint{3.907604in}{2.711777in}}%
\pgfusepath{clip}%
\pgfsetbuttcap%
\pgfsetroundjoin%
\pgfsetlinewidth{3.513125pt}%
\definecolor{currentstroke}{rgb}{0.000000,0.000000,0.000000}%
\pgfsetstrokecolor{currentstroke}%
\pgfsetstrokeopacity{0.800000}%
\pgfsetdash{}{0pt}%
\pgfpathmoveto{\pgfqpoint{0.853293in}{1.373979in}}%
\pgfpathlineto{\pgfqpoint{0.853293in}{2.458690in}}%
\pgfusepath{stroke}%
\end{pgfscope}%
\begin{pgfscope}%
\pgfpathrectangle{\pgfqpoint{0.675675in}{0.831623in}}{\pgfqpoint{3.907604in}{2.711777in}}%
\pgfusepath{clip}%
\pgfsetbuttcap%
\pgfsetroundjoin%
\pgfsetlinewidth{3.513125pt}%
\definecolor{currentstroke}{rgb}{0.000000,0.000000,0.000000}%
\pgfsetstrokecolor{currentstroke}%
\pgfsetstrokeopacity{0.800000}%
\pgfsetdash{}{0pt}%
\pgfpathmoveto{\pgfqpoint{1.445355in}{1.330590in}}%
\pgfpathlineto{\pgfqpoint{1.445355in}{1.373979in}}%
\pgfusepath{stroke}%
\end{pgfscope}%
\begin{pgfscope}%
\pgfpathrectangle{\pgfqpoint{0.675675in}{0.831623in}}{\pgfqpoint{3.907604in}{2.711777in}}%
\pgfusepath{clip}%
\pgfsetbuttcap%
\pgfsetroundjoin%
\pgfsetlinewidth{3.513125pt}%
\definecolor{currentstroke}{rgb}{0.000000,0.000000,0.000000}%
\pgfsetstrokecolor{currentstroke}%
\pgfsetstrokeopacity{0.800000}%
\pgfsetdash{}{0pt}%
\pgfpathmoveto{\pgfqpoint{2.037416in}{1.373979in}}%
\pgfpathlineto{\pgfqpoint{2.037416in}{1.373979in}}%
\pgfusepath{stroke}%
\end{pgfscope}%
\begin{pgfscope}%
\pgfpathrectangle{\pgfqpoint{0.675675in}{0.831623in}}{\pgfqpoint{3.907604in}{2.711777in}}%
\pgfusepath{clip}%
\pgfsetbuttcap%
\pgfsetroundjoin%
\pgfsetlinewidth{3.513125pt}%
\definecolor{currentstroke}{rgb}{0.000000,0.000000,0.000000}%
\pgfsetstrokecolor{currentstroke}%
\pgfsetstrokeopacity{0.800000}%
\pgfsetdash{}{0pt}%
\pgfpathmoveto{\pgfqpoint{2.629477in}{1.373979in}}%
\pgfpathlineto{\pgfqpoint{2.629477in}{2.458690in}}%
\pgfusepath{stroke}%
\end{pgfscope}%
\begin{pgfscope}%
\pgfpathrectangle{\pgfqpoint{0.675675in}{0.831623in}}{\pgfqpoint{3.907604in}{2.711777in}}%
\pgfusepath{clip}%
\pgfsetbuttcap%
\pgfsetroundjoin%
\pgfsetlinewidth{3.513125pt}%
\definecolor{currentstroke}{rgb}{0.000000,0.000000,0.000000}%
\pgfsetstrokecolor{currentstroke}%
\pgfsetstrokeopacity{0.800000}%
\pgfsetdash{}{0pt}%
\pgfpathmoveto{\pgfqpoint{3.221538in}{1.780745in}}%
\pgfpathlineto{\pgfqpoint{3.221538in}{2.458690in}}%
\pgfusepath{stroke}%
\end{pgfscope}%
\begin{pgfscope}%
\pgfpathrectangle{\pgfqpoint{0.675675in}{0.831623in}}{\pgfqpoint{3.907604in}{2.711777in}}%
\pgfusepath{clip}%
\pgfsetbuttcap%
\pgfsetroundjoin%
\pgfsetlinewidth{3.513125pt}%
\definecolor{currentstroke}{rgb}{0.000000,0.000000,0.000000}%
\pgfsetstrokecolor{currentstroke}%
\pgfsetstrokeopacity{0.800000}%
\pgfsetdash{}{0pt}%
\pgfpathmoveto{\pgfqpoint{3.813599in}{1.916334in}}%
\pgfpathlineto{\pgfqpoint{3.813599in}{2.458690in}}%
\pgfusepath{stroke}%
\end{pgfscope}%
\begin{pgfscope}%
\pgfpathrectangle{\pgfqpoint{0.675675in}{0.831623in}}{\pgfqpoint{3.907604in}{2.711777in}}%
\pgfusepath{clip}%
\pgfsetbuttcap%
\pgfsetroundjoin%
\pgfsetlinewidth{3.513125pt}%
\definecolor{currentstroke}{rgb}{0.000000,0.000000,0.000000}%
\pgfsetstrokecolor{currentstroke}%
\pgfsetstrokeopacity{0.800000}%
\pgfsetdash{}{0pt}%
\pgfpathmoveto{\pgfqpoint{4.405661in}{1.916334in}}%
\pgfpathlineto{\pgfqpoint{4.405661in}{2.458690in}}%
\pgfusepath{stroke}%
\end{pgfscope}%
\begin{pgfscope}%
\pgfpathrectangle{\pgfqpoint{0.675675in}{0.831623in}}{\pgfqpoint{3.907604in}{2.711777in}}%
\pgfusepath{clip}%
\pgfsetbuttcap%
\pgfsetroundjoin%
\definecolor{currentfill}{rgb}{0.000000,0.000000,0.000000}%
\pgfsetfillcolor{currentfill}%
\pgfsetfillopacity{0.800000}%
\pgfsetlinewidth{2.007500pt}%
\definecolor{currentstroke}{rgb}{0.000000,0.000000,0.000000}%
\pgfsetstrokecolor{currentstroke}%
\pgfsetstrokeopacity{0.800000}%
\pgfsetdash{}{0pt}%
\pgfsys@defobject{currentmarker}{\pgfqpoint{-0.055556in}{-0.000000in}}{\pgfqpoint{0.055556in}{0.000000in}}{%
\pgfpathmoveto{\pgfqpoint{0.055556in}{-0.000000in}}%
\pgfpathlineto{\pgfqpoint{-0.055556in}{0.000000in}}%
\pgfusepath{stroke,fill}%
}%
\begin{pgfscope}%
\pgfsys@transformshift{0.853293in}{1.373979in}%
\pgfsys@useobject{currentmarker}{}%
\end{pgfscope}%
\begin{pgfscope}%
\pgfsys@transformshift{1.445355in}{1.330590in}%
\pgfsys@useobject{currentmarker}{}%
\end{pgfscope}%
\begin{pgfscope}%
\pgfsys@transformshift{2.037416in}{1.373979in}%
\pgfsys@useobject{currentmarker}{}%
\end{pgfscope}%
\begin{pgfscope}%
\pgfsys@transformshift{2.629477in}{1.373979in}%
\pgfsys@useobject{currentmarker}{}%
\end{pgfscope}%
\begin{pgfscope}%
\pgfsys@transformshift{3.221538in}{1.780745in}%
\pgfsys@useobject{currentmarker}{}%
\end{pgfscope}%
\begin{pgfscope}%
\pgfsys@transformshift{3.813599in}{1.916334in}%
\pgfsys@useobject{currentmarker}{}%
\end{pgfscope}%
\begin{pgfscope}%
\pgfsys@transformshift{4.405661in}{1.916334in}%
\pgfsys@useobject{currentmarker}{}%
\end{pgfscope}%
\end{pgfscope}%
\begin{pgfscope}%
\pgfpathrectangle{\pgfqpoint{0.675675in}{0.831623in}}{\pgfqpoint{3.907604in}{2.711777in}}%
\pgfusepath{clip}%
\pgfsetbuttcap%
\pgfsetroundjoin%
\definecolor{currentfill}{rgb}{0.000000,0.000000,0.000000}%
\pgfsetfillcolor{currentfill}%
\pgfsetfillopacity{0.800000}%
\pgfsetlinewidth{2.007500pt}%
\definecolor{currentstroke}{rgb}{0.000000,0.000000,0.000000}%
\pgfsetstrokecolor{currentstroke}%
\pgfsetstrokeopacity{0.800000}%
\pgfsetdash{}{0pt}%
\pgfsys@defobject{currentmarker}{\pgfqpoint{-0.055556in}{-0.000000in}}{\pgfqpoint{0.055556in}{0.000000in}}{%
\pgfpathmoveto{\pgfqpoint{0.055556in}{-0.000000in}}%
\pgfpathlineto{\pgfqpoint{-0.055556in}{0.000000in}}%
\pgfusepath{stroke,fill}%
}%
\begin{pgfscope}%
\pgfsys@transformshift{0.853293in}{2.458690in}%
\pgfsys@useobject{currentmarker}{}%
\end{pgfscope}%
\begin{pgfscope}%
\pgfsys@transformshift{1.445355in}{1.373979in}%
\pgfsys@useobject{currentmarker}{}%
\end{pgfscope}%
\begin{pgfscope}%
\pgfsys@transformshift{2.037416in}{1.373979in}%
\pgfsys@useobject{currentmarker}{}%
\end{pgfscope}%
\begin{pgfscope}%
\pgfsys@transformshift{2.629477in}{2.458690in}%
\pgfsys@useobject{currentmarker}{}%
\end{pgfscope}%
\begin{pgfscope}%
\pgfsys@transformshift{3.221538in}{2.458690in}%
\pgfsys@useobject{currentmarker}{}%
\end{pgfscope}%
\begin{pgfscope}%
\pgfsys@transformshift{3.813599in}{2.458690in}%
\pgfsys@useobject{currentmarker}{}%
\end{pgfscope}%
\begin{pgfscope}%
\pgfsys@transformshift{4.405661in}{2.458690in}%
\pgfsys@useobject{currentmarker}{}%
\end{pgfscope}%
\end{pgfscope}%
\begin{pgfscope}%
\pgfpathrectangle{\pgfqpoint{0.675675in}{0.831623in}}{\pgfqpoint{3.907604in}{2.711777in}}%
\pgfusepath{clip}%
\pgfsetrectcap%
\pgfsetroundjoin%
\pgfsetlinewidth{2.509375pt}%
\definecolor{currentstroke}{rgb}{0.501961,0.501961,0.501961}%
\pgfsetstrokecolor{currentstroke}%
\pgfsetstrokeopacity{0.800000}%
\pgfsetdash{}{0pt}%
\pgfpathmoveto{\pgfqpoint{0.675675in}{1.916334in}}%
\pgfpathlineto{\pgfqpoint{4.583279in}{1.916334in}}%
\pgfusepath{stroke}%
\end{pgfscope}%
\begin{pgfscope}%
\pgfpathrectangle{\pgfqpoint{0.675675in}{0.831623in}}{\pgfqpoint{3.907604in}{2.711777in}}%
\pgfusepath{clip}%
\pgfsetbuttcap%
\pgfsetroundjoin%
\pgfsetlinewidth{3.513125pt}%
\definecolor{currentstroke}{rgb}{0.666667,0.200000,0.466667}%
\pgfsetstrokecolor{currentstroke}%
\pgfsetstrokeopacity{0.800000}%
\pgfsetdash{{12.950000pt}{5.600000pt}}{0.000000pt}%
\pgfpathmoveto{\pgfqpoint{0.853293in}{2.632243in}}%
\pgfpathlineto{\pgfqpoint{1.445355in}{2.534619in}}%
\pgfpathlineto{\pgfqpoint{2.037416in}{2.377336in}}%
\pgfpathlineto{\pgfqpoint{2.629477in}{2.409878in}}%
\pgfpathlineto{\pgfqpoint{3.221538in}{2.361066in}}%
\pgfpathlineto{\pgfqpoint{3.813599in}{2.377336in}}%
\pgfpathlineto{\pgfqpoint{4.405661in}{2.339371in}}%
\pgfusepath{stroke}%
\end{pgfscope}%
\begin{pgfscope}%
\pgfpathrectangle{\pgfqpoint{0.675675in}{0.831623in}}{\pgfqpoint{3.907604in}{2.711777in}}%
\pgfusepath{clip}%
\pgfsetbuttcap%
\pgfsetroundjoin%
\definecolor{currentfill}{rgb}{0.666667,0.200000,0.466667}%
\pgfsetfillcolor{currentfill}%
\pgfsetfillopacity{0.800000}%
\pgfsetlinewidth{2.007500pt}%
\definecolor{currentstroke}{rgb}{1.000000,1.000000,1.000000}%
\pgfsetstrokecolor{currentstroke}%
\pgfsetstrokeopacity{0.800000}%
\pgfsetdash{}{0pt}%
\pgfsys@defobject{currentmarker}{\pgfqpoint{-0.055556in}{-0.055556in}}{\pgfqpoint{0.055556in}{0.055556in}}{%
\pgfpathmoveto{\pgfqpoint{0.000000in}{-0.055556in}}%
\pgfpathcurveto{\pgfqpoint{0.014734in}{-0.055556in}}{\pgfqpoint{0.028866in}{-0.049702in}}{\pgfqpoint{0.039284in}{-0.039284in}}%
\pgfpathcurveto{\pgfqpoint{0.049702in}{-0.028866in}}{\pgfqpoint{0.055556in}{-0.014734in}}{\pgfqpoint{0.055556in}{0.000000in}}%
\pgfpathcurveto{\pgfqpoint{0.055556in}{0.014734in}}{\pgfqpoint{0.049702in}{0.028866in}}{\pgfqpoint{0.039284in}{0.039284in}}%
\pgfpathcurveto{\pgfqpoint{0.028866in}{0.049702in}}{\pgfqpoint{0.014734in}{0.055556in}}{\pgfqpoint{0.000000in}{0.055556in}}%
\pgfpathcurveto{\pgfqpoint{-0.014734in}{0.055556in}}{\pgfqpoint{-0.028866in}{0.049702in}}{\pgfqpoint{-0.039284in}{0.039284in}}%
\pgfpathcurveto{\pgfqpoint{-0.049702in}{0.028866in}}{\pgfqpoint{-0.055556in}{0.014734in}}{\pgfqpoint{-0.055556in}{0.000000in}}%
\pgfpathcurveto{\pgfqpoint{-0.055556in}{-0.014734in}}{\pgfqpoint{-0.049702in}{-0.028866in}}{\pgfqpoint{-0.039284in}{-0.039284in}}%
\pgfpathcurveto{\pgfqpoint{-0.028866in}{-0.049702in}}{\pgfqpoint{-0.014734in}{-0.055556in}}{\pgfqpoint{0.000000in}{-0.055556in}}%
\pgfpathlineto{\pgfqpoint{0.000000in}{-0.055556in}}%
\pgfpathclose%
\pgfusepath{stroke,fill}%
}%
\begin{pgfscope}%
\pgfsys@transformshift{0.853293in}{2.632243in}%
\pgfsys@useobject{currentmarker}{}%
\end{pgfscope}%
\begin{pgfscope}%
\pgfsys@transformshift{1.445355in}{2.534619in}%
\pgfsys@useobject{currentmarker}{}%
\end{pgfscope}%
\begin{pgfscope}%
\pgfsys@transformshift{2.037416in}{2.377336in}%
\pgfsys@useobject{currentmarker}{}%
\end{pgfscope}%
\begin{pgfscope}%
\pgfsys@transformshift{2.629477in}{2.409878in}%
\pgfsys@useobject{currentmarker}{}%
\end{pgfscope}%
\begin{pgfscope}%
\pgfsys@transformshift{3.221538in}{2.361066in}%
\pgfsys@useobject{currentmarker}{}%
\end{pgfscope}%
\begin{pgfscope}%
\pgfsys@transformshift{3.813599in}{2.377336in}%
\pgfsys@useobject{currentmarker}{}%
\end{pgfscope}%
\begin{pgfscope}%
\pgfsys@transformshift{4.405661in}{2.339371in}%
\pgfsys@useobject{currentmarker}{}%
\end{pgfscope}%
\end{pgfscope}%
\begin{pgfscope}%
\pgfpathrectangle{\pgfqpoint{0.675675in}{0.831623in}}{\pgfqpoint{3.907604in}{2.711777in}}%
\pgfusepath{clip}%
\pgfsetbuttcap%
\pgfsetroundjoin%
\pgfsetlinewidth{3.513125pt}%
\definecolor{currentstroke}{rgb}{0.200000,0.400000,0.466667}%
\pgfsetstrokecolor{currentstroke}%
\pgfsetstrokeopacity{0.800000}%
\pgfsetdash{{22.400000pt}{5.600000pt}{3.500000pt}{5.600000pt}}{0.000000pt}%
\pgfpathmoveto{\pgfqpoint{0.853293in}{0.858741in}}%
\pgfpathlineto{\pgfqpoint{1.149324in}{0.880435in}}%
\pgfpathlineto{\pgfqpoint{1.445355in}{1.086530in}}%
\pgfpathlineto{\pgfqpoint{1.741385in}{1.298049in}}%
\pgfpathlineto{\pgfqpoint{2.037416in}{1.336014in}}%
\pgfpathlineto{\pgfqpoint{2.333446in}{1.449908in}}%
\pgfpathlineto{\pgfqpoint{2.629477in}{1.547532in}}%
\pgfpathlineto{\pgfqpoint{2.925508in}{1.623462in}}%
\pgfusepath{stroke}%
\end{pgfscope}%
\begin{pgfscope}%
\pgfpathrectangle{\pgfqpoint{0.675675in}{0.831623in}}{\pgfqpoint{3.907604in}{2.711777in}}%
\pgfusepath{clip}%
\pgfsetbuttcap%
\pgfsetmiterjoin%
\definecolor{currentfill}{rgb}{0.200000,0.400000,0.466667}%
\pgfsetfillcolor{currentfill}%
\pgfsetfillopacity{0.800000}%
\pgfsetlinewidth{2.007500pt}%
\definecolor{currentstroke}{rgb}{1.000000,1.000000,1.000000}%
\pgfsetstrokecolor{currentstroke}%
\pgfsetstrokeopacity{0.800000}%
\pgfsetdash{}{0pt}%
\pgfsys@defobject{currentmarker}{\pgfqpoint{-0.055556in}{-0.055556in}}{\pgfqpoint{0.055556in}{0.055556in}}{%
\pgfpathmoveto{\pgfqpoint{-0.000000in}{-0.055556in}}%
\pgfpathlineto{\pgfqpoint{0.055556in}{0.055556in}}%
\pgfpathlineto{\pgfqpoint{-0.055556in}{0.055556in}}%
\pgfpathlineto{\pgfqpoint{-0.000000in}{-0.055556in}}%
\pgfpathclose%
\pgfusepath{stroke,fill}%
}%
\begin{pgfscope}%
\pgfsys@transformshift{0.853293in}{0.858741in}%
\pgfsys@useobject{currentmarker}{}%
\end{pgfscope}%
\begin{pgfscope}%
\pgfsys@transformshift{1.149324in}{0.880435in}%
\pgfsys@useobject{currentmarker}{}%
\end{pgfscope}%
\begin{pgfscope}%
\pgfsys@transformshift{1.445355in}{1.086530in}%
\pgfsys@useobject{currentmarker}{}%
\end{pgfscope}%
\begin{pgfscope}%
\pgfsys@transformshift{1.741385in}{1.298049in}%
\pgfsys@useobject{currentmarker}{}%
\end{pgfscope}%
\begin{pgfscope}%
\pgfsys@transformshift{2.037416in}{1.336014in}%
\pgfsys@useobject{currentmarker}{}%
\end{pgfscope}%
\begin{pgfscope}%
\pgfsys@transformshift{2.333446in}{1.449908in}%
\pgfsys@useobject{currentmarker}{}%
\end{pgfscope}%
\begin{pgfscope}%
\pgfsys@transformshift{2.629477in}{1.547532in}%
\pgfsys@useobject{currentmarker}{}%
\end{pgfscope}%
\begin{pgfscope}%
\pgfsys@transformshift{2.925508in}{1.623462in}%
\pgfsys@useobject{currentmarker}{}%
\end{pgfscope}%
\end{pgfscope}%
\begin{pgfscope}%
\pgfpathrectangle{\pgfqpoint{0.675675in}{0.831623in}}{\pgfqpoint{3.907604in}{2.711777in}}%
\pgfusepath{clip}%
\pgfsetrectcap%
\pgfsetroundjoin%
\pgfsetlinewidth{3.513125pt}%
\definecolor{currentstroke}{rgb}{0.000000,0.000000,0.000000}%
\pgfsetstrokecolor{currentstroke}%
\pgfsetstrokeopacity{0.800000}%
\pgfsetdash{}{0pt}%
\pgfpathmoveto{\pgfqpoint{0.853293in}{1.883793in}}%
\pgfpathlineto{\pgfqpoint{1.445355in}{1.352284in}}%
\pgfpathlineto{\pgfqpoint{2.037416in}{1.373979in}}%
\pgfpathlineto{\pgfqpoint{2.629477in}{1.872946in}}%
\pgfpathlineto{\pgfqpoint{3.221538in}{2.122429in}}%
\pgfpathlineto{\pgfqpoint{3.813599in}{2.122429in}}%
\pgfpathlineto{\pgfqpoint{4.405661in}{2.176665in}}%
\pgfusepath{stroke}%
\end{pgfscope}%
\begin{pgfscope}%
\pgfpathrectangle{\pgfqpoint{0.675675in}{0.831623in}}{\pgfqpoint{3.907604in}{2.711777in}}%
\pgfusepath{clip}%
\pgfsetbuttcap%
\pgfsetmiterjoin%
\definecolor{currentfill}{rgb}{0.000000,0.000000,0.000000}%
\pgfsetfillcolor{currentfill}%
\pgfsetfillopacity{0.800000}%
\pgfsetlinewidth{2.007500pt}%
\definecolor{currentstroke}{rgb}{1.000000,1.000000,1.000000}%
\pgfsetstrokecolor{currentstroke}%
\pgfsetstrokeopacity{0.800000}%
\pgfsetdash{}{0pt}%
\pgfsys@defobject{currentmarker}{\pgfqpoint{-0.078567in}{-0.078567in}}{\pgfqpoint{0.078567in}{0.078567in}}{%
\pgfpathmoveto{\pgfqpoint{-0.000000in}{-0.078567in}}%
\pgfpathlineto{\pgfqpoint{0.078567in}{0.000000in}}%
\pgfpathlineto{\pgfqpoint{0.000000in}{0.078567in}}%
\pgfpathlineto{\pgfqpoint{-0.078567in}{0.000000in}}%
\pgfpathlineto{\pgfqpoint{-0.000000in}{-0.078567in}}%
\pgfpathclose%
\pgfusepath{stroke,fill}%
}%
\begin{pgfscope}%
\pgfsys@transformshift{0.853293in}{1.883793in}%
\pgfsys@useobject{currentmarker}{}%
\end{pgfscope}%
\begin{pgfscope}%
\pgfsys@transformshift{1.445355in}{1.352284in}%
\pgfsys@useobject{currentmarker}{}%
\end{pgfscope}%
\begin{pgfscope}%
\pgfsys@transformshift{2.037416in}{1.373979in}%
\pgfsys@useobject{currentmarker}{}%
\end{pgfscope}%
\begin{pgfscope}%
\pgfsys@transformshift{2.629477in}{1.872946in}%
\pgfsys@useobject{currentmarker}{}%
\end{pgfscope}%
\begin{pgfscope}%
\pgfsys@transformshift{3.221538in}{2.122429in}%
\pgfsys@useobject{currentmarker}{}%
\end{pgfscope}%
\begin{pgfscope}%
\pgfsys@transformshift{3.813599in}{2.122429in}%
\pgfsys@useobject{currentmarker}{}%
\end{pgfscope}%
\begin{pgfscope}%
\pgfsys@transformshift{4.405661in}{2.176665in}%
\pgfsys@useobject{currentmarker}{}%
\end{pgfscope}%
\end{pgfscope}%
\begin{pgfscope}%
\pgfsetrectcap%
\pgfsetmiterjoin%
\pgfsetlinewidth{1.505625pt}%
\definecolor{currentstroke}{rgb}{0.000000,0.000000,0.000000}%
\pgfsetstrokecolor{currentstroke}%
\pgfsetdash{}{0pt}%
\pgfpathmoveto{\pgfqpoint{0.675675in}{0.831623in}}%
\pgfpathlineto{\pgfqpoint{0.675675in}{3.543400in}}%
\pgfusepath{stroke}%
\end{pgfscope}%
\begin{pgfscope}%
\pgfsetrectcap%
\pgfsetmiterjoin%
\pgfsetlinewidth{1.505625pt}%
\definecolor{currentstroke}{rgb}{0.000000,0.000000,0.000000}%
\pgfsetstrokecolor{currentstroke}%
\pgfsetdash{}{0pt}%
\pgfpathmoveto{\pgfqpoint{0.675675in}{0.831623in}}%
\pgfpathlineto{\pgfqpoint{4.583279in}{0.831623in}}%
\pgfusepath{stroke}%
\end{pgfscope}%
\end{pgfpicture}%
\makeatother%
\endgroup%

%% file: figures/synthetic-final/M3/delta6/size_asymmetric_errors.pgf
\begingroup%
\makeatletter%
\begin{pgfpicture}%
\pgfpathrectangle{\pgfpointorigin}{\pgfqpoint{4.757956in}{3.745236in}}%
\pgfusepath{use as bounding box, clip}%
\begin{pgfscope}%
\pgfsetbuttcap%
\pgfsetmiterjoin%
\definecolor{currentfill}{rgb}{1.000000,1.000000,1.000000}%
\pgfsetfillcolor{currentfill}%
\pgfsetlinewidth{0.000000pt}%
\definecolor{currentstroke}{rgb}{1.000000,1.000000,1.000000}%
\pgfsetstrokecolor{currentstroke}%
\pgfsetdash{}{0pt}%
\pgfpathmoveto{\pgfqpoint{0.000000in}{0.000000in}}%
\pgfpathlineto{\pgfqpoint{4.757956in}{0.000000in}}%
\pgfpathlineto{\pgfqpoint{4.757956in}{3.745236in}}%
\pgfpathlineto{\pgfqpoint{0.000000in}{3.745236in}}%
\pgfpathlineto{\pgfqpoint{0.000000in}{0.000000in}}%
\pgfpathclose%
\pgfusepath{fill}%
\end{pgfscope}%
\begin{pgfscope}%
\pgfsetbuttcap%
\pgfsetmiterjoin%
\definecolor{currentfill}{rgb}{1.000000,1.000000,1.000000}%
\pgfsetfillcolor{currentfill}%
\pgfsetlinewidth{0.000000pt}%
\definecolor{currentstroke}{rgb}{0.000000,0.000000,0.000000}%
\pgfsetstrokecolor{currentstroke}%
\pgfsetstrokeopacity{0.000000}%
\pgfsetdash{}{0pt}%
\pgfpathmoveto{\pgfqpoint{0.675675in}{0.799635in}}%
\pgfpathlineto{\pgfqpoint{4.657956in}{0.799635in}}%
\pgfpathlineto{\pgfqpoint{4.657956in}{3.545217in}}%
\pgfpathlineto{\pgfqpoint{0.675675in}{3.545217in}}%
\pgfpathlineto{\pgfqpoint{0.675675in}{0.799635in}}%
\pgfpathclose%
\pgfusepath{fill}%
\end{pgfscope}%
\begin{pgfscope}%
\pgfsetbuttcap%
\pgfsetroundjoin%
\definecolor{currentfill}{rgb}{0.000000,0.000000,0.000000}%
\pgfsetfillcolor{currentfill}%
\pgfsetlinewidth{1.505625pt}%
\definecolor{currentstroke}{rgb}{0.000000,0.000000,0.000000}%
\pgfsetstrokecolor{currentstroke}%
\pgfsetdash{}{0pt}%
\pgfsys@defobject{currentmarker}{\pgfqpoint{0.000000in}{-0.083333in}}{\pgfqpoint{0.000000in}{0.000000in}}{%
\pgfpathmoveto{\pgfqpoint{0.000000in}{0.000000in}}%
\pgfpathlineto{\pgfqpoint{0.000000in}{-0.083333in}}%
\pgfusepath{stroke,fill}%
}%
\begin{pgfscope}%
\pgfsys@transformshift{0.856688in}{0.799635in}%
\pgfsys@useobject{currentmarker}{}%
\end{pgfscope}%
\end{pgfscope}%
\begin{pgfscope}%
\definecolor{textcolor}{rgb}{0.000000,0.000000,0.000000}%
\pgfsetstrokecolor{textcolor}%
\pgfsetfillcolor{textcolor}%
\pgftext[x=0.856688in,y=0.667691in,,top]{\color{textcolor}{\rmfamily\fontsize{22.000000}{26.400000}\selectfont\catcode`\^=\active\def^{\ifmmode\sp\else\^{}\fi}\catcode`\%=\active\def
\end{pgfscope}%
\begin{pgfscope}%
\pgfsetbuttcap%
\pgfsetroundjoin%
\definecolor{currentfill}{rgb}{0.000000,0.000000,0.000000}%
\pgfsetfillcolor{currentfill}%
\pgfsetlinewidth{1.505625pt}%
\definecolor{currentstroke}{rgb}{0.000000,0.000000,0.000000}%
\pgfsetstrokecolor{currentstroke}%
\pgfsetdash{}{0pt}%
\pgfsys@defobject{currentmarker}{\pgfqpoint{0.000000in}{-0.083333in}}{\pgfqpoint{0.000000in}{0.000000in}}{%
\pgfpathmoveto{\pgfqpoint{0.000000in}{0.000000in}}%
\pgfpathlineto{\pgfqpoint{0.000000in}{-0.083333in}}%
\pgfusepath{stroke,fill}%
}%
\begin{pgfscope}%
\pgfsys@transformshift{2.430007in}{0.799635in}%
\pgfsys@useobject{currentmarker}{}%
\end{pgfscope}%
\end{pgfscope}%
\begin{pgfscope}%
\definecolor{textcolor}{rgb}{0.000000,0.000000,0.000000}%
\pgfsetstrokecolor{textcolor}%
\pgfsetfillcolor{textcolor}%
\pgftext[x=2.430007in,y=0.667691in,,top]{\color{textcolor}{\rmfamily\fontsize{22.000000}{26.400000}\selectfont\catcode`\^=\active\def^{\ifmmode\sp\else\^{}\fi}\catcode`\%=\active\def
\end{pgfscope}%
\begin{pgfscope}%
\pgfsetbuttcap%
\pgfsetroundjoin%
\definecolor{currentfill}{rgb}{0.000000,0.000000,0.000000}%
\pgfsetfillcolor{currentfill}%
\pgfsetlinewidth{1.505625pt}%
\definecolor{currentstroke}{rgb}{0.000000,0.000000,0.000000}%
\pgfsetstrokecolor{currentstroke}%
\pgfsetdash{}{0pt}%
\pgfsys@defobject{currentmarker}{\pgfqpoint{0.000000in}{-0.083333in}}{\pgfqpoint{0.000000in}{0.000000in}}{%
\pgfpathmoveto{\pgfqpoint{0.000000in}{0.000000in}}%
\pgfpathlineto{\pgfqpoint{0.000000in}{-0.083333in}}%
\pgfusepath{stroke,fill}%
}%
\begin{pgfscope}%
\pgfsys@transformshift{4.003327in}{0.799635in}%
\pgfsys@useobject{currentmarker}{}%
\end{pgfscope}%
\end{pgfscope}%
\begin{pgfscope}%
\definecolor{textcolor}{rgb}{0.000000,0.000000,0.000000}%
\pgfsetstrokecolor{textcolor}%
\pgfsetfillcolor{textcolor}%
\pgftext[x=4.003327in,y=0.667691in,,top]{\color{textcolor}{\rmfamily\fontsize{22.000000}{26.400000}\selectfont\catcode`\^=\active\def^{\ifmmode\sp\else\^{}\fi}\catcode`\%=\active\def
\end{pgfscope}%
\begin{pgfscope}%
\pgfsetbuttcap%
\pgfsetroundjoin%
\definecolor{currentfill}{rgb}{0.000000,0.000000,0.000000}%
\pgfsetfillcolor{currentfill}%
\pgfsetlinewidth{0.602250pt}%
\definecolor{currentstroke}{rgb}{0.000000,0.000000,0.000000}%
\pgfsetstrokecolor{currentstroke}%
\pgfsetdash{}{0pt}%
\pgfsys@defobject{currentmarker}{\pgfqpoint{0.000000in}{-0.027778in}}{\pgfqpoint{0.000000in}{0.000000in}}{%
\pgfpathmoveto{\pgfqpoint{0.000000in}{0.000000in}}%
\pgfpathlineto{\pgfqpoint{0.000000in}{-0.027778in}}%
\pgfusepath{stroke,fill}%
}%
\begin{pgfscope}%
\pgfsys@transformshift{0.704217in}{0.799635in}%
\pgfsys@useobject{currentmarker}{}%
\end{pgfscope}%
\end{pgfscope}%
\begin{pgfscope}%
\pgfsetbuttcap%
\pgfsetroundjoin%
\definecolor{currentfill}{rgb}{0.000000,0.000000,0.000000}%
\pgfsetfillcolor{currentfill}%
\pgfsetlinewidth{0.602250pt}%
\definecolor{currentstroke}{rgb}{0.000000,0.000000,0.000000}%
\pgfsetstrokecolor{currentstroke}%
\pgfsetdash{}{0pt}%
\pgfsys@defobject{currentmarker}{\pgfqpoint{0.000000in}{-0.027778in}}{\pgfqpoint{0.000000in}{0.000000in}}{%
\pgfpathmoveto{\pgfqpoint{0.000000in}{0.000000in}}%
\pgfpathlineto{\pgfqpoint{0.000000in}{-0.027778in}}%
\pgfusepath{stroke,fill}%
}%
\begin{pgfscope}%
\pgfsys@transformshift{0.784697in}{0.799635in}%
\pgfsys@useobject{currentmarker}{}%
\end{pgfscope}%
\end{pgfscope}%
\begin{pgfscope}%
\pgfsetbuttcap%
\pgfsetroundjoin%
\definecolor{currentfill}{rgb}{0.000000,0.000000,0.000000}%
\pgfsetfillcolor{currentfill}%
\pgfsetlinewidth{0.602250pt}%
\definecolor{currentstroke}{rgb}{0.000000,0.000000,0.000000}%
\pgfsetstrokecolor{currentstroke}%
\pgfsetdash{}{0pt}%
\pgfsys@defobject{currentmarker}{\pgfqpoint{0.000000in}{-0.027778in}}{\pgfqpoint{0.000000in}{0.000000in}}{%
\pgfpathmoveto{\pgfqpoint{0.000000in}{0.000000in}}%
\pgfpathlineto{\pgfqpoint{0.000000in}{-0.027778in}}%
\pgfusepath{stroke,fill}%
}%
\begin{pgfscope}%
\pgfsys@transformshift{1.330304in}{0.799635in}%
\pgfsys@useobject{currentmarker}{}%
\end{pgfscope}%
\end{pgfscope}%
\begin{pgfscope}%
\pgfsetbuttcap%
\pgfsetroundjoin%
\definecolor{currentfill}{rgb}{0.000000,0.000000,0.000000}%
\pgfsetfillcolor{currentfill}%
\pgfsetlinewidth{0.602250pt}%
\definecolor{currentstroke}{rgb}{0.000000,0.000000,0.000000}%
\pgfsetstrokecolor{currentstroke}%
\pgfsetdash{}{0pt}%
\pgfsys@defobject{currentmarker}{\pgfqpoint{0.000000in}{-0.027778in}}{\pgfqpoint{0.000000in}{0.000000in}}{%
\pgfpathmoveto{\pgfqpoint{0.000000in}{0.000000in}}%
\pgfpathlineto{\pgfqpoint{0.000000in}{-0.027778in}}%
\pgfusepath{stroke,fill}%
}%
\begin{pgfscope}%
\pgfsys@transformshift{1.607352in}{0.799635in}%
\pgfsys@useobject{currentmarker}{}%
\end{pgfscope}%
\end{pgfscope}%
\begin{pgfscope}%
\pgfsetbuttcap%
\pgfsetroundjoin%
\definecolor{currentfill}{rgb}{0.000000,0.000000,0.000000}%
\pgfsetfillcolor{currentfill}%
\pgfsetlinewidth{0.602250pt}%
\definecolor{currentstroke}{rgb}{0.000000,0.000000,0.000000}%
\pgfsetstrokecolor{currentstroke}%
\pgfsetdash{}{0pt}%
\pgfsys@defobject{currentmarker}{\pgfqpoint{0.000000in}{-0.027778in}}{\pgfqpoint{0.000000in}{0.000000in}}{%
\pgfpathmoveto{\pgfqpoint{0.000000in}{0.000000in}}%
\pgfpathlineto{\pgfqpoint{0.000000in}{-0.027778in}}%
\pgfusepath{stroke,fill}%
}%
\begin{pgfscope}%
\pgfsys@transformshift{1.803920in}{0.799635in}%
\pgfsys@useobject{currentmarker}{}%
\end{pgfscope}%
\end{pgfscope}%
\begin{pgfscope}%
\pgfsetbuttcap%
\pgfsetroundjoin%
\definecolor{currentfill}{rgb}{0.000000,0.000000,0.000000}%
\pgfsetfillcolor{currentfill}%
\pgfsetlinewidth{0.602250pt}%
\definecolor{currentstroke}{rgb}{0.000000,0.000000,0.000000}%
\pgfsetstrokecolor{currentstroke}%
\pgfsetdash{}{0pt}%
\pgfsys@defobject{currentmarker}{\pgfqpoint{0.000000in}{-0.027778in}}{\pgfqpoint{0.000000in}{0.000000in}}{%
\pgfpathmoveto{\pgfqpoint{0.000000in}{0.000000in}}%
\pgfpathlineto{\pgfqpoint{0.000000in}{-0.027778in}}%
\pgfusepath{stroke,fill}%
}%
\begin{pgfscope}%
\pgfsys@transformshift{1.956391in}{0.799635in}%
\pgfsys@useobject{currentmarker}{}%
\end{pgfscope}%
\end{pgfscope}%
\begin{pgfscope}%
\pgfsetbuttcap%
\pgfsetroundjoin%
\definecolor{currentfill}{rgb}{0.000000,0.000000,0.000000}%
\pgfsetfillcolor{currentfill}%
\pgfsetlinewidth{0.602250pt}%
\definecolor{currentstroke}{rgb}{0.000000,0.000000,0.000000}%
\pgfsetstrokecolor{currentstroke}%
\pgfsetdash{}{0pt}%
\pgfsys@defobject{currentmarker}{\pgfqpoint{0.000000in}{-0.027778in}}{\pgfqpoint{0.000000in}{0.000000in}}{%
\pgfpathmoveto{\pgfqpoint{0.000000in}{0.000000in}}%
\pgfpathlineto{\pgfqpoint{0.000000in}{-0.027778in}}%
\pgfusepath{stroke,fill}%
}%
\begin{pgfscope}%
\pgfsys@transformshift{2.080968in}{0.799635in}%
\pgfsys@useobject{currentmarker}{}%
\end{pgfscope}%
\end{pgfscope}%
\begin{pgfscope}%
\pgfsetbuttcap%
\pgfsetroundjoin%
\definecolor{currentfill}{rgb}{0.000000,0.000000,0.000000}%
\pgfsetfillcolor{currentfill}%
\pgfsetlinewidth{0.602250pt}%
\definecolor{currentstroke}{rgb}{0.000000,0.000000,0.000000}%
\pgfsetstrokecolor{currentstroke}%
\pgfsetdash{}{0pt}%
\pgfsys@defobject{currentmarker}{\pgfqpoint{0.000000in}{-0.027778in}}{\pgfqpoint{0.000000in}{0.000000in}}{%
\pgfpathmoveto{\pgfqpoint{0.000000in}{0.000000in}}%
\pgfpathlineto{\pgfqpoint{0.000000in}{-0.027778in}}%
\pgfusepath{stroke,fill}%
}%
\begin{pgfscope}%
\pgfsys@transformshift{2.186297in}{0.799635in}%
\pgfsys@useobject{currentmarker}{}%
\end{pgfscope}%
\end{pgfscope}%
\begin{pgfscope}%
\pgfsetbuttcap%
\pgfsetroundjoin%
\definecolor{currentfill}{rgb}{0.000000,0.000000,0.000000}%
\pgfsetfillcolor{currentfill}%
\pgfsetlinewidth{0.602250pt}%
\definecolor{currentstroke}{rgb}{0.000000,0.000000,0.000000}%
\pgfsetstrokecolor{currentstroke}%
\pgfsetdash{}{0pt}%
\pgfsys@defobject{currentmarker}{\pgfqpoint{0.000000in}{-0.027778in}}{\pgfqpoint{0.000000in}{0.000000in}}{%
\pgfpathmoveto{\pgfqpoint{0.000000in}{0.000000in}}%
\pgfpathlineto{\pgfqpoint{0.000000in}{-0.027778in}}%
\pgfusepath{stroke,fill}%
}%
\begin{pgfscope}%
\pgfsys@transformshift{2.277537in}{0.799635in}%
\pgfsys@useobject{currentmarker}{}%
\end{pgfscope}%
\end{pgfscope}%
\begin{pgfscope}%
\pgfsetbuttcap%
\pgfsetroundjoin%
\definecolor{currentfill}{rgb}{0.000000,0.000000,0.000000}%
\pgfsetfillcolor{currentfill}%
\pgfsetlinewidth{0.602250pt}%
\definecolor{currentstroke}{rgb}{0.000000,0.000000,0.000000}%
\pgfsetstrokecolor{currentstroke}%
\pgfsetdash{}{0pt}%
\pgfsys@defobject{currentmarker}{\pgfqpoint{0.000000in}{-0.027778in}}{\pgfqpoint{0.000000in}{0.000000in}}{%
\pgfpathmoveto{\pgfqpoint{0.000000in}{0.000000in}}%
\pgfpathlineto{\pgfqpoint{0.000000in}{-0.027778in}}%
\pgfusepath{stroke,fill}%
}%
\begin{pgfscope}%
\pgfsys@transformshift{2.358016in}{0.799635in}%
\pgfsys@useobject{currentmarker}{}%
\end{pgfscope}%
\end{pgfscope}%
\begin{pgfscope}%
\pgfsetbuttcap%
\pgfsetroundjoin%
\definecolor{currentfill}{rgb}{0.000000,0.000000,0.000000}%
\pgfsetfillcolor{currentfill}%
\pgfsetlinewidth{0.602250pt}%
\definecolor{currentstroke}{rgb}{0.000000,0.000000,0.000000}%
\pgfsetstrokecolor{currentstroke}%
\pgfsetdash{}{0pt}%
\pgfsys@defobject{currentmarker}{\pgfqpoint{0.000000in}{-0.027778in}}{\pgfqpoint{0.000000in}{0.000000in}}{%
\pgfpathmoveto{\pgfqpoint{0.000000in}{0.000000in}}%
\pgfpathlineto{\pgfqpoint{0.000000in}{-0.027778in}}%
\pgfusepath{stroke,fill}%
}%
\begin{pgfscope}%
\pgfsys@transformshift{2.903624in}{0.799635in}%
\pgfsys@useobject{currentmarker}{}%
\end{pgfscope}%
\end{pgfscope}%
\begin{pgfscope}%
\pgfsetbuttcap%
\pgfsetroundjoin%
\definecolor{currentfill}{rgb}{0.000000,0.000000,0.000000}%
\pgfsetfillcolor{currentfill}%
\pgfsetlinewidth{0.602250pt}%
\definecolor{currentstroke}{rgb}{0.000000,0.000000,0.000000}%
\pgfsetstrokecolor{currentstroke}%
\pgfsetdash{}{0pt}%
\pgfsys@defobject{currentmarker}{\pgfqpoint{0.000000in}{-0.027778in}}{\pgfqpoint{0.000000in}{0.000000in}}{%
\pgfpathmoveto{\pgfqpoint{0.000000in}{0.000000in}}%
\pgfpathlineto{\pgfqpoint{0.000000in}{-0.027778in}}%
\pgfusepath{stroke,fill}%
}%
\begin{pgfscope}%
\pgfsys@transformshift{3.180671in}{0.799635in}%
\pgfsys@useobject{currentmarker}{}%
\end{pgfscope}%
\end{pgfscope}%
\begin{pgfscope}%
\pgfsetbuttcap%
\pgfsetroundjoin%
\definecolor{currentfill}{rgb}{0.000000,0.000000,0.000000}%
\pgfsetfillcolor{currentfill}%
\pgfsetlinewidth{0.602250pt}%
\definecolor{currentstroke}{rgb}{0.000000,0.000000,0.000000}%
\pgfsetstrokecolor{currentstroke}%
\pgfsetdash{}{0pt}%
\pgfsys@defobject{currentmarker}{\pgfqpoint{0.000000in}{-0.027778in}}{\pgfqpoint{0.000000in}{0.000000in}}{%
\pgfpathmoveto{\pgfqpoint{0.000000in}{0.000000in}}%
\pgfpathlineto{\pgfqpoint{0.000000in}{-0.027778in}}%
\pgfusepath{stroke,fill}%
}%
\begin{pgfscope}%
\pgfsys@transformshift{3.377240in}{0.799635in}%
\pgfsys@useobject{currentmarker}{}%
\end{pgfscope}%
\end{pgfscope}%
\begin{pgfscope}%
\pgfsetbuttcap%
\pgfsetroundjoin%
\definecolor{currentfill}{rgb}{0.000000,0.000000,0.000000}%
\pgfsetfillcolor{currentfill}%
\pgfsetlinewidth{0.602250pt}%
\definecolor{currentstroke}{rgb}{0.000000,0.000000,0.000000}%
\pgfsetstrokecolor{currentstroke}%
\pgfsetdash{}{0pt}%
\pgfsys@defobject{currentmarker}{\pgfqpoint{0.000000in}{-0.027778in}}{\pgfqpoint{0.000000in}{0.000000in}}{%
\pgfpathmoveto{\pgfqpoint{0.000000in}{0.000000in}}%
\pgfpathlineto{\pgfqpoint{0.000000in}{-0.027778in}}%
\pgfusepath{stroke,fill}%
}%
\begin{pgfscope}%
\pgfsys@transformshift{3.529710in}{0.799635in}%
\pgfsys@useobject{currentmarker}{}%
\end{pgfscope}%
\end{pgfscope}%
\begin{pgfscope}%
\pgfsetbuttcap%
\pgfsetroundjoin%
\definecolor{currentfill}{rgb}{0.000000,0.000000,0.000000}%
\pgfsetfillcolor{currentfill}%
\pgfsetlinewidth{0.602250pt}%
\definecolor{currentstroke}{rgb}{0.000000,0.000000,0.000000}%
\pgfsetstrokecolor{currentstroke}%
\pgfsetdash{}{0pt}%
\pgfsys@defobject{currentmarker}{\pgfqpoint{0.000000in}{-0.027778in}}{\pgfqpoint{0.000000in}{0.000000in}}{%
\pgfpathmoveto{\pgfqpoint{0.000000in}{0.000000in}}%
\pgfpathlineto{\pgfqpoint{0.000000in}{-0.027778in}}%
\pgfusepath{stroke,fill}%
}%
\begin{pgfscope}%
\pgfsys@transformshift{3.654288in}{0.799635in}%
\pgfsys@useobject{currentmarker}{}%
\end{pgfscope}%
\end{pgfscope}%
\begin{pgfscope}%
\pgfsetbuttcap%
\pgfsetroundjoin%
\definecolor{currentfill}{rgb}{0.000000,0.000000,0.000000}%
\pgfsetfillcolor{currentfill}%
\pgfsetlinewidth{0.602250pt}%
\definecolor{currentstroke}{rgb}{0.000000,0.000000,0.000000}%
\pgfsetstrokecolor{currentstroke}%
\pgfsetdash{}{0pt}%
\pgfsys@defobject{currentmarker}{\pgfqpoint{0.000000in}{-0.027778in}}{\pgfqpoint{0.000000in}{0.000000in}}{%
\pgfpathmoveto{\pgfqpoint{0.000000in}{0.000000in}}%
\pgfpathlineto{\pgfqpoint{0.000000in}{-0.027778in}}%
\pgfusepath{stroke,fill}%
}%
\begin{pgfscope}%
\pgfsys@transformshift{3.759616in}{0.799635in}%
\pgfsys@useobject{currentmarker}{}%
\end{pgfscope}%
\end{pgfscope}%
\begin{pgfscope}%
\pgfsetbuttcap%
\pgfsetroundjoin%
\definecolor{currentfill}{rgb}{0.000000,0.000000,0.000000}%
\pgfsetfillcolor{currentfill}%
\pgfsetlinewidth{0.602250pt}%
\definecolor{currentstroke}{rgb}{0.000000,0.000000,0.000000}%
\pgfsetstrokecolor{currentstroke}%
\pgfsetdash{}{0pt}%
\pgfsys@defobject{currentmarker}{\pgfqpoint{0.000000in}{-0.027778in}}{\pgfqpoint{0.000000in}{0.000000in}}{%
\pgfpathmoveto{\pgfqpoint{0.000000in}{0.000000in}}%
\pgfpathlineto{\pgfqpoint{0.000000in}{-0.027778in}}%
\pgfusepath{stroke,fill}%
}%
\begin{pgfscope}%
\pgfsys@transformshift{3.850856in}{0.799635in}%
\pgfsys@useobject{currentmarker}{}%
\end{pgfscope}%
\end{pgfscope}%
\begin{pgfscope}%
\pgfsetbuttcap%
\pgfsetroundjoin%
\definecolor{currentfill}{rgb}{0.000000,0.000000,0.000000}%
\pgfsetfillcolor{currentfill}%
\pgfsetlinewidth{0.602250pt}%
\definecolor{currentstroke}{rgb}{0.000000,0.000000,0.000000}%
\pgfsetstrokecolor{currentstroke}%
\pgfsetdash{}{0pt}%
\pgfsys@defobject{currentmarker}{\pgfqpoint{0.000000in}{-0.027778in}}{\pgfqpoint{0.000000in}{0.000000in}}{%
\pgfpathmoveto{\pgfqpoint{0.000000in}{0.000000in}}%
\pgfpathlineto{\pgfqpoint{0.000000in}{-0.027778in}}%
\pgfusepath{stroke,fill}%
}%
\begin{pgfscope}%
\pgfsys@transformshift{3.931335in}{0.799635in}%
\pgfsys@useobject{currentmarker}{}%
\end{pgfscope}%
\end{pgfscope}%
\begin{pgfscope}%
\pgfsetbuttcap%
\pgfsetroundjoin%
\definecolor{currentfill}{rgb}{0.000000,0.000000,0.000000}%
\pgfsetfillcolor{currentfill}%
\pgfsetlinewidth{0.602250pt}%
\definecolor{currentstroke}{rgb}{0.000000,0.000000,0.000000}%
\pgfsetstrokecolor{currentstroke}%
\pgfsetdash{}{0pt}%
\pgfsys@defobject{currentmarker}{\pgfqpoint{0.000000in}{-0.027778in}}{\pgfqpoint{0.000000in}{0.000000in}}{%
\pgfpathmoveto{\pgfqpoint{0.000000in}{0.000000in}}%
\pgfpathlineto{\pgfqpoint{0.000000in}{-0.027778in}}%
\pgfusepath{stroke,fill}%
}%
\begin{pgfscope}%
\pgfsys@transformshift{4.476943in}{0.799635in}%
\pgfsys@useobject{currentmarker}{}%
\end{pgfscope}%
\end{pgfscope}%
\begin{pgfscope}%
\definecolor{textcolor}{rgb}{0.000000,0.000000,0.000000}%
\pgfsetstrokecolor{textcolor}%
\pgfsetfillcolor{textcolor}%
\pgftext[x=2.666815in,y=0.356068in,,top]{\color{textcolor}{\rmfamily\fontsize{22.000000}{26.400000}\selectfont\catcode`\^=\active\def^{\ifmmode\sp\else\^{}\fi}\catcode`\%=\active\def
\end{pgfscope}%
\begin{pgfscope}%
\pgfsetbuttcap%
\pgfsetroundjoin%
\definecolor{currentfill}{rgb}{0.000000,0.000000,0.000000}%
\pgfsetfillcolor{currentfill}%
\pgfsetlinewidth{1.505625pt}%
\definecolor{currentstroke}{rgb}{0.000000,0.000000,0.000000}%
\pgfsetstrokecolor{currentstroke}%
\pgfsetdash{}{0pt}%
\pgfsys@defobject{currentmarker}{\pgfqpoint{-0.083333in}{0.000000in}}{\pgfqpoint{-0.000000in}{0.000000in}}{%
\pgfpathmoveto{\pgfqpoint{-0.000000in}{0.000000in}}%
\pgfpathlineto{\pgfqpoint{-0.083333in}{0.000000in}}%
\pgfusepath{stroke,fill}%
}%
\begin{pgfscope}%
\pgfsys@transformshift{0.675675in}{0.799635in}%
\pgfsys@useobject{currentmarker}{}%
\end{pgfscope}%
\end{pgfscope}%
\begin{pgfscope}%
\definecolor{textcolor}{rgb}{0.000000,0.000000,0.000000}%
\pgfsetstrokecolor{textcolor}%
\pgfsetfillcolor{textcolor}%
\pgftext[x=0.411623in, y=0.699616in, left, base]{\color{textcolor}{\rmfamily\fontsize{22.000000}{26.400000}\selectfont\catcode`\^=\active\def^{\ifmmode\sp\else\^{}\fi}\catcode`\%=\active\def
\end{pgfscope}%
\begin{pgfscope}%
\pgfsetbuttcap%
\pgfsetroundjoin%
\definecolor{currentfill}{rgb}{0.000000,0.000000,0.000000}%
\pgfsetfillcolor{currentfill}%
\pgfsetlinewidth{1.505625pt}%
\definecolor{currentstroke}{rgb}{0.000000,0.000000,0.000000}%
\pgfsetstrokecolor{currentstroke}%
\pgfsetdash{}{0pt}%
\pgfsys@defobject{currentmarker}{\pgfqpoint{-0.083333in}{0.000000in}}{\pgfqpoint{-0.000000in}{0.000000in}}{%
\pgfpathmoveto{\pgfqpoint{-0.000000in}{0.000000in}}%
\pgfpathlineto{\pgfqpoint{-0.083333in}{0.000000in}}%
\pgfusepath{stroke,fill}%
}%
\begin{pgfscope}%
\pgfsys@transformshift{0.675675in}{1.486031in}%
\pgfsys@useobject{currentmarker}{}%
\end{pgfscope}%
\end{pgfscope}%
\begin{pgfscope}%
\definecolor{textcolor}{rgb}{0.000000,0.000000,0.000000}%
\pgfsetstrokecolor{textcolor}%
\pgfsetfillcolor{textcolor}%
\pgftext[x=0.411623in, y=1.386011in, left, base]{\color{textcolor}{\rmfamily\fontsize{22.000000}{26.400000}\selectfont\catcode`\^=\active\def^{\ifmmode\sp\else\^{}\fi}\catcode`\%=\active\def
\end{pgfscope}%
\begin{pgfscope}%
\pgfsetbuttcap%
\pgfsetroundjoin%
\definecolor{currentfill}{rgb}{0.000000,0.000000,0.000000}%
\pgfsetfillcolor{currentfill}%
\pgfsetlinewidth{1.505625pt}%
\definecolor{currentstroke}{rgb}{0.000000,0.000000,0.000000}%
\pgfsetstrokecolor{currentstroke}%
\pgfsetdash{}{0pt}%
\pgfsys@defobject{currentmarker}{\pgfqpoint{-0.083333in}{0.000000in}}{\pgfqpoint{-0.000000in}{0.000000in}}{%
\pgfpathmoveto{\pgfqpoint{-0.000000in}{0.000000in}}%
\pgfpathlineto{\pgfqpoint{-0.083333in}{0.000000in}}%
\pgfusepath{stroke,fill}%
}%
\begin{pgfscope}%
\pgfsys@transformshift{0.675675in}{2.172426in}%
\pgfsys@useobject{currentmarker}{}%
\end{pgfscope}%
\end{pgfscope}%
\begin{pgfscope}%
\definecolor{textcolor}{rgb}{0.000000,0.000000,0.000000}%
\pgfsetstrokecolor{textcolor}%
\pgfsetfillcolor{textcolor}%
\pgftext[x=0.411623in, y=2.072407in, left, base]{\color{textcolor}{\rmfamily\fontsize{22.000000}{26.400000}\selectfont\catcode`\^=\active\def^{\ifmmode\sp\else\^{}\fi}\catcode`\%=\active\def
\end{pgfscope}%
\begin{pgfscope}%
\pgfsetbuttcap%
\pgfsetroundjoin%
\definecolor{currentfill}{rgb}{0.000000,0.000000,0.000000}%
\pgfsetfillcolor{currentfill}%
\pgfsetlinewidth{1.505625pt}%
\definecolor{currentstroke}{rgb}{0.000000,0.000000,0.000000}%
\pgfsetstrokecolor{currentstroke}%
\pgfsetdash{}{0pt}%
\pgfsys@defobject{currentmarker}{\pgfqpoint{-0.083333in}{0.000000in}}{\pgfqpoint{-0.000000in}{0.000000in}}{%
\pgfpathmoveto{\pgfqpoint{-0.000000in}{0.000000in}}%
\pgfpathlineto{\pgfqpoint{-0.083333in}{0.000000in}}%
\pgfusepath{stroke,fill}%
}%
\begin{pgfscope}%
\pgfsys@transformshift{0.675675in}{2.858822in}%
\pgfsys@useobject{currentmarker}{}%
\end{pgfscope}%
\end{pgfscope}%
\begin{pgfscope}%
\definecolor{textcolor}{rgb}{0.000000,0.000000,0.000000}%
\pgfsetstrokecolor{textcolor}%
\pgfsetfillcolor{textcolor}%
\pgftext[x=0.411623in, y=2.758802in, left, base]{\color{textcolor}{\rmfamily\fontsize{22.000000}{26.400000}\selectfont\catcode`\^=\active\def^{\ifmmode\sp\else\^{}\fi}\catcode`\%=\active\def
\end{pgfscope}%
\begin{pgfscope}%
\pgfsetbuttcap%
\pgfsetroundjoin%
\definecolor{currentfill}{rgb}{0.000000,0.000000,0.000000}%
\pgfsetfillcolor{currentfill}%
\pgfsetlinewidth{1.505625pt}%
\definecolor{currentstroke}{rgb}{0.000000,0.000000,0.000000}%
\pgfsetstrokecolor{currentstroke}%
\pgfsetdash{}{0pt}%
\pgfsys@defobject{currentmarker}{\pgfqpoint{-0.083333in}{0.000000in}}{\pgfqpoint{-0.000000in}{0.000000in}}{%
\pgfpathmoveto{\pgfqpoint{-0.000000in}{0.000000in}}%
\pgfpathlineto{\pgfqpoint{-0.083333in}{0.000000in}}%
\pgfusepath{stroke,fill}%
}%
\begin{pgfscope}%
\pgfsys@transformshift{0.675675in}{3.545217in}%
\pgfsys@useobject{currentmarker}{}%
\end{pgfscope}%
\end{pgfscope}%
\begin{pgfscope}%
\definecolor{textcolor}{rgb}{0.000000,0.000000,0.000000}%
\pgfsetstrokecolor{textcolor}%
\pgfsetfillcolor{textcolor}%
\pgftext[x=0.411623in, y=3.445198in, left, base]{\color{textcolor}{\rmfamily\fontsize{22.000000}{26.400000}\selectfont\catcode`\^=\active\def^{\ifmmode\sp\else\^{}\fi}\catcode`\%=\active\def
\end{pgfscope}%
\begin{pgfscope}%
\definecolor{textcolor}{rgb}{0.000000,0.000000,0.000000}%
\pgfsetstrokecolor{textcolor}%
\pgfsetfillcolor{textcolor}%
\pgftext[x=0.356068in,y=2.172426in,,bottom,rotate=90.000000]{\color{textcolor}{\rmfamily\fontsize{22.000000}{26.400000}\selectfont\catcode`\^=\active\def^{\ifmmode\sp\else\^{}\fi}\catcode`\%=\active\def
\end{pgfscope}%
\begin{pgfscope}%
\pgfpathrectangle{\pgfqpoint{0.675675in}{0.799635in}}{\pgfqpoint{3.982281in}{2.745582in}}%
\pgfusepath{clip}%
\pgfsetbuttcap%
\pgfsetroundjoin%
\pgfsetlinewidth{3.513125pt}%
\definecolor{currentstroke}{rgb}{0.901961,0.623529,0.000000}%
\pgfsetstrokecolor{currentstroke}%
\pgfsetdash{}{0pt}%
\pgfpathmoveto{\pgfqpoint{0.856688in}{0.799635in}}%
\pgfpathlineto{\pgfqpoint{0.856688in}{2.858822in}}%
\pgfusepath{stroke}%
\end{pgfscope}%
\begin{pgfscope}%
\pgfpathrectangle{\pgfqpoint{0.675675in}{0.799635in}}{\pgfqpoint{3.982281in}{2.745582in}}%
\pgfusepath{clip}%
\pgfsetbuttcap%
\pgfsetroundjoin%
\pgfsetlinewidth{3.513125pt}%
\definecolor{currentstroke}{rgb}{0.901961,0.623529,0.000000}%
\pgfsetstrokecolor{currentstroke}%
\pgfsetdash{}{0pt}%
\pgfpathmoveto{\pgfqpoint{1.482775in}{2.172426in}}%
\pgfpathlineto{\pgfqpoint{1.482775in}{2.858822in}}%
\pgfusepath{stroke}%
\end{pgfscope}%
\begin{pgfscope}%
\pgfpathrectangle{\pgfqpoint{0.675675in}{0.799635in}}{\pgfqpoint{3.982281in}{2.745582in}}%
\pgfusepath{clip}%
\pgfsetbuttcap%
\pgfsetroundjoin%
\pgfsetlinewidth{3.513125pt}%
\definecolor{currentstroke}{rgb}{0.901961,0.623529,0.000000}%
\pgfsetstrokecolor{currentstroke}%
\pgfsetdash{}{0pt}%
\pgfpathmoveto{\pgfqpoint{1.956391in}{2.172426in}}%
\pgfpathlineto{\pgfqpoint{1.956391in}{2.378345in}}%
\pgfusepath{stroke}%
\end{pgfscope}%
\begin{pgfscope}%
\pgfpathrectangle{\pgfqpoint{0.675675in}{0.799635in}}{\pgfqpoint{3.982281in}{2.745582in}}%
\pgfusepath{clip}%
\pgfsetbuttcap%
\pgfsetroundjoin%
\pgfsetlinewidth{3.513125pt}%
\definecolor{currentstroke}{rgb}{0.901961,0.623529,0.000000}%
\pgfsetstrokecolor{currentstroke}%
\pgfsetdash{}{0pt}%
\pgfpathmoveto{\pgfqpoint{2.430007in}{2.172426in}}%
\pgfpathlineto{\pgfqpoint{2.430007in}{2.213610in}}%
\pgfusepath{stroke}%
\end{pgfscope}%
\begin{pgfscope}%
\pgfpathrectangle{\pgfqpoint{0.675675in}{0.799635in}}{\pgfqpoint{3.982281in}{2.745582in}}%
\pgfusepath{clip}%
\pgfsetbuttcap%
\pgfsetroundjoin%
\pgfsetlinewidth{3.513125pt}%
\definecolor{currentstroke}{rgb}{0.901961,0.623529,0.000000}%
\pgfsetstrokecolor{currentstroke}%
\pgfsetdash{}{0pt}%
\pgfpathmoveto{\pgfqpoint{3.529710in}{2.172426in}}%
\pgfpathlineto{\pgfqpoint{3.529710in}{2.199882in}}%
\pgfusepath{stroke}%
\end{pgfscope}%
\begin{pgfscope}%
\pgfpathrectangle{\pgfqpoint{0.675675in}{0.799635in}}{\pgfqpoint{3.982281in}{2.745582in}}%
\pgfusepath{clip}%
\pgfsetbuttcap%
\pgfsetroundjoin%
\pgfsetlinewidth{3.513125pt}%
\definecolor{currentstroke}{rgb}{0.901961,0.623529,0.000000}%
\pgfsetstrokecolor{currentstroke}%
\pgfsetdash{}{0pt}%
\pgfpathmoveto{\pgfqpoint{4.003327in}{2.172426in}}%
\pgfpathlineto{\pgfqpoint{4.003327in}{2.405801in}}%
\pgfusepath{stroke}%
\end{pgfscope}%
\begin{pgfscope}%
\pgfpathrectangle{\pgfqpoint{0.675675in}{0.799635in}}{\pgfqpoint{3.982281in}{2.745582in}}%
\pgfusepath{clip}%
\pgfsetbuttcap%
\pgfsetroundjoin%
\pgfsetlinewidth{3.513125pt}%
\definecolor{currentstroke}{rgb}{0.901961,0.623529,0.000000}%
\pgfsetstrokecolor{currentstroke}%
\pgfsetdash{}{0pt}%
\pgfpathmoveto{\pgfqpoint{4.476943in}{2.172426in}}%
\pgfpathlineto{\pgfqpoint{4.476943in}{2.858822in}}%
\pgfusepath{stroke}%
\end{pgfscope}%
\begin{pgfscope}%
\pgfpathrectangle{\pgfqpoint{0.675675in}{0.799635in}}{\pgfqpoint{3.982281in}{2.745582in}}%
\pgfusepath{clip}%
\pgfsetbuttcap%
\pgfsetroundjoin%
\definecolor{currentfill}{rgb}{0.901961,0.623529,0.000000}%
\pgfsetfillcolor{currentfill}%
\pgfsetlinewidth{2.007500pt}%
\definecolor{currentstroke}{rgb}{0.901961,0.623529,0.000000}%
\pgfsetstrokecolor{currentstroke}%
\pgfsetdash{}{0pt}%
\pgfsys@defobject{currentmarker}{\pgfqpoint{-0.055556in}{-0.000000in}}{\pgfqpoint{0.055556in}{0.000000in}}{%
\pgfpathmoveto{\pgfqpoint{0.055556in}{-0.000000in}}%
\pgfpathlineto{\pgfqpoint{-0.055556in}{0.000000in}}%
\pgfusepath{stroke,fill}%
}%
\begin{pgfscope}%
\pgfsys@transformshift{0.856688in}{0.799635in}%
\pgfsys@useobject{currentmarker}{}%
\end{pgfscope}%
\begin{pgfscope}%
\pgfsys@transformshift{1.482775in}{2.172426in}%
\pgfsys@useobject{currentmarker}{}%
\end{pgfscope}%
\begin{pgfscope}%
\pgfsys@transformshift{1.956391in}{2.172426in}%
\pgfsys@useobject{currentmarker}{}%
\end{pgfscope}%
\begin{pgfscope}%
\pgfsys@transformshift{2.430007in}{2.172426in}%
\pgfsys@useobject{currentmarker}{}%
\end{pgfscope}%
\begin{pgfscope}%
\pgfsys@transformshift{3.529710in}{2.172426in}%
\pgfsys@useobject{currentmarker}{}%
\end{pgfscope}%
\begin{pgfscope}%
\pgfsys@transformshift{4.003327in}{2.172426in}%
\pgfsys@useobject{currentmarker}{}%
\end{pgfscope}%
\begin{pgfscope}%
\pgfsys@transformshift{4.476943in}{2.172426in}%
\pgfsys@useobject{currentmarker}{}%
\end{pgfscope}%
\end{pgfscope}%
\begin{pgfscope}%
\pgfpathrectangle{\pgfqpoint{0.675675in}{0.799635in}}{\pgfqpoint{3.982281in}{2.745582in}}%
\pgfusepath{clip}%
\pgfsetbuttcap%
\pgfsetroundjoin%
\definecolor{currentfill}{rgb}{0.901961,0.623529,0.000000}%
\pgfsetfillcolor{currentfill}%
\pgfsetlinewidth{2.007500pt}%
\definecolor{currentstroke}{rgb}{0.901961,0.623529,0.000000}%
\pgfsetstrokecolor{currentstroke}%
\pgfsetdash{}{0pt}%
\pgfsys@defobject{currentmarker}{\pgfqpoint{-0.055556in}{-0.000000in}}{\pgfqpoint{0.055556in}{0.000000in}}{%
\pgfpathmoveto{\pgfqpoint{0.055556in}{-0.000000in}}%
\pgfpathlineto{\pgfqpoint{-0.055556in}{0.000000in}}%
\pgfusepath{stroke,fill}%
}%
\begin{pgfscope}%
\pgfsys@transformshift{0.856688in}{2.858822in}%
\pgfsys@useobject{currentmarker}{}%
\end{pgfscope}%
\begin{pgfscope}%
\pgfsys@transformshift{1.482775in}{2.858822in}%
\pgfsys@useobject{currentmarker}{}%
\end{pgfscope}%
\begin{pgfscope}%
\pgfsys@transformshift{1.956391in}{2.378345in}%
\pgfsys@useobject{currentmarker}{}%
\end{pgfscope}%
\begin{pgfscope}%
\pgfsys@transformshift{2.430007in}{2.213610in}%
\pgfsys@useobject{currentmarker}{}%
\end{pgfscope}%
\begin{pgfscope}%
\pgfsys@transformshift{3.529710in}{2.199882in}%
\pgfsys@useobject{currentmarker}{}%
\end{pgfscope}%
\begin{pgfscope}%
\pgfsys@transformshift{4.003327in}{2.405801in}%
\pgfsys@useobject{currentmarker}{}%
\end{pgfscope}%
\begin{pgfscope}%
\pgfsys@transformshift{4.476943in}{2.858822in}%
\pgfsys@useobject{currentmarker}{}%
\end{pgfscope}%
\end{pgfscope}%
\begin{pgfscope}%
\pgfpathrectangle{\pgfqpoint{0.675675in}{0.799635in}}{\pgfqpoint{3.982281in}{2.745582in}}%
\pgfusepath{clip}%
\pgfsetbuttcap%
\pgfsetroundjoin%
\pgfsetlinewidth{3.513125pt}%
\definecolor{currentstroke}{rgb}{0.000000,0.000000,0.000000}%
\pgfsetstrokecolor{currentstroke}%
\pgfsetdash{}{0pt}%
\pgfpathmoveto{\pgfqpoint{0.856688in}{1.362480in}}%
\pgfpathlineto{\pgfqpoint{0.856688in}{1.486031in}}%
\pgfusepath{stroke}%
\end{pgfscope}%
\begin{pgfscope}%
\pgfpathrectangle{\pgfqpoint{0.675675in}{0.799635in}}{\pgfqpoint{3.982281in}{2.745582in}}%
\pgfusepath{clip}%
\pgfsetbuttcap%
\pgfsetroundjoin%
\pgfsetlinewidth{3.513125pt}%
\definecolor{currentstroke}{rgb}{0.000000,0.000000,0.000000}%
\pgfsetstrokecolor{currentstroke}%
\pgfsetdash{}{0pt}%
\pgfpathmoveto{\pgfqpoint{1.482775in}{1.486031in}}%
\pgfpathlineto{\pgfqpoint{1.482775in}{2.172426in}}%
\pgfusepath{stroke}%
\end{pgfscope}%
\begin{pgfscope}%
\pgfpathrectangle{\pgfqpoint{0.675675in}{0.799635in}}{\pgfqpoint{3.982281in}{2.745582in}}%
\pgfusepath{clip}%
\pgfsetbuttcap%
\pgfsetroundjoin%
\pgfsetlinewidth{3.513125pt}%
\definecolor{currentstroke}{rgb}{0.000000,0.000000,0.000000}%
\pgfsetstrokecolor{currentstroke}%
\pgfsetdash{}{0pt}%
\pgfpathmoveto{\pgfqpoint{1.956391in}{2.090059in}}%
\pgfpathlineto{\pgfqpoint{1.956391in}{2.172426in}}%
\pgfusepath{stroke}%
\end{pgfscope}%
\begin{pgfscope}%
\pgfpathrectangle{\pgfqpoint{0.675675in}{0.799635in}}{\pgfqpoint{3.982281in}{2.745582in}}%
\pgfusepath{clip}%
\pgfsetbuttcap%
\pgfsetroundjoin%
\pgfsetlinewidth{3.513125pt}%
\definecolor{currentstroke}{rgb}{0.000000,0.000000,0.000000}%
\pgfsetstrokecolor{currentstroke}%
\pgfsetdash{}{0pt}%
\pgfpathmoveto{\pgfqpoint{2.430007in}{2.172426in}}%
\pgfpathlineto{\pgfqpoint{2.430007in}{2.227338in}}%
\pgfusepath{stroke}%
\end{pgfscope}%
\begin{pgfscope}%
\pgfpathrectangle{\pgfqpoint{0.675675in}{0.799635in}}{\pgfqpoint{3.982281in}{2.745582in}}%
\pgfusepath{clip}%
\pgfsetbuttcap%
\pgfsetroundjoin%
\pgfsetlinewidth{3.513125pt}%
\definecolor{currentstroke}{rgb}{0.000000,0.000000,0.000000}%
\pgfsetstrokecolor{currentstroke}%
\pgfsetdash{}{0pt}%
\pgfpathmoveto{\pgfqpoint{3.529710in}{2.172426in}}%
\pgfpathlineto{\pgfqpoint{3.529710in}{2.227338in}}%
\pgfusepath{stroke}%
\end{pgfscope}%
\begin{pgfscope}%
\pgfpathrectangle{\pgfqpoint{0.675675in}{0.799635in}}{\pgfqpoint{3.982281in}{2.745582in}}%
\pgfusepath{clip}%
\pgfsetbuttcap%
\pgfsetroundjoin%
\pgfsetlinewidth{3.513125pt}%
\definecolor{currentstroke}{rgb}{0.000000,0.000000,0.000000}%
\pgfsetstrokecolor{currentstroke}%
\pgfsetdash{}{0pt}%
\pgfpathmoveto{\pgfqpoint{4.003327in}{2.172426in}}%
\pgfpathlineto{\pgfqpoint{4.003327in}{2.213610in}}%
\pgfusepath{stroke}%
\end{pgfscope}%
\begin{pgfscope}%
\pgfpathrectangle{\pgfqpoint{0.675675in}{0.799635in}}{\pgfqpoint{3.982281in}{2.745582in}}%
\pgfusepath{clip}%
\pgfsetbuttcap%
\pgfsetroundjoin%
\pgfsetlinewidth{3.513125pt}%
\definecolor{currentstroke}{rgb}{0.000000,0.000000,0.000000}%
\pgfsetstrokecolor{currentstroke}%
\pgfsetdash{}{0pt}%
\pgfpathmoveto{\pgfqpoint{4.476943in}{2.172426in}}%
\pgfpathlineto{\pgfqpoint{4.476943in}{2.241066in}}%
\pgfusepath{stroke}%
\end{pgfscope}%
\begin{pgfscope}%
\pgfpathrectangle{\pgfqpoint{0.675675in}{0.799635in}}{\pgfqpoint{3.982281in}{2.745582in}}%
\pgfusepath{clip}%
\pgfsetbuttcap%
\pgfsetroundjoin%
\definecolor{currentfill}{rgb}{0.000000,0.000000,0.000000}%
\pgfsetfillcolor{currentfill}%
\pgfsetlinewidth{2.007500pt}%
\definecolor{currentstroke}{rgb}{0.000000,0.000000,0.000000}%
\pgfsetstrokecolor{currentstroke}%
\pgfsetdash{}{0pt}%
\pgfsys@defobject{currentmarker}{\pgfqpoint{-0.055556in}{-0.000000in}}{\pgfqpoint{0.055556in}{0.000000in}}{%
\pgfpathmoveto{\pgfqpoint{0.055556in}{-0.000000in}}%
\pgfpathlineto{\pgfqpoint{-0.055556in}{0.000000in}}%
\pgfusepath{stroke,fill}%
}%
\begin{pgfscope}%
\pgfsys@transformshift{0.856688in}{1.362480in}%
\pgfsys@useobject{currentmarker}{}%
\end{pgfscope}%
\begin{pgfscope}%
\pgfsys@transformshift{1.482775in}{1.486031in}%
\pgfsys@useobject{currentmarker}{}%
\end{pgfscope}%
\begin{pgfscope}%
\pgfsys@transformshift{1.956391in}{2.090059in}%
\pgfsys@useobject{currentmarker}{}%
\end{pgfscope}%
\begin{pgfscope}%
\pgfsys@transformshift{2.430007in}{2.172426in}%
\pgfsys@useobject{currentmarker}{}%
\end{pgfscope}%
\begin{pgfscope}%
\pgfsys@transformshift{3.529710in}{2.172426in}%
\pgfsys@useobject{currentmarker}{}%
\end{pgfscope}%
\begin{pgfscope}%
\pgfsys@transformshift{4.003327in}{2.172426in}%
\pgfsys@useobject{currentmarker}{}%
\end{pgfscope}%
\begin{pgfscope}%
\pgfsys@transformshift{4.476943in}{2.172426in}%
\pgfsys@useobject{currentmarker}{}%
\end{pgfscope}%
\end{pgfscope}%
\begin{pgfscope}%
\pgfpathrectangle{\pgfqpoint{0.675675in}{0.799635in}}{\pgfqpoint{3.982281in}{2.745582in}}%
\pgfusepath{clip}%
\pgfsetbuttcap%
\pgfsetroundjoin%
\definecolor{currentfill}{rgb}{0.000000,0.000000,0.000000}%
\pgfsetfillcolor{currentfill}%
\pgfsetlinewidth{2.007500pt}%
\definecolor{currentstroke}{rgb}{0.000000,0.000000,0.000000}%
\pgfsetstrokecolor{currentstroke}%
\pgfsetdash{}{0pt}%
\pgfsys@defobject{currentmarker}{\pgfqpoint{-0.055556in}{-0.000000in}}{\pgfqpoint{0.055556in}{0.000000in}}{%
\pgfpathmoveto{\pgfqpoint{0.055556in}{-0.000000in}}%
\pgfpathlineto{\pgfqpoint{-0.055556in}{0.000000in}}%
\pgfusepath{stroke,fill}%
}%
\begin{pgfscope}%
\pgfsys@transformshift{0.856688in}{1.486031in}%
\pgfsys@useobject{currentmarker}{}%
\end{pgfscope}%
\begin{pgfscope}%
\pgfsys@transformshift{1.482775in}{2.172426in}%
\pgfsys@useobject{currentmarker}{}%
\end{pgfscope}%
\begin{pgfscope}%
\pgfsys@transformshift{1.956391in}{2.172426in}%
\pgfsys@useobject{currentmarker}{}%
\end{pgfscope}%
\begin{pgfscope}%
\pgfsys@transformshift{2.430007in}{2.227338in}%
\pgfsys@useobject{currentmarker}{}%
\end{pgfscope}%
\begin{pgfscope}%
\pgfsys@transformshift{3.529710in}{2.227338in}%
\pgfsys@useobject{currentmarker}{}%
\end{pgfscope}%
\begin{pgfscope}%
\pgfsys@transformshift{4.003327in}{2.213610in}%
\pgfsys@useobject{currentmarker}{}%
\end{pgfscope}%
\begin{pgfscope}%
\pgfsys@transformshift{4.476943in}{2.241066in}%
\pgfsys@useobject{currentmarker}{}%
\end{pgfscope}%
\end{pgfscope}%
\begin{pgfscope}%
\pgfpathrectangle{\pgfqpoint{0.675675in}{0.799635in}}{\pgfqpoint{3.982281in}{2.745582in}}%
\pgfusepath{clip}%
\pgfsetrectcap%
\pgfsetroundjoin%
\pgfsetlinewidth{2.509375pt}%
\definecolor{currentstroke}{rgb}{0.501961,0.501961,0.501961}%
\pgfsetstrokecolor{currentstroke}%
\pgfsetstrokeopacity{0.800000}%
\pgfsetdash{}{0pt}%
\pgfpathmoveto{\pgfqpoint{0.675675in}{2.172426in}}%
\pgfpathlineto{\pgfqpoint{4.657956in}{2.172426in}}%
\pgfusepath{stroke}%
\end{pgfscope}%
\begin{pgfscope}%
\pgfpathrectangle{\pgfqpoint{0.675675in}{0.799635in}}{\pgfqpoint{3.982281in}{2.745582in}}%
\pgfusepath{clip}%
\pgfsetbuttcap%
\pgfsetroundjoin%
\pgfsetlinewidth{3.513125pt}%
\definecolor{currentstroke}{rgb}{0.901961,0.623529,0.000000}%
\pgfsetstrokecolor{currentstroke}%
\pgfsetdash{{3.500000pt}{5.775000pt}}{0.000000pt}%
\pgfpathmoveto{\pgfqpoint{0.856688in}{1.794909in}}%
\pgfpathlineto{\pgfqpoint{1.482775in}{2.515624in}}%
\pgfpathlineto{\pgfqpoint{1.956391in}{2.275386in}}%
\pgfpathlineto{\pgfqpoint{2.430007in}{2.193018in}}%
\pgfpathlineto{\pgfqpoint{3.529710in}{2.186154in}}%
\pgfpathlineto{\pgfqpoint{4.003327in}{2.289113in}}%
\pgfpathlineto{\pgfqpoint{4.476943in}{2.350889in}}%
\pgfusepath{stroke}%
\end{pgfscope}%
\begin{pgfscope}%
\pgfpathrectangle{\pgfqpoint{0.675675in}{0.799635in}}{\pgfqpoint{3.982281in}{2.745582in}}%
\pgfusepath{clip}%
\pgfsetbuttcap%
\pgfsetmiterjoin%
\definecolor{currentfill}{rgb}{0.901961,0.623529,0.000000}%
\pgfsetfillcolor{currentfill}%
\pgfsetlinewidth{2.007500pt}%
\definecolor{currentstroke}{rgb}{1.000000,1.000000,1.000000}%
\pgfsetstrokecolor{currentstroke}%
\pgfsetdash{}{0pt}%
\pgfsys@defobject{currentmarker}{\pgfqpoint{-0.052836in}{-0.044945in}}{\pgfqpoint{0.052836in}{0.055556in}}{%
\pgfpathmoveto{\pgfqpoint{0.000000in}{0.055556in}}%
\pgfpathlineto{\pgfqpoint{-0.052836in}{0.017168in}}%
\pgfpathlineto{\pgfqpoint{-0.032655in}{-0.044945in}}%
\pgfpathlineto{\pgfqpoint{0.032655in}{-0.044945in}}%
\pgfpathlineto{\pgfqpoint{0.052836in}{0.017168in}}%
\pgfpathlineto{\pgfqpoint{0.000000in}{0.055556in}}%
\pgfpathclose%
\pgfusepath{stroke,fill}%
}%
\begin{pgfscope}%
\pgfsys@transformshift{0.856688in}{1.794909in}%
\pgfsys@useobject{currentmarker}{}%
\end{pgfscope}%
\begin{pgfscope}%
\pgfsys@transformshift{1.482775in}{2.515624in}%
\pgfsys@useobject{currentmarker}{}%
\end{pgfscope}%
\begin{pgfscope}%
\pgfsys@transformshift{1.956391in}{2.275386in}%
\pgfsys@useobject{currentmarker}{}%
\end{pgfscope}%
\begin{pgfscope}%
\pgfsys@transformshift{2.430007in}{2.193018in}%
\pgfsys@useobject{currentmarker}{}%
\end{pgfscope}%
\begin{pgfscope}%
\pgfsys@transformshift{3.529710in}{2.186154in}%
\pgfsys@useobject{currentmarker}{}%
\end{pgfscope}%
\begin{pgfscope}%
\pgfsys@transformshift{4.003327in}{2.289113in}%
\pgfsys@useobject{currentmarker}{}%
\end{pgfscope}%
\begin{pgfscope}%
\pgfsys@transformshift{4.476943in}{2.350889in}%
\pgfsys@useobject{currentmarker}{}%
\end{pgfscope}%
\end{pgfscope}%
\begin{pgfscope}%
\pgfpathrectangle{\pgfqpoint{0.675675in}{0.799635in}}{\pgfqpoint{3.982281in}{2.745582in}}%
\pgfusepath{clip}%
\pgfsetrectcap%
\pgfsetroundjoin%
\pgfsetlinewidth{3.513125pt}%
\definecolor{currentstroke}{rgb}{0.000000,0.000000,0.000000}%
\pgfsetstrokecolor{currentstroke}%
\pgfsetdash{}{0pt}%
\pgfpathmoveto{\pgfqpoint{0.856688in}{1.424255in}}%
\pgfpathlineto{\pgfqpoint{1.482775in}{1.891004in}}%
\pgfpathlineto{\pgfqpoint{1.956391in}{2.131242in}}%
\pgfpathlineto{\pgfqpoint{2.430007in}{2.199882in}}%
\pgfpathlineto{\pgfqpoint{3.529710in}{2.199882in}}%
\pgfpathlineto{\pgfqpoint{4.003327in}{2.193018in}}%
\pgfpathlineto{\pgfqpoint{4.476943in}{2.206746in}}%
\pgfusepath{stroke}%
\end{pgfscope}%
\begin{pgfscope}%
\pgfpathrectangle{\pgfqpoint{0.675675in}{0.799635in}}{\pgfqpoint{3.982281in}{2.745582in}}%
\pgfusepath{clip}%
\pgfsetbuttcap%
\pgfsetmiterjoin%
\definecolor{currentfill}{rgb}{0.000000,0.000000,0.000000}%
\pgfsetfillcolor{currentfill}%
\pgfsetlinewidth{2.007500pt}%
\definecolor{currentstroke}{rgb}{1.000000,1.000000,1.000000}%
\pgfsetstrokecolor{currentstroke}%
\pgfsetdash{}{0pt}%
\pgfsys@defobject{currentmarker}{\pgfqpoint{-0.078567in}{-0.078567in}}{\pgfqpoint{0.078567in}{0.078567in}}{%
\pgfpathmoveto{\pgfqpoint{-0.000000in}{-0.078567in}}%
\pgfpathlineto{\pgfqpoint{0.078567in}{0.000000in}}%
\pgfpathlineto{\pgfqpoint{0.000000in}{0.078567in}}%
\pgfpathlineto{\pgfqpoint{-0.078567in}{0.000000in}}%
\pgfpathlineto{\pgfqpoint{-0.000000in}{-0.078567in}}%
\pgfpathclose%
\pgfusepath{stroke,fill}%
}%
\begin{pgfscope}%
\pgfsys@transformshift{0.856688in}{1.424255in}%
\pgfsys@useobject{currentmarker}{}%
\end{pgfscope}%
\begin{pgfscope}%
\pgfsys@transformshift{1.482775in}{1.891004in}%
\pgfsys@useobject{currentmarker}{}%
\end{pgfscope}%
\begin{pgfscope}%
\pgfsys@transformshift{1.956391in}{2.131242in}%
\pgfsys@useobject{currentmarker}{}%
\end{pgfscope}%
\begin{pgfscope}%
\pgfsys@transformshift{2.430007in}{2.199882in}%
\pgfsys@useobject{currentmarker}{}%
\end{pgfscope}%
\begin{pgfscope}%
\pgfsys@transformshift{3.529710in}{2.199882in}%
\pgfsys@useobject{currentmarker}{}%
\end{pgfscope}%
\begin{pgfscope}%
\pgfsys@transformshift{4.003327in}{2.193018in}%
\pgfsys@useobject{currentmarker}{}%
\end{pgfscope}%
\begin{pgfscope}%
\pgfsys@transformshift{4.476943in}{2.206746in}%
\pgfsys@useobject{currentmarker}{}%
\end{pgfscope}%
\end{pgfscope}%
\begin{pgfscope}%
\pgfsetrectcap%
\pgfsetmiterjoin%
\pgfsetlinewidth{1.505625pt}%
\definecolor{currentstroke}{rgb}{0.000000,0.000000,0.000000}%
\pgfsetstrokecolor{currentstroke}%
\pgfsetdash{}{0pt}%
\pgfpathmoveto{\pgfqpoint{0.675675in}{0.799635in}}%
\pgfpathlineto{\pgfqpoint{0.675675in}{3.545217in}}%
\pgfusepath{stroke}%
\end{pgfscope}%
\begin{pgfscope}%
\pgfsetrectcap%
\pgfsetmiterjoin%
\pgfsetlinewidth{1.505625pt}%
\definecolor{currentstroke}{rgb}{0.000000,0.000000,0.000000}%
\pgfsetstrokecolor{currentstroke}%
\pgfsetdash{}{0pt}%
\pgfpathmoveto{\pgfqpoint{0.675675in}{0.799635in}}%
\pgfpathlineto{\pgfqpoint{4.657956in}{0.799635in}}%
\pgfusepath{stroke}%
\end{pgfscope}%
\begin{pgfscope}%
\pgfsetbuttcap%
\pgfsetmiterjoin%
\definecolor{currentfill}{rgb}{1.000000,1.000000,1.000000}%
\pgfsetfillcolor{currentfill}%
\pgfsetfillopacity{0.400000}%
\pgfsetlinewidth{1.003750pt}%
\definecolor{currentstroke}{rgb}{0.000000,0.000000,0.000000}%
\pgfsetstrokecolor{currentstroke}%
\pgfsetstrokeopacity{0.400000}%
\pgfsetdash{}{0pt}%
\pgfpathmoveto{\pgfqpoint{1.569148in}{0.910746in}}%
\pgfpathlineto{\pgfqpoint{4.502400in}{0.910746in}}%
\pgfpathquadraticcurveto{\pgfqpoint{4.546845in}{0.910746in}}{\pgfqpoint{4.546845in}{0.955191in}}%
\pgfpathlineto{\pgfqpoint{4.546845in}{1.906348in}}%
\pgfpathquadraticcurveto{\pgfqpoint{4.546845in}{1.950792in}}{\pgfqpoint{4.502400in}{1.950792in}}%
\pgfpathlineto{\pgfqpoint{1.569148in}{1.950792in}}%
\pgfpathquadraticcurveto{\pgfqpoint{1.524704in}{1.950792in}}{\pgfqpoint{1.524704in}{1.906348in}}%
\pgfpathlineto{\pgfqpoint{1.524704in}{0.955191in}}%
\pgfpathquadraticcurveto{\pgfqpoint{1.524704in}{0.910746in}}{\pgfqpoint{1.569148in}{0.910746in}}%
\pgfpathlineto{\pgfqpoint{1.569148in}{0.910746in}}%
\pgfpathclose%
\pgfusepath{stroke,fill}%
\end{pgfscope}%
\begin{pgfscope}%
\pgfsetrectcap%
\pgfsetroundjoin%
\pgfsetlinewidth{2.509375pt}%
\definecolor{currentstroke}{rgb}{0.501961,0.501961,0.501961}%
\pgfsetstrokecolor{currentstroke}%
\pgfsetstrokeopacity{0.800000}%
\pgfsetdash{}{0pt}%
\pgfpathmoveto{\pgfqpoint{1.613592in}{1.773014in}}%
\pgfpathlineto{\pgfqpoint{1.835815in}{1.773014in}}%
\pgfpathlineto{\pgfqpoint{2.058037in}{1.773014in}}%
\pgfusepath{stroke}%
\end{pgfscope}%
\begin{pgfscope}%
\definecolor{textcolor}{rgb}{0.000000,0.000000,0.000000}%
\pgfsetstrokecolor{textcolor}%
\pgfsetfillcolor{textcolor}%
\pgftext[x=2.235815in,y=1.695237in,left,base]{\color{textcolor}{\rmfamily\fontsize{16.000000}{19.200000}\selectfont\catcode`\^=\active\def^{\ifmmode\sp\else\^{}\fi}\catcode`\%=\active\def
\end{pgfscope}%
\begin{pgfscope}%
\pgfsetbuttcap%
\pgfsetroundjoin%
\pgfsetlinewidth{3.513125pt}%
\definecolor{currentstroke}{rgb}{0.901961,0.623529,0.000000}%
\pgfsetstrokecolor{currentstroke}%
\pgfsetdash{}{0pt}%
\pgfpathmoveto{\pgfqpoint{1.835815in}{1.337444in}}%
\pgfpathlineto{\pgfqpoint{1.835815in}{1.559666in}}%
\pgfusepath{stroke}%
\end{pgfscope}%
\begin{pgfscope}%
\pgfsetbuttcap%
\pgfsetroundjoin%
\definecolor{currentfill}{rgb}{0.901961,0.623529,0.000000}%
\pgfsetfillcolor{currentfill}%
\pgfsetlinewidth{2.007500pt}%
\definecolor{currentstroke}{rgb}{0.901961,0.623529,0.000000}%
\pgfsetstrokecolor{currentstroke}%
\pgfsetdash{}{0pt}%
\pgfsys@defobject{currentmarker}{\pgfqpoint{-0.055556in}{-0.000000in}}{\pgfqpoint{0.055556in}{0.000000in}}{%
\pgfpathmoveto{\pgfqpoint{0.055556in}{-0.000000in}}%
\pgfpathlineto{\pgfqpoint{-0.055556in}{0.000000in}}%
\pgfusepath{stroke,fill}%
}%
\begin{pgfscope}%
\pgfsys@transformshift{1.835815in}{1.337444in}%
\pgfsys@useobject{currentmarker}{}%
\end{pgfscope}%
\end{pgfscope}%
\begin{pgfscope}%
\pgfsetbuttcap%
\pgfsetroundjoin%
\definecolor{currentfill}{rgb}{0.901961,0.623529,0.000000}%
\pgfsetfillcolor{currentfill}%
\pgfsetlinewidth{2.007500pt}%
\definecolor{currentstroke}{rgb}{0.901961,0.623529,0.000000}%
\pgfsetstrokecolor{currentstroke}%
\pgfsetdash{}{0pt}%
\pgfsys@defobject{currentmarker}{\pgfqpoint{-0.055556in}{-0.000000in}}{\pgfqpoint{0.055556in}{0.000000in}}{%
\pgfpathmoveto{\pgfqpoint{0.055556in}{-0.000000in}}%
\pgfpathlineto{\pgfqpoint{-0.055556in}{0.000000in}}%
\pgfusepath{stroke,fill}%
}%
\begin{pgfscope}%
\pgfsys@transformshift{1.835815in}{1.559666in}%
\pgfsys@useobject{currentmarker}{}%
\end{pgfscope}%
\end{pgfscope}%
\begin{pgfscope}%
\pgfsetbuttcap%
\pgfsetroundjoin%
\pgfsetlinewidth{3.513125pt}%
\definecolor{currentstroke}{rgb}{0.901961,0.623529,0.000000}%
\pgfsetstrokecolor{currentstroke}%
\pgfsetdash{{3.500000pt}{5.775000pt}}{0.000000pt}%
\pgfpathmoveto{\pgfqpoint{1.613592in}{1.448555in}}%
\pgfpathlineto{\pgfqpoint{2.058037in}{1.448555in}}%
\pgfusepath{stroke}%
\end{pgfscope}%
\begin{pgfscope}%
\pgfsetbuttcap%
\pgfsetmiterjoin%
\definecolor{currentfill}{rgb}{0.901961,0.623529,0.000000}%
\pgfsetfillcolor{currentfill}%
\pgfsetlinewidth{2.007500pt}%
\definecolor{currentstroke}{rgb}{1.000000,1.000000,1.000000}%
\pgfsetstrokecolor{currentstroke}%
\pgfsetdash{}{0pt}%
\pgfsys@defobject{currentmarker}{\pgfqpoint{-0.052836in}{-0.044945in}}{\pgfqpoint{0.052836in}{0.055556in}}{%
\pgfpathmoveto{\pgfqpoint{0.000000in}{0.055556in}}%
\pgfpathlineto{\pgfqpoint{-0.052836in}{0.017168in}}%
\pgfpathlineto{\pgfqpoint{-0.032655in}{-0.044945in}}%
\pgfpathlineto{\pgfqpoint{0.032655in}{-0.044945in}}%
\pgfpathlineto{\pgfqpoint{0.052836in}{0.017168in}}%
\pgfpathlineto{\pgfqpoint{0.000000in}{0.055556in}}%
\pgfpathclose%
\pgfusepath{stroke,fill}%
}%
\begin{pgfscope}%
\pgfsys@transformshift{1.835815in}{1.448555in}%
\pgfsys@useobject{currentmarker}{}%
\end{pgfscope}%
\end{pgfscope}%
\begin{pgfscope}%
\definecolor{textcolor}{rgb}{0.000000,0.000000,0.000000}%
\pgfsetstrokecolor{textcolor}%
\pgfsetfillcolor{textcolor}%
\pgftext[x=2.235815in,y=1.370777in,left,base]{\color{textcolor}{\rmfamily\fontsize{16.000000}{19.200000}\selectfont\catcode`\^=\active\def^{\ifmmode\sp\else\^{}\fi}\catcode`\%=\active\def
\end{pgfscope}%
\begin{pgfscope}%
\pgfsetbuttcap%
\pgfsetroundjoin%
\pgfsetlinewidth{3.513125pt}%
\definecolor{currentstroke}{rgb}{0.000000,0.000000,0.000000}%
\pgfsetstrokecolor{currentstroke}%
\pgfsetdash{}{0pt}%
\pgfpathmoveto{\pgfqpoint{1.835815in}{1.012984in}}%
\pgfpathlineto{\pgfqpoint{1.835815in}{1.235206in}}%
\pgfusepath{stroke}%
\end{pgfscope}%
\begin{pgfscope}%
\pgfsetbuttcap%
\pgfsetroundjoin%
\definecolor{currentfill}{rgb}{0.000000,0.000000,0.000000}%
\pgfsetfillcolor{currentfill}%
\pgfsetlinewidth{2.007500pt}%
\definecolor{currentstroke}{rgb}{0.000000,0.000000,0.000000}%
\pgfsetstrokecolor{currentstroke}%
\pgfsetdash{}{0pt}%
\pgfsys@defobject{currentmarker}{\pgfqpoint{-0.055556in}{-0.000000in}}{\pgfqpoint{0.055556in}{0.000000in}}{%
\pgfpathmoveto{\pgfqpoint{0.055556in}{-0.000000in}}%
\pgfpathlineto{\pgfqpoint{-0.055556in}{0.000000in}}%
\pgfusepath{stroke,fill}%
}%
\begin{pgfscope}%
\pgfsys@transformshift{1.835815in}{1.012984in}%
\pgfsys@useobject{currentmarker}{}%
\end{pgfscope}%
\end{pgfscope}%
\begin{pgfscope}%
\pgfsetbuttcap%
\pgfsetroundjoin%
\definecolor{currentfill}{rgb}{0.000000,0.000000,0.000000}%
\pgfsetfillcolor{currentfill}%
\pgfsetlinewidth{2.007500pt}%
\definecolor{currentstroke}{rgb}{0.000000,0.000000,0.000000}%
\pgfsetstrokecolor{currentstroke}%
\pgfsetdash{}{0pt}%
\pgfsys@defobject{currentmarker}{\pgfqpoint{-0.055556in}{-0.000000in}}{\pgfqpoint{0.055556in}{0.000000in}}{%
\pgfpathmoveto{\pgfqpoint{0.055556in}{-0.000000in}}%
\pgfpathlineto{\pgfqpoint{-0.055556in}{0.000000in}}%
\pgfusepath{stroke,fill}%
}%
\begin{pgfscope}%
\pgfsys@transformshift{1.835815in}{1.235206in}%
\pgfsys@useobject{currentmarker}{}%
\end{pgfscope}%
\end{pgfscope}%
\begin{pgfscope}%
\pgfsetrectcap%
\pgfsetroundjoin%
\pgfsetlinewidth{3.513125pt}%
\definecolor{currentstroke}{rgb}{0.000000,0.000000,0.000000}%
\pgfsetstrokecolor{currentstroke}%
\pgfsetdash{}{0pt}%
\pgfpathmoveto{\pgfqpoint{1.613592in}{1.124095in}}%
\pgfpathlineto{\pgfqpoint{2.058037in}{1.124095in}}%
\pgfusepath{stroke}%
\end{pgfscope}%
\begin{pgfscope}%
\pgfsetbuttcap%
\pgfsetmiterjoin%
\definecolor{currentfill}{rgb}{0.000000,0.000000,0.000000}%
\pgfsetfillcolor{currentfill}%
\pgfsetlinewidth{2.007500pt}%
\definecolor{currentstroke}{rgb}{1.000000,1.000000,1.000000}%
\pgfsetstrokecolor{currentstroke}%
\pgfsetdash{}{0pt}%
\pgfsys@defobject{currentmarker}{\pgfqpoint{-0.078567in}{-0.078567in}}{\pgfqpoint{0.078567in}{0.078567in}}{%
\pgfpathmoveto{\pgfqpoint{-0.000000in}{-0.078567in}}%
\pgfpathlineto{\pgfqpoint{0.078567in}{0.000000in}}%
\pgfpathlineto{\pgfqpoint{0.000000in}{0.078567in}}%
\pgfpathlineto{\pgfqpoint{-0.078567in}{0.000000in}}%
\pgfpathlineto{\pgfqpoint{-0.000000in}{-0.078567in}}%
\pgfpathclose%
\pgfusepath{stroke,fill}%
}%
\begin{pgfscope}%
\pgfsys@transformshift{1.835815in}{1.124095in}%
\pgfsys@useobject{currentmarker}{}%
\end{pgfscope}%
\end{pgfscope}%
\begin{pgfscope}%
\definecolor{textcolor}{rgb}{0.000000,0.000000,0.000000}%
\pgfsetstrokecolor{textcolor}%
\pgfsetfillcolor{textcolor}%
\pgftext[x=2.235815in,y=1.046317in,left,base]{\color{textcolor}{\rmfamily\fontsize{16.000000}{19.200000}\selectfont\catcode`\^=\active\def^{\ifmmode\sp\else\^{}\fi}\catcode`\%=\active\def
\end{pgfscope}%
\end{pgfpicture}%
\makeatother%
\endgroup%

%% file: figures/TwoStatevsUnif/h_laplace_loc.pgf
\begingroup%
\makeatletter%
\begin{pgfpicture}%
\pgfpathrectangle{\pgfpointorigin}{\pgfqpoint{5.344216in}{2.949100in}}%
\pgfusepath{use as bounding box, clip}%
\begin{pgfscope}%
\pgfsetbuttcap%
\pgfsetmiterjoin%
\definecolor{currentfill}{rgb}{1.000000,1.000000,1.000000}%
\pgfsetfillcolor{currentfill}%
\pgfsetlinewidth{0.000000pt}%
\definecolor{currentstroke}{rgb}{1.000000,1.000000,1.000000}%
\pgfsetstrokecolor{currentstroke}%
\pgfsetdash{}{0pt}%
\pgfpathmoveto{\pgfqpoint{0.000000in}{0.000000in}}%
\pgfpathlineto{\pgfqpoint{5.344216in}{0.000000in}}%
\pgfpathlineto{\pgfqpoint{5.344216in}{2.949100in}}%
\pgfpathlineto{\pgfqpoint{0.000000in}{2.949100in}}%
\pgfpathlineto{\pgfqpoint{0.000000in}{0.000000in}}%
\pgfpathclose%
\pgfusepath{fill}%
\end{pgfscope}%
\begin{pgfscope}%
\pgfsetbuttcap%
\pgfsetmiterjoin%
\definecolor{currentfill}{rgb}{1.000000,1.000000,1.000000}%
\pgfsetfillcolor{currentfill}%
\pgfsetlinewidth{0.000000pt}%
\definecolor{currentstroke}{rgb}{0.000000,0.000000,0.000000}%
\pgfsetstrokecolor{currentstroke}%
\pgfsetstrokeopacity{0.000000}%
\pgfsetdash{}{0pt}%
\pgfpathmoveto{\pgfqpoint{0.917618in}{0.831623in}}%
\pgfpathlineto{\pgfqpoint{5.244216in}{0.831623in}}%
\pgfpathlineto{\pgfqpoint{5.244216in}{2.849100in}}%
\pgfpathlineto{\pgfqpoint{0.917618in}{2.849100in}}%
\pgfpathlineto{\pgfqpoint{0.917618in}{0.831623in}}%
\pgfpathclose%
\pgfusepath{fill}%
\end{pgfscope}%
\begin{pgfscope}%
\pgfsetbuttcap%
\pgfsetroundjoin%
\definecolor{currentfill}{rgb}{0.000000,0.000000,0.000000}%
\pgfsetfillcolor{currentfill}%
\pgfsetlinewidth{1.505625pt}%
\definecolor{currentstroke}{rgb}{0.000000,0.000000,0.000000}%
\pgfsetstrokecolor{currentstroke}%
\pgfsetdash{}{0pt}%
\pgfsys@defobject{currentmarker}{\pgfqpoint{0.000000in}{-0.083333in}}{\pgfqpoint{0.000000in}{0.000000in}}{%
\pgfpathmoveto{\pgfqpoint{0.000000in}{0.000000in}}%
\pgfpathlineto{\pgfqpoint{0.000000in}{-0.083333in}}%
\pgfusepath{stroke,fill}%
}%
\begin{pgfscope}%
\pgfsys@transformshift{1.114281in}{0.831623in}%
\pgfsys@useobject{currentmarker}{}%
\end{pgfscope}%
\end{pgfscope}%
\begin{pgfscope}%
\definecolor{textcolor}{rgb}{0.000000,0.000000,0.000000}%
\pgfsetstrokecolor{textcolor}%
\pgfsetfillcolor{textcolor}%
\pgftext[x=1.114281in,y=0.699679in,,top]{\color{textcolor}{\rmfamily\fontsize{22.000000}{26.400000}\selectfont\catcode`\^=\active\def^{\ifmmode\sp\else\^{}\fi}\catcode`\%=\active\def
\end{pgfscope}%
\begin{pgfscope}%
\pgfsetbuttcap%
\pgfsetroundjoin%
\definecolor{currentfill}{rgb}{0.000000,0.000000,0.000000}%
\pgfsetfillcolor{currentfill}%
\pgfsetlinewidth{1.505625pt}%
\definecolor{currentstroke}{rgb}{0.000000,0.000000,0.000000}%
\pgfsetstrokecolor{currentstroke}%
\pgfsetdash{}{0pt}%
\pgfsys@defobject{currentmarker}{\pgfqpoint{0.000000in}{-0.083333in}}{\pgfqpoint{0.000000in}{0.000000in}}{%
\pgfpathmoveto{\pgfqpoint{0.000000in}{0.000000in}}%
\pgfpathlineto{\pgfqpoint{0.000000in}{-0.083333in}}%
\pgfusepath{stroke,fill}%
}%
\begin{pgfscope}%
\pgfsys@transformshift{2.687590in}{0.831623in}%
\pgfsys@useobject{currentmarker}{}%
\end{pgfscope}%
\end{pgfscope}%
\begin{pgfscope}%
\definecolor{textcolor}{rgb}{0.000000,0.000000,0.000000}%
\pgfsetstrokecolor{textcolor}%
\pgfsetfillcolor{textcolor}%
\pgftext[x=2.687590in,y=0.699679in,,top]{\color{textcolor}{\rmfamily\fontsize{22.000000}{26.400000}\selectfont\catcode`\^=\active\def^{\ifmmode\sp\else\^{}\fi}\catcode`\%=\active\def
\end{pgfscope}%
\begin{pgfscope}%
\pgfsetbuttcap%
\pgfsetroundjoin%
\definecolor{currentfill}{rgb}{0.000000,0.000000,0.000000}%
\pgfsetfillcolor{currentfill}%
\pgfsetlinewidth{1.505625pt}%
\definecolor{currentstroke}{rgb}{0.000000,0.000000,0.000000}%
\pgfsetstrokecolor{currentstroke}%
\pgfsetdash{}{0pt}%
\pgfsys@defobject{currentmarker}{\pgfqpoint{0.000000in}{-0.083333in}}{\pgfqpoint{0.000000in}{0.000000in}}{%
\pgfpathmoveto{\pgfqpoint{0.000000in}{0.000000in}}%
\pgfpathlineto{\pgfqpoint{0.000000in}{-0.083333in}}%
\pgfusepath{stroke,fill}%
}%
\begin{pgfscope}%
\pgfsys@transformshift{4.260898in}{0.831623in}%
\pgfsys@useobject{currentmarker}{}%
\end{pgfscope}%
\end{pgfscope}%
\begin{pgfscope}%
\definecolor{textcolor}{rgb}{0.000000,0.000000,0.000000}%
\pgfsetstrokecolor{textcolor}%
\pgfsetfillcolor{textcolor}%
\pgftext[x=4.260898in,y=0.699679in,,top]{\color{textcolor}{\rmfamily\fontsize{22.000000}{26.400000}\selectfont\catcode`\^=\active\def^{\ifmmode\sp\else\^{}\fi}\catcode`\%=\active\def
\end{pgfscope}%
\begin{pgfscope}%
\definecolor{textcolor}{rgb}{0.000000,0.000000,0.000000}%
\pgfsetstrokecolor{textcolor}%
\pgfsetfillcolor{textcolor}%
\pgftext[x=3.080917in,y=0.388056in,,top]{\color{textcolor}{\rmfamily\fontsize{22.000000}{26.400000}\selectfont\catcode`\^=\active\def^{\ifmmode\sp\else\^{}\fi}\catcode`\%=\active\def
\end{pgfscope}%
\begin{pgfscope}%
\pgfsetbuttcap%
\pgfsetroundjoin%
\definecolor{currentfill}{rgb}{0.000000,0.000000,0.000000}%
\pgfsetfillcolor{currentfill}%
\pgfsetlinewidth{1.505625pt}%
\definecolor{currentstroke}{rgb}{0.000000,0.000000,0.000000}%
\pgfsetstrokecolor{currentstroke}%
\pgfsetdash{}{0pt}%
\pgfsys@defobject{currentmarker}{\pgfqpoint{-0.083333in}{0.000000in}}{\pgfqpoint{-0.000000in}{0.000000in}}{%
\pgfpathmoveto{\pgfqpoint{-0.000000in}{0.000000in}}%
\pgfpathlineto{\pgfqpoint{-0.083333in}{0.000000in}}%
\pgfusepath{stroke,fill}%
}%
\begin{pgfscope}%
\pgfsys@transformshift{0.917618in}{0.831623in}%
\pgfsys@useobject{currentmarker}{}%
\end{pgfscope}%
\end{pgfscope}%
\begin{pgfscope}%
\definecolor{textcolor}{rgb}{0.000000,0.000000,0.000000}%
\pgfsetstrokecolor{textcolor}%
\pgfsetfillcolor{textcolor}%
\pgftext[x=0.443111in, y=0.731604in, left, base]{\color{textcolor}{\rmfamily\fontsize{22.000000}{26.400000}\selectfont\catcode`\^=\active\def^{\ifmmode\sp\else\^{}\fi}\catcode`\%=\active\def
\end{pgfscope}%
\begin{pgfscope}%
\pgfsetbuttcap%
\pgfsetroundjoin%
\definecolor{currentfill}{rgb}{0.000000,0.000000,0.000000}%
\pgfsetfillcolor{currentfill}%
\pgfsetlinewidth{1.505625pt}%
\definecolor{currentstroke}{rgb}{0.000000,0.000000,0.000000}%
\pgfsetstrokecolor{currentstroke}%
\pgfsetdash{}{0pt}%
\pgfsys@defobject{currentmarker}{\pgfqpoint{-0.083333in}{0.000000in}}{\pgfqpoint{-0.000000in}{0.000000in}}{%
\pgfpathmoveto{\pgfqpoint{-0.000000in}{0.000000in}}%
\pgfpathlineto{\pgfqpoint{-0.083333in}{0.000000in}}%
\pgfusepath{stroke,fill}%
}%
\begin{pgfscope}%
\pgfsys@transformshift{0.917618in}{1.638614in}%
\pgfsys@useobject{currentmarker}{}%
\end{pgfscope}%
\end{pgfscope}%
\begin{pgfscope}%
\definecolor{textcolor}{rgb}{0.000000,0.000000,0.000000}%
\pgfsetstrokecolor{textcolor}%
\pgfsetfillcolor{textcolor}%
\pgftext[x=0.443111in, y=1.538595in, left, base]{\color{textcolor}{\rmfamily\fontsize{22.000000}{26.400000}\selectfont\catcode`\^=\active\def^{\ifmmode\sp\else\^{}\fi}\catcode`\%=\active\def
\end{pgfscope}%
\begin{pgfscope}%
\pgfsetbuttcap%
\pgfsetroundjoin%
\definecolor{currentfill}{rgb}{0.000000,0.000000,0.000000}%
\pgfsetfillcolor{currentfill}%
\pgfsetlinewidth{1.505625pt}%
\definecolor{currentstroke}{rgb}{0.000000,0.000000,0.000000}%
\pgfsetstrokecolor{currentstroke}%
\pgfsetdash{}{0pt}%
\pgfsys@defobject{currentmarker}{\pgfqpoint{-0.083333in}{0.000000in}}{\pgfqpoint{-0.000000in}{0.000000in}}{%
\pgfpathmoveto{\pgfqpoint{-0.000000in}{0.000000in}}%
\pgfpathlineto{\pgfqpoint{-0.083333in}{0.000000in}}%
\pgfusepath{stroke,fill}%
}%
\begin{pgfscope}%
\pgfsys@transformshift{0.917618in}{2.445604in}%
\pgfsys@useobject{currentmarker}{}%
\end{pgfscope}%
\end{pgfscope}%
\begin{pgfscope}%
\definecolor{textcolor}{rgb}{0.000000,0.000000,0.000000}%
\pgfsetstrokecolor{textcolor}%
\pgfsetfillcolor{textcolor}%
\pgftext[x=0.443111in, y=2.345585in, left, base]{\color{textcolor}{\rmfamily\fontsize{22.000000}{26.400000}\selectfont\catcode`\^=\active\def^{\ifmmode\sp\else\^{}\fi}\catcode`\%=\active\def
\end{pgfscope}%
\begin{pgfscope}%
\definecolor{textcolor}{rgb}{0.000000,0.000000,0.000000}%
\pgfsetstrokecolor{textcolor}%
\pgfsetfillcolor{textcolor}%
\pgftext[x=0.387555in,y=1.840361in,,bottom,rotate=90.000000]{\color{textcolor}{\rmfamily\fontsize{22.000000}{26.400000}\selectfont\catcode`\^=\active\def^{\ifmmode\sp\else\^{}\fi}\catcode`\%=\active\def
\end{pgfscope}%
\begin{pgfscope}%
\pgfpathrectangle{\pgfqpoint{0.917618in}{0.831623in}}{\pgfqpoint{4.326598in}{2.017477in}}%
\pgfusepath{clip}%
\pgfsetbuttcap%
\pgfsetroundjoin%
\pgfsetlinewidth{4.516875pt}%
\definecolor{currentstroke}{rgb}{0.933333,0.466667,0.200000}%
\pgfsetstrokecolor{currentstroke}%
\pgfsetstrokeopacity{0.800000}%
\pgfsetdash{}{0pt}%
\pgfpathmoveto{\pgfqpoint{1.114281in}{1.388671in}}%
\pgfpathlineto{\pgfqpoint{1.114281in}{1.602748in}}%
\pgfusepath{stroke}%
\end{pgfscope}%
\begin{pgfscope}%
\pgfpathrectangle{\pgfqpoint{0.917618in}{0.831623in}}{\pgfqpoint{4.326598in}{2.017477in}}%
\pgfusepath{clip}%
\pgfsetbuttcap%
\pgfsetroundjoin%
\pgfsetlinewidth{4.516875pt}%
\definecolor{currentstroke}{rgb}{0.933333,0.466667,0.200000}%
\pgfsetstrokecolor{currentstroke}%
\pgfsetstrokeopacity{0.800000}%
\pgfsetdash{}{0pt}%
\pgfpathmoveto{\pgfqpoint{1.900935in}{1.258319in}}%
\pgfpathlineto{\pgfqpoint{1.900935in}{1.533705in}}%
\pgfusepath{stroke}%
\end{pgfscope}%
\begin{pgfscope}%
\pgfpathrectangle{\pgfqpoint{0.917618in}{0.831623in}}{\pgfqpoint{4.326598in}{2.017477in}}%
\pgfusepath{clip}%
\pgfsetbuttcap%
\pgfsetroundjoin%
\pgfsetlinewidth{4.516875pt}%
\definecolor{currentstroke}{rgb}{0.933333,0.466667,0.200000}%
\pgfsetstrokecolor{currentstroke}%
\pgfsetstrokeopacity{0.800000}%
\pgfsetdash{}{0pt}%
\pgfpathmoveto{\pgfqpoint{2.687590in}{1.320403in}}%
\pgfpathlineto{\pgfqpoint{2.687590in}{1.492805in}}%
\pgfusepath{stroke}%
\end{pgfscope}%
\begin{pgfscope}%
\pgfpathrectangle{\pgfqpoint{0.917618in}{0.831623in}}{\pgfqpoint{4.326598in}{2.017477in}}%
\pgfusepath{clip}%
\pgfsetbuttcap%
\pgfsetroundjoin%
\pgfsetlinewidth{4.516875pt}%
\definecolor{currentstroke}{rgb}{0.933333,0.466667,0.200000}%
\pgfsetstrokecolor{currentstroke}%
\pgfsetstrokeopacity{0.800000}%
\pgfsetdash{}{0pt}%
\pgfpathmoveto{\pgfqpoint{3.474244in}{1.201494in}}%
\pgfpathlineto{\pgfqpoint{3.474244in}{1.389792in}}%
\pgfusepath{stroke}%
\end{pgfscope}%
\begin{pgfscope}%
\pgfpathrectangle{\pgfqpoint{0.917618in}{0.831623in}}{\pgfqpoint{4.326598in}{2.017477in}}%
\pgfusepath{clip}%
\pgfsetbuttcap%
\pgfsetroundjoin%
\pgfsetlinewidth{4.516875pt}%
\definecolor{currentstroke}{rgb}{0.933333,0.466667,0.200000}%
\pgfsetstrokecolor{currentstroke}%
\pgfsetstrokeopacity{0.800000}%
\pgfsetdash{}{0pt}%
\pgfpathmoveto{\pgfqpoint{4.260898in}{1.181578in}}%
\pgfpathlineto{\pgfqpoint{4.260898in}{1.387981in}}%
\pgfusepath{stroke}%
\end{pgfscope}%
\begin{pgfscope}%
\pgfpathrectangle{\pgfqpoint{0.917618in}{0.831623in}}{\pgfqpoint{4.326598in}{2.017477in}}%
\pgfusepath{clip}%
\pgfsetbuttcap%
\pgfsetroundjoin%
\pgfsetlinewidth{4.516875pt}%
\definecolor{currentstroke}{rgb}{0.933333,0.466667,0.200000}%
\pgfsetstrokecolor{currentstroke}%
\pgfsetstrokeopacity{0.800000}%
\pgfsetdash{}{0pt}%
\pgfpathmoveto{\pgfqpoint{5.047552in}{1.067956in}}%
\pgfpathlineto{\pgfqpoint{5.047552in}{1.196210in}}%
\pgfusepath{stroke}%
\end{pgfscope}%
\begin{pgfscope}%
\pgfpathrectangle{\pgfqpoint{0.917618in}{0.831623in}}{\pgfqpoint{4.326598in}{2.017477in}}%
\pgfusepath{clip}%
\pgfsetbuttcap%
\pgfsetroundjoin%
\definecolor{currentfill}{rgb}{0.933333,0.466667,0.200000}%
\pgfsetfillcolor{currentfill}%
\pgfsetfillopacity{0.800000}%
\pgfsetlinewidth{2.007500pt}%
\definecolor{currentstroke}{rgb}{0.933333,0.466667,0.200000}%
\pgfsetstrokecolor{currentstroke}%
\pgfsetstrokeopacity{0.800000}%
\pgfsetdash{}{0pt}%
\pgfsys@defobject{currentmarker}{\pgfqpoint{-0.055556in}{-0.000000in}}{\pgfqpoint{0.055556in}{0.000000in}}{%
\pgfpathmoveto{\pgfqpoint{0.055556in}{-0.000000in}}%
\pgfpathlineto{\pgfqpoint{-0.055556in}{0.000000in}}%
\pgfusepath{stroke,fill}%
}%
\begin{pgfscope}%
\pgfsys@transformshift{1.114281in}{1.388671in}%
\pgfsys@useobject{currentmarker}{}%
\end{pgfscope}%
\begin{pgfscope}%
\pgfsys@transformshift{1.900935in}{1.258319in}%
\pgfsys@useobject{currentmarker}{}%
\end{pgfscope}%
\begin{pgfscope}%
\pgfsys@transformshift{2.687590in}{1.320403in}%
\pgfsys@useobject{currentmarker}{}%
\end{pgfscope}%
\begin{pgfscope}%
\pgfsys@transformshift{3.474244in}{1.201494in}%
\pgfsys@useobject{currentmarker}{}%
\end{pgfscope}%
\begin{pgfscope}%
\pgfsys@transformshift{4.260898in}{1.181578in}%
\pgfsys@useobject{currentmarker}{}%
\end{pgfscope}%
\begin{pgfscope}%
\pgfsys@transformshift{5.047552in}{1.067956in}%
\pgfsys@useobject{currentmarker}{}%
\end{pgfscope}%
\end{pgfscope}%
\begin{pgfscope}%
\pgfpathrectangle{\pgfqpoint{0.917618in}{0.831623in}}{\pgfqpoint{4.326598in}{2.017477in}}%
\pgfusepath{clip}%
\pgfsetbuttcap%
\pgfsetroundjoin%
\definecolor{currentfill}{rgb}{0.933333,0.466667,0.200000}%
\pgfsetfillcolor{currentfill}%
\pgfsetfillopacity{0.800000}%
\pgfsetlinewidth{2.007500pt}%
\definecolor{currentstroke}{rgb}{0.933333,0.466667,0.200000}%
\pgfsetstrokecolor{currentstroke}%
\pgfsetstrokeopacity{0.800000}%
\pgfsetdash{}{0pt}%
\pgfsys@defobject{currentmarker}{\pgfqpoint{-0.055556in}{-0.000000in}}{\pgfqpoint{0.055556in}{0.000000in}}{%
\pgfpathmoveto{\pgfqpoint{0.055556in}{-0.000000in}}%
\pgfpathlineto{\pgfqpoint{-0.055556in}{0.000000in}}%
\pgfusepath{stroke,fill}%
}%
\begin{pgfscope}%
\pgfsys@transformshift{1.114281in}{1.602748in}%
\pgfsys@useobject{currentmarker}{}%
\end{pgfscope}%
\begin{pgfscope}%
\pgfsys@transformshift{1.900935in}{1.533705in}%
\pgfsys@useobject{currentmarker}{}%
\end{pgfscope}%
\begin{pgfscope}%
\pgfsys@transformshift{2.687590in}{1.492805in}%
\pgfsys@useobject{currentmarker}{}%
\end{pgfscope}%
\begin{pgfscope}%
\pgfsys@transformshift{3.474244in}{1.389792in}%
\pgfsys@useobject{currentmarker}{}%
\end{pgfscope}%
\begin{pgfscope}%
\pgfsys@transformshift{4.260898in}{1.387981in}%
\pgfsys@useobject{currentmarker}{}%
\end{pgfscope}%
\begin{pgfscope}%
\pgfsys@transformshift{5.047552in}{1.196210in}%
\pgfsys@useobject{currentmarker}{}%
\end{pgfscope}%
\end{pgfscope}%
\begin{pgfscope}%
\pgfpathrectangle{\pgfqpoint{0.917618in}{0.831623in}}{\pgfqpoint{4.326598in}{2.017477in}}%
\pgfusepath{clip}%
\pgfsetbuttcap%
\pgfsetroundjoin%
\pgfsetlinewidth{4.516875pt}%
\definecolor{currentstroke}{rgb}{0.000000,0.466667,0.733333}%
\pgfsetstrokecolor{currentstroke}%
\pgfsetstrokeopacity{0.800000}%
\pgfsetdash{}{0pt}%
\pgfpathmoveto{\pgfqpoint{1.114281in}{1.902007in}}%
\pgfpathlineto{\pgfqpoint{1.114281in}{2.243857in}}%
\pgfusepath{stroke}%
\end{pgfscope}%
\begin{pgfscope}%
\pgfpathrectangle{\pgfqpoint{0.917618in}{0.831623in}}{\pgfqpoint{4.326598in}{2.017477in}}%
\pgfusepath{clip}%
\pgfsetbuttcap%
\pgfsetroundjoin%
\pgfsetlinewidth{4.516875pt}%
\definecolor{currentstroke}{rgb}{0.000000,0.466667,0.733333}%
\pgfsetstrokecolor{currentstroke}%
\pgfsetstrokeopacity{0.800000}%
\pgfsetdash{}{0pt}%
\pgfpathmoveto{\pgfqpoint{1.900935in}{1.886763in}}%
\pgfpathlineto{\pgfqpoint{1.900935in}{2.287233in}}%
\pgfusepath{stroke}%
\end{pgfscope}%
\begin{pgfscope}%
\pgfpathrectangle{\pgfqpoint{0.917618in}{0.831623in}}{\pgfqpoint{4.326598in}{2.017477in}}%
\pgfusepath{clip}%
\pgfsetbuttcap%
\pgfsetroundjoin%
\pgfsetlinewidth{4.516875pt}%
\definecolor{currentstroke}{rgb}{0.000000,0.466667,0.733333}%
\pgfsetstrokecolor{currentstroke}%
\pgfsetstrokeopacity{0.800000}%
\pgfsetdash{}{0pt}%
\pgfpathmoveto{\pgfqpoint{2.687590in}{1.801846in}}%
\pgfpathlineto{\pgfqpoint{2.687590in}{2.149402in}}%
\pgfusepath{stroke}%
\end{pgfscope}%
\begin{pgfscope}%
\pgfpathrectangle{\pgfqpoint{0.917618in}{0.831623in}}{\pgfqpoint{4.326598in}{2.017477in}}%
\pgfusepath{clip}%
\pgfsetbuttcap%
\pgfsetroundjoin%
\pgfsetlinewidth{4.516875pt}%
\definecolor{currentstroke}{rgb}{0.000000,0.466667,0.733333}%
\pgfsetstrokecolor{currentstroke}%
\pgfsetstrokeopacity{0.800000}%
\pgfsetdash{}{0pt}%
\pgfpathmoveto{\pgfqpoint{3.474244in}{1.768069in}}%
\pgfpathlineto{\pgfqpoint{3.474244in}{1.904248in}}%
\pgfusepath{stroke}%
\end{pgfscope}%
\begin{pgfscope}%
\pgfpathrectangle{\pgfqpoint{0.917618in}{0.831623in}}{\pgfqpoint{4.326598in}{2.017477in}}%
\pgfusepath{clip}%
\pgfsetbuttcap%
\pgfsetroundjoin%
\pgfsetlinewidth{4.516875pt}%
\definecolor{currentstroke}{rgb}{0.000000,0.466667,0.733333}%
\pgfsetstrokecolor{currentstroke}%
\pgfsetstrokeopacity{0.800000}%
\pgfsetdash{}{0pt}%
\pgfpathmoveto{\pgfqpoint{4.260898in}{1.671204in}}%
\pgfpathlineto{\pgfqpoint{4.260898in}{1.821739in}}%
\pgfusepath{stroke}%
\end{pgfscope}%
\begin{pgfscope}%
\pgfpathrectangle{\pgfqpoint{0.917618in}{0.831623in}}{\pgfqpoint{4.326598in}{2.017477in}}%
\pgfusepath{clip}%
\pgfsetbuttcap%
\pgfsetroundjoin%
\pgfsetlinewidth{4.516875pt}%
\definecolor{currentstroke}{rgb}{0.000000,0.466667,0.733333}%
\pgfsetstrokecolor{currentstroke}%
\pgfsetstrokeopacity{0.800000}%
\pgfsetdash{}{0pt}%
\pgfpathmoveto{\pgfqpoint{5.047552in}{1.632850in}}%
\pgfpathlineto{\pgfqpoint{5.047552in}{1.699859in}}%
\pgfusepath{stroke}%
\end{pgfscope}%
\begin{pgfscope}%
\pgfpathrectangle{\pgfqpoint{0.917618in}{0.831623in}}{\pgfqpoint{4.326598in}{2.017477in}}%
\pgfusepath{clip}%
\pgfsetbuttcap%
\pgfsetroundjoin%
\definecolor{currentfill}{rgb}{0.000000,0.466667,0.733333}%
\pgfsetfillcolor{currentfill}%
\pgfsetfillopacity{0.800000}%
\pgfsetlinewidth{2.007500pt}%
\definecolor{currentstroke}{rgb}{0.000000,0.466667,0.733333}%
\pgfsetstrokecolor{currentstroke}%
\pgfsetstrokeopacity{0.800000}%
\pgfsetdash{}{0pt}%
\pgfsys@defobject{currentmarker}{\pgfqpoint{-0.055556in}{-0.000000in}}{\pgfqpoint{0.055556in}{0.000000in}}{%
\pgfpathmoveto{\pgfqpoint{0.055556in}{-0.000000in}}%
\pgfpathlineto{\pgfqpoint{-0.055556in}{0.000000in}}%
\pgfusepath{stroke,fill}%
}%
\begin{pgfscope}%
\pgfsys@transformshift{1.114281in}{1.902007in}%
\pgfsys@useobject{currentmarker}{}%
\end{pgfscope}%
\begin{pgfscope}%
\pgfsys@transformshift{1.900935in}{1.886763in}%
\pgfsys@useobject{currentmarker}{}%
\end{pgfscope}%
\begin{pgfscope}%
\pgfsys@transformshift{2.687590in}{1.801846in}%
\pgfsys@useobject{currentmarker}{}%
\end{pgfscope}%
\begin{pgfscope}%
\pgfsys@transformshift{3.474244in}{1.768069in}%
\pgfsys@useobject{currentmarker}{}%
\end{pgfscope}%
\begin{pgfscope}%
\pgfsys@transformshift{4.260898in}{1.671204in}%
\pgfsys@useobject{currentmarker}{}%
\end{pgfscope}%
\begin{pgfscope}%
\pgfsys@transformshift{5.047552in}{1.632850in}%
\pgfsys@useobject{currentmarker}{}%
\end{pgfscope}%
\end{pgfscope}%
\begin{pgfscope}%
\pgfpathrectangle{\pgfqpoint{0.917618in}{0.831623in}}{\pgfqpoint{4.326598in}{2.017477in}}%
\pgfusepath{clip}%
\pgfsetbuttcap%
\pgfsetroundjoin%
\definecolor{currentfill}{rgb}{0.000000,0.466667,0.733333}%
\pgfsetfillcolor{currentfill}%
\pgfsetfillopacity{0.800000}%
\pgfsetlinewidth{2.007500pt}%
\definecolor{currentstroke}{rgb}{0.000000,0.466667,0.733333}%
\pgfsetstrokecolor{currentstroke}%
\pgfsetstrokeopacity{0.800000}%
\pgfsetdash{}{0pt}%
\pgfsys@defobject{currentmarker}{\pgfqpoint{-0.055556in}{-0.000000in}}{\pgfqpoint{0.055556in}{0.000000in}}{%
\pgfpathmoveto{\pgfqpoint{0.055556in}{-0.000000in}}%
\pgfpathlineto{\pgfqpoint{-0.055556in}{0.000000in}}%
\pgfusepath{stroke,fill}%
}%
\begin{pgfscope}%
\pgfsys@transformshift{1.114281in}{2.243857in}%
\pgfsys@useobject{currentmarker}{}%
\end{pgfscope}%
\begin{pgfscope}%
\pgfsys@transformshift{1.900935in}{2.287233in}%
\pgfsys@useobject{currentmarker}{}%
\end{pgfscope}%
\begin{pgfscope}%
\pgfsys@transformshift{2.687590in}{2.149402in}%
\pgfsys@useobject{currentmarker}{}%
\end{pgfscope}%
\begin{pgfscope}%
\pgfsys@transformshift{3.474244in}{1.904248in}%
\pgfsys@useobject{currentmarker}{}%
\end{pgfscope}%
\begin{pgfscope}%
\pgfsys@transformshift{4.260898in}{1.821739in}%
\pgfsys@useobject{currentmarker}{}%
\end{pgfscope}%
\begin{pgfscope}%
\pgfsys@transformshift{5.047552in}{1.699859in}%
\pgfsys@useobject{currentmarker}{}%
\end{pgfscope}%
\end{pgfscope}%
\begin{pgfscope}%
\pgfpathrectangle{\pgfqpoint{0.917618in}{0.831623in}}{\pgfqpoint{4.326598in}{2.017477in}}%
\pgfusepath{clip}%
\pgfsetrectcap%
\pgfsetroundjoin%
\pgfsetlinewidth{4.516875pt}%
\definecolor{currentstroke}{rgb}{0.933333,0.466667,0.200000}%
\pgfsetstrokecolor{currentstroke}%
\pgfsetstrokeopacity{0.800000}%
\pgfsetdash{}{0pt}%
\pgfpathmoveto{\pgfqpoint{1.114281in}{1.490217in}}%
\pgfpathlineto{\pgfqpoint{1.900935in}{1.392078in}}%
\pgfpathlineto{\pgfqpoint{2.687590in}{1.408988in}}%
\pgfpathlineto{\pgfqpoint{3.474244in}{1.306403in}}%
\pgfpathlineto{\pgfqpoint{4.260898in}{1.311472in}}%
\pgfpathlineto{\pgfqpoint{5.047552in}{1.152114in}}%
\pgfusepath{stroke}%
\end{pgfscope}%
\begin{pgfscope}%
\pgfpathrectangle{\pgfqpoint{0.917618in}{0.831623in}}{\pgfqpoint{4.326598in}{2.017477in}}%
\pgfusepath{clip}%
\pgfsetbuttcap%
\pgfsetroundjoin%
\definecolor{currentfill}{rgb}{0.933333,0.466667,0.200000}%
\pgfsetfillcolor{currentfill}%
\pgfsetfillopacity{0.800000}%
\pgfsetlinewidth{2.007500pt}%
\definecolor{currentstroke}{rgb}{1.000000,1.000000,1.000000}%
\pgfsetstrokecolor{currentstroke}%
\pgfsetstrokeopacity{0.800000}%
\pgfsetdash{}{0pt}%
\pgfsys@defobject{currentmarker}{\pgfqpoint{-0.055556in}{-0.055556in}}{\pgfqpoint{0.055556in}{0.055556in}}{%
\pgfpathmoveto{\pgfqpoint{0.000000in}{-0.055556in}}%
\pgfpathcurveto{\pgfqpoint{0.014734in}{-0.055556in}}{\pgfqpoint{0.028866in}{-0.049702in}}{\pgfqpoint{0.039284in}{-0.039284in}}%
\pgfpathcurveto{\pgfqpoint{0.049702in}{-0.028866in}}{\pgfqpoint{0.055556in}{-0.014734in}}{\pgfqpoint{0.055556in}{0.000000in}}%
\pgfpathcurveto{\pgfqpoint{0.055556in}{0.014734in}}{\pgfqpoint{0.049702in}{0.028866in}}{\pgfqpoint{0.039284in}{0.039284in}}%
\pgfpathcurveto{\pgfqpoint{0.028866in}{0.049702in}}{\pgfqpoint{0.014734in}{0.055556in}}{\pgfqpoint{0.000000in}{0.055556in}}%
\pgfpathcurveto{\pgfqpoint{-0.014734in}{0.055556in}}{\pgfqpoint{-0.028866in}{0.049702in}}{\pgfqpoint{-0.039284in}{0.039284in}}%
\pgfpathcurveto{\pgfqpoint{-0.049702in}{0.028866in}}{\pgfqpoint{-0.055556in}{0.014734in}}{\pgfqpoint{-0.055556in}{0.000000in}}%
\pgfpathcurveto{\pgfqpoint{-0.055556in}{-0.014734in}}{\pgfqpoint{-0.049702in}{-0.028866in}}{\pgfqpoint{-0.039284in}{-0.039284in}}%
\pgfpathcurveto{\pgfqpoint{-0.028866in}{-0.049702in}}{\pgfqpoint{-0.014734in}{-0.055556in}}{\pgfqpoint{0.000000in}{-0.055556in}}%
\pgfpathlineto{\pgfqpoint{0.000000in}{-0.055556in}}%
\pgfpathclose%
\pgfusepath{stroke,fill}%
}%
\begin{pgfscope}%
\pgfsys@transformshift{1.114281in}{1.490217in}%
\pgfsys@useobject{currentmarker}{}%
\end{pgfscope}%
\begin{pgfscope}%
\pgfsys@transformshift{1.900935in}{1.392078in}%
\pgfsys@useobject{currentmarker}{}%
\end{pgfscope}%
\begin{pgfscope}%
\pgfsys@transformshift{2.687590in}{1.408988in}%
\pgfsys@useobject{currentmarker}{}%
\end{pgfscope}%
\begin{pgfscope}%
\pgfsys@transformshift{3.474244in}{1.306403in}%
\pgfsys@useobject{currentmarker}{}%
\end{pgfscope}%
\begin{pgfscope}%
\pgfsys@transformshift{4.260898in}{1.311472in}%
\pgfsys@useobject{currentmarker}{}%
\end{pgfscope}%
\begin{pgfscope}%
\pgfsys@transformshift{5.047552in}{1.152114in}%
\pgfsys@useobject{currentmarker}{}%
\end{pgfscope}%
\end{pgfscope}%
\begin{pgfscope}%
\pgfpathrectangle{\pgfqpoint{0.917618in}{0.831623in}}{\pgfqpoint{4.326598in}{2.017477in}}%
\pgfusepath{clip}%
\pgfsetbuttcap%
\pgfsetroundjoin%
\pgfsetlinewidth{4.516875pt}%
\definecolor{currentstroke}{rgb}{0.000000,0.466667,0.733333}%
\pgfsetstrokecolor{currentstroke}%
\pgfsetstrokeopacity{0.800000}%
\pgfsetdash{{16.650000pt}{7.200000pt}}{0.000000pt}%
\pgfpathmoveto{\pgfqpoint{1.114281in}{2.063629in}}%
\pgfpathlineto{\pgfqpoint{1.900935in}{2.151456in}}%
\pgfpathlineto{\pgfqpoint{2.687590in}{1.962877in}}%
\pgfpathlineto{\pgfqpoint{3.474244in}{1.829602in}}%
\pgfpathlineto{\pgfqpoint{4.260898in}{1.777664in}}%
\pgfpathlineto{\pgfqpoint{5.047552in}{1.639478in}}%
\pgfusepath{stroke}%
\end{pgfscope}%
\begin{pgfscope}%
\pgfpathrectangle{\pgfqpoint{0.917618in}{0.831623in}}{\pgfqpoint{4.326598in}{2.017477in}}%
\pgfusepath{clip}%
\pgfsetbuttcap%
\pgfsetmiterjoin%
\definecolor{currentfill}{rgb}{0.000000,0.466667,0.733333}%
\pgfsetfillcolor{currentfill}%
\pgfsetfillopacity{0.800000}%
\pgfsetlinewidth{2.007500pt}%
\definecolor{currentstroke}{rgb}{1.000000,1.000000,1.000000}%
\pgfsetstrokecolor{currentstroke}%
\pgfsetstrokeopacity{0.800000}%
\pgfsetdash{}{0pt}%
\pgfsys@defobject{currentmarker}{\pgfqpoint{-0.055556in}{-0.055556in}}{\pgfqpoint{0.055556in}{0.055556in}}{%
\pgfpathmoveto{\pgfqpoint{-0.055556in}{-0.055556in}}%
\pgfpathlineto{\pgfqpoint{0.055556in}{-0.055556in}}%
\pgfpathlineto{\pgfqpoint{0.055556in}{0.055556in}}%
\pgfpathlineto{\pgfqpoint{-0.055556in}{0.055556in}}%
\pgfpathlineto{\pgfqpoint{-0.055556in}{-0.055556in}}%
\pgfpathclose%
\pgfusepath{stroke,fill}%
}%
\begin{pgfscope}%
\pgfsys@transformshift{1.114281in}{2.063629in}%
\pgfsys@useobject{currentmarker}{}%
\end{pgfscope}%
\begin{pgfscope}%
\pgfsys@transformshift{1.900935in}{2.151456in}%
\pgfsys@useobject{currentmarker}{}%
\end{pgfscope}%
\begin{pgfscope}%
\pgfsys@transformshift{2.687590in}{1.962877in}%
\pgfsys@useobject{currentmarker}{}%
\end{pgfscope}%
\begin{pgfscope}%
\pgfsys@transformshift{3.474244in}{1.829602in}%
\pgfsys@useobject{currentmarker}{}%
\end{pgfscope}%
\begin{pgfscope}%
\pgfsys@transformshift{4.260898in}{1.777664in}%
\pgfsys@useobject{currentmarker}{}%
\end{pgfscope}%
\begin{pgfscope}%
\pgfsys@transformshift{5.047552in}{1.639478in}%
\pgfsys@useobject{currentmarker}{}%
\end{pgfscope}%
\end{pgfscope}%
\begin{pgfscope}%
\pgfsetrectcap%
\pgfsetmiterjoin%
\pgfsetlinewidth{1.505625pt}%
\definecolor{currentstroke}{rgb}{0.000000,0.000000,0.000000}%
\pgfsetstrokecolor{currentstroke}%
\pgfsetdash{}{0pt}%
\pgfpathmoveto{\pgfqpoint{0.917618in}{0.831623in}}%
\pgfpathlineto{\pgfqpoint{0.917618in}{2.849100in}}%
\pgfusepath{stroke}%
\end{pgfscope}%
\begin{pgfscope}%
\pgfsetrectcap%
\pgfsetmiterjoin%
\pgfsetlinewidth{1.505625pt}%
\definecolor{currentstroke}{rgb}{0.000000,0.000000,0.000000}%
\pgfsetstrokecolor{currentstroke}%
\pgfsetdash{}{0pt}%
\pgfpathmoveto{\pgfqpoint{0.917618in}{0.831623in}}%
\pgfpathlineto{\pgfqpoint{5.244216in}{0.831623in}}%
\pgfusepath{stroke}%
\end{pgfscope}%
\begin{pgfscope}%
\pgfsetbuttcap%
\pgfsetmiterjoin%
\definecolor{currentfill}{rgb}{1.000000,1.000000,1.000000}%
\pgfsetfillcolor{currentfill}%
\pgfsetfillopacity{0.400000}%
\pgfsetlinewidth{1.003750pt}%
\definecolor{currentstroke}{rgb}{0.000000,0.000000,0.000000}%
\pgfsetstrokecolor{currentstroke}%
\pgfsetstrokeopacity{0.400000}%
\pgfsetdash{}{0pt}%
\pgfpathmoveto{\pgfqpoint{2.078800in}{2.022403in}}%
\pgfpathlineto{\pgfqpoint{5.088660in}{2.022403in}}%
\pgfpathquadraticcurveto{\pgfqpoint{5.133105in}{2.022403in}}{\pgfqpoint{5.133105in}{2.066847in}}%
\pgfpathlineto{\pgfqpoint{5.133105in}{2.693544in}}%
\pgfpathquadraticcurveto{\pgfqpoint{5.133105in}{2.737989in}}{\pgfqpoint{5.088660in}{2.737989in}}%
\pgfpathlineto{\pgfqpoint{2.078800in}{2.737989in}}%
\pgfpathquadraticcurveto{\pgfqpoint{2.034355in}{2.737989in}}{\pgfqpoint{2.034355in}{2.693544in}}%
\pgfpathlineto{\pgfqpoint{2.034355in}{2.066847in}}%
\pgfpathquadraticcurveto{\pgfqpoint{2.034355in}{2.022403in}}{\pgfqpoint{2.078800in}{2.022403in}}%
\pgfpathlineto{\pgfqpoint{2.078800in}{2.022403in}}%
\pgfpathclose%
\pgfusepath{stroke,fill}%
\end{pgfscope}%
\begin{pgfscope}%
\pgfsetbuttcap%
\pgfsetroundjoin%
\pgfsetlinewidth{4.516875pt}%
\definecolor{currentstroke}{rgb}{0.933333,0.466667,0.200000}%
\pgfsetstrokecolor{currentstroke}%
\pgfsetstrokeopacity{0.800000}%
\pgfsetdash{}{0pt}%
\pgfpathmoveto{\pgfqpoint{2.345467in}{2.449100in}}%
\pgfpathlineto{\pgfqpoint{2.345467in}{2.671322in}}%
\pgfusepath{stroke}%
\end{pgfscope}%
\begin{pgfscope}%
\pgfsetbuttcap%
\pgfsetroundjoin%
\definecolor{currentfill}{rgb}{0.933333,0.466667,0.200000}%
\pgfsetfillcolor{currentfill}%
\pgfsetfillopacity{0.800000}%
\pgfsetlinewidth{2.007500pt}%
\definecolor{currentstroke}{rgb}{0.933333,0.466667,0.200000}%
\pgfsetstrokecolor{currentstroke}%
\pgfsetstrokeopacity{0.800000}%
\pgfsetdash{}{0pt}%
\pgfsys@defobject{currentmarker}{\pgfqpoint{-0.055556in}{-0.000000in}}{\pgfqpoint{0.055556in}{0.000000in}}{%
\pgfpathmoveto{\pgfqpoint{0.055556in}{-0.000000in}}%
\pgfpathlineto{\pgfqpoint{-0.055556in}{0.000000in}}%
\pgfusepath{stroke,fill}%
}%
\begin{pgfscope}%
\pgfsys@transformshift{2.345467in}{2.449100in}%
\pgfsys@useobject{currentmarker}{}%
\end{pgfscope}%
\end{pgfscope}%
\begin{pgfscope}%
\pgfsetbuttcap%
\pgfsetroundjoin%
\definecolor{currentfill}{rgb}{0.933333,0.466667,0.200000}%
\pgfsetfillcolor{currentfill}%
\pgfsetfillopacity{0.800000}%
\pgfsetlinewidth{2.007500pt}%
\definecolor{currentstroke}{rgb}{0.933333,0.466667,0.200000}%
\pgfsetstrokecolor{currentstroke}%
\pgfsetstrokeopacity{0.800000}%
\pgfsetdash{}{0pt}%
\pgfsys@defobject{currentmarker}{\pgfqpoint{-0.055556in}{-0.000000in}}{\pgfqpoint{0.055556in}{0.000000in}}{%
\pgfpathmoveto{\pgfqpoint{0.055556in}{-0.000000in}}%
\pgfpathlineto{\pgfqpoint{-0.055556in}{0.000000in}}%
\pgfusepath{stroke,fill}%
}%
\begin{pgfscope}%
\pgfsys@transformshift{2.345467in}{2.671322in}%
\pgfsys@useobject{currentmarker}{}%
\end{pgfscope}%
\end{pgfscope}%
\begin{pgfscope}%
\pgfsetrectcap%
\pgfsetroundjoin%
\pgfsetlinewidth{4.516875pt}%
\definecolor{currentstroke}{rgb}{0.933333,0.466667,0.200000}%
\pgfsetstrokecolor{currentstroke}%
\pgfsetstrokeopacity{0.800000}%
\pgfsetdash{}{0pt}%
\pgfpathmoveto{\pgfqpoint{2.123244in}{2.560211in}}%
\pgfpathlineto{\pgfqpoint{2.567689in}{2.560211in}}%
\pgfusepath{stroke}%
\end{pgfscope}%
\begin{pgfscope}%
\pgfsetbuttcap%
\pgfsetroundjoin%
\definecolor{currentfill}{rgb}{0.933333,0.466667,0.200000}%
\pgfsetfillcolor{currentfill}%
\pgfsetfillopacity{0.800000}%
\pgfsetlinewidth{2.007500pt}%
\definecolor{currentstroke}{rgb}{1.000000,1.000000,1.000000}%
\pgfsetstrokecolor{currentstroke}%
\pgfsetstrokeopacity{0.800000}%
\pgfsetdash{}{0pt}%
\pgfsys@defobject{currentmarker}{\pgfqpoint{-0.055556in}{-0.055556in}}{\pgfqpoint{0.055556in}{0.055556in}}{%
\pgfpathmoveto{\pgfqpoint{0.000000in}{-0.055556in}}%
\pgfpathcurveto{\pgfqpoint{0.014734in}{-0.055556in}}{\pgfqpoint{0.028866in}{-0.049702in}}{\pgfqpoint{0.039284in}{-0.039284in}}%
\pgfpathcurveto{\pgfqpoint{0.049702in}{-0.028866in}}{\pgfqpoint{0.055556in}{-0.014734in}}{\pgfqpoint{0.055556in}{0.000000in}}%
\pgfpathcurveto{\pgfqpoint{0.055556in}{0.014734in}}{\pgfqpoint{0.049702in}{0.028866in}}{\pgfqpoint{0.039284in}{0.039284in}}%
\pgfpathcurveto{\pgfqpoint{0.028866in}{0.049702in}}{\pgfqpoint{0.014734in}{0.055556in}}{\pgfqpoint{0.000000in}{0.055556in}}%
\pgfpathcurveto{\pgfqpoint{-0.014734in}{0.055556in}}{\pgfqpoint{-0.028866in}{0.049702in}}{\pgfqpoint{-0.039284in}{0.039284in}}%
\pgfpathcurveto{\pgfqpoint{-0.049702in}{0.028866in}}{\pgfqpoint{-0.055556in}{0.014734in}}{\pgfqpoint{-0.055556in}{0.000000in}}%
\pgfpathcurveto{\pgfqpoint{-0.055556in}{-0.014734in}}{\pgfqpoint{-0.049702in}{-0.028866in}}{\pgfqpoint{-0.039284in}{-0.039284in}}%
\pgfpathcurveto{\pgfqpoint{-0.028866in}{-0.049702in}}{\pgfqpoint{-0.014734in}{-0.055556in}}{\pgfqpoint{0.000000in}{-0.055556in}}%
\pgfpathlineto{\pgfqpoint{0.000000in}{-0.055556in}}%
\pgfpathclose%
\pgfusepath{stroke,fill}%
}%
\begin{pgfscope}%
\pgfsys@transformshift{2.345467in}{2.560211in}%
\pgfsys@useobject{currentmarker}{}%
\end{pgfscope}%
\end{pgfscope}%
\begin{pgfscope}%
\definecolor{textcolor}{rgb}{0.000000,0.000000,0.000000}%
\pgfsetstrokecolor{textcolor}%
\pgfsetfillcolor{textcolor}%
\pgftext[x=2.745467in,y=2.482433in,left,base]{\color{textcolor}{\rmfamily\fontsize{16.000000}{19.200000}\selectfont\catcode`\^=\active\def^{\ifmmode\sp\else\^{}\fi}\catcode`\%=\active\def
\end{pgfscope}%
\begin{pgfscope}%
\pgfsetbuttcap%
\pgfsetroundjoin%
\pgfsetlinewidth{4.516875pt}%
\definecolor{currentstroke}{rgb}{0.000000,0.466667,0.733333}%
\pgfsetstrokecolor{currentstroke}%
\pgfsetstrokeopacity{0.800000}%
\pgfsetdash{}{0pt}%
\pgfpathmoveto{\pgfqpoint{2.345467in}{2.124640in}}%
\pgfpathlineto{\pgfqpoint{2.345467in}{2.346862in}}%
\pgfusepath{stroke}%
\end{pgfscope}%
\begin{pgfscope}%
\pgfsetbuttcap%
\pgfsetroundjoin%
\definecolor{currentfill}{rgb}{0.000000,0.466667,0.733333}%
\pgfsetfillcolor{currentfill}%
\pgfsetfillopacity{0.800000}%
\pgfsetlinewidth{2.007500pt}%
\definecolor{currentstroke}{rgb}{0.000000,0.466667,0.733333}%
\pgfsetstrokecolor{currentstroke}%
\pgfsetstrokeopacity{0.800000}%
\pgfsetdash{}{0pt}%
\pgfsys@defobject{currentmarker}{\pgfqpoint{-0.055556in}{-0.000000in}}{\pgfqpoint{0.055556in}{0.000000in}}{%
\pgfpathmoveto{\pgfqpoint{0.055556in}{-0.000000in}}%
\pgfpathlineto{\pgfqpoint{-0.055556in}{0.000000in}}%
\pgfusepath{stroke,fill}%
}%
\begin{pgfscope}%
\pgfsys@transformshift{2.345467in}{2.124640in}%
\pgfsys@useobject{currentmarker}{}%
\end{pgfscope}%
\end{pgfscope}%
\begin{pgfscope}%
\pgfsetbuttcap%
\pgfsetroundjoin%
\definecolor{currentfill}{rgb}{0.000000,0.466667,0.733333}%
\pgfsetfillcolor{currentfill}%
\pgfsetfillopacity{0.800000}%
\pgfsetlinewidth{2.007500pt}%
\definecolor{currentstroke}{rgb}{0.000000,0.466667,0.733333}%
\pgfsetstrokecolor{currentstroke}%
\pgfsetstrokeopacity{0.800000}%
\pgfsetdash{}{0pt}%
\pgfsys@defobject{currentmarker}{\pgfqpoint{-0.055556in}{-0.000000in}}{\pgfqpoint{0.055556in}{0.000000in}}{%
\pgfpathmoveto{\pgfqpoint{0.055556in}{-0.000000in}}%
\pgfpathlineto{\pgfqpoint{-0.055556in}{0.000000in}}%
\pgfusepath{stroke,fill}%
}%
\begin{pgfscope}%
\pgfsys@transformshift{2.345467in}{2.346862in}%
\pgfsys@useobject{currentmarker}{}%
\end{pgfscope}%
\end{pgfscope}%
\begin{pgfscope}%
\pgfsetbuttcap%
\pgfsetroundjoin%
\pgfsetlinewidth{4.516875pt}%
\definecolor{currentstroke}{rgb}{0.000000,0.466667,0.733333}%
\pgfsetstrokecolor{currentstroke}%
\pgfsetstrokeopacity{0.800000}%
\pgfsetdash{{16.650000pt}{7.200000pt}}{0.000000pt}%
\pgfpathmoveto{\pgfqpoint{2.123244in}{2.235751in}}%
\pgfpathlineto{\pgfqpoint{2.567689in}{2.235751in}}%
\pgfusepath{stroke}%
\end{pgfscope}%
\begin{pgfscope}%
\pgfsetbuttcap%
\pgfsetmiterjoin%
\definecolor{currentfill}{rgb}{0.000000,0.466667,0.733333}%
\pgfsetfillcolor{currentfill}%
\pgfsetfillopacity{0.800000}%
\pgfsetlinewidth{2.007500pt}%
\definecolor{currentstroke}{rgb}{1.000000,1.000000,1.000000}%
\pgfsetstrokecolor{currentstroke}%
\pgfsetstrokeopacity{0.800000}%
\pgfsetdash{}{0pt}%
\pgfsys@defobject{currentmarker}{\pgfqpoint{-0.055556in}{-0.055556in}}{\pgfqpoint{0.055556in}{0.055556in}}{%
\pgfpathmoveto{\pgfqpoint{-0.055556in}{-0.055556in}}%
\pgfpathlineto{\pgfqpoint{0.055556in}{-0.055556in}}%
\pgfpathlineto{\pgfqpoint{0.055556in}{0.055556in}}%
\pgfpathlineto{\pgfqpoint{-0.055556in}{0.055556in}}%
\pgfpathlineto{\pgfqpoint{-0.055556in}{-0.055556in}}%
\pgfpathclose%
\pgfusepath{stroke,fill}%
}%
\begin{pgfscope}%
\pgfsys@transformshift{2.345467in}{2.235751in}%
\pgfsys@useobject{currentmarker}{}%
\end{pgfscope}%
\end{pgfscope}%
\begin{pgfscope}%
\definecolor{textcolor}{rgb}{0.000000,0.000000,0.000000}%
\pgfsetstrokecolor{textcolor}%
\pgfsetfillcolor{textcolor}%
\pgftext[x=2.745467in,y=2.157973in,left,base]{\color{textcolor}{\rmfamily\fontsize{16.000000}{19.200000}\selectfont\catcode`\^=\active\def^{\ifmmode\sp\else\^{}\fi}\catcode`\%=\active\def
\end{pgfscope}%
\end{pgfpicture}%
\makeatother%
\endgroup%

%% file: figures/SE_est_match/h_laplace_SEmatch.pgf
\begingroup%
\makeatletter%
\begin{pgfpicture}%
\pgfpathrectangle{\pgfpointorigin}{\pgfqpoint{5.378809in}{2.976518in}}%
\pgfusepath{use as bounding box, clip}%
\begin{pgfscope}%
\pgfsetbuttcap%
\pgfsetmiterjoin%
\definecolor{currentfill}{rgb}{1.000000,1.000000,1.000000}%
\pgfsetfillcolor{currentfill}%
\pgfsetlinewidth{0.000000pt}%
\definecolor{currentstroke}{rgb}{1.000000,1.000000,1.000000}%
\pgfsetstrokecolor{currentstroke}%
\pgfsetdash{}{0pt}%
\pgfpathmoveto{\pgfqpoint{0.000000in}{0.000000in}}%
\pgfpathlineto{\pgfqpoint{5.378809in}{0.000000in}}%
\pgfpathlineto{\pgfqpoint{5.378809in}{2.976518in}}%
\pgfpathlineto{\pgfqpoint{0.000000in}{2.976518in}}%
\pgfpathlineto{\pgfqpoint{0.000000in}{0.000000in}}%
\pgfpathclose%
\pgfusepath{fill}%
\end{pgfscope}%
\begin{pgfscope}%
\pgfsetbuttcap%
\pgfsetmiterjoin%
\definecolor{currentfill}{rgb}{1.000000,1.000000,1.000000}%
\pgfsetfillcolor{currentfill}%
\pgfsetlinewidth{0.000000pt}%
\definecolor{currentstroke}{rgb}{0.000000,0.000000,0.000000}%
\pgfsetstrokecolor{currentstroke}%
\pgfsetstrokeopacity{0.000000}%
\pgfsetdash{}{0pt}%
\pgfpathmoveto{\pgfqpoint{0.777311in}{0.705277in}}%
\pgfpathlineto{\pgfqpoint{5.278809in}{0.705277in}}%
\pgfpathlineto{\pgfqpoint{5.278809in}{2.876518in}}%
\pgfpathlineto{\pgfqpoint{0.777311in}{2.876518in}}%
\pgfpathlineto{\pgfqpoint{0.777311in}{0.705277in}}%
\pgfpathclose%
\pgfusepath{fill}%
\end{pgfscope}%
\begin{pgfscope}%
\pgfpathrectangle{\pgfqpoint{0.777311in}{0.705277in}}{\pgfqpoint{4.501497in}{2.171241in}}%
\pgfusepath{clip}%
\pgfsetbuttcap%
\pgfsetroundjoin%
\pgfsetlinewidth{3.513125pt}%
\definecolor{currentstroke}{rgb}{0.000000,0.000000,0.000000}%
\pgfsetstrokecolor{currentstroke}%
\pgfsetstrokeopacity{0.800000}%
\pgfsetdash{}{0pt}%
\pgfpathmoveto{\pgfqpoint{0.981925in}{1.304781in}}%
\pgfpathlineto{\pgfqpoint{0.981925in}{1.535174in}}%
\pgfusepath{stroke}%
\end{pgfscope}%
\begin{pgfscope}%
\pgfpathrectangle{\pgfqpoint{0.777311in}{0.705277in}}{\pgfqpoint{4.501497in}{2.171241in}}%
\pgfusepath{clip}%
\pgfsetbuttcap%
\pgfsetroundjoin%
\pgfsetlinewidth{3.513125pt}%
\definecolor{currentstroke}{rgb}{0.000000,0.000000,0.000000}%
\pgfsetstrokecolor{currentstroke}%
\pgfsetstrokeopacity{0.800000}%
\pgfsetdash{}{0pt}%
\pgfpathmoveto{\pgfqpoint{3.028060in}{1.231310in}}%
\pgfpathlineto{\pgfqpoint{3.028060in}{1.416852in}}%
\pgfusepath{stroke}%
\end{pgfscope}%
\begin{pgfscope}%
\pgfpathrectangle{\pgfqpoint{0.777311in}{0.705277in}}{\pgfqpoint{4.501497in}{2.171241in}}%
\pgfusepath{clip}%
\pgfsetbuttcap%
\pgfsetroundjoin%
\pgfsetlinewidth{3.513125pt}%
\definecolor{currentstroke}{rgb}{0.000000,0.000000,0.000000}%
\pgfsetstrokecolor{currentstroke}%
\pgfsetstrokeopacity{0.800000}%
\pgfsetdash{}{0pt}%
\pgfpathmoveto{\pgfqpoint{5.074195in}{1.081904in}}%
\pgfpathlineto{\pgfqpoint{5.074195in}{1.304039in}}%
\pgfusepath{stroke}%
\end{pgfscope}%
\begin{pgfscope}%
\pgfpathrectangle{\pgfqpoint{0.777311in}{0.705277in}}{\pgfqpoint{4.501497in}{2.171241in}}%
\pgfusepath{clip}%
\pgfsetbuttcap%
\pgfsetroundjoin%
\definecolor{currentfill}{rgb}{0.000000,0.000000,0.000000}%
\pgfsetfillcolor{currentfill}%
\pgfsetfillopacity{0.800000}%
\pgfsetlinewidth{2.007500pt}%
\definecolor{currentstroke}{rgb}{0.000000,0.000000,0.000000}%
\pgfsetstrokecolor{currentstroke}%
\pgfsetstrokeopacity{0.800000}%
\pgfsetdash{}{0pt}%
\pgfsys@defobject{currentmarker}{\pgfqpoint{-0.055556in}{-0.000000in}}{\pgfqpoint{0.055556in}{0.000000in}}{%
\pgfpathmoveto{\pgfqpoint{0.055556in}{-0.000000in}}%
\pgfpathlineto{\pgfqpoint{-0.055556in}{0.000000in}}%
\pgfusepath{stroke,fill}%
}%
\begin{pgfscope}%
\pgfsys@transformshift{0.981925in}{1.304781in}%
\pgfsys@useobject{currentmarker}{}%
\end{pgfscope}%
\begin{pgfscope}%
\pgfsys@transformshift{3.028060in}{1.231310in}%
\pgfsys@useobject{currentmarker}{}%
\end{pgfscope}%
\begin{pgfscope}%
\pgfsys@transformshift{5.074195in}{1.081904in}%
\pgfsys@useobject{currentmarker}{}%
\end{pgfscope}%
\end{pgfscope}%
\begin{pgfscope}%
\pgfpathrectangle{\pgfqpoint{0.777311in}{0.705277in}}{\pgfqpoint{4.501497in}{2.171241in}}%
\pgfusepath{clip}%
\pgfsetbuttcap%
\pgfsetroundjoin%
\definecolor{currentfill}{rgb}{0.000000,0.000000,0.000000}%
\pgfsetfillcolor{currentfill}%
\pgfsetfillopacity{0.800000}%
\pgfsetlinewidth{2.007500pt}%
\definecolor{currentstroke}{rgb}{0.000000,0.000000,0.000000}%
\pgfsetstrokecolor{currentstroke}%
\pgfsetstrokeopacity{0.800000}%
\pgfsetdash{}{0pt}%
\pgfsys@defobject{currentmarker}{\pgfqpoint{-0.055556in}{-0.000000in}}{\pgfqpoint{0.055556in}{0.000000in}}{%
\pgfpathmoveto{\pgfqpoint{0.055556in}{-0.000000in}}%
\pgfpathlineto{\pgfqpoint{-0.055556in}{0.000000in}}%
\pgfusepath{stroke,fill}%
}%
\begin{pgfscope}%
\pgfsys@transformshift{0.981925in}{1.535174in}%
\pgfsys@useobject{currentmarker}{}%
\end{pgfscope}%
\begin{pgfscope}%
\pgfsys@transformshift{3.028060in}{1.416852in}%
\pgfsys@useobject{currentmarker}{}%
\end{pgfscope}%
\begin{pgfscope}%
\pgfsys@transformshift{5.074195in}{1.304039in}%
\pgfsys@useobject{currentmarker}{}%
\end{pgfscope}%
\end{pgfscope}%
\begin{pgfscope}%
\pgfpathrectangle{\pgfqpoint{0.777311in}{0.705277in}}{\pgfqpoint{4.501497in}{2.171241in}}%
\pgfusepath{clip}%
\pgfsetrectcap%
\pgfsetroundjoin%
\pgfsetlinewidth{3.513125pt}%
\definecolor{currentstroke}{rgb}{0.000000,0.000000,0.000000}%
\pgfsetstrokecolor{currentstroke}%
\pgfsetstrokeopacity{0.800000}%
\pgfsetdash{}{0pt}%
\pgfpathmoveto{\pgfqpoint{0.981925in}{1.414067in}}%
\pgfpathlineto{\pgfqpoint{3.028060in}{1.326647in}}%
\pgfpathlineto{\pgfqpoint{5.074195in}{1.221698in}}%
\pgfusepath{stroke}%
\end{pgfscope}%
\begin{pgfscope}%
\pgfpathrectangle{\pgfqpoint{0.777311in}{0.705277in}}{\pgfqpoint{4.501497in}{2.171241in}}%
\pgfusepath{clip}%
\pgfsetbuttcap%
\pgfsetroundjoin%
\definecolor{currentfill}{rgb}{0.000000,0.000000,0.000000}%
\pgfsetfillcolor{currentfill}%
\pgfsetfillopacity{0.800000}%
\pgfsetlinewidth{2.007500pt}%
\definecolor{currentstroke}{rgb}{1.000000,1.000000,1.000000}%
\pgfsetstrokecolor{currentstroke}%
\pgfsetstrokeopacity{0.800000}%
\pgfsetdash{}{0pt}%
\pgfsys@defobject{currentmarker}{\pgfqpoint{-0.055556in}{-0.055556in}}{\pgfqpoint{0.055556in}{0.055556in}}{%
\pgfpathmoveto{\pgfqpoint{0.000000in}{-0.055556in}}%
\pgfpathcurveto{\pgfqpoint{0.014734in}{-0.055556in}}{\pgfqpoint{0.028866in}{-0.049702in}}{\pgfqpoint{0.039284in}{-0.039284in}}%
\pgfpathcurveto{\pgfqpoint{0.049702in}{-0.028866in}}{\pgfqpoint{0.055556in}{-0.014734in}}{\pgfqpoint{0.055556in}{0.000000in}}%
\pgfpathcurveto{\pgfqpoint{0.055556in}{0.014734in}}{\pgfqpoint{0.049702in}{0.028866in}}{\pgfqpoint{0.039284in}{0.039284in}}%
\pgfpathcurveto{\pgfqpoint{0.028866in}{0.049702in}}{\pgfqpoint{0.014734in}{0.055556in}}{\pgfqpoint{0.000000in}{0.055556in}}%
\pgfpathcurveto{\pgfqpoint{-0.014734in}{0.055556in}}{\pgfqpoint{-0.028866in}{0.049702in}}{\pgfqpoint{-0.039284in}{0.039284in}}%
\pgfpathcurveto{\pgfqpoint{-0.049702in}{0.028866in}}{\pgfqpoint{-0.055556in}{0.014734in}}{\pgfqpoint{-0.055556in}{0.000000in}}%
\pgfpathcurveto{\pgfqpoint{-0.055556in}{-0.014734in}}{\pgfqpoint{-0.049702in}{-0.028866in}}{\pgfqpoint{-0.039284in}{-0.039284in}}%
\pgfpathcurveto{\pgfqpoint{-0.028866in}{-0.049702in}}{\pgfqpoint{-0.014734in}{-0.055556in}}{\pgfqpoint{0.000000in}{-0.055556in}}%
\pgfpathlineto{\pgfqpoint{0.000000in}{-0.055556in}}%
\pgfpathclose%
\pgfusepath{stroke,fill}%
}%
\begin{pgfscope}%
\pgfsys@transformshift{0.981925in}{1.414067in}%
\pgfsys@useobject{currentmarker}{}%
\end{pgfscope}%
\begin{pgfscope}%
\pgfsys@transformshift{3.028060in}{1.326647in}%
\pgfsys@useobject{currentmarker}{}%
\end{pgfscope}%
\begin{pgfscope}%
\pgfsys@transformshift{5.074195in}{1.221698in}%
\pgfsys@useobject{currentmarker}{}%
\end{pgfscope}%
\end{pgfscope}%
\begin{pgfscope}%
\pgfpathrectangle{\pgfqpoint{0.777311in}{0.705277in}}{\pgfqpoint{4.501497in}{2.171241in}}%
\pgfusepath{clip}%
\pgfsetbuttcap%
\pgfsetroundjoin%
\definecolor{currentfill}{rgb}{0.200000,0.733333,0.933333}%
\pgfsetfillcolor{currentfill}%
\pgfsetfillopacity{0.800000}%
\pgfsetlinewidth{2.509375pt}%
\definecolor{currentstroke}{rgb}{1.000000,1.000000,1.000000}%
\pgfsetstrokecolor{currentstroke}%
\pgfsetstrokeopacity{0.800000}%
\pgfsetdash{}{0pt}%
\pgfsys@defobject{currentmarker}{\pgfqpoint{-0.064550in}{-0.107583in}}{\pgfqpoint{0.064550in}{0.107583in}}{%
\pgfpathmoveto{\pgfqpoint{-0.000000in}{-0.107583in}}%
\pgfpathlineto{\pgfqpoint{0.064550in}{0.000000in}}%
\pgfpathlineto{\pgfqpoint{0.000000in}{0.107583in}}%
\pgfpathlineto{\pgfqpoint{-0.064550in}{0.000000in}}%
\pgfpathlineto{\pgfqpoint{-0.000000in}{-0.107583in}}%
\pgfpathclose%
\pgfusepath{stroke,fill}%
}%
\begin{pgfscope}%
\pgfsys@transformshift{0.981925in}{1.460386in}%
\pgfsys@useobject{currentmarker}{}%
\end{pgfscope}%
\begin{pgfscope}%
\pgfsys@transformshift{3.028060in}{1.359018in}%
\pgfsys@useobject{currentmarker}{}%
\end{pgfscope}%
\begin{pgfscope}%
\pgfsys@transformshift{5.074195in}{1.307546in}%
\pgfsys@useobject{currentmarker}{}%
\end{pgfscope}%
\end{pgfscope}%
\begin{pgfscope}%
\pgfsetbuttcap%
\pgfsetroundjoin%
\definecolor{currentfill}{rgb}{0.000000,0.000000,0.000000}%
\pgfsetfillcolor{currentfill}%
\pgfsetlinewidth{1.505625pt}%
\definecolor{currentstroke}{rgb}{0.000000,0.000000,0.000000}%
\pgfsetstrokecolor{currentstroke}%
\pgfsetdash{}{0pt}%
\pgfsys@defobject{currentmarker}{\pgfqpoint{0.000000in}{-0.083333in}}{\pgfqpoint{0.000000in}{0.000000in}}{%
\pgfpathmoveto{\pgfqpoint{0.000000in}{0.000000in}}%
\pgfpathlineto{\pgfqpoint{0.000000in}{-0.083333in}}%
\pgfusepath{stroke,fill}%
}%
\begin{pgfscope}%
\pgfsys@transformshift{0.981925in}{0.705277in}%
\pgfsys@useobject{currentmarker}{}%
\end{pgfscope}%
\end{pgfscope}%
\begin{pgfscope}%
\definecolor{textcolor}{rgb}{0.000000,0.000000,0.000000}%
\pgfsetstrokecolor{textcolor}%
\pgfsetfillcolor{textcolor}%
\pgftext[x=0.981925in,y=0.573333in,,top]{\color{textcolor}{\rmfamily\fontsize{14.000000}{16.800000}\selectfont\catcode`\^=\active\def^{\ifmmode\sp\else\^{}\fi}\catcode`\%=\active\def
\end{pgfscope}%
\begin{pgfscope}%
\pgfsetbuttcap%
\pgfsetroundjoin%
\definecolor{currentfill}{rgb}{0.000000,0.000000,0.000000}%
\pgfsetfillcolor{currentfill}%
\pgfsetlinewidth{1.505625pt}%
\definecolor{currentstroke}{rgb}{0.000000,0.000000,0.000000}%
\pgfsetstrokecolor{currentstroke}%
\pgfsetdash{}{0pt}%
\pgfsys@defobject{currentmarker}{\pgfqpoint{0.000000in}{-0.083333in}}{\pgfqpoint{0.000000in}{0.000000in}}{%
\pgfpathmoveto{\pgfqpoint{0.000000in}{0.000000in}}%
\pgfpathlineto{\pgfqpoint{0.000000in}{-0.083333in}}%
\pgfusepath{stroke,fill}%
}%
\begin{pgfscope}%
\pgfsys@transformshift{3.028060in}{0.705277in}%
\pgfsys@useobject{currentmarker}{}%
\end{pgfscope}%
\end{pgfscope}%
\begin{pgfscope}%
\definecolor{textcolor}{rgb}{0.000000,0.000000,0.000000}%
\pgfsetstrokecolor{textcolor}%
\pgfsetfillcolor{textcolor}%
\pgftext[x=3.028060in,y=0.573333in,,top]{\color{textcolor}{\rmfamily\fontsize{14.000000}{16.800000}\selectfont\catcode`\^=\active\def^{\ifmmode\sp\else\^{}\fi}\catcode`\%=\active\def
\end{pgfscope}%
\begin{pgfscope}%
\pgfsetbuttcap%
\pgfsetroundjoin%
\definecolor{currentfill}{rgb}{0.000000,0.000000,0.000000}%
\pgfsetfillcolor{currentfill}%
\pgfsetlinewidth{1.505625pt}%
\definecolor{currentstroke}{rgb}{0.000000,0.000000,0.000000}%
\pgfsetstrokecolor{currentstroke}%
\pgfsetdash{}{0pt}%
\pgfsys@defobject{currentmarker}{\pgfqpoint{0.000000in}{-0.083333in}}{\pgfqpoint{0.000000in}{0.000000in}}{%
\pgfpathmoveto{\pgfqpoint{0.000000in}{0.000000in}}%
\pgfpathlineto{\pgfqpoint{0.000000in}{-0.083333in}}%
\pgfusepath{stroke,fill}%
}%
\begin{pgfscope}%
\pgfsys@transformshift{5.074195in}{0.705277in}%
\pgfsys@useobject{currentmarker}{}%
\end{pgfscope}%
\end{pgfscope}%
\begin{pgfscope}%
\definecolor{textcolor}{rgb}{0.000000,0.000000,0.000000}%
\pgfsetstrokecolor{textcolor}%
\pgfsetfillcolor{textcolor}%
\pgftext[x=5.074195in,y=0.573333in,,top]{\color{textcolor}{\rmfamily\fontsize{14.000000}{16.800000}\selectfont\catcode`\^=\active\def^{\ifmmode\sp\else\^{}\fi}\catcode`\%=\active\def
\end{pgfscope}%
\begin{pgfscope}%
\definecolor{textcolor}{rgb}{0.000000,0.000000,0.000000}%
\pgfsetstrokecolor{textcolor}%
\pgfsetfillcolor{textcolor}%
\pgftext[x=3.028060in,y=0.340000in,,top]{\color{textcolor}{\rmfamily\fontsize{18.000000}{21.600000}\selectfont\catcode`\^=\active\def^{\ifmmode\sp\else\^{}\fi}\catcode`\%=\active\def
\end{pgfscope}%
\begin{pgfscope}%
\pgfsetbuttcap%
\pgfsetroundjoin%
\definecolor{currentfill}{rgb}{0.000000,0.000000,0.000000}%
\pgfsetfillcolor{currentfill}%
\pgfsetlinewidth{1.505625pt}%
\definecolor{currentstroke}{rgb}{0.000000,0.000000,0.000000}%
\pgfsetstrokecolor{currentstroke}%
\pgfsetdash{}{0pt}%
\pgfsys@defobject{currentmarker}{\pgfqpoint{-0.083333in}{0.000000in}}{\pgfqpoint{-0.000000in}{0.000000in}}{%
\pgfpathmoveto{\pgfqpoint{-0.000000in}{0.000000in}}%
\pgfpathlineto{\pgfqpoint{-0.083333in}{0.000000in}}%
\pgfusepath{stroke,fill}%
}%
\begin{pgfscope}%
\pgfsys@transformshift{0.777311in}{0.705277in}%
\pgfsys@useobject{currentmarker}{}%
\end{pgfscope}%
\end{pgfscope}%
\begin{pgfscope}%
\definecolor{textcolor}{rgb}{0.000000,0.000000,0.000000}%
\pgfsetstrokecolor{textcolor}%
\pgfsetfillcolor{textcolor}%
\pgftext[x=0.395138in, y=0.635833in, left, base]{\color{textcolor}{\rmfamily\fontsize{14.000000}{16.800000}\selectfont\catcode`\^=\active\def^{\ifmmode\sp\else\^{}\fi}\catcode`\%=\active\def
\end{pgfscope}%
\begin{pgfscope}%
\pgfsetbuttcap%
\pgfsetroundjoin%
\definecolor{currentfill}{rgb}{0.000000,0.000000,0.000000}%
\pgfsetfillcolor{currentfill}%
\pgfsetlinewidth{1.505625pt}%
\definecolor{currentstroke}{rgb}{0.000000,0.000000,0.000000}%
\pgfsetstrokecolor{currentstroke}%
\pgfsetdash{}{0pt}%
\pgfsys@defobject{currentmarker}{\pgfqpoint{-0.083333in}{0.000000in}}{\pgfqpoint{-0.000000in}{0.000000in}}{%
\pgfpathmoveto{\pgfqpoint{-0.000000in}{0.000000in}}%
\pgfpathlineto{\pgfqpoint{-0.083333in}{0.000000in}}%
\pgfusepath{stroke,fill}%
}%
\begin{pgfscope}%
\pgfsys@transformshift{0.777311in}{1.573773in}%
\pgfsys@useobject{currentmarker}{}%
\end{pgfscope}%
\end{pgfscope}%
\begin{pgfscope}%
\definecolor{textcolor}{rgb}{0.000000,0.000000,0.000000}%
\pgfsetstrokecolor{textcolor}%
\pgfsetfillcolor{textcolor}%
\pgftext[x=0.395138in, y=1.504329in, left, base]{\color{textcolor}{\rmfamily\fontsize{14.000000}{16.800000}\selectfont\catcode`\^=\active\def^{\ifmmode\sp\else\^{}\fi}\catcode`\%=\active\def
\end{pgfscope}%
\begin{pgfscope}%
\pgfsetbuttcap%
\pgfsetroundjoin%
\definecolor{currentfill}{rgb}{0.000000,0.000000,0.000000}%
\pgfsetfillcolor{currentfill}%
\pgfsetlinewidth{1.505625pt}%
\definecolor{currentstroke}{rgb}{0.000000,0.000000,0.000000}%
\pgfsetstrokecolor{currentstroke}%
\pgfsetdash{}{0pt}%
\pgfsys@defobject{currentmarker}{\pgfqpoint{-0.083333in}{0.000000in}}{\pgfqpoint{-0.000000in}{0.000000in}}{%
\pgfpathmoveto{\pgfqpoint{-0.000000in}{0.000000in}}%
\pgfpathlineto{\pgfqpoint{-0.083333in}{0.000000in}}%
\pgfusepath{stroke,fill}%
}%
\begin{pgfscope}%
\pgfsys@transformshift{0.777311in}{2.442270in}%
\pgfsys@useobject{currentmarker}{}%
\end{pgfscope}%
\end{pgfscope}%
\begin{pgfscope}%
\definecolor{textcolor}{rgb}{0.000000,0.000000,0.000000}%
\pgfsetstrokecolor{textcolor}%
\pgfsetfillcolor{textcolor}%
\pgftext[x=0.395138in, y=2.372825in, left, base]{\color{textcolor}{\rmfamily\fontsize{14.000000}{16.800000}\selectfont\catcode`\^=\active\def^{\ifmmode\sp\else\^{}\fi}\catcode`\%=\active\def
\end{pgfscope}%
\begin{pgfscope}%
\definecolor{textcolor}{rgb}{0.000000,0.000000,0.000000}%
\pgfsetstrokecolor{textcolor}%
\pgfsetfillcolor{textcolor}%
\pgftext[x=0.339583in,y=1.790898in,,bottom,rotate=90.000000]{\color{textcolor}{\rmfamily\fontsize{18.000000}{21.600000}\selectfont\catcode`\^=\active\def^{\ifmmode\sp\else\^{}\fi}\catcode`\%=\active\def
\end{pgfscope}%
\begin{pgfscope}%
\pgfsetrectcap%
\pgfsetmiterjoin%
\pgfsetlinewidth{1.505625pt}%
\definecolor{currentstroke}{rgb}{0.000000,0.000000,0.000000}%
\pgfsetstrokecolor{currentstroke}%
\pgfsetdash{}{0pt}%
\pgfpathmoveto{\pgfqpoint{0.777311in}{0.705277in}}%
\pgfpathlineto{\pgfqpoint{0.777311in}{2.876518in}}%
\pgfusepath{stroke}%
\end{pgfscope}%
\begin{pgfscope}%
\pgfsetrectcap%
\pgfsetmiterjoin%
\pgfsetlinewidth{1.505625pt}%
\definecolor{currentstroke}{rgb}{0.000000,0.000000,0.000000}%
\pgfsetstrokecolor{currentstroke}%
\pgfsetdash{}{0pt}%
\pgfpathmoveto{\pgfqpoint{0.777311in}{0.705277in}}%
\pgfpathlineto{\pgfqpoint{5.278809in}{0.705277in}}%
\pgfusepath{stroke}%
\end{pgfscope}%
\begin{pgfscope}%
\pgfsetbuttcap%
\pgfsetmiterjoin%
\definecolor{currentfill}{rgb}{1.000000,1.000000,1.000000}%
\pgfsetfillcolor{currentfill}%
\pgfsetfillopacity{0.400000}%
\pgfsetlinewidth{1.003750pt}%
\definecolor{currentstroke}{rgb}{0.000000,0.000000,0.000000}%
\pgfsetstrokecolor{currentstroke}%
\pgfsetstrokeopacity{0.400000}%
\pgfsetdash{}{0pt}%
\pgfpathmoveto{\pgfqpoint{2.113393in}{2.049821in}}%
\pgfpathlineto{\pgfqpoint{5.123253in}{2.049821in}}%
\pgfpathquadraticcurveto{\pgfqpoint{5.167697in}{2.049821in}}{\pgfqpoint{5.167697in}{2.094265in}}%
\pgfpathlineto{\pgfqpoint{5.167697in}{2.720962in}}%
\pgfpathquadraticcurveto{\pgfqpoint{5.167697in}{2.765407in}}{\pgfqpoint{5.123253in}{2.765407in}}%
\pgfpathlineto{\pgfqpoint{2.113393in}{2.765407in}}%
\pgfpathquadraticcurveto{\pgfqpoint{2.068948in}{2.765407in}}{\pgfqpoint{2.068948in}{2.720962in}}%
\pgfpathlineto{\pgfqpoint{2.068948in}{2.094265in}}%
\pgfpathquadraticcurveto{\pgfqpoint{2.068948in}{2.049821in}}{\pgfqpoint{2.113393in}{2.049821in}}%
\pgfpathlineto{\pgfqpoint{2.113393in}{2.049821in}}%
\pgfpathclose%
\pgfusepath{stroke,fill}%
\end{pgfscope}%
\begin{pgfscope}%
\pgfsetbuttcap%
\pgfsetroundjoin%
\definecolor{currentfill}{rgb}{0.200000,0.733333,0.933333}%
\pgfsetfillcolor{currentfill}%
\pgfsetfillopacity{0.800000}%
\pgfsetlinewidth{2.509375pt}%
\definecolor{currentstroke}{rgb}{1.000000,1.000000,1.000000}%
\pgfsetstrokecolor{currentstroke}%
\pgfsetstrokeopacity{0.800000}%
\pgfsetdash{}{0pt}%
\pgfsys@defobject{currentmarker}{\pgfqpoint{-0.064550in}{-0.107583in}}{\pgfqpoint{0.064550in}{0.107583in}}{%
\pgfpathmoveto{\pgfqpoint{-0.000000in}{-0.107583in}}%
\pgfpathlineto{\pgfqpoint{0.064550in}{0.000000in}}%
\pgfpathlineto{\pgfqpoint{0.000000in}{0.107583in}}%
\pgfpathlineto{\pgfqpoint{-0.064550in}{0.000000in}}%
\pgfpathlineto{\pgfqpoint{-0.000000in}{-0.107583in}}%
\pgfpathclose%
\pgfusepath{stroke,fill}%
}%
\begin{pgfscope}%
\pgfsys@transformshift{2.380059in}{2.568185in}%
\pgfsys@useobject{currentmarker}{}%
\end{pgfscope}%
\end{pgfscope}%
\begin{pgfscope}%
\definecolor{textcolor}{rgb}{0.000000,0.000000,0.000000}%
\pgfsetstrokecolor{textcolor}%
\pgfsetfillcolor{textcolor}%
\pgftext[x=2.780059in,y=2.509851in,left,base]{\color{textcolor}{\rmfamily\fontsize{16.000000}{19.200000}\selectfont\catcode`\^=\active\def^{\ifmmode\sp\else\^{}\fi}\catcode`\%=\active\def
\end{pgfscope}%
\begin{pgfscope}%
\pgfsetbuttcap%
\pgfsetroundjoin%
\pgfsetlinewidth{3.513125pt}%
\definecolor{currentstroke}{rgb}{0.000000,0.000000,0.000000}%
\pgfsetstrokecolor{currentstroke}%
\pgfsetstrokeopacity{0.800000}%
\pgfsetdash{}{0pt}%
\pgfpathmoveto{\pgfqpoint{2.380059in}{2.152058in}}%
\pgfpathlineto{\pgfqpoint{2.380059in}{2.374280in}}%
\pgfusepath{stroke}%
\end{pgfscope}%
\begin{pgfscope}%
\pgfsetbuttcap%
\pgfsetroundjoin%
\definecolor{currentfill}{rgb}{0.000000,0.000000,0.000000}%
\pgfsetfillcolor{currentfill}%
\pgfsetfillopacity{0.800000}%
\pgfsetlinewidth{2.007500pt}%
\definecolor{currentstroke}{rgb}{0.000000,0.000000,0.000000}%
\pgfsetstrokecolor{currentstroke}%
\pgfsetstrokeopacity{0.800000}%
\pgfsetdash{}{0pt}%
\pgfsys@defobject{currentmarker}{\pgfqpoint{-0.055556in}{-0.000000in}}{\pgfqpoint{0.055556in}{0.000000in}}{%
\pgfpathmoveto{\pgfqpoint{0.055556in}{-0.000000in}}%
\pgfpathlineto{\pgfqpoint{-0.055556in}{0.000000in}}%
\pgfusepath{stroke,fill}%
}%
\begin{pgfscope}%
\pgfsys@transformshift{2.380059in}{2.152058in}%
\pgfsys@useobject{currentmarker}{}%
\end{pgfscope}%
\end{pgfscope}%
\begin{pgfscope}%
\pgfsetbuttcap%
\pgfsetroundjoin%
\definecolor{currentfill}{rgb}{0.000000,0.000000,0.000000}%
\pgfsetfillcolor{currentfill}%
\pgfsetfillopacity{0.800000}%
\pgfsetlinewidth{2.007500pt}%
\definecolor{currentstroke}{rgb}{0.000000,0.000000,0.000000}%
\pgfsetstrokecolor{currentstroke}%
\pgfsetstrokeopacity{0.800000}%
\pgfsetdash{}{0pt}%
\pgfsys@defobject{currentmarker}{\pgfqpoint{-0.055556in}{-0.000000in}}{\pgfqpoint{0.055556in}{0.000000in}}{%
\pgfpathmoveto{\pgfqpoint{0.055556in}{-0.000000in}}%
\pgfpathlineto{\pgfqpoint{-0.055556in}{0.000000in}}%
\pgfusepath{stroke,fill}%
}%
\begin{pgfscope}%
\pgfsys@transformshift{2.380059in}{2.374280in}%
\pgfsys@useobject{currentmarker}{}%
\end{pgfscope}%
\end{pgfscope}%
\begin{pgfscope}%
\pgfsetrectcap%
\pgfsetroundjoin%
\pgfsetlinewidth{3.513125pt}%
\definecolor{currentstroke}{rgb}{0.000000,0.000000,0.000000}%
\pgfsetstrokecolor{currentstroke}%
\pgfsetstrokeopacity{0.800000}%
\pgfsetdash{}{0pt}%
\pgfpathmoveto{\pgfqpoint{2.157837in}{2.263169in}}%
\pgfpathlineto{\pgfqpoint{2.602281in}{2.263169in}}%
\pgfusepath{stroke}%
\end{pgfscope}%
\begin{pgfscope}%
\pgfsetbuttcap%
\pgfsetroundjoin%
\definecolor{currentfill}{rgb}{0.000000,0.000000,0.000000}%
\pgfsetfillcolor{currentfill}%
\pgfsetfillopacity{0.800000}%
\pgfsetlinewidth{2.007500pt}%
\definecolor{currentstroke}{rgb}{1.000000,1.000000,1.000000}%
\pgfsetstrokecolor{currentstroke}%
\pgfsetstrokeopacity{0.800000}%
\pgfsetdash{}{0pt}%
\pgfsys@defobject{currentmarker}{\pgfqpoint{-0.055556in}{-0.055556in}}{\pgfqpoint{0.055556in}{0.055556in}}{%
\pgfpathmoveto{\pgfqpoint{0.000000in}{-0.055556in}}%
\pgfpathcurveto{\pgfqpoint{0.014734in}{-0.055556in}}{\pgfqpoint{0.028866in}{-0.049702in}}{\pgfqpoint{0.039284in}{-0.039284in}}%
\pgfpathcurveto{\pgfqpoint{0.049702in}{-0.028866in}}{\pgfqpoint{0.055556in}{-0.014734in}}{\pgfqpoint{0.055556in}{0.000000in}}%
\pgfpathcurveto{\pgfqpoint{0.055556in}{0.014734in}}{\pgfqpoint{0.049702in}{0.028866in}}{\pgfqpoint{0.039284in}{0.039284in}}%
\pgfpathcurveto{\pgfqpoint{0.028866in}{0.049702in}}{\pgfqpoint{0.014734in}{0.055556in}}{\pgfqpoint{0.000000in}{0.055556in}}%
\pgfpathcurveto{\pgfqpoint{-0.014734in}{0.055556in}}{\pgfqpoint{-0.028866in}{0.049702in}}{\pgfqpoint{-0.039284in}{0.039284in}}%
\pgfpathcurveto{\pgfqpoint{-0.049702in}{0.028866in}}{\pgfqpoint{-0.055556in}{0.014734in}}{\pgfqpoint{-0.055556in}{0.000000in}}%
\pgfpathcurveto{\pgfqpoint{-0.055556in}{-0.014734in}}{\pgfqpoint{-0.049702in}{-0.028866in}}{\pgfqpoint{-0.039284in}{-0.039284in}}%
\pgfpathcurveto{\pgfqpoint{-0.028866in}{-0.049702in}}{\pgfqpoint{-0.014734in}{-0.055556in}}{\pgfqpoint{0.000000in}{-0.055556in}}%
\pgfpathlineto{\pgfqpoint{0.000000in}{-0.055556in}}%
\pgfpathclose%
\pgfusepath{stroke,fill}%
}%
\begin{pgfscope}%
\pgfsys@transformshift{2.380059in}{2.263169in}%
\pgfsys@useobject{currentmarker}{}%
\end{pgfscope}%
\end{pgfscope}%
\begin{pgfscope}%
\definecolor{textcolor}{rgb}{0.000000,0.000000,0.000000}%
\pgfsetstrokecolor{textcolor}%
\pgfsetfillcolor{textcolor}%
\pgftext[x=2.780059in,y=2.185392in,left,base]{\color{textcolor}{\rmfamily\fontsize{16.000000}{19.200000}\selectfont\catcode`\^=\active\def^{\ifmmode\sp\else\^{}\fi}\catcode`\%=\active\def
\end{pgfscope}%
\end{pgfpicture}%
\makeatother%
\endgroup%